\pdfoutput=1 
\documentclass[a4paper, twoside, openright]{book}

\usepackage{amsmath, amsthm, amssymb, amsfonts, mathtools}

\usepackage[english]{babel}

\usepackage{stix} 
\pdfmapfile{+stix.map} 
\usepackage{multirow} 
\usepackage{bbold} 
\usepackage[group-separator={\,}]{siunitx} 

\usepackage{booktabs}
\usepackage{colortbl, color}
\usepackage{afterpage} 
\usepackage{fancyhdr}
\usepackage[yyyymmdd,hhmmss]{datetime}
\usepackage{enumerate} 

\usepackage{etoolbox}
\definecolor{light-gray}{gray}{0.80}
\newcommand{\headrulecolor}[1]{\patchcmd{\headrule}{\hrule}{\color{#1}\hrule}{}{}}
\headrulecolor{light-gray}
\usepackage{graphicx} 
\graphicspath{ {Figures/} } 

\allowdisplaybreaks[1]

\usepackage[symbol*]{footmisc}
\DefineFNsymbols*{lamport}{*\dagger\ddagger\S\P\|{**}{\dagger\dagger}{\ddagger\ddagger}}
\setfnsymbol{lamport}
\DefineFNsymbols*{lamportnostar}[math]{\star\dagger\ddagger\S\P\|\sharp{\star\star}{\dagger\dagger}{\ddagger\ddagger}{\sharp\sharp}}
\setfnsymbol{lamportnostar}
\usepackage{chngcntr}
\counterwithin*{footnote}{section} 

\usepackage[backend=bibtex, bibstyle=nature, doi=false, eprint=false, doi=false, isbn=false, url=false, hyperref=false, backref=false]{biblatex}

\def\ep{\varepsilon}

\def\ra{\rightarrow}
\def\Chi{\mathcal{X}}
\def\Sphere{\mathbf{S}}

\definecolor{bblue}{RGB}{0,81,158}
\definecolor{bred}{RGB}{230,0,27}
\definecolor{bgold}{RGB}{205,181,144}
\definecolor{bgray}{gray}{0.80}

\DeclareMathAlphabet{\mathpzc}{OT1}{pzc}{m}{it}
\def\de{\mathpzc{d}}
\def\a{\mathpzc{A}}
\def\f{\mathpzc{f}}

\DeclarePairedDelimiterX\innerr[2]{(}{)}{\,#1 \;\delimsize\vert\; #2\,}
\newcommand{\inner}[2]{\innerr*{#1}{#2}}
\newcommand{\coef}[1]{\left[\ep^{#1}\right]}

\def\2F1{\mbox{$_2${F}$_1$}} 
\def\id{\mathbb{1}} 

\newcommand{\fracn}[2]{\frac{\num{#1}}{\num{#2}}}
\def\const{\mathrm{const}}

\def\?{{\bf (?)}}

\providecommand{\prognamestyle}[1]{\textit{#1}}
\providecommand{\codenamestyle}[1]{\texttt{#1}}
\def\mathematica{\prognamestyle{Mathematica}}

\def\Si{\text{Si}}
\def\Ci{\text{Ci}}
\def\sinc{\text{sinc}}

\def\lwidth{0.98\columnwidth}
\def\mwidth{0.8\columnwidth}
\def\swidth{0.7\columnwidth}

\newtheorem{mydef}{Definition}
\newtheorem{mythe}{Theorem}

\newcommand*\diff{\mathop{}\!\mathrm{d}}

\def\eulergamma{\boldsymbol{\gamma}}

\begin{document}

\pagestyle{empty}
\begin{titlepage}
  \thispagestyle{empty}
\begin{center}
\begin{minipage}{0.75\linewidth}
    \centering
    \vspace{3cm}
    {{\huge Dynamics of Nonlinear Waves on Bounded
        Domains\par}} \vspace{3cm}
    {\Large Maciej Maliborski\par}
    \vspace{3cm}
    {A Dissertation Presented to the Faculty of Physics, Astronomy and
      Applied Computer Science of the Jagiellonian University in
      Candidacy for the Degree of Doctor of Philosophy\par}
    \vspace{3cm}
    {\Large Advisor: Prof. Piotr Bizo\'n \\
      Co-advisor: Dr. Andrzej Rostworowski \par} \vspace{3cm}
    {\Large November 2014}
  \end{minipage}
\end{center}
\clearpage

\end{titlepage}
\pagestyle{empty}

\newpage
\begin{center}
  \copyright \hspace{1ex} Copyright by Maciej Maliborski, 2014.
  \\
  All rights reserved.
\end{center}
\newpage

\pagestyle{empty}

\begin{flushleft}
  Wydzia\l{} Fizyki, Astronomii i Informatyki Stosowanej
  \\
  Uniwersytet Jagiello\'nski
\end{flushleft}

\vspace{4ex}

\begin{center}
  {\large \textbf{O\'swiadczenie}}
\end{center}

Ja ni\.zej podpisany Maciej Maliborski (nr indeksu: 417) doktorant
Wydzia\l{}u Fizyki, Astronomii i Informatyki Stosowanej Uniwersytetut
Jagiello\'n{}skiego o\'swiadczam, \.ze przed\l{}o\.zona przezemnie
rozprawa doktorska pt. ,,Dynamika nieliniowych fal na \linebreak{}
zwartych rozmaito\'sciach'' (ang. ,,Dynamics of Nonlinear Waves on
Bounded Domains'') jest oryginalna i przedstawia wyniki bada\'n
wykonanych przeze mnie osobi\'scie, pod kierunkiem prof. dr hab Piotra
Bizonia oraz dr Andrzeja Rostworowskiego.  Prac\k{e} napisa\l{}em
samodzielnie.

O\'swiadczam, \.ze moja rozprawa doktorska zosta\l{}a opracowana
zgodnie z Ustaw\k{a} o~prawie autorskim i prawach pokrewnych z dnia 4
lutego 1994 r. (Dziennik Ustaw 1994 nr 24 poz. 83 wraz z
p\'o\'zniejszymi zmianami).

Jestem \'swiadom, \.ze niezgodno\'s\'c niniejszego o\'swiadczenia z
prawd\k{a} ujawniona w~dowolnym czasie, niezale\.znie od skutk\'ow
prawnych wynikaj\k{a}cych z ww. ustawy, mo\.ze spowodowa\'c
uniewa\.znienie stopnia nabytego na podstawie tej rozprawy.

\vspace{16ex}

\begin{center}
  \begin{tabular}{lcr}
    Krak\'ow, dnia ..............................  &
    \hspace{16ex}  &
    ...........................................
  \end{tabular}
\end{center}

\clearpage
\selectlanguage{english}
\pagestyle{empty}
\mbox{}
\newpage

\pagestyle{empty}
\vspace*{\fill}
\begin{center}
  \begin{minipage}{.667\textwidth}
    \begin{flushright}
      \textit{To my wife Natalia.}
      \vspace{6.67cm}
    \end{flushright}
  \end{minipage}
\end{center}
\vspace*{\fill}

\newpage
\pagestyle{empty}
\mbox{}
\newpage

\pagestyle{empty}
\mbox{}
\vspace{14ex}

\noindent {\Huge\bf Acknowledgments}
\vspace{9ex}

\noindent I would like to thank Prof. Piotr Bizo\'n and Dr. Andrzej
Rostworowski, my supervisors, for their continuous help, support and
encouragement during my studies.  I would also like to express
gratitude to my colleagues, Dr. Patryk Mach and Pawe\l{} Biernat, for
valuable discussions and work on common projects (also those not
finished yet) and for sharing the office.  Finally, I am indebted to
Prof. Leszek Hadasz and Dr. Sebastian Szybka for their discreet and
valuable advice, even though they may not be aware of their important
contribution.

The hospitality of Max Planck Institute for Gravitational Physics
(Albert Einstein Institute) and the Erwin Schr\"odinger Institute of
the University of Vienna is gratefully acknowledged.

Moreover, I would like to acknowledge financial support of the NCN
Grant No.~DEC-2012/06/A/ST2/00397, and series of the Polish Ministry
of Science and Higher Education Grants (Dean's Grants)
No. 7150/E-338/M/2012, No.~7150/E-338/M/2013 and
No.~7150/E-338/M/2014.

The computations were carried out with the supercomputer ``Deszno'' at
the Institute of Physics of the Jagiellonian University and with the
supercomputers ``Mars'' and ``Zeus'' maintained by Academic Computer
Centre CYFRONET AGH (through computational Grants
No.~MNiSW/IBM\_BC\_HS21/UJ/071/2013 and
No.~MNiSW/\linebreak{}Zeus\_lokalnie/UJ/027/2014).

\newpage
\pagestyle{empty}
\mbox{}
\newpage

\pagestyle{empty}
\mbox{}
\vspace{14ex}

\noindent {\Huge\bf Abstract}
\vspace{9ex}

This thesis is concerned with dynamics of conservative nonlinear waves
on bounded domains.  In general, there are two scenarios of evolution.
Either the solution behaves in an oscillatory, quasiperiodic manner or
the nonlinear effects cause the energy to concentrate on smaller
scales leading to a turbulent behaviour.  Which of these two
possibilities occurs depends on a model and the initial conditions.

In the quasiperiodic scenario there exist very special time-periodic
solutions.  They result for a delicate balance between dispersion and
nonlinear interaction.  The main body of this dissertation is
concerned with construction (by means of perturbative and numerical
methods) of time-periodic solutions for various nonlinear wave
equations on bounded domains.

While turbulence is mainly associated with hydrodynamics, recent
research in General Relativity has also revealed turbulent phenomena.
Numerical studies of a self-gravitating massless scalar field in
spherical symmetry gave evidence that anti-de Sitter space is unstable
against black hole formation.  On the other hand there appeared many
examples of asymptotically anti-de Sitter solutions which evade
turbulent behaviour and appear almost periodic for long times.  We
discuss here these two contrasting scenarios putting special attention
to the construction and properties of strictly time-periodic
solutions.  We analyze different models where solutions of this type
exist.  Moreover, we describe similarities and differences among these
models concerning properties of time-periodic solutions and methods
used for their construction.

\newpage
\pagestyle{empty}
\mbox{}
\newpage

\pagestyle{fancy}
\frontmatter

\tableofcontents

\mainmatter
\chapter*{Preface}
\addcontentsline{toc}{chapter}{Preface}

\tolerance=1200
\interfootnotelinepenalty=10000

This thesis is concerned with dynamics of conservative nonlinear waves
on bounded domains.  In general, there are two scenarios of evolution.
Either the solution behaves in an oscillatory, quasiperiodic manner or
the nonlinear effects cause the energy to concentrate on smaller
scales leading to a turbulent behaviour.  Which of these two
possibilities occurs depends on a model and the initial conditions.

In the quasiperiodic scenario there exist very special time-periodic
solutions.  They result for a delicate balance between dispersion and
nonlinear interaction.  The main body of this dissertation is
concerned with construction of time-periodic solutions for various
nonlinear wave equations on bounded domains.

While turbulence is mainly associated with hydrodynamics, recent
research in General Relativity has also revealed turbulent phenomena.
Numerical studies of a self-gravitating massless scalar field in
spherical symmetry gave evidence that anti-de Sitter (AdS) space is
unstable against black hole formation.  It was demonstrated that there
exists a large class of small perturbations of AdS which grow in time
and eventually lead to collapse---it is in stark contrast to the
behaviour of small perturbations of Minkowski space which disperse to
infinity.  This mechanism attributed to resonant energy transfer which
is seen in weakly nonlinear perturbative calculations.

On the other hand there appeared many examples of asymptotically AdS
solutions which evade turbulent behaviour and appear almost periodic
for long times.  We discuss here these two contrasting scenarios
putting special attention to the construction (by means of
perturbative and numerical methods) and properties of strictly
time-periodic solutions.  We analyze different models where solutions
of this type exist.  We describe similarities and differences among
these models concerning properties of time-periodic solutions and
methods used for their construction.

Although studies in this thesis are directly connected to the problem
of stability of AdS space, of course we did not expect to answer to
this difficult question.  We give details of numerical algorithms and
explain steps of perturbative calculations pointing out strengths and
weaknesses of chosen approaches.  We hope this text will serve as a
reference for further studies in this rapidly growing area of
research.

This text is organized as follows.  In Part~\ref{part:Preliminaries},
composed of three chapters, we introduce the subject of our studies.
In Chapter~\ref{cha:Introduction} we give a motivation and introduce
some basic concepts.  Next, in Chapter~\ref{cha:Methods}, we give an
abstract description of techniques (both perturbative and numerical)
used to construct time-periodic solutions.  Variants of described
methods are used to study concrete systems in the following chapters.
In Chapter~\ref{cha:Models} we introduce the models that we aim to
study.  We motivate their consideration, derive equations of motion
and formulate initial or initial-boundary value problems.  We also
analyze the spectra of linear operators arising in the study of linear
stability.

These models are studied in detail in Part~\ref{part:Studies}, which
is composed of two chapters.  In Chapter~\ref{cha:Turbulence} we study
turbulent phenomena in nonlinear evolution systems and the
(in)stability problem for generic perturbations.  We focus on the
question of how the dispersive and nondispersive spectrum of linear
perturbations affects the nonlinear dynamics.  We investigate this
issue by studying self-gravitating massless scalar field in a
perfectly reflecting cavity (Section~\ref{sec:BoxTurbulence}) and the
Yang-Mills (YM) field propagating on the Einstein Universe
(Section~\ref{sec:YMTurbulence}).  In these models we have a freedom
of changing the character of the eigenfrequencies by imposing
different boundary conditions (the scalar field case) or by
considering perturbations in different topological sectors (the YM
field model).  In addition, we give the details of numerical methods
used to solve the evolution equations.  For the YM model we present
the results of perturbative methods used to describe a single linear
mode initial data, which also serve as a starting point in the
construction of time-periodic solutions.  We point out that the
nondispersive spectrum does not forbid the resonances to occur.

In Chapter~\ref{cha:Periodic} we study in detail time-periodic
solutions for the systems of equations derived in
Chapter~\ref{cha:Models}.  In Section~\ref{sec:AdSPeriodic} we discuss
the methods of constructing time-periodic solutions for the real
self-gravitating massless scalar field in spherical symmetry (in any
number of spatial dimensions).  We point out the differences between
even and odd spatial dimensions and propose alternative methods for
both cases.  These techniques are then adapted (in
Section~\ref{sec:Standing}) to the construction of standing waves for
the complex scalar field.  For both real and complex field cases we
find both stable and unstable solutions.

In analogy to the scalar field case we analyze the Bianchi IX
cohomogenity-two biaxial ansatz which allows for pure gravitational
dynamical degrees of freedom in the $1+1$ setting, and possesses
time-periodic solutions (Section~\ref{sec:BCS}).  The construction of
these solutions is much more demanding and their stability analysis is
not conclusive.

In Sections~\ref{sec:BoxPeriodic} and \ref{sec:YMPeriodic} we continue
the studies initiated in Chapter~\ref{cha:Turbulence} on the spherical
cavity and the YM models, respectively.  We construct time-periodic
solutions in both cases for dispersive and nondispersive linear
spectrum and discuss how the character of the spectrum affects the
structure of solutions and also the methods to construct them.

We conclude and discuss some directions for future work in
Chapter~\ref{cha:Summary}.

In appendices we give additional details of numerical and analytical
techniques.  In Appendix~\ref{cha:AppOrthogonalPolynomials} we state
the most important properties of orthogonal polynomials and give a
list of useful identities used in this thesis.
Appendix~\ref{cha:polyn-pseud-meth} contains a detailed description of
pseudospectral spatial discretization based on Chebyshev polynomials,
in particular their adaptation to spherically symmetric problems.
Additionally in Appendix~\ref{cha:runge-kutta-methods} we present the
Runge-Kutta time integration methods and state their most important
properties.  These include symplectic methods mainly used in this
work.  Finally, in Appendix~\ref{cha:inter-coeff} we explain the
method we use to calculate integrals appearing in perturbative
calculations which are crucial for efficient symbolic manipulation.

\part{Preliminaries}
\label{part:Preliminaries}

\chapter{Introduction}
\label{cha:Introduction}

In this chapter we introduce the subject of our studies.  After giving
the motivation (Section~\ref{sec:Motivation}), we review the anti-de
Sitter space (Section~\ref{sec:AdS}).  In Section~\ref{sec:Waves} we
shortly review mathematical studies of time-periodic solutions and
weak-turbulence phenomena.

\section{Motivation---(in)stability of AdS space}
\label{sec:Motivation}

Asymptotically anti-de Sitter (aAdS) spacetimes have come to play a
central role in theoretical physics, prominently due to the AdS/CFT
correspondence which conjectures a duality between gravity in the AdS
bulk and a quantum conformal field theory with a large number of
strongly interacting degrees of freedom living in the spacetime
corresponding to the AdS conformal boundary
\cite{10.1023/A:1026654312961, Witten:1998qj}.  Despite on important
role of AdS space plays in these theories, the question of its
stability remains unanswered till now, see
\cite{10.1007/s10714-014-1724-0} for a review.  In contrast, the
questions of stability of Minkowski and de~Sitter spacetimes have been
answered in affirmative in \cite{christodoulou2014global} and
\cite{10.1007/BF01205488} respectively.

Recent numerical and analytical studies of spherically symmetric
self-gravitating massless scalar field system with negative
cosmological constant indicated that AdS space may be unstable against
the formation of a black hole under arbitrarily small perturbations
\cite{br, jrb, Buchel2012} (under reflecting boundary conditions).
Although gravitational collapse seems to be a generic fate of small
perturbations of AdS, it was suggested in \cite{br} that there may
also exist a set of initial data for which the evolutions remains
globally regular in time.  This conjecture was substantiated in
\cite{MRPRL}, where the evidence for the existence of globally
regular, nonlinearly stable, time-periodic solutions within the same
model was given.  A similar class of aAdS solutions was studied in
\cite{Buchel2013}.  A similar behaviour is expected for the pure
vacuum case with simplifying symmetry assumptions (cohomogenity-two
biaxial Bianchi~IX ansatz) and with no symmetry assumptions
\cite{Dias2012a,dhms,Horowitz2012}.  These studies indicate that the
structure of phase space for AdS gravity is complex and poorly
understood.  Most recent analytical studies concentrate on the
analysis of wave equations on the fixed AdS \cite{0264-9381-21-12-012}
or the AdS-Schwarzschild backgrounds \cite{Warnick2013}.  Therefore
the numerical simulations will play a key role in further
investigations of these problems with major emphasis on assistance of
analytic attempts.

The existence of (nontrivial) time-periodic solutions of ordinary and
partial differential equations is a fundamental problem.  The
existence proofs of time-periodic solutions to simple semi-linear PDEs
require advanced mathematical techniques (we give some references in
the following section).  Thus finding time-periodic solutions to
complicated elliptic-hyperbolic systems of PDEs is challenging and
thus particularly interesting on its own regardless of the stability
problem of AdS.  On the other hand it is well known that, in
asymptotically flat spacetimes, there are no nontrivial time-periodic
solutions to Einstein equations \cite{0264-9381-27-5-055007,
  0264-9381-27-17-175011}.  This nonexistence proof points the
difference between the bounded and unbounded domains---the mechanism
that allows for a nontrivial time-periodic solutions in aAdS cases is
the lack of dissipation of energy.  In this thesis we give numerical
and analytical studies of numerous systems of PDEs on bounded domains
which may guide and stimulate any further more rigorous efforts.

Additionally, these studies where intended to develop numerical
methods for efficient integration of the Einstein equations with
negative cosmological constant in spherical symmetry.  Applied space
discretization methods are the core of our numerical algorithm used to
find time-periodic solutions, they are also used in the time
evolution.  We intend to develop methods which are robust and general,
in particular we apply two different methods for the Einstein-massless
scalar field system depending on a parity of space dimension.  Our
studies demonstrate the efficiency of using spectral methods in space
discretization for the aAdS spaces.  Initially developed for studies
of turbulence phenomena, spectral methods increased their wide
applicability in numerical solutions to the Einstein equations, see
e.g. \cite{PhysRevD.65.064029, PhysRevD.82.104023} for an application
of the Galerkin approach in the studies of gravitational collapse of
self-gravitating scalar field with spherical symmetry assumption in
asymptotically flat case.  For review on spectral methods in numerical
relativity we refer to \cite{lrr-2009-1}.  Besides spatial
discretization we also emphasize particular properties of symplectic
integration methods \cite{hairer2006geometric}.  Time stepping methods
are usually selected to achieve some prescribed accuracy at the least
cost, that is to minimize the product of the cost per time step and
the number of time steps needed.  For general systems of ODEs the
symplectic method is necessarily implicit (which is more costly than
the explicit one) but their long time near conservation of the systems
invariants and large stability domains (compared to explicit methods)
make them especially advantageous in the studies of Hamiltonian
systems.  In practice, the ODE methods that we use are either explicit
or implicit depending on a structure of the system under study.

Moreover, efficient implementation of these techniques may be regarded
as a good exercise for those willing to master their programming
skills in \mathematica{} \cite{mathematica}.  Writing codes for
numerical solution of evolutionary PDEs in \mathematica{} when using
semi-discrete method of lines approach (MOL), is reduced to
implementation of the spatial discretization\footnote{Simple problems
  can be handled entirely by \mathematica{}, these more advanced need
  to be written by the user.} due to modular structure of the
\codenamestyle{NDSolve} function \cite{NDSolve} and plenty of ODE
integration algorithms already implemented \cite{NDSolveTutorial1,
  NDSolveTutorial2}.  Functional and pattern matching programming
facilities of \mathematica{} were particularly advantageous in concise
and readable implementation of our perturbative construction of
time-periodic solutions.

While most of the numerical methods where implemented in
\mathematica{}, some were coded from scratch in
\prognamestyle{FORTRAN}.  Besides \mathematica{} we also acknowledge
other plotting software, \prognamestyle{Pyxplot} \cite{Pyxplot} and
\prognamestyle{CustomTicks} package \cite{CustomTicks}, used to
produce figures placed in this thesis.  We further acknowledge
\prognamestyle{GRQUICK} package \cite{GRQUICK} used to derive
necessary field equations.

\section{An overview of AdS space}
\label{sec:AdS}

\begin{figure}[!p]
  \centering
  \includegraphics[width=\swidth]{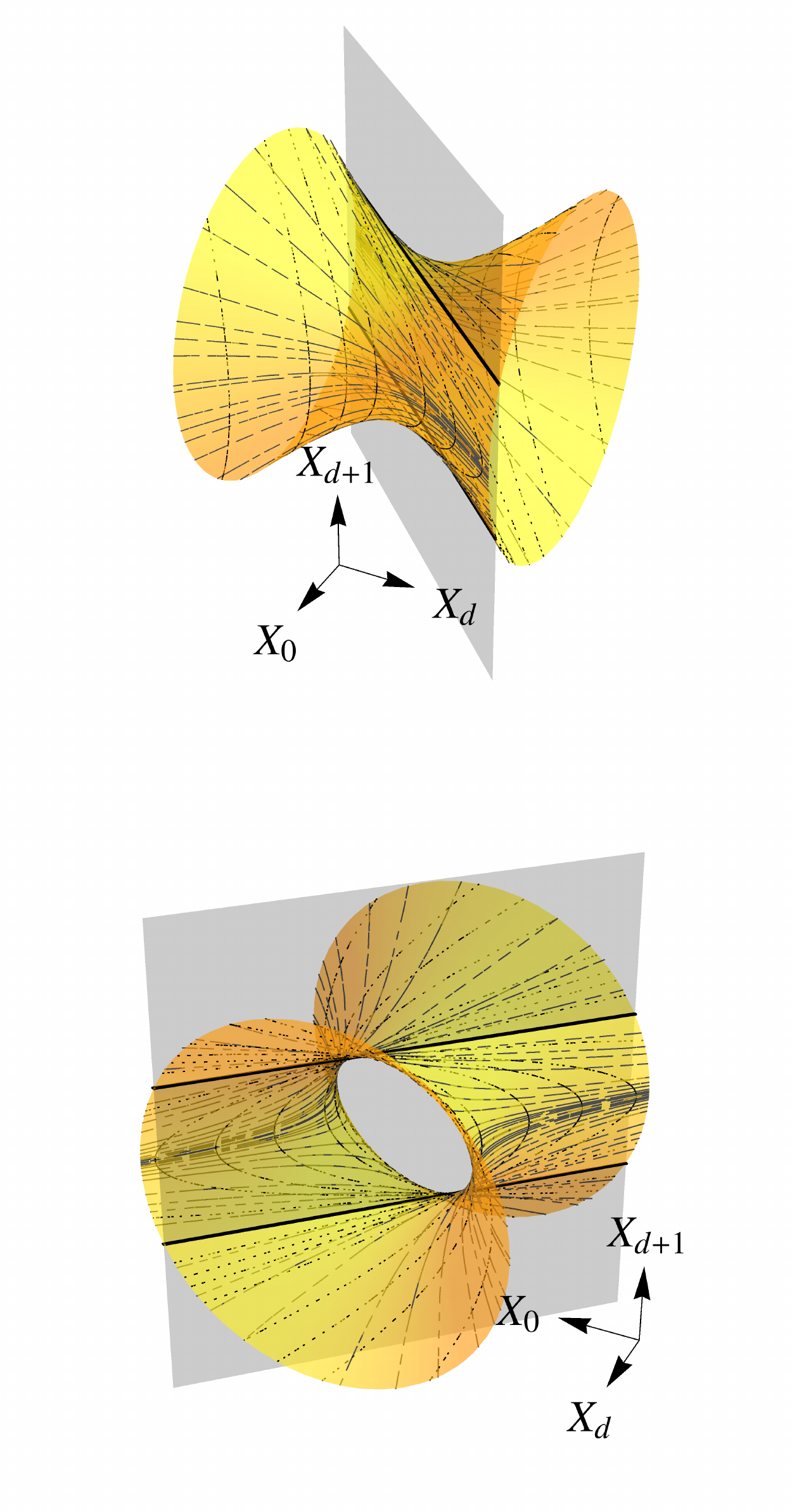}
  \caption{The AdS space (the hyperboloid
    $X_{0}^2-X_{d}^2+X_{d+1}^{2}=\ell^{2}$, $X_{i}=0$,
    $i=1,\dots,d-1$) cut by the hyperplane $X_{0}=X_{d}$, separating
    two patches of Poincar\'e coordinates (\ref{eq:10}) $z>0$ and
    $z<0$.  The intersection of hyperboloid with the cutting plane
    corresponds to the boundary $z\ra\pm\infty$ in this coordinate
    system (solid black lines). The black dotted lines are the
    $z=\const$, while gray dashed lines denote $t=\const$.}
  \label{fig:AdSHyperboloid}
\end{figure}

Anti-de Sitter (AdS) spacetime is the maximally symmetric solution of
the vacuum Einstein equations
\begin{equation}
  \label{eq:1}
  R_{\alpha\beta} - \frac{1}{2}Rg_{\alpha\beta} + \Lambda g_{\alpha\beta}=0,
\end{equation}
with negative cosmological constant $\Lambda<0$.  This solution
appeared for the first time in the paper
\cite{10.1023/A:1026755309811} as cosmological solution regarded as a
four-dimensional spacetime model of the Universe, though the name AdS
appeared much later \cite{calabi1962relativistic}.  Here we generalize
AdS solution to arbitrary dimensions.  Geometrically, AdS$_{d+1}$ can
be thought of as hyperboloid of radius $\ell>0$ (which is related to
the cosmological constant by $\Lambda=-d(d-1)/(2\ell^{2})$)
\begin{equation}
  \label{eq:2}
  -X_{0}^{2} + \sum_{k=1}^dX_{k}^{2} - X_{d+1}^{2} = -\ell^{2},
\end{equation}
embedded in $(d+2)$-dimensional flat, $O(d,2)$ invariant space, with a
line element
\begin{equation}
  \label{eq:3}
  \diff s^2 = - \diff X_{0}^{2} + \sum_{k=1}^{d}\diff X_{k}^{2} - \diff X_{d+1}^{2},
\end{equation}
see Fig.~\ref{fig:AdSHyperboloid}.  One of many possible
parametrizations of (\ref{eq:2}) is
\begin{subequations}
  \label{eq:4}
  \begin{align}
    X_{0} &= \ell\cosh{\rho}\cos{\tau},
    \\
    X_{d+1} &= \ell\cosh{\rho}\sin{\tau},
    \\
    X_{k} &= \ell\sinh{\rho}\,n_{k}, \quad k=1,\ldots,d,
  \end{align}
\end{subequations}
with $-\pi\leq\tau<\pi$, $0\leq\rho<\infty$, and $n_{k}$'s such that
the condition (\ref{eq:2}) holds, i.e. $\sum_{k=1}^{d}n_{k}^{2}=1$.
In fact, $n_{k}$ parametrize $\Sphere^{d-1}$, so in spherical
coordinates $(\theta_{1},\ldots,\theta_{d-2},\varphi)$
\begin{subequations}
  \label{eq:5}
  \begin{align}
    n_{1} &= \sin{\theta_{1}}\sin{\theta_{2}}\cdots
    \sin{\theta_{d-2}}\sin{\varphi},
    \\
    n_{2} &= \sin{\theta_{1}}\sin{\theta_{2}}\cdots
    \sin{\theta_{d-2}}\cos{\varphi},
    \\
    &\ \vdots \nonumber
    \\
    n_{d-1} &= \sin{\theta_{1}}\cos{\theta_{2}}
    \\
    n_{d} &= \cos{\theta_{1}},
  \end{align}
\end{subequations}
with the angle ranges $\phi\in[0,2\pi)$ and $\theta_{k}\in[0,\pi]$ for
$k=1,\ldots,d-2$.  Using (\ref{eq:4}) and (\ref{eq:5}) the induced
metric on the hyperboloid (\ref{eq:2}) is
\begin{equation}
  \label{eq:6}
  \diff s^{2} = \ell^{2}\left(-\cosh^{2}\!\rho\,\diff\tau^{2} + \diff\rho^{2}
    + \sinh^{2}\!\rho\,\diff\Omega_{d-1}^{2}\right),
\end{equation}
with $\diff\Omega_{d-1}^{2}$ denoting the round metric on unit
$\Sphere^{d-1}$.  In fact, this is one of many possible
parametrizations of AdS$_{d+1}$ space, see \cite{griffiths2009exact}.
For convenience we introduce a new radial coordinate, $x$, by setting
\begin{equation}
  \label{eq:7}
  \tan{x} = \sinh{\rho},
\end{equation}
with $x\in[0,\pi/2)$; then the metric (\ref{eq:6}) takes the following
form
\begin{equation}
  \label{eq:8}
  \diff s^{2} = \frac{\ell^{2}}{\cos^{2}\!{x}}
  \left( -\diff\tau^{2} + \diff x^{2} +
    \sin^{2}\!{x}\,\diff\Omega^{2}_{d-1}\right).
\end{equation}
We compactify the AdS space by extending the range of the radial
coordinate to $x\in[0,\pi/2]$.

The AdS$_{d+1}$ has the topology $\Sphere\times\mathbb{R}^{d}$ so
there are closed timelike curves, parametrized by, e.g.
$X_{0}=\ell\cos{\tau}$, $X_{d+1}=\ell\sin{\tau}$ and $X_{i}=0$,
$i=1,\ldots,d$.  This feature is lost if we unwrap the hyperboloid by
considering the covering space of AdS (abbreviated as CAdS).  We
denote the unwrapped timelike coordinate by $t\in\mathbb{R}$.  The
CAdS$_{d+1}$ has the topology $\mathbb{R}^{d+1}$ and contains no
closed timelike curves.  Hereafter we consider only CAdS$_{d+1}$,
i.e. we identify the AdS$_{d+1}$ with its covering space.

\begin{figure}[!t]
  \centering
  \begin{tabular}{lc}
    \includegraphics[width=0.45\textwidth]{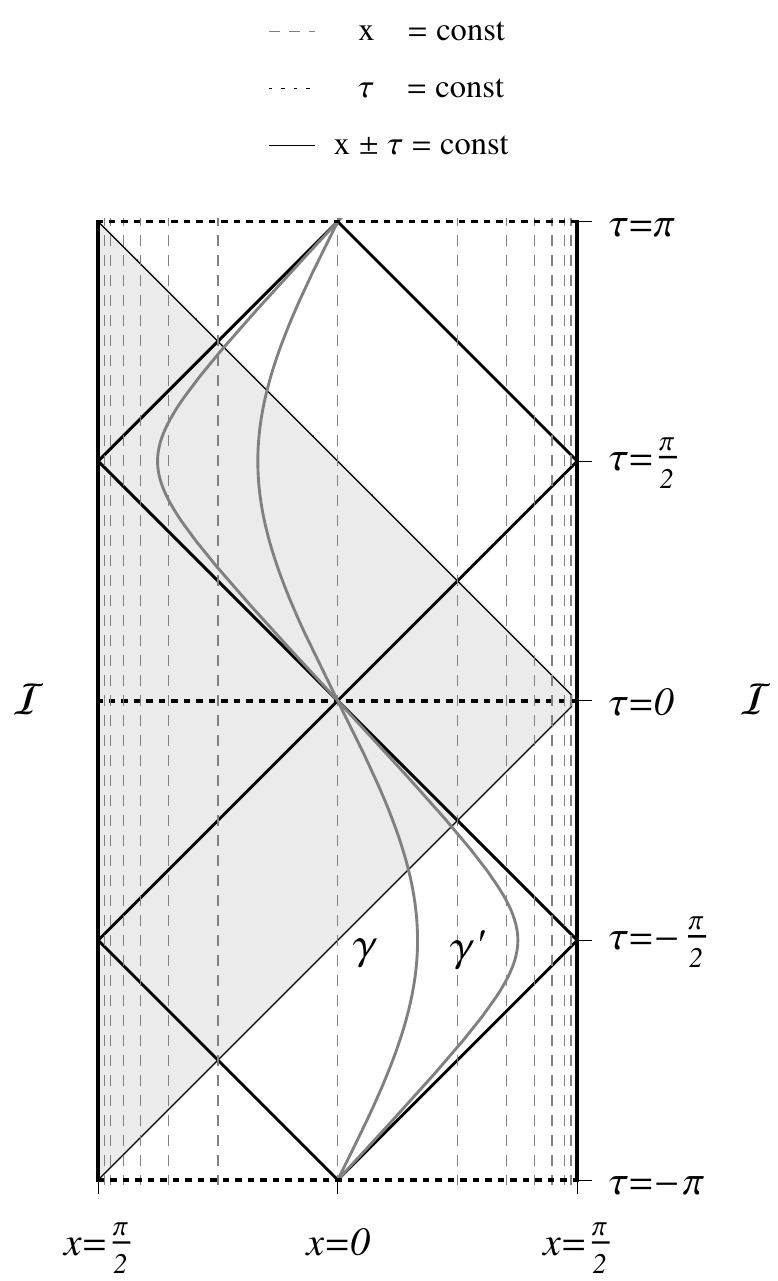} &
    \raisebox{.22\height}{\includegraphics[width=0.45\textwidth]
      {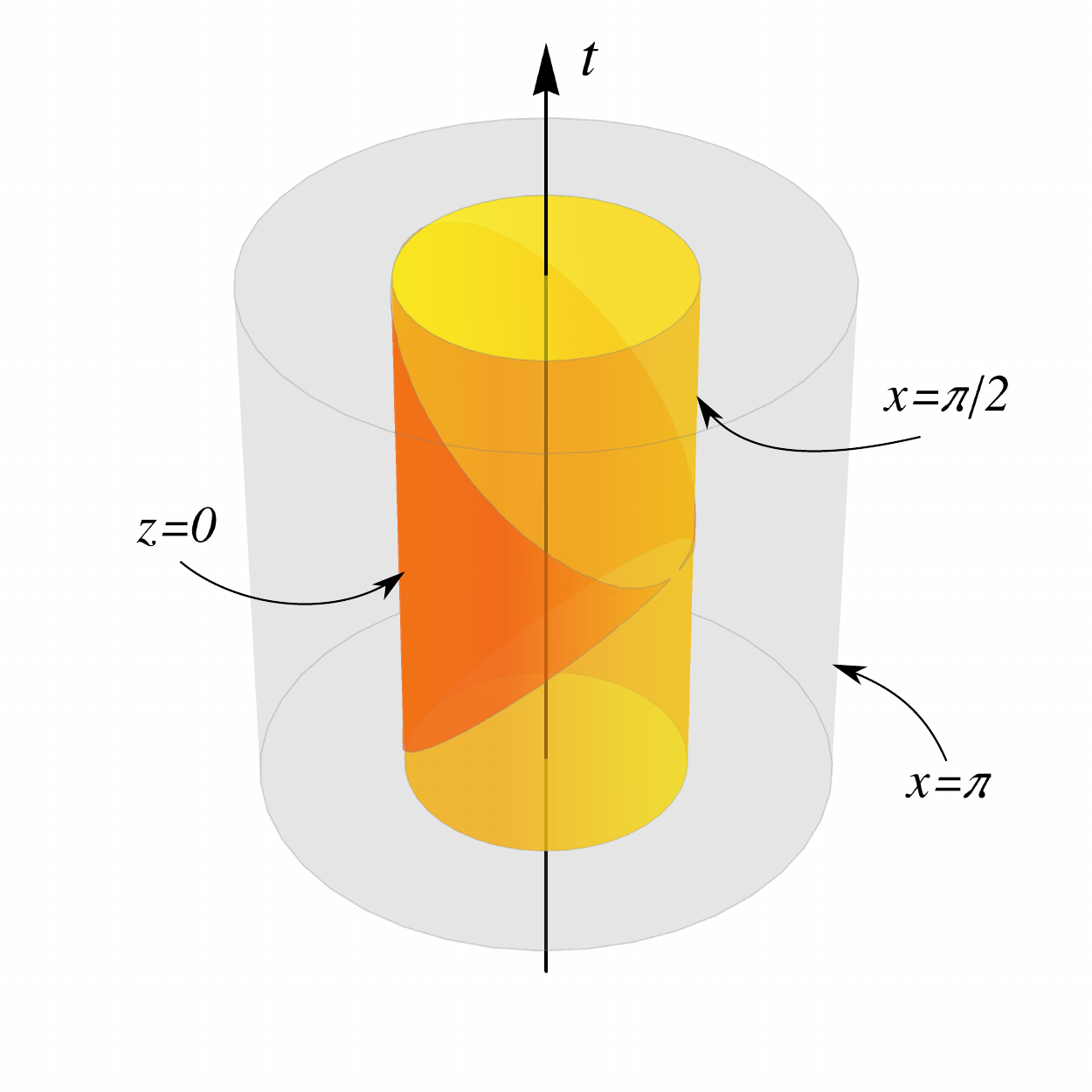}}
  \end{tabular}
  \caption{\textit{Left panel}. Penrose diagram for AdS space with top
    ($\tau=\pi$) and bottom ($\tau=-\pi$) surfaces identified.  Each
    point on this diagram represents $\Sphere^{d-1}$. Timelike
    geodesics $\gamma$ and $\gamma'$ are indicated, plotted with gray
    solid lines, manifest time-periodicity.  The null geodesics
    (diagonal black lines) reach the conformal boundary with finite
    coordinate time.  The shaded region corresponds to the part of AdS
    space covered by the Poincar\'e charts. \textit{Right panel}. The
    diagram showing the Einstein Universe and the AdS space as its
    portion; with $(d-2)$ dimensions suppressed these represent
    concentric cylinders with diameters $x=\pi$ and $x=\pi/2$
    respectively.}
  \label{fig:AdSPenroseDiagram}
\end{figure}

The metrics given in (\ref{eq:6}) and (\ref{eq:8}) are indeed
solutions to the vacuum Einstein equations (\ref{eq:1}) with
$\Lambda=-d(d-1)/(2\ell^2)$.  In coordinates (\ref{eq:8}) we see that
the conformal boundary of AdS, corresponding to spatial and null
infinity, is the hypersurface $x=\pi/2$.  It is the timelike cylinder
$\mathcal{I}=\mathbb{R}\times\Sphere^{d-1}$ with the boundary metric
\begin{equation}
  \label{eq:9}
  \diff s^2_{\mathcal{I}}=-\diff t^2 +
  \sin^{2}\!{x}\,\diff\Omega^2_{\Sphere^{d-1}}.
\end{equation}
Since the spatial and null infinity is a timelike hypersurface the
information may be lost to or gained from this surface in finite
coordinate time $t$.  The consequence of this is that there exists no
complete Cauchy surface---the AdS space is not globally hyperbolic.
This means that given initial data on a constant $t$ hypersurface, it
is not possible to prescribe evolution in the region beyond the Cauchy
development of this surface.  This is seen on
Fig.~\ref{fig:AdSPenroseDiagram} showing the Penrose diagram of AdS in
global coordinates (\ref{eq:8}).  The unique determination of
evolution is possible only if the data on $\mathcal{I}$ is also
prescribed \cite{Friedrich1995125, hawking1973large}.

Important property of AdS space is its conformal structure.  From
(\ref{eq:8}) we see that the AdS metric $\hat{g}$ is conformally
related to the Einstein Universe\footnote{Known also as the Einstein
  cylinder, because of its topology $\mathbb{R}\times\Sphere^{3}$, is
  on of the solutions to the Friedman equations with dust matter
  content and positive cosmological constant.} (its generalizations to
higher dimensions) metric via $g=\Omega^{2}\hat{g}$, with conformal
factor\footnote{Which should not be confused with $\diff\Omega^2$.}
$\Omega=\cos{x}/\ell$, see also Fig.~\ref{fig:AdSPenroseDiagram}.

Another commonly used parametrization of (\ref{eq:2}) is given below
\begin{subequations}
  \label{eq:10}
  \begin{align}
    X_{0} &= \frac{1}{2z}\left(z^{2} + \ell^{2} + \bar{x}^{2} -
      t^{2}\right),
    \\
    X_{i} &= \frac{x_{i}}{z}\ell, \quad i=1,\ldots,d-1,
    \\
    X_{d} &= \frac{1}{2z}\left(z^{2} - \ell^{2} + \bar{x}^{2} -
      t^{2}\right),
    \\
    X_{d+1} &= \frac{t}{z}\ell,
  \end{align}
\end{subequations}
where we use the shorthand notation for $\bar{x}^{2} \equiv
\sum_{i=1}^{d-1}x_{i}^{2}$.  The coordinates $z$, $x_{i}$, $t$ are
known as Poincar\'e coordinates, and for them the metric of
AdS$_{d+1}$ takes the form
\begin{equation}
  \label{eq:11}
  \diff s^{2} = \frac{\ell^{2}}{z^{2}}\left(-\diff t^{2} + \diff z^{2}
      + \diff\bar{x}^{2}\right),
\end{equation}
which apparently is conformally flat.  In fact these coordinates give
us two different charts ($z>0$ or $z<0$) separated by the hypersurface
$X_{0}=X_{d}$ which corresponds to taking the limits $z\ra\pm\infty$
(see Fig.~\ref{fig:AdSHyperboloid}).  On the other hand the limit
$z\ra 0$ corresponds to the Minkowski space-time.

The detailed analysis of both global and Poincar\'e coordinates shows
that the Poincar\'e AdS boundary contains both the points of the AdS
boundary ($x=\pi/2$) and the points of the AdS bulk in global
coordinates (see Fig.~\ref{fig:AdSPenroseDiagram}).  For the details
of mapping of Poincar\'e AdS boundary to the global AdS we refer to
\cite{doi:10.1007/s10714-007-0446-y}.  In contrast to the global
coordinates both of the Poincar\'e charts $z<0$ and $z>0$ do not cover
the whole AdS space.  In fact, these coordinate patches cover only the
shaded part of the Penrose diagram depicted on
Fig.~\ref{fig:AdSPenroseDiagram}.  Though widely used in the AdS/CFT
(because of the flatness of the boundary metric), the Poincar\'e
coordinates are not suitable for studies of global properties of the
AdS space.  Therefore we use the global coordinates and consider
AdS$_{d+1}$ with metric given in Eq.~(\ref{eq:8}).

It is also useful to derive geometric quantities of AdS as these serve
as singularity formation indicators in the dynamical aAdS spaces.  The
AdS$_{d+1}$, as being maximally symmetric space, has a constant scalar
curvature
\begin{equation}
  \label{eq:12}
  R = -\frac{d(d+1)}{\ell^{2}},
\end{equation}
while the Riemann tensor takes the form
\begin{equation}
  \label{eq:13}
  R_{\mu\nu\rho\sigma} = - \frac{1}{\ell^2}\left(g_{\mu\rho}g_{\nu\sigma}
  - g_{\nu\rho}g_{\mu\sigma}\right),
\end{equation}
and so the Kretschmann invariant reads
\begin{equation}
  \label{eq:14}
  K = R^{\mu\nu\rho\sigma}R_{\mu\nu\rho\sigma} = 2\frac{d(d+1)}{\ell^{4}}.
\end{equation}

\section{Waves on bounded domains---time-periodic solutions and weak turbulence}
\label{sec:Waves}

Model problems in the studies of evolutionary Hamiltonian PDEs on
bounded domains\footnote{Here by bounded (compact) domain we mean
  finite interval or finite cube with appropriate boundary conditions
  imposed, e.g. with periodic boundary conditions for problems on a
  $d$-dimensional torus $\mathbf{T}^{d}$.} are the semi-linear wave
equation (NLW)
\begin{equation}
  \label{eq:15}
  u_{tt} - \Delta u + f(u) = 0,
\end{equation}
(with $\Delta$ denoting the $d$-dimensional flat Laplacian) and the
nonlinear Schr\"odinger equation (NLS)
\begin{equation}
  \label{eq:16}
  iu_t - \Delta u + f\left(\left|u\right|^{2}\right)u = 0,
\end{equation}
Studies on the NLS are also motivated by its importance in modeling
various physical systems: the Bose-Einstein condensates, nonlinear
optics or plasma physics.  These problems are usually posed on a
$d$-dimensional cube $x\in\mathcal{D}=(0,\pi)^{d}$, $t\in\mathbb{R}$
and typically with Dirichlet boundary conditions
$\left. u(t,x)\right|_{x\in\partial\mathcal{D}}=0$.  Techniques
developed for these toy models (especially for $d=1$) are further
applied to more complex systems.

Over decades mathematicians considered various modifications of
(\ref{eq:15}) and (\ref{eq:16}) including other boundary conditions
(like periodic boundary conditions then the domain is
$\mathbf{T}^{d}$), generalization of the nonlinear term, and also the
extension to higher dimensions; also same results are often derived or
refined using different techniques.  We give here only few results of
vast mathematical literature on the existence of time-periodic and
quasi-periodic solutions to the nonlinear wave and Schr\"odinger
equations.

First studies of the one-dimensional problem (\ref{eq:15}) reach back
to the seminal work \cite{CPA:CPA3160310103} where using variational
method it was shown that (under some restrictions on $f$) there exists
(not necessarily small) time-periodic solutions.  Moreover, these form
very special class of solutions with period being rational multiple of
$\pi$.  Further proofs of the existence of time-periodic solutions of
NLW based on a Kolmogorov-Arnold-Moser (KAM) type techniques were
developed independently in \cite{10.1007/BF02566000} and in
\cite{10.1007/BF02104499}.  First of all, these give solutions when
$f'(0)\neq 0$ (a problem with the mass term---the nonresonant case)
and complement the result of \cite{CPA:CPA3160310103} in a sense that
here solutions period is non-rational multiple of $\pi$.  Using other
techniques (method based on the Nash-Moser generalized implicit
function theorem and on the Lyapunov-Schmidt decomposition) these
results were extended to include generic class of nonlinearities
\cite{CPA:CPA3160461102}.  A very special result, the proof of
existence of small amplitude time-periodic solutions to the
one-dimensional NLW with cubic defocusing term, i.e. for
$f(u)=-u^{3}$, was given in \cite{10.1007/BF01077432}.
Generalizations of former proofs to higher dimensional problems were
given in \cite{10.1007/BF01902055} (for massive case on
$\mathbf{T}^{d}$).

Not always theorems guarantee existence of continuous family of
time-periodic solutions (the case studied in this thesis), rather the
existence of Cantor-like sets (of measure zero) of such solutions were
proved, see e.g. \cite{CPA:CPA3160461102}; while these results are
constantly refined to include more solutions
\cite{10.1007/s00220-004-1255-8}.  For more recent results, including
generalizations to higher dimensions and other types of evolution
equations see \cite{berti2011, 10.1007/s00220-009-0817-1} and
references therein.

The complicated proofs of existence of time-periodic solutions give
little information about their form and properties (including their
stability).  An exception is \cite{10.1007/BF02557130,
  Khrustalev2001239} where the analytical construction of
time-periodic solutions (based on the Poincar\'e-Lindstedt
perturbative approach) for the massless one-dimensional NLW with cubic
nonlinearity and periodic boundary conditions is explicitly
demonstrated.  The authors of \cite{10.1007/BF02557130,
  Khrustalev2001239} construct solutions bifurcating from a
fundamental frequency of linear mode and in particular they resolve
the leading-order expansion term which is a nontrivial infinite
superposition of linear modes rather than just a single mode (feature
present in one of the models considered below).

Very little is known about the behaviour of generic initial
conditions.  The results on the existence of small amplitude solutions
(starting from finite number of linear modes) which stay almost
periodic for exponentially long times are given in
\cite{Bambusi199873, 10.1023/A:1013943111479}.  Such behaviour was
confirmed recently in the numerical studies of a one-dimensional NLW
with positive mass term (the nonresonant case) and quadratic
nonlinearity with periodic boundary conditions, see
\cite{10.1007/s00205-007-0095-z, 10.1080/03605302.2012.683503}.
Authors consider small amplitude initial data concentrated in one
Fourier mode whose evolution demonstrates that the energy remains
essentially localized in the initial Fourier mode over time scales
that are much longer than predicted by standard perturbation theory.
These results are then accompanied with proofs relaying on a modulated
Fourier expansions in time \cite{hairer2006geometric}.

The theorem stating the existence of solutions with growing in time
higher Sobolev norms\footnote{Sobolev norm
  $\left\|\,\cdot\,\right\|_{s,p}$ is a natural norm of the Sobolev
  space $W^{s,p}$---space of functions such that their weak
  derivatives of order up to $s$ have finite $L^{p}$ norm.  In one
  dimensional case such norm is defined as: $\left\|f\right\|_{s,p} :=
  \left(\sum_{i=0}^{s}\left\|f^{(i)}\right\|_{p}^{p}\right)^{1/p}$.
  For special case $p=2$ the space $W^{s,2}$ is denoted by $H^{s}$
  \cite{SobolevSpace}.} for the NLS on $\mathbf{T}^{2}$ with cubic
defocusing nonlinearity was proved in \cite{Colliander2010}.  Theses
particular solutions exhibit energy transfer from low to high modes
which in turn induce a growth in time of higher Sobolev norms (the
conservation of energy associated with (\ref{eq:16}) implies that
$H^{1}$ norm of solution stays bounded).  This statement was further
refined in \cite{Guardia2012} where the existence of solutions with
polynomial time estimates was proved.

Numerical studies in \cite{Colliander2013} of a toy model (the one
proposed in \cite{Colliander2010} to approximate dynamics of the
original problem) have shown that indeed a simplified finite
dimensional dynamical system approximates the original problem, and
more importantly that the energy cascade is a generic phenomenon (a
conclusion supported by statistical studies).

Studies of the energy cascade of cubic NLS on multidimensional torus
where reported in \cite{10.3934/dcds.2012.32.2063} where authors study
a special small amplitude $\ep$ initial conditions and give precise
description of growth of higher modes.  This construction is valid
only up to time of order $1/\ep$ and does not depend on the focusing
or defocusing character of nonlinearity.  Authors provide numerical
illustration of that phenomena and observe stabilization in higher
Sobolev norms (despite the fact they consider resonant case); after a
sufficiently long time all modes are present but the energy flow is
less active (defocusing case).

In contrast to the instability results (the one showing that starting
with initial data with small $H^{s}$ norms these grow arbitrarily in
time) there are results showing the long-time (orbital) stability of
plane waves (a single mode initial conditions), see e.g.
\cite{doi:10.1080/03605302.2013.785562, Wilson2014}.

All these studies concern relatively simple cases namely the equations
(\ref{eq:15}) and (\ref{eq:16}) have a very simple linear modes
(especially for problems posed on $\mathbf{T}^{d}$ these are
exponents) for which any product (the nonlinearity) can be expressed
as finite sums easily.  This stays in contrast to the equations
considered in this thesis, where the explicit form of eigenfunctions
is complicated so this makes the analysis particularly involved.
Moreover, in most of the considered cases we deal with coupled
elliptic-hyperbolic systems which makes the problem still more
challenging.

\section{Notation and conventions}
\label{sec:Notation}

Below we list applied conventions:

\begin{enumerate}[\itshape i\upshape)]
\item Einstein summation convention only applies when used with Greek
  indices. Sums involving latin indices are always given with
  summation $\sum$ sign.

\item The letter $d$ is reserved to denote the number of spatial
  dimensions hence the number of spacetime dimensions is $d+1$.

\item By $\mathbb{N}_{0}$ we denote the set of non-negative integer
  numbers $\mathbb{N}_{0}=\mathbb{N}\cup\{0\}=\{0,1,2,\ldots\}$.

\item For later convenience it is useful to introduce the following
  notation: let $\coef{\lambda} f$ denote the coefficient at
  $\ep^{\lambda}$ in the (formal) power series expansion of
  $f=\sum_{\lambda} f_{\lambda} \ep^{\lambda}$.  The operator
  $\coef{\lambda}$ is known as the coefficient extraction operator
  \cite{Knuth1994}.

\item The letter $\delta$ is used both as Kronecker delta and as
  metric function.  We state explicitly when $\delta$ symbol refers to
  the former.

\item We use overdots and primes to denote differentiation with
  respect to temporal and spatial variables respecitvely for functions
  of two arguments; for derivatives of functions of single argument we
  use primes regardless of character of its argument.  Additionally we
  use the subscript notation where it does not lead to confusion.

\item The $\Sphere^{n}$ stands for $n$-dimensional sphere.

\item The letter $L$ is reserved to denote the linear differential
  operator; by $L^{2}$ we denote the Hilbert space of square
  integrable functions.

\item For a smooth real function $f(x)$ the Pad\'e approximant of
  order $[L/M]$ is defined as
  $\left[L/M\right]_{f}(x):=P_{L}(x)/Q_{M}(x)$ where $P_{L}$ and
  $Q_{M}$ are real polynomials of orders $L$ and $M$ accordingly such
  that $f(x)-P_{L}(x)/Q_{M}(x)=\mathcal{O}\left(x^{L+M+1}\right)$.

\item $\Re$ stands for real while $\Im$ stands for imaginary parts of
  complex quantity.

\item When two small quantities are involved we use two typographical
  variants of lowercase Greek ``epsilon'', namely $\varepsilon$ and
  its lunate form $\epsilon$.

\item We use the Landau notation \cite{LandauSymbols} and write
  $f(x)=\mathcal{O}\left(g(x)\right)$ when $x\ra{}x_{0}$ for any such
  function $f$ that $f(x)\in\mathcal{O}\left(g(x)\right)$ (for
  functions of discrete argument we use the symbol $O$ instead of
  $\mathcal{O}$ and we assume the discrete variable tending to
  infinity).
\end{enumerate}

\vspace{2ex}

We also give a list of frequently used abbreviations:

\vspace{3ex}

\begin{tabular}{l p{10cm}}
  aAds & asymptotically anti-de Sitter \\
  AdS & anti-de Sitter \\
  AdS$_{d+1}$ & is identified with a $(d+1)$ dimensional anti-de Sitter
  space and is used when the space dimension $d$ is explicitly
  mentioned \\
  CAdS &  covering space of anti-de Sitter. In the text after the
  discussion of AdS solution we implicitly assume while referring to AdS
  space its covering \\
  CAS & computer algebra system \\
  EKG & Einstein-Klein-Gordon \\
  FD & finite difference \\
  FDA & finite difference approximation \\
  MOL & method of lines \\
  NLS & nonlinear Schr\"odinger equation \\
  NLW & nonlinear wave equation \\
  ODE & ordinary differential equation \\
  PDE & partial differential equation \\
  RHS & right-hand side \\
  RK & Runge-Kutta method \\
  YM & Yang-Mills
\end{tabular}

\chapter{Constructing time-periodic solutions}
\label{cha:Methods}

In this chapter we review methods we use to find time-periodic
solutions for studied PDEs in more abstract setting.  A more
exhaustive discussion, regarding necessary modifications and details
of the construction, is postponed until concrete models are analyzed.
After introducing the problem (Section~\ref{sec:MethodsProblem}) we
describe in Section~\ref{sec:MethodsMethods} both of used methods: the
analytical approach, which is based on Poincar\'e-Lindstedt
perturbative expansion, and the numerical root-finding approach with
pseudospectral discretization.  We also comment on other applicable
techniques, in particular we briefly review the minimization approach
(Section~\ref{sec:MethodsOther}).  This chapter is also intended to
set necessary notation and conventions used throughout this thesis.

\section{Formulation of the problem}
\label{sec:MethodsProblem}

Let us consider an initial-boundary value problem\footnote{We restrict
  this consideration to the simplest case of a second order in time
  one dimensional wave equation which appears in all of the considered
  problems here.  Generalization to other types of equations, e.g. to
  NLS, or to higher dimensions is straightforward.}
\begin{equation}
  \label{eq:17}
    u_{tt} = F(u),
\end{equation}
$(t,x)\in[0,\infty)\times [a,b]$, with $-\infty<a<b<\infty$ where $F$
is a smooth function of $u$ and its spatial derivatives only, in
particular it is assumed to be explicitly independent of time so that
(\ref{eq:17}) is an autonomous NLW.  For simplicity we take the
homogeneous Dirichlet boundary conditions $u(t,a)=u(t,b)=0$ (other
combination of non-dissipative conditions are permitted).  When
Eq.~(\ref{eq:17}) is considered as an initial boundary value problem,
the initial conditions $u(0,x)$ and $u_{t}(0,x)$ are assumed to be
smooth, in particular to avoid corner singularities they have to
fulfill boundary conditions accompanied to (\ref{eq:17}).

We are looking for nontrivial solutions to Eq.~(\ref{eq:17})
satisfying
\begin{equation}
  \label{eq:18}
  u(t+T,x) = u(t,x), \quad \forall t\in\mathbb{R},
\end{equation}
with some finite period $T>0$.  First, we consider the linear
perturbation of the static solutions to this equation.  For any static
solution $u(t,x)=S(x)$ satisfying
\begin{equation}
  \label{eq:19}
  F(S) = 0,
\end{equation}
we consider a small perturbations, i.e. we set
\begin{equation}
  \label{eq:20}
  u = S + v,
\end{equation}
with $\left\|v\right\|\ll 1$ (in some suitable norm).  When we perform
the Taylor expansion of the RHS of (\ref{eq:17})
\begin{equation}
  \label{eq:21}
  F(S + v) = F(S) + F'(S) v + \ldots,
\end{equation}
and neglect higher order terms, we obtain the linear, homogeneous wave
equation for the perturbation $v$, namely
\begin{equation}
  \label{eq:22}
  v_{tt} + Lv = 0,
\end{equation}
where by $L=-F'(S)$ we denote the linear differential operator.  We
use the standard approach to solve Eq.~(\ref{eq:22}) and separate
variables, by taking the ansatz $v(t,x) = e^{-i\omega t}\tilde{v}(x)$,
which transforms Eq.~(\ref{eq:22}) to the eigenvalue problem for
$\tilde{v}$.  Then, the solution to the eigenequation
\begin{equation}
  \label{eq:23}
  L e_{j}(x) = \omega_{j}^{2} e_{j}(x), \quad j=0,1,\ldots,
\end{equation}
yields a complete set of eigenvectors
$\left\{e_{j}(x)\,\middle|\ j\in\mathbb{N}_{0}\right\}$ (the
eigenbasis) and real eigenvalues $\omega_{j}^{2}$ of the self-adjoint
Sturm-Liouville operator.  For linearly stable static solution $S$ the
linear operator $L$ is positive definite and each
$\omega_{j}^{2}\ge 0$ (which we assume here).  Using $e_{j}(x)$ any
solution to Eq.~(\ref{eq:22}) can be written as
\begin{equation}
  \label{eq:24}
  v(t,x) = \sum_{j\geq 0} \hat{v}_{j}\cos\left(\omega_{j}\,t
    + \varphi_{j}\right)e_{j}(x),
\end{equation}
which is parametrized by real constants: the amplitudes $\hat{v}_{j}$
and the phases $\varphi_{j}$, which are uniquely determined by the
initial conditions $v(0,x)$ and $v_{t}(0,x)$.

The spectrum of an operator $L$ gives the eigenfrequencies\footnote{We
  commonly use the term 'spectrum of $L$' when referring to the square
  roots of its eigenvalues---the eigenfrequencies of $L$.} which are
\textit{nondispersive} when $\diff \omega_j/\diff j=\text{const}$, and
\textit{dispersive} in the opposite.  These are also referred as being
\textit{resonant} (as these producing resonances) and
\textit{nonresonant} (the case when no resonances are expected)
respectively.  There is also a term \textit{completely resonant} (or
\textit{fully resonant}) which refers to the cases when all the
eigenfrequencies are rational multiples of one another (this the most
common for the considered case).  We will show that this distinction
(between resonant and noresonant spectrum) is sometimes misleading
since resonances are equally common also in nondispersive cases.  We
argue that the presence of resonances or their lack should be
attributed to the structure of equations rather than to the character
of the linear spectrum.

\section{Techniques}
\label{sec:MethodsMethods}

\subsection{Perturbative approach}
\label{sec:MethodsPerturbative}

The analysis of the linearized equation (\ref{eq:22}) and the
eigenvalue problem (\ref{eq:23}) is the first step toward the
construction of time-periodic solutions to the nonlinear equation
(\ref{eq:17}).  Clearly, for unstable static solutions no
time-periodic perturbations are expected to exist, though
time-periodic solutions may be unstable on their own.

The linear equation (\ref{eq:22}) has (infinitely) many nontrivial
time-periodic solutions.  In particular, the very special
time-periodic solution is just a single eigenmode
\begin{equation}
  \label{eq:25}
  v(t,x) = \hat{v}_{\gamma}\cos\left(\omega_{\gamma}\,t
    + \varphi_{\gamma}\right)e_{\gamma}(x),
\end{equation}
for any $\gamma\in\mathbb{N}_{0}$.  One can add more eigenmodes to
(\ref{eq:25}) in such a way that the solution still be periodic with
fundamental period $T_{\gamma}=2\pi/\omega_{\gamma}$ or one can modify
formula (\ref{eq:25}) so that it will no longer be periodic with that
period.  In particular, for any fixed frequency $\omega_{\gamma}$
($\gamma\in\mathbb{N}_{0}$) the solutions to Eq.~(\ref{eq:22}) in the
following form
\begin{equation}
  \label{eq:26}
  v(t,x) = \sum_{j\in O_{\gamma}} \hat{v}_{j}\cos\left(\omega_{j}\,t
    + \varphi_{j}\right)e_{j}(x),
\end{equation}
where
\begin{equation}
  \label{eq:27}
  O_{\gamma} = \{ k\in\mathbb{N}_{0}\,|\ m\,\omega_{\gamma}=\omega_{k},
  \ m\in\mathbb{N} \},
\end{equation}
is the set of resonant indices associated with $\gamma$, are periodic
with period $T_{\gamma}$ for any $\hat{v}_{j}\in\mathbb{R}$,
$j\in O_{\gamma}$.  If $v(t,x)$ contains other eigenfrequencies absent
in $O_{\gamma}$ then it is no longer periodic with period
$T_{\gamma}$.  There is a distinction between finite number of such
additional frequencies present in $v$, we have then a
\textit{quasi-periodic} solution, while solutions with infinite number
of additional frequencies present we call an \textit{almost-periodic}.

The observation that the linear equation possesses time-periodic
solutions (in fact many of them) is the starting point of our
construction of time-periodic solutions around the equilibrium $S$ to
the corresponding nonlinear equation.  It is natural to expect that a
small amplitude solutions to the nonlinear problem smoothly converge
to the solutions of the linear equation.  Therefore for each
eigenfrequency $\omega_{\gamma}$ ($\gamma\in\mathbb{N}_{0}$) there
should be a bifurcation branch of a family of time-periodic solutions
for the nonlinear system.  Of course with this approach one can
construct the bifurcating solutions only; the question about the
existence of other types of solutions like the non-bifurcating
time-periodic solutions found in \cite{CPA:CPA3160310103} stays open.

The perturbative procedure proposed in this thesis relies on the
Poincar\'e-Lindstedt method \cite{bender1999advanced,
  kevorkian1996multiple} in which both the solution profile function
and its oscillation frequency are expanded in powers of a small
parameter $\ep$ which measures the size of the solution (or
equivalently the magnitude of nonlinearity).  Since the equation
(\ref{eq:17}) is autonomous the oscillation frequency of time-periodic
solution is \textit{a priori} unknown.  For convenience we change
variables
\begin{equation}
  \label{eq:28}
  \tau = \Omega\,t,
\end{equation}
and rewrite Eq.~(\ref{eq:17}) for $v=S-u$ as
\begin{equation}
  \label{eq:29}
  \Omega^{2}v_{\tau\tau} + Lv + f(v) = 0,
\end{equation}
where $f$ is the nonlinear part of $F$ such that $f(0)=0$.  Then, we
define perturbative series expansion in $\ep$ for both the solution
$v$ and the frequency $\Omega$
\begin{align}
  \label{eq:30}
  v(t,x;\ep) &= \sum_{\lambda\geq 1}\ep^{\lambda}\,v_{\lambda}(\tau,x),
  \\
  \label{eq:31}
  \Omega(\ep) &= \sum_{\lambda\geq 0}\ep^{\lambda}\,\xi_{\lambda}.
\end{align}
We plug (\ref{eq:30}) and (\ref{eq:31}) into (\ref{eq:29}), collect
terms with the same powers of $\ep$ and require the resulting
equations to be satisfied order by order.  In addition we demand the
solution to be uniformly bounded and at the same periodic with the
frequency $\Omega$.  Any continuous family of solutions bifurcating
from the eigenfrequency $\omega_{\gamma}$ ($\gamma \in
\mathbb{N}_{0}$) is constructed as follows.  At the lowest order the
equation for $v_{1}$ has the form
\begin{equation}
  \label{eq:32}
  \xi_{0}^{2}\,\frac{\partial^{2}v_{1}}{\partial\tau^{2}} + Lv_{1} = 0,
\end{equation}
which is solved by setting $\xi_{0}=\omega_{\gamma}$, and by taking
the following superposition
\begin{equation}
  \label{eq:33}
  v_{1}(\tau,x) =
  \sum_{j\in O_{\gamma}}
  \hat{v}_{1,j}\,\cos\left( \frac{\omega_{j}}{\omega_{\gamma}}\tau
    + \varphi_{1,j}\right) e_{j}(x),
\end{equation}
as a time-periodic function with a fundamental period $2\pi$ in
$\tau$, with a set of free parameters: $\hat{v}_{1,j}$ and
$\varphi_{1,j}$.\footnote{The time translation symmetry of
  (\ref{eq:29}) allows us to fix the phase of solutions such that
  their time derivative vanishes at $\tau=0$, which in particular
  implies $\varphi_{1,j}=0$.}  If we neglect the higher order terms in
the perturbative expansion (\ref{eq:30}) and (\ref{eq:31}) we would
get only the linear approximation to the time-periodic solution,
cf. (\ref{eq:26}), which would be a good approximation only for
$|\ep|\ll 1$.  Therefore, we solve also higher order perturbative
equations, enforcing time periodicity, order by order, to get better
and better approximations to the time-periodic solution of
Eq.~(\ref{eq:29}).

At higher orders, in particular for $\lambda=2$ (or first nontrivial
order), the perturbative equation takes the form of an inhomogeneous
wave equation
\begin{equation}
  \label{eq:34}
  \omega_{\gamma}^{2}\,\frac{\partial^{2}v_{2}}{\partial\tau^{2}}
  + Lv_{2} = S_{2},
\end{equation}
with the source term depending on all lower order terms, here
$S_{2}\equiv S_{2}(v_{1},\xi_{1})$.  For generic choice of free
parameters in (\ref{eq:33}) there will be no chance to obtain bounded
solution to (\ref{eq:34}).  This is because the source term, in
general, would contain resonant terms, i.e. terms like
$\cos(p\tau)\,e_{n}(x)$ with $p=\omega_{n}/\omega_{\gamma}$,
$n\in\mathbb{N}_{0}$, $p\in\mathbb{N}$; these will be present even for
dispersive (nonresonant) spectrum of linear
perturbations.\footnote{There is always possibility for one resonance
  to appear---that for the fundamental mode frequency.}  When not
removed, they produce secular terms in $v_{2}$, terms proportional to
$\tau$, which would spoil the periodicity and result in unbounded
solution.  Therefore, we use a freedom we have in setting the
parameters $\xi_{1}$ and $\hat{v}_{1,j}$ in (\ref{eq:33}) to remove
all of the resonant terms present in $S_{2}$.  It may happen (and is
common in considered cases) that only one resonance appears at the
lowest nontrivial order, here at $\lambda=2$, which is then removed by
the frequency shift $\xi_{1}$, so instead of (\ref{eq:33}) it suffices
to take a single mode in $v_{1}$, i.e.
\begin{equation}
  \label{eq:35}
  v_{1}(\tau,x) =
  \hat{v}_{1,\gamma}\,\cos\left( \tau + \varphi_{1,\gamma}\right) e_{\gamma}(x),
\end{equation}
(the amplitude $\hat{v}_{1,\gamma}$ will be fixed by the
\textit{normalization condition} which will define expansion parameter
$\ep$).  When more (typically infinite number) resonances are present
at second order then using (\ref{eq:33}) is necessary and the
amplitudes $\hat{v}_{1,j}$ are used to produce a uniformly bounded
solution to (\ref{eq:34}).  This then takes the general form
\begin{equation}
  \label{eq:36}
  v_{2}(\tau,x) = \sum_{j\geq 0}\hat{v}_{2,j}(\tau)e_{j}(x),
\end{equation}
with $\hat{v}_{2,j}(\tau)$ being $2\pi$-periodic trigonometric
polynomials with initial values unspecified at this order.  Whether
the sum in (\ref{eq:36}) is finite or not depends on the
compatibility of boundary expansion of eigenbasis with that of smooth
solutions to the nonlinear equation (\ref{eq:29}).  Higher order terms
are derived in similar way using the integration constants and
frequency expansion coefficient we have from the previous order.

This method is a variant of the Poincar\'e-Lindsted perturbative
approach, since we also relax the initial conditions to remove all of
the resonant terms appearing at successive perturbative orders
$\lambda$.  This may be also regarded as a shooting method since our
aim is to find periodic and bounded solution by suitably tuning the
initial values such that no secular terms appear and
$v(2\pi,x)=v(0,x)$.

When only single eigenmode is present in (\ref{eq:33}) and higher
order terms in (\ref{eq:30}) can be written as a finite linear
combination of eigenfunctions we can reduce the problem of solving
PDEs to a much simpler task of integrating ODEs.  Then, taking the
advantages of CAS in manipulating lengthy expressions, the
perturbative calculation can be carried up to very high order.
Additionally, if the structure of the problem allows for some sort of
algorithmization, then the whole procedure of constructing of
time-periodic solutions can be efficiently implemented yielding 'a
black box' procedure (for example this is the case for the EKG system
in even spatial dimensions).

Analysis of concrete examples shows that the above method requires
some modifications.  Details of perturbative construction of
time-periodic solutions will be given on the case by case manner,
however the idea behind many technical issues remain the same, with
some minor modifications coming from special forms of analyzed
equations.  In particular, for the Einstein equations, in addition to
the wave equation for the dynamical degrees of freedom, one needs to
solve also the constraints, which are relatively easy to work out.
Also, special form of governing equations may force some specific form
of perturbative expansions, for example, with only odd coefficients
present.  These and other details are addressed in the subsequent
chapters.

\subsection{Numerical approach}
\label{sec:MethodsNumeric}

For the numerical construction of time-periodic solutions we also
rewrite (\ref{eq:17}) as (\ref{eq:29}) with the change of variables
$\tau=\Omega\,t$.  This is crucial since in the original time
coordinate the size of the domain is not know \textit{a priori} since
the period (equivalently the frequency $\Omega$) is one of the
unknowns.  This transformation fixes the size of temporal domain, in
effect the problem (\ref{eq:29}) with $2\pi$-periodicity condition
\begin{equation}
  \label{eq:37}
  v(\tau+2\pi,x) = v(\tau,x), \quad \tau\in\mathbb{R},
\end{equation}
is posed on the rectangular domain $(\tau,x)\in[0,2\pi)\times[a,b]$.
Any smooth periodic (in $\tau$) function satisfying boundary
conditions associated with (\ref{eq:17}), e.g.
$v(\tau,a)=v(\tau,b)=0$, can be written in a following
way\footnote{The expansion in terms of eigenbasis functions is
  advantageous when the boundary expansion of smooth solution conforms
  with that of $e_{j}(x)$ otherwise the eigenbasis functions should be
  replaced by other complete set of functions, e.g. by Chebyshev
  polynomials which are \textit{the standard} polynomials for generic
  boundary conditions \cite{boyd2001chebyshev}.}
\begin{equation}
  \label{eq:38}
  v(\tau,x) = \sum_{k\geq 0}\sum_{j\geq 0}\left(
    \hat{v}_{k,j}\cos(k\tau) + \hat{w}_{k,j}\sin(k\tau)
  \right) e_{j}(x),
\end{equation}
with arbitrary real parameters $\hat{v}_{k,j}$ and $\hat{w}_{k,j}$.
We can fix the phase of the time-periodic solution and eliminate half
of the coefficients present in (\ref{eq:38}) thanks to the time
reflection symmetry $\tau\ra-\tau$ of Eq.~(\ref{eq:29}).  Without loss
of generality we set $\partial_{\tau}v|_{\tau=0}=0$, which implies
$\hat{w}_{k,j} = 0$ for $k,j\in\mathbb{N}_{0}$.

Following the numerical spectral approach we consider finite
dimensional truncation of (\ref{eq:38}).  Let $\mathsf{B}_{K,N}$ be a
finite dimensional subspace of
$L^{2}\bigl(\left[0,2\pi\right) \times \left[a,b\right]\bigr)$
\begin{equation}
  \label{eq:39}
  \mathsf{B}_{K,N} =
  \text{span}\left\{\left.\cos(k\tau)\,e_j(x)\,\right|\ k=0,1,\ldots,K,
    \ j=0,1,\ldots,N \right\},
\end{equation}
of dimension $\text{dim}\left(\mathsf{B}_{K,N}\right)=(K+1)(N+1)$.
Next, we approximate a smooth function $v(\tau,x)$ by considering its
finite expansion in terms of $(N+1)$ eigenbasis functions and $(K+1)$
trigonometric polynomials
\begin{equation}
  \label{eq:40}
  \mathcal{I}_{K,N}v(\tau,x) =
  \sum_{k=0}^{K}\sum_{j=0}^{N}\hat{v}_{k,j}\cos(k\tau)\,e_{j}(x),
\end{equation}
where $\mathcal{I}_{K,N}$ is the orthogonal projection from
$L^{2}\bigl([0,2\pi)\times[a,b]\bigr)$ onto $\mathsf{B}_{K,N}$ (to
simplify the notation in the following chapters we omit the explicit
projection operator and simply identify approximated function with its
interpolant remembering that we always consider its finite-dimensional
approximation).  If the expansion coefficients $\hat{v}_{k,j}$ decay
exponentially fast with growing mode index,\footnote{We use the
  numerical jargon phrase and say about the 'spectral convergence'
  when the generalized Fourier coefficients $\hat{f}_j$ of smooth
  function $f(x)$ decay faster than any negative integer power of
  $j$.} then the truncated sum (\ref{eq:40}) should serve as good
approximation to the continuous solution.  Using finite dimensional
representation (\ref{eq:40}) it is easy to compute its time derivative
\begin{equation}
  \label{eq:41}
  \frac{\partial^{2}}{\partial \tau^{2}}\mathcal{I}_{K,N}v(t,x) =
  \sum_{k=0}^{K}\sum_{j=0}^{N}\left(-k^{2}\right)
  \hat{v}_{k,j}\cos(k\tau)\,e_{j}(x),
\end{equation}
and the action of linear operator $L$
\begin{equation}
  \label{eq:42}
  L\bigl(\mathcal{I}_{K,N}v(t,x)\bigr) =
  \sum_{k=0}^{K}\sum_{j=0}^{N}\omega_{j}^{2}\,\hat{v}_{k,j}\cos(k\tau)\,e_{j}(x),
\end{equation}
To determine the expansion coefficients $\hat{v}_{k,j}$ of
time-periodic solution we plug (\ref{eq:40}) into (\ref{eq:29}) and
using relations (\ref{eq:41}) and (\ref{eq:42}) we get
\begin{multline}
  \label{eq:43}
  \sum_{k=0}^{K}\sum_{j=0}^{N}\bigl(-\Omega^{2} k^{2} +
  \omega_{j}^{2}\bigr) \hat{v}_{k,j}\cos\left(k\tau\right)\,e_{j}(x)
  \\
  +
  f\left(\,\sum_{k=0}^{K}\sum_{j=0}^{N}\hat{v}_{k,j}\cos(k\tau)\,e_{j}(x)\right)
  = 0.
\end{multline}
Then using the collocation approach \cite{boyd2001chebyshev,
  hesthaven2007spectral, shen2011spectral, trefethen2000spectral} we
require for (\ref{eq:43}) to be identically satisfied\footnote{In fact
  we require the residuals to vanish identically at the collocation
  points.} on a discrete set of points---suitably chosen collocation
points for the eigenbasis in space and trigonometric polynomials in
time
\begin{equation}
  \label{eq:44}
  \bigl\{(\tau_{k},\,x_{j})\,\bigl|\ k=0,\ldots,K,\ j=0,\ldots,N\bigr\},
\end{equation}
(thus the action of $f$ is computed in physical space).  In this way
we get a system of $(K+1)\times(N+1)$ nonlinear algebraic equations
for $(K+1)\times(N+1)+1$ unknowns.  To close the system we add the
condition fixing the amplitude of the solution, e.g. by setting
\begin{equation}
  \label{eq:45}
  g\left(\mathcal{I}_{K,N}v(0,\,\cdot\,)\right) = \ep,
\end{equation}
to some prescribed value $\ep\in\mathbb{R}$ (not necessarily small)
which identifies a particular solution.  The choice of $g$ depends on
a used parametrization of time-periodic solutions, one of the possible
choices is to control the amplitude of dominant mode (we comment on
the issue of taking proper parametrization throughout the subsequent
parts of the text).  This gives another constraint on the expansion
coefficients $\hat{v}_{k,j}$ and allows to uniquely determine the
approximate time-periodic solution to Eq.~(\ref{eq:29}).

The resulting set of equations is solved with the Newton-Raphson
algorithm using an approximate Jacobian matrix of (\ref{eq:43}) and
(\ref{eq:45}) computed numerically using first order forward
FDA.\footnote{Which is default option for the \codenamestyle{FindRoot}
  function when no Jacobian option is passed \cite{FindRoot}.} It is
not only the most strightforward approach, but for dense systems that
we encounter this is also the most effective method, since even with
an analytic formula computing numerical value of the Jacobian would be
very expensive.

The above method is a modified version of numerical method (without
explicit derivation of iterative scheme) introduced in
\cite{0951-7715-3-1-010} intended to derive spatially periodic
breathers of classical $\phi^{4}$ field theory.  Alternative approach
was proposed in \cite{PhysRevD.89.065027} to study self-interacting
scalar field on a fixed AdS background where the spatial
discretization was realized by using Chebyshev polynomial
approximation on divided radial domain.

\section{Alternative methods}
\label{sec:MethodsOther}

In this section we review the variational method for computation of
time-periodic solutions which was used by the author at the very
initial stage of his studies.  Most of the methods developed to
find-time periodic solutions of PDEs are generalizations of shooting
method for ODEs.  The method of \cite{10.1007/s00332-009-9058-x} is
also an extension of multi-shooting method, it was originally used for
computation of time-periodic solution to the Benjamin-Ono equation (a
dispersive PDE modelling the evolution of waves on deep water, with
periodic boundary conditions); see also \cite{Ambrose23022010,
  Ambrose201415} and references therein.

Such minimization procedure can be applied to a wide variety of PDEs,
for simplicity we illustrate it and consider (\ref{eq:17}), which we
rewrite as a system of two first order equations using vector notation
\begin{equation}
  \label{eq:46}
  \mathbf{U}_t(t,x) =
  \begin{pmatrix}
    0 & 1 \\
    F & 0
  \end{pmatrix}
  \mathbf{U}(t,x),
\end{equation}
where
\begin{equation}
  \label{eq:47}
  \mathbf{U}(t,x) :=
  \begin{pmatrix}
    u(t,x) \\
    v(t,x)
  \end{pmatrix}, \quad \text{and} \quad v(t,x):=u_{t}(t,x).
\end{equation}
For such two component vectors $\mathbf{U}_1(x), \mathbf{U}_2(x)\in
L^2\times L^2$, we define the scalar product
\begin{equation}
  \label{eq:48}
  \left<\mathbf{U}_1,\,\mathbf{U}_2\right> := \int_{a}^{b}
  \bigl( u_1(x)u_2(x) + v_1(x)v_2(x) \bigr) \diff x.
\end{equation}
To find nontrivial time periodic solutions we define the functional
\begin{equation}
  \label{eq:49}
  \Gamma(\mathbf{U}_0,T) := G(\mathbf{U}_0,T) + \Phi(\mathbf{U}_0),
\end{equation}
of a vector of initial conditions $\mathbf{U}(0,x):=\mathbf{U}_{0}$
and a period $T$, with
\begin{multline}
  \label{eq:50}
  G(\mathbf{U}_0,T) := \frac{1}{2} \left\| \mathbf{U}(T,\,\cdot\,) -
    \mathbf{U}_0 \right\|^2 =
  \\
  \frac{1}{2} \int_{a}^{b} \left[ \bigl( u(T,x)-u_0(x) \bigr)^2 +
    \bigl( v(T,x)-v_0(x) \bigr)^2 \right] \diff x,
\end{multline}
and look for minimizers of $\Gamma$ hoping that such minimum exist and
the minimum value of $G$ is zero.  The $\Phi$ part in (\ref{eq:49}),
called the penalty function, is arbitrary and can be defined to fix
the phase of time-periodic solution or its size (the amplitude).  The
inclusion of penalty functional is very important since it fixes the
values of the free parameters that describe the manifold of nontrivial
time-periodic solutions (without this part in $G$ only trivial
solutions are found, like constant or traveling waves).

The minimization is performed by the Broyden-Fletcher-Goldfarb-Shanno
(BFGS) algorithm \cite{nocedal2006numerical}, which requires not only
the value of the functional $\Gamma$ at specified point but also its
gradients $\frac{\delta\Gamma}{\delta\mathbf{U}_0}$ and
$\frac{\delta\Gamma}{\delta T}$ evaluated at each of its internal
step.  Computing $\frac{\delta\Gamma}{\delta\mathbf{U}_0}$ can be very
costly numerically, e.g. by straightforward use of finite
differencing.  The advantage of the method of
\cite{10.1007/s00332-009-9058-x} is that it allows to compute gradient
of (\ref{eq:50}) with respect to the initial data vector
$\mathbf{U}_{0}$ in as little computational time as it takes to get
value of $\Gamma$ itself (typically variational derivatives of penalty
functional $\Phi$ are easy to obtain and are computationally
inexpensive).  We shortly review this approach below.

The derivative of $G$ with respect to $T$ is easy to obtain, indeed
from Eq.~(\ref{eq:50}) we have
\begin{equation}
  \label{eq:51}
  \frac{\partial}{\partial T} G(\mathbf{U}_0,T) =
  \left< \mathbf{U}(T,\,\cdot\,) - \mathbf{U}_0,\,
    \mathbf{U}_t(T,\,\cdot\,)\right>.
\end{equation}
The variational derivative of $G$ with respect to initial conditions is
\begin{equation}
  \label{eq:52}
  \dot G = \frac{\diff}{\diff\epsilon} G(\mathbf{U}_0+\epsilon\,
  \widetilde{\mathbf{U}}_0,T)\Big|_{\epsilon=0} =
  \left< \mathbf{U}(T,\,\cdot\,) - \mathbf{U}_0,\,
    \widetilde{\mathbf{U}}(T,\,\cdot\,) - \widetilde{\mathbf{U}}_0 \right>,
\end{equation}
with $\widetilde{\mathbf{U}}(T,x):=\frac{\diff}{\diff\epsilon}
\mathbf{U}(T,x)\bigr|_{\epsilon=0}$.  To eliminate the unknown
function $\widetilde{\mathbf{U}}(T,x)$ from Eq.~(\ref{eq:52}) we would
like to rewrite $\dot{G}$ as
\begin{equation}
  \label{eq:53}
  \dot{G} = \left<\frac{\delta G}{\delta\mathbf{U}_{0}},\,
    \widetilde{\mathbf{U}}_{0}\right>.
\end{equation}
Therefore for $s=T-t$ we define an auxiliary quantity
$\mathbf{Q}(s,x)$ such that the following condition holds
\begin{equation}
  \label{eq:54}
  \left<\mathbf{Q}(s,\,\cdot\,),\,
    \widetilde{\mathbf{U}}(T-s,\,\cdot\,)\right>
  =\const,
\end{equation}
and at $s=0$
\begin{equation}
  \label{eq:55}
  \mathbf{Q}(0,x) = \mathbf{U}(T,x) - \mathbf{U}_{0}(x),
\end{equation}
is satisfied (so $\mathbf{Q}$ measures the deviation from
periodicity).  Linearizing (\ref{eq:17}) around any solution
$\mathbf{U}(t,x)$ we get the linear evolution equation for the
perturbation of initial conditions $\widetilde{\mathbf{U}}$
\begin{equation}
  \label{eq:56}
  \widetilde{\mathbf{U}}_{t} =
  F_{0}\left(\mathbf{U}\right)\widetilde{\mathbf{U}}.
\end{equation}
Then, differentiating (\ref{eq:54}) with respect to $s$ and using
(\ref{eq:56}) we get
\begin{equation}
  \label{eq:57}
  \biggl<\frac{\partial}{\partial s}\mathbf{Q}(s,\,\cdot\,),\,
  \widetilde{\mathbf{U}}(T-s\,\cdot\,)\biggr>
  -  \biggl<\mathbf{Q}(s,\,\cdot\,),\,
  F_{0}\bigl(\mathbf{U}(T-s,\,\cdot\,)\bigr)
  \widetilde{\mathbf{U}}(T-s,\,\cdot\,)\biggr>
  = 0,
\end{equation}
and whence we derive the linear evolution PDE for $\mathbf{Q}(s,x)$
\begin{equation}
  \label{eq:58}
  \frac{\partial}{\partial s}\mathbf{Q} =
  F_{0}^{*}\bigl(\mathbf{U}(T-s,\,\cdot\,)\bigr)\mathbf{Q}\,,
\end{equation}
where $F_{0}^{*}$ is adjoint of $F_{0}$ with respect to the inner
product (\ref{eq:48}).  The initial conditions for the above adjoint
PDE are these given in Eq.~(\ref{eq:55}).  In this way we get the
gradient of $G$ with respect to the vector of initial conditions
$\mathbf{U}_{0}$, i.e.
\begin{equation}
  \label{eq:59}
  \frac{\delta G}{\delta\mathbf{U}_{0}} =
  \mathbf{Q}(T,x) - \mathbf{Q}(0,x),
\end{equation}
which can be obtained at the cost of solving Eq.~(\ref{eq:58}), which
is comparable to the cost of solving Eq.~(\ref{eq:46}) itself.

Unfortunately, this approach (at least in the form used by the author)
has a serious drawback, namely slow convergence.  Solving both
evolution equations (\ref{eq:46}) and (\ref{eq:58}) using the MOL
approach with spectral discretization in space together with
Runge-Kutta (RK) time-integration algorithm will lead to the overall
polynomial convergence.  For very precise calculations and for very
complex PDEs, like the Einstein equations, this would be very
inefficient.  For this reason we prefer the method presented in the
previous section.  The spectral decomposition method, for the problems
at hand, gives the overall fast spectral convergence, and as a
consequence more efficient algorithm.

\chapter{Models}
\label{cha:Models}

In this chapter we introduce and motivate models studied in this
thesis.  We derive equations of motion and analyze their linearization
and the associated solution to the eigenvalue problems.  We emphasize
the regularity and boundary conditions for the fields as well as
particular coordinate and gauge choices.

\section{Einstein-Klein-Gordon system}
\label{sec:AdSModel}

The Einstein-Klein-Gordon (EKG) system is described by an action
\begin{equation}
  \label{eq:60}
  \mathcal{S} = \int \diff^{d+1}x \sqrt{-g}\left(\mathcal{L}_{g}
    + \mathcal{L}_{\phi}\right),
\end{equation}
where
\begin{equation}
  \label{eq:61}
  \mathcal{L}_{g} = \frac{1}{16\pi G}(R - 2\Lambda),
\end{equation}
is the Lagrangian of the Hilbert-Einstein action with the cosmological
constant term, while
\begin{equation}
  \label{eq:62}
  \mathcal{L}_{\phi} =
  -\frac{1}{2}\left(\nabla^{\mu}\phi\nabla_{\mu}\phi^{*}\right)
  + V\left(|\phi|^{2}\right),
\end{equation}
is the general Lagrangian of a scalar field $\phi$ minimally coupled
to gravity, $^{*}$ stands for complex conjugation, $V$ is a
self-interaction term.  Variation of the action with respect to the
metric $g_{\mu\nu}$ yields the Einstein equations
\begin{equation}
  \label{eq:63}
  R_{\mu\nu} - \frac{1}{2}g_{\mu\nu}R + \Lambda g_{\mu\nu} = 8\pi G\, T_{\mu\nu},
\end{equation}
with the stress-energy tensor
\begin{equation}
  \label{eq:64}
  T_{\mu\nu} = \frac{1}{2}\left(\nabla_{\mu}\phi\nabla_{\nu}\phi^{*} +
    \nabla_{\nu}\phi\nabla_{\mu}\phi^{*} -
    g_{\mu\nu}\nabla^{\alpha}\phi\nabla_{\alpha}\phi^{*}\right)
  - V\left(|\phi|^{2}\right)g_{\mu\nu}.
\end{equation}
Variation with respect to the scalar field gives the Klein-Gordon
equation
\begin{equation}
  \label{eq:65}
  \square\phi + V'\left(|\phi|^{2}\right)\phi = 0.
\end{equation}
The system of equations (\ref{eq:63})-(\ref{eq:65}) is the EKG system,
one of the PDE systems studied in this thesis.

In the asymptotically flat situation the above system (with
$V\equiv{}0$) has been extensively studied in the past leading to
important insights about the dynamics of gravitational collapse.  In
particular, Christodoulou showed that for small initial data the
fields disperse to infinity \cite{ch1}, while for large initial data
black holes are formed \cite{ch2}.  A borderline between these two
generic outcomes of evolution was explored numerically by Choptuik
leading to the discovery of critical phenomena at the threshold of
black hole formation \cite{ch}.  Following studies explored other
matter models which resulted in deeper understanding of critical
behaviour in gravitational collapse, see e.g. \cite{lrr-2007-5} and
references therein.

The numerical studies of critical phenomena in the presence of
negative cosmological constant were initiated in
\cite{PhysRevD.62.124012} studying self-gravitating EKG system in
(2+1) dimensions under the assumption of axial symmetry.  These
studies concentrated on the threshold phenomena and asymptotic
behaviour after the black hole formation.  Later studies were
concentrating on identification and understanding of the critical
solution \cite{PhysRevD.63.044007, PhysRevD.66.044015}.  For a long
time the evolution of small amplitude perturbations of AdS within the
EKG system stayed unexplored (see \cite{10.1007/s10714-014-1724-0} for
a comment) until the works \cite{br, jrb}, which were later
independently confirmed and further extended to the complex scalar
field case \cite{Buchel2012, Buchel2013}.  These studies were
accompanied with analytical considerations \cite{Dias2012a, dhms} and
most recently \cite{Holzegel20142436, Holzegel:2011rk, Holzegel2013}
including the proofs of stability of the Schwarzschild-AdS solution
and the global well-posedness of the EKG system with $\Lambda<0$
(these studies were restricted to spherical symmetry).  Finally,
recent numerical studies \cite{Balasubramanian2014} of the system
(\ref{eq:63})-(\ref{eq:65}) support connection of AdS dynamics with
the famous Fermi-Pasta-Ulam (FPU) problem, see
e.g. \cite{10.1063/1.1855036}.

Regarding the complexity of the dynamics of Einstein equations with
negative cosmological constant recent numerical studies are mainly
restricted to spherical symmetry (therefore the scalar field models
are considered---as being the simplest matter models---in order to
evade Birkhoff theorem \cite{d1992introducing} and generate dynamics).
A notable exception is \cite{PhysRevD.85.084038} which in $2+1$
setting ($SO(3)$ symmetry imposed in five spacetime dimensions)
demonstrate prompt black hole formation from large amplitude scalar
perturbations and its relaxation through quasinormal ringing.

Complex scalar field studies include both the construction and
properties of stationary configurations (standing waves/boson stars)
and their dynamics.  Different variants of the EKG system, with
relaxed symmetry assumptions as well as with numerous forms of
self-interaction term, were studied in the past both with and without
cosmological constant.  Among the results \cite{Buchel2012,
  Buchel2013} directly connected to the subject of this thesis the
most recent analysis of the EKG with $\Lambda<0$ include
configurations (boson stars and boson shells) with V-shaped
($V(|\phi|^{2})=|\phi|$) potential studied in \cite{Hartmann2013} and
solutions with helical symmetry \cite{Stotyn2014a, Stotyn2014b}.

Here, to reduce the complexity of the EKG system, we impose spherical
symmetry and consider self-gravitating scalar field only by putting
the potential term $V$ to zero (so in particular we exclude massive
fields).  We regard the space dimension $d$ appearing in the EKG
system as a discrete parameter so in particular we do not prefer any
of $d=3$, $d=28$ or $d=2014$ cases.

\subsection{Equations of motion}
\label{sec:AdSScalarEqations}

We parametrize the $(d+1)$-dimensional asymptotically AdS metric
(compare with (\ref{eq:8})) by the spherically symmetric ansatz
\begin{equation}
  \label{eq:66}
  \diff s^2 = \frac{\ell^{2}}{\cos^{2}x}
  \left( -A e^{-2\delta}\diff t^2 + A^{-1}\diff x^2
    + \sin^2{x}\diff\Omega^{2}_{d-1}\right),
\end{equation}
where $\ell^2=-d(d-1)/(2\Lambda)$, $\diff\Omega^2_{d-1}$ is the round
metric on $\Sphere^{d-1}$, $-\infty<t<\infty$, $0\leq x<\pi/2$, and
$A$, $\delta$ are functions of $(t,x)$.  For this ansatz, the
evolution of a self-gravitating massless scalar field $\phi(t,x)$ is
governed by the following system (using units where $8\pi G=d-1$)
\begin{equation}
  \label{eq:67}
  \dot\Phi = \left( Ae^{-\delta}\Pi \right)', \quad
  \dot\Pi = \frac{1}{\tan^{d-1}{x}}\left( \tan^{d-1}{x}Ae^{-\delta}\Phi \right)',
\end{equation}
\begin{align}
  \label{eq:68}
  \delta' &= -\sin{x}\cos{x}\left( \left|\Phi\right|^{2} +
    \left|\Pi\right|^{2} \right),
  \\
  \label{eq:69}
  A' &= \frac{d-2+2\sin^{2}{x}}{\sin{x}\cos{x}}(1-A) + A\delta',
  \\
  \label{eq:70}
  \dot{A} &= -2\sin{x}\cos{x}A^{2}e^{-\delta}\Re\left(\Phi\,\Pi^{*}\right),
\end{align}
where ${}^{\cdot}=\partial_t$, ${}'=\partial_x$, and
\begin{equation}
  \label{eq:71}
  \Phi = \phi', \qquad \Pi = A^{-1}e^{\delta}\dot\phi.
\end{equation}
The set of equations (\ref{eq:67})-(\ref{eq:69}) has the same form for
both real and complex valued scalar field $\phi$, compare
\cite{jrb,MRPRL} with the equivalent set of equations given in
\cite{Buchel2012} which differ by the absence of the scaling factor
$\cos^{d-1}{x}$.  Note that the length scale $\ell$ drops out from the
equations.  For the vacuum case $\phi\equiv 0$, this system has a
one-parameter family of static solutions
\begin{equation}
  \label{eq:72}
  \delta = \const, \quad A = 1 - M \frac{\cos^{2}{x}}{\tan^{d-2}x},
\end{equation}
which are the Schwarzschild-AdS black holes when $M>0$ (a $d+1$
dimensional Kottler metric \cite{3906}, derived also independently in
\cite{4065}) and the pure AdS for $M=0$.  In analogy to (\ref{eq:72})
we define the mass function for the system (\ref{eq:67})-(\ref{eq:69})
with dynamical matter content
\begin{equation}
  \label{eq:73}
  m(t,x) = \bigl( 1 - A(t,x)\bigr)\frac{\tan^{d-2}x}{\cos^{2}x},
\end{equation}
and from the Hamiltonian constraint (\ref{eq:69}) we get the
expression for the mass density
\begin{equation}
  \label{eq:74}
  m'(t,x) = -\frac{\sin^{d-2}{x}}{\cos^{d}{x}}A\delta'.
\end{equation}
Integrating this equation and using the slicing constraint
(\ref{eq:68}) we obtain the total mass
\begin{equation}
  \label{eq:75}
  M = \lim_{x\ra\pi/2} m(t,x) = \int_{0}^{\pi/2}A\left(\left|\Phi\right|^{2}
    + \left|\Pi\right|^{2}\right)\tan^{d-1}x\diff x,
\end{equation}
which, if finite (we assume this here), is also constant of
motion.\footnote{This is easy to see by imposing the mildest
  assumption on the falloff at $x=\pi/2$ for the scalar field, i.e.
  with $\phi\sim (\pi/2-x)^{\alpha}$ the integral (\ref{eq:75}) is
  finite when $\alpha\geq d/2$ (for smooth solutions there is
  $A(t,\pi/2)=1$ and $\delta(t,\pi/2)=\const$, see discussion in the
  following section).  Then from the definition (\ref{eq:73}) and the
  momentum constraint (\ref{eq:70}) we get
  $\lim_{x\ra\pi/2}\dot{m}(t,x)=0$.}  Within the adapted polar-areal
coordinate system (\ref{eq:66}) the Hamiltonian constraint
(\ref{eq:69}) can be transformed as follow.  Using the identity
\begin{equation}
  \label{eq:76}
  \frac{d-2+2\sin^{2}x}{\sin{x}\cos{x}} = \frac{\cos^{d}x}{\sin^{d-2}x}
  \left(\frac{\sin^{d-2}x}{\cos^{d}x}\right)',
\end{equation}
the Eq.~(\ref{eq:69}) can be rewritten as
\begin{equation}
  \label{eq:77}
  \left( \frac {\sin^{d-2} x} {\cos^d x} A \right)' -
  \left( \frac {\sin^{d-2} x} {\cos^d x} A \right) \delta' =
  \left( \frac {\sin^{d-2} x} {\cos^d x} \right)',
\end{equation}
which multiplied by $\exp(-\delta)$ and further integrated by parts
yields
\begin{equation}
  \label{eq:78}
  1 - A = e^{\delta}\frac{\cos^{d}x} {\sin^{d-2}x}\int_0^xe^{-\delta(t,y)}\left(
    \left|\Phi(t,y)\right|^{2} + \left|\Pi(t,y)\right|^{2}\right)
  \tan^{d-1}y\diff y.
\end{equation}

In the case of the complex scalar field Lagrangian (\ref{eq:62}) is
manifestly $U(1)$ invariant and the associated conserved current is
\begin{equation}
  \label{eq:79}
  J^{\mu} = \frac{i}{2}\bigl(\phi^{*}\nabla^{\mu}\phi -
    \phi\nabla^{\mu}\phi^{*}\bigr),
\end{equation}
while the conserved charge $Q$ is given by the integral
\begin{equation}
  \label{eq:80}
  Q = - \Im\int_{0}^{\pi/2}\phi\,\Pi^{*}\tan^{d-1}x\diff x,
\end{equation}
which is finite whenever $M<\infty$.  For the real scalar field the
charge vanishes identically while for complex scalar field it is the
second constant of motion.

The Ricci scalar computed directly from the definition for the metric
(\ref{eq:66}) expressed in terms of $A(t,x)$ and $\delta(t,x)$ is
fairly complicated, while using the Einstein equations it can be
reduced to the following form
\begin{equation}
  \label{eq:81}
  R = -\frac{d(d+1)}{\ell^{2}} +
  \frac{d-1}{\ell^{2}} A\cos^{2}{x}\left(\left|\Phi\right|^{2}
    + \left|\Pi\right|^{2}\right).
\end{equation}
Similarly the Kretschmann scalar takes a particularly simple form
\begin{multline}
  \label{eq:82}
  K = \frac{(d-1)(d+1)}{\ell^{4}}\left(\left|\Phi\right|^{2} -
    \left|\Pi\right|^{2}\right)^{2}A^{2}\cos^{4}{x}
  \\
  + 2\frac{d-1}{\ell^{4}}A\cot^{2}{x}
  \Bigl(d(d-3)-(d-1)(d-2)A+2\cos^{2}{x}\Bigr)
  \left(\left|\Phi\right|^{2} - \left|\Pi\right|^{2}\right)
  \\
  + 2\frac{d(d+1)}{\ell^{4}} +
  \frac{d(d-1)^{2}(d-2)}{\ell^{4}}\frac{(1-A)^{2}}{\sin^{4}{x}},
\end{multline}

\subsection{Regularity and boundary conditions}
\label{sec:AdSScalarBoundary}

Smoothness at the origin $x=0$ implies
\begin{subequations}
  \label{eq:83}
  \begin{align}
    \phi(t,x) &= \check{\phi}_{0}(t) + \sum_{i\geq 1}\check{\phi}_{2i}(t)x^{2i},
    \\
    \delta(t,x) &= \sum_{i\geq 1}\check{\delta}_{2i}(t)x^{2i},
    \\
    A(t,x) &= 1 + \sum_{i\geq 1}\check{A}_{2i}(t)x^{2i},
  \end{align}
\end{subequations}
so that only even powers of $x$ appear in these power series, and the
higher order terms in (\ref{eq:83}) are uniquely determined by
$\check{\phi}_{0}(t)$, e.g.
\begin{equation}
  \label{eq:84}
  \check{\phi}_{2}(t) = \frac{\check{\phi}_{0}''(t)}{2d},
  \quad \check{\delta}_{2}(t) = -\frac{\check{\phi}_{0}(t)^{2}}{2},
  \quad \check{A}_{2}(t) = -\frac{\check{\phi}_{0}'(t)^{2}}{d}.
\end{equation}
To fix the remaining gauge freedom in this system we set
$\delta(t,0)=0$, making the coordinate time $t$ to be the proper time
of the central observer.  It is easy to check\footnote{For example
  using the \mathematica{} notebook \texttt{boundary.nb} available at
  the NR/HEP 2 Spring School website \url{http://goo.gl/sb4qjZ}.}
that smoothness at spatial infinity and the finiteness of the total
mass (\ref{eq:75}) imply power series expansion near $x=\pi/2$ (using
$z=\pi/2-x$) depending on parity of space dimension $d$; explicitly
for even $d\geq 2$
\begin{subequations}
  \label{eq:85}
  \begin{align}
    \phi(t,x) &= \breve{\phi}_{0} + \breve{\phi}_{d}(t)z^{d} +
    \sum_{i\geq 1}\breve{\phi}_{d+2i}(t)z^{d+2i},
    \\[1ex]
    \delta(t,x) &= \breve{\delta}_{0}(t) + \sum_{i\geq 0}
    \breve{\delta}_{2d+2i}(t)z^{2d+2i},
    \\[1ex]
    A(t,x) &= 1 - Mz^{d} + \sum_{i\geq 1}\breve{A}_{d+2i}(t)z^{d+2i},
  \end{align}
\end{subequations}
while for odd $d\geq 3$
\begin{subequations}
  \label{eq:86}
  \begin{align}
    \phi(t,x) &= \breve{\phi}_{0} + \breve{\phi}_{d}(t)z^{d} +
    \sum_{i\geq 1}\breve{\phi}_{d+2i}(t)z^{d+2i} + M\sum_{i\geq
      0}\breve{\phi}_{2d+2i}(t)z^{2d+2i},
    \\[1ex]
    \delta(t,x) &= \breve{\delta}_{0}(t) + \sum_{i\geq
      0}\breve{\delta}_{2d+2i}(t)z^{2d+2i} + M\sum_{i\geq
      0}\breve{\delta}_{3d+2i}(t)z^{3d+2i},
    \\[1ex]
    A(t,x) &= 1 - Mz^{d} + \sum_{i\geq 1}\breve{A}_{d+2i}(t)z^{d+2i} +
    M\sum_{i\geq 0} \breve{A}_{2d+2i}(t)z^{2d+2i}.
  \end{align}
\end{subequations}
The subsequent terms are expressed by the constants $\breve{\phi}_{0}$
and $M$ specified by the initial data and functions
$\breve{\phi}_{d}(t)$ and $\breve{\delta}_{0}(t)$ which are determined
by the evolution.  In particular for compactly supported initial
perturbations localized at the origin we have
$\phi(0,\pi/2)=\breve{\phi}_{0}\equiv 0$ which implies homogeneous
Dirichlet conditions.  From the momentum constraint it follows that
the mass $M$ of the system is conserved.  For even $d$ the series
expansion (\ref{eq:85}) is always even regardless of the total mass
$M$ which nevertheless is present (implicitly) in higher order terms.
However, for odd $d$ the expansion (\ref{eq:86}) is neither even nor
odd for $M\neq 0$ (only when $M\equiv 0$ this expansion is odd in
$z$).  This has an important consequence---as we refer to this point
few times in this thesis---which causes us to use different variants
of methods depending on parity of $d$.

Local well-posedness, the first step toward a solution of a global
Cauchy problem, was proved in \cite{Friedrich1995125,
  Holzegel:2011rk}.

Using the Taylor series expansion at $x=0$, Eqs.~(\ref{eq:83}) and
(\ref{eq:84}), we find that Ricci (\ref{eq:81}) and Kretschmann
(\ref{eq:82}) scalars evaluated at the origin are polynomials in
$\left|\Pi(t,0)\right|$ only, i.e.
\begin{equation}
  \label{eq:87}
  R\bigl|_{x=0} = -\frac{d(d+1)}{\ell^{2}}
  - \frac{d-1}{\ell^{2}}\left|\Pi(t,0)\right|^{2},
\end{equation}
\begin{equation}
  \label{eq:88}
  K\bigl|_{x=0} = 2 \frac{d(d+1)}{\ell^{4}}
  + 4 \frac{d-1}{\ell^{4}}\left|\Pi(t,0)\right|^{2}
  + 2 \frac{(d-1)(2d-1)}{d\,\ell^{4}}\left|\Pi(t,0)\right|^{4}.
\end{equation}
It follows from here that whenever $\left|\Pi(t,0)\right|$ stays
bounded, so are Ricci and Kretschmann scalars at $x=0$.

\subsection{Linear perturbations---the eigenvalue problem}
\label{sec:AdSScalarEigen}

The extended studies of linear perturbations of AdS space with scalar,
electromagnetic and gravitational fields were presented in
\cite{0264-9381-21-12-012}, where linear stability of AdS with respect
to scalar, electromagnetic and gravitational perturbations was
demonstrated.  Here we consider in detail only the scalar case for the
problem at hand.  The spectrum of the linear essentially self-adjoint
operator, which governs linearized perturbations of AdS${}_{d+1}$
\begin{equation}
  \label{eq:89}
  \ddot\phi + L\phi = 0,
\end{equation}
$L = -(\tan x)^{1-d} \partial_x \left((\tan x)^{d-1} \partial_x
\right)$, is given by $\omega_j^2=(d+2j)^2$, $j\in\mathbb{N}_{0}$. The
eigenfunctions read
\begin{equation}
  \label{eq:90}
  e_j(x) = 2 \frac{\sqrt{j!\,(j+d-1)!}}{\Gamma\left(j+\frac{d}{2}\right)}
  \cos^{d}{x} P_j^{(d/2-1,d/2)} (\cos 2x)\,,
\end{equation}
where $P_j^{(\alpha,\beta)} (x)$ are the Jacobi polynomials (see
Appendix~\ref{sec:AppJacobiPolynomials}).  These eigenfunctions form
an orthonormal basis in the Hilbert space of functions
$L^2\left([0,\pi/2],\,\tan^{d-1}\!{x}\,\diff x\right)$.  Below we
denote the inner product on this Hilbert space by
\begin{equation}
  \label{eq:91}
  \inner{f}{g}:=\int_0 ^{\pi/2} f(x) g(x) \tan^{d-1} x\,\diff x.
\end{equation}
Note that with the choice of normalization constant in (\ref{eq:90})
we have $\inner{e_{i}}{e_{j}} = \delta_{ij}$ and $\inner{e_i'}{e_j'} =
\omega_j^2\,\delta_{ij}$ for $i,j\in\mathbb{N}_{0}$ (here
$\delta_{ij}$ stands for Kronecker delta).  These eigenfunctions are
regular at the origin, having an even Taylor expansion at $x=0$
\begin{multline}
  \label{eq:92}
  e_{j}(x) = \frac{1}{\Gamma(d/2)} \sqrt{\frac{\Gamma(d+j)}{j!}}
  \Biggl( 2 - \frac{\omega _j^2}{d}\,x^{2} +
  \frac{\omega_j^2\left(4(d-1) + 3\omega_j^2\right)}{12d(d+2)}\,x^{4}
  \\
  + \mathcal{O}\left(x^{6}\right)\Biggr).
\end{multline}
While parity of (\ref{eq:92}) is independent of $d$ near $x=\pi/2$ we
have (using $z=\pi/2-x$)
\begin{multline}
  \label{eq:93}
  e_{j}(x) = (-1)^j z^d \sqrt{\frac{\Gamma (d+j)}{j!}}  \Biggl(
  \frac{\omega_{j}}{\Gamma(d/2+1)} - \frac{3\omega_j^3 -
    2(d-1)d\,\omega_j}{12\,\Gamma(d/2+2)}\,z^{2}
  \\
  + \frac{\omega_j \left( - 60\left(d^2-1\right)\omega_j^2 + 4(d-1) d
      (d+2) (5d+2) + 45 \omega_j^4
    \right)}{1440\,\Gamma(d/2+3)}\,z^{4}
  \\
  + \mathcal{O}\left(z^{6}\right) \Biggr),
\end{multline}
which is even or odd depending on the space dimension $d$ only
(independently on the eigenmode index $j\in\mathbb{N}_{0}$).  This is
in contrast to the series expansion (\ref{eq:86}) of the scalar field
$\phi(t,x)$ for odd $d$ and $M\neq 0$, where the scalar field has no
definite symmetry.  Therefore prescribing initial data such as a
superposition of finite number of eigenmodes (in particular a single
eigenmode) for odd $d$ will lead to a corner singularity
(incompatibility of initial and boundary data) and consequently to
non-smooth evolution.  This incompatibility has a direct consequence
in the design of numerical method for time-evolution and construction
of time-periodic solutions (both perturbativelly and numerically);
this issue will be addressed and discussed in the subsequent chapters
of this work.

\section[Cohomogenity-two biaxial Bianchi~IX ansatz]{Cohomogenity-two
  biaxial Bianchi~IX\newline ansatz}
\label{sec:BCSModel}

Due to the Birkhoff theorem \cite{d1992introducing}, Einstein's
equations in spherical symmetry have no dynamical degrees of
freedom. Thus, in order to generate dynamics in spherical symmetry one
needs to include matter.  In the previous section we have considered a
very simple matter model---a minimally coupled self-gravitating scalar
field.  One can evade the Birkhoff theorem at the expense of going to
higher (odd spacetime) dimensions. Such model which admits radially
symmetric gravitational waves was introduced in
\cite{PhysRevLett.95.071102} (see also \cite{bizon2006bypass}).  It
was numerically studied in the context of critical behaviour in vacuum
gravitational collapse \cite{PhysRevLett.95.071102,
  PhysRevLett.97.131101}.  Later, the nine-dimensional version was
used for the numerical studies of stability of the
Schwarzschild-Tangherlini black hole \cite{PhysRevD.72.121502}.

Here we adapt cohomogenity-two biaxial Bianchi~IX ansatz to the
studies of aAdS spaces.  It appears that this provides the simplest
$1+1$ dimensional system with pure gravitational degrees of freedom
where the turbulent dynamics was also observed \cite{Bizon2014bcs}
supporting the AdS instability conjecture \cite{br}.  However in this
thesis we apply this model to study purely gravitational time-periodic
solutions to the vacuum Einstein equations.  These solutions which are
held exclusively by the gravitational field, appear to be nonlinearly
stable (against perturbations within this ansatz) thus may be
considered as an realization of the Wheeler geon \cite{PhysRev.97.511,
  PhysRev.135.B271},

Another motivation for these studies is the special role that
AdS$_{5}$ plays in AdS/CFT correspondence, which states that Type IIB
string theory on the product space AdS$_{5}\times\Sphere^{5}$ is
equivalent to $N=4$ super Yang-Mills theory on the four-dimensional
conformal boundary \cite{10.1023/A:1026654312961, Witten:1998qj}.

\subsection{Equations of motion}
\label{sec:BCS-equations-motion}

Following \cite{bizon2006bypass} we take the metric ansatz for aAdS
space in $(4+1)$ dimensions (\ref{eq:66}) and replace the round metric
on $\Sphere^{3}$ with the homogeneously squashed metric, thereby
breaking the $SO(4)$ isometry to $SO(3)\times U(1)$
\begin{equation}
  \label{eq:94}
  \diff s^{2} = \frac{\ell^{2}}{\cos^{2}{x}} \left[ -Ae^{-2\delta}\diff t^{2}
    + A^{-1}\diff x^{2} + \frac{\sin^{2}{x}}{4}\left( e^{2B}\left(
        \sigma_{1}^{2} + \sigma_{2}^{2}\right) + e^{-4B}\sigma_{3}^{2}
    \right)\right].
\end{equation}
The $A$, $\delta$ and $B$ are functions of $(t,x)\in
(-\infty,\infty)\times[0,\pi/2)$ and $\sigma_{i}$ are left invariant
one-forms on $SU(2)$, which in terms of the Euler angles
($0\leq\theta\leq\pi$, $0\leq\phi,\psi\leq 2\pi$) take the form
\begin{equation}
  \label{eq:95}
    \sigma_{1} + i\sigma_{2} = e^{i\psi}\left( \cos{\theta}\diff\phi +
      i\diff\theta \right), \quad
    \sigma_{3} = \diff\psi - \sin{\theta}\diff\phi.
\end{equation}
In this way the squashing parameter $B$ becomes a dynamical degree of
freedom and the Birkhoff theorem is evaded.  Substituting the ansatz
(\ref{eq:94}) into the vacuum Einstein equations in $(4+1)$ dimensions
\begin{equation}
  \label{eq:96}
  R_{\alpha\beta} = \frac{2}{3}\Lambda g_{\alpha\beta},
\end{equation}
with $\ell^{2}=-6/\Lambda$, one gets the following system of PDEs (in
the following we use overdots and primes to denote $\partial_{t}$ and
$\partial_{x}$ respectively)
\begin{equation}
  \label{eq:97}
  \dot B = Ae^{-\delta}\Pi, \quad \dot\Pi = \frac{1}{\tan^{3}{x}}\left(
    \tan^{3}{x}Ae^{-\delta}\beta\right)'
  - \frac{4}{3}\frac{e^{-\delta}}{\sin^{2}{x}} \left( e^{-2B}
    - e^{-8B} \right),
\end{equation}
\begin{align}
  \label{eq:98}
  \delta' &= -2\sin{x}\cos{x} \left( \beta^{2} + \Pi^{2} \right),
  \\
  \label{eq:99}
  A' &= \frac{3-\cos{2x}}{\sin{x}\cos{x}} \left(1-A\right) + \delta'A
  + \frac{2}{3}\frac{4e^{-2B}-e^{-8B}-3}{\tan{x}},
  \\
  \label{eq:100}
  \dot A &= -4\sin{x}\cos{x} A^{2}e^{-\delta}\beta\Pi,
\end{align}
where for convenience we introduce the fields
\begin{equation}
  \label{eq:101}
  \beta = B', \quad \Pi = A^{-1}e^{\delta}\dot B.
\end{equation}
This system has a very similar structure to the EKG system of
Section~\ref{sec:AdSScalarEqations},
cf. Eqs.~(\ref{eq:67})-(\ref{eq:71}) with $d=4$.  However here the
Einstein equations are explicitly nonlinear (in terms of the $B$
field), and due to the exponential nonlinearity, the system
(\ref{eq:97})-(\ref{eq:101}) does not have the reflection symmetry
$B\ra -B$ which is present for the self-gravitating minimally coupled
scalar field.

The mass function (defined in analogy to the mass in Schwarzschild-AdS
solution) for this system reads
\begin{equation}
  \label{eq:102}
  m(t,x) = \bigl(1-A(t,x)\bigr)\frac{\sin^{2}{x}}{\cos^{4}{x}},
\end{equation}
(which is the Eq.~(\ref{eq:73}) with $d=4$).  From the Hamiltonian
constraint (\ref{eq:99}) it follows that
\begin{equation}
  \label{eq:103}
  m'(t,x) = 2 \tan^{3}{x} \left[ \left(\beta^{2} + \Pi^{2}\right)A +
    \frac{3 + e^{-8B} - 4e^{-2B}}{3\sin^{2}{x}} \right],
\end{equation}
thus the conserved mass of the system reads\footnote{Which is constant
  of motion by analogous argument as for the EKG system.}
\begin{multline}
  \label{eq:104}
  M = \lim_{x\ra\pi/2}m(t,x) = \\
  = 2\int_{0}^{\pi/2}\left[ \left(\beta^{2} + \Pi^{2}\right)A +
    \frac{3 + e^{-8B} - 4e^{-2B}}{3\sin^{2}{x}} \right]\tan^{3}x \diff x.
\end{multline}
We intend to solve the system (\ref{eq:97})-(\ref{eq:99}) for smooth
initial data, prescribed on a $t=0$ slice, with finite total mass
(\ref{eq:104}).  Smooth solutions have the following Taylor series
expansion at the origin ($x=0$)
\begin{subequations}
  \label{eq:105}
  \begin{align}
    B(t,x) &= \check{B}_{0}(t)x^{2} + \mathcal{O}\left(x^{4}\right),
    \\
    \delta(t,x) &= \mathcal{O}\left(x^{4}\right),
    \\
    A(t,x) &= 1 + \mathcal{O}\left(x^{4}\right).
  \end{align}
\end{subequations}
We fix the gauge freedom by setting $\delta(t,0)=0$, so that $t$ be
the proper time at origin.  The power series expansions in
(\ref{eq:105}) are given in terms of $\check{B}_{0}(t)$ and its higher
derivatives.  Smoothness at spatial infinity $x=\pi/2$ and finiteness
of the total mass $M$ imply (using $z=x-\pi/2$)
\begin{subequations}
  \label{eq:106}
  \begin{align}
    B(t,x) &= \breve{B}_{4}(t)z^{4} + \mathcal{O}\left(z^{6}\right),
    \\
    \delta(t,x) &= \breve{\delta}_{0}(t) +
    \mathcal{O}\left(z^{8}\right),
    \\
    A(t,x) &= 1 - M z^{4} + \mathcal{O}\left(z^{6}\right),
  \end{align}
\end{subequations}
where the functions $\breve{B}_{4}(t)$, $\breve{\delta}_{0}(t)$ and a
constant $M$ uniquely determine the power series expansions.  It
follows from (\ref{eq:106}) that the behaviour of fields at the
conformal boundary of AdS$_{5}$ (at $x=\pi/2$) is completely fixed by
the smoothness assumption so there is no freedom in imposing the
boundary data.  Thus, the Cauchy problem for smooth initial data is
well defined without the need of specifying boundary data at
$x=\pi/2$, here again they are reflecting boundary conditions.  For
zero initial data $B=0$, $\dot B=0$ (no squashing) there is no
dynamics (due to the Birkhoff theorem) and the only solution is the
AdS$_{5}$ space ($A=1$, $\delta=\textrm{const}$) or the
Schwarzschild-AdS ($A=1+M\cos^{4}{x}/\sin^{2}{x}$, $M=\textrm{const}$
$\delta=\textrm{const}$).  As for the scalar field case
(Section~\ref{sec:AdSScalarEqations}) we rewrite the constraint
equations in the form suitable for numerical integration.  For given
functions $\beta$ and $\Pi$ (treated as independent dynamical
variables) the remaining metric functions, namely $\delta$ and $A$,
are expressed by the integrals; the metric function $\delta$ is given
by
\begin{equation}
  \label{eq:107}
  \delta(t,x) = -2\int_{0}^{x}\sin{y}\cos{y}\left(\beta(t,y)^{2}
    + \Pi(t,y)^{2}\right) \diff y,
\end{equation}
while (\ref{eq:99}) can be rewriten as
\begin{equation}
  \label{eq:108}
  \left(\frac{\sin^{2}{x}}{\cos^{4}{x}}A\right)' -
  \frac{\sin^{2}{x}}{\cos^{4}{x}}A\delta' =
  \left(\frac{\sin^{2}{x}}{\cos^{4}{x}}\right)' +
  \frac{2}{3}\frac{\sin^{2}{x}}{\cos^{4}{x}}\left(
    \frac{4e^{-2B}-e^{-8B}-3}{\tan{x}}\right),
\end{equation}
which multiplied by $e^{-\delta}$ and integrated by parts, taking into
account (\ref{eq:105}), yields
\begin{multline}
  \label{eq:109}
  1- A(t,x) = \\
  2\frac{\cos^{4}{x}}{\sin^{2}{x}}e^{\delta}\int_{0}^{x}
  e^{-\delta}\left( \beta^{2} + \Pi^{2} -
    \frac{1}{3}\frac{4e^{-2B} - e^{-8B} - 3}{\sin^{2}{y}} \right)
  \tan^{3}y \diff y.
\end{multline}

The Ricci scalar for pure vacuum case is constant $R=-20/\ell^{2}$,
cf.~(\ref{eq:96}), while the Kretschmann scalar is a complicated
function when expressed in terms of the metric functions $\delta$, $A$
and $B$.  For the diagnostics we monitor a value of the Kretschmann
scalar evaluated at the origin which, using boundary expansion
(\ref{eq:105}) reduces to
\begin{equation}
  \label{eq:110}
  K\bigl|_{x=0} = \frac{40}{\ell^{4}}
  \left(1 + \frac{108}{5}\,B''(t,0)^{2}\right).
\end{equation}

\subsection{Linear perturbations---the eigenvalue problem}
\label{sec:BCSEigenvalue}

The linear perturbations of AdS solution within the ansatz
(\ref{eq:94}) are governed by the equation
\begin{equation}
  \label{eq:111}
  \ddot B + L B = 0, \quad L =
  -\frac{1}{\tan^{3}{x}}\partial_{x}\left(\tan^{3}{x}\partial_{x}\right)
  + \frac{8}{\sin^{2}{x}},
\end{equation}
which is a particular case of the master equation describing the
evolution of linear perturbations of AdS space
\cite{0264-9381-21-12-012}.  The operator $L$ is essentially
self-adjoint on the Hilbert space $L^{2} \left([0,\pi/2],
  \tan^{3}x\diff x\right)$ equipped with the inner product denoted as
\begin{equation}
  \label{eq:112}
  \inner{f}{g}:=\int_{0}^{\pi/2}f(x)g(x)\tan^{3}x \diff x.
\end{equation}
The eigenvalues of $L$ are $\omega_{j}^{2} = (6+2j)^{2}$ and the
eigenfunctions read
\begin{equation}
  \label{eq:113}
  e_{j}(x) = 2\sqrt{\frac{(j+3)(j+4)(j+5)}{(j+1)(j+2)}}
  \sin^{2}{x}\cos^{4}{x}\,P^{(3,2)}_{j}(\cos{2x}),
\end{equation}
$j\in\mathbb{N}_{0}$, and $P^{(\alpha,\beta)}_{n}(x)$ are the Jacobi
polynomials (see Appendix~\ref{sec:AppJacobiPolynomials}).  The
constant factor in (\ref{eq:113}) is fixed by the normalization
condition $\inner{e_{i}}{e_{j}} = \delta_{ij}$ for
$i,j\in\mathbb{N}_{0}$ (where $\delta_{ij}$ stands for Kronecker
delta).  Here again as for the scalar field case the spectrum of
linear operator $L$ is nondispersive.  The important property of the
basis functions (\ref{eq:113}) is that their behaviour near the origin
$x=0$
\begin{multline}
  \label{eq:114}
  e_{j}(x) = \frac{1}{3}\sqrt{(j+1)(j+2)(j+3)^3(j+4)(j+5)} \biggl(
  x^{2} + \frac{1}{48} \left(3\omega_{j}^{2} + 4\right)x^{4}
  \\
  + \frac{9\omega_{j}^{4} + 60\omega_{j}^{2} - 128}{5760}\,x^6 +
  \mathcal{O}\left(x^{8}\right) \biggr),
\end{multline}
and near the outer boundary $x=\pi/2$
\begin{multline}
  \label{eq:115}
  e_{j}(x) = (-1)^j \sqrt{(j+1)(j+2)(j+3)(j+4)(j+5)} \biggl( z^4 -
  \frac{1}{12}\left(\omega_{j}^{2}-16\right)z^{6}
  \\
  + \left(\frac{1}{384}\left(\omega_{j}^{2} - 36\right)\omega_{j}^{2}
    + \frac{6}{5}\right)z^8 + \mathcal{O}\left(z^{10}\right) \biggr),
\end{multline}
where $z=x-\pi/2$, is compatible with the regularity conditions of the
full problem with $M\neq 0$ (cf. (\ref{eq:114}) and (\ref{eq:115})
with the corresponding conditions (\ref{eq:105}) and (\ref{eq:106})).
For that reason the basis functions (\ref{eq:113}) are natural
candidates for the expansion of the metric functions $B$ and
$A$,\footnote{In fact for $1-A$ because $e_{j}(0)=e_{j}(\pi/2)=0$.}
which can be used in both numerical and perturbative calculations.

\section{Spherical cavity model}
\label{sec:BoxModel}

We consider small perturbations of a portion of (3+1) dimensional
Minkowski spacetime enclosed inside a timelike worldtube
$\mathbb{R}\times\Sphere^3$.  Admittedly, this problem is somewhat
artificial geometrically, yet we think that it sheds some new light on
the results of \cite{br,jrb}.  The aim of the work \cite{mPRL} was to
see how these findings are affected by placing a reflecting mirror at
some finite radius $r=R$.  In other words, instead of asymptotically
flat boundary conditions, we consider the interior problem inside a
ball of radius $R$ with either Dirichlet $\phi(t,R)=0$ or Neumann
$\phi'(t,R)=0$ boundary condition.

The idea of putting the scalar field in a box is not completely new.
The similarity of quantization of scalar field theory in AdS and in
the box with reflecting boundary conditions was pointed out in
\cite{PhysRevD.18.3565}.  The first attempt toward simulations of
black holes (a binary system) in AdS space was carried out in
\cite{PhysRevD.82.104037, 1742-6596-229-1-012072} with a reflecting
boundary conditions imposed on a spherical box (though the
cosmological constant was set to zero).  The simulations showed
stability of the numerical scheme confirmed by the convergence tests
for times up to two reflections of gravitational signal produced in
the merger off the boundary.

Following the work \cite{br} there appeared papers
\cite{PhysRevD.84.066006, Garfinkle2012therm} where the instability of
AdS was studied.  Because the radial coordinate was not compactified
the authors prescribed the reflecting boundary conditions at some
fixed radial distance from the origin to mimic the reflective property
of AdS space.  Since they studied the system with both cosmological
constant and perfectly reflecting mirror their results were not
conclusive as far as the role of the $\Lambda$ term is concerned.
Despite the use of the nonuniform spatial grid to track the steep
gradients of the scalar field profiles, the results where not
conclusive for small perturbations---finally the sharp scalar peak
travels over whole spatial domain (see comments in \cite{jrb}).  A
similar setup, restricting the radial domain of AdS (so also with
$\Lambda<0$) and putting the reflecting boundary conditions for the
complex scalar field, was studied \cite{Buchel2012,Buchel2013}.

Here we set the cosmological constant to zero and study the
self-gravitating EKG system with reflecting boundary conditions
imposed at the spherical cavity.  We present extended studies of this
system, by supplementing \cite{mPRL, MR2014}.  In particular we
perform more detailed analysis of the Neumann boundary problem since
it substantially differs form the Dirichlet case as pointed out in
\cite{MR2014} (see \cite{Okawa2014} with opposite conclusions).
Furthermore we construct and describe small amplitude time-periodic
solutions in this system, both for Dirichlet and Neumann boundary
conditions, mainly to point out on the differences among other systems
we consider.

The well-posedness of this is system is not proved, however we
demonstrate this showing convergence tests (supporting the consistency
and stability) of our numerical methods.  On a well-posedness proof of
certain initial boundary value problems for the vacuum Einstein
equations see \cite{10.1007/s002200050571} and also
\cite{0264-9381-29-11-113001}.

\subsection{Equations of motion}
\label{sec:BoxEquations}

We restrict ourselves to spherical symmetry and to $d=3$ and consider
a minimally coupled self-gravitating real massless scalar field
$\phi(t,r)$ as a matter source with stress-energy tensor (\ref{eq:64})
(again we study the simplest case and set potential $V\equiv 0$).  It
is convenient to parametrize the general spherically symmetric metric
of (3+1)-dimensional spacetime in the following way
\begin{equation}
  \label{eq:116}
  \diff s^{2} = - A e^{-2\delta} \diff t^2 + A^{-1} \diff r^2 +
  r^2 \diff\Omega^{2}_{2}\,,
\end{equation}
where $r$ is the areal radial coordinate, $\diff\Omega^{2}_{2}$ is the
round metric on $\Sphere^{2}$, and both metric functions $A$ and
$\delta$ depend on $t$ and $r$.  We introduce auxiliary variables
\begin{equation}
  \label{eq:117}
  \Phi = \phi', \quad \Pi = A^{-1}e^{\delta}\dot{\phi},
\end{equation}
(hereafter primes and dots denote $\partial_{r}$ and $\partial_{t}$,
respectively) and write the wave equation (\ref{eq:65}) in first order
(in time) form
\begin{equation}
  \label{eq:118}
  \dot\Phi = \left( A e^{-\delta}\Pi \right)'\,, \quad
  \dot\Pi = \frac{1}{r^2}\left( r^2 A e^{-\delta}\Phi \right)'\,.
\end{equation}
The Einstein equations (\ref{eq:63}) take a form (in units where $4\pi
G=1$)
\begin{align}
  \label{eq:119}
  \delta' &= - r \left( \Phi^2 + \Pi^2 \right),
  \\
  \label{eq:120}
  A' &= \frac{1-A}{r} - A r \left( \Phi^2 + \Pi^2 \right),
  \\
  \label{eq:121}
  \dot A &= - 2 r e^{-\delta} A^{2}\Phi\Pi,
\end{align}
(in fact, this system can be derived from the set of equations
(\ref{eq:67})-(\ref{eq:70}) with $d=3$ by change of variables
$r/\ell=\tan{x}$ in the the limit $\ell\ra\infty$, which corresponds
to $\Lambda=0$).  Both the slicing condition (\ref{eq:119}) and the
Hamiltonian constraint (\ref{eq:120}) for given boundary conditions
(which we discuss below) determine the geometry of spacetime for given
matter content prescribed by the $\Phi(t,r)$ and $\Pi(t,r)$ scalar
fields.  The momentum constraint (\ref{eq:121}) can be used to monitor
the accuracy of numerical solution.

The mass function, $m(t,r)$, defined as
\begin{equation}
  \label{eq:122}
  m(t,r) = \frac{1}{2}r\left(1 - A(t,r)\right),
\end{equation}
gives a measure of mass contained within the 2-sphere of radius $r$ at
time $t$.  Using this the total mass (the energy) of the system can be
expressed as the volume integral
\begin{equation}
  \label{eq:123}
  M = \frac{1}{2} \int_0^R A \left( \Phi^2 + \Pi^2 \right)r^2\diff r,
\end{equation}
which is constant of motion when reflecting boundary conditions are
imposed at $r=R$ (see discussion below).  Additionally the quantity
(\ref{eq:122}) is useful to monitor the formation of apparent horizon.
In the adapted coordinates if an apparent horizon forms at point
$(t,r)$ in spacetime then $2m(t,r)/r=1$ (equivalently then the metric
function $A(t,r)$ drops to zero).  The drawback of the coordinate
system (\ref{eq:116}) is that whenever a black-hole forms the
evolution freezes while a finite precision in numerical calculations
causes the code to break.

\subsection{Regularity and boundary conditions}
\label{sec:BoxBoundary}

To ensure regularity at the origin $r=0$ we require that
\begin{subequations}
  \label{eq:124}
  \begin{align}
    \phi(t,r) &= \check{\phi}_{0}(t) + \mathcal{O}\left( r^{2}
    \right),
    \\
    A(t,r) &= 1 + \mathcal{O}\left(r^2\right),
    \\
    \delta(t,r) &= \mathcal{O}\left(r^2\right),
  \end{align}
\end{subequations}
with expansion coefficients uniquely given in terms of
$\check{\phi}_{0}(t)$.  In fact all the odd powers of $r$ are absent
in this Taylor expansion, which implies that both the metric functions
and the scalar field are even functions of $r$.  We set
$\delta(t,0)=0$ so that $t$ is the proper time at the center of
adopted coordinate system.

The boundary condition and the requirement of smoothness imply that
the coefficients of the power series expansions at $r=R$ (here
$z=1-r/R$)
\begin{subequations}
  \label{eq:125}
  \begin{align}
    \breve{\phi}(t,r) &= \sum_{k\geq 0}\breve{\phi}_k(t)z^k,
    \\
    \breve{\delta}(t,r) &= \sum_{k\geq 0}\breve{\delta}_k(t)z^k,
    \\
    \breve{A}(t,r) &= \sum_{k\geq 0}\breve{A}_k(t)z^k,
  \end{align}
\end{subequations}
are determined recursively by the leading order terms. For example, at
the lowest order for Dirichlet condition ($\breve{\phi}_{0}(t)\equiv
0$) we get
\begin{subequations}
  \label{eq:126}
  \begin{align}
    \breve{\phi}_{2}(t) &=
    \frac{1+\breve{A}_{0}(t)}{2\breve{A}_{0}(t)}\breve{\phi}_{1}(t),
    \\
    \breve{A}_{1}(t) &=
    \breve{A}_{0}(t)\left(1+\breve{\phi}_{1}(t)^{2}\right) - 1,
    \\
    \breve{\delta}_{1}(t) &= \breve{\phi}_{1}(t)^{2},
  \end{align}
\end{subequations}
and for Neumann condition ($\breve{\phi}_{1}(t)\equiv 0$)
\begin{subequations}
  \label{eq:127}
  \begin{align}
    \breve{\phi}_{2}(t) &=
    \frac{e^{2\breve{\delta}_{0}(t)}R^{2}}{2\breve{A}_{0}(t)^{2}}
    \left( \breve{\delta}_{0}'(t) \breve{\phi}_{0}'(t) +
      \breve{\phi}_{0}''(t) \right),
    \\
    \breve{A}_{1}(t) &= \breve{A}_{0}(t) - 1 +
    \frac{e^{2\breve{\delta}_{0}(t)}R^{2}}{\breve{A}_{0}(t)}
    \breve{\phi}_{0}'(t)^{2},
    \\
    \breve{\delta}_{1}(t) &=
    \frac{e^{2\breve{\delta}_{0}(t)}R^{2}}{\breve{A}_{0}(t)^{2}}
    \breve{\phi}_{0}'(t)^{2}.
  \end{align}
\end{subequations}
Taken at $t=0$, the expansion (\ref{eq:125}) express the compatibility
conditions between initial and boundary values.  It follows
immediately from equation (\ref{eq:121}) and the definition
(\ref{eq:122}) that for both boundary conditions, Dirichlet and
Neumann, we have
\begin{equation}
  \label{eq:128}
  \breve{A}_0'(t) = 0,
\end{equation}
which implies that the total energy (\ref{eq:123}) is constant of
motion.  We solve the initial-boundary value problem for the system
(\ref{eq:118})-(\ref{eq:121}) with boundary conditions compatible
smooth initial data, i.e. fulfilling the conditions either
(\ref{eq:126}) or (\ref{eq:127}), for both of the boundary conditions
imposed at the cavity located at $R=1$ (we can fix its position
without loosing generality).

\subsection{Linear perturbations---the eigenvalue problem}
\label{sec:BoxEigenvalue}

The evolution of linearized perturbations (propagating on the fixed
Minkowski background) is governed by the linear radial wave equation
\begin{equation}
  \label{eq:129}
  \ddot \phi + L\phi = 0, \quad
  L = - \frac{1}{r^{2}} \partial_{r}\left( r^{2} \partial_{r} \right).
\end{equation}
After separation of variables one obtains the eigenvalue problem---the
spherical Bessel equation \cite{Fun:SphBesselEq}.  Solving the
eigenvalue problem $Le_{j}(r)=\omega_{j}^{2}e_{j}(r)$ for the operator
$L$ we find its eigenfunctions
\begin{equation}
  \label{eq:130}
  e_{j}(r) =
  \sqrt{2}\left(1-\frac{\sin{2\omega_{j}}}{2\omega_{j}}\right)^{-1/2}
  \frac{\sin{\omega_{j}r}}{r}, \quad j\in\mathbb{N}_{0},
\end{equation}
which form an orthogonal basis on a Hilbert space
$L^{2}([0,1],r^{2}\diff r)$ with respect to the inner product
\begin{equation}
  \label{eq:131}
  \inner{f}{g} := \int_{0}^{1}f(r) g(r) r^{2} \diff r.
\end{equation}
The eigenvalues of $L$ for the Dirichlet boundary condition,
$\phi(t,1)=0$, are given explicitly
\begin{equation}
  \label{eq:132}
  \omega_{j}^{2} = (j+1)^{2}\pi^{2}, \quad j\in\mathbb{N}_{0},
\end{equation}
while for the Neumann boundary condition $\phi'(t,1)=0$, the
eigenvalues $\omega_{j}^{2}$ are determined from the transcendental
equation
\begin{equation}
  \label{eq:133}
  \omega_{j} = \tan\omega_{j}, \quad \omega_{j}>0, \quad j\in\mathbb{N}_{0},
\end{equation}
The first few roots of this equation are listed in
Table~\ref{tab:BoxEigenval}.
\begin{table}[t]
  \centering
  \begin{tabular}[h]{c|cccccc}
    \toprule
    $j$ & 0 & 1 & 2 & 3 & 4 & $\cdots$ \\ \midrule
    $\omega_{j}$ & \num{4.493409} & \num{7.725252} & \num{10.90412} &
    \num{14.06619} & \num{17.22076} & $\cdots$ \\
    \bottomrule
  \end{tabular}
  \caption{The numerical values of first few eigenfrequencies of the
    linear operator (\ref{eq:129}) governing the linear
    scalar perturbations in a cavity with Neumann boundary
    condition $\phi'(t,1)=0$.}
  \label{tab:BoxEigenval}
\end{table}
It is worth noting that the first derivatives of the eigenfunctions
(\ref{eq:130}) are also orthogonal with respect to (\ref{eq:131})
\begin{equation}
  \label{eq:134}
  \inner{e_{i}'}{e_{j}'} = \omega_{i}^{2}\,\delta_{ij},
  \quad i,\,j\in\mathbb{N}_{0},
\end{equation}
for each of the boundary condition individually (here $\delta_{ij}$
stands for the Kronecker delta).

Clearly, in the Dirichlet case the spectrum is nondispersive
(frequencies $\omega_{j}$ are equidistant), as for AdS within
considered models.  For the Neumann boundary condition the
eigenfrequencies are only asymptotically nondispersive (equidistant)
\begin{equation}
  \label{eq:135}
  \omega_{j} = \pi\left( j + \frac{1}{2} \right) - \frac{1}{\pi j} +
  \frac{1}{2\pi j^{2}} +\mathcal{O}\left(j^{-3}\right),
  \quad j\ra\infty, \quad j\in\mathbb{N}.
\end{equation}
The question of how the character of the spectrum of the linear
operator (\ref{eq:129}) (whether it is resonant or only
asymptotically resonant) affects the dynamics is addressed in
Chapter~\ref{cha:Turbulence}.

\section{Yang-Mills on Einstein Universe}
\label{sec:YMModel}

A disadvantage of previous models is their complexity---the Einstein
equations, even when restricted to spherical symmetry, form a set of
coupled elliptic-hyperbolic PDEs.  Thus their analytical analysis is
quite involved, but not completely impossible as we demonstrate in
this thesis.  Therefore we intent to study the system describing
evolution of nonlinear waves on a curved background---the spherically
symmetric YM field propagating on the Einstein Universe.  This not
only reduces the problem to a single NLW equation but also under some
restrictions stays in direct connection with the studies of AdS as we
demonstrate below.

We consider the YM field with $SU(2)$ gauge group on a $(d+1)$
dimensional spacetime $(\mathcal{M},g)$.  Given the YM potential
$A_{\mu}=\sum_{a=1}^{3}A^{a}_{\mu}T_{a}$ (where $T_a$ the are usual
generators of the $\mathfrak{su}(2)$ algebra) the field-strength
two-form is
\begin{equation}
  \label{eq:136}
  F_{\mu\nu}=\partial_{\mu}A_{\nu}-\partial_{\nu}A_{\mu}+[A_{\mu},A_{\nu}].
\end{equation}
The YM action written in terms of $F$ is
\begin{equation}
  \label{eq:137}
  \mathcal{S} = \int\text{Tr}\left(F_{\mu\nu}F^{\mu\nu}\right)
  \sqrt{-g}\diff^{d+1}x,
\end{equation}
where trace is computed over the gauge group indices. The
Euler-Lagrange equation for the action is
\begin{equation}
  \label{eq:138}
  \nabla_{\mu}F^{\mu\nu}+[A_{\mu},F^{\mu\nu}]=0.
\end{equation}
We note that if we consider a manifold $(\hat{\mathcal{M}},\hat{g})$
with the conformally related metric
\begin{equation}
  \label{eq:139}
  g = \Omega^{2}\,\hat{g},
\end{equation}
and the YM potential $\hat{A}_{\mu}=A_{\mu}$ on
$(\hat{\mathcal{M}},\hat{g})$ then the integrand (\ref{eq:137})
transforms as
\begin{equation}
  \label{eq:140}
  F_{\mu\nu}F^{\mu\nu}\sqrt{-g} =
  \Omega^{d-3}\hat{F}_{\mu\nu}\hat{F}^{\mu\nu}\sqrt{-\hat{g}}.
\end{equation}
This shows that the YM theory is conformally invariant only for $d=3$,
and then the field equations (\ref{eq:138}) have the same form when
expressed in terms of $\hat{F}$ and $F$.  Therefore taking into
account conformal structure of AdS instead of studying the YM field
propagating on AdS$_{4}$ we study this problem posed in four
dimensional Einstein Universe which technically translates to
extending the domain from upper hemisphere to the whole 3-sphere.  In
this way we overcome the ambiguity of boundary conditions at the
conformal boundary of AdS which is regular for the YM field in $d=3$
space dimensions.

Global existence for the YM field equation on AdS$_{4}$ was proved in
\cite{choquet1989global} (with the 'no flux' boundary conditions at
timelike infinity), while extension to general globally hyperbolic
Lorentzian manifolds was carried in \cite{Chrusciel:1997zb}.

\subsection{Equations of motion}
\label{sec:YMEquations}

We consider the YM field propagating on the Einstein Universe
\begin{equation}
  \label{eq:141}
  \diff s^2 = -\diff t^2 + \diff \Omega^2_3,
\end{equation}
where the metric on a round $\Sphere^{3}$ is
\begin{equation}
  \label{eq:142}
  \diff\Omega^2_3 =
  \diff x^2 + \sin^2\!{x}\,(\diff\vartheta^2 + \sin^2\vartheta
  \diff\varphi^2 ),
\end{equation}
with coordinate ranges $x,\vartheta \in [0,\pi]$ and
$\varphi\in[0,2\pi)$.

The most general, spherically symmetric SU(2) connection in $(3+1)$
dimensions can be written as \cite{PhysRevLett.38.121}
\begin{multline}
  \label{eq:143}
  A = W_{1}\,\tau_3\,\diff t + W_{2}\,\tau_3\,\diff x + (W_{3}\,\tau_1
  + W_{4}\,\tau_2)\diff\vartheta
  \\
  + (\cot\vartheta\,\tau_3 + W_{3}\,\tau_2 -
  W_{4}\,\tau_1)\sin\vartheta\,\diff\varphi,
\end{multline}
where $W_{1}$, $W_{2}$, $W_{3}$ and $W_{4}$ are real functions
depending on $t$, $x$ and $(\tau_1,\tau_2,\tau_3)$ is the standard
basis of su(2) Lie algebra.  To simplify (\ref{eq:143}) further we
take purely magnetic ansatz \cite{PhysRevD.60.124011,
  0264-9381-30-9-095009}.  Then making the suitable gauge choices we
can set all $W_{i}$ but $W_{3}$ to zero.  The reduced YM connection
with $W_{3}\equiv W$ is
\begin{equation}
  \label{eq:144}
  A = W\,\tau_1\,\diff\vartheta
  + (\cot\vartheta\,\tau_3 + W\,\tau_2)\sin\vartheta\,\diff\varphi.
\end{equation}
For this particular ansatz the YM curvature (\ref{eq:136}) is
\begin{multline}
  \label{eq:145}
  F = \left(\dot{W}\,\tau_1\,\diff t + W'\,\tau_1\,\diff
    x\right)\wedge \diff\vartheta
  \\
  + \left(\dot{W}\,\tau_2\,\diff t + W'\,\tau_2\,\diff x -
    \left(1-W^2\right)\tau_3 \,\diff\vartheta\right)
  \wedge\sin\vartheta\,\diff\varphi,
\end{multline}
where we use the notation $\dot{}\equiv \partial_{t}$ and
$'\equiv\partial_{x}$.  The action functional (\ref{eq:137}) reduces
to
\begin{equation}
  \label{eq:146}
  \mathcal{S} = 4\pi\int \left( -\dot{W}^2 + W'^2 +
    \frac{(1-W^2)^2}{2\sin^2x}\right) \diff x \diff t.
\end{equation}
The Euler-Lagrange equation for this action gives the YM equation for
the potential $W$
\begin{equation}
  \label{eq:147}
  -\ddot{W}+W''+\frac{W(1-W^2)}{\sin^2x} = 0.
\end{equation}
For this equation the conserved energy is given by the integral
\begin{equation}
  \label{eq:148}
  E = \frac{1}{2}\int_{0}^{\pi} \left( \dot{W}^2 + W'^2 +
    \frac{\left(1-W^2\right)^2}{2\sin^2x} \right) \diff x.
\end{equation}
The equation (\ref{eq:147}) has a reflection symmetry which means that
if $W$ is a solution of (\ref{eq:147}) then also $(-W)$ is a solution.

\subsection{Static solutions}
\label{sec:YMStatic}

Let us discuss the static solutions $W(t,x) = S(x)$ of equation
(\ref{eq:147}).  Regular solutions of
\begin{equation}
  \label{eq:149}
  S''+ \frac{S(1 - S^{2})}{\sin^{2}{x}} = 0,
\end{equation}
behave near $x=0$ as follows
\begin{equation}
  \label{eq:150}
  S(x) = \pm \left[1 - \check{S}_{2} x^{2}
    + \frac{\check{S}_{2}}{30}(9\check{S}_{2}-2)x^{4}
    + \mathcal{O}\left(x^{6}\right)\right],
\end{equation}
where $\check{S}_{2}$ is a free parameter.  Regular solutions near the
opposite pole $x=\pi$ behave as
\begin{equation}
  \label{eq:151}
  S(x) = \pm \left[1 - \breve{S}_{2}(x-\pi)^{2}
    + \frac{\breve{S}_{2}}{30}(9\breve{S}_{2}-2)(x-\pi)^{6}
    + \mathcal{O}\left((x-\pi)^{6}\right)\right],
\end{equation}
with $\breve{S}_{2}$ being a free parameter.  All higher order terms
in the expansions (\ref{eq:150}) and (\ref{eq:151}) are uniquely
determined by the leading order expansion coefficients $\check{S}_{2}$
and $\breve{S}_{2}$ respectively.  Using shooting method we have found
only two smooth static solutions to (\ref{eq:149}) (up to the
reflection symmetry $S \rightarrow -S$): the trivial solution $S(x)=1$
(vacuum) with vanishing YM curvature (\ref{eq:145}) and zero total
energy $E=0$ and the nontrivial configuration $S(x)=\cos{x}$ (kink)
with the total energy $E=3\pi/8$ (the solution found in
\cite{hosotani1984exact} for the Einstein-Yang-Mills system).  In the
following, without loose of generality, we assume $W(t,0)=1$.  These
static solutions separate the phase space of solutions of the equation
(\ref{eq:147}) into two topologically distinct sectors:\footnote{These
  should not be confused with the YM connection ansatz sectors.}
solutions with $W(t,0)=W(t,\pi)=1$ and solutions with
$W(t,0)=-W(t,\pi)=1$.  This means, that any solution to equation
(\ref{eq:147}) starting in one of the topological sectors must stay in
that sector during smooth time evolution.  We begin the analysis of
solutions for each of the topological sectors by solving the equation
governing linear perturbations around static solutions.

\subsection{Linear perturbations---the eigenvalue problem}
\label{sec:YMEigenvalue}

We consider smooth solutions of (\ref{eq:147}) of the form
\begin{equation}
  \label{eq:152}
  W(t,x) = S(x) + u(t,x),
\end{equation}
where $S(x)$ is one of the static solutions of (\ref{eq:149}) and
$u(t,x)$ is a perturbation fulfilling regularity conditions at both
poles of $\Sphere^{3}$, i.e. the $u(t,x)$ is even function of $x$ at
$x=0$ and $x=\pi$.  Substituting (\ref{eq:152}) into (\ref{eq:147}) we
obtain the evolution equation for the perturbation $u(t,x)$ of the
static solution $S(x)$
\begin{equation}
  \label{eq:153}
  \ddot{u} - u'' + \frac{3S^{2}-1}{\sin^{2}{x}}u
  + \frac{3S+u}{\sin^{2}{x}}u^{2}=0,
\end{equation}
which can be written in a following canonical form
\begin{equation}
  \label{eq:154}
  \ddot u + L u + f(u) = 0,
\end{equation}
and $L$ is a linear operator
\begin{equation}
  \label{eq:155}
  L = - \partial^{2}_{x} + \frac{3S^{2}-1}{\sin^{2}{x}},
\end{equation}
and $f(u)$ denotes the nonlinear part of (\ref{eq:153})
\begin{equation}
  \label{eq:156}
  f(u) = \frac{3S + u}{\sin^{2}{x}} u^{2}.
\end{equation}
The total energy (\ref{eq:148}) of (\ref{eq:152}) can be written as a
sum of two components
\begin{equation}
  \label{eq:157}
  E[S+u] = E[S] + E[u; S],
\end{equation}
the energy of static solution ($E[S=1]=0$ or $E[S=\cos{x}]=3\pi/8$)
and the energy associated with perturbation $u$.  The former one is
given by the integral
\begin{equation}
  \label{eq:158}
  E[u; S] = \int_{0}^{\pi}\left(\frac{1}{2}\dot u + \frac{1}{2} u' +
    \frac{3S^{2}-1}{2\sin^{2}{x}}u^{2} +
    u^{3}\frac{4S+u}{4\sin^{2}{x}}\right)\,\diff x.
\end{equation}

Next, we consider linear perturbations of $S$. Dropping the nonlinear
term $f(u)$ in (\ref{eq:154}) and performing separation of variables
we get the eigenvalue problem for essentially self-adjoint operator
$L$ on Hilbert space $L^{2}\left([0,\pi],\diff x\right)$ equipped with
the inner product
\begin{equation}
  \label{eq:159}
  \inner{f}{g} := \int_{0}^{\pi}f(x)g(x)\diff x.
\end{equation}
The eigenfunctions of $L$ are given by
\begin{equation}
  \label{eq:160}
  e_j(x) =
  \frac{(j+1)\sqrt{j(j+2)}\Gamma(j)}
  {2\sqrt{2}\Gamma\left(j+\frac{3}{2}\right)}
  \sin^{2}{x}\,P_{j-1}^{\left(3/2,\,3/2\right)}(\cos{x}),
  \quad j\in\mathbb{N},
\end{equation}
(both for $S=1$ and $S=\cos{x}$).  Alternatively (\ref{eq:160}) can be
written in a more compact form
\begin{equation}
  \label{eq:161}
  e_j(x) = \frac{\csc{x}}{\sqrt{2\pi}}
  \left(\sqrt{\frac{j+2}{j}}\sin(j x) -
    \sqrt{\frac{j}{j+2}}\sin((j+2)x)\right),
  \quad j\in\mathbb{N},
\end{equation}
which is particularly useful when implementing numerical
routines\footnote{Still the former can be applied by using, e.g.
  \prognamestyle{FORTRAN} subroutines available at
  \url{http://goo.gl/MV57oj}.}.  The corresponding eigenvalues are
\begin{equation}
  \label{eq:162}
  \omega_{j}^{2} =
  \begin{cases}
    (j+1)^2, & \text{for } S=1, \\
    (j+1)^2 - 3, & \text{for } S=\cos{x},
  \end{cases}
  \quad j\in\mathbb{N},
\end{equation}
(Note as opposed to other models we start the numbering of the
eigenfunctions and eigenvalues with $j=1$.)  Since for both cases the
eigenvalues $\omega_{j}^{2}$ are positive the corresponding static
solutions are linearly stable.

The eigenfunctions (\ref{eq:161}) for vacuum and kink static solutions
are identical, which is a direct consequence of the fact that the
linear operators (\ref{eq:155}) for both of the static solutions
commute.  The normalization constant in (\ref{eq:160}) and so in
(\ref{eq:161}) was chosen such that
$\inner{e_{i}}{e_{j}}=\delta_{ij}$.  The eigenfunctions (\ref{eq:161})
have exactly $(j-1)$ zeroes, they are alternately even and odd, with
respect to the equator of the three-sphere, for $j=1, 3, \ldots$ they
are even functions of argument $x$, whereas for $j=2,4,\ldots$ they
are odd functions, i.e. we have $e_j(x)=(-1)^{j+1}e_j(\pi-x)$.  Their
Taylor series expansions at the poles are, for $x=0$
\begin{equation}
  \label{eq:163}
  e_{j}(x) = (j+1)\sqrt{\frac{j(j+2)}{2\pi}} \Biggl(\frac{2}{3}x^{2}
  - \frac{1}{45}\bigl(3j(j+2)+1\bigr)x^{4}
  + \mathcal{O}\left(x^{6}\right) \Biggr),
\end{equation}
and similarly for $x=\pi$ using the symmetry of $e_{j}(x)$.  Therefore
the expansions of $e_{j}(x)$ conform with the regularity conditions
for the nonlinear problem (\ref{eq:153}) and whence can be used as an
expansion functions of smooth solutions in both the numerical and
perturbative calculations.

Another important feature concerns the derived spectrum of linear
perturbations.  The eigenvalues for vacuum and kink static solutions
correspond to nondispersive and dispersive cases respectively.  In the
former case, frequencies of the linear problem are equally spaced
integer numbers $\omega_j=j+1$ (starting with $\omega_{1}=2$).  For
the kink static solution, frequencies are only asymptotically
equidistant, i.e. for $j\rightarrow\infty$ we have $\omega_j = j + 1
- 3/(2j) + \mathcal{O}\left(j^{-2}\right)$.  This difference has a
direct consequence on the nonlinear dynamics, which will be a subject
for the subsequent sections of this thesis.

\part{Studies}
\label{part:Studies}
\chapter{Turbulence, resonances and (in)stability}
\label{cha:Turbulence}

In this chapter we consider generic behaviour of nonlinear waves
propagating on bounded domains.  We focus on the problem how the
dispersive and nondispersive spectrum of linear perturbations affects
the nonlinear dynamics.  We investigate this issue by studying models
of Sections~\ref{sec:BoxModel} and \ref{sec:YMModel}, which give us a
possibility to change the character of the eigenvalues by considering
different boundary conditions (the scalar field case) or by
considering perturbations in different topological sectors (the YM
field model).

In addition we give the details of numerical methods used to solve the
evolution equations.  For the YM model we present the results of
perturbative methods which serve as a starting point in the
construction of time-periodic solutions.  We point out that the
nondispersive spectrum does not forbid the resonances to occur.

The first part of this chapter is based mainly on the paper
\cite{mPRL} and in part on more recent work presented in
\cite{MR2014}.

\section{Spherical cavity model}
\label{sec:BoxTurbulence}

Defining a new metric function $B(t,r)$ by the relation\footnote{This
  should not be confused with the squashing field of
  Section~\ref{sec:BCSModel}.}
\begin{equation}
  \label{eq:164}
  B := A e^{-\delta},
\end{equation}
the field equations (\ref{eq:118})-(\ref{eq:120}) can be rewritten as
\begin{align}
  \label{eq:165}
  \delta' &= - r\left( \Phi^{2} + \Pi^{2} \right),
  \\
  \label{eq:166}
  \left(r\,B\right)' &= e^{-\delta},
  \\
  \label{eq:167}
  \dot{\Phi} &= (B\,\Pi)',
  \\
  \label{eq:168}
  \dot{\Pi} &= r^{-2}\left(r^{2}B\,\Phi\right)',
\end{align}
(we skip here the momentum constraint (\ref{eq:121}) since we are
using the constrained evolution scheme).  It was emphasized in
\cite{mPRL} that this form of equations greatly reduces the complexity
of a numerical algorithm and is particularly useful for the numerical
integration of the Hamiltonian constraint (\ref{eq:120}).

In the following we present two approaches used to solve the
initial-boundary value problem for the system
(\ref{eq:165})-(\ref{eq:168}).  The first one, based on the finite
difference approximation (FDA), is used to investigate the behaviour
of solutions starting from generic initial conditions---for which the
solutions develop huge gradients and an apparent horizon forms---as
was discussed in \cite{mPRL} (to be precise, this is a slightly
modified and improved version of the code that was used in
\cite{mPRL}).  The second one, using pseudospectral discretization,
has an advantage of spectral convergence when the solution stays
smooth during time evolution.  This approach was used to solve the
initial-boundary value problem for the system
(\ref{eq:165})-(\ref{eq:168}) for the first time by the author in
\cite{MR2014}; it is also a core of the numerical algorithm applied to
construct time-periodic solutions, which is described in
Section~\ref{sec:BoxPeriodicNumeric}.

\subsection{Numerical evolution scheme}
\label{sec:BoxEvolution}

\subsubsection{Finite difference method}
\label{sec:BoxFDA}

We take a numerical grid for the radial coordinate $r\in[0,1]$ with
$N$ equally spaced points (grid nodes)
\begin{align}
  \label{eq:169}
  r_{i} = (i-1)h, \quad i=1,2,\ldots, N,
\end{align}
where $h = 1/(N-1)$ is the grid spacing constant.  We discretize the
equations (\ref{eq:165})-(\ref{eq:168}) with finite difference method
of fourth order in grid spacing $h$
\cite{doi:10.1137/1.9780898717839}.  In this way we obtain a system of
$2N$ coupled ODEs for the $\phi_{i}(t)\equiv\phi(t,r_{i})$,
$\Pi_{i}(t)\equiv\Pi(t,r_{i})$, $i=1,\ldots,N$ dependent variables
(the discretized version of equations (\ref{eq:167}), (\ref{eq:168})),
with $t$ being a continuous independent variable, subject to a
discrete version of elliptic constraint equations (discussed below).
Resulting system is solved by using 'standard' numerical algorithms
for the integration of ODEs; This is commonly known the MOL approach
(or semi-discretization \cite{iserles2009first}).  Concerning the
Einstein equations, the problem we are facing, we use the constrained
evolution scheme \cite{AlcubierreBook}.  Instead of using the momentum
constraint to update the metric function $A(t,r)$, we solve the
Hamiltonian constraint, which is particularly advantageous in
spherical symmetry when using polar-areal coordinates.  This, together
with a need to solve the slicing condition, implies that for any
numerical method used to advance solution in time the constraint
equations have to be solved very often, precisely at each intermediate
step of the time integration algorithm.  This is the reason why the
Einstein equations, viewed as a coupled system of hyperbolic and
elliptic PDEs, are very expensive to solve and with no symmetry
assumptions a free evolution scheme is commonly used (though
application of advanced techniques like multi-grid
methods\footnote{Which should not be confused with adaptive mesh
  refinement techniques so successful in numerical relativity see,
  e.g. \cite{Pretorius2006246}.} may substantially reduce the
complexity of solving the constraint equations
\cite{10.1007/BF00770203}, these require much more effort to be
adopted and implemented to the problems at hand than the use of free
evolution).  Therefore any feasible enhancement of the algorithm
solving the constraints will result in a significant gain of
performance of the overall algorithm used to solve the time dependent
Einstein equations.  This is the reason why we prefer to solve the
system (\ref{eq:165})-(\ref{eq:168}) instead of
(\ref{eq:118})-(\ref{eq:121}).

A discrete version of the slicing condition (\ref{eq:165}) together
with the boundary condition $\delta(t,0)=0$ (\ref{eq:124}) reads
\begin{align}
  \label{eq:170}
  \delta_{1} &= 0,
  \\
  \label{eq:171}
  \sum_{j=1}^{N}D^{(1)}_{ij}\delta_{j} &= - r_{i} \left(\Phi_{i}^{2} +
    \Pi_{i}^{2}\right), \quad i=2,\ldots,N,
\end{align}
(to simplify the notation in the following we drop an explicit time
dependence of the scalar field and metric functions).  The $D^{(1)}$
is the first order FDA derivative operator (whose explicit form is
given below).  For a given matter content, represented by the two
vectors of length $N$ in the FDA representation ($\Phi_{i}$ and
$\Pi_{i}$, $i=1,\ldots,N$), the vector representing metric function
$\delta(t,r)$ ($\delta_{i}$, $i=1,\ldots,N$) is then given as a
solution to the linear algebraic equation with the banded main matrix
\begin{equation}
  \label{eq:172}
  \frac{1}{12h}\begin{pmatrix}
    12h \\
    -8 & 1 & 8 & -1 \\
    1 & -8 & 0 & 8 & -1 \\
    & 1 & -8 & 0 & 8 & -1 \\
    & & \ddots & \ddots & \ddots & \ddots & \ddots &  \\
    & & & 1 & -8 & 0 & 8 & -1 \\
    & & & -1 & 6 & -18 & 10 & 3 \\
    & & & 3 & -16 & 36 & -48 & 25
  \end{pmatrix}.
\end{equation}
This system is solved using banded version of the LU factorization
algorithm \cite{iserles2009first}.  Since the matrix (\ref{eq:172}) is
constant (time-independent), the factorization is performed only once,
during the initialization, and the cost of solving the system
(\ref{eq:170}), (\ref{eq:171}) is $O\left(20N\right)$ (not including
computing the RHS), while the factorization cost is between $O(24N)$
and $O(56N)$.\footnote{The factorization and then the solution
  procedure are carried by referencing to the \prognamestyle{LAPACK}
  routines \codenamestyle{gbtrf} and \codenamestyle{gbtrs}
  respectively \cite{laug}.}  In a very similar way we solve the
Hamiltonian constraint (\ref{eq:166}) where $\delta$ plays the role of
a source (whence the slicing condition has to be solved first).  The
discrete version of this equation take the form of the algebraic
system (after performing differentiation on the LHS, which saves an
additional $N$ floating-point division operations)
\begin{align}
  \label{eq:173}
  B_{1} &= 1,
  \\
  \label{eq:174}
  \sum_{j=1}^{N}\left(\id_{ij} + r_{i}D^{(1)}_{ij}\right)B_{j} &=
  e^{-\delta_{i}}, \quad i=2,\ldots,N,
\end{align}
where $\id_{ij}$ is the $(i,j)$-th element of the identity
matrix.  The main matrix of this system is
\begin{equation}
  \label{eq:175}
  {\small
  \begin{pmatrix}
    1 \\[1ex]
    -\frac{2}{3} & \frac{1}{12} + 1 & \frac{2}{3} & -\frac{1}{12} \\[1ex]
    \frac{1}{6} & -\frac{4}{3} & 1 & \frac{4}{3} & -\frac{1}{6} \\[1ex]
    & \frac{1}{4} & -2 & 1 & 2 & -\frac{1}{4} \\[1ex]
    & & \ddots & \ddots & \ddots & \ddots & \ddots &  \\[1ex]
    & & & \frac{N-3}{12} & -\frac{2(N-3)}{3} & 1 & \frac{2(N-3)}{3} &
    -\frac{N-3}{12} \\[1ex]
    & & & -\frac{N-2}{12} & \frac{N-2}{2} & -\frac{3(N-2)}{2} &
    \frac{5(N-2)}{6} + 1 & \frac{N-2}{4} \\[1ex]
    & & & \frac{N-1}{4} & -\frac{4(N-1)}{3} & 3(N-1) & -4(N-1) &
    \frac{25(N-1)}{12} + 1
  \end{pmatrix}.
  }
\end{equation}
This again is solved with the use of the LU factorization, so the
complexity of the algorithm to compute a solution to the Hamiltonian
constraint is of the same order as for solving the slicing condition.
This should be compared with the cost of solving (\ref{eq:120})
directly where the resulting main matrix (after performing the FDA
discretization), being time-dependent through its dependence on the
scalar field, would have to be factorized at each time which would at
least double an overall complexity.

The RHS of the FDA discretized evolution equations (\ref{eq:167}) and
(\ref{eq:168}) are written in terms of the $\phi$ field instead of its
spatial derivative, the $\Phi$ field introduced in (\ref{eq:117}).
After performing dozens of numerical experiments, using different
schemes imposing the boundary conditions at the cavity, we found that
the problem of solving (\ref{eq:165})-(\ref{eq:168}) or equivalently
(\ref{eq:118})-(\ref{eq:120}) is unstable when $\Phi$ is used instead
of $\phi$ (this is also the case for a free wave equation written in
terms of corresponding quantities to $\Phi$ and $\Pi$).  For that
reason, in the FDA approach we are forced to evolve in time the scalar
field $\phi$ instead of its gradient.

At the interior of the grid, i.e. for the grid points
$i=2,\ldots,N-1$ the discrete version of (\ref{eq:167}) and
(\ref{eq:168}) reads
\begin{align}
  \label{eq:176}
  \dot\phi_{i} &= B_{i}\Pi_{i},
  \\
  \label{eq:177}
  \dot\Pi_{i} &= B_{i} \left(D^{(2)}\phi\right)_{i} + \left( B_{i} +
    e^{-\delta_{i}}\right) \frac{\left(D^{(1)}\phi\right)_{i}}{r_{i}}.
\end{align}
For the node $i=1$ located at the origin $r=0$, where equation
(\ref{eq:168}) is singular, we calculate the RHS using l'Hopital's
rule which together with regularity conditions (\ref{eq:124}) gives
\begin{align}
  \label{eq:178}
  \dot\phi_{1} &= \Pi_{1},
  \\
  \label{eq:179}
  \dot\Pi_{1} &= 3 \left(D^{(2)}\phi\right)_{1},
\end{align}
At the outer boundary $r=1$ the Dirichlet condition $\phi(t,1)=0$ is
straightforward to impose by setting
\begin{align}
  \label{eq:180}
  \dot\phi_{N} &= 0,
  \\
  \label{eq:181}
  \dot\Pi_{N} &= 0,
\end{align}
A stable scheme for the Neumann boundary condition $\phi'(t,1)=0$,
which does not produce any spurious oscillations, is
\begin{align}
  \label{eq:182}
  \dot\phi_{N} &= \frac{1}{25} \left( -3\dot\phi_{N-4} +
    16\dot\phi_{N-3} - 36\dot\phi_{N-2} + 48\dot\phi_{N-1} \right),
  \\
  \label{eq:183}
  \dot\Pi_{N} &= B_{N} \left(D^{(2)}\phi\right)_{N},
\end{align}
which we derive from the condition $\left(D^{(1)}\phi\right)_{N}=0$,
by taking its time derivative and then solving for $\dot{\phi}_{N}$.

To filter out high frequencies, inevitably present in the FDA
discretization approach, the Kreiss-Oliger type artificial dissipation
\cite{kreiss1973methods} (see also \cite{AlcubierreBook, lrr-2012-9}
for more details) is added to the RHS of the dynamical equations
\begin{align}
  \label{eq:184}
  \dot\phi_{i} &\ra \dot\phi_{i} - \epsilon_{d}
  \left(Q_{d}\phi\right)_{i},
  \\
  \label{eq:185}
  \dot\Pi_{i} &\ra \dot\Pi_{i} - \epsilon_{d}
  \left(Q_{d}\Pi\right)_{i},
\end{align}
at the $i=1,\ldots,N-3$ grid points.  We deliberately do not modify
the time derivatives at the last three grid points ($i=N-2,N-1,N$) in
order not to affect imposed boundary conditions.  The correct choice
of the order of the dissipation operator $Q_{d}$ and its strength, the
free parameter $\epsilon_{d}$, which from linear stability analysis
has to be $0\leq\epsilon_{d}<1$, ensures that the order of accuracy of
the FDA scheme is not affected.  In practice, it was sufficient to set
$\epsilon_{d}=0.01$ or $\epsilon_{d}=0.1$ in this problem.\footnote{In
  general whenever larger values of $\epsilon_{d}$ are needed to
  stabilize the evolution this signals that the discretization scheme
  is unstable. In that case the resolution is to change the scheme
  rather than to increase the strength of dissipation.}  Due to the
used scheme (especially (\ref{eq:173}) and (\ref{eq:174})) and the
sparsity of discrete FDA operators the overall cost of computing RHS
of evolution equations (\ref{eq:167}) and (\ref{eq:168}) is $O(N)$.
Since for large and moderate amplitudes of initial perturbations high
gradients appear and finally the black hole forms we take very large
number of grid points, usually $N$ is of order $2^{14}-2^{16}$, to
resolve fine structures of the solution.  Therefore the use of an
energy conserving time-integration method, the Gauss-Runge-Kutta
method (discussed in Appendix~\ref{sec:AppIRK}), would greatly affect
the overall performance.  Such method requires a solution to the
nonlinear algebraic system of equations of size $O\left(N^{2}\right)$,
which for such large grids would be a significant additional cost.
Hence, we prefer to use an explicit method and the time-integration of
evolution equations is done using an adaptive---self adjusting time
step size, explicit Runge-Kutta-Dormand-Prince algorithm
\cite{Dormand198019} (see also the Appendix~\ref{sec:AppERK}).  For
stability reasons time step is of order $1/N$, thus an overall
complexity of the evolution scheme (cost of integrating the equations
per unit time interval) is $O\left(N^{2}\right)$.  In the following
sections we present results of energy conservation tests confirming
our choice and robustness of this approach.

The fourth order finite difference operators used in (\ref{eq:171}),
(\ref{eq:172}), (\ref{eq:174}), (\ref{eq:175}), (\ref{eq:179}),
(\ref{eq:177}), (\ref{eq:182}) and (\ref{eq:183}) are constructed to
utilize symmetry of differentiated function at the origin $r=0$
\begin{equation}
  \label{eq:186}
  f(-r)=f(r),
\end{equation}
i.e. whenever the symmetric stencils applied at the grid points near
the origin involve the function values with negative indices, these
are replaced with corresponding values with positive indices
\begin{equation}
  \label{eq:187}
  f(-jh) = f(jh)\ \Rightarrow\ f_{-j+1} = f_{j+1}, \quad j=1,2,\ldots\,.
\end{equation}
These FD operators are listed below: the first order derivative
\begin{align}
  & \left(D^{(1)}f\right)_{1} &=&\ \ \ 0,
  \\[1ex]
  & \left(D^{(1)}f\right)_{2} &=&\ \ \ \frac{1}{12h}\Bigl( - 8f_{1} + f_{2} +
    8f_{3} - f_{4} \Bigr),
  \\[1ex]
  & \left(D^{(1)}f\right)_{i} &=&\ \ \ \frac{1}{12h}\Bigl( f_{i-2} - 8f_{i-1}
    + 8f_{i+1} - f_{i+2} \Bigr),
  \\[1ex]
  & \left(D^{(1)}f\right)_{N-1}\!\!\!\! &=&\ \ \ \frac{1}{12h}\Bigl( - f_{N-4} +
    6f_{N-3} - 18f_{N-2} + 10f_{N-1} + 3f_{N} \Bigr),
  \\[1ex]
  & \left(D^{(1)}f\right)_{N} &=&\ \ \ \frac{1}{12h}\Bigl( 3f_{N-4} -
    16f_{N-3} + 36f_{N-2} - 48f_{N-1} + 25f_{N} \Bigr),
\end{align}
and the second order derivative
\begin{align}
  & \left(D^{(2)}f\right)_{1} &=&\ \ \ \frac{1}{12h^{2}}\Bigl(
  -30f_{1} + 32f_{2} -2f_{3} \Bigr),
  \\[1ex]
  & \left(D^{(2)}f\right)_{2} &=&\ \ \ \frac{1}{12h^{2}}\Bigl( 16f_{1}
  - 31f_{2} + 16f_{3} - f_{4} \Bigr),
  \\[1ex]
  & \left(D^{(2)}f\right)_{i} &=&\ \ \ \frac{1}{12h^{2}}\Bigl( -f_{i-2} +
  16f_{i-1} - 30f_{i} + 16f_{i+1} - f_{i+2} \Bigr),
  \\[1ex]
  & \left(D^{(2)}f\right)_{N-1}\!\!\!\! &=&\ \
  \begin{multlined}[t]
    \frac{1}{12h^{2}}\Bigl( f_{N-5} - 6f_{N-4} + 14f_{N-3} - 4f_{N-2}
    \\
    \qquad\qquad\qquad\qquad - 15f_{N-1} + 10f_{N} \Bigr),
  \end{multlined}
  \\[1ex]
  & \left(D^{(2)}f\right)_{N} &=&\ \
  \begin{multlined}[t]
    \frac{1}{12h^{2}}\Bigl( -10f_{N-5} + 61f_{N-4} - 156f_{N-3} +
    214f_{N-2}
    \\
    - 154f_{N-1} + 45f_{N} \Bigr),
  \end{multlined}
\end{align}
where $i=3,\ldots,N-2$.  At the outer boundary $r=1$ we use
nonsymmetric stencils, taking into account more grid points than for
the symmetric stencils, to retain an overall fourth order of
convergence.  The sixth order dissipation operator $Q_{d}$ (see
e.g. \cite{AlcubierreBook}), also derived for use with symmetric
function (\ref{eq:186})-(\ref{eq:187}) reads
\begin{align}
  & \left(Q_{d}f\right)_{1}&=&\ \ \frac{1}{h}\Bigl( 20f_{1} - 30f_{2} +
  12f_{3} - 2f_{4} \Bigr),
  \\[1ex]
  & \left(Q_{d}f\right)_{2} &=&\ \ \frac{1}{h}\Bigl( -15f_{1} + 26f_{2} -
  16f_{3} + 6f_{4} - f_{5} \Bigr),
  \\[1ex]
  & \left(Q_{d}f\right)_{3} &=&\ \ \frac{1}{h}\Bigl( 6f_{1} - 16f_{2} +
  20f_{3} - 15f_{4} + 6f_{5} - f_{6} \Bigr),
  \\[1ex]
  & \left(Q_{d}f\right)_{i} &=&\ \
  \begin{multlined}[t]
    \frac{1}{h}\Bigl( -f_{i-3} + 6f_{i-2} - 15f_{i-1} + 20f_{i}
    \\
    \qquad\qquad\qquad\qquad - 15f_{i+1} + 6f_{i+2} - f_{i+3} \Bigr),
  \end{multlined}
  \\[1ex]
  & \left(Q_{d}f\right)_{N-2}\!\!\!\! &=&\ \
  \begin{multlined}[t]
    \frac{1}{h}\Bigl( f_{N-7} - 8f_{N-6} + 27f_{N-5} - 50f_{N-4} +
    55f_{N-3}
    \\
    - 36f_{N-2} + 13f_{N-1} - 2f_{N} \Bigr),
  \end{multlined}
  \\[1ex]
  & \left(Q_{d}f\right)_{N-1}\!\!\!\! &=&\ \
  \begin{multlined}[t]
    \frac{1}{h}\Bigl(\, 2f_{N-7} - 15f_{N-6}
    + 48f_{N-5} - 85f_{N-4} + 90f_{N-3}
    \\
    - 57f_{N-2} + 20f_{N-1} -3f_{N} \Bigr),
  \end{multlined}
  \\[1ex]
  & \left(Q_{d}f\right)_{N} &=&\ \
  \begin{multlined}[t]
    \frac{1}{h} \Bigl(\, 3f_{N-7} - 22f_{N-6} + 69f_{N-5} - 120f_{N-4}
    + 125f_{N-3}
    \\
    - 78f_{N-2} + 27f_{N-1} - 4f_{N} \Bigr),
  \end{multlined}
\end{align}
where $i=4,\ldots,N-3$ (even though we do not use them, we list the
$N-2$, $N-1$, $N$ schemes for completeness).
\subsubsection{Pseudospectral method}
\label{sec:BoxCheb}

To discretize the system of equations (\ref{eq:165})-(\ref{eq:168}) we
use the Chebyshev pseudospectral method adapted to spherical symmetry,
which is described in detail in Appendix~\ref{cha:polyn-pseud-meth}.
We take $N+1$ radial Chebyshev points, given in Eq.~(\ref{eq:635}) and
evolve in time the values of the dynamical fields at the grid nodes;
we use the following notation $\Phi_{i}(t)\equiv\Phi(t,r=x_{i})$ and
$\Pi_{i}(t)\equiv\Pi(t,r=x_{i})$ for $i=0,1,\ldots,N$ (where we drop
the time dependence for convenience).  As in the FDA approach, the
constraint equations, after performing disretization in space, become
algebraic equations for the function values at the grid points.  The
discrete version of the slicing condition (\ref{eq:165}) is then
\begin{equation}
  \label{eq:188}
  \sum_{j=0}^{N}D_{ij}^{(1,+)}\delta_{j} =
  -x_{i}\left(\Phi_{i}^{2} + \Pi_{i}^{2}\right), \quad i=1,\ldots,N,
\end{equation}
while the remaining equation ($i=0$) is the gauge condition
$\delta(t,0)=0$ which with use of (\ref{eq:641}) simply is
\begin{equation}
  \label{eq:189}
  \sum_{i=0}^{N}\frac{w_{i}}{x_{i}}\delta_{i} = 0,
\end{equation}
where $w_{i}$ are associated weights to $x_{i}$ in the barycentric
representation of interpolating polynomial (see
\cite{doi:10.1137/S0036144502417715} and the discussion in the
Appendix~\ref{cha:polyn-pseud-meth}).  Similarly, a discrete version
of the Hamiltonian constraint equation (\ref{eq:166}) is
\begin{equation}
  \label{eq:190}
  \sum_{j=0}^{N}\left(\id_{ij} + x_{i}D_{ij}^{(1,+)}\right)B_{j}
  = e^{-\delta_{i}}, \quad i=1,\ldots,N,
\end{equation}
together with the discrete boundary condition $B(t,0)=1$ given as
\begin{equation}
  \label{eq:191}
  \frac{{\displaystyle\sum_{i=0}^{N}\frac{w_{i}}{x_{i}}B_{i}}}
  {\displaystyle\sum_{i=0}^{N}\frac{w_{i}}{x_{i}}} = 1.
\end{equation}
For both, the slicing condition and the Hamiltonian constraints, the
resulting algebraic equations are solved using the LU factorization as
for the FDA case.\footnote{Here the appearing differentiation matrices
  are dense so the proper \prognamestyle{LAPACK} routines,
  \codenamestyle{getrf} and \codenamestyle{getrs}, used for
  factorization and for solving the resulting system perform
  $O(2/3N^{3})$ and $O(2N^{2})$ floating-point operations
  respectively.}  The pseudospectral approximation to the dynamical
equations (\ref{eq:167}) and (\ref{eq:168}) is
\begin{align}
  \label{eq:192}
  \dot{\Phi}_{i} &= \sum_{j=0}^{N}D_{ij}^{(1,+)}B_{j}\Pi_{j},
  \\
  \label{eq:193}
  \dot{\Pi}_{i} &= B_{i}\sum_{j=0}^{N}D_{ij}^{(1,-)}\Phi_{j} +
  \left(B_{i} + e^{-\delta_{i}}\right)\frac{\Phi_{i}}{x_{i}},
\end{align}
for $i=0,\ldots,N$.  Thus the overall cost of computing the RHSs of
dynamical equations is $O(N^{2})$.  The boundary condition at $r=1$ is
imposed by replacing the $i=0$ equation for one of the dynamical
variables by setting either $\Pi_{0}=0$ for Dirichlet boundary
condition or by $\Phi_{0}=0$ for Neumann boundary condition
respectively.

Because of the stiffness of the resulting ODE system (the property of
Chebyshev pseudospectral method) we are forced to use an implicit
time-integration algorithm here \cite{hesthaven2007spectral,
  hairer1996solving}.  The implicit methods are more costly per time
step than the explicit ones, but their stability properties put less
stringent restrictions of the magnitude of integration step
\cite{hairer1996solving} (thus it suffices to take time step size of
order $O(1/N)$ with implicit RK methods, while explicit RK restricts
integration step to $O(1/N^2)$).  Also due to the spectral convergence
of the spatial discretization, usually we do not need to take $N$ to
be large (in practice not greater than $2^{8}$) in order to provide an
accurate approximation to the solution for this problem, at least for
smooth solutions.  For ODEs of moderate sizes the use of an implicit
time integrator does not cause significant increase of computational
cost compared to the cost of computing the RHS of the equations. In
addition, using the implicit method we can benefit from applying the
symplectic, energy conserving algorithm, such as the Gauss-RK method
discussed in Appendix~\ref{sec:AppIRK}, for the time integration.

\subsection{Dirichlet boundary condition}
\label{sec:BoxDirichlet}

\begin{figure}[!t]
  \centering
  \includegraphics[width=\swidth]
  {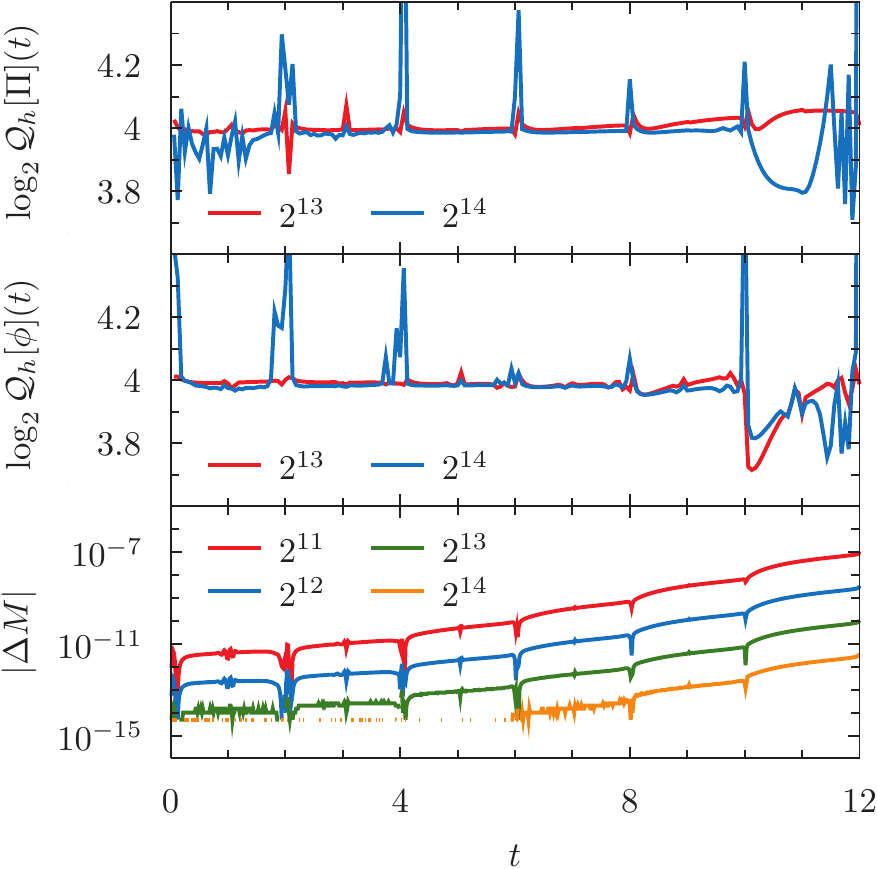}
  \caption{The results of convergence and mass conservation tests of
    the FDA method of Section~\ref{sec:BoxFDA}, derived for the
    Dirichlet boundary condition at the cavity ($\phi(t,1)=0$) and the
    initial data (\ref{eq:194}) with the amplitude $\ep=16$.  The
    curves are labelled by values of $h^{-1}$ for grid spacing $h$.
    For both tests we use the fixed time-stepping integration with
    step size $\Delta t=h/4$.  \textit{Top and middle panels}.  The
    convergence factor (defined in Eq.~(\ref{eq:195}) and computed
    with respect to discrete $\ell_{2}$ norm) is close to the expected
    value for the fourth order FDA scheme employed here (the fifth
    order accurate time-integration algorithm does not improve the
    convergence since an overall error is dominated by the lower order
    spatial discretization).  The variation of convergence factor for
    late times, before the black hole formation at
    $\mbox{$t\simeq{}12$}$, is caused by the significant errors in
    approximation of functions with huge gradients.  \textit{Bottom
      panel}.  The total mass conservation test, the absolute
    difference $\mbox{$\Delta{}M:=M(t)-M(0)$}$, for the same initial
    and boundary data.  The continuous loss of the mass (the
    difference is negative) for low resolution runs is reduced when
    denser grids are used.  For the highest resolution shown here,
    $h=2^{-14}$, for early times $\mbox{$t\simeq{}5$}$ the total mass
    is conserved up to the machine precision.}
  \label{fig:BoxDirichletConvergence}
\end{figure}

Numerical results presented below were generated from Gaussian-type
initial data of the form\footnote{The same as the one used in
  \cite{mPRL}.  There is a typo in the width of the Gaussian, $32$
  should be replaced with $64$, in the Eq.~(13) in \cite{mPRL} to
  reproduce results of that paper.}
\begin{equation}
  \label{eq:194}
  \Phi(0,r)=0,\quad \Pi(0,r)=\varepsilon \exp\left( -64
    \tan^{2}\frac{\pi}{2} r \right).
\end{equation}
These initial data vanish exponentially as $r\ra 1$ so compatibility
conditions (\ref{eq:126}) or (\ref{eq:127}) are not an issue.  The
convergence test for the FDA method where performed together with the
total mass (\ref{eq:123}) conservation test.  The results for the
initial data (\ref{eq:194}) are presented in
Fig.~\ref{fig:BoxDirichletConvergence} where we plot the convergence
factor which for a quantity $f$ is defined as
\begin{equation}
  \label{eq:195}
  \mathcal{Q}_{h}[f] :=
  \frac{\left\|f^{(4h)}-f^{(2h)}\right\|}{\left\|f^{(2h)}-f^{(h)}\right\|},
\end{equation}
where by $f^{(h)}$ we mean the function $f$ approximated on a FDA grid
with the spacing $h$.  For convergent FD symmetric scheme of order $p$
the Richardson approximation states that $\mathcal{Q}_{h}[f]=2^{p}$ in
the limit $h\ra 0$.

\begin{figure}[p]
  \includegraphics[width=\lwidth]
  {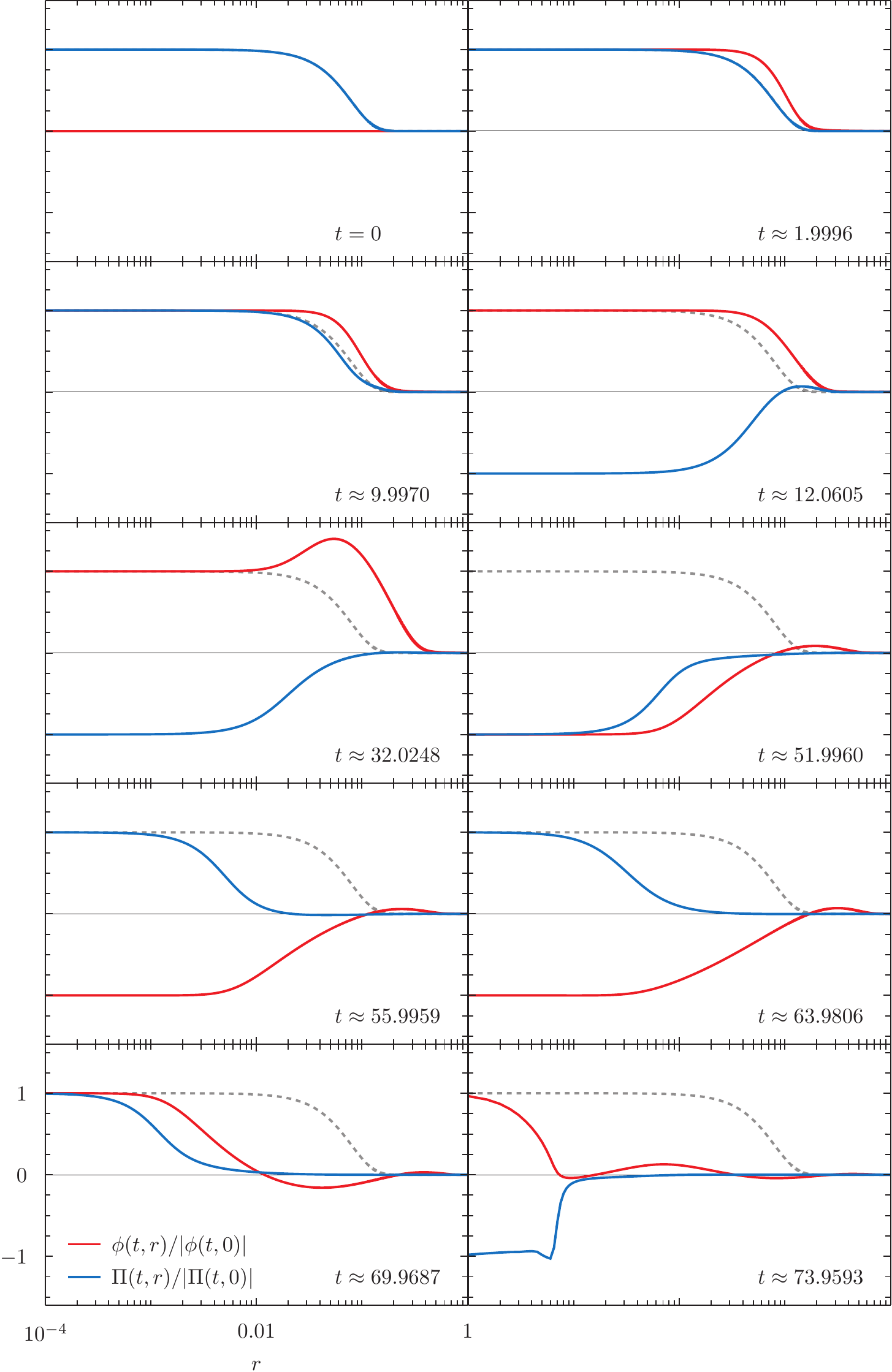}
  \caption{The snapshots from time-evolution of initial data
    (\ref{eq:194}) with $\ep=8$ (gray dashed line) for which the
    collapse occurs at $t_{AH}\approx 73.9503$ (the last frame).  To
    suppress the varying amplitude of the scalar fields we normalize
    $\phi(t,r)$ (red lines) and $\Pi(t,r)$ (blue lines) by dividing
    them by their values at the origin (the magnitude $\Pi(t,r=0)$
    ranges from $8$ at $t=0$ up to $\sim 4212$ at $t=t_{AH}$).  Note
    the logarithmic scale on the horizontal axis, and the successive
    narrowing of the scalar field pulse with increasing time.}
  \label{fig:BoxDirichletEvolutionMulti}
\end{figure}

\begin{figure}[t]
  \centering
  \includegraphics[width=\swidth]{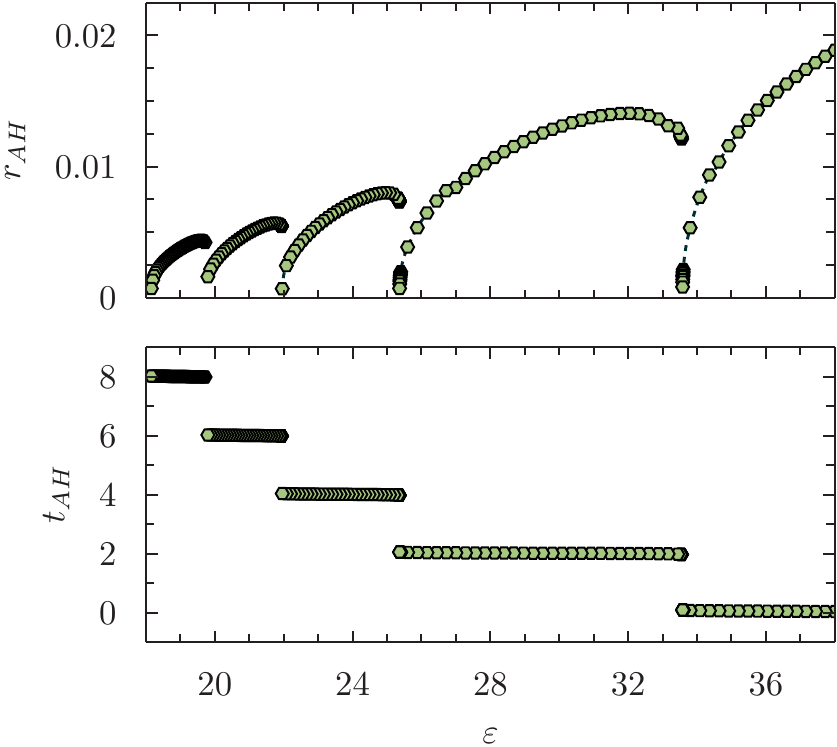}
  \caption{An apparent horizon radius $r_{AH}$ and a corresponding
    formation time $t_{AH}$ as a function of the amplitude of initial
    data (\ref{eq:194}). At critical points
    $\lim_{\ep\rightarrow\ep_{n}^+} r_{AH}(\ep)=0$, while the horizon
    formation time exhibits jumps of size
    $t_{AH}(\ep_{n+1})-t_{AH}(\ep_{n})\approx 2$ (time in which the
    pulse traverses the cavity back and forth).}
  \label{fig:BoxTeeth}
\end{figure}

\begin{figure}[!th]
  \centering
  \includegraphics[width=\swidth]
  {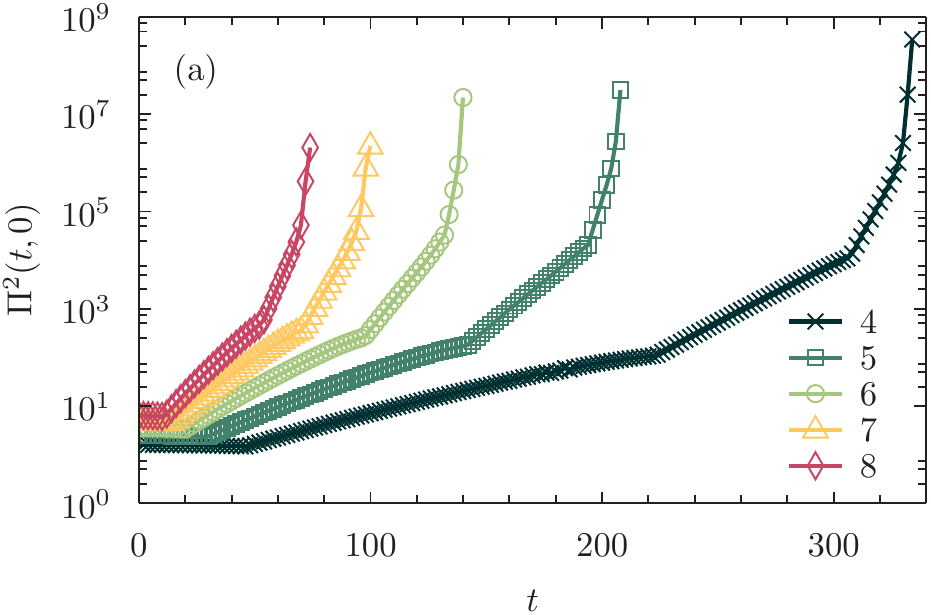}
  \\[2.5ex]
  \includegraphics[width=\swidth]
  {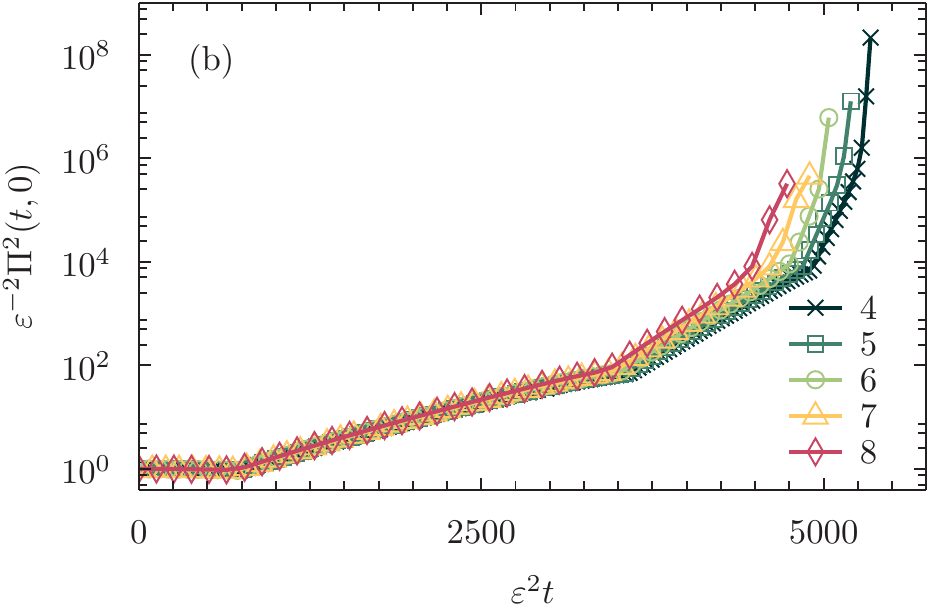}
  \caption{\textit{Top panel}.  The function $\Pi^{2}(t,0)$ for
    solutions with initial data (\ref{eq:194}) for several moderately
    small amplitudes. For clarity of the plot only the envelopes of
    rapid oscillations are depicted. \textit{Bottom panel}.  The
    curves from the plot top panel scaled according to
    $\ep^{-2}\Pi^{2}(\ep^{2} t,0)$. Plotted curves are labelled by the
    value of initial data amplitude $\ep$.}
  \label{fig:BoxRicci}
\end{figure}

The results are very similar to those of \cite{br}, as can be seen by
comparing Figs.~\ref{fig:BoxTeeth} and \ref{fig:BoxRicci} with the
analogues figures in \cite{br}.  For large amplitudes the evolution is
not affected by the mirror; the wave packet rapidly collapses, forming
an apparent horizon at a point where the metric function $A(t,r)$ goes
to zero.  However, a wave packet which is marginally too weak to form
a horizon on the first implosion, does so on the second implosion
after being reflected back by the mirror.  As in the AdS case, this
leads to a sequence of critical amplitudes $\ep_n$ for which the
solutions, after making $n$ bounces, asymptote locally Choptuik's
critical solution (see Fig.~\ref{fig:BoxTeeth}).  The snapshots of the
time-evolution showing continuous narrowing of scalar field profiles
with increasing time is shown on
Fig.~\ref{fig:BoxDirichletEvolutionMulti} (for $\ep=8$ in
(\ref{eq:194}) the black hole is formed after 36 reflections).
To track the steepening of the wave packet for very small amplitudes,
we follow \cite{br} and monitor the Ricci scalar at the origin, i.e.
we plot $R(t,0)=-2\Pi(t,0)^{2}$.  This function oscillates with
approximate period $2$ (for the cavity of unit radius).  Initially,
the amplitude stays almost constant but after some time it begins to
grow exponentially and eventually a horizon forms (see
Fig.~\ref{fig:BoxRicci}(a)).  As shown in Fig.~\ref{fig:BoxRicci}(b),
the time of onset of exponential growth $T$ scales with the amplitude
of initial data as $T\sim\ep^{-2}$, which indicates that arbitrarily
small perturbations (for which it is impossible numerically to track
the formation of a horizon) eventually start growing.

In \cite{br} the numerical results were corroborated by a nonlinear
perturbation analysis which demonstrated that the instability of AdS
is caused by the resonant transfer of energy from low to high
frequencies.  For the problem at hand, as in AdS, the spectrum is
fully resonant (that is, the frequencies $\omega_{j}$ are rational
multiples of one another), so the entire perturbation analysis of
\cite{br} can be formally repeated in our case.  We say 'formally'
because, in contrast to the AdS case, the eigenmodes (\ref{eq:130})
violate the compatibility conditions at $r=1$ (see
Eqs.~(\ref{eq:126}), the same holds for Neumann boundary conditions
where also Eqs.~(\ref{eq:127}) are not satisfied by the eigenmodes)
and therefore they cannot be taken as smooth initial data.

\begin{figure}[!th]
  \centering
  \includegraphics[width=\swidth]
  {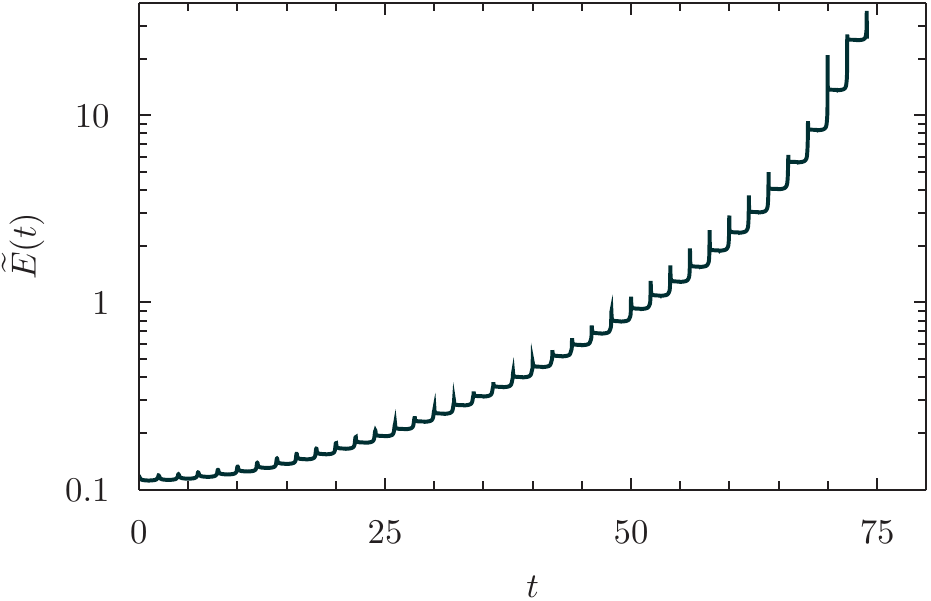}
  \caption{Plot of the weighted energy norm $\widetilde E(t)$
    (\ref{eq:194}) for the solution with initial amplitude
    $\ep=8$.  The steep bursts of growth occur when the pulse implodes
    through the center.}
  \label{fig:BoxSobolevNorm}
\end{figure}

The transfer of energy to higher modes (which is equivalent to the
concentration of energy on smaller scales) can be quantified by
monitoring the energy contained in the linear modes
\begin{equation}
  \label{eq:196}
  E_{j}=\Pi_{j}^{2}+\omega_{j}^{-2}\Phi_{j}^{2}\,,
\end{equation}
where $\Phi_{j}:=\inner{\sqrt{A}\,\Phi}{e'_{j}}$ and
$\Pi_{j}:=\inner{\sqrt{A}\,\Pi}{e_{j}}$, with the inner product
defined in (\ref{eq:131}).  Then, the total energy (\ref{eq:123}) can
be expressed as the Parseval sum $M=\sum_{j\geq 0}E_j(t)$.  The
evidence for the energy transfer is shown in
Fig.~\ref{fig:BoxSobolevNorm} which depicts a Sobolev-type weighted
energy norm
\begin{equation}
  \label{eq:197}
  \widetilde E(t) \equiv \left\|\phi(t,\cdot)\right\|_{1}^{2}
  = \sum_{j\geq 0}(1+j)^{2}E_{j}(t).
\end{equation}
The growth of $\widetilde E(t)$ in time means that the distribution of
energy shifts from low to high frequencies.  The characteristic
staircase shape of $\widetilde E(t)$ indicates that the energy
transfer occurs mainly during the subsequent implosions through the
center.  This observation leads to the conclusion that the only role
of the mirror is to reflect the pulse so that it can be focused during
the next implosion.

\begin{figure}[!th]
  \centering
  \includegraphics[width=\swidth]
  {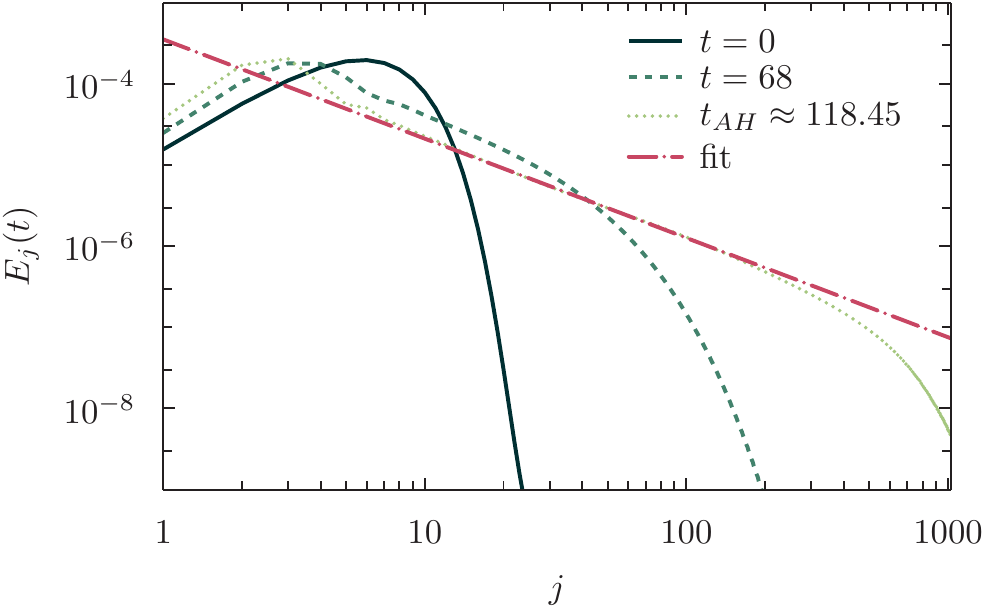}
  \caption{Energy spectra (\ref{eq:196}) at three moments of time
    (initial, intermediate, and just before collapse) for the solution
    with initial amplitude $\ep=6$.  The fit of the power law
    $E_{j}\sim j^{-\alpha}$ performed on the interval $j\in[16,128]$
    gives the slope $\alpha\approx 1.2$.}
  \label{fig:BoxSpectra}
\end{figure}

Another aspect of the turbulent cascade is shown in
Fig.~\ref{fig:BoxSpectra} which depicts the spectrum of energy (that
is, the distribution of the total energy over the linear modes) for
the solution with initial data (\ref{eq:194}) and $\ep=6$.  Initially,
the energy is concentrated in low modes; the exponential decay of the
spectrum expresses the smoothness of initial data.  During the
evolution the range of excited modes increases and the spectrum
becomes broader.  Just before horizon formation an intermediate range
of the spectrum exhibits the power-law scaling $E_{j}\sim j^{-\alpha}$
with exponent $\alpha = (1.2\pm 0.1)\,$.\footnote{Approximately the
  same value was observed for a perturbed AdS$_{4}$ space
  \cite{MRIJMPA}.}  Energy spectra in evolutions of different families
of small initial data exhibit the same slope (up to a numerical error)
which indicates that the exponent $\alpha$ is universal.  We note that
the power-law spectrum with a similar exponent was also observed in
the AdS case. As pointed out in \cite{br}, the black hole formation
provides a cut-off for the turbulent energy cascade for solutions of
Einstein's equations (in analogy to viscosity in the case of the
Navier-Stokes equation). It is natural to conjecture that the
power-law decay is a consequence of the loss of smoothness of the
solution during collapse;\footnote{This is in not the case in $d=2$
  space dimensions where for small perturbations the apparent horizon
  cannot form and solution stays smooth for arbitrary times to the
  future \cite{PhysRevLett.111.041102, 2013AcPPB..44.2603J}.} however
we have not been able to compute the exponent $\alpha$ analytically.

\begin{figure}[t]
  \centering
  \includegraphics[width=\swidth]
  {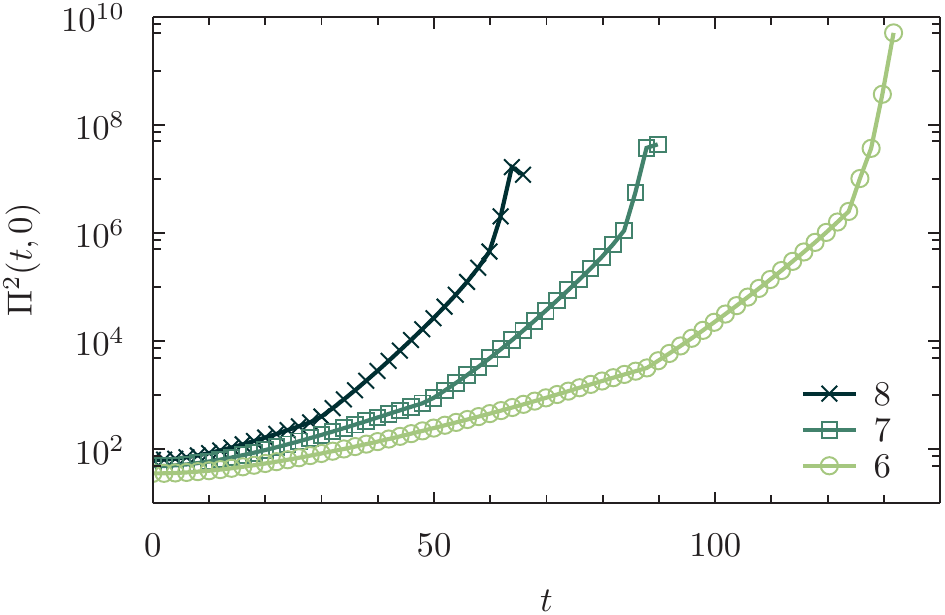}
  \\[2.5ex]
  \includegraphics[width=\swidth]
  {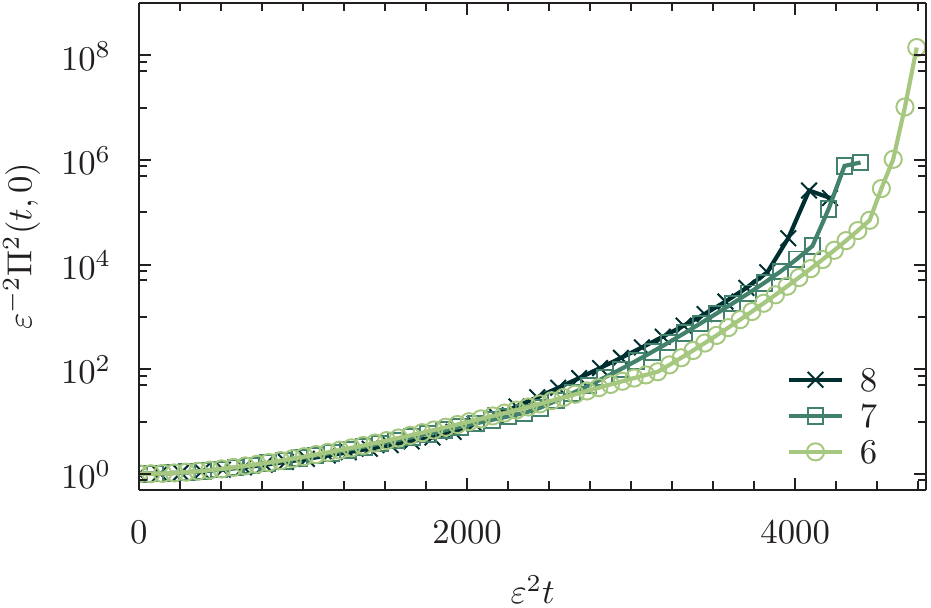}
  \caption{The analogue of Fig.~\ref{fig:BoxRicci} for the Neumann
    boundary condition (for the same initial data (\ref{eq:194})). For
    considered values of amplitudes $\ep$ of initial perturbation the
    time of apparent horizon formation exhibits the same type of
    scaling $t_{AH}\sim\ep^{-2}$.}
  \label{fig:BoxRicciNeumann}
\end{figure}

Close parallels between the results presented here (published in
\cite{mPRL}) and \cite{br, jrb} indicate that the turbulent behaviour
is not an exclusive domain of asymptotically AdS spacetimes but a
typical feature of 'confined' Einstein's gravity with reflecting
boundary conditions.  This answers the question about the role of the
negative cosmological constant $\Lambda$ posed at the end of
\cite{br}: the only role of $\Lambda$ is to generate the timelike
boundary at spatial and null infinity.

\subsection{Neumann boundary condition}
\label{sec:BoxNeumann}

The resonant case, being a close analogue of the AdS case, showed a
perfect scaling with the amplitude of the initial perturbation
(compare the Fig.~\ref{fig:BoxRicci}, with the key numerical evidence
for AdS instability, the Fig.~2 in \cite{br}) and the similar behaviour
of energy spectra to the AdS case (compare the
Fig.~\ref{fig:BoxSpectra} with the Fig.~2 in \cite{MRIJMPA}) and
strengthened the evidence for a robust mechanism of instability
sketched in \cite{br}.  Despite the fact that the analogous scaling in
the Neumann boundary case, depicted on Fig.~\ref{fig:BoxRicciNeumann},
did not seem compelling enough, it was concluded by the author in
\cite{mPRL} in quest of further robustness that \textit{(...)~the
  spectrum of linearized perturbations need not be fully resonant for
  triggering the instability}.

\begin{figure}[!t]
  \centering
  \includegraphics[width=\mwidth]
  {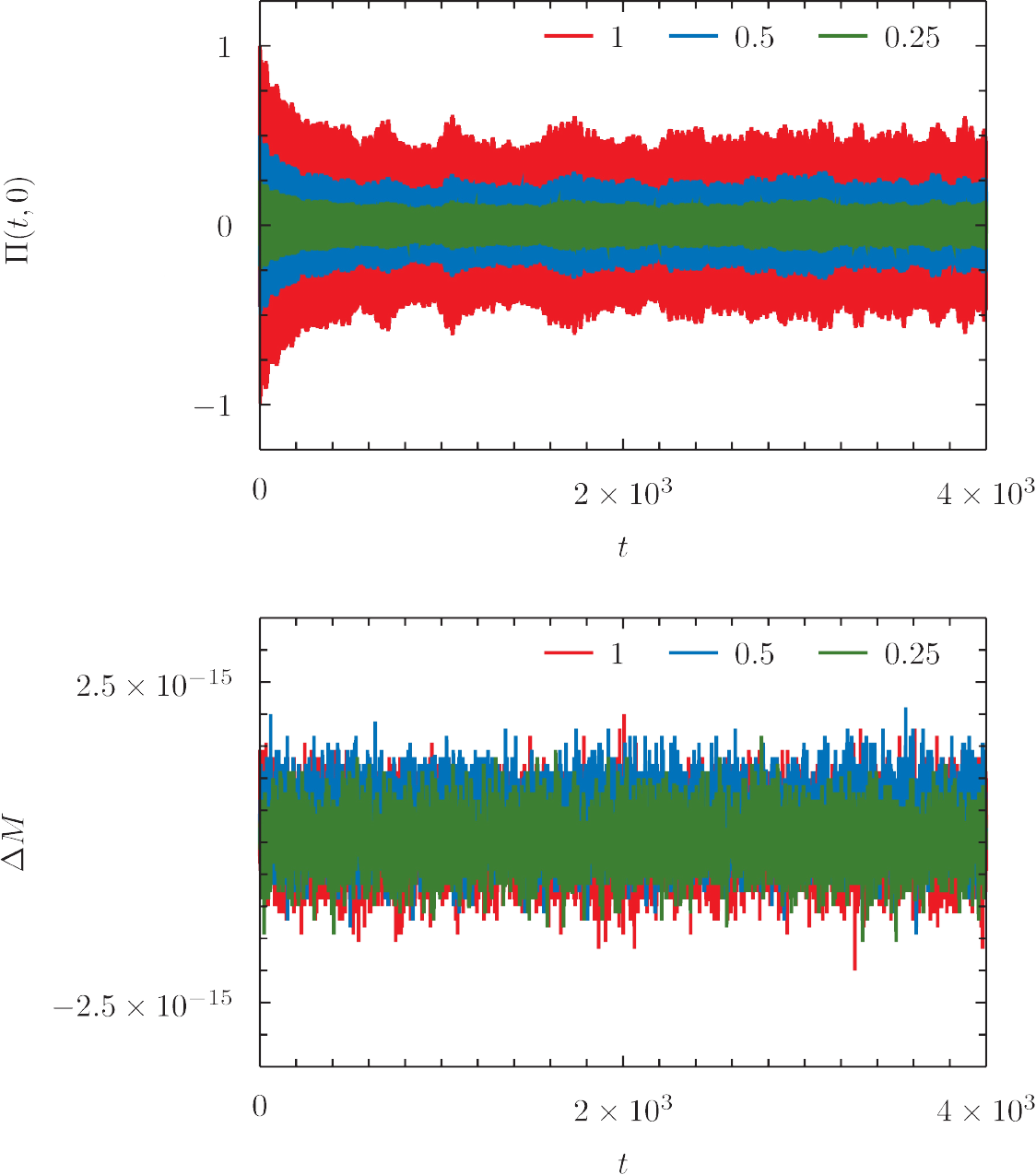}
  \caption{\textit{Top panel}. The function $\Pi(t,0)$ for solutions
    with Neumann boundary condition at the cavity with initial data
    (\ref{eq:194}) for small amplitude shows very different behaviour
    as opposed to moderate and large perturbations, compare with
    Fig.~\ref{fig:BoxRicciNeumann}. \textit{Bottom panel}.  The
    spectral code (described in Section~\ref{sec:BoxCheb}) together
    with symplectic Gauss-RK integration method (see
    Appendix~\ref{sec:AppIRK}) with small enough time step is able to
    conserve the total mass (\ref{eq:123}) up to $\sim 2.5\times
    10^{-15}$ (the absolute error) over long integration times. The
    256 radial Chebyshev points were used to produces these results.}
  \label{fig:BoxNeumannRicciAndEnergy}
\end{figure}

\begin{figure}[!t]
  \centering
  \includegraphics[width=\mwidth]
  {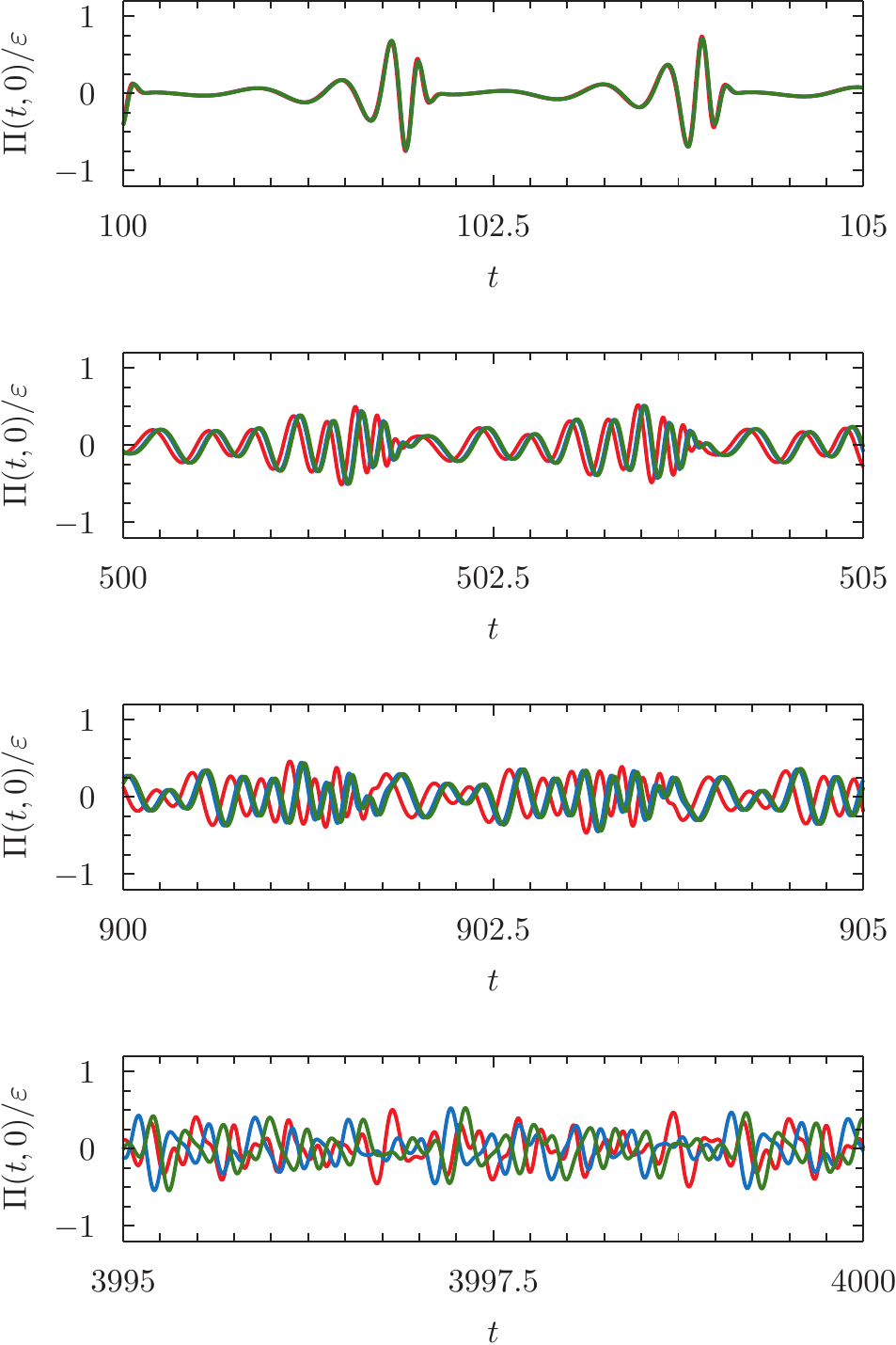}
  \caption{The closeup of Fig.~\ref{fig:BoxNeumannRicciAndEnergy}
    showing scaling of $\Pi(t,0)$ function with an amplitude of the
    perturbation $\ep$ (with the same color coding).  Due to the
    dispersive spectra for the Neumann boundary condition the
    initially localized preturbation spreads over the entire spatial
    domain which prevents the collapse.  For late time there is also
    phase shift between the signals of different amplitudes.}
  \label{fig:BoxNeumannRicciCloseup}
\end{figure}

On the other hand the authors of \cite{dhms} came to the opposite
conclusion based on the nonlinear perturbation analysis.  The clash
between those two statements became even more prominent with the
discovery of concrete examples of (nonlinearly) stable aAdS solutions
\cite{MRPRL}, previously advocated in \cite{dhms} and the question
what makes them immune to the instability discovered in \cite{br}.
Thus we developed Chebyshev pseudospectral spatial discretization for
the system (\ref{eq:165})-(\ref{eq:168}) and used symplectic
time-integrator with the aim of performing high precision, long-time,
stable, energy conserving evolution with an even smaller amplitudes
than considered before.  The solutions for the same family of initial
data (\ref{eq:194}) are depicted in
Fig.~\ref{fig:BoxNeumannRicciAndEnergy}.  For the Neumann boundary
condition we found that the scaling (shown at
Fig.~\ref{fig:BoxRicciNeumann}) does not improve as we decrease the
amplitude, while for $\ep\lesssim 1$ (for this concrete family of
initial perturbations, given in (\ref{eq:194})) the instability is not
triggered at all.  However, the scalar field $\Pi(t,r)$ exhibits
direct scaling with an amplitude $\ep$ of initial perturbation, as is
shown on Fig.~\ref{fig:BoxNeumannRicciCloseup}, and no growth of the
Ricci scalar occurs over relatively long times---in fact the magnitude
of oscillations slightly decreases.  There is no indication of scaling
previously observed for the Dirichlet case (see
Fig.~\ref{fig:BoxRicci}), after the rescaling by the initial amplitude
the signal registered by the central observer almost converges for
early times, while for late times we observe phase shift between the
signals of different amplitudes (bottom panels on
Fig.~\ref{fig:BoxNeumannRicciCloseup}).

The quantitative difference in the long-time behaviour for different
boundary conditions imposed at the cavity is shown on
Fig.~\ref{fig:BoxDirichletNeumannEvolutionProfilesMulti}, where the
Dirichlet and Neumann evolutions starting from the same initial
configuration are shown together.  For the Dirichlet boundary
condition the pulse stays approximately compact and gets more
compressed after each implosion through the center (as was illustrated
in the previous section).  In contrast, for Neumann boundary condition
the evolution is totally different (the difference is seen immediately
after the first reflection).  Not only the signal gets shifted in
phase (with respect to Dirichlet case, as shown on $t=2$ panel) after
each reflection of the cavity, it also spreads over the entire spatial
domain very fast (this is clearly visible on the $t=200$ panel;
compare red (Dirichlet) and blue (Neumann) profiles).  This feature is
also apparent on a signal registered at the center, shown on
Figs.~\ref{fig:BoxNeumannRicciAndEnergy} and
\ref{fig:BoxNeumannRicciCloseup}.  Because waves with different
lengths are reaching the center at different times (lack of coherence)
the signal is no longer a sharp peak, it gets more wavy at late
time. Moreover the energy conservation implies that the signal has to
have smaller amplitude.  The results of performed simulations
illustrate the main differences between the evolutions with different
reflecting boundary conditions imposed on the cavity.

The spreading (or its lack) is caused by the dispersive (or
nondisperive) character of the spectrum of linear perturbation
operator (which was derived in Section~\ref{sec:BoxEigenvalue}).  On
the other hand, there is a gravitational focusing present (the
nonlinear effect) acting over very long time (many successive
implosions).  The fate of initial perturbation depends on whether the
focusing effect dominates the dispersion---the signal gets compressed
and finally the black hole forms---or focusing by gravity cannot
overcome the spreading and the evolution stays smooth.  This can be
controlled by the size of initial perturbation (the amplitude $\ep$ of
gaussian perturbation in our case).  This explains the behaviour we
observe for large and moderate amplitudes, see
Fig.~\ref{fig:BoxRicciNeumann}, when the gravitational focusing
dominates the dispersion (which is inevitably present when the Neumann
boundary condition is imposed on cavity), while for small
perturbations, Fig.~\ref{fig:BoxNeumannRicciAndEnergy}, the evolution
is dominated by the dispersion.  In contrast, for the Dirichlet
boundary condition, where the dispersion is absent, all generic
perturbations are expected to collapse regardless of their size.  This
conclusion is based on an extrapolation of the observed scaling shown
on Fig.~\ref{fig:BoxRicci}, which improves when we decrease amplitude.
There exists also non-generic perturbations which do not trigger black
hole formation, as for the AdS case \cite{MRPRL}, these are
time-periodic solutions and will be the subject of
Section~\ref{sec:BoxPeriodic}.

\afterpage{%
    \clearpage
    \ifodd\value{page}
        \expandafter\afterpage
    \fi
    {%
    \begin{figure}[p]
      \includegraphics[width=\lwidth]
      {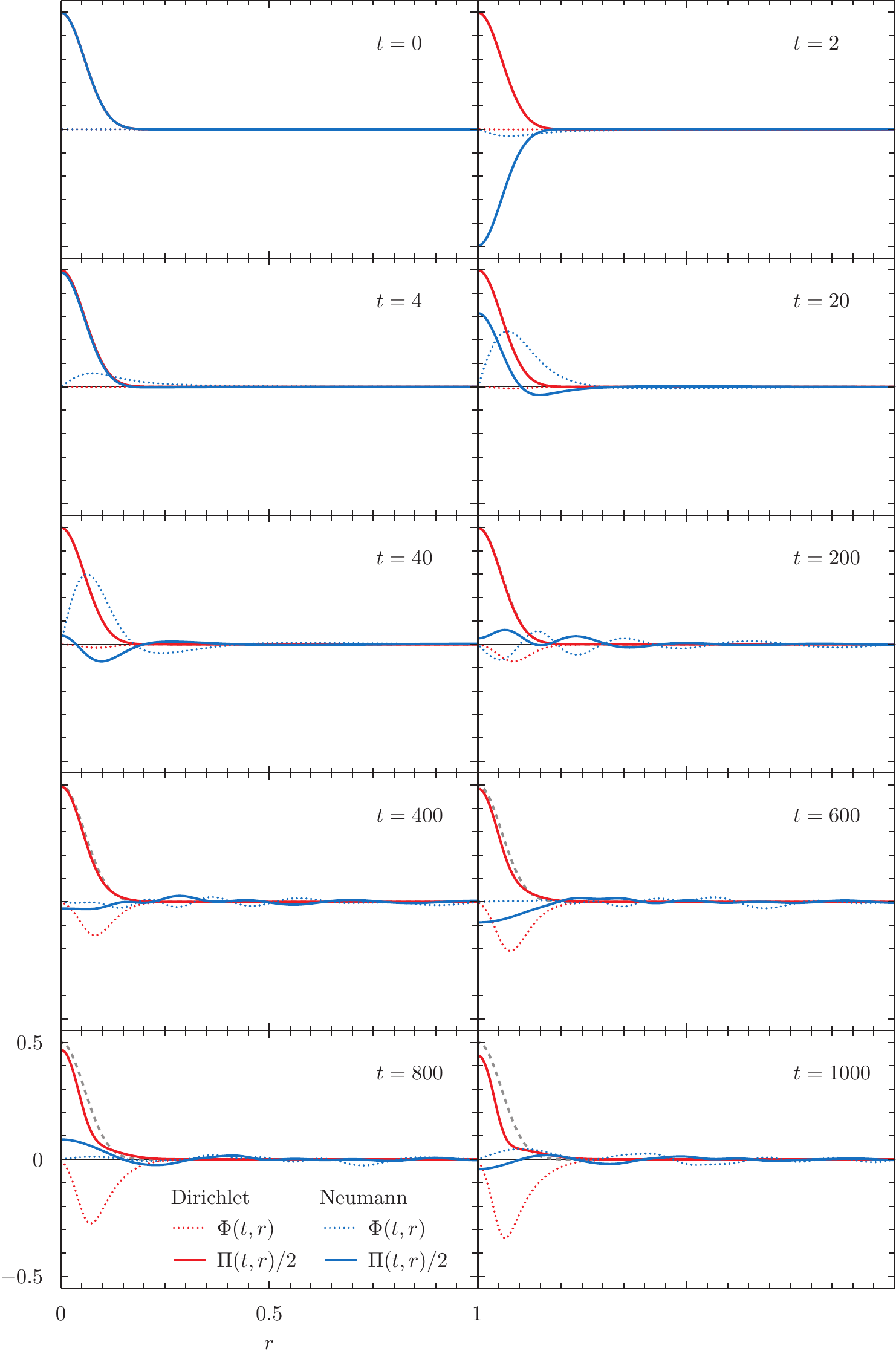}
      \caption{The comparison of time-evolution for both Dirichlet
        (red) and Neumann (blue) boundary conditions.  The initial
        conditions were given by (\ref{eq:194}) with amplitude below
        the collapse threshold for nonresonant case, namely $\ep=1$
        (gray dashed line), for which the collapse for Dirichlet case
        is expected to occur at time $t\approx 4736$.  The scalar
        field $\Phi(t,r)$ plotted with dotted line and the $\Pi(t,r)$
        with solid line.  (For better visibility the scalar field
        $\Pi$ is divided by the factor of $2$.)  The difference in
        evolution is clearly visible.}
        \label{fig:BoxDirichletNeumannEvolutionProfilesMulti}
    \end{figure}
    \clearpage
    \begin{figure}[p]
      \includegraphics[width=\lwidth]
      {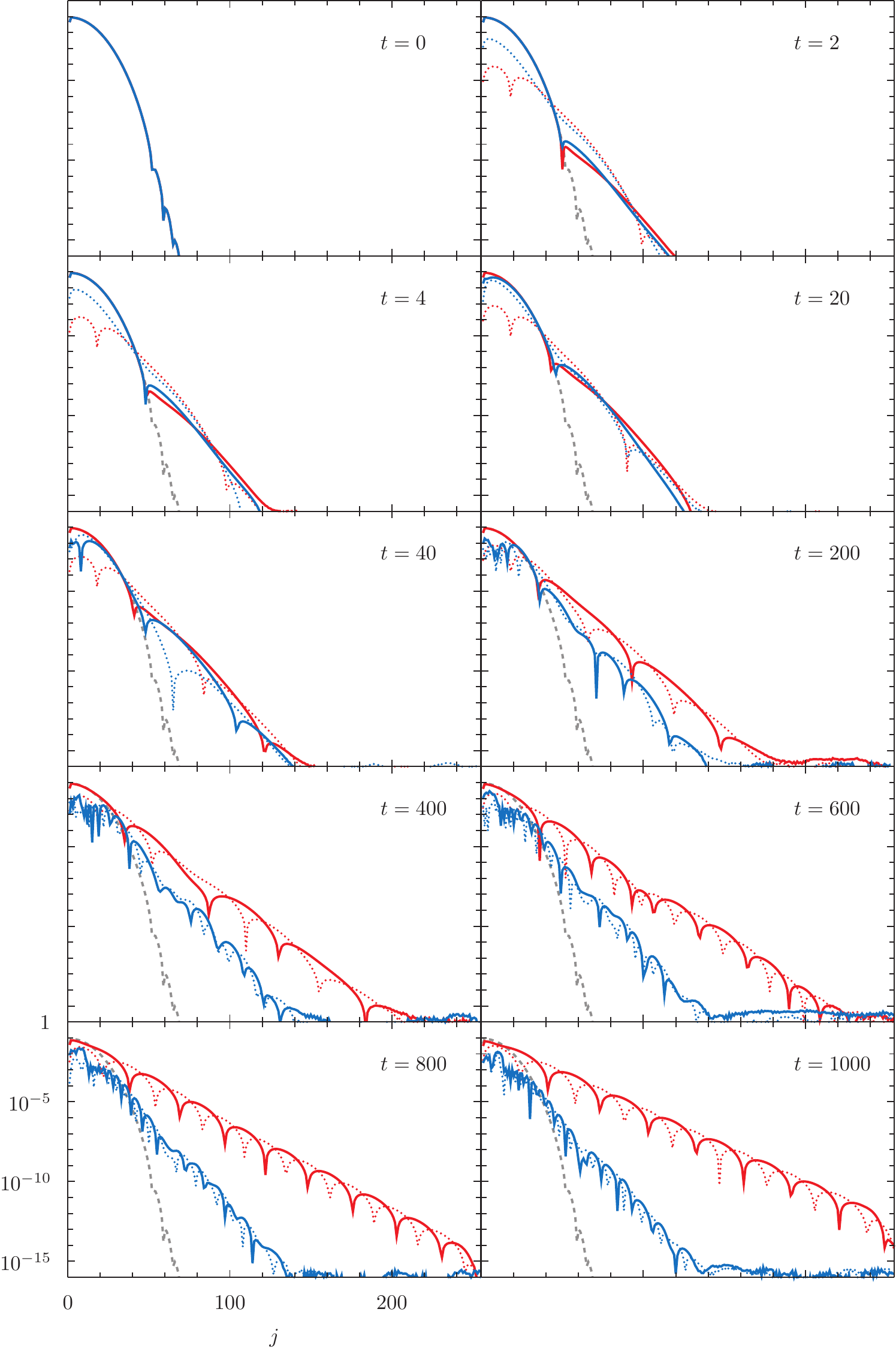}
      \caption{The time-evolution of Chebyshev coefficients
        corresponding to the profiles of $\Phi(t,r)$ and $\Pi(t,r)$
        (dashed and solid lines respectively) shown on
        Fig.~\ref{fig:BoxDirichletNeumannEvolutionProfilesMulti}.
        While for the Neumann boundary condition (blue lines) the
        spectra equilibrate at an almost constant slope, for the
        Dirichlet boundary condition (red lines) higher modes are
        being excited (energy concentrates on smaller spatial scales).
        This is also visible on function profiles, the scalar field
        spreads out over entire domain for Neumann case, while for
        Dirichlet case it gets constantly steeper after each implosion
        through the center.}
        \label{fig:BoxDirichletNeumannEvolutionCoefficientsMulti}
    \end{figure}
    \clearpage
    }%
}

The competition of dispersion and focusing is also visible on
Fig.~\ref{fig:BoxDirichletNeumannEvolutionCoefficientsMulti} showing
Chebyshev decomposition of the function profiles, $\Phi$ and $\Pi$,
plotted on accompanying
Fig.~\ref{fig:BoxDirichletNeumannEvolutionProfilesMulti}.  The
broadening range of excited Chebyshev coefficients in the resonant
case signals the need to use larger number of polynomials to represent
fine structures of approximated functions; the focusing effect of
gravity causes the energy to concentrate on smaller scales.  This
continuous transfer of energy to progressively smaller scales is the
main difficulty in numerical simulations.

Clearly the attractive property of spectral methods, the
infinite-order convergence (also called spectral accuracy), is lost
when the approximated functions are not smooth (lose derivatives,
become discontinuous or form shocks).  Such a loss of regularity
seriously degrades the rate of convergence of spectral approximations
and makes the method ineffective.  Sometimes the postprocessing
methods like filtering or the Gibbs complementary basis technique can
be used to deal with Gibbs phenomena \cite{hesthaven2007spectral} and
recover high-order accuracy, however the applicability of these
methods to the problem at hand is questionable.  Since the FD methods
are computationally less costly (the linear versus quadratic
complexity in our problem) the number of degrees of freedom can be
greately increased to achieve accurate results by resolving fine
features of the solution.  This 'brute force' approach has a natural
limitation due to finite computational resources.  However for
solutions staying smooth, which is the case for the Neumann boundary
conditions, the spectral methods greatly outperform FD discretization.

Since in the nonresonant case the dispersion dominates over the
focusing effects, the late time spectra of Chebyshev coefficients
oscillate around some equilibrium state for arbitrarily long times and
higher order polynomials are not excited above the threshold
determined by finite numerical precision (here the double
floating-point precision).  Therefore, there is a finite range of
Chebyshev polynomials needed to accurately represent solution, and
thanks to the spectral convergence this number is not very large (in
practice we have used at most $256$ Chebyshev modes).  This makes the
method presented in Section \ref{sec:BoxCheb} a very robust tool
(justified by the total mass conservation test and its result shown on
Fig.~\ref{fig:BoxNeumannRicciAndEnergy}) for solving the Cauchy
problem with Neumann boundary condition and small initial
amplitudes.  Additionally, this approach for the spatial
discretization of the field equations constitutes the main component
of the numerical methods used to find time-periodic solutions in
Section~\ref{sec:BoxPeriodic}.

The differences of evolutions for the problems with dispersive and
nondispersive character of the linear spectrum are also discussed in
the following section for the YM model, as well as in
Section~\ref{sec:Standing}, where we study the stability problem of
standing waves in AdS.

\section{Yang-Mills on Einstein Universe}
\label{sec:YMTurbulence}
Here we deal with the Cauchy problem for the YM model introduced in
Section~\ref{sec:YMModel}.  We show the qualitative difference in
dynamics of solutions when the model admits the dispersive or
nondispersive spectra of linear perturbations.

\subsection{Numerical evolution scheme}
\label{sec:YMEvolution}

To solve the Cauchy problem for the perturbations of a static solution
$S$ ($S=1$ or $S=\cos{x}$) we rewrite equation (\ref{eq:153}) as a
first order system
\begin{align}
  \label{eq:198}
  \dot{u} &= v,
  \\
  \label{eq:199}
  \dot{v} &= u'' - \csc^{2}{x}\left(3S^{2}-1\right)u
  - \csc^{2}{x}\left(3S+u\right)u^{2}.
\end{align}
We solve this system numerically using the MOL approach with a
pseudospectral spatial discretization which goes as follows.  We
assume the truncated approximation
\begin{equation}
  \label{eq:200}
  u(t,x) = \sum_{i=1}^{N} \hat{u}_{i}(t)e_{i}(x), \quad
  v(t,x) = \sum_{i=1}^{N} \hat{v}_{i}(t)e_{i}(x),
\end{equation}
where $e_{i}(x)$ are the eigenfunctions (\ref{eq:161}) of the linear
operator (\ref{eq:155}).  The time derivatives of $\hat{u}_{i}$ and
$\hat{v}_{i}$ are computed by plugging (\ref{eq:200}) to
(\ref{eq:198}) and (\ref{eq:199}) and then projecting onto the
eigenmodes $e_{k}(x)$.  This yields the system of $2N$ coupled ODEs
\begin{align}
  \label{eq:201}
  \frac{\diff}{\diff t}\hat{u}_{k} &= \hat{v}_{k},
  \\
  \label{eq:202}
  \frac{\diff}{\diff t}\hat{v}_{k} &= - \omega_{k}^{2}\hat{u}_{k}
  + \inner{e_{k}}{\csc^{2}{x}\left(3S+u\right)u^{2}},
\end{align}
$k=1,\ldots,N$, with the inner product defined in (\ref{eq:159}). The
nonlinear term in (\ref{eq:202}) is complicated when expressed in
terms of $\hat{u}_{i}$ therefore it is calculated numerically.  Since
both sides of (\ref{eq:199}) have an even Taylor expansion at both
poles of the three-sphere, we approximate the nonlinear term by
\begin{equation}
  \label{eq:203}
  \csc^{2}{x}\left(3S+u\right)u^{2} = \sum_{i=1}^{N} \tilde{a}_{i}e_{i}(x).
\end{equation}
Equating both sides of (\ref{eq:203}) at the set of $N$ collocation
points
\begin{equation}
  \label{eq:204}
  x_{i} = \frac{2i-1}{2N}\pi, \quad i=1,\ldots,N,
\end{equation}
suited for the eigenbasis (\ref{eq:161}) (chosen as the best analytic
approximation for the zeros of $e_{N+1}(x)$, \cite{Boyd199935}) we get
a system of $N$ linear equations for $N$ unknown coefficients
$\tilde{a}_{i}$, $i=1,\ldots,N$.  Therefore, the used approach is the
so-called Galerkin method with numerical integration
\cite{shen2011spectral}.  The algebraic equations (\ref{eq:203}) are
solved using LU factorization of the eigenbasis matrix which appears
on the RHS.  Thus the overall theoretical cost of computing the RHS is
of order $O(2N^{2})$.  The advantage of using the eigenbasis
$e_{i}(x)$ as expansion functions in (\ref{eq:200}) is threefold.
First, the boundary conditions at the poles of three-sphere are
automatically satisfied (see also discussion in
Appendix~\ref{cha:polyn-pseud-meth}); second, the linear part of the
equation (\ref{eq:202}) can be computed exactly; and finally, this
allows for a direct comparison with perturbative calculations.

The total energy of a perturbation $u$, expressed by the integral
(\ref{eq:158}), can be computed in the following convenient way.
Given a function decomposition into the $N$ eigenbasis functions, as
in (\ref{eq:200}), the last term in the integrand (\ref{eq:158}) can
be written as\footnote{This is easy to show noting the form of
  eigenbasis expressed in terms of Jacobi polynomials (\ref{eq:160}).
  The $P^{(\alpha,\beta)}_{j}(x)$ is a polynomial of order $j$ in $x$
  so $e_{j}(x)\sim \sin^{2}(x)\cos\left((j-1)x\right)$, and the
  highest Fourier mode present in $j$-th eigenmode is $e_{j}(x)\sim
  \cos\left((j+1)x\right)$. Thus for $u$ with the eigenmode
  decomposition (\ref{eq:200}) the highest Fourier mode of the
  nonlinear term in (\ref{eq:205}) would be $u^{4}/\sin^{2}{x} \sim
  \cos\left(3(N+1)x\right)\cos\left((N-1)x\right) \sim
  \cos\left((4N+2)x\right)$.}
\begin{equation}
  \label{eq:205}
  \csc^{2}{x}\left( S + \frac{u}{4} \right)u^{3} =
  \sum_{j=0}^{4N+2}\tilde{b}_{j}\cos(jx).
\end{equation}
Since
\begin{equation}
  \label{eq:206}
  \int_{0}^{\pi} \cos(jx) \diff x =
  \begin{cases}
    \pi, & j=0, \\
    0, & j\neq 0,
  \end{cases}
\end{equation}
and because the energy associated with the linear part of governing
equation can be easily calculated for $u$ given as the expansion
(\ref{eq:200}), we get a numerically convenient representation for the
energy
\begin{equation}
  \label{eq:207}
  E[u; S] =
  \frac{1}{2}\sum_{i=1}^{N}E_{i} + \pi\tilde{b}_{0},
\end{equation}
where
\begin{equation}
  \label{eq:208}
  E_{i} := \hat{v}_{i}^{2}+\omega_{i}^{2}\,\hat{u}_{i}^{2},
\end{equation}
are eigenmode energies and
\begin{equation}
  \label{eq:209}
  \tilde{b}_{0} = \frac{1}{\pi} \int_{0}^{\pi}
  \csc^{2}{x}\left( S + \frac{u}{4} \right)u^{3}\,\diff x.
\end{equation}
The Fourier coefficient $\tilde{b}_{0}$ can be easily computed by
solving the linear system of equations from (\ref{eq:205})
\begin{equation}
  \label{eq:210}
  \csc^{2}{x}\left( S + \frac{u}{4} \right)u^{3}\biggr|_{x=x_{k}} =
  \sum_{j=0}^{4N+2}\tilde{b}_{j}\cos (j x_{k}),
  \quad x_{k}=\pi\frac{2k-1}{2K+1},
\end{equation}
$k=1,\ldots ,K=4N+3$.  We integrate the system of equations
(\ref{eq:201}) and (\ref{eq:202}) using the partitioned Runge-Kutta
method (PRK) (see Appendix~\ref{sec:AppPRK}).  The symplectic methods
for general Hamiltonian systems are necessarily implicit.
Nevertheless, for systems with separable Hamiltonians, which is the
case for the problem at hand, there exist a class of explicit PRK.
Therefore, here we can have benefits of using the symplectic
integrator with little computational cost (or with no cost at all
since stable evolution it suffices to take integration step of order
$O(1/N)$ for the problem at hand) to have an energy conservation over
very long time-integration intervals.  The superiority of symplectic
time-integration algorithms is advocated in the following sections.

\subsection{Weakly nonlinear perturbations}
\label{sec:YMweaklyNonlPert}

In this section we concentrate on analytic methods which we use to
describe solutions to (\ref{eq:153}) with small inital data ($0<
|\ep|\ll 1$)
\begin{equation}
  \label{eq:211}
  u(0,x) = \ep f(x), \quad \dot{u}(0,x)=\ep g(x),
\end{equation}
where $f,\,g:\;[0,\pi]\mapsto\mathbb{R}$ are smooth functions,
fulfilling the regularity conditions---being even functions of $x$ at
both poles of $\Sphere^{3}$.  We consider initial conditions with
single eigenmode only.  For such restricted initial conditions the
derivation of perturbative solutions is straightforward and the
systematic analysis is relatively easy---for more generic data this
would be hardly possible.  The results of such analysis gives insight
into the dynamics because they reveal the structure of interactions
between the eigenmodes coupled through nonlinearity.  In particular it
shows that for the dispersive case resonances are equally common as
for the nondispersive case.  Moreover, these perturbative calculations
will serve as a starting point in the construction of time-periodic
solutions.  Motivated by this we impose the following initial data
\begin{equation}
  \label{eq:212}
  u(0,x) = \ep\,e_{\gamma}(x), \quad \dot{u}(0,x)=0,
\end{equation}
where $\gamma\in\mathbb{N}$ is a fixed eigenmode index and the
amplitude $\ep$ will serve as an expansion parameter.

\subsubsection{Poincar\'e-Lindstedt method}
\label{sec:poinc-lindst-appr}

Due to the nonlinearity of the Eq.~(\ref{eq:153}), it is natural to
expect that the coefficient of the eigenmode $e_{\gamma}(x)$ will no
longer be a harmonic function oscillating with the eigenmode natural
frequency $\omega_{\gamma}$.  We assume that its oscillation frequency
$\Omega(\ep)$ will depend on the magnitude of initial data and will
reduce to $\omega_{\gamma}$ in the limit $\ep\ra 0$.  To simplify the
perturbative calculation we introduce the new time variable by the
simple rescaling
\begin{equation}
  \label{eq:213}
  \tau = \Omega(\ep)\,t.
\end{equation}
Eq.~(\ref{eq:153}) written in terms of $\tau$ is
\begin{equation}
  \label{eq:214}
  \Omega^{2}\frac{\partial^{2} u}{\partial\tau^{2}} + Lu
  + \csc^{2}{x}\left(3S+u\right)u^{2} = 0,
\end{equation}
(we use notation set in (\ref{eq:155}) for the linear differential
operator $L$).  Then, we expand both, the solution $u$ and the
frequency $\Omega$, in power series in $\ep$
\begin{equation}
  \label{eq:215}
  u(\tau, x;\ep) = \sum_{\lambda\geq 1}\ep^{\lambda}u_{\lambda}(\tau,x),
\end{equation}
\begin{equation}
  \label{eq:216}
  \Omega(\ep) = \omega_{\gamma} +
  \sum_{\lambda\geq 1}\ep^{\lambda}\xi_{\lambda}.
\end{equation}
The $\xi_{\lambda}$ are unknown constants to be determined by the
requirement that all $u_{\lambda}$ are uniformly bounded for
$t\ra\infty$.  Plugging (\ref{eq:215}) and (\ref{eq:216}) into
(\ref{eq:213}) and performing Taylor expansion in $\ep$ we get a
series of initial value problems for the $u_{\lambda}$; we give here
the lowest order equations which are essential in these considerations
\begin{align}
  \label{eq:217}
  \omega_{\gamma}^{2} \frac{\partial^{2}u_{1}}{\partial\tau^{2}} + Lu_{1} &= 0,
  \\
  \label{eq:218}
  \omega_{\gamma}^{2} \frac{\partial^{2}u_{2}}{\partial\tau^{2}} +
  Lu_{2} &= - 3\csc^{2}{x}\,S\,u_{1}^{2} - 2
  \xi_{1}\omega_{\gamma}\frac{\partial^{2}u_{1}}{\partial\tau^{2}},
  \\
  \label{eq:219}
  \begin{split}
    \omega_{\gamma}^{2} \frac{\partial^{2}u_{3}}{\partial\tau^{2}} +
    Lu_{3} &= - \csc^{2}{x}\,u_{1}\left(u_{1}^{2} + 6S\,u_{2}\right) -
    \left(\xi_{1}^{2} +
      2\xi_{2}\omega_{\gamma}\right)\frac{\partial^{2}u_{1}}{\partial\tau^{2}}
    \\
    & \quad -
    2\xi_{1}\omega_{\gamma}\frac{\partial^{2}u_{2}}{\partial\tau^{2}},
  \end{split}
  \\
  \label{eq:220}
  \begin{split}
    \omega_{\gamma}^{2} \frac{\partial^{2}u_{4}}{\partial\tau^{2}} +
    Lu_{4} &= - 3\csc^{2}{x}\,\left(u_{1}^{2}u_{2} + S\,u_{2}^{2} +
      2S\,u_{1}u_{3}\right)
    \\
    & \quad - 2\left(\xi_{1}\xi_{2}+\xi_{3}\omega_{\gamma}\right)
    \frac{\partial^{2}u_{1}}{\partial\tau^{2}} -
    \left(\xi_{1}^{2}+2\xi_{2}\omega_{\gamma}\right)
    \frac{\partial^{2}u_{2}}{\partial\tau^{2}}
    \\
    & \quad -
    2\xi_{1}\omega_{\gamma}\frac{\partial^{2}u_{3}}{\partial\tau^{2}}.
  \end{split}
\end{align}
The solution of (\ref{eq:217}) with inital conditions (\ref{eq:212})
is
\begin{equation}
  \label{eq:221}
  u_{1}(\tau,x) = \cos{\tau}\,e_{\gamma}(x).
\end{equation}
The higher order equations, in particular
(\ref{eq:218})-(\ref{eq:220}), are solved as follows.  We assume that
at the order $\lambda\geq 2$ the solution is given as a linear
combination of the eigenbasis functions
\begin{equation}
  \label{eq:222}
  u_{\lambda}(\tau,x) = \sum_{j\geq 1}\hat{u}_{\lambda,j}(\tau)e_{j}(x).
\end{equation}
Then, projecting a particular perturbative equation onto the
successive eigenmodes we find that the evolution of the expansion
coefficients $\hat{u}_{\lambda,j}(\tau)$ in (\ref{eq:222}) is governed
by the system of second order inhomogeneous ODEs
\begin{equation}
  \label{eq:223}
  \omega_{\gamma}^{2}\,\frac{\diff^2\hat{u}_{\lambda,k}(\tau)}{\diff\tau^2}
  + \omega_k^2\,\hat{u}_{\lambda,k}(\tau) =
  \inner{e_{k}}{s_{\lambda}(\tau,\,\cdot\,)}, \quad k\in\mathbb{N},
\end{equation}
where $s_{\lambda}(\tau,x)$ denote the source terms of the
perturbative equations (the first three of them are the RHSs of
Eqs.~(\ref{eq:218})-(\ref{eq:220})).  Solving these with the zero
initial conditions
\begin{equation}
  \label{eq:224}
  \hat{u}_{\lambda,k}(0)=\frac{\diff \hat{u}_{\lambda,k}}{\diff\tau}(0)=0,
\end{equation}
(coming from our choice (\ref{eq:212})) we determine the unique
solution.  It turns out that at each perturbative order
$\lambda\geq 2$ the system (\ref{eq:223}) is finite, since the RHS
vanishes for $k>k_{\ast}(\gamma,\lambda)$, this implies that the
solution (\ref{eq:222}) is a finite combination of the eigenmodes.

It may happen that the projection on a given mode $k$ contains the
frequency $\omega_{k}/\omega_{\gamma}$.  Such terms give rise the
secular terms in solution $\hat{u}_{\lambda,k}$, i.e. terms which
grow linearly with time $\tau$ (the ratio $\omega_{k}/\omega_{\gamma}$
does not have to be an integer number).  The idea behind the
transformation (\ref{eq:213}) and the associated perturbative
expansion (\ref{eq:216}) is to use the free parameters $\xi_{i}$ to
systematically remove secular terms appearing in successive orders of
perturbative calculation.  Since, there is only one such parameter
available at each order, whenever there appear more then one secular
terms, due to the resonant interactions between consecutive orders of
perturbative expansion, the scheme breaks down.

We show in the following that the Poincar\'e-Lindstedt method applied
to the considered problem yields the $\mathcal{O}(\ep^{4})$ accurate,
uniformly bounded result for both the resonant and nonresonant cases
but it breaks down at the 4th order due to an unremovable resonance.
There is one special case for the perturbations of the kink solution
with the fundamental mode, $\gamma=1$ case, where the perturbative
procedure can be continued indefinitely with no secular terms
appearing at all.  Below we present the results of the perturbative
calculations and point out their drawbacks by comparing them with the
numerical solution.  Since the resonant and nonresonant cases are
distinct we consider them separately.  In the following section we
show how using a multiple-scale approach get one can perturbative
formulae which provide better approximations of solutions.

\paragraph{Vacuum sector perturbations}
\label{sec:vacuum-resonant-case}

A careful analysis of generated perturbative series solutions, for
several choices of $\gamma$ in the first order approximation
(\ref{eq:221}), gives the following eigenmode decomposition of the
lowest terms in perturbative approximation (\ref{eq:215})
\begin{align}
  \label{eq:225}
  u_{2}(\tau,x) &= \sum_{j=1}^{\gamma}\hat{u}_{2,2j-1}(\tau)\,e_{2j-1}(x),
  \\
  \label{eq:226}
  u_{3}(\tau,x) &=
  \begin{cases}
    {\displaystyle
     \ \sum_{j=1}^{(3\gamma+1)/2}\hat{u}_{3,2j-1}(\tau)\,e_{2j-1}(x)},
    & \text{for}\ \gamma\ \text{odd},
    \\[3ex]
    {\displaystyle
      \ \sum_{j=1}^{3\gamma/2}\hat{u}_{3,2j}(\tau)\,e_{2j}(x)}, &
    \text{for}\ \gamma\ \text{even},
  \end{cases}
  \\
  \label{eq:227}
  u_{4}(\tau,x) &= \sum_{j=1}^{2\gamma}\hat{u}_{4,2j-1}(\tau)\,e_{2j-1}(x).
\end{align}
For odd $\gamma$ there are only odd eigenmodes present in the
solution, whereas the even modes are excited only for even $\gamma$ at
odd perturbative orders.  The expansion of $\Omega$ also depends on
the parity of $\gamma$, namely for even $\gamma$ there are only even
powers of $\ep$ in (\ref{eq:216}) present, while for $\gamma$ odd this
perturbative series has no fixed parity.

It turns out that at the fourth order there are always more than one
resonances present for any choice of $\gamma$.  By inspection we find
that the secular terms in (\ref{eq:227}) appear for the eigenmodes
\begin{equation}
  \label{eq:228}
  \{e_{2j-1}(x)\,|\ j=1,\ldots,\gamma+1\}\cup\{e_{\gamma}(x)\},
\end{equation}
The resonance for $e_{\gamma}(x)$ can always be removed by a suitable
choice of the frequency shift parameter $\xi_{3}$.  Thus the number of
resonances at the fourth perturbative order is equal $(\gamma+1)$ or
$\gamma$, depending of the parity of $\gamma$.

\begin{figure}[!t]
  \centering
  \includegraphics[width=\mwidth]
  {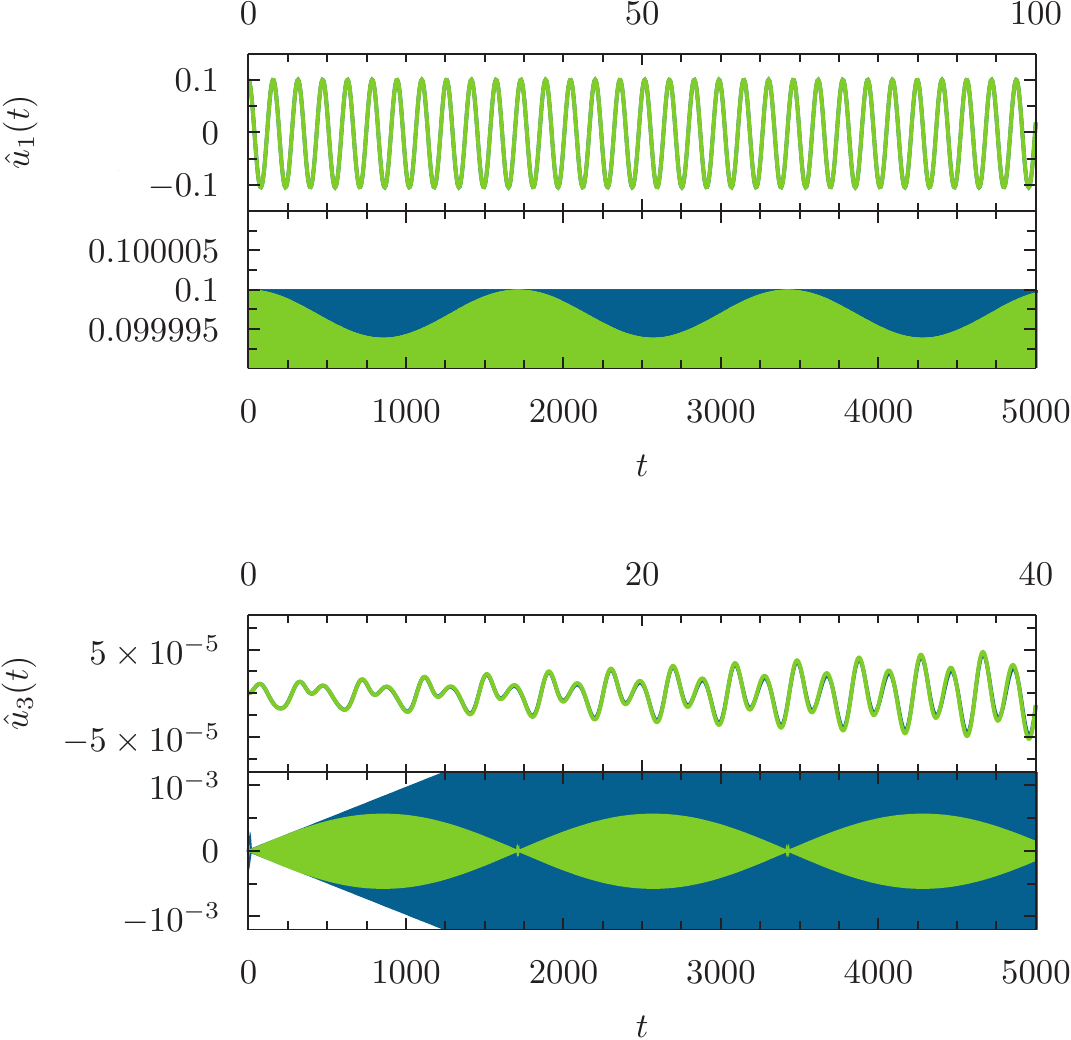}
  \caption{Comparison of perturbative solution given in (\ref{eq:229})
    and (\ref{eq:230}) derived using the Poincar\'e-Lindstedt method
    (blue line) with full numerical evolution (green line).  The
    single mode $\gamma=1$ initial conditions (\ref{eq:212}) were
    imposed with the amplitude $\ep=1/10$.  \textit{Top panel}.  The
    perturbative solution gives uniformly bounded and relatively good
    approximation for the coefficient $\hat{u}_{1}(t)$, however it
    does not describe its small amplitude modulation. \textit{Bottom
      panel}.  Due to an unremovable resonance the perturbative
    solution for $\hat{u}_{3}(t)$ contains the secular term (see
    Eq.~(\ref{eq:230})).  The perturbative solution is thus expected
    to break down at time $t\sim 1/(2\,\ep^{3})$.  This is indeed
    visible on the bottom graph, while for early times ($t\lesssim
    500$) the amplitude of the mode grows linearly in time, it then
    decreases---the long-time evolution of $\hat{u}_{3}(t)$ exhibits
    beating oscillations.  This indicates some kind of recurrence in
    this system, the energy initially deposed in fundamental mode
    moves to higher modes and then returns almost periodically.  The
    regular perturbation expansion fails to predict this effect.}
  \label{fig:YMPoincareOneE1}
\end{figure}

\begin{figure}[!t]
  \centering
  \includegraphics[width=\swidth]
  {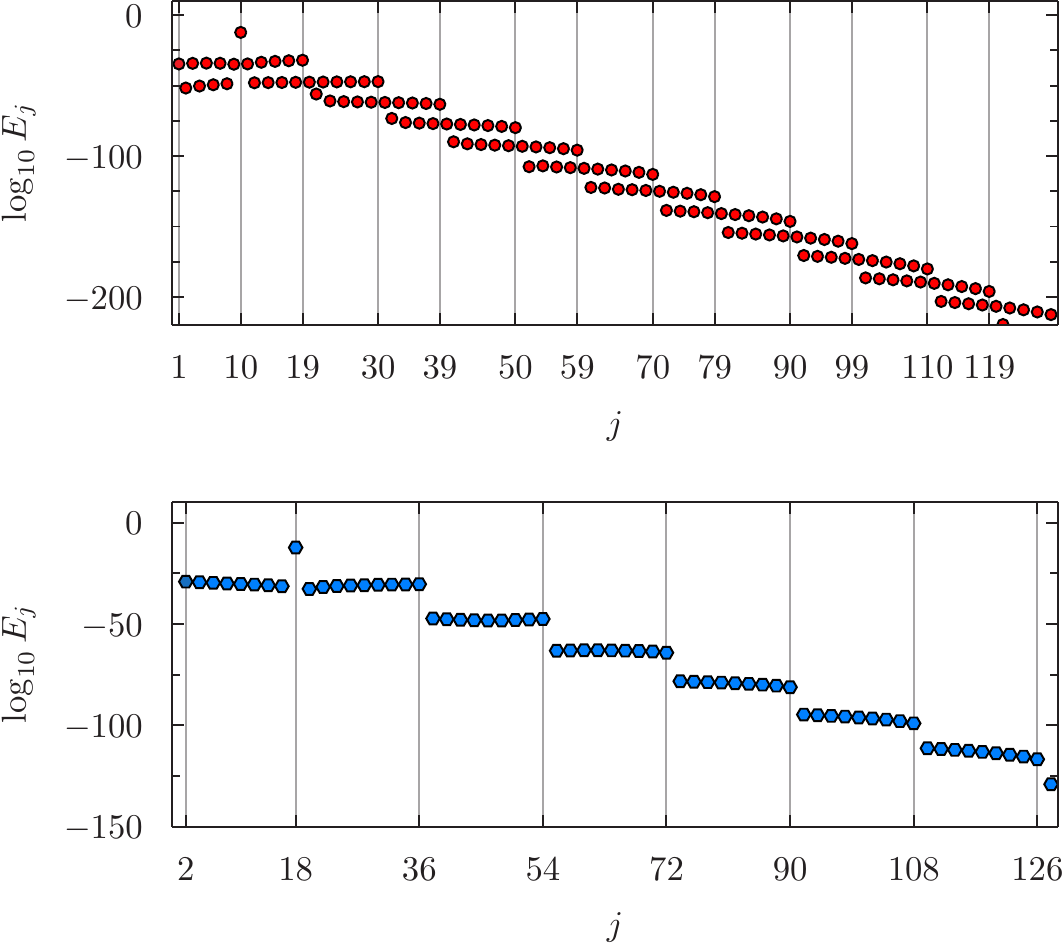}
  \caption{The energy spectra (with energy of eigenmode $j$ defined in
    Eq.~(\ref{eq:208})) for a single eigenmode initial data
    (\ref{eq:212}) exhibits characteristic staircase form.  Due to the
    nonlinear interactions, the eigenmodes are excited in groups of
    comparable amplitudes.  This structure is correctly predicted by
    the perturbative calculation.  To emphasize the separation between
    the eigenmode clusters we have taken small initial amplitude
    $\ep=10^{-6}$.  This forced us to use the extended precision
    arithmetic calculations, with up to $200$ significant digits,
    carried by the \mathematica.  Here we plot the snapshot at
    $t=2\pi$, at later times the character of the spectra stays
    unaltered.  \textit{Top panel}.  The vacuum perturbation by
    eigenmode $\gamma=10$.  At the leading order, excluding
    $e_{\gamma}(x)$ which dominates the evolution, there are
    $e_{1}(x), e_{3}(x), \ldots, e_{19}(x)$ present, a lower order
    contribution is due to eigenmodes $e_{2}(x), e_{4}(x),\ldots,
    e_{30}(x)$, and $e_{1}(x), e_{3}(x), \ldots, e_{39}(x)$ are
    excited at fourth order, as stated in
    (\ref{eq:225})-(\ref{eq:227}). \textit{Bottom panel}.  The kink
    perturbation with eigenmode $\gamma=18$.  Due to the symmetry,
    only even modes are excited for that case.  Here again the
    perturbative result (\ref{eq:231})-(\ref{eq:233}) predicts
    observed spectra; the highest index of excited eigenmode is $36$
    in second, $54$ in third and $72$ in fourth order.}
  \label{fig:YMEnergySpectraStairs}
\end{figure}

Here we give the fourth order accurate result of calculations for
$\gamma=1$ (the simplest case), which includes the frequency expansion
\begin{equation}
  \label{eq:229}
  \Omega(\ep) = 2 - \frac{5}{6\pi}\ep^{2}
  - \frac{5}{3\sqrt{6}\pi^{3/2}}\ep^{3} + \mathcal{O}\left(\ep^{4}\right),
\end{equation}
and the solution profile
\begin{multline}
  \label{eq:230}
  u(\tau,x;\ep) = \ep\,\cos{\tau}\,e_{1}(x) +
  \ep^{2}\,\frac{1}{2\sqrt{6\pi}}\bigl(-3 + 2\cos{\tau} -
  \cos{2\tau}\bigr)e_1(x)
  \\
  + \ep^{3}\Biggl[ \frac{1}{144 \pi}\bigl(-72 + 41\cos{\tau} +
  24\cos{2\tau} + 7\cos{3\tau}\bigr)e_{1}(x)
  \\
  +\frac{1}{72\sqrt{5}\pi}\bigl(5 \cos{\tau} - 4\cos{2\tau} -
  \cos{3\tau}\bigr)e_{3}(x) \Biggr]
  \\
  + \ep^{4} \Biggl[ \frac{\sqrt{6}}{864\pi^{3/2}}\bigl(-180 +
  91\cos{\tau} + 64\cos{2\tau} + 21\cos{3\tau} +
  4\cos{4\tau}\bigr)e_{1}(x)
  \\
  + \frac{1}{2880\sqrt{30}\pi^{3/2}} \bigl(- 525 + 760\cos{\tau} +
  16\cos{2\tau} - 216\cos{3\tau} - 35\cos{4\tau}
  \\
  -420\tau\sin{2\tau} \bigr)e_3(x) \Biggr] +
  \mathcal{O}\left(\ep^{5}\right),
\end{multline}
which up to the fifth order is approximated with only two eigenmodes
$e_{1}(x)$ and $e_{3}(x)$.  Note the presence of the secular term in
the coefficient of the $e_{3}(x)$ mode, as given by (\ref{eq:228}).
Fig.~\ref{fig:YMPoincareOneE1} shows the comparison of (\ref{eq:229})
and (\ref{eq:230}) with the numerical solution.  Whereas the
perturbative solution provides good approximation to the true solution
for moderate values of $\ep$ and early times, it fails to predict the
long-time behaviour of the solution.  The secular term in
(\ref{eq:230}) destroys the approximation to $\hat{u}_{3}(t)$, where
we observe beating oscillations in numerical solution.  The similar
structure, the amplitude modulation of $\hat{u}_{1}(t)$, as read off
from Eqs.~(\ref{eq:229}) and (\ref{eq:230}), is also not recovered by
Poincar\'e-Lindstedt method.\footnote{Even if we continue the
  perturbative calculation in order to obtain better approximation to
  the true solution, allowing in the same time for secular terms, we
  would get only the secular term $\ep^{6}\tau$ in $\hat{u}_{1}(t)$
  and no terms which recover the fine structure of time evolution of
  $\hat{u}_{1}(t)$.}  To summarize, the derived expansion explains
characteristic staircase energy spectra for a single mode initial
conditions, see Fig.~\ref{fig:YMEnergySpectraStairs}, and provides
reasonable approximation for times of order
$t\lesssim\left(\omega_{\gamma}\ep^{3}\right)^{-1}$.  Due to large
number of unremovable resonances the Poincar\'e-Lindstedt method does
not give a uniformly bounded approximation; it also fails to reproduce
observed amplitude modulations of the eigenmodes.

In order to obtain a better approximation we will use a multiple-scale
approach and show here the results for this particular case
($\gamma=1$), but first we apply the Poincar\'e-Lindstedt method to
the dispersive case.

\paragraph{Kink sector perturbations}
\label{sec:kink-nonr-case}
The analogous analysis for the perturbations of the kink solution
$S=\cos{x}$ leads to very similar result for the eigenmode
decomposition
\begin{align}
  \label{eq:231}
  u_{2}(\tau,x) &= \sum_{j=1}^{\gamma}\hat{u}_{2,2j}(\tau)\,e_{2j}(x),
  \\
  \label{eq:232}
  u_{3}(\tau,x) &=
  \begin{cases}
    {\displaystyle
      \ \sum_{j=1}^{(3\gamma+1)/2}\hat{u}_{3,2j-1}(\tau)\,e_{2j-1}(x)},
    & \text{for}\ \gamma\ \text{odd},
    \\[3ex]
    {\displaystyle\ \sum_{j=1}^{3\gamma/2}\hat{u}_{3,2j}(\tau)\,e_{2j}(x)},
    & \text{for}\ \gamma\ \text{even},
  \end{cases}
  \\
  \label{eq:233}
  u_{4}(\tau,x) &= \sum_{j=1}^{2\gamma}\hat{u}_{4,2j}(\tau)\,e_{2j}(x).
\end{align}
Here in contrast to the previous case, for even $\gamma$ the solution
consist of even eigenmodes only (the frequency expansion has both even
and odd terms), while for odd $\gamma$ th solution contains also the
odd modes present at odd perturbative orders (the frequency is an even
function in $\ep$).

The secular terms at fourth order are produced for the eigenmodes (for
$\gamma\neq 1$)
\begin{equation}
  \label{eq:234}
  \{e_{2j}(x)\,|\ j=1,\ldots,\gamma\}\cup\{e_{\gamma}(x)\},
\end{equation}
and the secular term for the eigenmode $e_{\gamma}$ can always be
eliminated by correctly setting the value of the parameter $\xi_{3}$.
So there are $\gamma$ or $(\gamma-1)$ resonant terms (depending of the
parity of $\gamma$) at fourth order.  The case $\gamma=1$ is special
in the sense that there appears only one resonance (for the eigenmode
$e_{1}(x)$) at any order $\lambda\geq 2$.  This resonant term can be
removed by setting $\xi_{\lambda-1}=0$.  Therefore, for the
fundamental eigenmode perturbation of the kink solution the
nonlinearity does not affect the oscillation frequency.  Secondly,
because of lack of additional resonances, since we can remove
resonances at each perturbative order, the Poincar\'e-Lindstedt method
gives a uniformly bounded approximation up to an arbitrarily high
order.

Here, as an example, we give the fourth order accurate solution for
$\gamma=1$ (for $\gamma>1$ the formulae are much more complex)
\begin{equation}
  \label{eq:235}
  \Omega(\ep) = 1 + \mathcal{O}\left(\ep^{5}\right),
\end{equation}
\begin{multline}
  \label{eq:236}
  u(\tau,x;\ep) = \ep\,\cos{\tau}\,e_{1}(x) +
  \ep^{2}\frac{1}{6\sqrt{\pi}}\biggl(-1 - 3\cos\left(2\tau\right) +
  4\cos\left(\sqrt{6}\tau\right)\biggr)e_{2}(x)
  \\
  + \ep^{3} \Biggl[ - \frac{1}{18\pi}\biggl(23 \cos (\tau )+\cos (3
  \tau )+\left(-12-4 \sqrt{6}\right) \cos \left(\tau -\sqrt{6} \tau
  \right)
  \\
  +\left(4 \sqrt{6}-12\right) \cos \left(\sqrt{6} \tau +\tau
  \right)\biggr)e_1(x)
  \\
  + \frac{\sqrt{5}}{36\pi}\biggl(3 \cos (\tau )+5 \cos (3 \tau )+16
  \cos \left(\sqrt{13} \tau \right)
  \\
  +\left(4 \sqrt{6}-12\right) \cos \left(\tau -\sqrt{6} \tau
  \right)+\left(-12-4 \sqrt{6}\right) \cos \left(\sqrt{6} \tau +\tau
  \right)\biggr)e_3(x) \Biggr]
  \\
  + \ep^{4} \Biggl[ \frac{1}{864\pi^{3/2}}\biggl(215 + 1068 \cos (2
  \tau )+45 \cos (4 \tau )-3504 \cos \left(\sqrt{6} \tau \right)+48
  \cos \left(2 \sqrt{6} \tau \right)
  \\
  -216 \cos \left(2 \tau -\sqrt{6} \tau \right)-216 \cos
  \left(\sqrt{6} \tau +2 \tau \right)+\left(1280+320 \sqrt{13}\right)
  \cos \left(\tau -\sqrt{13} \tau \right)
  \\
  +\left(1280-320 \sqrt{13}\right) \cos \left(\sqrt{13} \tau +\tau
  \right)\biggr)e_{2}(x)
  \\
  -\frac{1}{3168\pi^{3/2}}\biggl(147\sqrt{3} + 220 \sqrt{3} \cos (2
  \tau )+385 \sqrt{3} \cos (4 \tau )-528 \sqrt{3} \cos \left(\sqrt{6}
    \tau \right)
  \\
  -528 \sqrt{3} \cos \left(2 \sqrt{6} \tau \right)-1632 \sqrt{3} \cos
  \left(\sqrt{22} \tau \right)
  \\
  +\left(2112 \sqrt{2}-1848 \sqrt{3}\right) \cos \left(2 \tau
    -\sqrt{6} \tau \right)
  \\
  +\left(-2112 \sqrt{2}-1848 \sqrt{3}\right) \cos \left(\sqrt{6} \tau
    +2 \tau \right)
  \\
  +\left(2816 \sqrt{3}-704 \sqrt{39}\right) \cos \left(\tau -\sqrt{13}
    \tau \right)
  \\
  +\left(2816 \sqrt{3}+704 \sqrt{39}\right) \cos \left(\sqrt{13} \tau
    +\tau \right)\biggr)e_{4}(x) \Biggr] +
  \mathcal{O}\left(\ep^{5}\right).
\end{multline}
Note the presence of both rational and irrational frequencies in the
above formula.
\begin{figure}[!th]
  \centering
  \includegraphics[width=\mwidth]
  {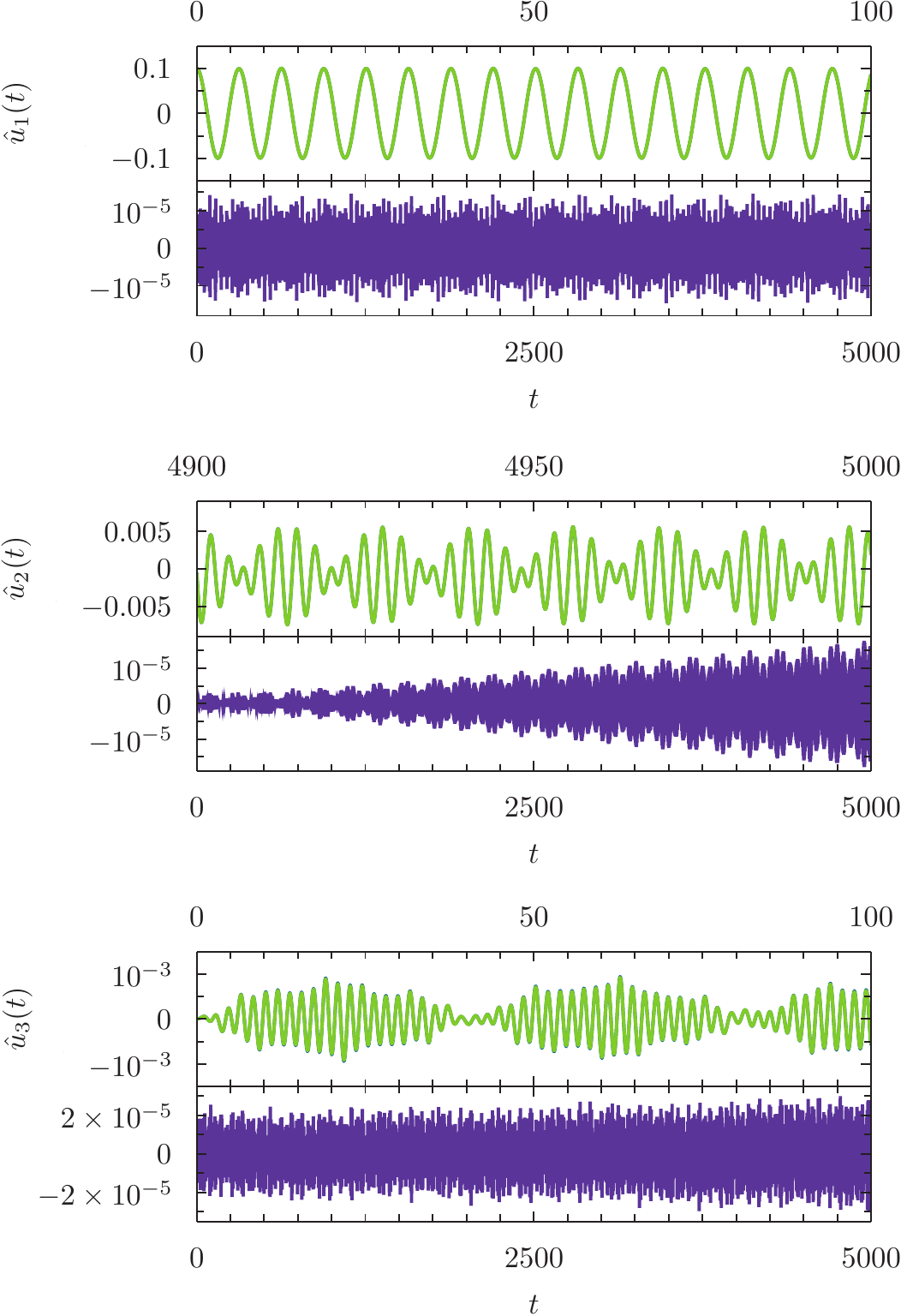}
  \caption{The plot showing the comparison of the perturbative
    solution derived using Poincar\'e-Lindstedt method (blue line)
    with numerical solution (green line) and the difference of both
    (purple line).  The single mode $\gamma=1$ initial conditions
    (\ref{eq:212}) were imposed with the amplitude $\ep=1/10$.
    \textit{Top panel}.  The fourth order accurate series expansion
    (\ref{eq:236}) provides good approximation for the coefficient
    $\hat{u}_{1}(t)$.  \textit{Middle panel}.  The divergence of
    perturbative approximation for $\hat{u}_{2}(t)$ slowly grows with
    time.  This variance is caused by systematic phase change between
    the numerical and analytical solutions.  \textit{Bottom
      panel}. The beating oscillations of eigenmode coefficients are
    predicted by the low order perturbative calculation.}
  \label{fig:YMPoincareCosE1}
\end{figure}
\begin{figure}[!th]
  \centering
  \includegraphics[width=\mwidth]
  {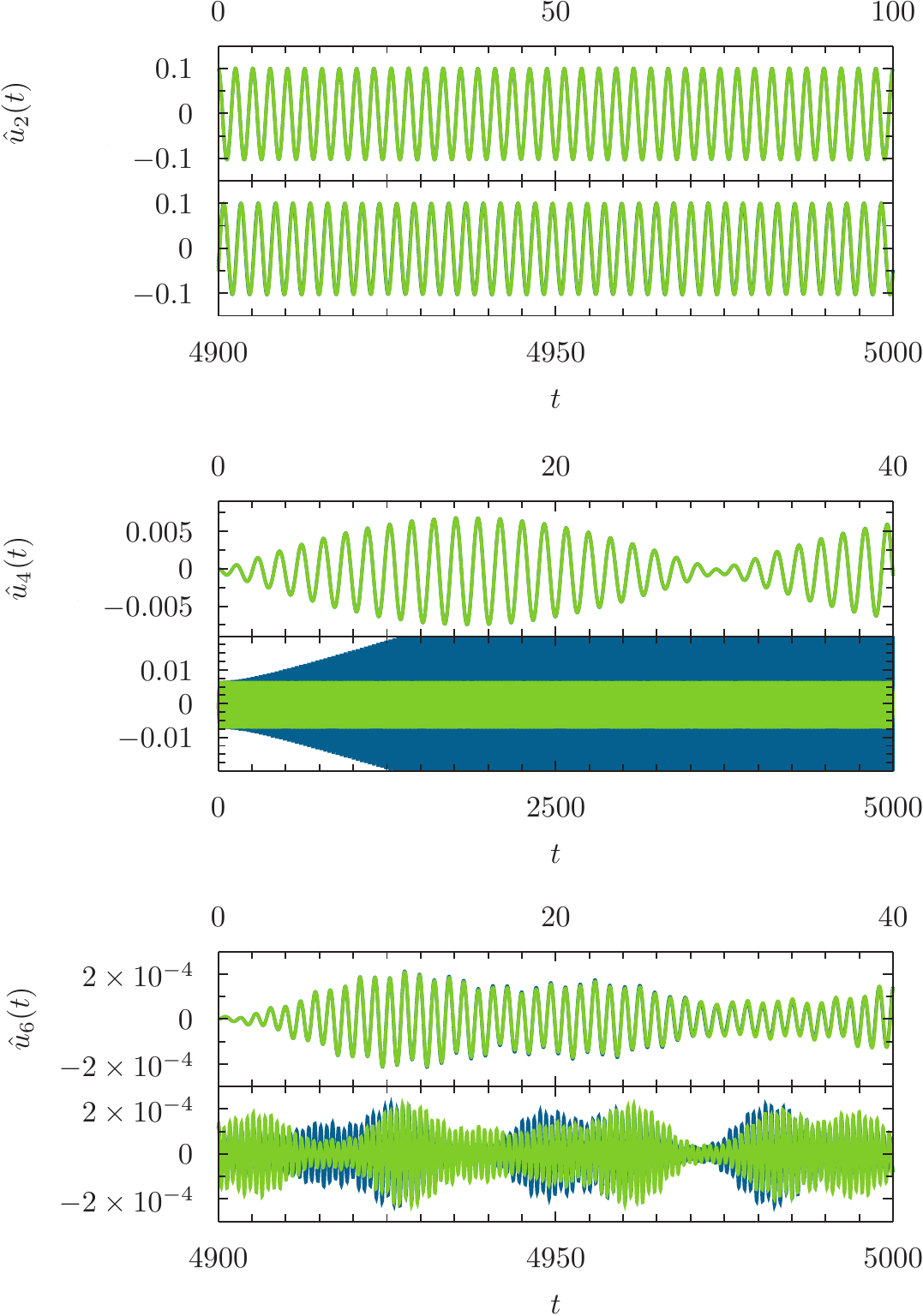}
  \caption{The plot showing the comparison of the perturbative
    solution derived using Poincar\'e-Lindstedt method (blue line)
    with numerical solution (green line).  The single mode $\gamma=2$
    initial conditions (\ref{eq:212}) were imposed with the amplitude
    $\ep=1/10$.  \textit{Top panel}. The fourth order accurate series
    expansion provides good approximation for the coefficient
    $\hat{u}_{2}(t)$. \textit{Middle panel}.  The presence of the
    resonance at fourth perturbative order causes the eigenmode
    coefficient $\hat{u}_{4}(t)$ to grow unboundedly when
    $t\ra\infty$.  The break down of perturbative expansion at time
    $\mbox{$t\sim{}1/\sqrt{6}\,\ep^{3}$}$ is clearly visible.
    \textit{Bottom panel}. Despite the fact that for early times the
    perturbative result conforms with numerical solution, for late
    times there is significant phase shift between the signals of
    $\hat{u}_{6}(t)$.  Similar behaviour is observed for the
    coefficient $\hat{u}_{8}(t)$ which is excited at fourth order (not
    shown here).}
  \label{fig:YMPoincareCosE2}
\end{figure}
In this special case (lack of secular terms) the perturbative result
(which is uniformly bounded) provides satisfactory approximation to
the numerical data, see Fig.~\ref{fig:YMPoincareCosE1}. A very slow
growth of difference between the numerical and analytical solutions is
due to the systematic change in phase of the signals (both numerical
and analytical solutions stay bounded).  This error is reduced with
order of perturbative approximation.  We have noted that phase error
increases with $\gamma$.  We show the $\gamma=2$ case on
Fig.~\ref{fig:YMPoincareCosE2}, which is qualitatively similar to
$\gamma=1$ except that we get one secular term in the $\hat{u}_{4}(t)$
coefficient, see Eq.~(\ref{eq:234}).  The fourth order accurate
approximation contains only even modes, up to $e_{8}(x)$, as given in
(\ref{eq:231})-(\ref{eq:233}).  For the coefficient $\hat{u}_{2}(t)$
the perturbative result is sufficient, while for higher modes, the
$\hat{u}_{6}(t)$ shown here, we observe a significant phase shift with
respect to numerical solution.  In that case, the Poincar\'e-Lindstedt
method is capable to capture the slow modulation of fast oscillations,
especially for early times.
\subsubsection{Multiple-scale approach}
\label{sec:mult-scale-appr}
We employ an alternative to the Poincar\'e-Lindstedt method namely the
multiple-scale approach \cite{bender1999advanced,
  kevorkian1996multiple} to derive perturbative approximation to the
numerical solution shown in Section~\ref{sec:vacuum-resonant-case} on
Fig.~\ref{fig:YMPoincareOneE1}.  In order to simplify the analysis and
shorten the presentation we proceed as follows.  As we have seen in
the previous section, for the case of $S=1$ and $\gamma=1$ the
solution (\ref{eq:230}) is approximated (up to the fourth perturbative
expansion) with only two eigenmodes $e_{1}(x)$ and $e_{3}(x)$.
Therefore assuming $u(t,x)=\hat{u}_{1}(t)e_{1}(x) +
\hat{u}_{3}(t)e_{3}(x)$, and evaluating the eigenmode projections as
in (\ref{eq:202}) we derived the evolution equations for the expansion
coefficients
\begin{subequations}
  \label{eq:237}
  \begin{align}
    \frac{\diff^{2}\hat{u}_{1}}{\diff t^{2}} + 4\hat{u}_{1} &=
    -\frac{20}{9\pi}\hat{u}_{1}^{3}
    -\frac{2}{3\pi}\left(3\sqrt{6\pi}-\sqrt{5}\hat{u}_3\right)\hat{u}_{1}^{2}
    \nonumber \\
    & \quad -\frac{68}{15\pi}\hat{u}_{3}^{2}\,\hat{u}_{1} -
    \frac{2}{225\pi} \left(46\sqrt{5}\hat{u}_3+225\sqrt{6\pi}\right)
    \hat{u}_{3}^{2},
    \\
    \frac{\diff^{2}\hat{u}_{3}}{\diff t^{2}} + 16\hat{u}_{3} &= -
    \frac{872}{225\pi}\hat{u}_{3}^{2} -
    \frac{4}{15\sqrt{5}\pi}
    \left(23\hat{u}_1+24\sqrt{6\pi}\right)\hat{u}_{3}^{2}
    \nonumber \\
    & \quad - \frac{4}{15\pi}\left(17\hat{u}_1+15\sqrt{6\pi}
    \right)\hat{u}_{3} + \frac{2\sqrt{5}}{9\pi}\hat{u}_{1}^{3}.
  \end{align}
\end{subequations}
This system of coupled ODEs is supplied with the following initial
conditions
\begin{equation}
  \label{eq:238}
  \begin{aligned}
    \hat{u}_{1}(0) &= \ep, \quad \frac{\diff \hat{u}_{1}}{\diff t}(0) = 0,
    \\
    \hat{u}_{3}(0) &= 0, \quad \frac{\diff \hat{u}_{3}}{\diff t}(0) = 0,
  \end{aligned}
\end{equation}
to match the condition (\ref{eq:212}) with $\gamma=1$.  Of course the
solution of (\ref{eq:237}) will not be a solution to the original
problem (\ref{eq:153}), however as we have seen in perturbative
calculation these problems are equivalent up to and including fourth
order.  Thus the approximate solution to (\ref{eq:237}) will be a good
approximation to (\ref{eq:153}) (for this specific initial conditions)
and at the same time the analysis of the system (\ref{eq:237}) is much
easier.

Following the multiple-scale approach we introduce two times: the fast
time $T_{0}:=t$ and the slow time $T_{2}:=\ep^{2}t$, and treat them as
independent variables.  We assume
\begin{align}
  \label{eq:239}
  \hat{u}_{1}(t) &= \sum_{i\geq 1}\ep^{i}\hat{u}_{1,i}(T_{0},T_{2}),
  \\
  \label{eq:240}
  \hat{u}_{3}(t) &= \sum_{i\geq 1}\ep^{i}\hat{u}_{3,i}(T_{0},T_{2}),
\end{align}
where the expansion coefficients are functions of both slow and fast
times.  The second derivative in (\ref{eq:237}) is then replaced by
\begin{equation}
  \label{eq:241}
  \frac{\diff^{2}}{\diff t^{2}} = \frac{\partial^{2}}{\partial T_{0}^{2}}
  + 2 \ep^{2} \frac{\partial^{2}}{\partial T_{0}\partial T_{2}}
  + \ep^{4} \frac{\partial^{2}}{\partial T_{2}^{2}}.
\end{equation}
The general solution to the linear order approximation of
(\ref{eq:237}) is
\begin{equation}
  \label{eq:242}
  \begin{aligned}
    \hat{u}_{1,1}(T_{0},T_{2}) &= a_{1,1}(T_{2})\cos{2T_{0}} +
    b_{1,1}(T_{2})\sin{2T_{0}},
    \\
    \hat{u}_{3,1}(T_{0},T_{2}) &= a_{3,1}(T_{2})\cos{4T_{0}} +
    b_{3,1}(T_{2})\sin{4T_{0}},
  \end{aligned}
\end{equation}
where the integration constants are now functions of $T_{2}$.  Next,
we solve the second order perturbative equations imposing generic
initial conditions
\begin{equation}
  \label{eq:243}
  \begin{aligned}
    \hat{u}_{1,2}(0,T_{2}) &= a_{1,2}(T_{2}), \quad
    \frac{\partial\hat{u}_{1,2}}{\partial T_{0}}(0,T_{2}) = b_{1,2}(T_{2}),
    \\
    \hat{u}_{3,2}(0,T_{2}) &= a_{3,2}(T_{2}), \quad
    \frac{\partial\hat{u}_{3,2}}{\partial T_{0}}(0,T_{2}) = b_{3,2}(T_{2}),
  \end{aligned}
\end{equation}
with free functions $a_{1,2}(T_{2})$, $b_{1,2}(T_{2})$,
$a_{3,2}(T_{2})$ and $b_{3,2}(T_{2})$ to be determined later. Since no
resonances occur at second order the solution, $\hat{u}_{1,2}$ and
$\hat{u}_{3,2}$, stay bounded.  The secular terms appear at the third
order.  The condition that their coefficients vanish is given by the
following system of first order ODEs
\begin{subequations}
  \label{eq:244}
  \begin{align}
    \label{eq:245}
    b_{1,1}' &=
    a_{1,1}\left[\frac{5}{6\pi}\left(b_{1,1}^{2}+a_{1,1}^{2}\right) +
      \frac{17}{15\pi}\left(b_{3,1}^{2}+a_{3,1}^{2}\right)\right],
    \\
    \label{eq:246}
    a_{1,1}' &=
    -b_{1,1}\left[\frac{5}{6\pi}\left(a_{1,1}^{2}+b_{1,1}^{2}\right) +
      \frac{17}{15\pi}\left(a_{3,1}^{2} + b_{3,1}^{2}\right)\right],
    \\
    \label{eq:247}
    b_{3,1}' &=
    a_{3,1}\left[\frac{17}{30\pi}\left(b_{1,1}^{2}+a_{1,1}^{2}\right)
      + \frac{409}{600\pi}\left(b_{3,1}^{2}+a_{3,1}^{2}\right)\right],
    \\
    \label{eq:248}
    a_{3,1}' &=
    -b_{3,1}\left[\frac{17}{30\pi}\left(a_{1,1}^{2}+b_{1,1}^{2}\right)
      + \frac{409}{600\pi}\left(a_{3,1}^{2}+b_{3,1}^{2}\right)\right],
  \end{align}
\end{subequations}
whose solution determines $\hat{u}_{1,1}$ and $\hat{u}_{3,1}$
uniquely.  The structure of (\ref{eq:244}) admits solutions of the
form\footnote{This is easy to show that $a_{1,1}^{2}+b_{1,1}^{2}$ and
  $a_{3,1}^{2}+b_{3,1}^{2}$ are constant.  From this we get
  (\ref{eq:250})-(\ref{eq:253}).}
\begin{subequations}
  \label{eq:249}
  \begin{align}
    \label{eq:250}
    a_{1,1}(T_{2}) &= \alpha_{1}\cos(\beta_{1}T_{2}+\theta_{1}),
    \\
    \label{eq:251}
    b_{1,1}(T_{2}) &= \alpha_{1}\sin(\beta_{1}T_{2}+\theta_{1}),
    \\
    \label{eq:252}
    a_{3,1}(T_{2}) &= \alpha_{3}\cos(\beta_{3}T_{2}+\theta_{3}),
    \\
    \label{eq:253}
    b_{3,1}(T_{2}) &= \alpha_{3}\sin(\beta_{3}T_{2}+\theta_{3}).
  \end{align}
\end{subequations}
The parameters $\alpha_{i}$ and $\beta_{i}$ are not independent,
plugging (\ref{eq:249}) into (\ref{eq:244}) gives
\begin{equation}
  \label{eq:254}
  \beta_{1} =
  \frac{1}{30\pi}\left(25\alpha_1^2 + 34\alpha_3^2\right), \quad
  \beta_{3} =
  \frac{1}{600\pi}\left(340\alpha_1^2 + 409\alpha_3^2\right).
\end{equation}
We take the amplitudes $\alpha_{1}=1$, $\alpha_{3}=0$ and the phases
$\theta_{1}=\theta_{3}=0$ in order to match the initial conditions
(\ref{eq:238}).  From this we see that the first order approximation
(\ref{eq:242}) reduces to
\begin{equation}
  \label{eq:255}
  \hat{u}_{1,1}(T_{0},T_{2}) = \cos\left(2T_{0}-\frac{5}{6\pi}T_{2}\right),
  \quad \hat{u}_{3,1}(T_{0},T_{2})=0,
\end{equation}
which is exactly the same what we get using the Poincar\'e-Lindstedt
approach, cf. Eqs.~(\ref{eq:229}) and (\ref{eq:230}).  Going to the
fourth perturbative order we find the conditions for the absence of
secular terms, which are two pairs of linear ODEs, namely
\begin{figure}[pt]
  \centering
  \includegraphics[width=\mwidth]
  {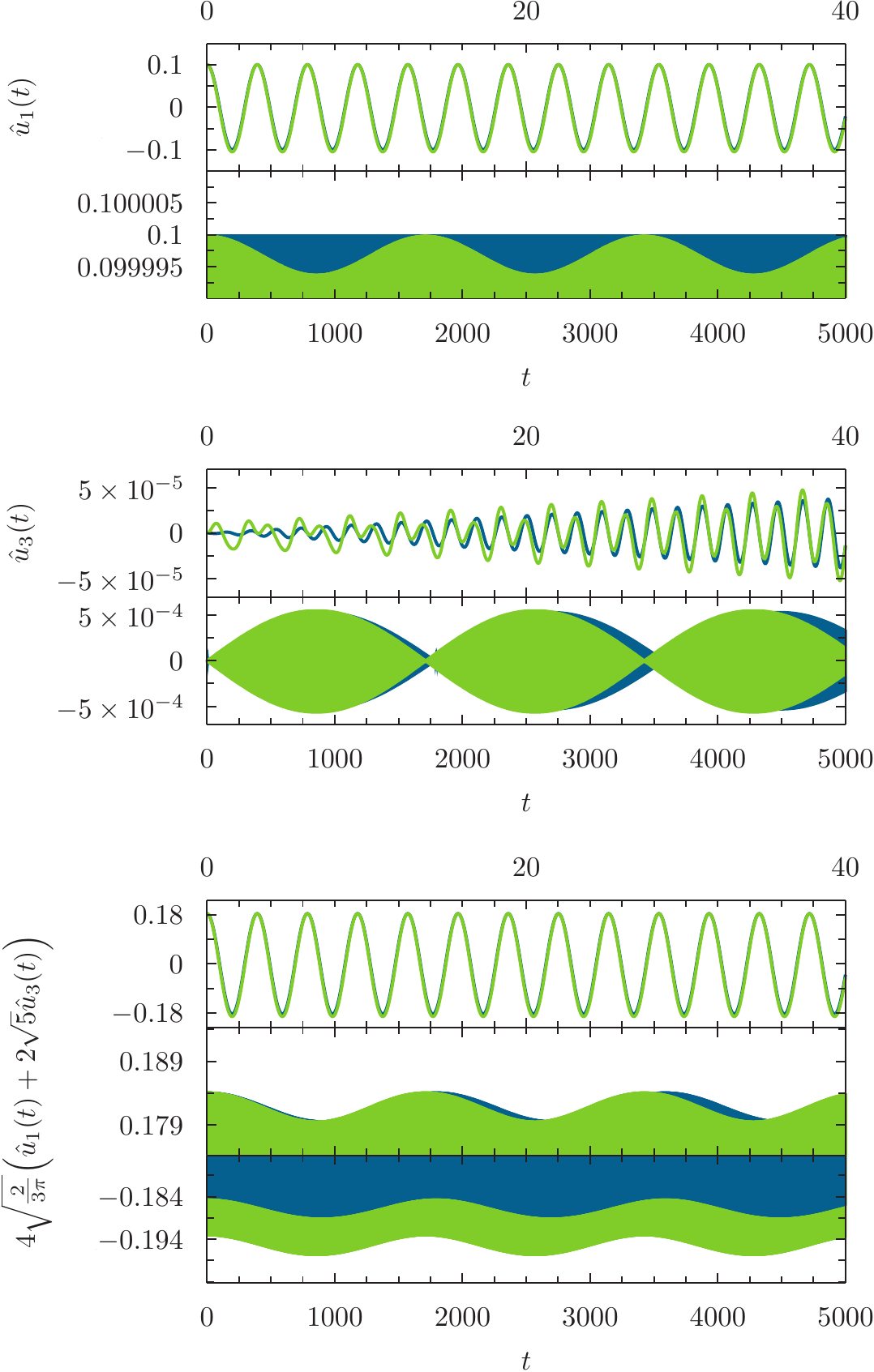}
  \caption{The comparison of the multiple-scale perturbative
    calculations with numerical solution of
    (\ref{eq:237})-(\ref{eq:238}) with $\ep=1/10$.  The perturbative
    solution (blue line) recovers qualitatively subtle features of
    numerical solution (green line) as opposed to Poincar\'e-Lindstedt
    approach, cf. Fig.~\ref{fig:YMPoincareOneE1}.  \textit{Top and
      middle panels}.  The approximation (\ref{eq:255}) contains only
    phase change, and does not predict amplitude modulation to
    $\hat{u}_{1}(t)$, whereas the third order accurate (\ref{eq:259})
    formula gives qualitatively satisfactory result for the beating
    oscillations of $\hat{u}_{3}(t)$.  \textit{Bottom panel}.  Despite
    the fact that the perturbative solution does not contain the term
    reproducing the $\hat{u}_{1}(t)$ modulation the approximation to
    $u''(t,0)$ (see discussion in text) resembles numerical result
    quite well.}
  \label{fig:YMMultiscale}
\end{figure}
\begin{align}
  \label{eq:256}
  \begin{split}
    b_{1,2}' -
    \frac{5}{6\pi}\sin\left(\frac{5}{3\pi}T_{2}\right)b_{1,2} &=
    \frac{5}{3\pi}\left[2 +
      \cos\left(\frac{5}{3\pi}T_{2}\right)\right]a_{1,2} +
    \frac{5}{4\pi^{3/2}}\sqrt{\frac{3}{2}}
    \\
    & \quad + \frac{5}{6\sqrt{6}\pi^{3/2}}
    \left[5\cos\left(\frac{5}{3\pi}T_{2}\right) +
      \frac{1}{2}\cos\left(\frac{10}{3\pi}T_{2}\right)\right],
    \\
    a_{1,2}' +
    \frac{5}{6\pi}\sin\left(\frac{5}{3\pi}T_{2}\right)a_{1,2} &=
    \frac{5}{6\pi}\left[ -1 +
      \frac{1}{2}\cos\left(\frac{5}{3\pi}T_{2}\right) \right]b_{1,2}
    \\
    & \quad - \frac{5}{12\sqrt{6}\pi^{3/2}} \left[
      \sin\left(\frac{5}{3\pi}T_{2}\right) +
      \frac{1}{2}\sin\left(\frac{10}{3\pi}T_{2}\right) \right],
  \end{split}
  \\
  \intertext{and}
  \label{eq:257}
  \begin{split}
    b_{3,2}' - \frac{34}{15\pi}a_{3,2} &=
    \frac{7}{6\sqrt{30}\pi^{3/2}}\cos\left(\frac{5}{3\pi}T_{2}\right),
    \\
    a_{3,2}' + \frac{17}{120\pi}b_{3,2} &=
    \frac{7}{24\sqrt{30}\pi^{3/2}}\sin\left(\frac{5}{3\pi}T_{2}\right).
  \end{split}
\end{align}
The solution to (\ref{eq:257}) with the initial conditions
$a_{3,2}(0)=b_{3,2}(0)=0$ is
\begin{equation}
  \label{eq:258}
  \begin{aligned}
    a_{3,2}(T_{2}) &= \frac{7}{132} \sqrt{\frac{5}{6\pi}}
    \left(\cos\left(\frac{17}{30\pi}T_2\right) -
      \cos\left(\frac{5}{3\pi}T_2\right)\right),
    \\
    b_{3,2}(T_{2}) &= \frac{7}{33}\sqrt{\frac{5}{6\pi}}
    \left(\sin\left(\frac{17}{30\pi}T_2\right) -
      \sin\left(\frac{5}{3\pi}T_2\right)\right).
  \end{aligned}
\end{equation}
This gives an unique second order accurate approximation to
$\hat{u}_{3}$, which has the following form
\begin{equation}
  \label{eq:259}
  \hat{u}_{3,2}(T_{0},T_{2}) = -\frac{7}{66}\sqrt{\frac{5}{6\pi}}
  \sin\left(\frac{11}{20\pi}T_{2}\right)
  \sin\left(4T_{0}-\frac{67}{60\pi}T_{2}\right).
\end{equation}
Unfortunately this approach also gives $\hat{u}_{1,2}(T_{0},T_{2})\sim
T_{2}$ (the solution to (\ref{eq:256}) with the initial conditions
$a_{1,2}(0)=b_{1,2}(0)=0$ contains secular term $T_{2}$).  This
inconsistency should be removed when even slower time scales (like
$\ep^{4}t$) were included.  Nonetheless, using multiple-scale approach
we where able to provide the approximation to the system
(\ref{eq:237}), which reproduces beating oscillations of the amplitude
$\hat{u}_{3}(x)$.  This is seen on Fig.~\ref{fig:YMMultiscale}, where
we compare the perturbative result, including only the leading order
terms, i.e. first order for $\hat{u}_{1}(t)$ as given in
Eq.~(\ref{eq:255}), and the second order approximation to
$\hat{u}_{3}(t)$, the formula (\ref{eq:259}).  The beating oscillation
of the former is quite well recovered, as opposed to the
Poincar\'e-Lindstedt method, while the conformity of fast oscillations
is modest.  Neglecting second order approximation to $\hat{u}_{1}(t)$,
because of presence of the secular term, we have been able to provide
only the phase frequency shift, as in previous section.  However, if
we look back at the starting problem of finding approximate solution
to (\ref{eq:153}) (with $S=1$) and for initial conditions
(\ref{eq:212}) with $\gamma=1$ and regard the time evolution of
solution profile $u(t,x)$, not its eigenmode decomposition
coefficients, we find this perturbative result quite satisfactory.  If
we look e.g. at the time evolution of the second spatial derivative
at $x=0$, i.e. if we compare the following linear combination
\begin{equation}
  \label{eq:260}
  u''(t,0) \approx 4\sqrt{\frac{2}{3\pi}}\Big(\hat{u}_1(t) +
  2\sqrt{5}\,\hat{u}_3(t)\Big),
\end{equation}
(from (\ref{eq:163}) $e_{j}''(0)=\frac{2}{3}(j+1)\sqrt{2j(j+2)/\pi}$),
then we see that the multiple-scale method provides good enough
approximation.  (As we have seen in perturbative calculations the two
mode approximation to $u(t,x)$ is adequate for fundamental mode
initial conditions.)  This agreement is easy to understand because the
long-time modulation of $\hat{u}_{1}(t)$ is subdominant with respect
to the modulation of $\hat{u}_{3}(t)$ (an effect two orders of
magnitude smaller in this case, see~Fig.~\ref{fig:YMMultiscale}),
additionally the numerical factor $2\sqrt{5}$ in this combination
amplifies this difference.

\subsection{Generic initial perturbations}
\label{sec:YMLargeData}

In the previous section we have compared the dynamics for
nondispersive and dispersive cases (for perturbations of the static
solutions $S=1$ and $S=\cos{x}$ respectively) for very simple initial
conditions, namely a single mode perturbations.  In both cases
dynamics looks very similar, the remaining eigenmodes are
excited---the energy flows from the mode excited initially to other
modes.  They cluster in groups with similar amount of energy as
predicted by the perturbative calculations---a finite range of modes
is excited at each perturbative order, see
Eqs.~(\ref{eq:225})-(\ref{eq:227}) and (\ref{eq:231})-(\ref{eq:233}).
A characteristic feature of the evolution seen in the eigenmode
coefficients is their nontrivial modulated oscillation.  In
perturbative calculation there appear a number of resonances, for both
nondispersive and dispersive cases.  Due to the resulting secular
term, perturbative expansion method fails to produce a uniformly
bounded approximation to the solution.  With multiple-scale approach
we were able to predict the observed amplitude modulation, at least
for special case shown previously, a fundamental mode initial
conditions.  Due to the nonlinear coupling between the modes dynamics
is fairly complicated and hard to describe analytically.  Clearly,
full understanding of mode interactions is beyond our reach,
nevertheless some more general statements concerning the dynamics can
be made.

\begin{figure}[!t]
  \centering
  \includegraphics[width=\swidth]
  {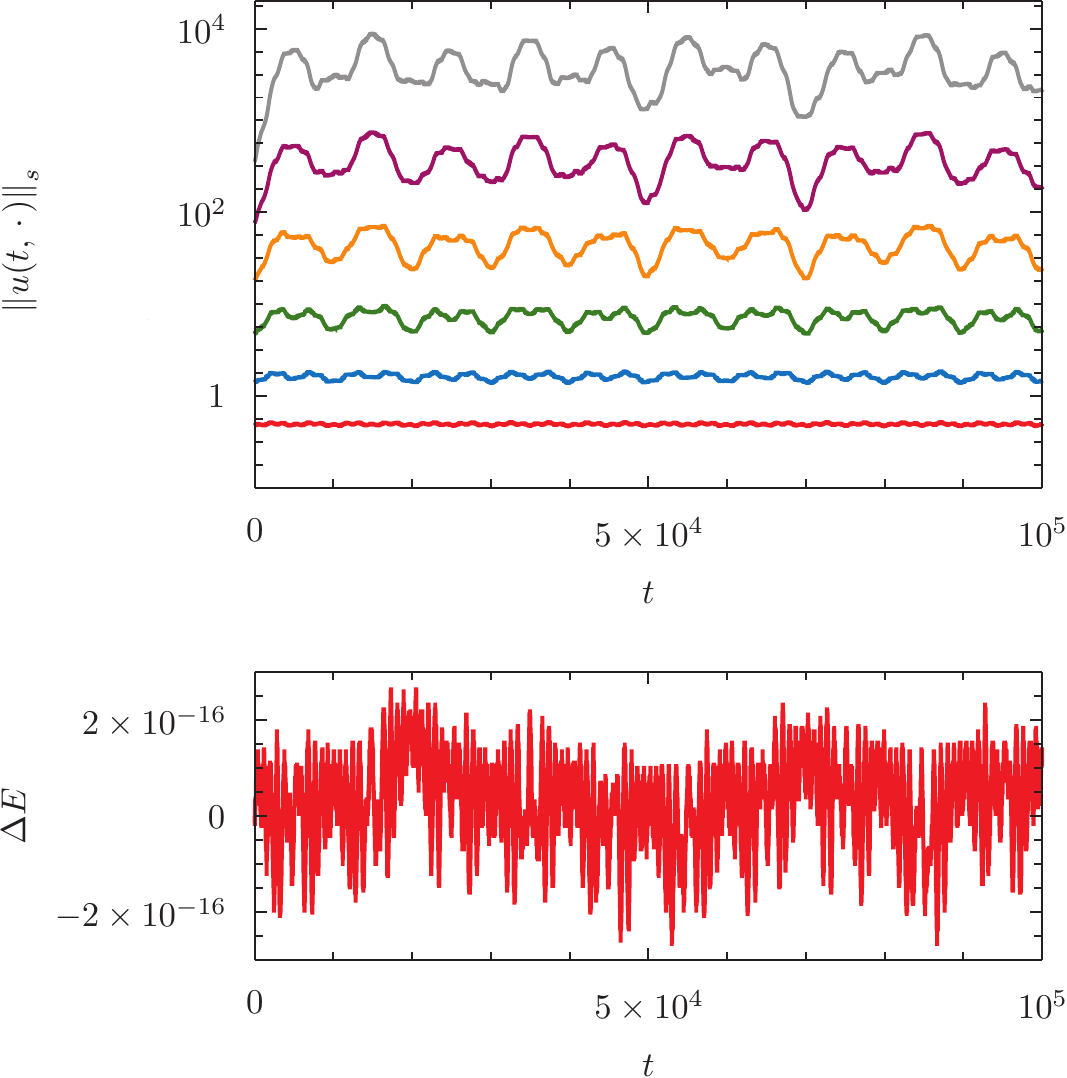}
  \caption{The long-time evolution for the gaussian type perturbation
    (\ref{eq:261}) with $\ep=2$ in vacuum topological sector
    (nondispersive case).  \textit{Top panel}.  Higher Sobolev type
    norms (\ref{eq:262}) with $s=1,\ldots,6$, plotted with different
    line colors, from bottom to top.  These norms tend to grow
    monotonically for early times, however at long time scales they
    seem to saturate---no further growth is observed.  \textit{Bottom
      panel}.  The absolute error of conserved energy
    $\Delta{}E:=E(t)-E(0)$ shows reliability of spectral
    discretization and symplectic PRK.  Here we have taken $N=128$
    eigenmodes, PRK method of order 6 and the time-step
    $\Delta{}t=\pi/(4N)$.}
  \label{fig:YMEnergyNormsOne}
\end{figure}

In order to examine this problem further we analyze the dynamics for a
generic initial conditions.  We choose the Gaussian like localized
distribution as the initial condition
\begin{equation}
  \label{eq:261}
  u(0,x) = 0, \quad
  v(0,x) =
  \ep\,\frac{\sin^{2}{x}}{2}
  \exp\left(-4\sin^{2}\left(\frac{x}{2}\right)\right),
\end{equation}
with the amplitude, $\ep$ playing the role of the control parameter.
Smoothness of chosen function profile implies that the generalized
Fourier coefficients of $v(0,x)$ fall off exponentially.  Due to the
global existence results for the Cauchy problem in this model, such
solution stays smooth for all times, so its generalized Fourier
coefficients will always exhibit exponential fall off with no
polynomial tail, in contrast to the model of previous section.  The
nonlinearity would cause this spectrum to evolve in time.
\begin{figure}[!t]
  \centering
  \includegraphics[width=\swidth]
  {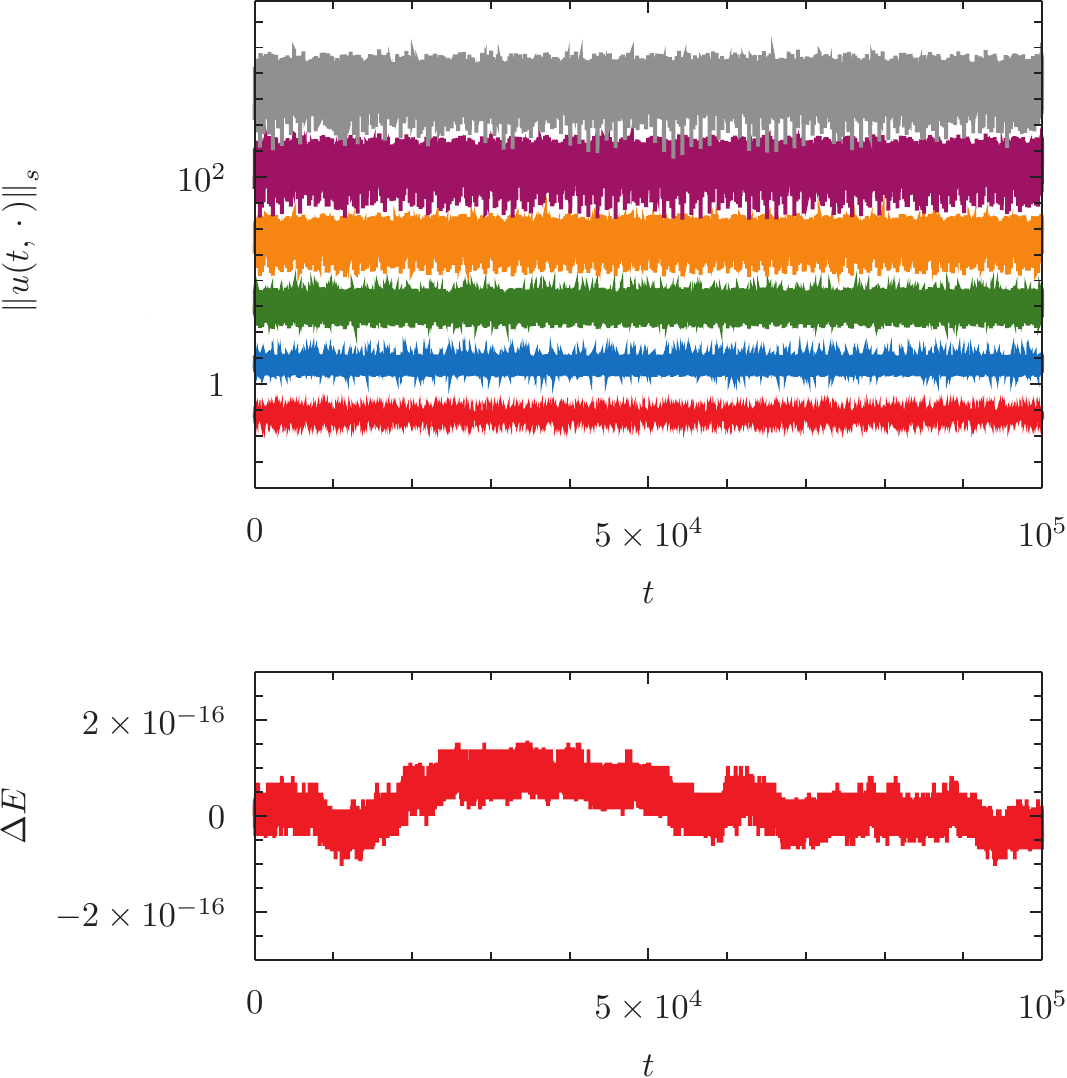}
  \caption{An analogue of Fig.~(\ref{fig:YMEnergyNormsOne}) for
    dispersive case (a kink topological sector).  \textit{Top panel}.
    Higher Sobolev type norms (\ref{eq:262}) with $s=1,\ldots,6$,
    plotted with different line colors, from bottom to top, clearly
    stay bounded for relatively long times.  \textit{Bottom panel}.
    Similarly to the previous case the energy is conserved up to the
    rounding errors.}
  \label{fig:YMEnergyNormsCos}
\end{figure}
In this case a perturbative calculation, similar to the previous
sections, would be hardly possible, thus here we rely only on a
numerical approach.  To analyze the energy transfer, similarly to the
model of previous section, we define the Sobolev type energy norms
\begin{equation}
  \label{eq:262}
  \left\| u(t,\,\cdot\,) \right\|_{s} :=
  \left( \sum_{j\geq 1}\left(1+j\right)^{2s}E_{j}(t) \right)^{1/2},
  \quad s\in\mathbb{N}_{0}\,,
\end{equation}
with the eigenmode energies defined in Eq.~(\ref{eq:208}) for the
solution represented as in (\ref{eq:200}).  The outcomes of the
long-time numerical integration, carried out with the method presented
in Section~\ref{sec:YMEvolution}, for the initial conditions
(\ref{eq:261}), with $\ep=2$, for both nondispersive and dispersive
cases are shown on Fig.~\ref{fig:YMEnergyNormsOne} and
Fig.~\ref{fig:YMEnergyNormsCos} respectively.

\begin{figure}[!t]
  \centering
  \includegraphics[width=\swidth]
  {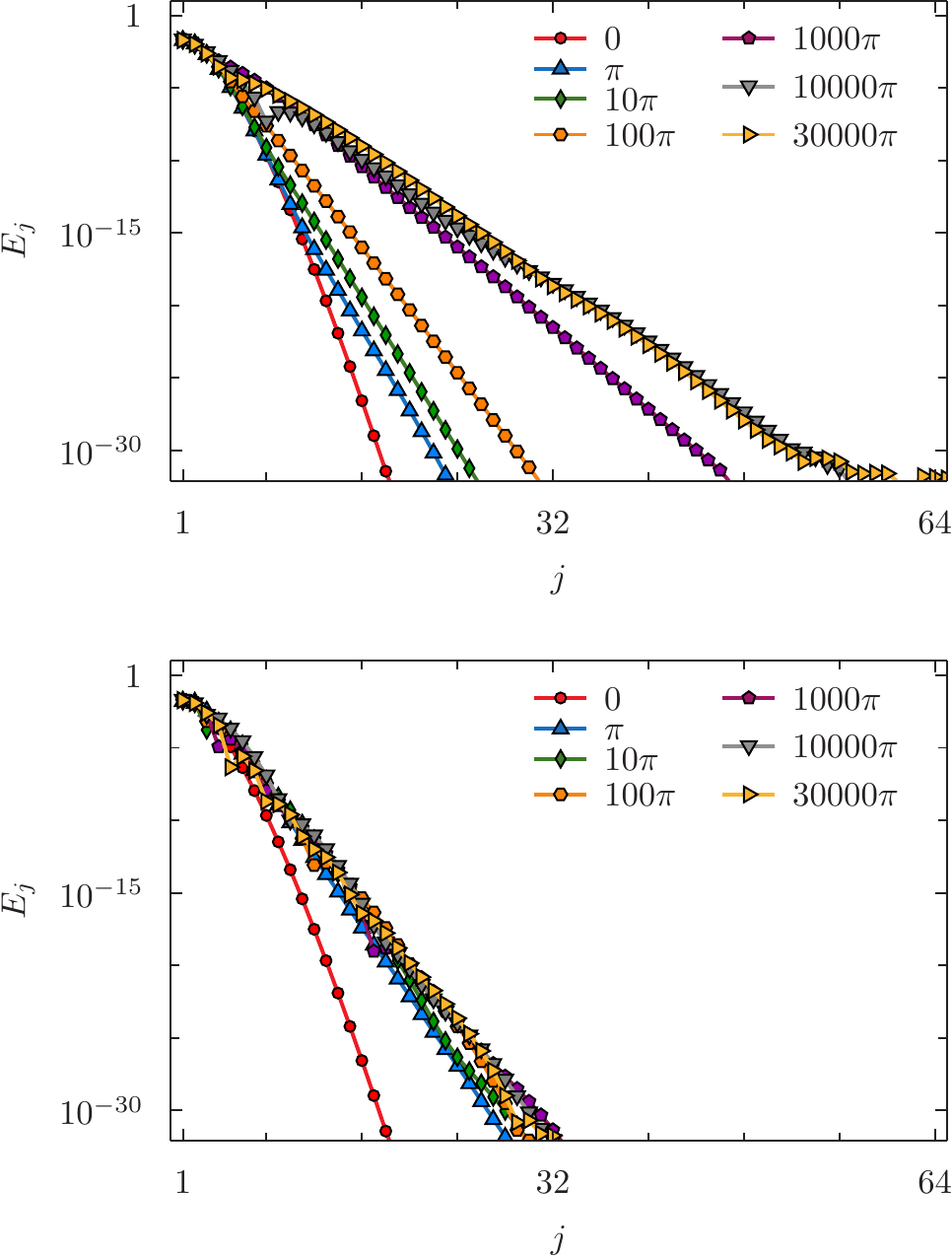}
  \caption{The time evolution of energy spectra of solution with the
    gaussian like initial conditions (\ref{eq:261}), with the
    amplitude $\ep=2$.  The snapshots at successive times are labelled
    by different point types.  \textit{Top panel}.  The $S=1$ case.
    Note the equilibration of energy spectra.  \textit{Bottom panel}.
    In the $S=\cos{x}$ case the slope of energy spectra is greater
    than for the vacuum case.  Also the equilibrium distribution is
    attained much faster.}
  \label{fig:YMEnergySpectraOneCos}
\end{figure}

Clearly the dynamics in these two cases are completely different.  At
first glance, a noisy character of plotted quantities for the
dispersive case is clearly visible, whereas in the nondispersive case
the curves are smooth.  This is the sampling effect which causes the
time dependence of norms for the dispersive case look irregular.  It
is an artifact of probing with constant intervals the functions with
irrational and rational frequencies which is precisely the case here.
A fundamental difference between these two cases concerns the growth
and oscillatory behaviour of corresponding norms.

\begin{figure}[!t]
  \centering
  \includegraphics[width=\swidth]
  {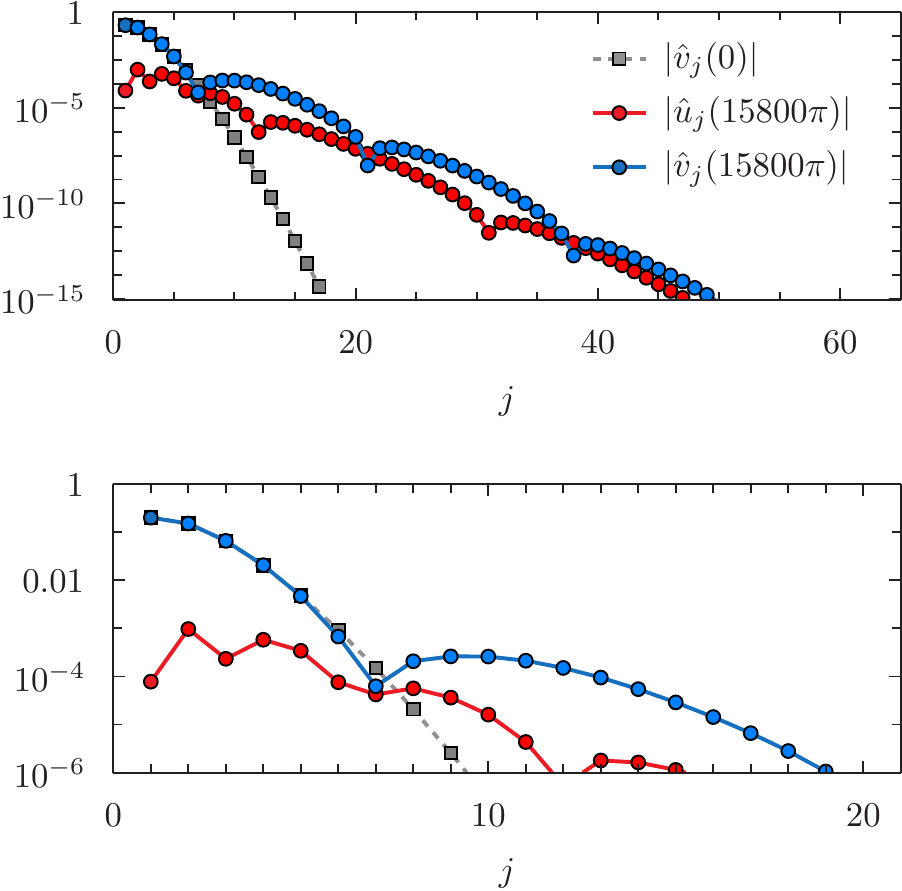}
  \caption{\textit{Top panel}.  The almost reccurence of the solution
    for nondispersive case (vacuum perturbations).  The absolute
    values of expansion coefficients of $u(t,x)$ and $v(t,x)$ for
    initial conditions (\ref{eq:261}) with $\ep=2$ (gray; at $t=0$
    $\hat{u}_{j}\equiv 0$) and at time $t=15800\pi$ (red and blue
    respectively).  Not exact reccurence is observed, the higher modes
    are still excited.  \textit{Bottom panel}.  The lowest
    $j=1,\ldots,5$ eigenmode coefficients of $v(t,x)$ at $t=15800\pi$
    almost coincide with the initial configuration decomposition (the
    difference of the fundamental mode coefficient is of order
    $10^{-8}$).  Since these dominate the solution profiles almost
    coincide while $u(15800\pi,x)$ stays close to zero.  We do not
    observe such a recurrence in the dispersive (kink perturbations).}
  \label{fig:YMSpectraOneReccurence}
\end{figure}

For the nondispersive case the norms grow monotonically for early
times, however at later times they saturate and no further growth is
observed.  Nevertheless a variation of norms is evident.  This
suggests the energy flow from low to high frequencies and back (see
discussion below).  Such behaviour is absent in the dispersive case.
For perturbations around kink the norms saturate very fast and stay
bounded.  After the initial phase of energy transfer the energy flow
is limited, this is also visible on a energy spectrum given in
Fig.~\ref{fig:YMEnergySpectraOneCos} where we plot the energy
distribution among the modes for few instants of time for solutions in
both topological sectors.

Moreover, for the nondispersive case we observe the almost reccurence
to the initial configuration.  If we plot the solution profile at
times corresponding to the local minimum of higher norms
(\ref{eq:262}), shown on Fig.~\ref{fig:YMEnergyNormsOne}, e.g. at time
$t=15800\pi\approx 5\times 10^{4}$, the solution profiles almost
coincide with the initial data (\ref{eq:261}).  We say almost, because
not all energy deposed in higher modes returns to the lower modes, as
the slope of the energy spectra is almost constant for large times
(see Fig.~\ref{fig:YMEnergySpectraOneCos}) and the higher norms do not
drop to their initial values, but they decrease sufficiently
Fig.~\ref{fig:YMEnergyNormsOne}.  This recurrence is shown on
Fig.~\ref{fig:YMSpectraOneReccurence} where we plot the eigenbasis
expansion coefficients (\ref{eq:200}) at $t=0$ and $t=15800\pi$ (for
solution with $\ep=2$).  The amplitudes of first five eigenmodes are
close to their initial values (with relative difference not larger
than $10^{-2}$, while the fundamental mode, which dominates in the
$v(t=15800\pi,x)$ expansion, is approximately equal its initial value
with difference of order $10^{-8}$), so the $L^{2}$ norm of difference
is
$\|v(t=15800\pi,\,\cdot\,)-v(t=0,\,\cdot\,)\|_{L^{2}}\approx 10^{-3}$.
Additionally, since the expansion coefficients of $u(t=15800\pi,x)$
are not greater in magnitude than $10^{-3}$ this function is close to
$0$ (with $\|u(t=15800\pi,\,\cdot\,)\|_{L^{2}}\approx 10^{-3}$).  We
observe a similar behaviour when looking at the solution for different
times (even with very sparse probing in time) corresponding to
successive local minimum of higher norms
(Fig.~\ref{fig:YMEnergyNormsOne}).  While looking at solutions for the
kink case we could not find any such recurrence.

\begin{figure}[!t]
  \centering
  \includegraphics[width=\mwidth]
  {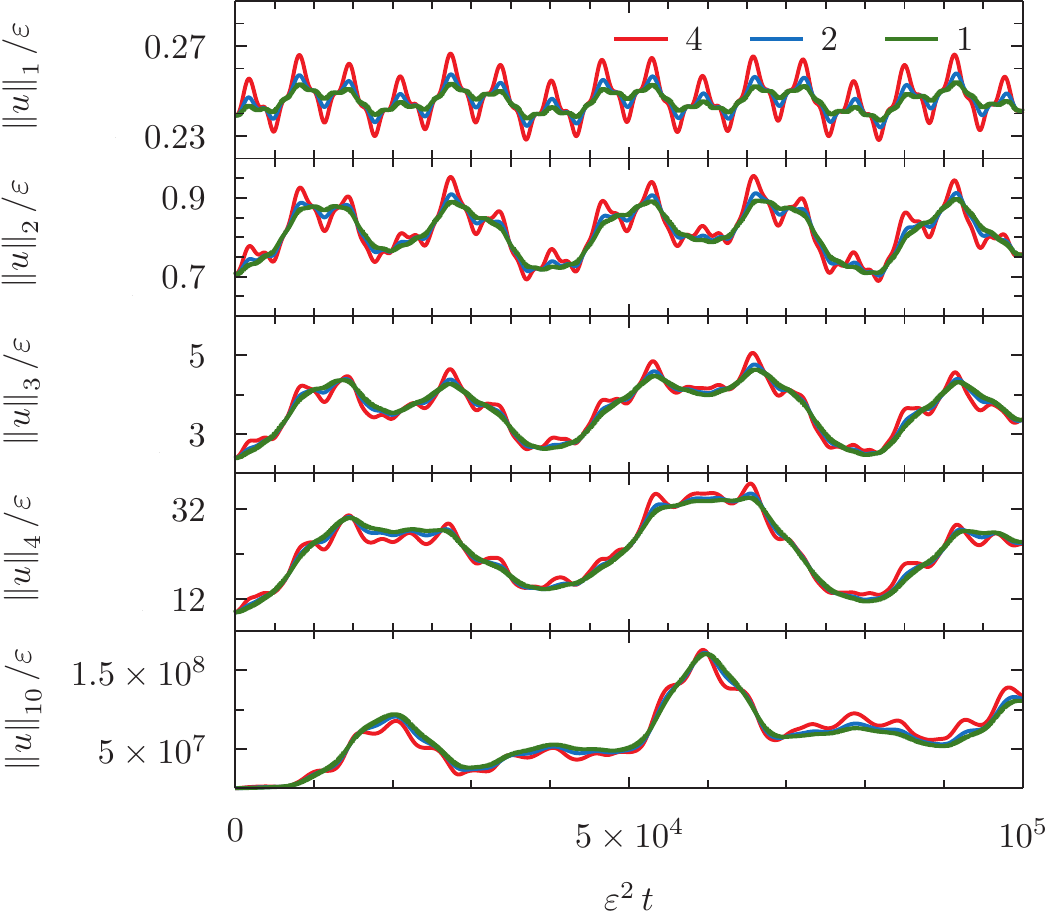}
  \caption{The scaling of Sobolev norms (\ref{eq:262}) for
    nondispersive case (vacuum perturbations) with the amplitude $\ep$
    of initial perturbation (\ref{eq:261}) (the line color codes the
    amplitude).  This scaling indicates the convergence to the
    universal curve, distinct for each family of initial data, as
    $\ep\ra{}0$.  We observe such convergence for all norms $s\geq 1$
    and for other families of initial conditions.  No such scaling
    appear for dispersive case (kink perturbations).}
  \label{fig:YMNormsOneScaling}
\end{figure}

Most remarkably, for the solutions in the vacuum topological sector,
with nondispersive spectrum of linear perturbations the norms shown on
Fig.~\ref{fig:YMEnergyNormsOne} indicate the scaling property with the
amplitude of initial data as shown on
Fig.~(\ref{fig:YMNormsOneScaling}).  The norms defined in
(\ref{eq:262}) for any $s$ almost overlap after rescaling
$\ep^{-1}\left\| u\left(\ep^{2}t,\,\cdot\,\right) \right\|_{s}$ to
universal curves for each family of initial conditions.  This scaling
improves as $\ep\ra 0$.  We do not observe this type of scaling for
the dispersive case.

Finally, we stress out that using symplectic integration algorithm,
the partitioned RK method, gives an enormous advantage of having a
numerical solution with constant (up to the double-precision
floating-point rounding errors) total energy.

\section{Conclusions}
\label{sec:TurbulenceConclusions}

In this chapter we have considered two models, namely the EKG system
confined in a spherical perfectly reflecting cavity and the YM field
propagating on the Einstein Universe, to study generic turbulent
dynamics of nonlinear waves on bounded domains.  These studies
concerned mainly the effect the spectrum of linear perturbations
(nondispersive and dispersive character) has on the global dynamics
(we continue these studies in the next chapter).  The spectrum was
controlled either by changing boundary conditions (the EKG system) or
by considering perturbations of various static solutions (the YM
equation).

Studying the EKG system in spherical cavity with Dirichlet boundary
condition we observed the growth of Ricci scalar (monitored at the
origin) with time for arbitrarily small amplitudes $\ep$ of initial
perturbations.  The characteristic growth of this quantity scales with
$\ep$ in the same ways as in the aAdS case.  This scaling and its
extrapolation for $\ep\ra 0$ supports the conjecture for the
instability of AdS space.  We draw the same conclusion in the case of
the minimally coupled scalar field confined in spherical cavity
without cosmological term.  We stress that most of the energy is
transferred during the implosion of the scalar field packet through
the origin.  We observe different behaviour when the Neumann boundary
condition is imposed at cavity.  For large and moderate amplitudes of
perturbation we note the similar scaling.  However this does not
improve when we further decrease $\ep$ (in contrary to Dirichlet
case).  This suggests the existence of a threshold in initial data
amplitude below which black hole formation is not triggered---further
energy transfer from large to small scales is stopped.

For the YM field propagating on the Einstein Universe the
singularities cannot occur (due to the global existence; this is no
longer true for higher dimensional spheres as in higher dimensional
flat spaces \cite{PhysRevD.64.121701}), however the difference in the
dynamics for dispersive and nondispersive cases (considering solutions
in different topological sectors) is also observed.  The analysis of
generic initial perturbations for the YM system shows another
difference in the dynamics for the problems with different character
of the spectrum.  The energy spectra slope in both cases equilibrates,
but for dispersive spectra this happens much faster than for the
nondispersive case.  This suggests the difference in the process of
energy transfer among the modes.  Indeed this was observed while
looking at the time evolution of higher Sobolev type norms of these
solutions.  While for the perturbations of kink norms stay bounded for
long time evolution and does not show any particular structure the
perturbations around vacuum reveal some kind of recurrent patterns.
Further investigation showed that the almost reccurence is actually
taking place.  This indicates that the energy transfer is not
unidirectional since this flows from high to low frequency modes also.
Moreover, each of the norms scale with the size of initial
perturbation which suggests some universal behaviour and supports the
conjecture on nonlinear stability of the vacuum static solution.

In order to gain more insight into the mechanism of the energy
transfer among the modes of linearized problem we have performed
series of perturbative calculations by considering simple---single
mode---initial conditions.  Using naive perturbative approach we have
shown that, in contrast to the expectation, the resonances are equally
present for both dispersive and nondispersive cases.  Further with
these simple techniques we where able to explain the characteristic
staircase energy spectrum.  However the Poincar\'e-Lindstedt approach
failed to predict the longtime mode amplitude modulation (which is
caused by the energy reccurence effect).  Due to the presence of
unremovable resonances secular terms arise and we get an unbounded
approximation valid only on short time intervals covering only early
phases of slow modulation.  Evident exception is the fundamental mode
perturbation of the kink solution, which is the nonresonant case in
the sense that there are no resonances present in the naive
perturbative calculation thus we can produce a bounded approximation
of arbitrary order.  We have analyzed the simple two dimensional
system of ODEs approximating the fundamental mode perturbation of
vacuum static solution.  With multiple-scale approach we were able to
predict beating oscillations for that ODEs.  Further studies of this
simple system may give us more informations about the full problem.
In particular a time-periodic solutions may be constructed for that
system and these, in a small amplitude limit, give an approximate
time-periodic solutions of the considered PDE.  This issue requires
further studies both analytical and numerical.

\chapter{Time-periodic and stationary solutions}
\label{cha:Periodic}

In this chapter, complementary to the studies in the previous chapter,
we focus on time-periodic solutions to the equations derived in
Chapter~\ref{cha:Models}.

We start with an analysis of the EKG model (in
Section~\ref{sec:AdSPeriodic}) for which the perturbative construction
of time-periodic solutions seems to be the simplest one (conceptually)
among the models considered in this thesis.  We develop methods
applicable both to even and odd space dimensions $d$ (these cases
differ in the asymptotic behaviour of the fields near the conformal
boundary of AdS$_{d+1}$ and need to be studied separately).  Next, in
Section~\ref{sec:Standing} we study standing waves---complex analogues
of time-periodic solutions of the EKG system.  We construct them by
the same of methods as time-periodic solutions.  Further, we analyze
in detail the stability of standing waves, in particular we derive the
spectrum of their linear perturbations. In Section~\ref{sec:BCS} we
investigate pure gravitational time-periodic solutions within
cohomogenity-two biaxial Bianchi~IX ansatz.  The exponential
nonlinearity of field equations requires number of modifications in
perturbative approach in comparison with the EKG system.
Sections~\ref{sec:BoxPeriodic} and \ref{sec:YMPeriodic} are devoted to
studies of systems allowing for both resonant and nonresonant
spectrum.  We examine a self-interacting scalar field in a cavity with
Dirichlet and Neumann boundary conditions imposed and emphasize a very
different structure of the time-periodic solutions in both cases.  A
very special structure of resonances for the YM system makes the
perturbative construction particularly involved especially for the
solutions in the kink topological sector.  Each of the considered
models illustrates different aspects of time-periodic solutions which
gives us a broader perspective.

Each section ends with the presentation of derived solutions, analysis
of methods and their detailed discussion.  Additionally we give a
general summary of this chapter in the closing
Section~\ref{sec:PeriodicConclusions}.

The Sections~\ref{sec:AdSPeriodic} and \ref{sec:Standing} are based on
\cite{MRPRL} and \cite{MR2014} respectively.

\section{Einstein-Klein-Gordon system---time-periodic solutions}
\label{sec:AdSPeriodic}

In this section we construct and analyze the time-periodic solutions
to the EKG system with real scalar field in $d+1$ spacetime dimensions
(we treat $d$ solely as a parameter in governing PDE system).  We
describe in detail the perturbative
(Section~\ref{sec:AdSPeriodicPerturbative}) and numerical
(Section~\ref{sec:AdSPeriodicNumeric-Even}) methods for their
construction.  We stress the differences between even and odd $d$
cases---we develop separate methods for even and odd space dimensions.
In perturbative calculations we stress the appearance of the
exceptional cancellation of resonant terms, which greately simplifies
the construction.  Additionally, in Section~\ref{sec:AdSEvolution} we
give details of spatial discretization method used in numerical
evolution scheme.  Results are presented and discussed in
Section~\ref{sec:AdSPeriodicResults}.

\subsection{Perturbative construction}
\label{sec:AdSPeriodicPerturbative}

We seek for time-periodic solutions of the system
(\ref{eq:67})-(\ref{eq:71}) for the real scalar field in the following
form ($0<\left|\ep\right|\ll 1$)
\begin{align}
  \label{eq:263}
  \phi(\tau,x;\ep) &= \ep\,\cos{\tau}\,e_{\gamma}(x)\,\,+
  \!\!\!\sum_{\mbox{{\small odd }}\lambda\geq 3}\!\!\!
  \ep^{\lambda}\,\phi_{\lambda}(\tau,x),
  \\
  \label{eq:264}
  \delta(\tau,x;\ep) &= \!\!\sum_{\mbox{{\small even }} \lambda\geq 2}
  \!\!\!\ep^{\lambda}\,\delta_{\lambda}(\tau,x),
  \\
  \label{eq:265}
  A(\tau,x;\ep) &= 1 \,\, - \!\!\!\sum_{\mbox{{\small even
      }}\lambda\geq 2} \!\!\!\ep^{\lambda}\,A_{\lambda}(\tau,x),
\end{align}
where $e_{\gamma}(x)$ is a dominant mode in the solution
(\ref{eq:263}) in the limit $\ep\ra 0$, and $\tau=\Omega\,t$ is the
rescaled time variable with
\begin{equation}
  \label{eq:266}
  \Omega(\ep) = \omega_{\gamma}\,\, + \!\!\!\sum_{\mbox{{\small even }}\lambda\geq 2}
  \!\!\!\ep^{\lambda}\,\xi_{\lambda}.
\end{equation}
We define the perturbative parameter $\ep$ to be the amplitude of the
mode $\gamma$ present at time $t=0$, which means we enforce
\begin{equation}
  \label{eq:267}
  \left.\inner{e_{\gamma}}{\phi}\right|_{t=0} = \ep, \quad
  \left.\inner{e_{\gamma}}{\partial_{t}\phi}\right|_{t=0} = 0,
\end{equation}
the normalization and phase fixing conditions (the same condition as
adapted in \cite{MRPRL}; another also convenient choice would be to
fix the value of the scalar field at the origin by setting
$\phi(0,0)=\ep$ at $t=0$, or the one used in
\cite{PhysRevD.89.065027}).

Due to the incompatibility of Taylor expansions at $x=\pi/2$ of the
scalar field and the eigenbasis functions (\ref{eq:90}) for odd $d$
(as stressed in Section~\ref{sec:AdSScalarEigen}), the use of this
basis would not be practical both for perturbative (because it
produces infinite sums) and for numerical calculations (since it
degrades the convergence rate).  For this reason the methods for even
$d$ presented in \cite{MRPRL} (whose extended description we provide
in the next section) cannot be applied for odd $d$, in particular for
$d=3$.  This of course does not imply that there are no time-periodic
solutions to the system (\ref{eq:67})-(\ref{eq:70}) for odd space
dimensions, there do exist time-periodic solutions for odd $d$ and, as
for even $d$, they are also bifurcating from a single eigenmode.  We
first discuss a general approach applicable for any $d\geq 2$, which
is the only way, so far, to obtain the perturbatively time-periodic
solutions for this model when $d$ is odd.\footnote{Such perturbative
  approach was applied for the first time to the $3+1$ dimensional EKG
  system by J.E.~Santos (private communication).}

\subsubsection{General approach for any space dimension}
\label{sec:AdSPeriodicPerturbative-Odd}

We take a single mode $e_{\gamma}(x)$, the solution to the linear wave
equation on a fixed AdS$_{d+1}$ background, as the first order
solution to the coupled system (\ref{eq:67})-(\ref{eq:69})
\begin{equation}
  \label{eq:268}
  \phi_{1}(\tau,x) = \cos{\tau}\,e_{\gamma}(x),
\end{equation}
as is already set in (\ref{eq:263}).  At each even order $\lambda$ the
constraint equations are solved by use of equations (\ref{eq:68}) and
(\ref{eq:78}), and their perturbative expansions, namely
\begin{equation}
  \label{eq:269}
  \delta_{\lambda} =
  - \int_{0}^{x}\sin{y}\cos{y}\coef{\lambda}\left(\Phi^{2}+\Pi^{2}\right)
  \diff y,
\end{equation}
and
\begin{equation}
  \label{eq:270}
  A_{\lambda} = - \frac{\cos^{d}{x}}{\sin^{d-2}{x}}\coef{\lambda}
  e^{\delta}\int_{0}^{x}e^{-\delta}\delta'\frac{\sin^{d-2}{y}}{\cos^{d}{y}}
  \diff y,
\end{equation}
where we fix the gauge condition by taking $\delta(t,0)=0$ which
implies $\delta_{\lambda}(t,0)=0$ for any $\lambda$.  The perturbative
version of Hamiltonian constraint equation can be further simplified
integrating (\ref{eq:270}) by parts, which then yields
\begin{equation}
  \label{eq:271}
  A_{\lambda} = - \frac{\cos^{d}{x}}{\sin^{d-2}{x}}
  \coef{\lambda}e^{\delta}
  \int_{0}^{x}e^{-\delta}\left(\frac{\sin^{d-2}{y}}{\cos^{d}{y}}\right)'
  \diff y.
\end{equation}
The evolution equations (\ref{eq:67}), at each odd order $\lambda\geq
3$, reduce to a linear inhomogeneous wave equation on the AdS$_{d+1}$
background
\begin{equation}
  \label{eq:272}
  \omega_{\gamma}^{2}\ddot{\phi}_{\lambda} + L\phi_{\lambda} = S_{\lambda},
\end{equation}
with $L$ as given after equation (\ref{eq:89}) and a source term,
$S_{\lambda}$, depending on all lower than $\lambda$ order terms of
the perturbative expansion (\ref{eq:263})-(\ref{eq:266}), which is
hard to write in a concise form for general $\lambda$ as we did for
the constraints.  In particular, for $\lambda=2$ we have the
backreaction formulae
\begin{equation}
  \label{eq:273}
  \delta_{2} = - \int_{0}^{x}\sin{y}\cos{y}
  \left(\phi_{1}'^{2} + \omega_{\gamma}^{2}\dot{\phi}_{1}^{2}\right)\diff y,
\end{equation}
\begin{equation}
  \label{eq:274}
  A_{2} = -\delta_{2} + \frac{\cos^{d}{x}}{\sin^{d-2}{x}}
  \int_{0}^{x}\delta_{2}\left(\frac{\sin^{d-2}{y}}{\cos^{d}{y}}\right)'
  \diff y.
\end{equation}
A third order wave equation has the following form
\begin{multline}
  \label{eq:275}
  -\omega_{\gamma}^{2}\ddot{\phi}_{3} - L\phi_{3} = S_{3} =
  \frac{-d+1+\cos{2x}}{\sin{x}\cos{x}}A_{2}\phi_{1}' +
  \omega_{\gamma}^{2}\left(\dot{A}_{2} +
    \dot{\delta}_{2}\right)\dot{\phi}_{1}
  \\
  + 2\omega_{\gamma}\left(\xi_{2} +
    \omega_{\gamma}\left(A_{2}+\delta_{2}\right)\right)\ddot{\phi}_{1},
\end{multline}
(compare with formula (16) in \cite{br} for $d=3$, noting also that
from (\ref{eq:274}) we have $\left(A_{2}+\delta_{2}\right)' =
A_{2}\left(-d + 1 + \cos{2x}\right) / \left(\sin{x}\cos{x}\right)$).
Integrating (\ref{eq:273}) and (\ref{eq:274}) for $\phi_{1}$ given in
(\ref{eq:268}) and plugging the results into the RHS of equation
(\ref{eq:275}) we compute the source function $S_{3}$.  After some
elementary trigonometric manipulations this can be rearranged to the
following form
\begin{equation}
  \label{eq:276}
  S_{3}(t,x) = \cos{\tau}\,S_{3,1}(x) + \cos{3\tau}\,S_{3,3}(x),
\end{equation}
which is independent on the choice of $d$ and $\gamma$.  Therefore as
a solution to (\ref{eq:275}) we assume
\begin{equation}
  \label{eq:277}
  \phi_{3}(t,x) = \cos{\tau}\,\phi_{3,1}(x) + \cos{3\tau}\,\phi_{3,3}(x),
\end{equation}
which (by linear independence of cosines) reduces the PDE
(\ref{eq:275}) to a system of two decoupled second order ODEs for the
Fourier modes $\phi_{3,1}(x)$ and $\phi_{3,3}(x)$.  We solve them with
Dirichlet boundary conditions $\phi_{3,1}(\pi/2)=\phi_{3,3}(\pi/2)=0$,
and since these are second order equations we are left with two
integration constants, denoted as $c_{3,1}$ and $c_{3,3}$, one for
each of the two equations (we adopt the convention where
$c_{\lambda,j}$ stands for the integration constant appearing at order
$\lambda$ for the $j$-th Fourier mode).

The appearing resonances could break our construction.  Since by the
ansatz (\ref{eq:277}) we have forced the solution to be uniformly
bounded when $t\ra\infty$ the secular terms could not occur.  However
these affect the spatial profile and generically cause the solution to
be unbounded at $x=0$ which violates the assumption on regularity of
the solution.  The idea behind this perturbative construction (as
stated in Section~\ref{sec:MethodsPerturbative}) is to use available
integration constants $c_{\lambda,j}$ and frequency $\ep$-expansion
parameters $\xi_{\lambda}$ to remove all of the resonant terms.  Here
we can require $\inner{e_{\gamma}}{S_{3,1}}=0$ (or equivalently
$\lim_{x\ra 0}\left|\phi_{3,1}(x)\right| < \infty$) by fixing
$\xi_{2}$, thus $c_{3,1}$ stays undetermined.  Further examination
shows that, first of all, the possible resonant term is absent in
$S_{3,3}$, i.e. the projection $\inner{e_{k_{*}}}{S_{3,3}}$ is zero
for $k_{*}=d+3\gamma$ (the $k_{*}$ is derived from the condition
$3\omega_{\gamma}=\omega_{k_{*}}$; see \cite{Craps2014} for rigorous
proof of this statement).  Cancellation of this resonance implies that
$\phi_{3,3}(x)$ is regular at the origin, therefore $c_{3,3}$ is also
an undetermined parameter at this perturbative order.  Secondly, for
any odd $d$ the function $S_{3,3}(x)$, in contrast to $S_{3,1}(x)$,
can be written as a finite linear combination eigenmodes, alike the
$\phi_{3,3}(x)$.  To sum up, we solve (\ref{eq:275}) by assuming
(\ref{eq:277}) and integrating the resulting ODEs in $x$ with
Dirichlet boundary conditions at $x=\pi/2$.  To remove a resonance (or
equivalently to ensure regularity at the origin) we fix the frequency
correction while, the two remaining integration constants stay
undetermined (as we will see one of them remains a free parameter,
which we fix by the normalization condition (\ref{eq:267}), while the
other one will be used to remove a resonance appearing at the higher
perturbative order).

The solution to the constraint equations at fourth and at any higher
even order $\lambda$, is in principle straightforward to get with
integral formulae (\ref{eq:269}) and (\ref{eq:271}).  Nevertheless,
for odd $d$ this turns out to be very time and resource consuming task
for \mathematica{} especially when we take $d$ and $\gamma$ to be
large (in practice for $d\geq 5$ and $\gamma\geq 3$).  At fifth order,
the structure of the source function in (\ref{eq:272}) is very similar
to that from lower order (\ref{eq:277}), so are the steps we take in
solving for $\phi_{5}(\tau,x)$.  Explicitly, the RHS of (\ref{eq:272})
at $\lambda=5$ has the following form
\begin{equation}
  \label{eq:278}
  S_{5}(t,x) = \cos{\tau}\,S_{5,1}(x) + \cos{3\tau}\,S_{5,3}(x)
  + \cos{5\tau}\,S_{5,5}(x),
\end{equation}
whence we assume
\begin{equation}
  \label{eq:279}
  \phi_{5}(t,x) = \cos{\tau}\,\phi_{5,1}(x) + \cos{3\tau}\,\phi_{5,3}(x)
  + \cos{5\tau}\,\phi_{5,5}(x).
\end{equation}
As in lower $\lambda=3$ order, the Fourier mode being the highest
multiple of the fundamental frequency $\omega_{\gamma}$, i.e. here
the function $\phi_{5,5}(x)$ stays regular at the origin.  The
regularity conditions for the remaining coefficients in (\ref{eq:279})
can be satisfied by setting properly the frequency correction
$\xi_{4}$ and the integration constant $c_{3,3}$.  The free parameters
would be therefore $c_{5,1}$, $c_{5,3}$ and $c_{5,5}$ (with $c_{5,1}$
fixed by the normalization condition).  The $c_{5,3}$, $c_{5,5}$ and
$\xi_{6}$ will be used to remove three resonances present at the next
order $\lambda=7$.

In general, for any odd $\lambda\geq 3$, the source function to the
wave equation (\ref{eq:272}) has a form of finite sum of
$(\lambda+1)/2$ terms
\begin{equation}
  \label{eq:280}
  S_{\lambda}(\tau,x) = \cos{\tau}\,S_{\lambda,1}(x) + \cos{3\tau}\,S_{\lambda,3}(x)
  + \cdots + \cos{\lambda\tau}\,S_{\lambda,\lambda}(x),
\end{equation}
where the highest Fourier mode $S_{\lambda,\lambda}(x)$ is always a
finite combination of eigenmodes (\ref{eq:90}) (this would not be the
case for the lower modes $S_{\lambda,1},\ldots,S_{\lambda,\lambda-2}$
in odd $d$; for $d$ even all Fourier modes of $S_{\lambda}$ are finite
combinations of eigenmodes---we use this fact in the following
section).  Also, the exceptional cancellation
\begin{equation}
  \label{eq:281}
  \inner{e_{k_{*}}}{S_{\lambda,\lambda}} \equiv 0,
\end{equation}
occurs at each order $\lambda$ for the eigenmode $k_{*}$ whose
frequency $\omega_{k_{*}}$ is a $\lambda$-th multiple of the
fundamental frequency $\omega_{\gamma}$ of a given time-periodic
solution we are constructing, i.e.
\begin{equation}
  \label{eq:282}
  \lambda\omega_{\gamma} = \omega_{k_{*}} \quad \Rightarrow \quad
  k_{*} = \frac{(\lambda - 1)d}{2} + \lambda\gamma.
\end{equation}
This regular structure implies the following.  The form of the source
(\ref{eq:280}) suggests
\begin{equation}
  \label{eq:283}
  \phi_{\lambda}(\tau,x) =
  \sum_{i=1}^{(\lambda+1)/2}\cos\left((2i-1)\tau\right)\phi_{\lambda,2i-1}(x),
\end{equation}
where there are $(\lambda+1)/2$ integration constants present in
$\phi_{\lambda}(\tau,x)$ one for each of $(\lambda+1)/2$ Fourier modes
(when we force the boundary condition $\phi_{\lambda}(t,\pi/2)=0$).
Additionally forcing the normalization condition (\ref{eq:267}) fixes
one of them and we are left with $(\lambda+1)/2-1$ integration
constants after solving the order $\lambda$.  In the next odd
perturbative order $\lambda+2$ there will be $(\lambda+1)/2+1$ terms
in (\ref{eq:283}) present, therefore we should expect the same number
of resonances to appear (one resonance for each Fourier component).
At first glance it appears that we have not enough parameters to
remove all the resonances, but it turns out not to be the case.  At
each order we have also the frequency expansion parameter
(\ref{eq:266}), and moreover, as pointed earlier, the resonance to the
highest Fourier mode is always absent so the number of available
parameters matches exactly the number of resonances and is equal
$(\lambda+1)/2$ in any perturbative order $\lambda+2$.  The freedom we
have to set the integration constants gives us the possibility to
cancel all appearing resonances order by order.

We note that this scheme applies at the lowest nontrivial perturbative
order $\lambda=3$, where we have only one free parameter, namely
$\xi_{2}$, while there are two possible resonances to occur, for
frequencies $\omega_{\gamma}$ and $3\omega_{\gamma}$.  Because of the
exceptional resonance cancellation (\ref{eq:281}) for (\ref{eq:282})
actually there is only one resonance present and we can continue the
construction procedure by solving the higher order equations.  Without
such a cancellation we would have to introduce additional parameters,
for example by modifying the linear order approximation (\ref{eq:268})
(which is the case for the model with cavity discussed in
Section~\ref{sec:BoxPeriodicPerturbative-Dirichlet}, with some
variants present also in other considered models) in some nontrivial
way.  The exceptional cancellation greatly simplifies the whole
construction procedure and moreover it actually makes the even $d$
case particularly simple to be carried up to a very high order of
$\ep$, which we discuss in detail below.

\subsubsection{The algorithmic approach for even number of space
  dimensions}
\label{sec:AdSPeriodicPerturbative-Even}

Due to the boundary expansion of the solutions of the system
(\ref{eq:67})-(\ref{eq:71}), as was discussed in
Section~\ref{sec:AdSScalarEigen}, the even $d$ case admits a special
form of the solution, since the following expansions
\begin{align}
  \label{eq:284}
  \phi_{\lambda}(\tau,x) &= \sum_{j\geq 0}
  \hat{\phi}_{\lambda,j}(\tau)e_j(x),
  \\
  \label{eq:285}
  \delta_{\lambda}(\tau,x) &= \sum_{j\geq 0}
  \hat{\delta}_{\lambda,j}(\tau)\bigl(e_{j}(x)-e_{j}(0)\bigr),
  \\
  \label{eq:286}
  A_{\lambda}(\tau,x) &= \sum_{j\geq 0}
  \hat{A}_{\lambda,j}(\tau)e_j(x)\,,
\end{align}
are finite at each order $\lambda$ of the perturbative expansions
(\ref{eq:263})-(\ref{eq:265}) (with expansion coefficients
$\hat{\phi}_{\lambda,j}(\tau)$, $\hat{\delta}_{\lambda,j}(\tau)$,
$\hat{A}_{\lambda,j}(\tau)$ being periodic in $\tau$).  This allows
for a straightforward algorithmization of building up the successive
terms in (\ref{eq:263})-(\ref{eq:265}).  The gauge condition
$\delta(t,0)=0$ is fixed by the form of expansion in (\ref{eq:285}).

For any even $\lambda\geq 2$ we solve the constraint equations
(\ref{eq:68}) and (\ref{eq:69}) in the following way.  Inserting the
series (\ref{eq:264}) and (\ref{eq:285}) into (\ref{eq:68}) and
projecting onto $e_k'(x)$, we get
\begin{equation}
  \label{eq:287}
  \hat{\delta}_{\lambda, k} = - \frac{1}{2 \omega_k^2}
  \inner{e_k'}{\coef{\lambda} \sin(2x) \left( \Phi^2 + \Pi^2 \right)},
  \quad k\in\mathbb{N}_{0},
\end{equation}
an explicit formula for the expansion coefficients of the
$\delta_{\lambda}$ function.  Solving Hamiltonian constraint is not as
strightforward since it involves a solution of a linear algebraic
system.  Inserting the series (\ref{eq:265}) and (\ref{eq:286}) into
(\ref{eq:69}) and projecting onto $e_k(x)$, we get a linear system of
equations for the coefficients $\hat{A}_{\lambda,j}(\tau)$
\begin{multline}
  \label{eq:288}
  \sum_{j\geq 0} \left[ (d-1) \delta_{kj} + \inner{e_k}{\frac {1}{2}
      \sin2x \, e_j' - \cos 2x\, e_j}\right] \hat{A}_{\lambda,j} =
  \\
  \frac{1}{4} \inner{e_k}{\coef{\lambda}(\sin 2x)^2 A \left(\Phi^2 +
      \Pi^2\right)}, \quad k\in\mathbb{N}_{0},
\end{multline}
(here $\delta_{ij}$ stands for the Kronecker delta).  It is useful to
note that the principal matrix of this system is tridiagonal. This
system, supplied with the boundary condition coming from (\ref{eq:83})
\begin{equation}
  \label{eq:289}
  \left. \coef{\lambda}(1-A)
  \right|_{x=0} = \sum_{j\geq 0} \hat{A}_{\lambda,j} e_j(0) = 0,
\end{equation}
allows for a unique solution for the coefficients
$\hat{A}_{\lambda,j}(\tau)$.

Then, for odd $\lambda \geq 3$, $\phi_{\lambda}$ fulfills an
inhomogeneous wave equation on the pure AdS$_{d+1}$ background
(\ref{eq:272}).  Projecting this equation onto $e_k(x)$ one finds that
the coefficients $\hat{\phi}_{\lambda,k}$ in (\ref{eq:284}) behave as
forced harmonic oscillators
\begin{equation}
  \label{eq:290}
  \left(\omega_{\gamma}^2\,\partial^{2}_{\tau} +
    \omega_{k}^2\right) \hat{\phi}_{\lambda,k} =
  \inner{e_k}{S_{\lambda}},
  \quad k\in\mathbb{N}_{0}.
\end{equation}
Solving these, we get two integration constants for each of the
equations
\begin{equation}
  \label{eq:291}
  \hat{\phi}_{\lambda,k}(0) = c_{\lambda,k},
  \quad \partial_{\tau}\hat{\phi}_{\lambda,k}(0) = \tilde{c}_{\lambda,k},
  \quad k\in\mathbb{N}_{0}.
\end{equation}
Because of the form of the lowest order perturbative expansion
(\ref{eq:263}) and because of the freedom we have to define the
perturbative parameter $\ep$, to meet (\ref{eq:267}) we set
\begin{equation}
  \label{eq:292}
  \hat{\phi}_{\lambda,\gamma}(0) = 0, \quad
  \partial_{\tau}\hat{\phi}_{\lambda,\gamma}(0) = 0,
\end{equation}
which fixes two integration constants in (\ref{eq:291}), namely
\begin{equation}
  \label{eq:293}
  c_{\lambda,\gamma} = \tilde{c}_{\lambda,\gamma}=0.
\end{equation}
In this way the dominant mode $e_{\gamma}(x)$ is present only at the
linear order approximation (\ref{eq:263}).  It turns out that at each
perturbative order all $\hat{\phi}_{\lambda,k}$ tune in phase to the
dominant mode, i.e. the following holds
\begin{equation}
  \label{eq:294}
  \partial_{\tau}\hat{\phi}_{\lambda,k}\bigr|_{\tau=0}
  \sim \partial_{\tau}\hat{\phi}_{\lambda,\gamma}\bigr|_{\tau=0},
  \quad k\in\mathbb{N}_{0}, \quad k\neq\gamma,
\end{equation}
so the choice (\ref{eq:293}) reduces by the factor of two the number
of integration constants by setting all of $\tilde{c}_{\lambda,k}$ in
(\ref{eq:291}) to zero.

Now, if $\inner{e_k}{S_{\lambda}}$ in (\ref{eq:290}) contains the
resonant terms $\cos(\tau\,\omega_{k}/\omega_{\gamma})$ (there are no
terms like $\sin(\tau\,\omega_{k}/\omega_{\gamma})$ present in
$\inner{e_k}{S_{\lambda}}$, because we have set all
$\tilde{c}_{\lambda,k}$ to zero already) this gives rise to secular
terms $\tau\sin(\tau\,\omega_{k}/\omega_{\gamma})$ in
$\hat{\phi}_{\lambda,k}(\tau)$.  Such terms would spoil the
periodicity and lead to the unbounded solution; thus, they have to be
removed.  This fixes the correction to the frequency
$\omega_{\gamma, \lambda-1}$, and the integration constants
$c_{\lambda,k}$ in (\ref{eq:291}).  Namely, it turns out that in order
not to produce spurious resonant terms in higher perturbative orders,
all but odd (in $\tau$) frequencies in the solutions for
$\hat{\phi}_{\lambda,k}$ have to be removed.  Therefore the
coefficients in front of $\cos(2i\,\tau)$, $i\in\mathbb{N}$, and
$\cos(i/j\,\tau)$ with $i,j\in\mathbb{N}$, $(i \bmod j)\neq 0$, have
to be set to zero.  These conditions fix uniquely most of the
integration constants $c_{\lambda,k}$ in (\ref{eq:291}).

The number of the essential integration constants in (\ref{eq:291})
remaining after removal of all spurious resonances is easy to
determine.  Since at any order $\lambda\geq 3$ the source function
$S_{\lambda}$ in (\ref{eq:272}) is a linear combination of finite
number of eigenmodes which means that we have
\begin{equation}
  \label{eq:295}
  \inner{e_{k}}{S_{\lambda}} \equiv 0, \quad k>k_{*},
\end{equation}
with $k_{*}=(\lambda-1)(d+1)/2 + \lambda\gamma$, the number of
possible resonances is also finite. The resonant frequencies are those
which are odd multiples of the fundamental frequency
$\omega_{\gamma}$, i.e. frequencies
\begin{equation}
  \label{eq:296}
  \{(2k+1)\omega_{\gamma} \ |\ k=0,1,\ldots, K\},
\end{equation}
where $K$ is determined from the condition for the largest possible
frequency present in $S_{\lambda}$
\begin{equation}
  \label{eq:297}
  (2K+1)\omega_{\gamma} \leq \omega_{k_{*}},
\end{equation}
(finite number of eigenmodes implies a finite number of frequencies
present in $S_{\lambda}$).  Therefore the expected number of
resonances is $K+1$ with (from the condition (\ref{eq:297}))
\begin{equation}
  \label{eq:298}
  K =
  \left\lfloor\frac{\omega_{k_{*}}}{2\omega_{\gamma}}-\frac{1}{2}\right\rfloor
  = \left\lfloor\frac{\lambda - 1}{2}
    \left(\frac{1}{\omega_{\gamma}}+1\right)\right\rfloor.
\end{equation}
Actually the number of integration constants is $K$ since always for
the $k=0$ case in (\ref{eq:296}) the corresponding integration
constant is fixed by the normalization condition, in our case by the
requirement (\ref{eq:267}).  At the same time there is an exceptional
cancellation present in $S_{\lambda}$ at any $\lambda$ (as pointed out
in Section~\ref{sec:AdSPeriodicPerturbative-Odd}) and thus the number
of available constants is exactly equal to the number of resonant
frequencies appearing at order $\lambda+2$.  These undetermined
integration constants will be fixed together with
$\omega_{\gamma,\lambda+1}$ to remove $(\lambda+1)/2 + \lfloor
(\lambda-1)/(2\omega_{\gamma})\rfloor$ secular terms present in
$\phi_{\lambda+2}$.  Therefore we can continue this procedure to
follow the same steps in higher order of perturbative expansion,
removing all of appearing resonances by fixing all of the available
parameters leading to a unique time-periodic solution.
\subsubsection{Integrals}
The advantage of using the decomposition (\ref{eq:284})-(\ref{eq:286})
is clearly visible when performing actual calculations.  In fact, all
the projections onto $e_k$ (or $e'_k$) appearing at any order of the
perturbative procedure described above, can be reduced to just a few
inner products: $\inner{e_k}{e_i\,e_j}$, $\inner{e_k}{\cos{2x}\,e_i}$,
$\inner{e_k}{\sin{2x}\,e_i'}$,
$\inner{e_k}{\csc{x}\,\sec{x}\,\cos{2x}\,e_i\,e_j'}$,
$\inner{e_k}{\csc{x}\,\sec{x}\,e_i\,e_j'}$.  Thus, the whole procedure
of building up such a perturbative solution is relatively easy to
implement.

The products appearing in (\ref{eq:287}), (\ref{eq:288}) and
(\ref{eq:290}) can be expressed in terms of finite sums (for even $d$)
\begin{equation}
  \label{eq:299}
  e_{i}(x)e_{j}'(x) = \sum_{k=\max(0,|j-i|-d/2)}^{i+j+d/2}
  \inner{e_{k}'}{e_{i}\,e_{j}'} \frac{e_{k}'(x)}{\omega_{k}^{2}},
\end{equation}

\begin{equation}
  \label{eq:300}
  \sin{2x}\,e_{i}(x)e_{j}(x) = \sum_{k=\max(0,|i-j|-d/2-1)}^{i+j+d/2+1}
  \inner{e_{k}'}{\sin{2x}\,e_{i}\,e_{j}}\frac{e_{k}'(x)}{\omega_{k}^{2}},
\end{equation}

\begin{equation}
  \label{eq:301}
  \sin{2x}\,e_{i}'(x) = \sum_{j=\max(0,i-1)}^{i+1}
  \inner{e_{j}}{\sin{2x}\,e_{i}'}e_{j}(x),
\end{equation}

\begin{equation}
  \label{eq:302}
  \sin{2x}\,e_{i}'(x)e_{j}'(x) = \sum_{k=\max(0,|i-j|-d/2-1)}^{i+j+d/2+1}
  \inner{e_{k}'}{\sin{2x}\,e_{i}'\,e_{j}'} \frac{e_{k}'(x)}{\omega_{k}^{2}},
\end{equation}

\begin{multline}
  \label{eq:303}
  \cos{2x}\csc{x}\sec{x}\,e_{i}(x)e_{j}'(x) =
  \\
  \sum_{k=\max(0,i-j-d/2)}^{i+j+d/2}
  \inner{e_{k}}{\cos{2x}\,\csc{x}\,\sec{x}\,e_{i}\,e_{j}'}e_{k}(x),
\end{multline}

\begin{equation}
  \label{eq:304}
  \csc{x}\sec{x}\,e_{i}(x)e_{j}'(x) = \sum_{k=\max(0,i-j-d/2+1)}^{i+j+d/2-1}
  \inner{e_{k}}{\csc{x}\,\sec{x}\,e_{i}\,e_{j}'} e_{k}(x)
\end{equation}
and
\begin{equation}
  \label{eq:305}
  e_{i}(x)e_{j}(x) = \sum_{k=\max(0,|i-j|-d/2)}^{i+j+d/2}
  \inner{e_{k}}{e_{i}\,e_{j}}e_{k}(x),
\end{equation}
with the expansion coefficients calculated in a way presented in
Appendix~\ref{cha:inter-coeff}.  Explicitly, the projections in
(\ref{eq:299})-(\ref{eq:305}) and these present in (\ref{eq:288})
computed with respect to the inner product (\ref{eq:91})
are\footnote{The integral (\ref{eq:306}) is an exception here, since
  instead of using the method of Appendix~\ref{cha:inter-coeff}
  directly (which would give a very complicated sum) it can be reduced
  to $\inner{e_{k}}{e_{i}\,e_{j}}$ by integration by parts and using
  the orthogonality property of $e_{j}'(x)$ and symmetry with respect
  to indices.}
\begin{equation}
  \label{eq:306}
  \inner{e_{k}}{e_{i}'\,e_{j}'} = \frac{1}{2}\left(\omega_{i}^{2} +
    \omega_{j}^{2} - \omega_{k}^{2}\right) \inner{e_{k}}{e_{i}\,e_{j}},
\end{equation}

\begin{multline}
  \label{eq:307}
  \inner{e_{k}}{e_{i}\,e_{j}} =
  \mathcal{N}_{i}\,\mathcal{N}_{j}\,\mathcal{N}_{k} \sum_{s=0}^i
  \sum_{r=0}^j \sum_{q=0}^k \Biggl[ (-1)^{i+j+k-(q+r+s)}
  \\
  \times \binom{i+\alpha}{s} \binom{i+\beta}{i-s} \binom{j+\alpha}{r}
  \binom{j+\beta}{j-r} \binom{k+\alpha}{q} \binom{k+\beta}{l-q}
  \\
  \times \frac{\Gamma (d+q+r+s+1)
    \Gamma\left(\frac{d}{2}+i+j+k-q-r-s\right)}{2
    \Gamma\left(\frac{3d}{2}+i+j+k+1\right)} \Biggr],
\end{multline}

\begin{equation}
  \label{eq:308}
  \inner{e_{k}'}{\sin{2x}\,e_{i}\,e_{j}} = \sum_{l=\max(0,k-1)}^{k+1}
  \inner{\sin{2x}\,e_{k}'}{e_{l}}\inner{e_{l}}{e_{i}\,e_{j}},
\end{equation}

\begin{multline}
  \label{eq:309}
  \inner{e_{j}}{\sin{2x}\,e_{i}'} =
  \mathcal{N}_{i}\,\mathcal{N}_{j}\,\Biggl\{ -d \sum_{s=0}^i
  \sum_{r=0}^j \Biggl[ (-1)^{i+j-(s+r)}
  \\
  \times \binom{i+\alpha}{s} \binom{i+\beta}{i-s} \binom{j+\alpha}{r}
  \binom{j+\beta}{j-r}
  \\
  \times \frac{\Gamma\left(\frac{d}{2}+r+s+1\right)
    \Gamma\left(\frac{d}{2}+i+j-r-s+1\right)}{\Gamma(d+i+j+2)} \Biggr]
  \\
  - 2 (\alpha +\beta +i+1) \sum_{s=0}^{i-1} \sum_{r=0}^j \Biggl[
  (-1)^{i+j-(s+r+1)}
  \\
  \times \binom{i+\alpha+1-1}{s} \binom{i+\beta+1-1}{i-s-1}
  \binom{j+\alpha}{r} \binom{j+\beta}{j-r}
  \\
  \times \frac{\Gamma\left(\frac{d}{2}+r+s+2\right)
    \Gamma\left(\frac{d}{2}+i+j-r-s\right)}{\Gamma(d+i+j+2)} \Biggr]
  \Biggr\},
\end{multline}

\begin{equation}
  \label{eq:310}
  \inner{e_{k}'}{\sin{2x}\,e_{i}'\,e_{j}'} = \sum_{l=\max(0,i-1)}^{i+1}
  \inner{\sin{2x}\,e_{i}'}{e_{l}}\inner{e_{l}}{e_{j}'\,e_{k}'},
\end{equation}

\begin{multline}
  \label{eq:311}
  \inner{e_{k}}{\cos{2x}\,\csc{x}\,\sec{x}\,e_{i}\,e_{j}'} =
  \mathcal{N}_{i}\,\mathcal{N}_{j}\,\mathcal{N}_{k} \Biggl\{ -d
  \sum_{s=0}^i \sum_{r=0}^j \sum_{q=0}^k \Biggl[ (-1)^{i+j+k-(s+r+q)}
  \\
  \times \binom{i+\alpha}{s} \binom{i+\beta}{i-s} \binom{j+\alpha }{r}
  \binom{j+\beta }{j-r} \binom{k+\alpha}{q} \binom{k+\beta}{k-q}
  \\
  \times \Bigl(d-2 (i+j+k-2(q+r+s))\Bigr)
  \\
  \times
  \frac{\Gamma(d+q+r+s)\Gamma\left(\frac{d}{2}+i+j+k-q-r-s\right)}
  {4\Gamma\left(\frac{3d}{2}+i+j+k+1\right)} \Biggr]
  \\
  + 2 (\alpha +\beta +j+1) \sum_{s=0}^i \sum_{r=0}^{j-1} \sum_{q=0}^k
  \Biggl[ (-1)^{i+j+k-(s+r+q)}
  \\
  \times \binom{i+\alpha}{s} \binom{i+\beta}{i-s}
  \binom{j+\alpha+1-1}{r} \binom{j+\beta+1-1}{j-r-1}
  \binom{k+\alpha}{q} \binom{k+\beta}{k-q}
  \\
  \times \Bigl(d-2 (i+j+k-2 (q+r+s+1))\Bigr)
  \\
  \times \frac{\left(\Gamma(d+q+r+s+1)
      \Gamma\left(\frac{d}{2}+i+j+k-q-r-s-1\right)\right)}{4
    \Gamma\left(\frac{3d}{2}+i+j+k+1\right)} \Biggr] \Biggr\},
\end{multline}

\begin{multline}
  \label{eq:312}
  \inner{e_{k}}{\csc{x}\,\sec{x}\,e_{i}\,e_{j}'} =
  \mathcal{N}_{i}\,\mathcal{N}_{j}\,\mathcal{N}_{k} \Biggl\{ -d
  \sum_{s=0}^i \sum_{r=0}^j \sum_{q=0}^k \Biggl[ (-1)^{i+j+k-(s+r+q)}
  \\
  \times \binom{i+\alpha}{s} \binom{i+\beta}{i-s} \binom{j+\alpha}{r}
  \binom{j+\beta}{j-r} \binom{k+\alpha}{q} \binom{k+\beta}{k-q}
  \\
  \times \frac{\Gamma(d+q+r+s)
    \Gamma\left(\frac{d}{2}+i+j+k-q-r-s\right)}{2
    \Gamma\left(\frac{3d}{2}+i+j+k\right)} \Biggr]
  \\
  -2 (\alpha +\beta +j+1) \sum_{s=0}^i \sum_{r=0}^{j-1} \sum_{q=0}^k
  \Biggl[ (-1)^{i+j+k-(s+r+q+1)}
  \\
  \times \binom{i+\alpha}{s} \binom{i+\beta}{i-s}
  \binom{j+\alpha+1-1}{r} \binom{j+\beta+1-1}{j-r-1}
  \binom{k+\alpha}{q} \binom{k+\beta}{k-q}
  \\
  \times \frac{\Gamma(d+q+r+s+1)
    \Gamma\left(\frac{d}{2}+i+j+k-q-r-s-1\right)}{2
    \Gamma\left(\frac{3d}{2}+i+j+k\right)} \Biggr] \Biggr\},
\end{multline}

\begin{multline}
  \label{eq:313}
  \inner{e_{j}}{\cos{2x}\,e_{i}} = \mathcal{N}_{i}\,\mathcal{N}_{j}
  \sum_{s=0}^{i} \sum_{r=0}^{j} \Biggl[ (-1)^{i+j-(r+s+1)}
  \\
  \times \binom{i+\alpha}{s} \binom{i+\beta}{i-s} \binom{j+\alpha}{r}
  \binom{j+\beta}{j-r}
  \\
  \times \frac{1}{2 \Gamma(d+i+j+2)} \Bigl(\Gamma(r+s+\beta+1)
  \Gamma(i+j-(r+s)+\beta+1)
  \\
  - \Gamma(r+s+\beta+2) \Gamma(i+j-(r+s)+\beta)\Bigr) \Biggr],
\end{multline}
where by $\mathcal{N}_{j}$ we denote the normalization factor coming
from the eigenbasis functions (\ref{eq:90})
\begin{equation}
  \label{eq:314}
  \mathcal{N}_{j} = 2 \frac{\sqrt{j!(j+d-1)!}}
  {\gamma\left(j+\frac{d}{2}\right)},
\end{equation}
and we also use $\alpha=d/2-1$, $\beta=d/2$ to shorten the notation.

The important properties of the integrals
(\ref{eq:306})-(\ref{eq:313}) are their symmetries like e.g.
\begin{equation}
  \label{eq:315}
  \inner{e_{k}}{e_{i}'e_{j}'} = \inner{e_{k}}{e_{j}'e_{i}'}
  \quad \forall\,i,j,k\in\mathbb{N}_{0},
\end{equation}
which are worth noting when performing calculations.

\subsection{Numerical evolution scheme}
\label{sec:AdSEvolution}

The evolution code presented in this section can be also used for the
complex field case, therefore to retain generality throughout this
section we keep the absolute values of complex quantities where they
are necessary, and we treat the scalar field and its momenta as
complex valued functions.

The substantial parts of methods presented here are used not only to
solve the Cauchy problem but also for finding the time-periodic and
standing wave solutions to the system (\ref{eq:67})-(\ref{eq:70}) with
real and complex scalar field respectively.

For the clarity of presentation we discuss first the method relying on
the use of eigenbasis functions (\ref{eq:90}), since it is simpler
than the second method which is based on the Chebyshev polynomials.
The drawback of using the eigenfunctions is that this approach is
applicable only in even number of space dimensions.  This is due to
the incompatibility of boundary expansion as we stressed in
Section~\ref{sec:AdSScalarEigen}.  Only for even $d$ the boundary
expansion of eigenfunctions and dynamical fields are compatible and
only then the use of such expansion is justified.  For this reason we
also develop a pseudospectral method, which is not limited in use by
the boundary behaviour of the approximated functions.

We use the MOL approach with the pseudospectral discretization in
space to solve the initial-value problem of the system
(\ref{eq:67})-(\ref{eq:69}) by using constrained evolution scheme,
i.e. we do not use explicitly the momentum constraint equation
(\ref{eq:70}) to advance metric function $A$.  Therefore, the
substantial part of the description of our methods is devoted to the
discussion of how to solve the constraint equations effectively.

\subsubsection{Eigenbasis expansion}
\label{sec:AdSEvolutionEigenbasis}

We expand both scalar fields $\phi(t,x)$ and $\Pi(t,x)$ into $N$
eigenmodes of the linear problem (\ref{eq:90})
\begin{equation}
  \label{eq:316}
  \phi(t,x) = \sum_{j=0}^{N-1}\hat{\phi}_{j}(t)\,e_{j}(x),
  \qquad \Pi(t,x) = \sum_{j=0}^{N-1}\hat{\Pi}_{j}(t)\,e_j(x),
\end{equation}
(this scheme is identical for both complex and real field, in the
former case the expansion coefficients are complex valued functions,
in effect the size of resulting ODE system is two times larger
compared to the real case, since we need to evolve in time both real
and imaginary parts of the dynamical fields).  We require for the
equations (\ref{eq:67}) to be identically satisfied at the set of $N$
collocation points chosen to be
\begin{equation}
  \label{eq:317}
  \Bigl\{x_{i}\in (0,\pi/2)\,\Bigl|\ e_{N}(x_{i})=0,
  \ i=0,1,\ldots N-1\Bigr\}.
\end{equation}
For convenience, instead of evolving in time the values of the
dynamical fields at the discrete set of spatial grid points, we evolve
their eigenbasis expansion coefficients.  To calculate time
derivatives of $\hat{\phi}_{j}(t)$ and $\hat{\Pi}_{j}(t)$ instead of
using (\ref{eq:67}) we take the following equivalent pair of dynamical
equations
\begin{align}
  \label{eq:318}
  \dot{\phi} &= e^{-\delta}A\Pi\,,
  \\
  \label{eq:319}
  \dot{\Pi} &= e^{-\delta}\left( A\Phi' +
    \frac{d-1-(1-A)\cos{2x}}{\sin{x}\cos{x}}\Phi\right),
\end{align}
where we have eliminated spatial derivatives of the metric functions
$\delta$ and $A$ by using the constraint equations (it is essential
since we know how to differentiate quantities which can be decomposed
in a chosen set of basis functions, to retain spectral convergence
these have to be in the same class of functions having the same
boundary behaviour).  We know, \textit{a posteriori}, that the metric
function $\delta$ and the integrand in (\ref{eq:78}) can be
efficiently approximated\footnote{By efficiency we mean that for
  smooth function expanded in a basis with compatible boundary
  behaviour the the expansion coefficients decay rapidly
  (exponentially) with wave number.} as follows
\begin{align}
  \label{eq:320}
  \delta(t,x) &= \sum_{j=0}^{N-1}\hat{\delta}_{j}(t)\cos(2 j x),
  \\
  \label{eq:321}
  e^{-\delta}\left(\left|\Phi\right|^2 + \left|\Pi\right|^2\right) &=
  \sum_{j=0}^{N-1}\tilde{A}_{j}(t)e_{j}(x).
\end{align}
Substituting (\ref{eq:320}) into (\ref{eq:68}) we get
\begin{equation}
  \label{eq:322}
  \sum_{j=1}^{N-1} (-2 j) \sin(2 j x) \hat{\delta}_j(t) =
  -\sin{x}\cos{x}\left(\left|\Phi\right|^2 + \left|\Pi\right|^2\right).
\end{equation}
Then, equation (\ref{eq:322}) evaluated at the set of collocation
points (\ref{eq:317}) together with one extra condition for the
remaining unknown $\hat{\delta}_{0}(t)$ (as it is absent in this
system, reflecting the gauge freedom) which is $\delta(t,0) =
\sum_{j=0}^{N-1}\hat{\delta}_{j}(t) = 0$, forms a closed linear system
for the Fourier coefficients of the $\delta$ function.  Similarly,
evaluating the sides of (\ref{eq:321}) at the same set of collocation
points we get the linear system of equations to be solved for the
expansion coefficients $\tilde{A}_{j}(t)$.  This allows us to solve
for the metric function $A$, which using (\ref{eq:78}) is approximated
as
\begin{equation}
  \label{eq:323}
  A(t,x) = 1 - e^{\delta}\frac{\cos^{d}x}{\sin^{d-2}x}
  \sum_{j=0}^{N-1}w^{(d)}_{j}(x)\tilde{A}_{j}(t),
\end{equation}
where the weight functions $w^{(d)}_{j}(x)$ read\footnote{This
  integral can be calculated by the change of variables $z=\cos{2y}$
  and use of the integral of Jacobi polynomials (\ref{eq:596}).  By
  the relation (\ref{eq:595}) the result can be expressed in terms of
  Jacobi polynomials again.}
\begin{multline}
  \label{eq:324}
  w^{(d)}_j(x) = \int_0^{x}e_{j}(y)\tan^{d-1}y\diff y =
  \\
  \frac{\sqrt{j!(d+j-1)!}}{(d/2+j)!}
  \sin^{d}\!x\,P^{(d/2,d/2-1)}_{j}(\cos{2x}).
\end{multline}
Finally, substituting the expansions (\ref{eq:316}) into the wave
equation (\ref{eq:318}), (\ref{eq:319}) and evaluating both sides at
the collocation points we get the linear system of equations to be
solved for the time derivatives of $\hat{\phi}_j(t)$ and
$\hat{\Pi}_j(t)$.

This scheme differs slightly from the one presented in \cite{MRPRL} by
the choice of collocation points and the form of expansion in
(\ref{eq:321}).  The first modification is motivated by the
universality of our method.  Previously used collocation points where
simple to obtain (given by an analytic formula) and good enough for
only moderate values of space dimension $d$ (note the explicit
dependence of basis functions (\ref{eq:90}) on this parameter), while
for large $d$ the matrices which appear in this pseudospectral
approach are ill-conditioned since have a large spectral condition
number (\cite{ConditionNumber}, see also
\cite[431]{ralston2001first}).  The desire to increase of numerical
stability and accuracy for large $d$ induced us to use the formula
(\ref{eq:317}) (which does not affect the small $d$ cases, it slightly
lowers the condition number though).  The second change, the expansion
of the integrand (\ref{eq:321}), is very convenient here, since then
the weights (\ref{eq:324}) are easy to get (using the integral
relation for the Jacobi functions, see
Section~\ref{sec:AppJacobiPolynomials} in Appendix), as opposed to the
cosines used in \cite{MRPRL}, where the integrals have to be computed
for each $d$ separately and in fact are rather hard to get.  Together,
these modifications lead to a more universal (in terms of $d$) and
more stable numerical scheme.

The time integration of the system (\ref{eq:318})-(\ref{eq:319})
together with the above method for solving the constraints
(\ref{eq:320})-(\ref{eq:324}) is carried by use of the implicit
Gauss-Runge-Kutta method.  The choice between explicit and implicit RK
time integration method is dictated by the intended application of
presented pseudospectral discretization method.  For the long
time-evolution of initial conditions which do not develop
singularities (whence such for which we do not expect black hole
formation) moderate number of eigenmodes in (\ref{eq:316}) is needed
to ensure precise approximation in such cases.  Therefore, the cost of
solving implicit equations for the internal stages of RK process is
moderate compared to the overall cost of determining the RHS of
dynamical equations (\ref{eq:318}) and (\ref{eq:319}).  Moreover, this
cost is compensated by the properties of the symplectic
Guass-Legendre-RK methods as stressed in the Appendix
\ref{sec:AppIRK}.  The conservation of integrals of motions (conserved
mass and charge in the case of the complex scalar field) makes these
integrators especially applicable for the problems being considered.

The total conserved mass, given by the integral (\ref{eq:75}), can be
expressed\footnote{This is possible due to the linearity (in terms of
  dynamical variables) of the field equations.} as the Parseval sum,
i.e.
\begin{equation}
  \label{eq:325}
  M = \sum_{j=0}^{N-1} E_{j},
\end{equation}
where the $E_{j}$ is defined as
\begin{equation}
  \label{eq:326}
  E_{j} = \left|\tilde{\Pi}_{j}\right|^{2} +
  \omega_{j}^{-2}\left|\tilde{\Phi}_{j}\right|^{2},
\end{equation}
with
\begin{equation}
  \label{eq:327}
  \tilde{\Phi}_{j} := \inner{e_{j}'}{\sqrt{A}\,\Phi},
  \qquad \tilde{\Pi}_{j} := \inner{e_{j}}{\sqrt{A}\,\Pi}.
\end{equation}
The quantity $E_{j}$ can be interpreted as the energy of the mode
$e_{j}(x)$.  The projections (\ref{eq:327}) can be easily computed
numerically since the matrices which need to be inverted are constant
and can be factorized in advance (using LU algorithm).  For the
complex scalar field case the conserved charge defined in
(\ref{eq:80}) can be calculated in the following way.  Noting that the
integrand (the $\phi\,\Pi^{*}$ term) has the same structure of Taylor
expansion at $x=\pi/2$ as the eigenbasis functions in the case of even
$d$, it can be written as
\begin{equation}
  \label{eq:328}
  - \Im \phi\,\Pi^{*} = \sum_{j=0}^{N-1} \hat{q}_{j} e_{j}(x).
\end{equation}
Then, plugging (\ref{eq:328}) into (\ref{eq:80}) we obtain
\begin{equation}
  \label{eq:329}
  Q = \int_{0}^{\pi} \sum_{j=0}^{N-1}\hat{q}_{j} e_{j}(x) \tan^{d-1}{x}\diff x =
  \sum_{j=0}^{N-1} \hat{q}_{j} \int_{0}^{\pi} e_{j}(x) \tan^{d-1}{x}\diff x,
\end{equation}
where the weighted integral of the eigenfunctions can be easily
calculated using the integral of the Jacobi polynomials
(\ref{eq:324}), and yields
\begin{equation}
  \label{eq:330}
  \int_{0}^{\pi} e_{j}(x) \tan^{d-1}{x}\diff x =
  \frac{2(-1)^j}{(d+2j)\Gamma\left(d/2\right)}\sqrt{\frac{(d+j-1)!}{j!}}.
\end{equation}
The conservation of the discrete versions of the total mass and charge
by the proposed method is demonstrated in subsequent sections.

\subsubsection{Polynomial expansion}
\label{sec:AdSEvolutionCheb}

Due to incompatibility of eigenbasis functions with regularity
conditions at $x=\pi/2$ for odd $d$, as stressed in
Section~\ref{sec:AdSScalarEigen}, the method of the previous section
is not applicable.  In principle, it can be applied to the odd $d$
cases but then it results in slow (only polynomial) convergence and
thus is highly inefficient.  Therefore we use a polynomial
pseudospectral method where instead of the eigenfunctions we take
Chebyshev polynomials rescaled to the interval $[-\pi/2,\pi/2]$ as
basis functions for the expansion of the scalar field (we are using
double covering method which is described in detail in the
Appendix~\ref{cha:polyn-pseud-meth}).  The property of having no
specified symmetry at the boundary makes Chebyshev polynomials
specially useful for problems with generic boundary conditions
\cite{boyd2001chebyshev}.  Therefore, the proposed discretization
method in space for the system (\ref{eq:67})-(\ref{eq:69}) is
applicable for both even and odd number of space dimensions.

The straightforward replacement of expansion functions in the method
of previous section, although possible and at the first sight obvious
step, leads to an unstable numerical scheme.  The first problem we
encounter is solving the Hamiltonian constraint.  Using the analogues
of (\ref{eq:321}), (\ref{eq:323}) and (\ref{eq:324}) with $e_{j}(x)$
replaced by even or odd Chebyshev polynomials
$T_{j}\left(2/\pi\,x\right)$ (taking care of the symmetry of the
integrand in (\ref{eq:321})) generates spurious oscillations near the
origin, especially for large values of $d$ ($d>4$); even if it does
not give easily seen effects for smaller values of $d$ this does not
mean that there are no serious problems with this approach.  Therefore
we use the formula (\ref{eq:77}) and write
\begin{equation}
  \label{eq:331}
  \left(\frac{\sin^{d-2}{x}}{\cos^{d}{x}}Ae^{-\delta}\right)' =
  \left(\frac{\sin^{d-2}{x}}{\cos^{d}{x}}\right)'e^{-\delta}.
\end{equation}
Then, introducing the following combination of metric functions (noted
previously to be useful in Section~\ref{sec:BoxEvolution})
\begin{equation}
  \label{eq:332}
  B := A e^{-\delta},
\end{equation}
instead of the function $A$ itself, we rewrite (\ref{eq:331}) as
equivalent of Hamiltonian constraint (\ref{eq:69})
\begin{equation}
  \label{eq:333}
  B + \frac{\sin{x}\cos{x}}{d-2 + 2\sin^{2}{x}}B' = e^{-\delta}.
\end{equation}
The use of metric variable $B$ not only reduce complexity of the
overall algorithm, since this is the only way the metric functions
enter the evolution equations (\ref{eq:67}), it also removes the
numerical instability otherwise occurring near the origin.

The second and more serious problem concerns the instability appearing
near the outer boundary ($x=\pi/2$) which is especially manifest for
large values of $d$ and which gets greatly amplified with the
increasing number of grid points (an effect indicating a numerical
instability).  This can be understood by looking at the equations we
are solving and knowing the properties of methods we are using.  Due
to the regularity conditions the scalar field falls as $x$ goes to
$\pi/2$ more rapidly with increasing $d$, with the falloff rate given
in (\ref{eq:85}) and (\ref{eq:86}).  Then the appearing divisions of
small numbers when $x\ra \pi/2$ and the clustering of Chebyshev grid
points near the boundary is the reason for observed spurious
oscillations with rapidly growing amplitude.  To overcome this we
rescale dependent variables---both of the scalar fields $\Phi$ and
$\Pi$---by taking into account their fall off behaviour near $x=\pi/2$
(as in \cite{Buchel2012}) and we define
\begin{equation}
  \label{eq:334}
  \Psi := \frac{\Phi}{\cos^{d-2}x}, \quad \Xi := \frac{\Pi}{\cos^{d-1}x}.
\end{equation}
In terms of these new variables the evolution equations (\ref{eq:67})
read
\begin{align}
  \label{eq:335}
  \dot{\Psi} &= \frac{1}{\cos^{d-2}{x}}\left(\cos^{d-1}{x}B\Xi\right)',
  \\
  \label{eq:336}
  \dot{\Xi} &= \frac{1}{\sin^{d-1}{x}}
  \left(\frac{\sin^{d-1}{x}}{\cos{x}}B\Psi\right)',
\end{align}
while the slicing condition (\ref{eq:68}) takes the form
\begin{equation}
  \label{eq:337}
  \delta' = -\sin{x}\cos^{2d-3}{x}\left(\left|\Psi\right|^{2}
    + \cos^{2}{x}\left|\Xi\right|^{2}\right).
\end{equation}
The Dirichlet boundary condition at $x=\pi/2$
\begin{equation}
  \label{eq:338}
  \Phi(t,\pi/2) = 0, \quad \Pi(t,\pi/2) = 0,
\end{equation}
(following from regularity conditions (\ref{eq:85}) and (\ref{eq:86}))
translates to the same conditions for new fields
\begin{equation}
  \label{eq:339}
  \Psi(t,\pi/2) = 0, \quad \Xi(t,\pi/2) = 0.
\end{equation}
The rescaling (\ref{eq:334}) does not affect regularity conditions at
the origin (\ref{eq:83}), i.e. $\Psi$ and $\Xi$ are odd and even
functions at $x=0$ respectively.  This redefinition of the dynamical
fields is crucial for stable numerical scheme with Chebyshev
pseudospectral discretization, and together with use of (\ref{eq:333})
as Hamiltonian constraint equation leads to a stable (for long
evolution times) numerical scheme.

In the case of local methods, such as FDA, the numerical instabilities
discussed above can be eliminated without using the rescaling
(\ref{eq:334}).  Instead, the l'Hopital rule is used on a portion of
computational grid where the equations are most singular.  This
approach was used in the previous works \cite{br,jrb} and it is
described in detail in \cite{MRIJMPA}.  For the global interpolation
methods, the overhead of computing the additional quantities is large
compared to local approach, therefore such rescaling is preferred.

We discretize equations (\ref{eq:335})-(\ref{eq:337}) and
(\ref{eq:333}) using Chebyshev pseudospectral method in spherical
symmetry together with the barycentric interpolation formula (as is
described in Appendix~\ref{cha:polyn-pseud-meth}; using the notation
for the Chebyshev weights $w_{i}$ and differentiation matrices
$D^{(n,\pm)}$) as follows.  The constraint equations are given as a
solution to the algebraic systems, which are derived similarly as for
the cavity model derived in Section~\ref{sec:BoxEvolution}.  Using
$N+1$ radial Chebyshev points (\ref{eq:635}) scaled to the range of
global radial coordinate of AdS
\begin{equation}
  \label{eq:340}
  x_{i} = \frac{\pi}{2}\cos\left(\frac{i\pi}{2N+1}\right), \quad
  i=0,1,\ldots,N,
\end{equation}
and introducing the shorthand notation $\delta_{i}$ for function value
at $i$-th grid point $\delta(t,x_{i})$, similarly for $B$, we have
\begin{equation}
  \label{eq:341}
  \sum_{j=0}^{N}D^{(1,+)}_{ij}\delta_{j}
  = -\sin{x_{i}}\cos^{2d-3}{x_{i}}\left(\left|\Psi_{i}\right|^{2} +
    \cos^{2}{x_{i}}\left|\Xi_{i}\right|^{2} \right),
\end{equation}
\begin{equation}
  \label{eq:342}
  \sum_{j=0}^{N}\left(\id_{ij} +
    \frac{\sin{x_{j}}\cos{x_{j}}}{d-2+2\sin^{2}{x_{j}}}D^{(1,+)}_{ij}\right)B_{j}
  = e^{-\delta_{i}},
\end{equation}
($i = 1,\ldots,N$) accompanied with a discrete versions of boundary
conditions $\delta(t,0)=0$
\begin{equation}
  \label{eq:343}
  \sum_{i=0}^{N}\frac{w_{i}}{x_{i}}\delta_{i} = 0,
\end{equation}
and $B(t,0)=1$
\begin{equation}
  \label{eq:344}
  \frac{{\displaystyle\sum_{i=0}^{N}\frac{w_{i}}{x_{i}}B_{i}}}
  {\displaystyle\sum_{i=0}^{N}\frac{w_{i}}{x_{i}}} = 1.
\end{equation}
The time derivatives of the scalar fields $\Psi$ and $\Xi$ are coded
using the expanded forms of (\ref{eq:335}) and (\ref{eq:336})
\begin{align}
  \label{eq:345}
  \dot{\Psi}_{i} &= \cos{x_{i}}\left(\sum_{j=0}^{N}
    D^{(1,+)}_{ij}B_{j}\Xi_{j}\right) - (d-1)\sin{x_{i}} B_{i}\Xi_{i},
  \\
  \label{eq:346}
  \dot{\Xi}_{i} &=
  \frac{1}{\cos{x_{i}}}\left(\sum_{j=0}^{N}D^{(1,-)}_{ij}B_{j}\Psi_{j}\right)
  + \left(\frac{\tan{x_{i}}}{\cos{x_{i}}} +
    \frac{d-1}{\sin{x_{i}}}\right)B_{i}\Psi_{i},
\end{align}
for $i=1,\ldots,N$, with boundary conditions (\ref{eq:339}) forced by
setting
\begin{equation}
  \label{eq:347}
  \dot{\Psi}_{0} = 0, \quad \dot{\Xi}_{0} = 0,
\end{equation}
(note the inverse ordering of Chebyshev grid points when using formula
(\ref{eq:340})).  The resulting equations are then integrated in time
using the Gauss-Legendre RK method (see Appendix~\ref{sec:AppIRK}).
The sample results obtained by using (\ref{eq:341})-(\ref{eq:347}), in
particular construction of time-periodic solutions are presented in
the following sections.

The total conserved mass of the system given by the integral
(\ref{eq:75}) with use of this approach is computed as follows
\begin{equation}
  \label{eq:348}
  \begin{aligned}
    M &= \int_{0}^{\pi/2}\cos^{2(d-2)}{x} A\left(\left|\Psi\right|^{2}
      + \cos^{2}{x}\left|\Xi\right|^{2}\right)\tan^{d-1}x \diff x
    \\
    & = \int_{0}^{\pi/2}\sum_{j}T_{j}\left(\frac{2}{\pi}x\right)m_{j}
    \diff x
    \\
    & = \sum_{j}m_{j}\int_{0}^{\pi/2}T_{j}\left(\frac{2}{\pi}x\right)
    \diff x,
  \end{aligned}
\end{equation}
where we use decomposition of the integrand in Chebyshev polynomials
\begin{equation}
  \label{eq:349}
  \sin^{d-1}{x}\cos^{d-3}{x}\,A\left(\left|\Psi\right|^{2} +
    \cos^{2}{x}\left|\Xi\right|^{2}\right) =
  \sum_{j} m_{j}T_{j}\left(\frac{2}{\pi}x\right),
\end{equation}
and the integrals of Chebyshev polynomials are given in
Appendix~\ref{sec:AppJacobiPolynomials}.  It should be noted that the
integrand (\ref{eq:348}) has a fixed parity near the center depending
on $d$ (it is even function of $x$ variable if $d$ is odd, while it is
even when $d$ is odd).  For that reason the expansion in
(\ref{eq:349}) is carried in terms of even or odd Chebyshev
polynomials only, depending on the space dimension $d$.

Similarly we calculate the conserved charge for the complex case.
Since in the definition there is the $\phi$ function present and not
its spatial derivative $\Phi$, which is the evaluated quantity, it
introduces minor complication here.  From the definition
(\ref{eq:334}) we can obtain Chebyshev expansion of the $\Phi$ field
\begin{equation}
  \label{eq:350}
  \cos^{d-2}{x}\,\Psi(t,x) = \Phi(t,x) =
  \sum_{j\geq 0} \hat{\Phi}_{2j+1}(t)\,T_{2j+1}\left(\frac{2}{\pi}x\right),
\end{equation}
which then we integrate in radial direction using relations
(\ref{eq:605})-(\ref{eq:607}) and get
\begin{equation}
  \label{eq:351}
  \phi(t,x) = \sum_{j\geq 0} \hat{\phi}_{2j}(t) \biggl(
  T_{2j}\left(\frac{2}{\pi}x\right) - 1\biggr),
\end{equation}
where the integration constant is set to match the boundary condition
$\phi(t,\pi/2)=0$.  Knowing that, the charge $Q$ can be computed, in
analogy to (\ref{eq:348}), as follows
\begin{equation}
  \label{eq:352}
  \begin{aligned}
    Q &= - \Im \int_{0}^{\pi/2} \phi\,\Xi^{*} \sin^{d-1}{x} \diff x
    \\
    & = - \Im \int_{0}^{\pi/2} \sum_{j}
    T_{j}\left(\frac{2}{\pi}x\right) \hat{q}_{j} \diff x
    \\
    & = - \sum_{j} \Im \hat{q}_{j} \int_{0}^{\pi/2}
    T_{j}\left(\frac{2}{\pi}x\right) \diff x,
  \end{aligned}
\end{equation}
with the expansion coefficients $\hat{q}_{j}$ determined from
\begin{equation}
  \label{eq:353}
  \phi\,\Xi^{*} \sin^{d-1}{x} = \sum_{j}
  T_{j}\left(\frac{2}{\pi}x\right) \hat{q}_{j},
\end{equation}
equated at the set of collocation points (\ref{eq:340}).

\subsection{Numerical construction}
\label{sec:AdSPeriodicNumeric-Even}

Numerical construction of time-periodic solutions relies heavily on
the time evolution code mainly because both use the same spatial
discretization method.  For the same reasons as
Section~\ref{sec:AdSEvolution}, here we also apply two approaches to
the spatial discretization of the field equations, depending on a
parity of space dimension $d$.  Before giving details of numerical
algorithms we first discuss their common aspects.

Seeking for time-periodic solutions numerically it is convenient to
use the rescaled time coordinate $\tau=\Omega\,t$ where, as in the
perturbative construction, $\Omega$ denotes the frequency of the
solution we are looking for.  In this way the numerical grid in
temporal direction has fixed size and we are looking for
$2\pi$-periodic functions.  We construct numerically an algebraic
system for the expansion coefficients (or equivalently for their
values at numerical grid, depending on applied method) of the
dynamical fields.  As an output we get the time-periodic configuration
of these variables.  The corresponding metric functions $\delta$ and
$A$ (the space-time geometry) can be determined, at each instant of
time, by solving the constraint equations for the time-periodic
sources.

The resulting system of algebraic equations is solved with the
Newton-Raphson algorithm.  To initialize this iterative procedure we
take the data corresponding to a single mode configuration of
dynamical fields (as a first order approximation to the time-periodic
solution), i.e.
\begin{equation}
  \label{eq:354}
  \phi(\tau,x) \sim \cos{\tau}\,e_{\gamma}(x),
  \quad \Omega=\omega_{\gamma}, \quad \gamma\in\mathbb{N}_{0},
\end{equation}
while looking for solution bifurcating from the eigenmode
$e_{\gamma}(x)$.  These provide a good guess for small and moderate
amplitudes only.  For larger absolute values of $\ep$, in the
nonlinear regime, the convergence may be restricted and very slow when
we start Newton's iteration far from the true solution.  So, instead
of taking a single mode approximation (\ref{eq:354}) we use the local
polynomial extrapolation of previously derived solutions of smaller
amplitudes.  This slightly speeds up the convergence of Newton's
algorithm, and makes possible to find large amplitude time-periodic
solutions.

The proportionality constant in (\ref{eq:354}) is fixed by a
particular normalization condition.  As in \cite{MRPRL} we can define
a parameter $\ep$ to be an amplitude of dominant mode
\begin{equation}
  \label{eq:355}
  \left.\inner{e_{\gamma}}{\phi}\right|_{t=0} = \ep,
\end{equation}
as we set in perturbative approach, cf. Eq.~(\ref{eq:267}).  The phase
of time-periodic solution is already fixed by (\ref{eq:354}).  This
choice is particularly straightforward to implement in the numerical
code and also easy to force in perturbative calculation but it may not
be the best choice in determining large amplitude solution (as we will
see below).  Another equally simple parametrization of the solutions
we get by controlling a central value of the scalar field at some
instant of time, set for convenience to $t=0$, i.e.
\begin{equation}
  \label{eq:356}
  \phi(0,0) = \ep.
\end{equation}
This being easy to set in the code using the eigenbasis expansion is
not convenient for rescaled variables (\ref{eq:334}) since there we
operate on a gradient of $\phi$ which for smooth solutions vanishes at
$x=0$.  Thus in order to have an universal parametrization (suitable
for both numerical approaches) alternatively we choose to control the
magnitude of the $\Pi$ field, or $\Xi$ respectively (since
$\Xi(t,0)=\Pi(t,0)$), at $\tau=\pi/2$
\begin{equation}
  \label{eq:357}
  \Pi(\pi/2,0) = \ep,
\end{equation}
while leaving the same phase of solutions, i.e. setting $\Pi(0,x)=0$,
in both cases.

Since these parametrizations may vary among the formulae and figures
presented in this work, in order to avoid a confusion we stress
explicitly to which parametrization particular results are referring
to.
\subsubsection{Eigenbasis expansion}
\label{sec:AdSPeriodicEigenbasis}

We expand both dynamical fields $\phi$ and $\Pi$ into eigenmodes of
the linearized problem in space and Fourier modes in time as follows
\begin{align}
  \label{eq:358}
  \phi(\tau, x) &= \sum_{k=0}^{K-1}\sum_{j=0}^{N-1}
  \hat{\phi}_{k,j} \cos\left(\left(2k+1\right)\tau_{\phantom{}}\right)\,e_j(x),
  \\
  \label{eq:359}
  \Pi(\tau, x) &= \sum_{k=0}^{K-1}\sum_{j=0}^{N-1}
  \hat{\Pi}_{k,j} \sin\left(\left(2k+1\right)\tau_{\phantom{}}\right)\,e_j(x).
\end{align}
Then to solve necessary equations by means of pseudospectral method we
choose a compatible grid points, i.e. $K$ collocation points in time
$\tau_i = \pi(i-1/2)/(2K+1)$, $i=1,\ldots,K$ and $N$ collocation
points in space (\ref{eq:317}).  Next, at each instant of time
$\tau_i$ we calculate the coefficients
\begin{align}
  \label{eq:360}
  \hat{\phi}_j(\tau_{i}) &= \sum_{k=0}^{K-1}
  \hat{\phi}_{k,j}\cos\left(\left(2k+1\right)\tau_{i}\right),
  \\
  \hat{\Pi}_j(\tau_{i}) &= \sum_{k=0}^{K-1}
  \hat{\Pi}_{k,j}\sin\left(\left(2k+1\right)\tau_{i}\right),
\end{align}
and put them as an input for our spectral procedure (see
Section~\ref{sec:AdSEvolutionEigenbasis}), getting as the output their
time derivatives.  Equating those to the time derivatives of
(\ref{eq:358}) and (\ref{eq:359}) (remembering that $\partial_t =
\Omega\,\partial_{\tau}$) at the set of $K \times N$ grid points
$(\tau_{k}, x_{j})$, $k=1,\ldots,K$, $j=1,\ldots,N$ (so we require for
the residuals to vanish identically at the collocation points),
together with the additional equation, either setting the amplitude of
the dominant mode $\gamma$ in the initial data to $\ep$
\begin{equation}
  \label{eq:361}
  \sum_{k=0}^{K-1}\hat{\phi}_{k,\gamma}=\ep,
\end{equation}
or the one corresponding to (\ref{eq:356})
\begin{equation}
  \label{eq:362}
  \sum_{k=0}^{K-1}\sum_{j=0}^{N-1}\hat{\phi}_{k,j}\,e_{j}(0) = \ep,
\end{equation}
we get a nonlinear system of $2 \times K \times N + 1$ equations for
$2 \times K \times N + 1$ unknowns: $\hat{\phi}_{k,j}$,
$\hat{\Pi}_{k,j}$ and $\Omega$ ($k=0,1,\ldots,K-1$,
$j=0,1,\ldots,N-1$).

As a single mode approximation for the starting values of the Newton
algorithm we take
\begin{align}
  \label{eq:363}
  \hat{\phi}_{0,\gamma} &= \ep,
  \\
  \label{eq:364}
  \hat{\Pi}_{0,\gamma} &= -\ep\,\omega_{\gamma},
  \\
  \label{eq:365}
  \Omega &= \omega_{\gamma},
\end{align}
while using (\ref{eq:361}) or
\begin{align}
  \label{eq:366}
  \hat{\phi}_{0,\gamma} &= \ep\,\frac{1}{e_{\gamma}(0)},
  \\
  \hat{\Pi}_{0,\gamma} &= -\ep\,\omega_{\gamma}\,\frac{1}{e_{\gamma}(0)},
  \\
  \Omega &= \omega_{\gamma},
\end{align}
for the parametrization defined by (\ref{eq:356}).

\subsubsection{Chebyshev polynomials expansion}
\label{sec:AdSPeriodicChebyshev}

Using the Chebyshev polynomials in pseudospectral discretization in
space we are able to construct time-periodic solutions in any space
dimension $d\geq 2$.  With this approach we expand the scalar fields
$\Psi(\tau,x)$ and $\Xi(\tau,x)$, introduced in
Section~\ref{sec:AdSEvolutionCheb}, in the Fourier basis in time
\begin{align}
  \label{eq:367}
  \Psi(\tau,x) &=
  \sum_{k=0}^{K-1}\cos\bigl((2k+1)\tau\bigr)\hat{\Psi}_{k}(x),
  \\
  \label{eq:368}
  \Xi(\tau,x) &=
  \sum_{k=0}^{K-1}\sin\bigl((2k+1)\tau\bigr)\hat{\Xi}_{k}(x).
\end{align}
For spatial discretization we use the nodal representation, i.e. we
operate on the function values at the grid points $\hat{\Psi}_{ki}
\equiv \hat{\Psi}_{k}(x_{i})$ and $\hat{\Xi}_{ki} \equiv
\hat{\Xi}_{k}(x_{i})$.

With $N$ radial Chebyshev collocation points in space (\ref{eq:340}),
and $K$ collocation points in time $\tau_{k} = \pi(k-1/2)/(2K+1)$,
$k=1,\ldots,K$, at each instant of time $\tau_{k}$ we calculate values
of the fields $\Psi(\tau_{k},x_{i})$ and $\Xi(\tau_{k},x_{i})$ at grid
points $x_{i}$.  Similarly as in the eigenbasis code we us the time
evolution procedure to get as the output their time derivatives.
Equating those to the time derivatives of (\ref{eq:367}) and
(\ref{eq:368}) (by the chain rule $\partial_t =
\Omega\,\partial_{\tau}$) at the set of $K \times N$ tensor product
grid $(\tau_k, x_i)$, together with the additional equation
\begin{equation}
  \label{eq:369}
  \Xi(\pi/2, 0) = \sum_{k=0}^{K-1}(-1)^{k}\,\hat{\Xi}_{k}(0) = \ep,
\end{equation}
(a discrete version of Eq.~(\ref{eq:357})) we close the system of $2
\times K \times N + 1$ nonlinear equations for $2 \times K \times N +
1$ unknowns: $\hat{\Psi}_{ki}$, $\hat{\Xi}_{ki}$ and $\Omega$
($k=0,1,\ldots,K-1$, $i=0,1,\ldots,N-1$).

As a starting configuration for the numerical root-finding algorithm
for a solution bifurcating from eigenmode $\gamma\in\mathbb{N}_{0}$
and fulfilling the normalization condition (\ref{eq:369}) we take
\begin{align}
  \label{eq:370}
  \hat{\Psi}_{0}(x) &=
  -\frac{\ep}{\omega_{\gamma}}\,
  \frac{e_{\gamma}'(x)}{e_{\gamma}(0)}\,\cos^{2-d}{x},
  \\
  \label{eq:371}
  \hat{\Xi}_{0}(x) &=
  \ep\,\frac{e_{\gamma}(x)}{e_{\gamma}(0)}\,\cos^{1-d}{x},
  \\
  \label{eq:372}
  \Omega &= \omega_{\gamma},
\end{align}
with higher Fourier harmonics in (\ref{eq:367}) and (\ref{eq:368}) set
to zero.

\subsection{Results}
\label{sec:AdSPeriodicResults}

We present and analyze results obtained by using the methods developed
in the preceding sections, concentrating on solutions bifurcating from
fundamental mode $\gamma=0$ in $d=3$ and $d=4$ space dimensions in
order to present all of the techniques we have derived together with
their outcomes.\footnote{We deliberately exclude the $d=2$ case from
  these considerations and study $d=4$ instead because of peculiar
  properties of three-dimensional gravity \cite{2013AcPPB..44.2603J}.}
Properties of excited ($\gamma>0$) time-periodic solutions correspond
to these of the fundamental family ($\gamma=0$); also solutions of
different space dimensions share similar features.  For completeness
we comment on other $\gamma$ and $d$ cases when necessary.

Using the perturbative approach we have derived approximation to the
solutions with different $d$ and $\gamma$ of high orders (in even
$d$).  Because of length and complexity of generated formulae we
restrict their presentation and give fourth order accurate results
only.  The most compact expressions among the odd $d$ are these for
$d=3$ and $\gamma=0$ which we give below
\begin{equation}
  \label{eq:373}
  \phi_{1}(\tau,x) = \cos{\tau}\,e_{0}(x) =
  \sqrt{\frac{2}{\pi}}\cos{\tau}\,(3\cos{x} + \cos{3x}),
\end{equation}
\begin{multline}
  \label{eq:374}
  \delta_{2}(\tau,x) = \frac{3}{4\pi}\Bigl(15\cos{2x} + 6\cos{4x} +
  \cos{6x} - 22\Bigr)
  \\
  - \frac{3}{32\pi}\cos{2\tau}\,\Bigl(48\cos{2x} + 36\cos{4x} +
  16\cos{6x} + 3\cos{8x} - 103\Bigr),
\end{multline}
\begin{multline}
  \label{eq:375}
  A_{2}(\tau,x) = \frac{9}{2\pi}x\csc{x}\Bigl(3\cos{x} +
  \cos{3x}\Bigr)
  \\
  - \frac{9}{16\pi}\Bigl(\csc{x}\bigl(3\sin{3x} + 3\sin{5x} +
  \sin{7x})\bigr) + 1\Bigr)
  \\
  - \frac{3}{8\pi}\cos{2\tau}\,\Bigl(4\cos{2x} - 4\cos{4x} - 4\cos{6x}
  - \cos{8x} + 5\Bigr),
\end{multline}
\begin{multline}
  \label{eq:376}
  \phi_{3}(\tau,x) = \cos{\tau} \Bigl[-\frac{27\sqrt{2}x^2}{\pi^{3/2}}
  (3\cos{x} + \cos{3x})
  \\
  - \frac{27x}{4\sqrt{2}\pi^{3/2}} \csc{x} (17 \cos{2x} + 2\cos{4x} -
  \cos{6x} + 14)
  \\
  +\frac{3}{896\sqrt{2}\pi^{3/2}} \bigl( 18\left(2207 +
    224\pi^2\right) \cos{x} + 6\left(2071 + 224\pi^2\right) \cos{3x}
  \\
  - 657\cos{5x} - 195\cos{7x} - 55\cos{9x} + 3\cos{11x} \bigr) \Bigr]
  \\
  -\frac{3}{12928\sqrt{2}\pi^{3/2}}\cos{3\tau} \Bigl(12774 \cos{x} -
  21566\cos{3x} - 23283\cos{5x}
  \\
  - 4497\cos{7x} + 459\cos{9x} - 303\cos{11x}\Bigr)
\end{multline}
\begin{multline}
  \label{eq:377}
  \delta_{4}(\tau,x) = -\frac{81x^2}{2\pi^2}\Bigl( 15\cos{2x} +
  6\cos{4x} + \cos{6x} - 25 \Bigr)
  \\
  + \frac{27x}{8\pi^2}\Bigl( 588\sin{2x} + 84\sin{4x} - 4\sin{6x} -
  3\sin{8x} \Bigr)
  \\
  +\frac{9}{401408\pi^2}\Bigl( \bigl(2257920\pi^2 +
  68231520\bigr)\cos{2x}
  \\
  + \bigl(903168\pi^2 + 12972456\bigr)\cos{4x} + \bigl(150528\pi^2 +
  1720096\bigr)\cos{6x}
  \\
  + 29820\cos{8x} + 6048\cos{10x} - 4200\cos{12x} - 1056\cos{14x}
  \\
  - 231\cos{16x} - 3311616\pi^2 - 82954453 \Bigr)
  \\
  \cos{2\tau} \biggl[\frac{81x^2}{16\pi^{2}} \bigl( 48\cos{2x} +
  36\cos{4x} + 16\cos{6x} + 3\cos{8x} - 10 \bigr)
  \\
  - \frac{27x}{64\pi^{2}} \bigl( 840\sin{2x} + 264\sin{4x} - 4\sin{6x}
  - 45\sin{8x} - 12\sin{10x} \bigr)
  \\
  -\frac{9}{5067776\pi^2} \Bigl( \bigl(11402496\pi^2 +
  47866728\bigr)\cos{2x} + \bigl(8551872\pi^2 + 19485144\bigr)\cos{4x}
  \\
  + \bigl(3800832\pi^2 + 15722392\bigr)\cos{6x} + \bigl(712656\pi^2 +
  3532137\bigr)\cos{8x}
  \\
  - 482664\cos{10x} - 317604\cos{12x} - 89688\cos{14x} - 24467856\pi^2
  - 85716445 \Bigr)\biggr]
  \\
  + \frac{9}{827392\pi^2}\cos{4\tau}\Bigl(1948896\cos{2x} +
  661752\cos{4x} - 552160\cos{6x}
  \\
  - 731724\cos{8x} - 290976\cos{10x} - 33720\cos{12x}
  \\
  - 3936\cos{14x} - 3333\cos{16x} - 994799\Bigr),
\end{multline}
\begin{multline}
  \label{eq:378}
  A_{4}(\tau,x) =
  -\frac{81x^3}{\pi^{2}}\bigl(3\cos{x}+\cos{3x}\bigr)\csc{x}
  \\
  + \frac{243 x^2}{4\pi^{2}}\bigl(7\cos{2x}+4\cos{4x}+\cos{6x}+4\bigr)
  \\
  + \frac{27x}{64\pi^2}\Bigl( \bigl(144\pi^2 + 3261\bigr)\cos{x} +
  \bigl(48\pi^2 + 2023\bigr)\cos{3x}
  \\
  + 684\cos{5x} + 72\cos{7x} - 12\cos{9x} \Bigr)\csc{x}
  \\
  -\frac{9}{250880\pi^2}\Bigl( \bigl(987840\pi^2 +
  32843296\bigr)\cos{2x} + \bigl(564480\pi^2 + 13266816\bigr)\cos{4x}
  \\
  + \bigl(141120\pi^2 + 2509356\bigr)\cos{6x} + 256644\cos{8x} +
  89008\cos{10x}
  \\
  + 16912\cos{12x} + 2980\cos{14x} + 255\cos{16x} + 564480\pi^2 +
  21904013\Bigr)
  \\
  \cos{2\tau} \biggl[\frac{81x^2}{4\pi^2} \Bigl( 4\cos{2x} - 4\cos{4x}
  - 4\cos{6x} - \cos{8x} + 5\Bigr)
  \\
  - \frac{9}{180992\pi^2}\Bigl( \bigl(135744\pi^2 +
  5377082\bigr)\cos{2x} -\bigl(135744\pi^2 - 628730\bigr)\cos{4x}
  \\
  - \bigl(135744\pi^2 + 930322\bigr)\cos{6x} - \bigl(33936\pi^2 +
  458887\bigr)\cos{8x}
  \\
  - 103154\cos{10x} - 20486\cos{12x} + 202\cos{14x} + 169680\pi^2 +
  4194451 \Bigr)
  \\
  + \frac{27x}{32\pi^2}\bigl( 462\cos{x} + 150\cos{3x} - 57\cos{5x} -
  45\cos{7x}
  \\
  - \cos{9x} + 3\cos{11x} \Bigr) \csc{x} \biggr]
  \\
  - \frac{9}{103424\pi^2}\cos{4\tau} \Bigl(50372\cos{2x} +
  83904\cos{4x} - 19068\cos{6x}
  \\
  - 74676\cos{8x} - 32548\cos{10x} + 128\cos{12x} + 1244\cos{14x}
  \\
  - 303\cos{16x} - 9053\Bigr),
\end{multline}
together with frequency expansion coefficients
\begin{equation}
  \label{eq:379}
  \xi_{2} = \frac{153}{4\pi}, \quad
  \xi_{4} = 324 - \frac{\num{9843147}}{\num{6272}\pi^{2}}.
\end{equation}
\begin{figure}[ph]
  \centering
  \includegraphics[width=\lwidth]{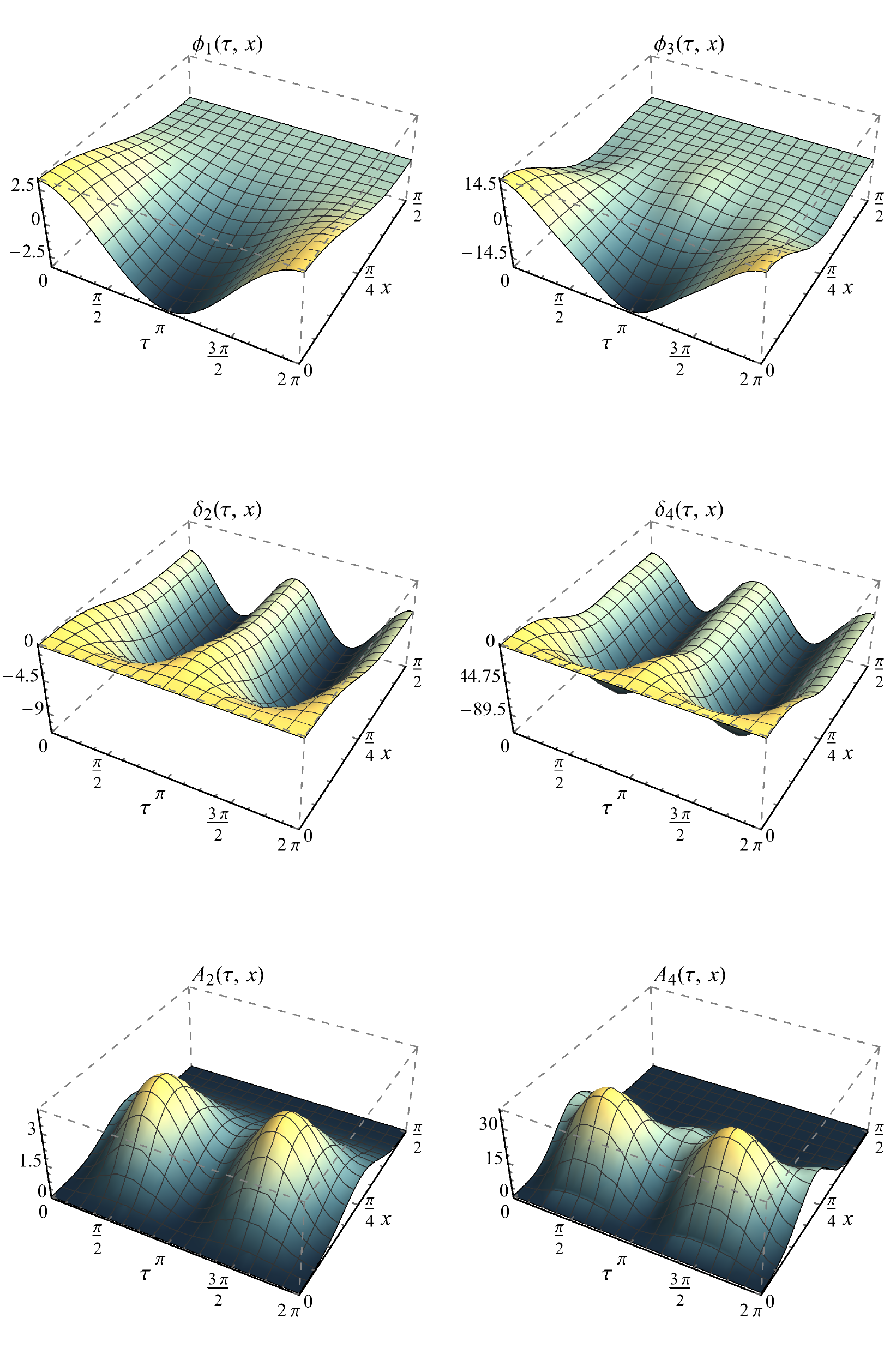}
  \caption{The perturbative result of the fundamental $\gamma=0$
    time-periodic solution in $d=3$ space dimensions derived up to the
    fourth order Eqs.~(\ref{eq:373})-(\ref{eq:378}).  \textit{Top
      panels}.  Profiles of the scalar field in the first (left) and
    the third (right) perturbative order.  \textit{Middle and bottom
      panels}.  Note that the metric functions have frequency two
    times larger than the scalar field since the source is quadratic
    in $\phi$, cf. Eqs.~(\ref{eq:68}) and (\ref{eq:69}).}
  \label{fig:AdSPerturbativeD3Gamma0Plot3D}
\end{figure}
A spatio-temporal plots of the expressions
(\ref{eq:373})-(\ref{eq:378}) are given on
Fig.~\ref{fig:AdSPerturbativeD3Gamma0Plot3D}.  Note the presence of
terms like $x^{n}$ and $x^{n}\csc{x}$ ($n\in\mathbb{N}$) in the
expressions given above, which would lead to an infinite decomposition
in terms of eigenbasis (\ref{eq:90}).  These appear due to
incompatibility of Taylor series expansion of $e_{j}(x)$ at $x=\pi/2$
with (\ref{eq:86}) for odd $d$.  This is not an issue of the even $d$
case when both the scalar field and the metric functions at each
perturbative order can be expressed in terms of finite sum of the
eigenmodes.  Below we list the fourth order accurate result for the
$d=4$ and $\gamma=0$ case
\begin{equation}
  \label{eq:380}
  \phi_{1}(\tau,x) = \cos{\tau}\,e_{0}(x),
\end{equation}
\begin{multline}
  \label{eq:381}
  \delta_{2}(\tau,x) = \frac{24}{5} (-5+3\cos{2\tau}) -
  \frac{4}{35}\sqrt{6}(-7+\cos{2\tau})\,e_0(x)
  \\
  - \frac{8}{35}\sqrt{6}(-2+\cos{2\tau})\,e_1(x) -
  \frac{2}{7\sqrt{15}}(-3+5\cos{2\tau})\,e_2(x)
  \\
  - \frac{8}{35}\sqrt{\frac{2}{15}}\cos{2\tau}\,e_3(x),
\end{multline}
\begin{multline}
  \label{eq:382}
  A_{2}(\tau,x) = -\frac{16}{35}\sqrt{\frac{2}{3}}
  (-7+3\cos{2\tau})\,e_0(x) -
  \frac{2}{35}\sqrt{6}(6+\cos{2\tau})\,e_1(x)
  \\
  +\frac{8 (-1+\cos{2 \tau })}{7\sqrt{15}}\,e_2(x)
  +\frac{2}{7}\sqrt{\frac{2}{15}}\cos{2\tau}\,e_3(x),
\end{multline}
\begin{multline}
  \label{eq:383}
  \phi_{3}(\tau,x) = \frac{20}{7}(\cos{\tau}-\cos{3\tau})\,e_0(x) +
  \frac{8}{105}(125\cos{\tau}+11\cos{3\tau})\,e_1(x)
  \\
  +\frac{4}{231}\sqrt{\frac{2}{5}}(101\cos{\tau}+51\cos{3\tau})\,e_2(x)
  -\frac{8}{1617\sqrt{5}}(13\cos{\tau}-105\cos{3\tau})\,e_3(x)
  \\
  -\frac{1}{4063917\sqrt{35}}(27471\cos{\tau}-207172\cos{3\tau})\,e_4(x)
  \\
  +\frac{8}{6435}\sqrt{\frac{2}{7}}(\cos{\tau}+9 \cos{3\tau})\,e_5(x),
\end{multline}
\begin{multline}
  \label{eq:384}
  \delta_{4}(\tau,x) =
  \frac{32}{104439825}
  (-3479309751+1982598091\cos{2\tau}-45954594\cos{4\tau})
  \\
  +\frac{32\sqrt{6}}{94325}(20923+3904\cos{2\tau}+3969\cos{4\tau})\,e_0(x)
  \\
  -\frac{16 \sqrt{\frac{2}{3}} (-94251367202+12586063269 \cos{2 \tau
    }-12234777366 \cos{4 \tau })}{34848088275}\,e_1(x)
  \\
  -\frac{4 (-536248711287+285801253865 \cos{2 \tau }-37360161720
    \cos{4 \tau })}{20908852965 \sqrt{15}}\,e_2(x)
  \\
  -\frac{16 \sqrt{\frac{2}{15}} (-154937773452+203668775089 \cos{2
      \tau }+15741275382 \cos{4 \tau })}{104544264825}\,e_3(x)
  \\
  -\frac{8 \sqrt{\frac{2}{105}} (-4063917+12979559 \cos{2 \tau
    }+4220796 \cos{4 \tau })}{4063917}\,e_4(x)
  \\
  -\frac{32 (-849235504+2934075730 \cos{2 \tau }+5845218939 \cos{4
      \tau })}{84631071525 \sqrt{21}}\,e_5(x)
  \\
  +\frac{2 \sqrt{\frac{2}{7}} (31611401+396965205 \cos{2 \tau
    }-1161665316 \cos{4 \tau })}{12090153075}\,e_6(x)
  \\
  +\frac{32 (-795732+6839506 \cos{2 \tau }-10011303 \cos{4 \tau
    })}{11813806719 \sqrt{5}}\,e_7(x)
  \\
  -\frac{448}{314925}\sqrt{\frac{2}{55}}(1+9\cos{4\tau})\,e_8(x),
\end{multline}
\begin{multline}
  \label{eq:385}
  A_{4}(\tau,x) =
  -\frac{64}{282975}\sqrt{\frac{2}{3}}
  (-339325+68748\cos{2\tau}-25137\cos{4\tau})\,e_0(x)
  \\
  -\frac{4 \sqrt{\frac{2}{3}} (-61190369850+72040451743 \cos{2 \tau
    }+6709214358 \cos{4 \tau })}{11616029425}\,e_1(x)
  \\
  -\frac{16 (72825307383+18196770701 \cos{2 \tau }+14841306396 \cos{4
      \tau })}{20908852965 \sqrt{15}}\,e_2(x)
  \\
  +\frac{4 \sqrt{\frac{2}{15}} (-125671015736+109046406409 \cos{2 \tau
    }-1102779594 \cos{4 \tau })}{20908852965}\,e_3(x)
  \\
  +\frac{32 \sqrt{\frac{2}{105}} (-368281821+949123105 \cos{2 \tau
    }+193006296 \cos{4 \tau })}{995659665}\,e_4(x)
  \\
  +\frac{4 (-11375093143+27542568360 \cos{2 \tau }+16550711478 \cos{4
      \tau })}{36270459225 \sqrt{21}}\,e_5(x)
  \\
  +\frac{16 \sqrt{\frac{2}{7}} (-4799751221+6131264895 \cos{2 \tau
    }+6368712966 \cos{4 \tau })}{761679643725}\,e_6(x)
  \\
  +\frac{4 (-31346157+3599740 \cos{2 \tau }+8365842 \cos{4 \tau
    })}{6563225955 \sqrt{5}}\,e_7(x)
  \\
  +\frac{64 \sqrt{\frac{2}{55}} (-31+81 \cos{4 \tau
    })}{566865}\,e_8(x),
\end{multline}
and
\begin{equation}
  \label{eq:386}
  \xi_{2} = \frac{464}{7}, \quad \xi_{4} = \frac{45614896}{11319}.
\end{equation}
Analysis of gathered perturbative formulae lead us to the following
observation.  For any even $d\geq 2$ and any choice of $\gamma$ the
solution at a given perturbative order $\lambda\geq 2$, with series
expansion (\ref{eq:284})-(\ref{eq:286}), has the following structure
\begin{equation}
  \label{eq:387}
  \begin{aligned}
    \phi_{\lambda}(\tau,x) &=
    \sum_{j=0}^{(\lambda-1)(d+1)/2+\lambda\gamma}
    \hat{\phi}_{\lambda,j}(\tau)e_j(x),
    \\
    \hat{\phi}_{\lambda,j}(\tau) &= \sum_{k=0}^{(\lambda-1)/2}
    \hat{\phi}_{\lambda,j,2k+1}\cos(2k+1)\tau,
  \end{aligned}
\end{equation}
for $\lambda$ odd, and for $\lambda$ even
\begin{equation}
  \label{eq:388}
  \begin{aligned}
    \delta_{\lambda}(\tau,x) &= \sum_{j\geq
      0}^{(\lambda(d+1)-d)/2+\lambda\gamma}
    \hat{\delta}_{\lambda,j}(\tau)\bigl(e_{j}(x)-e_{j}(0)\bigr),
    \\
    \hat{\delta}_{\lambda,j}(\tau) &=
    \sum_{k=0}^{\lambda/2}\hat{\delta}_{\lambda,j,2k}\cos{2k\tau},
  \end{aligned}
\end{equation}
analogously
\begin{equation}
  \label{eq:389}
  \begin{aligned}
    A_{\lambda}(\tau,x) &= \sum_{j\geq
      0}^{(\lambda(d+1)-d)/2+\lambda\gamma}
    \hat{A}_{\lambda,j}(\tau)e_j(x)\,,
    \\
    \hat{A}_{\lambda,j}(\tau) &=
    \sum_{k=0}^{\lambda/2}\hat{A}_{\lambda,j,2k}\cos{2k\tau},
  \end{aligned}
\end{equation}
while the frequency expansion contains only even powers of $\ep$, as
in Eq.~(\ref{eq:266}).  Additionally, independently of parity of $d$,
the perturbative solutions share the following symmetries
\begin{equation}
  \label{eq:390}
  \begin{aligned}
    \phi(\tau,x;\ep) &= -\phi(\tau,x;-\ep), \\
    \delta(\tau,x;\ep) &= \delta(\tau,x;-\ep), \\
    A(\tau,x;\ep) &= A(\tau,x;-\ep), \\
    \Omega(\ep) &= \Omega(-\ep),
  \end{aligned}
\end{equation}
and
\begin{equation}
  \label{eq:391}
  \begin{aligned}
    \phi(\tau,x;\ep) &= -\phi(\tau+\pi,x;\ep), \\
    \delta(\tau,x;\ep) &= \delta(\tau+\pi,x;\ep), \\
    A(\tau,x;\ep) &= A(\tau+\pi,x;\ep),
  \end{aligned}
\end{equation}
with $\tau\in[0,2\pi]$, $x\in[0,\pi/2]$.

\begin{figure}[!t]
  \centering
  \includegraphics[width=\swidth]
  {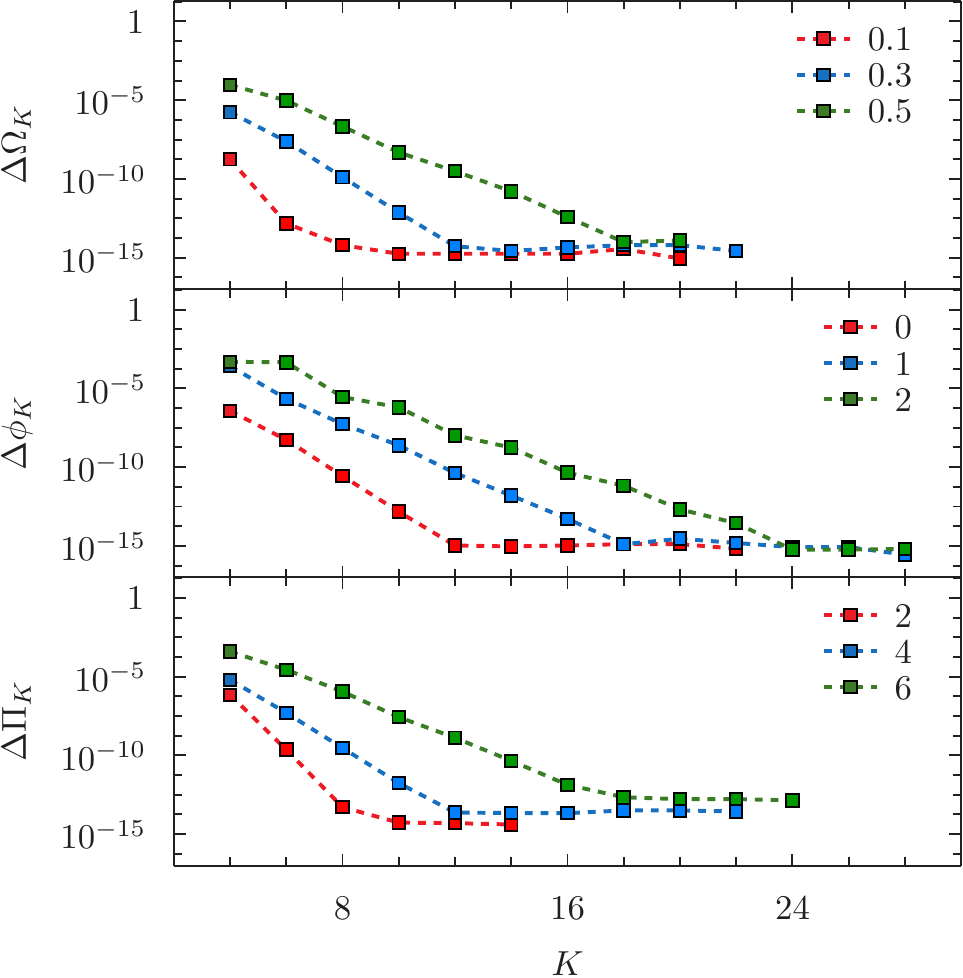}
  \caption{The results of convergence tests of the eigenbasis
    expansion numerical method used to find time-periodic solutions.
    In these tests we varied $K$ in the truncated expansion
    (\ref{eq:358}) and (\ref{eq:359}) with $N=2K$.  The reference
    solution is the one with large $K=\tilde{K}$ (which ranges from
    $16$ to $30$ on presented plots).  All tests were performed with
    with machine double precision arithmetics.  \textit{Top panel}.
    The absolute frequency error $\Delta
    \Omega_{K}:=\left|\Omega_{K}-\Omega_{\tilde{K}}\right|$ for $d=4$
    and $\gamma=0$ case for different amplitude solutions with
    $\ep=\phi(0,0)$.  \textit{Middle panel}.  The $\phi(\tau,x)$
    function absolute error $\Delta
    \phi_{K}:=\left\|\phi_{K}-\phi_{\tilde{K}}\right\|_{2}$ for $d=4$
    and $\ep=3/10$ with different $\gamma$.  \textit{Bottom panel}.
    The absolute error of the $\Pi(\tau,x)$ function $\Delta
    \phi_{K}:=\left\|\Pi_{K}-\Pi_{\tilde{K}}\right\|_{2}$ of the
    fundamental ($\gamma=0$) solution with $\ep=3/10$ in different
    $d$.  The discrete $l^{2}$-norm was calculated on a set of equally
    spaced grid points $x_{i}=i\,\pi/128$, $i=1,\ldots,63$ and
    $\tau_{j}=j\,\pi/128$, $j=1,\ldots,127$.}
  \label{fig:AdSPeriodicNumericConvergence}
\end{figure}
\begin{figure}[!t]
  \centering
  \includegraphics[width=\swidth]
  {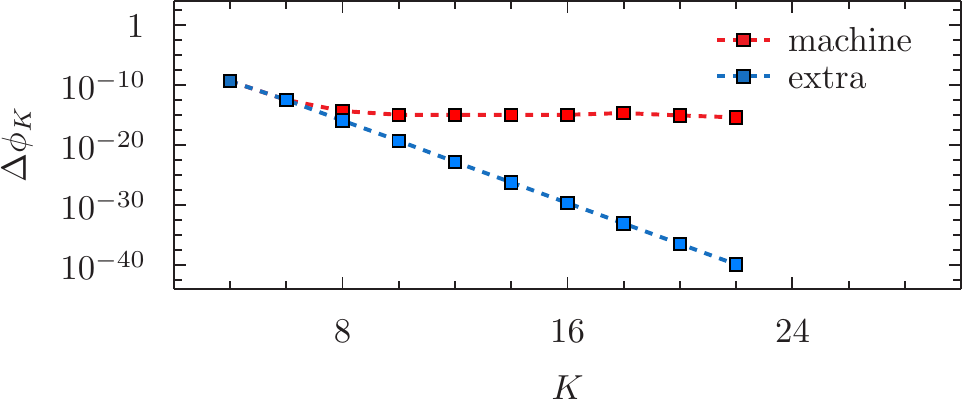}
  \caption{The analogue of
    Fig.~\ref{fig:AdSPeriodicNumericConvergence} (middle panel)
    comparing results with machine precision and with extended fixed
    80 digits precision---the absolute error of $\phi(\tau,x)$
    function in $d=4$ with $\gamma=0$ and $\ep=1/10$.  For machine
    precision calculations the absolute error saturates at the level
    $\sim 10^{-15}$ already for $K=8$ the results wit extended
    precision arithmetics demonstrate that the error tends
    exponentially to zero with $K\ra\infty$ (same rate of convergence
    is seen for both $\Delta\Omega_{K}$ and $\Delta\Pi_{K}$).}
  \label{fig:AdSPeriodicNumericConvergenceExtra}
\end{figure}
\begin{figure}[!t]
  \centering
  \includegraphics[width=\swidth]
  {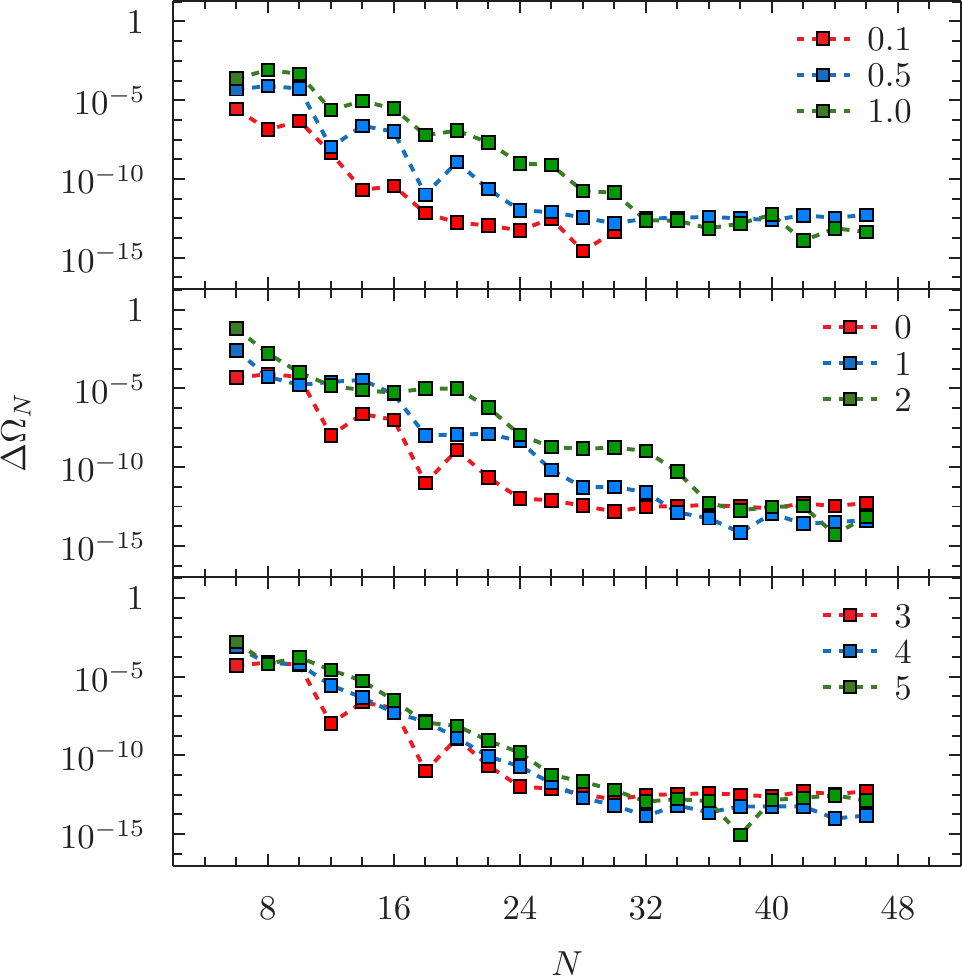}
  \caption{The results of convergence tests of the Chebyshev
    polynomials expansion numerical method used to find time-periodic
    solutions.  (analogue of
    Fig.~\ref{fig:AdSPeriodicNumericConvergence}).  Here we show the
    absolute frequency error $\Delta
    \Omega_{N}:=\left|\Omega_{N}-\Omega_{N=48}\right|$ as a function
    of number of Chebyshev grid points with the number of Fourier
    modes in (\ref{eq:367}) and (\ref{eq:367}) fixed (typically with
    $K=8$ for $\gamma=0$ up to $K=12$ for $\gamma=2$).  \textit{Top
      panel}.  The $d=3$, $\gamma=0$ case for different amplitude
    $\ep=\Pi(\pi/2,0)$.  \textit{Middle panel}.  The $d=3$, $\ep=0.5$
    case with varying $\gamma=0,1$ and $2$.  \textit{Bottom panel}.
    The convergence rate for $\gamma=0$ and $\ep=0.5$ and $d=3,4$ and
    $5$.}
  \label{fig:AdSPeriodicNumericConvergenceChebyshev}
\end{figure}

Using numerical techniques of
Section~\ref{sec:AdSPeriodicNumeric-Even} we have derived hundreds of
solutions in various combinations of $d$ and $\gamma$.  A sample of
results of extensive convergence tests, we have performed to analyze
and verify used numerical methods, are presented on
Figs.~\ref{fig:AdSPeriodicNumericConvergence} and
\ref{fig:AdSPeriodicNumericConvergenceExtra} for the eigenbasis
expansion and Fig.~\ref{fig:AdSPeriodicNumericConvergenceChebyshev} of
the Chebyshev polynomial spatial discretization approach.  First of
all they show fast exponential (spectral) convergence; secondly they
indicate that in order to accurately resolve solutions with large
$\ep$ we need to increase the number of expansion coefficients (grid
points) both in space and time.  The same holds when we increase
either $\gamma$ or $d$ with $\ep$ fixed.

\begin{figure}[!p]
  \centering
    \includegraphics[width=\swidth]
    {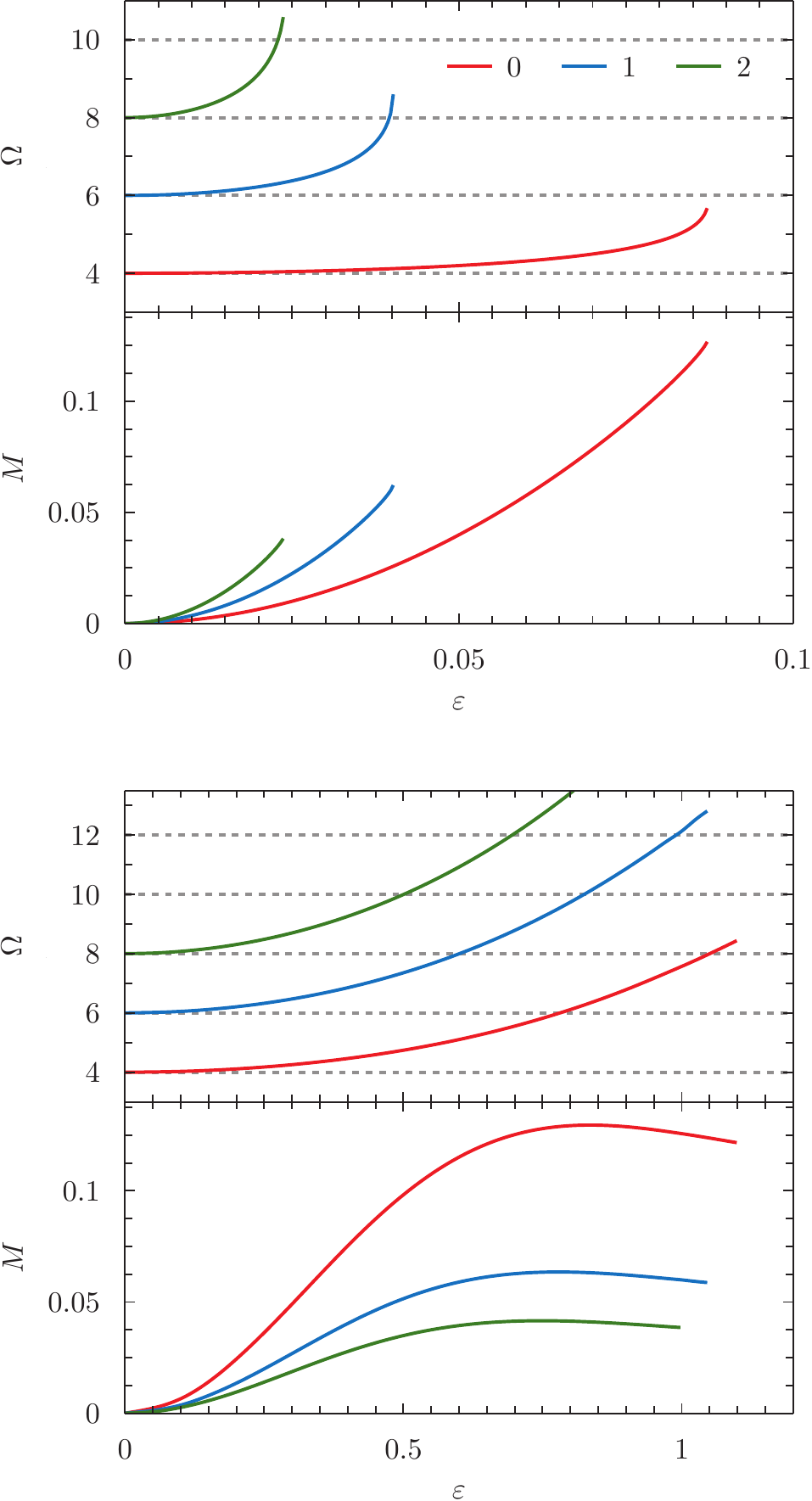}
    \caption{Bifurcation diagrams for the fundamental ($\gamma=0$) and
      first excited ($\gamma=1,2$) time-periodic solutions in $d=4$
      with different parametrizations used (with $\gamma$ color
      coded).  \textit{Top panel}.  For
      $\ep=\left.\inner{e_{\gamma}}{\phi}\right|_{\tau=0}$ we where
      unable to find solutions beyond a finite range of $\ep$.  For
      these limiting values, the frequency $\Omega$ grows rapidly but
      stays finite (so does the mass $M$ of the solutions).
      \textit{Bottom panel}.  When expressed in terms of
      $\ep=\phi(0,0)$, the frequency increases monotonically
      (unboundedly) for solutions in highly nonlinear regime while the
      mass of solutions stays bounded with local maxima decreasing and
      moving toward $\ep=0$ with increasing $\gamma$.}
  \label{fig:AdSPeriodicOmegaMassNew}
\end{figure}

\begin{figure}[!t]
  \centering
  \includegraphics[width=\swidth]
  {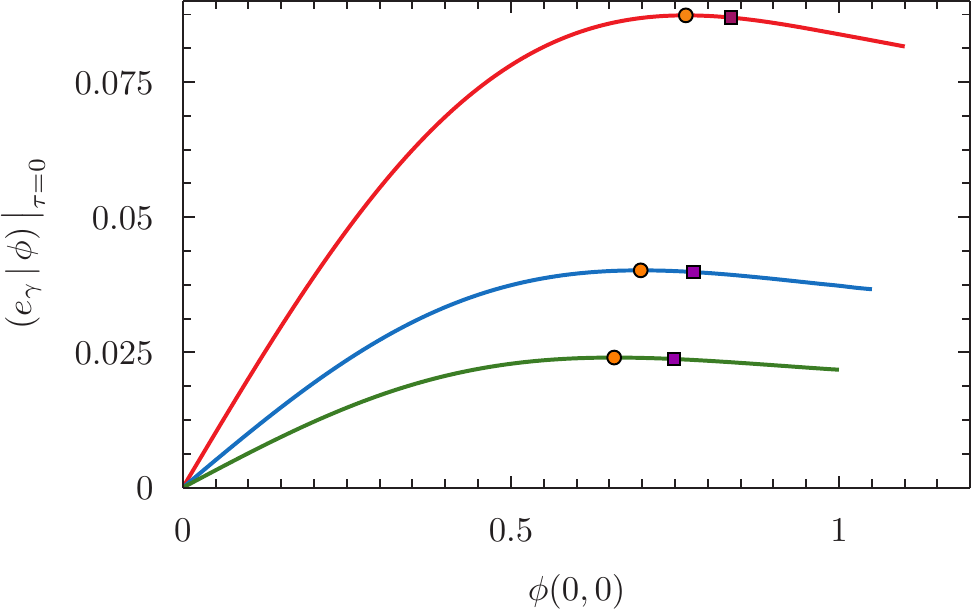}
  \caption{The dependence of the dominant mode amplitude on a value of
    the scalar field at the origin.  The curves are labeled by the
    index $\gamma$.  This dependence explains why we were unable to
    find large amplitude time-periodic solutions.  Dots mark the
    extrema while squares correspond to stationary points of the total
    mass, cf. Fig.~\ref{fig:AdSPeriodicOmegaMassNew}.}
  \label{fig:AdSPeriodicDominant}
\end{figure}

When using the parameter $\ep =
\left.\inner{e_{\gamma}}{\phi}\right|_{\tau=0}$ in the numerical
procedure (for even $d$), the same as in \cite{MRPRL}, we where able
to find time-periodic solutions within a finite range of $\ep$ only,
for any $\gamma$.  Additionally the range of allowed amplitudes
shrinks with increasing $\gamma$, e.g.
$\left|\ep\right|\lesssim\num{0.087342}$ for $\gamma=0$,
$\left|\ep\right|\lesssim\num{0.040182}$ for $\gamma=1$ and
$\left|\ep\right|\lesssim\num{0.024052}$ for $\gamma=2$ (in $d=4$).
At these limiting values both the frequency $\Omega$ and mass $M$ of
the solutions stay finite, see Fig.~\ref{fig:AdSPeriodicOmegaMassNew},
and solution profiles do not indicate any signs of that limitation.
However, with the condition $\ep=\phi(0,0)$ implemented, our numerical
procedure is able to converge for arbitrarily large values of $\ep$
(of course with the limitation on large enough $N$ and $K$ used and
with good initial guess for the Newton method provided).  This feature
of analyzed solutions can be understood when we recover from
$\ep=\phi(0,0)$ parametrized data the value of dominant eigenmode
amplitudes $\left.\inner{e_{\gamma}}{\phi}\right|_{\tau=0}$.  Such
results for considered cases are shown on
Fig.~\ref{fig:AdSPeriodicDominant}, which explains this difference by
showing that the dominant mode amplitude stays bounded and no
solutions of given family exist with amplitudes above certain limit.
We observe a similar effect in other considered models admitting
time-periodic solutions.

\begin{figure}[hp]
  \centering
  \includegraphics[width=\lwidth]
  {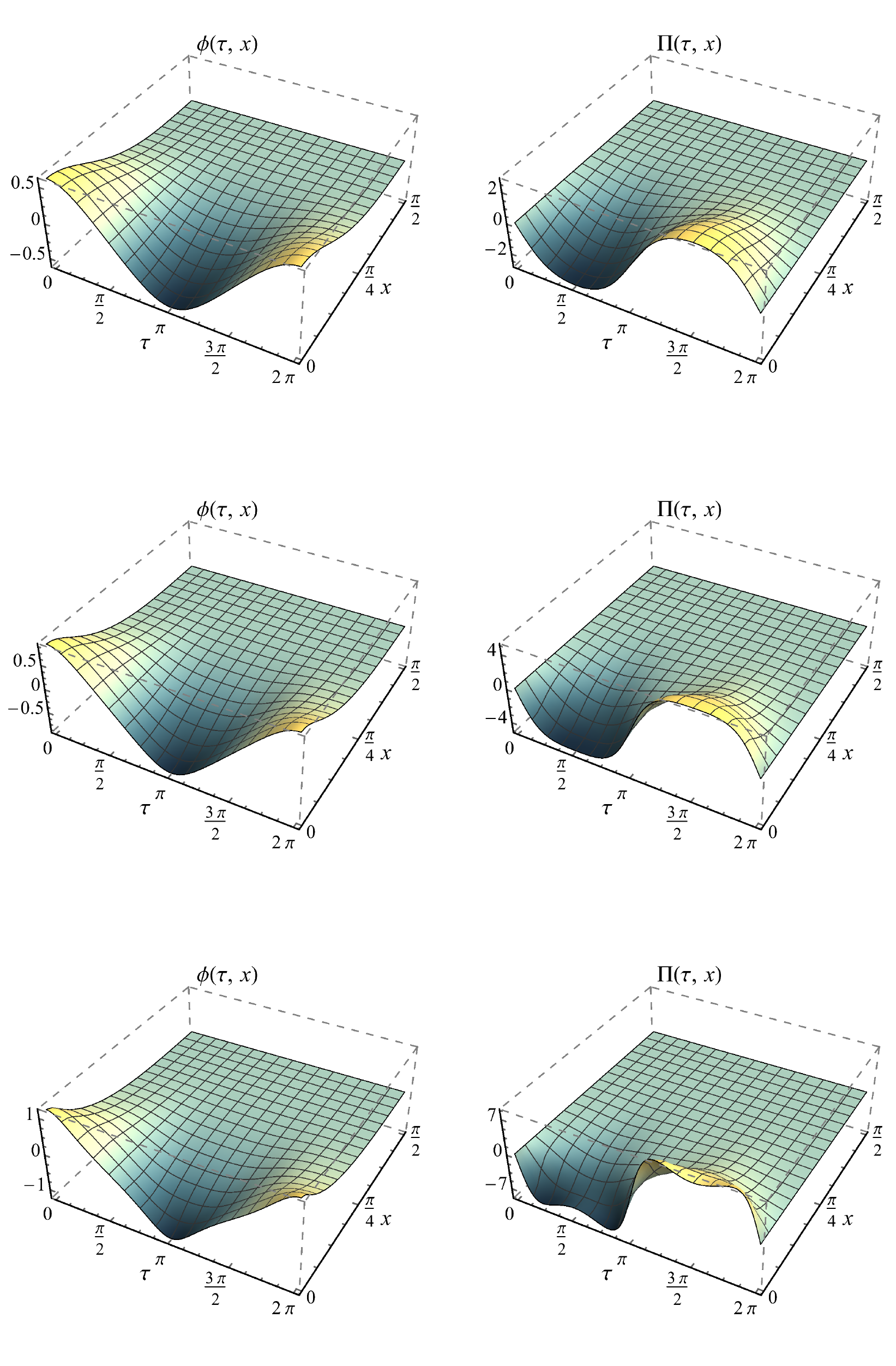}
  \caption{The spatio-temporal plots of fundamental $\gamma=0$
    time-periodic solutions in $d=4$ of increasing amplitudes
    $\phi(0,0)=0.6,\,0.9$ and $1.2$ (from top to bottom).  The
    solution changes its profile from almost harmonic oscillation to a
    v-shape and square oscillation for $\phi$ and $\Pi$ respectively.
    Solutions presented here were obtained on a grid of $28\times 96$
    points.}
  \label{fig:AdSPeriodicD4E0Multi3D}
\end{figure}

From the data shown on Fig.~\ref{fig:AdSPeriodicOmegaMassNew} it is
clear that the frequency of large amplitude solutions grows
monotonically with $\ep=\phi(0,0)$, while the total mass of these
solutions stays bounded, $M\leq M_{\ast}\equiv M(\ep_{\ast})$, where
$\ep=\ep_{\ast}$ are stationary points of $M(\ep)$ (with currently
available data we report on one such point on each bifurcation curve
of time-periodic solutions).  This is analogous to boson star
solutions in asymptotically flat case, see e.g. \cite{LaiPhD}.  For
configurations considered here, in $d=4$, these limiting values are:
$M_{\ast}\approx\num{0.129566}$ at $\ep_{\ast}\approx\num{0.835152}$
for $\gamma=0$, $M_{\ast}\approx\num{0.063512}$ at
$\ep_{\ast}\approx\num{0.778368}$ for $\gamma=1$ and
$M_{\ast}\approx\num{0.041587}$ at $\ep_{\ast}\approx\num{0.748553}$
for $\gamma=3$.  In the boson star models closest to origin stationary
point on a $M(\ep)$ graph plays a fundamental role in the linear
stability analysis of these solutions, separating stable and unstable
configurations.  Also, discussed here, time-periodic solutions share
similar properties, namely solutions with amplitudes $\ep<\ep_{\ast}$
are stable while such with amplitudes $\ep>\ep_{\ast}$ are not (we
look into their stability by a direct time evolution, and give more
details on that later in this section).  We also note that
$\ep_{\ast}$ and $M_{\ast}$ are both decreasing functions of $\gamma$
and $d$ (with an exception in $d=2$ where $M_{\ast}$ increases with
$\gamma$ and $M_{\ast}>1$).

Even though the dominant mode amplitude retains its maximal value this
does not imply that time-periodic solutions of even larger values of
scalar field at the origin $\phi(0,0)$ are no longer solutions
bifurcating from that mode (suggesting that it is no longer dominant
mode).  Although with $\phi(0,0)$ increasing, other modes also
increase their amplitudes and dominant mode amplitude decreases that
mode still dominates over remaining Fourier components.  Therefore all
of the solutions on a given curve shown on
Fig.~\ref{fig:AdSPeriodicOmegaMassNew} are bifurcating from single
(liner) mode.  Also, because the critical points of the mass function
are located to the right of the extremum points of dominant mode
amplitudes, all of the solutions derived in \cite{MRPRL} are stable
(see discussion above).

On Fig. \ref{fig:AdSPeriodicD4E0Multi3D} we plot profiles of
time-periodic solutions bifurcating from the fundamental mode in $d=4$
with increasing amplitudes $\ep=\phi(0,0)=0.6,\,0.9$ and $1.2$.  These
change smoothly with $\ep$ from an almost harmonic oscillation
dominated by $e_{0}(x)$ mode to a v-shape and square like oscillation
for $\phi$ and $\Pi$ fields in the nonlinear regime.  Qualitatively
similar profiles we get when looking at different space dimensions
$d\geq 2$ with tendency for the solutions to become more and more
compact near the origin as $d$ increases (which is a typical feature
of higher dimensional gravity).

\begin{figure}[!t]
  \centering
  \includegraphics[width=\swidth]
  {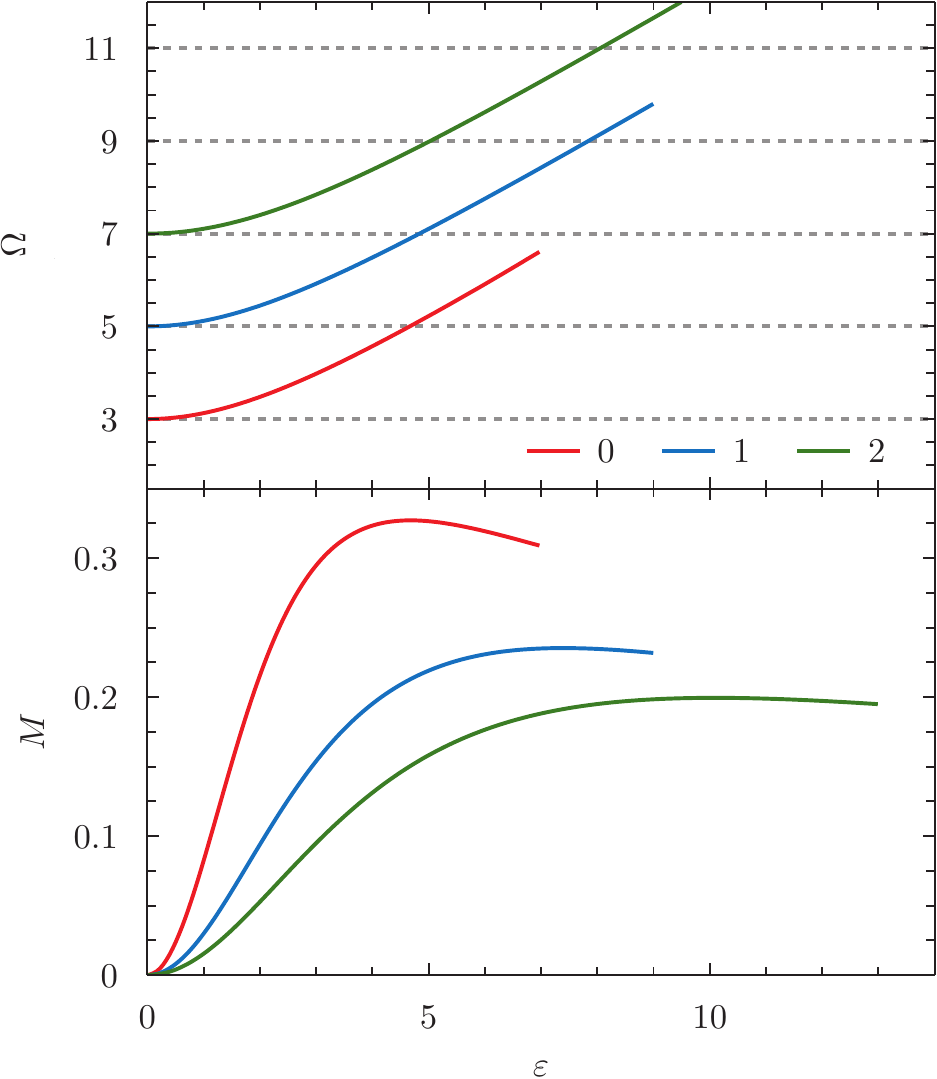}
  \caption{\textit{Top panel}.  The frequency of time-periodic
    solutions in $d=3$ bifurcating from the lowest eigenmodes
    $\gamma=0,1$ and $2$ (with $\gamma$ color coded) parametrized by
    $\ep=\Pi(\pi/2,0)$. \textit{Bottom panel}.  The corresponding
    total mass of the solutions derived using the Chebyshev
    polynomials in space discretization.}
  \label{fig:AdSPeriodicOmegaMassD3}
\end{figure}

While for odd $d$ the expansion of dynamical variables in terms of
eigenmodes makes less sense and since we were using $\Pi(\pi/2,0)$ as
a parameter in our Chebyshev pseudospectral code
(Section~\ref{sec:AdSPeriodicNumeric-Even}), we observe a similar
frequency and mass behaviour as for even $d$ when using the
$\phi(0,0)$ parametrization.  For completeness we give the results for
$d=3$ case on Fig.~\ref{fig:AdSPeriodicOmegaMassD3}.  While the
presence of extremum points of mass function is clear, their position
for different families $\gamma$ appears to be different from what we
have stated before (here $\ep_{\ast}$ increases with $\gamma$).
Again, this is an effect of parametrization being used, when expressed
in terms of the same parameter, e.g. with
$\Omega^{-1}\,\Pi(\pi/2,0)$, the bifurcation diagrams do not differ
qualitatively among the even and odd $d$.

\subsubsection{Consistency of perturbative and numerical results}

With two independent methods, aimed to construct time-periodic
solutions, we have performed series of comparisons to verify our
results,\footnote{These were also very useful at the early stages of
  implementation of numerical routines.} beside the convergence tests
of the numerical approach, shown above.

\begin{figure}[!t]
  \centering
  \includegraphics[width=\swidth]
  {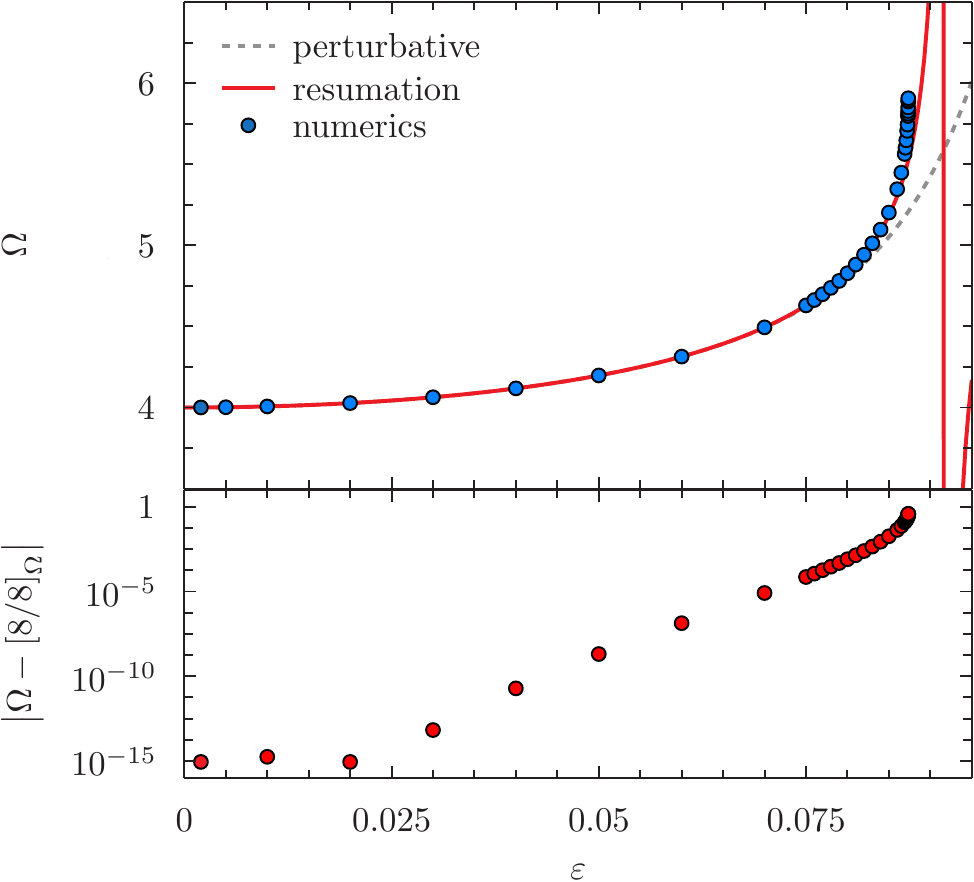}
  \caption{The comparison of oscillation frequency of time-periodic
    solution bifurcating from eigenmode $e_{0}(x)$ obtained from
    numerical procedure and perturbative calculation (parametrized be
    the $e_{0}(x)$ amplitude).  \textit{Top panel}. The numerical data
    (circles) align on a smooth curve which is well approximated by
    perturbative expansion (dashed gray line) only for small $\ep$.
    The Pad\'e resumation accelerates the convergence (solid red
    line). \textit{Bottom panel}.  The absolute difference between
    numerical data and $[8/8]_{\Omega}$ Pad\'e resumation of $\Omega$
    series (\ref{eq:392}).}
  \label{fig:AdSPeriodicOmegaPade}
\end{figure}

Using the perturbative method, for small values of $\ep$,
independently of its definition, we get fast convergence to the
numerical solution.  In \cite{MRPRL} we compared the numerical and
perturbative results by looking at the frequency of time-periodic
solutions bifurcating from $\gamma=0$ mode, whose explicit
perturbative series expansion reads
\begin{multline}
  \label{eq:392}
  \Omega_{\gamma=0}(\ep) = 4 + \fracn{464}{7}\,\ep^2 +
  \fracn{45614896}{11319}\,\ep^4 +
  \fracn{173158711507904383595696}{533797475350414275}\,\ep^6
  \\[0.8ex]
  + \fracn{19627018631453126466665156076805265104}
  {662148921092395909349993941125}\,\ep^8
  \\[0.8ex]
  \scalebox{0.95}{$ +
    \fracn{11072083972904297030696081837311640731963665649719
      830983726203187072}
    {37987752877653278511560882826978473140548594404054292393
      75}\,\ep^{10}$}
  \\[0.8ex]
  \scalebox{0.82}{$ +
    \fracn{20613105125948158448048760017030229079689748556646
      392089437804474799052885011038708688}
    {68546459749514836602259551946772532940114419607536660561
      179461719235846875}\,\ep^{12}$}
  \\
  \scalebox{0.52}{$+
    \fracn{24216542819360089111564790288089436774302059204922
      8107610368936295482539712220892052818656526424727746942
      2007330764469053088}
    {75437310986074872081884732335514765729460118981804160056
      314432853260497637542014881219556515583717903326984375}
    \,\ep^{14}$}
  \\
  \scalebox{0.48}{$ +
    \fracn{12184796673481160409607205110094880611473197401052
      3689696856780657632912793694941145734596660794921287901
      133445732356417910379756334718715484623394013328}
    {34663657226073755946021842599809006931016008158183111178
      3200326162192008896279909208258037781601796306594030016
      75300805518566174084765625}\,\ep^{16}$}
  \\
  + \mathcal{O}\left(\ep^{18}\right),
\end{multline}
(where $\ep=\inner{e_{0}}{\phi}\left.\right|_{\tau=0}$).  The results
are in excellent agreement, especially when we refine both the
numerical data, by using extended precision arithmetics, and the
perturbative series, by performing Pad\'e resumation
\cite{bender1999advanced}.  While a direct summation of (\ref{eq:392})
gives satisfactory approximation for small values of $\ep$, a Pad\'e
approximation of $\Omega$ greatly improves convergence, see
Fig.~\ref{fig:AdSPeriodicOmegaPade}.  Moreover, Pad\'e approximation
can be used to estimate the radius of convergence of the series
(\ref{eq:392}) or equivalently maximal allowed value of dominant
eigenmode amplitude.
The zeros of the denominators of $[n/n]_{\Omega}$ nearest to the
origin are: $\num{0.128251}$, $\num{0.101469}$, $\num{0.094708}$ and
$\num{0.091904}$ for $n=2,4,6$ and $8$ respectively.  These values
converge to the limiting value of the dominant eigenmode $e_{0}(x)$
amplitude $\num{0.087342}$, above which no time-periodic solutions
exist, this is also illustrated on
Fig.~\ref{fig:AdSPeriodicOmegaPade}.

\begin{table}[!t]
  \centering
  \begin{tabular}{c ll ll}
    \toprule
    \multirow{2}{*}{$\ \gamma\ $} &
    \multicolumn{2}{c}{$\coef{2}\Omega$}  &
    \multicolumn{2}{c}{$\coef{4}\Omega$} \\
    & \multicolumn{1}{c}{perturbative} & \multicolumn{1}{c}{numeric} &
    \multicolumn{1}{c}{perturbative} & \multicolumn{1}{c}{numeric}
    \\ \midrule
    0 & \num{0.1328125} & \num{0.132812496} &
    \num{-0.00456582} & \num{-0.00456541}
    \\
    1 & \num{0.1223958(3)} & \num{0.12239583341} &
    \num{-0.0031294310} & \num{-0.0031294322}
    \\
    \bottomrule
  \end{tabular}
  \caption{The comparison of frequency of small amplitude
    ($\ep=\Pi(\pi/2,0)$) time-periodic solutions in $d=3$ space
    dimensions.  The numerically derived (extracted from even polynomial
    fit) expansion coefficients agree, within the numerical and fitting
    errors (the quality of fit was intentionally reduced to show
    difference in numbers; the fit was performed on interval
    $\ep\in [0.01,1.5]$
    using sample of $7$ points).  The numerical solutions were
    derived using Chebyshev polynomials code with $N=12$ and $K=24$
    grid points.}
  \label{tab:AdSFrequencyCompareOdd3}
\end{table}
\begin{table}[!t]
  \centering
  \begin{tabular}{c ll ll}
    \toprule
    \multirow{2}{*}{$\ \gamma\ $} &
    \multicolumn{2}{c}{$\coef{2}\Omega$}  &
    \multicolumn{2}{c}{$\coef{4}\Omega$}
    \\
    & \multicolumn{1}{c}{perturbative} & \multicolumn{1}{c}{numeric} &
    \multicolumn{1}{c}{perturbative} & \multicolumn{1}{c}{numeric}
    \\ \midrule
    0 & \num{0.19921875} & \num{0.199218749948} &
    \num{-0.007024971} & \num{-0.007024966}
    \\
    1 & \num{0.158953125} & \num{0.15895312493} &
    \num{-0.00393161} & \num{-0.003931596}
    \\
    \bottomrule
  \end{tabular}
  \caption{The analogue of Tab.~\ref{tab:AdSFrequencyCompareOdd3} in $d=5$
    space dimensions.}
  \label{tab:AdSFrequencyCompareOdd5}
\end{table}

Because of difficulties that \mathematica{} encounters in manipulating
complex expressions, appearing in odd $d$, perturbative expansions in
odd $d$ are of lower order than those for even $d$.  Therefore,
instead of so detailed comparison as above we have checked that the
frequency $\Omega$ of numerically constructed solutions has the same
form, as a function of $\ep$, as given by the perturbative expansion.
Since we are using different parametrizations in numerical code and in
perturbative calculations, the expansion (\ref{eq:379}) cannot be
compared directly but has to be transformed accordingly.  The
frequency expansion (\ref{eq:379}) expressed in terms of
$\ep=\Pi(\pi/2,0)$ reads
\begin{equation}
  \label{eq:393}
  \Omega_{\gamma=0}(\ep) = 3 + \frac{17}{128}\ep^{2}
  + \left(\fracn{15}{8192}\pi^2-\fracn{7342227}{324337664}\right)\ep^{4}
  + \mathcal{O}\left(\ep^{6}\right).
\end{equation}
Then the expansion coefficients, like these in (\ref{eq:393}), can be
directly compared with a least-square fit of even polynomial to a
series of numerically derived solutions.  In that way we have compared
the frequencies of the ground state $\gamma=0$ and first excited state
$\gamma=1$ in $d=3, 5$; the results, which are presented in
Tab.~\ref{tab:AdSFrequencyCompareOdd3} and
\ref{tab:AdSFrequencyCompareOdd5}, show an excellent agreement of
perturbative approach and numerical scheme with spatial discretization
based Chebyshev pseudospectral method.  Furthermore we observe the
expected convergence, i.e. the difference of the two decreases when
refining discretization (which extends beyond machine precision).

Lastly, we have verified that for even $d$ cases, where both of
presented numerical approaches (the eigenbasis and Chebyshev spatial
expansion) are applicable these produce consistent results (within
discretization errors).

\subsubsection{Stability of time-periodic solutions}

\begin{figure}[pt]
  \centering
  \begin{tabular}{cc}
    \includegraphics[width=0.46\columnwidth]
    {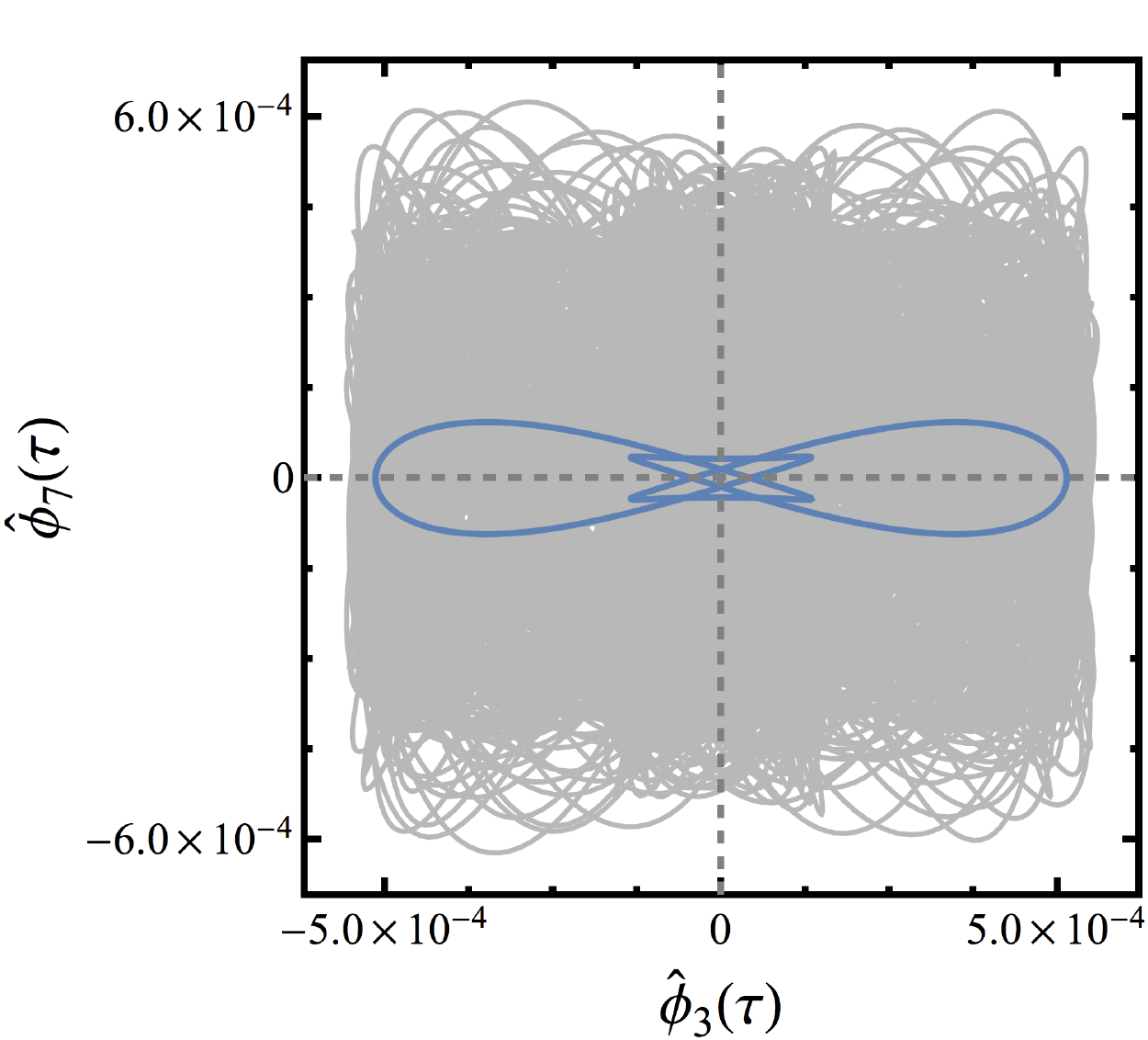} &
    \includegraphics[width=0.46\columnwidth]
    {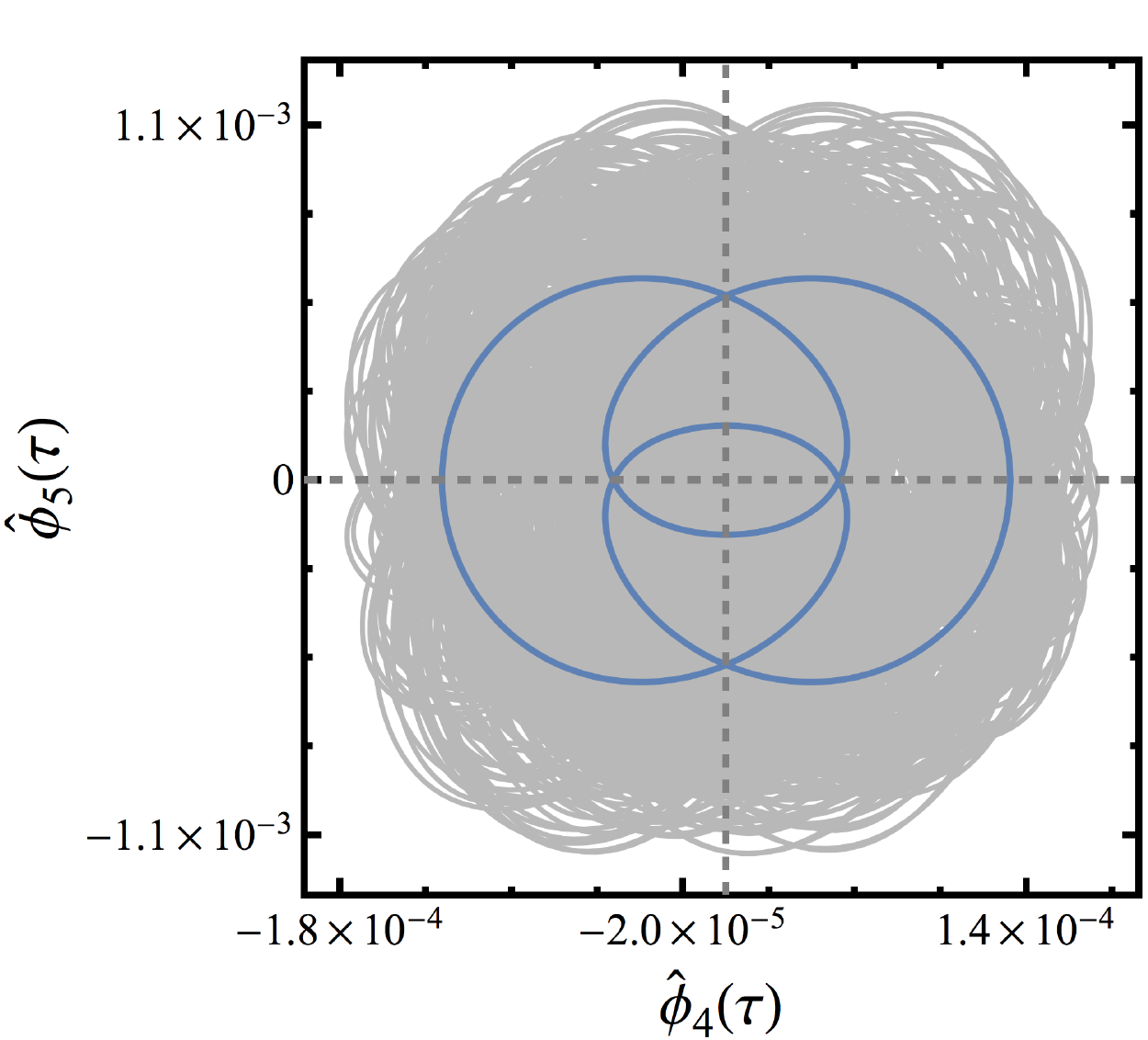}
    \\[4ex]
    \includegraphics[width=0.46\columnwidth]
    {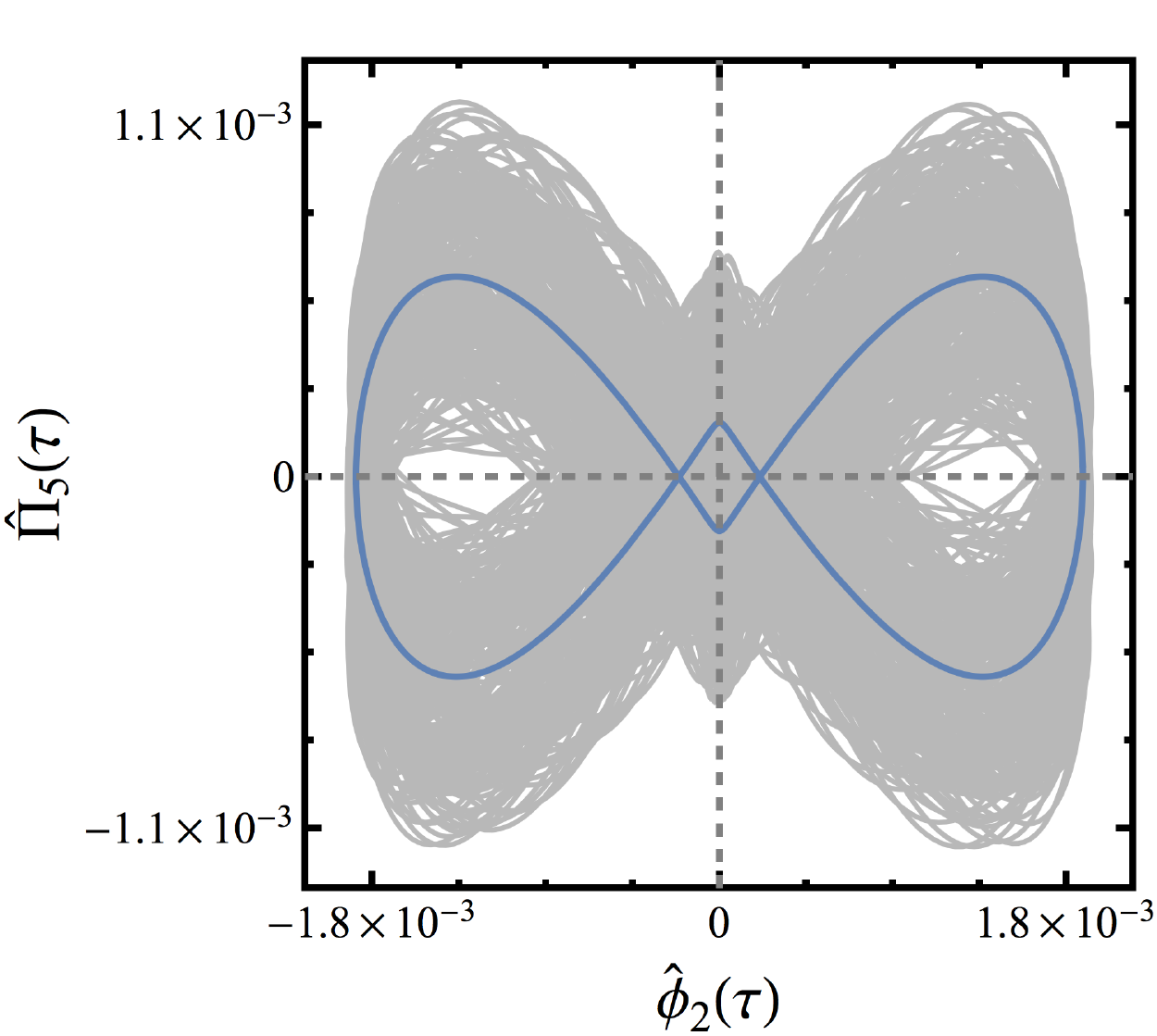} &
    \includegraphics[width=0.46\columnwidth]
    {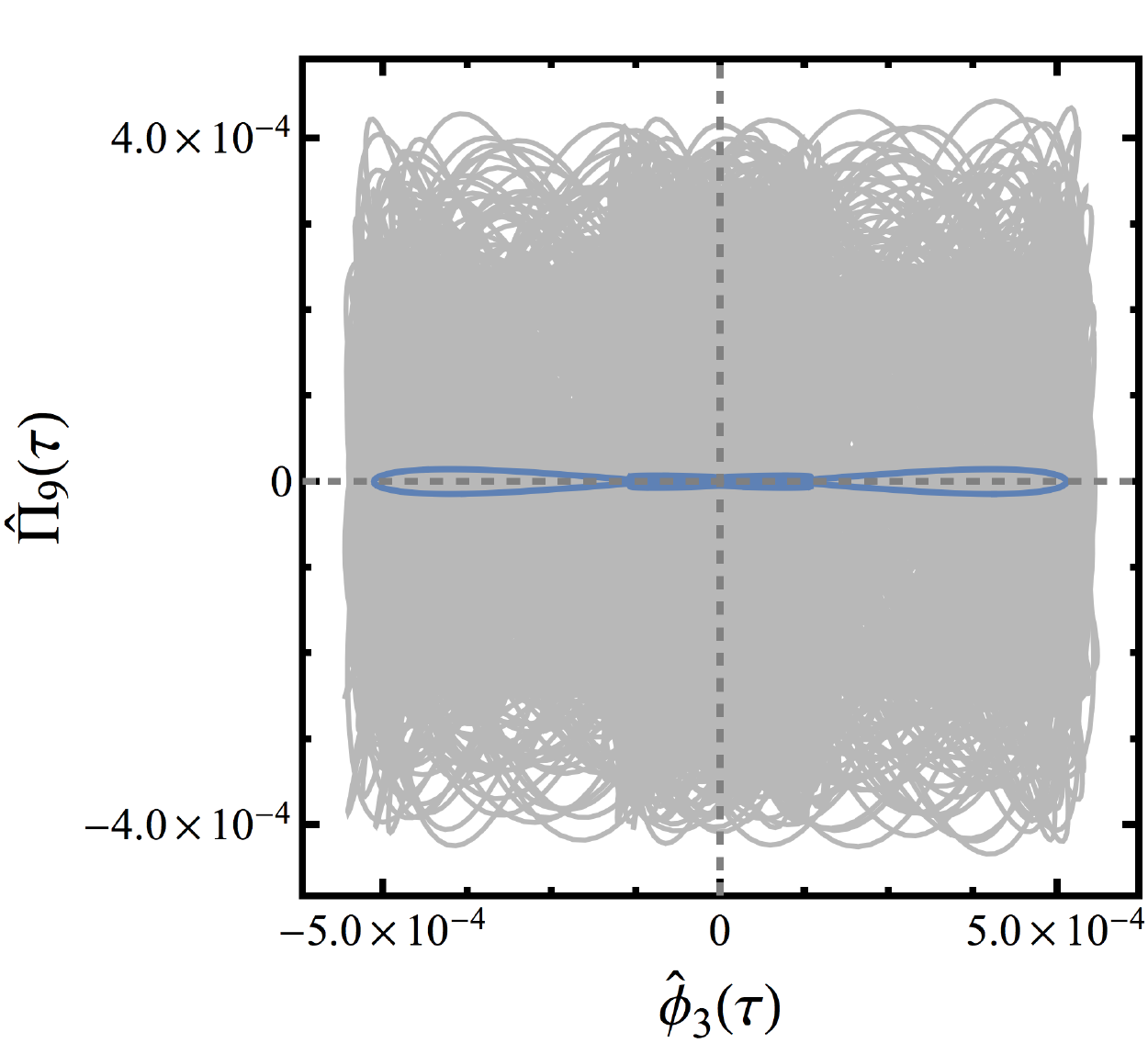}
    \\[4ex]
    \includegraphics[width=0.46\columnwidth]
    {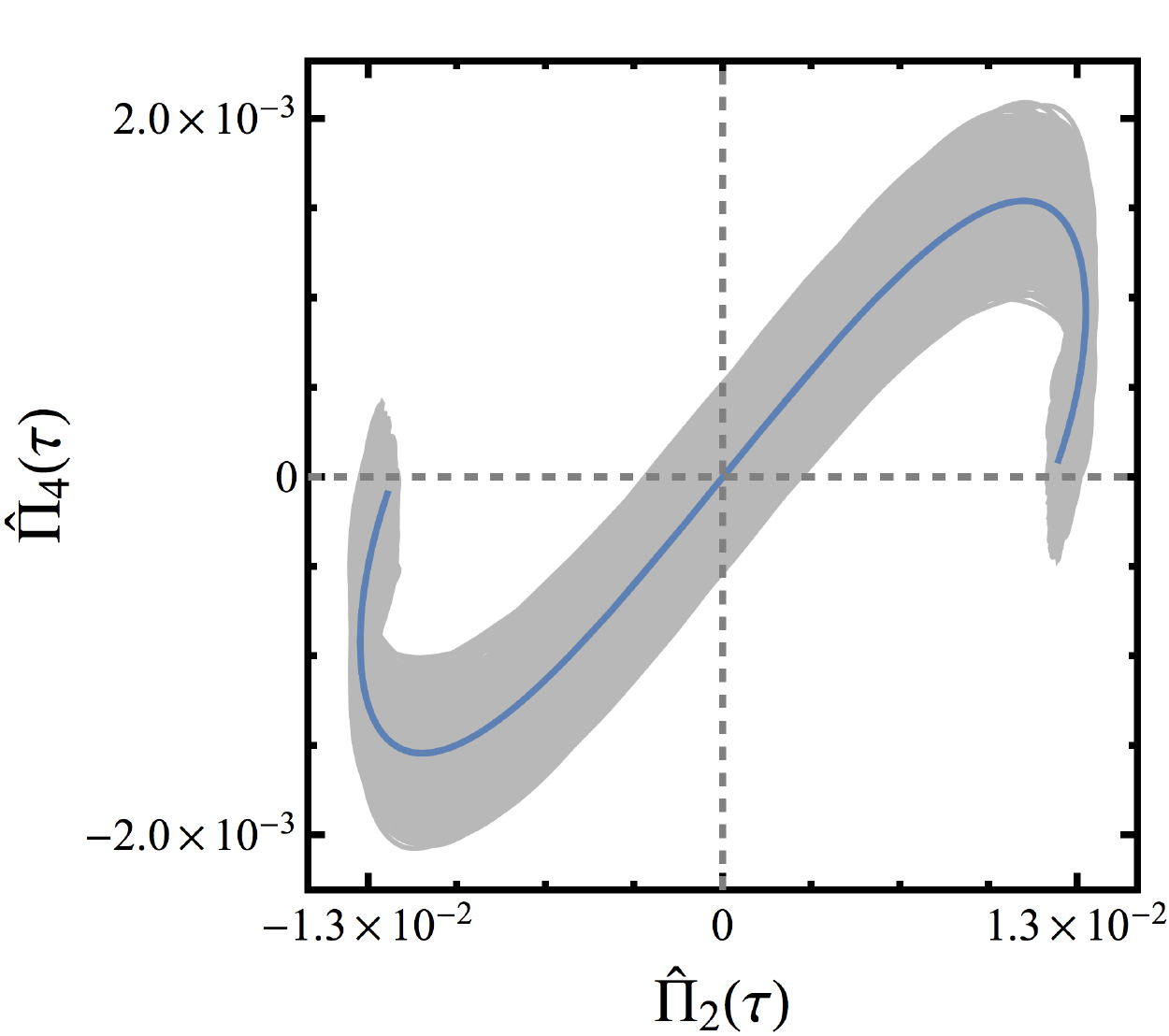} &
    \includegraphics[width=0.46\columnwidth]
    {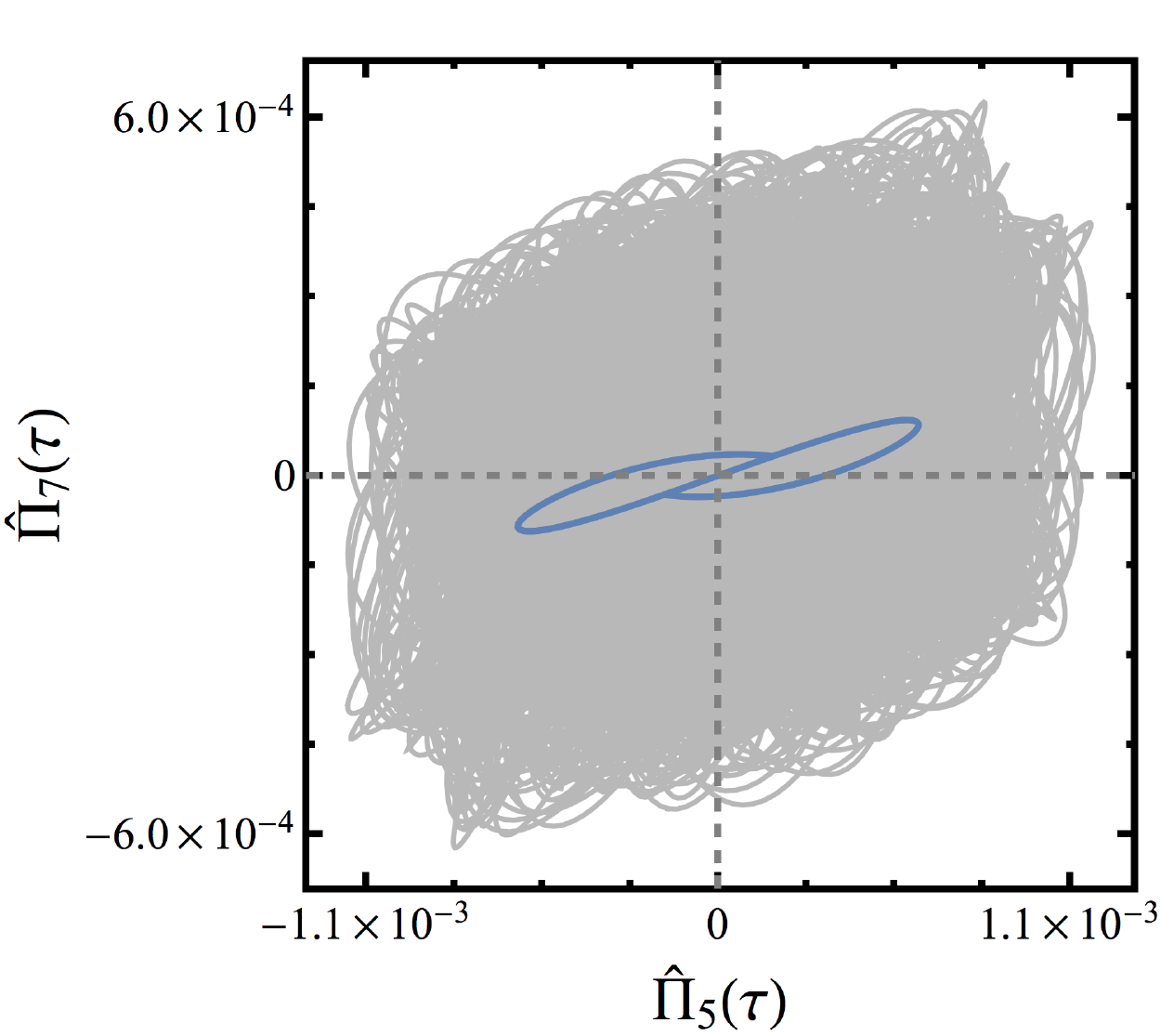}
  \end{tabular}
  \caption{The plots of projections of a phase space of perturbed, by
    a Gaussian pulse (\ref{eq:394}) with $\epsilon=1/2$, fundamental
    ($\gamma=0$) time-periodic solution in $d=4$ of central amplitude
    $\ep=1/2$ ($\Omega\approx\num{4.742050}$) plotted with gray lines
    together with unperturbed trajectories overlaid (blue lines).  The
    'spaghetti lines' show the explored trajectories in a phase space
    for $\tau\in[0,500\pi]$ and they do not expand any further during
    long time evolution.  The evolution was performed using $N=128$
    eigenmodes with fourth order Gauss-RK with time step $\Delta
    t=2^{-11}\pi$.}
  \label{fig:AdSPeriodicLoopsE0}
\end{figure}

\begin{figure}[pt]
  \centering
  \begin{tabular}{cc}
    \includegraphics[width=0.46\columnwidth]
    {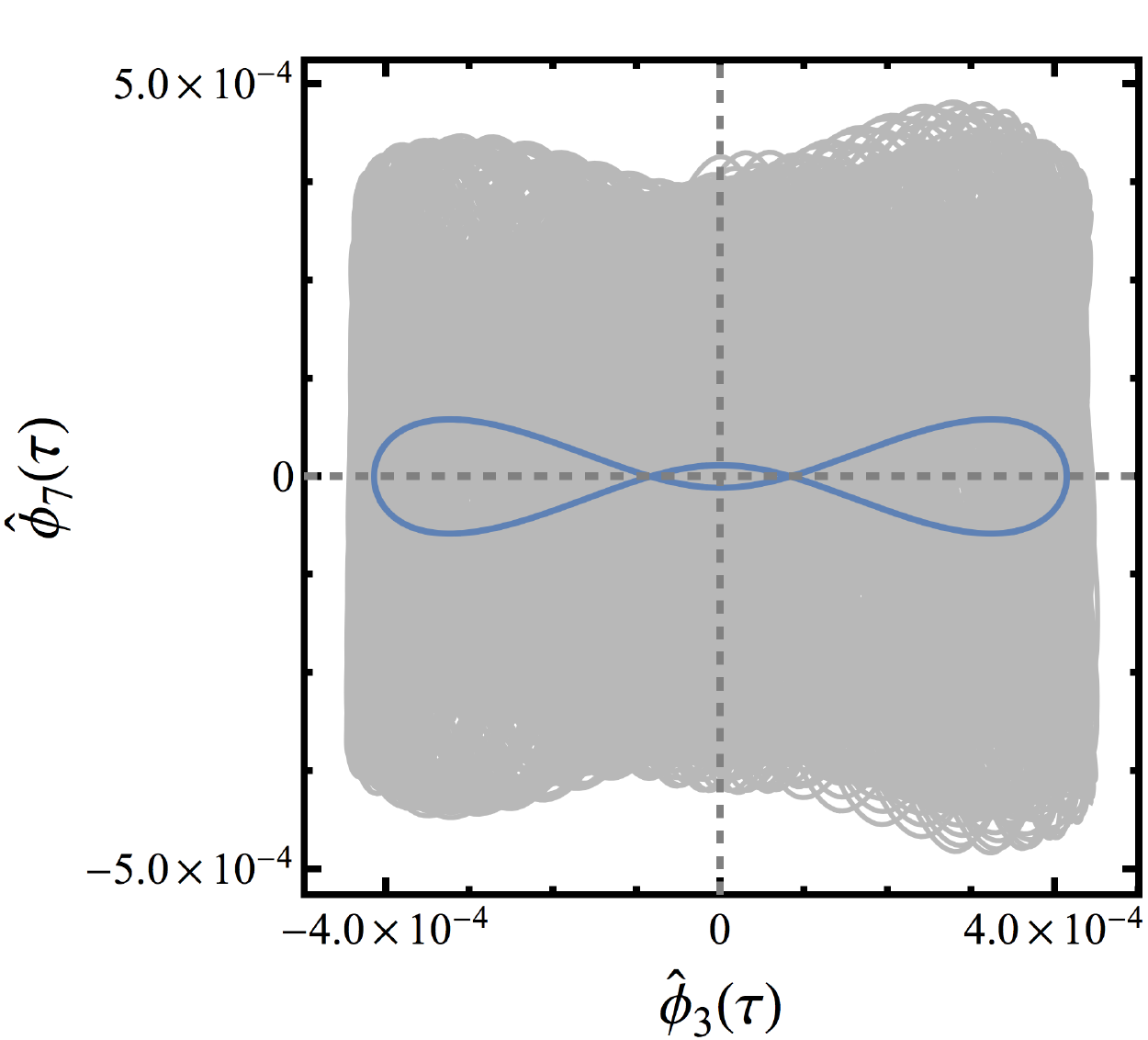} &
    \includegraphics[width=0.46\columnwidth]
    {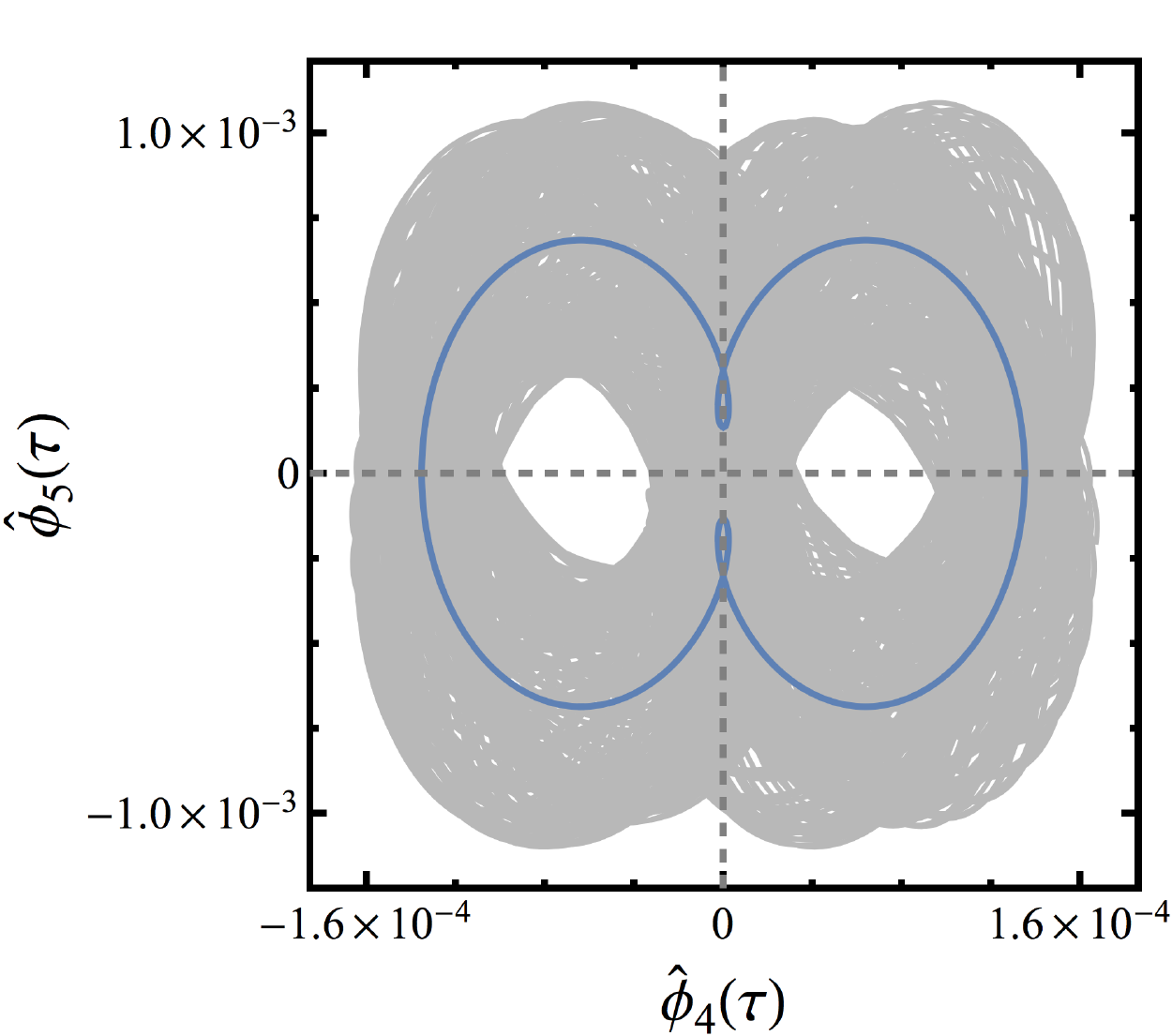}
    \\[4ex]
    \includegraphics[width=0.46\columnwidth]
    {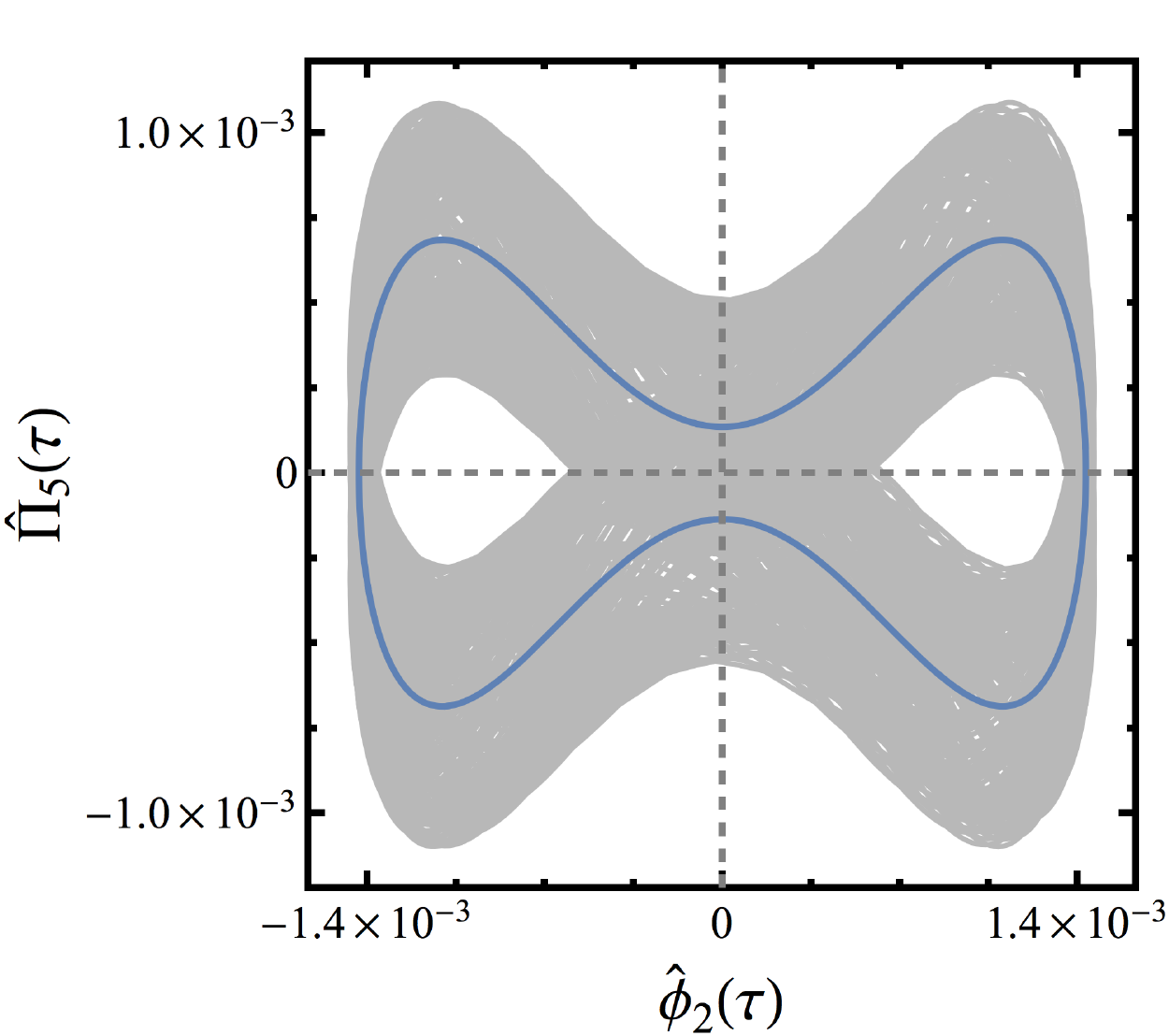} &
    \includegraphics[width=0.46\columnwidth]
    {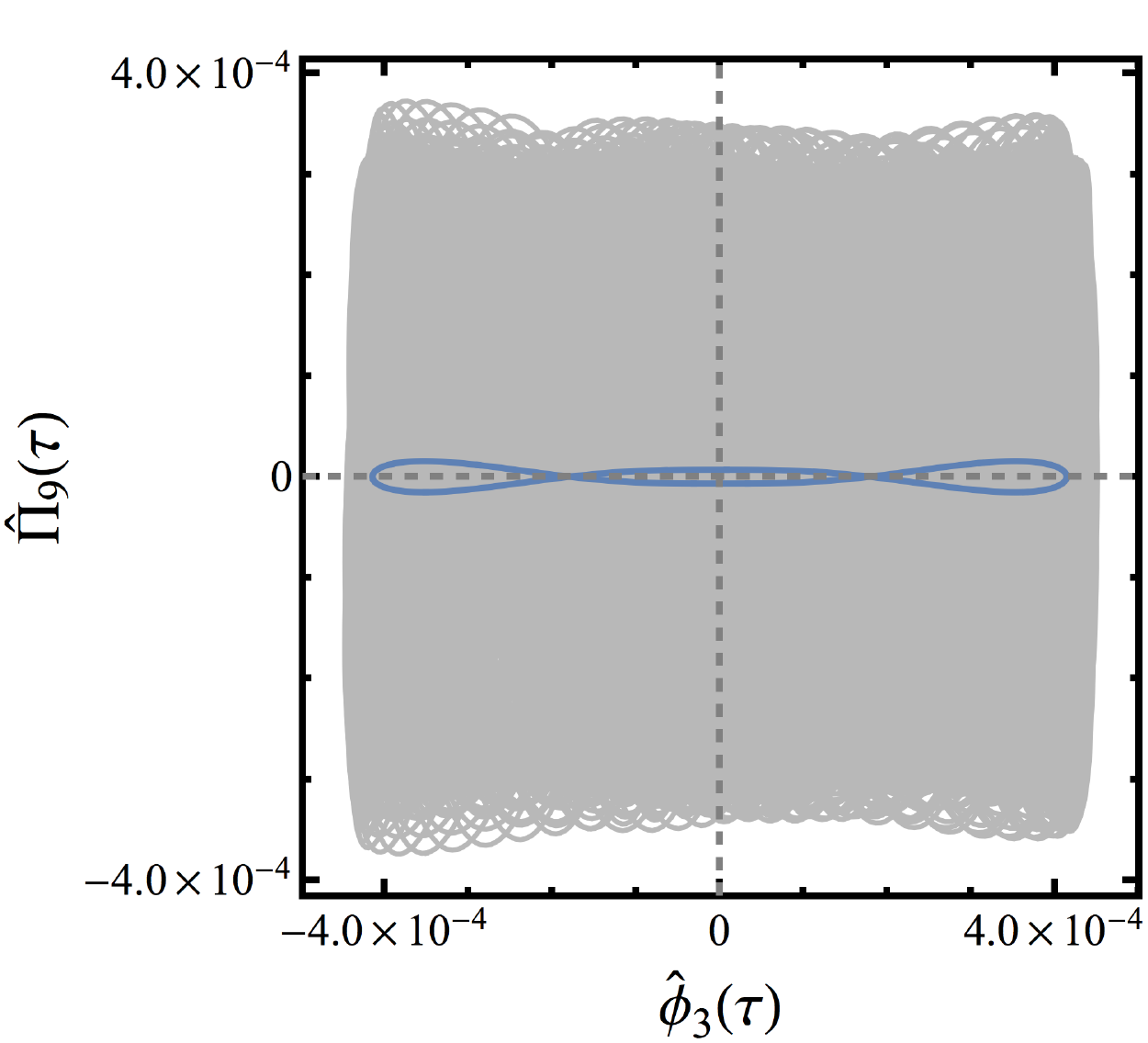}
    \\[4ex]
    \includegraphics[width=0.46\columnwidth]
    {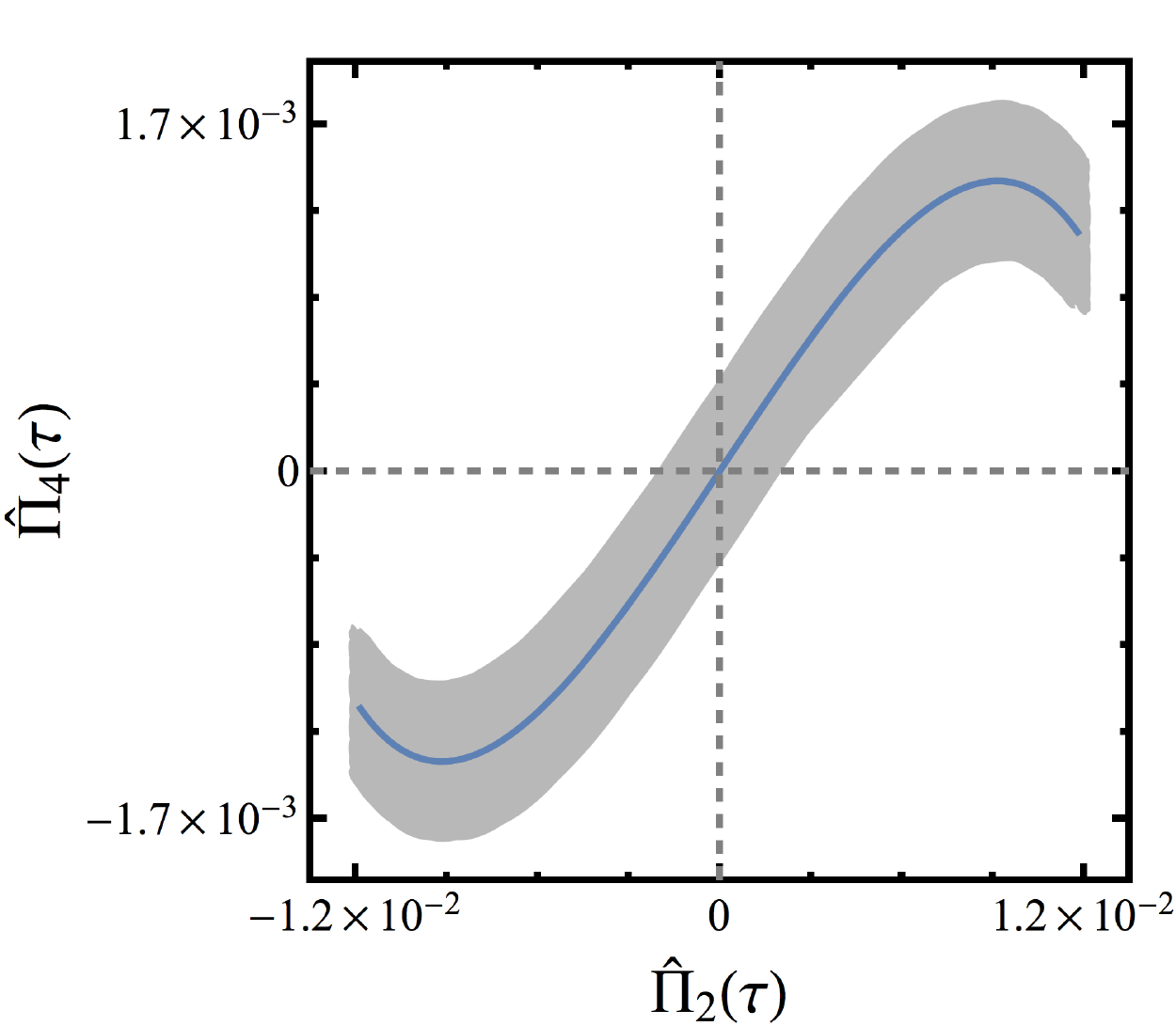} &
    \includegraphics[width=0.46\columnwidth]
    {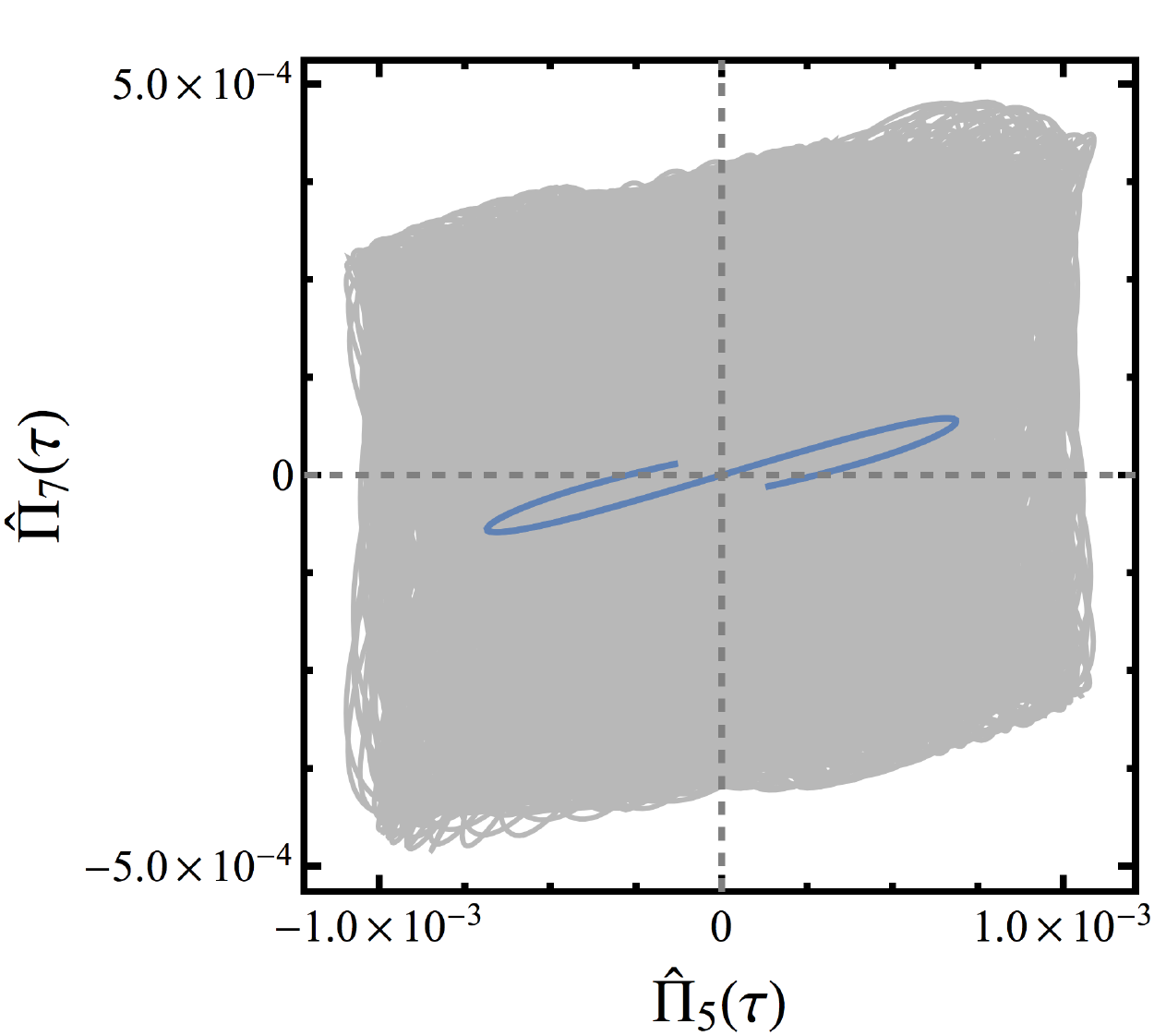}
  \end{tabular}
  \caption{The analogue of Fig.~\ref{fig:AdSPeriodicLoopsE0} with the
    same parameters used except that here we show first excited
    time-periodic solution ($\gamma=1$) of central amplitude
    $\ep=3/10$ ($\Omega\approx\num{6.470658}$).}
  \label{fig:AdSPeriodicLoopsE1}
\end{figure}

\begin{figure}[!t]
  \centering
  \includegraphics[width=\swidth]
  {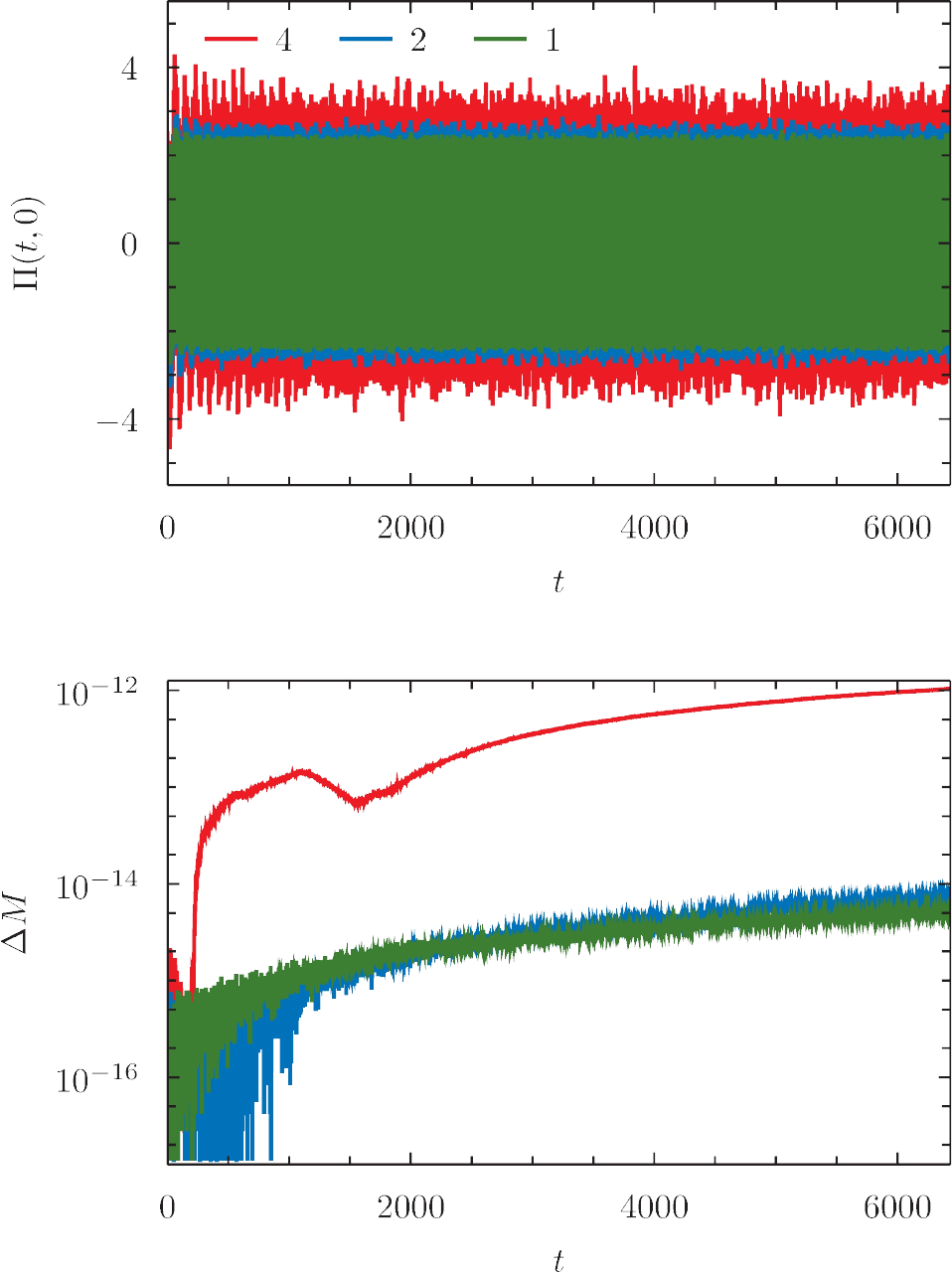}
  \caption{The results of time evolution of stable fundamental
    time-periodic solution ($\gamma=0$, $\ep=0.5$) in $d=4$ space
    dimensions.  \textit{Top panel}.  The evolution of perturbed
    time-periodic solution with Gaussian initial perturbation
    (\ref{eq:394}) of small amplitudes (color coded).  While for large
    amplitudes ($\epsilon\gtrsim 8$) we observe growth of $\Pi(t,0)$,
    for smaller perturbation the solutions stay bounded over long
    times.  \textit{Bottom panel}.  The absolute error of total mass
    $\Delta M:=M(t)-M(0)$ of the dynamical solution.  The time
    evolution was performed with Gauss-RK of order $4$ with time step
    $\Delta t=2^{-8}\pi/N$, with $N=160$.  Growth of $\Delta M$ for
    $\epsilon=4$ indicates that the number of modes $N$ is too small
    (we verified that this not affect our conclusions).}
  \label{fig:AdSPeriodicEvolutionD4E0Stable}
\end{figure}

To study the stability of constructed time-periodic solutions we read
off the coefficients of the expansion, either (\ref{eq:358}) and
(\ref{eq:359}) or (\ref{eq:367}) and (\ref{eq:368}), depending on the
method we use, at the time $t=0$ and put them as the initial data into
spectral evolution code.  For small amplitude solutions, those with
$\ep<\ep_{\ast}$, time evolution is periodic in time despite of the
presence of truncation errors and some amount of numerical noise in
prepared initial data.  This is depicted by closed loops in
Figs.~\ref{fig:AdSPeriodicLoopsE0} and \ref{fig:AdSPeriodicLoopsE1}
representing different sections of the phase space, spanned by the set
of coefficients $\{\hat{\phi}_i(t),\,\hat{\Pi}_j(t)\}$.  This provides
strong evidence not only for the existence of the time-periodic
solutions but also for their (nonlinear) stability.  This argument for
the stability is strengthen by the fact that if we perturb these
solutions slightly, e.g. by setting the nonzero initial
momenta\footnote{The phase of time-periodic solutions was set such
  that $\Pi(t=0,x)\equiv 0$.}
\begin{equation}
  \label{eq:394}
  \Pi(0,x) = \epsilon\,\frac{2}{\pi}
  \exp\left( -\frac{4}{\pi^{2}}\frac{\tan^{2}{x}}{\sigma^{2}} \right),
\end{equation}
(with $\sigma=1/16$) then its evolution is no longer periodic, but
stays close to the periodic orbit,
cf. Figs.~\ref{fig:AdSPeriodicLoopsE0} and
\ref{fig:AdSPeriodicLoopsE1}.  The Ricci scalar (\ref{eq:81})
evaluated at the origin stays bounded over integrated time intervals,
and in contrary to pure AdS case \cite{br,jrb} we do not observe any
such scaling with initial data amplitude.  While for larger amplitudes
we note that after several reflections initial perturbation starts to
grow, the energy begins to flow rapidly from high to low modes (for
$\epsilon=8$ this starts at $t\approx 75$) which causes the mass to
leak out of the system (due to the fact that we are evolving in time
fixed number of modes, which are sufficient to represent large
amplitude solutions only at a very early phase of the evolution) but
this is not triggered for smaller initial perturbations, see
Fig.~\ref{fig:AdSPeriodicEvolutionD4E0Stable}.  Actually what we
observe here is analogous to what we have seen in the spherical cavity
model with Neumann boundary condition (Section~\ref{sec:BoxNeumann}),
and this appear to be a common feature of the models we studied.

\begin{figure}[!t]
  \centering
  \includegraphics[width=\swidth]
  {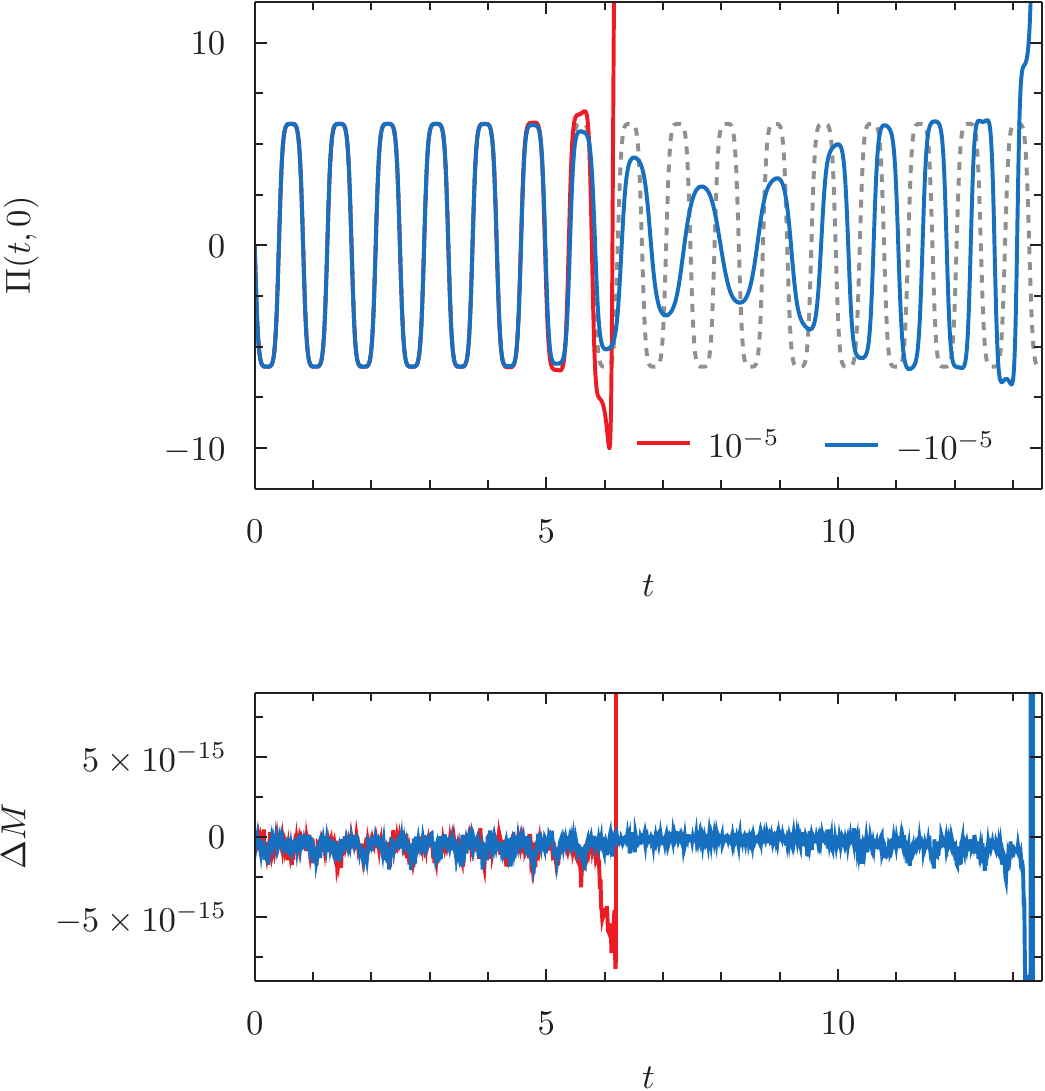}
  \caption{The results of time evolution of unstable fundamental
    time-periodic solution ($\gamma=0$, $\ep=1$) in $d=4$ space
    dimensions.  The time-periodic solution was derived on grid with
    $28\times 96$ points.  The scalar field $\Pi$ of time-periodic
    solution evaluated at the origin is shown with gray dashed line.
    \textit{Top panel}.  The evolution of perturbed time-periodic
    solution with Gaussian initial perturbation (\ref{eq:394}) of
    amplitude $\epsilon=\pm{}10^{-5}$.  \textit{Bottom panel}.  The
    absolute error of total mass $\Delta M:=M(t)-M(0)$ of the
    dynamical solution.  The time evolution was performed with
    Gauss-RK of order $4$ with time step $\Delta t=2^{-8}\pi/N$, with
    $N=160$, which prior to the black hole formation conserves the
    total mass up to the machine precision.}
  \label{fig:AdSPeriodicEvolutionD4E0Unstable}
\end{figure}

\begin{figure}[!th]
  \centering
  \includegraphics[width=\swidth]
  {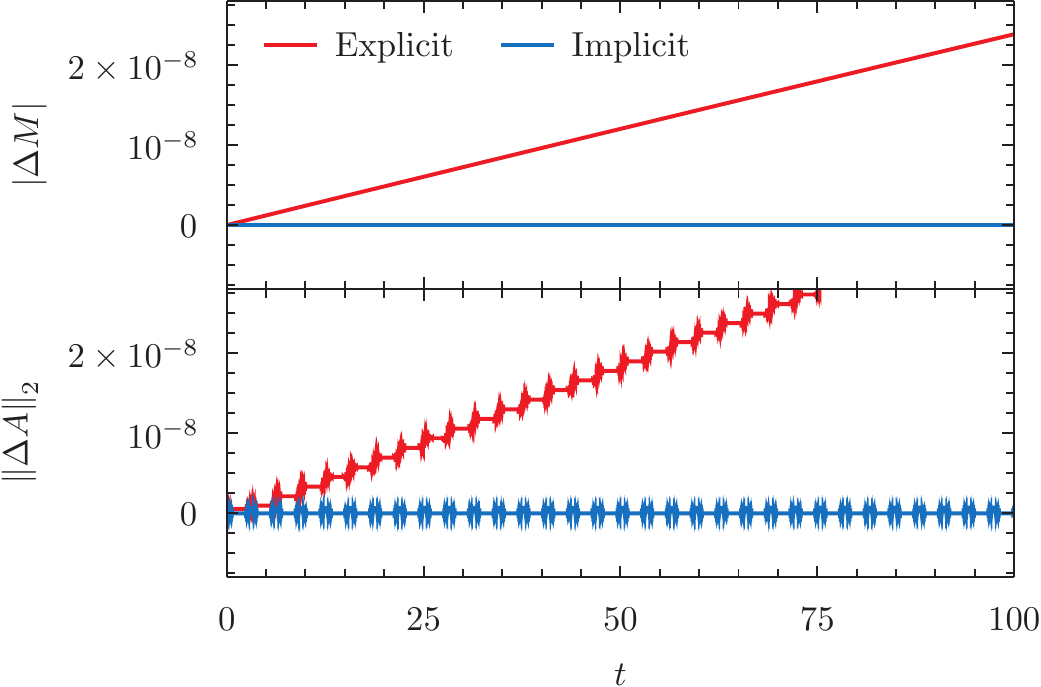}
  \caption{Comparison of explicit and implicit RK methods used to
    integrate system of equations (\ref{eq:67})-(\ref{eq:70}).  For
    this test we use the eigenbasis expansion code with $N=112$ in
    $d=4$ with initial data $\phi(0,x)=0$ and (\ref{eq:394}) with
    $\ep=1$.  In both cases the fixed step method was chosen with
    $\Delta t=\pi/448$.  \textit{Top panel}.  The mass conservation
    error increases linearly with time in the case of explicit RK
    method (we take default fourth order integrator of \mathematica{},
    i.e. the one derived in \cite{Sofroniou20041157}).  In contrast,
    with fourth order Gauss-RK method and exactly the same parameters
    the mass is almost conserved.  \textit{Bottom panel}.  The
    constraint violation norm
    $\left\|\Delta A\right\|_{2} :=
    \left\|A_{\textrm{free}}(t,\,\cdot\,) -
      A_{\textrm{cnst}}(t,\,\cdot\,)\right\|_{2}$
    shows qualitatively similar behaviour (we use the constrained
    evolution and get $A_{\textrm{cnst}}$, the $A_{\textrm{free}}$ was
    derived by solving independently (\ref{eq:70})).}
  \label{fig:AdSEvolutionD4ExplicitVsImplicit}
\end{figure}
\begin{figure}[!th]
  \centering
  \includegraphics[width=\swidth]
  {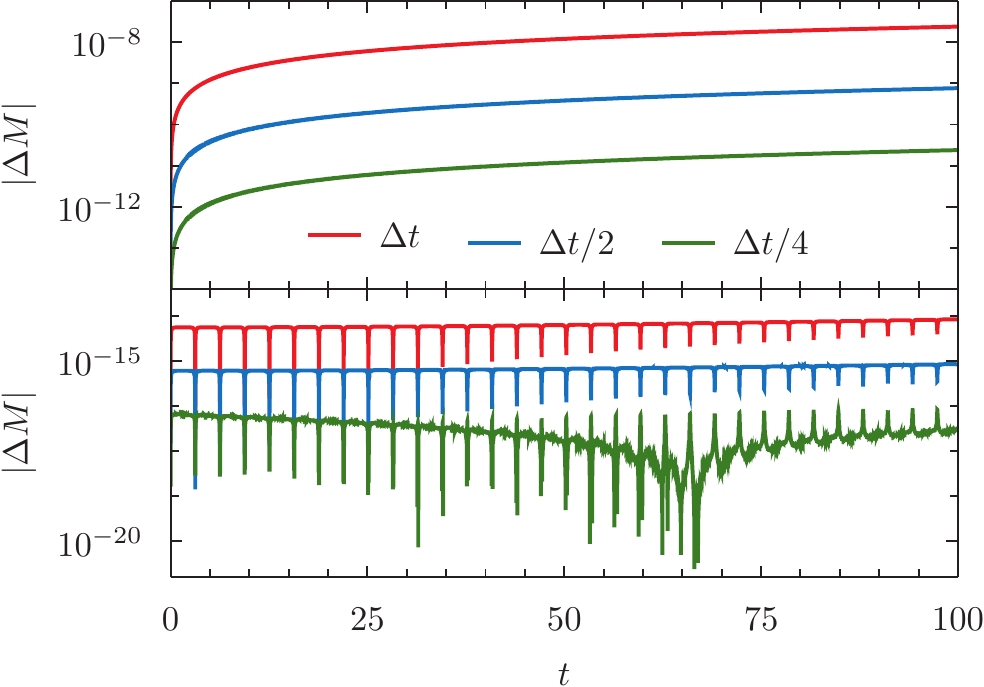}
  \caption{The convergence of total mass error during time-evolution
    with the time step size (for evolutions with the same parameters
    as on Fig.~\ref{fig:AdSEvolutionD4ExplicitVsImplicit}).
    \textit{Top panel}.  The fourth order explicit RK method.
    \textit{Bottom panel}.  For used step sizes the error of the
    fourth order implicit Gauss-RK is at the level of machine
    precision.  Small drift of error is inevitably caused by the
    accumulation of rounding errors.}
  \label{fig:AdSEvolutionD4ExplicitVsImplicitConvergence}
\end{figure}

It turns out, as already mentioned at the beginning of this section,
that not all of the solutions on a bifurcation branch are stable.
Evolution of initial conditions corresponding to time-periodic
solutions with $\ep>\ep_{\ast}$ shows that these are indeed unstable
with respect to small perturbations.  The results of one of such
evolutions in $d=4$ is shown on
Fig.~\ref{fig:AdSPeriodicEvolutionD4E0Unstable}.  We have taken a
solution with amplitude $\ep=1$ bifurcating from the fundamental mode
$\gamma=0$ ($\ep_{\ast}\approx\num{0.835152}$).  We plot the time
evolution of $\Pi$ evaluated at the origin for both positive and
negative initial perturbations (\ref{eq:394}) with amplitudes
$\epsilon=\pm{}10^{-5}$ along with the periodic oscillation.  While
one of the solutions (here the one with positive initial momenta)
collapses after performing just few oscillations then diverges
indicating black hole formation the other (the one with negative
perturbation) stays smooth for almost three times longer and also
collapse.\footnote{Because of numerical errors present in
  time-periodic data we where not able to extend the evolution much
  further by decreasing the Gaussian amplitude.}  The later solution,
showing a delayed collapse, follows a slower oscillation interfering
with its natural frequency ($7.5<t<9$).  This behaviour, which we do
not understand so far, together with full linear stability analysis of
constructed solutions clearly deserve further extended studies.

\begin{figure}[!t]
  \centering
  \includegraphics[width=\swidth]
  {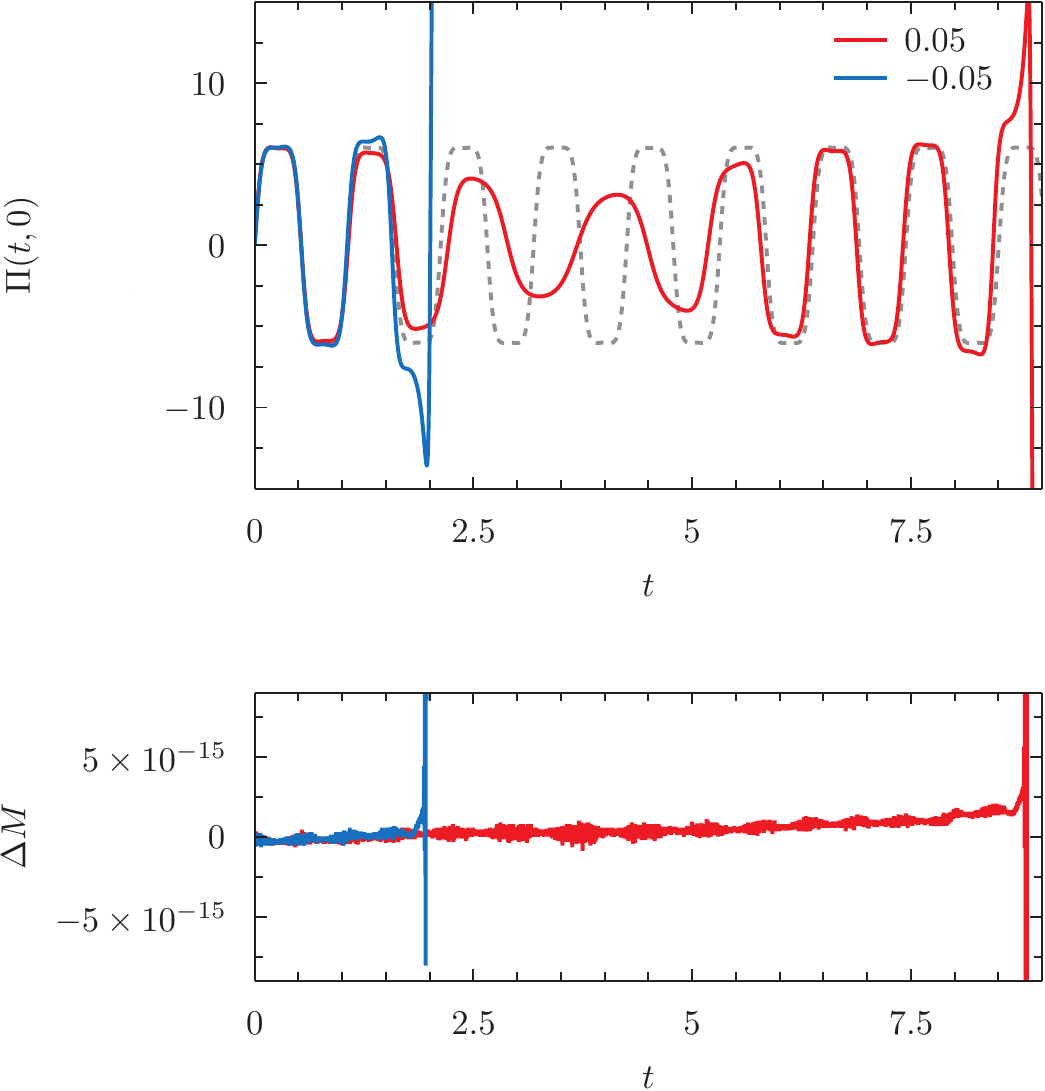}
  \caption{The analogue of
    Fig.~\ref{fig:AdSPeriodicEvolutionD4E0Unstable} in $d=3$ space
    dimensions. The time-periodic solution ($\gamma=0$, $\ep=6$) was
    derived with Chebyshev space discretization method on grid with
    $24\times 192$ points.  The scalar field $\Pi$ of time-periodic
    solution evaluated at the origin is shown in gray dashed line
    (note that $\Pi(t,0)=\Xi(t,0)$).  \textit{Top panel}.  The
    evolution of perturbed time-periodic solution with Gaussian
    initial perturbation (\ref{eq:395}) of amplitude
    $\epsilon=\pm{}1/20$.  \textit{Bottom panel}.  The absolute error
    of total mass $\Delta M:=M(t)-M(0)$ of the dynamical solution.
    The time evolution was performed with Gauss-RK of order $4$ with
    time step $\Delta t=\pi/N^{2}$, which prior to the black hole
    formation conserves the total mass up to the machine precision.}
  \label{fig:AdSPeriodicEvolutionD3E0Unstable}
\end{figure}

Additionally, on Fig.~\ref{fig:AdSPeriodicEvolutionD4E0Unstable} we
show an absolute error in the conservation of the total mass to verify
that the observed behaviour is not caused by the lack of resolution
and to point out the robustness of our numerical methods (the loss of
conservation of mass is caused by the fact that when the solution
approaches collapse, a scalar field profile develops a steep gradient
and the number of modes used in the truncation does not suffice to
represent the solution accurately).  Moreover, on
Figs.~\ref{fig:AdSEvolutionD4ExplicitVsImplicit} and
\ref{fig:AdSEvolutionD4ExplicitVsImplicitConvergence} we compare
evolutions of Gaussian perturbations (\ref{eq:394}) of AdS with
explicit and implicit (symplectic) time integrators.  These, and also
other tests we performed which are not included here, indicate the
superiority of Gauss-RK in long time energy conservation.  We note
that the use of symplectic integrators does not necessarily guarantee
preservation of the constraints for a free evolution
scheme,\footnote{Not shown here; all of the results given in this
  thesis were obtained by solving constrained system.} see
\cite{1751-8121-41-38-382005}.

In order to test the numerical methods based on Chebyshev
pseudospectral approach and also to verify that such behaviour is not
an exclusive feature of even $d$, we have performed a series of tests
in $d=3$ case.  We observed qualitatively the same behaviour.  While
time-periodic solutions to the left of the mass local maximum
(Fig.~\ref{fig:AdSPeriodicOmegaMassD3}) appear to be stable (not shown
here, we observe qualitatively similar to the results shown on
Fig.~\ref{fig:AdSPeriodicEvolutionD4E0Stable}), those on the unstable
part of bifurcation branch collapse to a black hole.  On
Fig.~\ref{fig:AdSPeriodicEvolutionD3E0Unstable} we plot the
time-evolution of $\gamma=0$, $\ep=6$ solution perturbed by
\begin{equation}
  \label{eq:395}
  \Pi(0,x) = \epsilon\,\exp\left(-16\tan^{2}{x}\right),
\end{equation}
with amplitudes $\epsilon=\pm{}1/20$.  Here again with symplectic
Gauss-RK method with sufficiently many grid points and sufficiently
small time integration step taken, the method we are using allows for
stable total mass conserving evolution.

\section{Einstein-Klein-Gordon system---standing waves}
\label{sec:Standing}

The system (\ref{eq:67})-(\ref{eq:69}) for a complex scalar field
$\phi$ admits a particular class of solutions called standing waves.
These are configurations with harmonic scalar field dependence and
time-independent metric
\begin{equation}
  \label{eq:396}
  \phi(t,x) = e^{i \Omega t}\f(x), \ \
  \delta(t,x)=\de(x), \
  A(t,x)=\a(x),
\end{equation}
where we assume the frequency of scalar oscillation to be positive
$\Omega>0$.  We refer to solutions of (\ref{eq:67})-(\ref{eq:69}) with
this particular structure (\ref{eq:396}) as standing waves rather than
boson stars.\footnote{The later being used to name models of star-like
  configurations, which are expected to be localized in space
  (represented as functions of compact support), while the former one,
  as we will see below, are not localized solutions instead these fill
  out a whole space.}  For a review of different models of boson star
solutions, their dynamics and possible astrophysical and cosmological
relevance see \cite{Jetzer1992163, 0264-9381-20-20-201, lrr-2012-6}
and references therein.

With a stationarity ansatz (\ref{eq:396}) the system
(\ref{eq:67})-(\ref{eq:71}) is reduced to the following set of ODEs
The mass (\ref{eq:75}) and the charge (\ref{eq:80}) of the standing
wave solution take the following form
\begin{equation}
  \label{eq:400}
  M = \int_{0}^{\pi/2}\a\left[\,\f'^{2} +
    \left(\frac{\Omega\, e^{\de}}{\a}\f\right)^{2}\right]
  \tan^{d-1}{x}\,\diff x,
\end{equation}
and
\begin{equation}
  \label{eq:401}
  Q = \int_{0}^{\pi/2}\frac{\Omega\, e^{\de}}{\a}\f^{2}
  \tan^{d-1}{x}\,\diff x,
\end{equation}
respectively.

In the subsequent section we construct the solutions of the system
(\ref{eq:397})-(\ref{eq:399}) both perturbativelly
(Section~\ref{sec:StandingPerturbative}) and numerically
(Section~\ref{sec:StandingNumerical}) applying methods developed for
studies of the time-periodic solutions with self-gravitating real
scalar field (see Sections~\ref{sec:AdSPeriodicPerturbative} and
\ref{sec:AdSPeriodicNumeric-Even}).  Next, in
Section~\ref{sec:StandingLinearStab} we study linear stability of
small amplitude solutions---these studies extend recent works
\cite{Buchel2012,Buchel2013}.  The results are analyzed in
Section~\ref{sec:StandingResults}.

Since the issue of incompatibility of eigenbasis functions, given in
Eq.~(\ref{eq:90}), with the regularity conditions for odd space
dimension $d$ also manifests here we restrict the following analysis
to even $d$ only.  For the odd $d$ the adaptation of techniques of
previous section is strightforward, while properties of the solutions
are analogous to these of even $d$.

\subsection{Perturbative construction}
\label{sec:StandingPerturbative}

The method of solving the standing wave equations
(\ref{eq:397})-(\ref{eq:399}) for small amplitude solutions follows
the same steps as for the time-periodic solution in
Section~\ref{sec:AdSPeriodicPerturbative-Even}.  Using perturbative
approach we seek for solution in a form
\begin{align}
  \label{eq:402}
  \f(x;\ep) &= \sum_{\mbox{{\small odd }}\lambda \geq
    1}\ep^{\lambda}\,\f_{\lambda}(x), \quad \f_{1}(x) =
  \frac{e_{\gamma}(x)}{e_{\gamma}(0)}
  \\
  \label{eq:403}
  \de(x;\ep) &= \sum_{\mbox{{\small even }} \lambda \geq
    2}\ep^{\lambda}\,\de_{\lambda}(x),
  \\
  \label{eq:404}
  \a(x;\ep) &= 1 - \sum_{\mbox{{\small even }} \lambda \geq
    2}\ep^{\lambda}\,\a_{\lambda}(x),
  \\
  \label{eq:405}
  \Omega(\ep) &= \omega_{\gamma} + \sum_{\mbox{{\small even }} \lambda
    \geq 2}\ep^{\lambda}\,\xi_{\lambda}.
\end{align}
where $e_{\gamma}(x)$ is a dominant mode in the solution in the limit
$\ep\ra 0$ ($e_{j}(x)$ as before denotes the eigenfunction see
Section~\ref{sec:AdSScalarEigen}).

Since the $e_{\gamma}(x)$ function has exactly $\gamma$ nodes we refer
to the solution with dominant mode $\gamma=0$ as a ground state
solution while for solutions with $\gamma>0$ as excited states (as in
the boson star nomenclature).  This particular choice of $\f_{1}(x)$
in (\ref{eq:402}) together with a requirement $\f_{\lambda}(0)=0$ for
$\lambda\geq 3$ fixes a value of scalar field at the origin to
$\f(0)=\ep$.  We refer to that value as a central density of the
standing wave solution which parametrize given family of solutions
(solutions with fixed number of nodes).

At each perturbative order $\lambda\geq 2$ we decompose scalar field
and metric functions in the eigenbasis $e_{j}(x)$ in a following way
\begin{align}
  \label{eq:406}
  \f_{\lambda}(x) &= \sum_{j} \hat{\f}_{\lambda,j} e_j(x),
  \\
  \label{eq:407}
  \de_{\lambda}(x) &= \sum_j\hat{\de}_{\lambda,j}
  \bigl(e_j(x)-e_{j}(0)\bigr),
  \\
  \label{eq:408}
  \a_{\lambda}(x) &= \sum_j\hat{\a}_{\lambda,j}e_j(x).
\end{align}
(With this form of expansion we fixed the gauge choice by setting
$\de(0)=0$.)  It is important to note that, for any even space
dimension $d$, the sums in (\ref{eq:406})-(\ref{eq:408}) are finite at
each order $\lambda$ of the perturbative expansions
(\ref{eq:402})-(\ref{eq:405}).  We plug the expansion
(\ref{eq:402})-(\ref{eq:405}) into (\ref{eq:397})-(\ref{eq:399}),
perform the Taylor series expansion around $\ep=0$ and require for the
coefficients of the resulting polynomial in $\ep$ to vanish
identically.  A relative simplicity of the solution procedure in this
case is a consequence of the form of the ansatz (\ref{eq:396}) which
reduces a PDE system to an ODE system.  This, with use of
(\ref{eq:406})-(\ref{eq:408}), allow us to obtain the solution by
solving the linear algebraic system for Fourier coefficients instead
of solving a coupled ODEs (such idea is also explored in the numerical
construction).  The perturbative procedure starts at $\lambda=2$ (the
first order equation is already satisfied by $\f_{1}(x)$ defined in
(\ref{eq:402})).  Inserting (\ref{eq:403}) and (\ref{eq:407}) into
(\ref{eq:403}) and projecting onto $e_k'(x)$, we get a solution
\begin{equation}
  \label{eq:409}
  \hat{\de}_{\lambda, k} = - \frac{1}{2\omega_k^2}
  \inner{e_k'}{\coef{\lambda}\sin 2x\left[\,\f'^{2} + \left(\frac{\Omega
          e^{\de}}{\a}\f\right)^{2}\right]}\,.
\end{equation}
Similarly, inserting the series (\ref{eq:404}) and (\ref{eq:408}) into
(\ref{eq:399}) and projecting onto $e_k(x)$ (after multiplication by a
trigonometric factor $\sin x\cos x$), we get a linear system of
equations for the coefficients $\hat{\a}_{\lambda,j}$
\begin{multline}
  \label{eq:410}
  \sum_j \left[ (d-1) \delta_{kj} + \inner{e_k}{\frac{1}{2} \sin2x \,
      e_j' - \cos 2x\, e_j}\right] \hat{\a}_{\lambda,j} =
  \\
  \frac{1}{4} \inner{e_k}{\coef{\lambda}(\sin 2x)^2 \a\left[\f'^{2} +
      \left(\frac{\Omega e^{\de}}{\a}\f\right)^{2}\right]}\,.
\end{multline}
It is useful to note that the principal matrix of this system is
tridiagonal.  This system supplied with the $\a_{\lambda}(0) =
\sum_j\hat{\a}_{\lambda,j}e_j(0) = 0$ condition allows for the unique
solution for the coefficients $\hat{\a}_{\lambda,j}$.  In this way we
solve constraints equations in each even perturbative order
$\lambda\geq 2$.

For odd $\lambda\geq 3$ the series expansion of (\ref{eq:397}) has a
form of linear inhomogeneous second order ODE
\begin{equation}
  \label{eq:411}
  \omega_{\gamma}^{2}\,\f_{\lambda} - L\f_{\lambda} = S_{\lambda},
\end{equation}
with a source function $S_{\lambda}$ depending on all lower order
expansion coefficients: $\f_{1}(x),\ldots,\f_{\lambda-2}(x)$,
$\a_{2}(x),\ldots,\a_{\lambda-1}(x)$,
$\de_{2}(x),\ldots,\de_{\lambda-1}(x)$ and
$\xi_{2},\ldots,\xi_{\lambda-1}$.  Using the orthogonality property of
basis functions we get
\begin{equation}
  \label{eq:412}
  \hat{\f}_{\lambda,j} =
  \frac{\inner{e_{j}}{S_{\lambda}}}{\omega_{\gamma}^{2} - \omega_{j}^{2}},
  \quad j\in\mathbb{N}_{0},\ j\neq\gamma.
\end{equation}
For $j=\gamma$ we use frequency correction $\xi_{\lambda-1}$ to
satisfy an itegrability condition
\begin{equation}
  \label{eq:413}
  \inner{e_{\gamma}}{S_{\lambda}} = 0,
\end{equation}
then the free coefficient of $\f_{\lambda}(x)$ namely
$\hat{f}_{\lambda,\gamma}$ is determined form normalization condition
\begin{equation}
  \label{eq:414}
  \f_{\lambda}(0) = 0.
\end{equation}
In this way we get a unique solution for any perturbative order
$\lambda$ with $d$ and $\gamma$ serving as the only parameters.

\subsection{Numerical construction}
\label{sec:StandingNumerical}

As for the time-periodic solutions we also construct standing wave
solution by solving the system (\ref{eq:397})-(\ref{eq:399})
numerically.  Here we represent the solution by a set of $3N$ Fourier
coefficients We expand the scalar field $\f(x)$ and metric functions
$\de(x)$, $\a(x)$ into $N$ eigenmodes of linearized problem
(\ref{eq:90}) in a following way
\begin{align}
  \label{eq:415}
  \f(x) &= \sum_{j=0}^{N-1}\hat{\f}_{j} e_j(x),
  \\
  \label{eq:416}
  \de(x) &= \sum_{j=0}^{N-1}\hat{\de}_{j}\left(e_{j}(x) - e_{j}(0)\right),
  \\
  \label{eq:417}
  \a(x) & = 1 - \sum_{j=0}^{N-1} \hat{\a}_{j} \left(e_j(x) - e_{j}(0)\right),
\end{align}
With this form of expansion both the boundary conditions at $x=\pi/2$
(for even $d$) and regularity conditions at the origin are satisfied.
The gauge condition $\de(0)=0$ is satisfied identically by this
expansion.  Then we require (\ref{eq:397})-(\ref{eq:399}) together
with (\ref{eq:415})-(\ref{eq:417}) to be satisfied at the set of $N$
collocation points (\ref{eq:317}).  We supply this system with an
additional equation fixing the central value of scalar field
$\f(0)=\ep$.  In this way we get a nonlinear system of $3N+1$
equations for $3N+1$ unknowns the expansion coefficients
$\{\hat{\f}_{j}, \hat{\a}_{j}, \hat{\de}_{j}\}_{0\leq j \leq N-1}$ and
the frequency $\Omega$.  Solving this system for a fixed $\ep$ with
Newton-Raphson root-finding algorithm yields the standing wave
solution.  To find a standing wave solution with $\gamma$ nodes in
$\f(x)$ profile we start Newton's method with the following initial
data
\begin{equation}
  \label{eq:418}
  \hat{\f}_{\gamma} = \frac{\ep}{e_{j}(0)}, \quad \Omega=\omega_{\gamma},
\end{equation}
and with all other Fourier coefficients in
(\ref{eq:415})-(\ref{eq:417}) set to zero.

\subsubsection{Remark on alternative numerical methods}
\label{sec:StandingNumericalRemark}

To construct standing wave solutions we could use an alternative
approach, e.g. the shooting method or the FD discretization of
Eqs.~(\ref{eq:415})-(\ref{eq:417}) (see \cite{LaiPhD} for application
of FDA to an analogous problem in asymptotically flat spacetime).  The
use of such local methods is particularly advantageous when looking
for solutions with large central density.  This is related to the
character of the solution; when we increase $\ep$ the solution profile
$\f(x)$ gets steeper and steeper, while the metric function $\a(x)$
develops a narrow local minimum.  These features are hard to resolve
with relatively small values of $N$.  Additionally local methods do
not suffer from the boundary behaviour issue thus work equally well in
odd and even dimensions.

Using a shooting approach we integrate
Eqs.~(\ref{eq:415})-(\ref{eq:417}) outward starting at $x=0$ with the
boundary conditions
\begin{equation}
  \label{eq:419}
  \f(0)=\ep, \quad \f'(0)=0, \quad \de(0)=0, \quad \a(0)=0,
\end{equation}
which follow from regularity conditions.  The condition for the
eigenvalue $\Omega$ is the regularity at the conformal boundary of AdS
\begin{equation}
  \label{eq:420}
  \f(\pi/2) = 0.
\end{equation}
Alternatively, we can fix the value of $\Omega$ and shoot for $\ep$
(it turns out that $\Omega(\ep)$ is monotonic function).  For general
value of $\ep$ (with fixed value of $\Omega>d$ in $d$ space
dimensions) the solution will not satisfy the condition
(\ref{eq:420}).  Only for specific values of $\ep$ this condition can
be fulfilled.  Moreover, being positive function at the origin with
negative slope near the center $\f(x)$ may cross the real axis many
times or may have no zero at all.  So, if we fix
$\Omega>\omega_{\gamma}$ then there will be an infinite number of
solutions regular at $x=\pi/2$ each with at least $\gamma$ nodes.  In
a very similar way one can solve numerically equations governing
linear perturbations which are discussed below.

\subsection{Linear stability}
\label{sec:StandingLinearStab}

To study the linear stability we make the perturbative ansatz
$(0<|\mu|\ll 1)$
\begin{align}
  \label{eq:421}
  \phi(t,x) & = e^{i\Omega t}\Bigl(\,\f(x) + \mu\,\psi(t,x) +
  \mathcal{O}\left(\mu^{2}\right) \Bigr),\\
  \delta(t,x) & = \de(x) + \mu\Bigl(\alpha(t,x) - \beta(t,x)\Bigr) +
  \mathcal{O}\left(\mu^{2}\right), \\
  A(t,x) & = \a(x) \Bigl( 1 + \mu\,\alpha(t,x) +
  \mathcal{O}\left(\mu^{2}\right) \Bigr),
\end{align}
and we neglect higher order terms in $\mu$.  Next, we assume harmonic
time dependence of the perturbation\footnote{Note the change of signs
  in the exponents with respect to \cite{MR2014}, which is a typo in
  that paper.  The easiest way to correct that misprint is to use
  (\ref{eq:422}) instead of (22) in \cite{MR2014} with $\psi_{+}$ and
  $\psi_{-}$ interchanged in the equations and the following
  discussion. }
\begin{subequations}
  \label{eq:422}
  \begin{align}
    \psi(t,x) & = \psi_{+}(x)e^{-i\Chi t} + \psi_{-}(x)e^{i\Chi t}, \\
    \alpha(t,x) & = \alpha(x)\cos\Chi t, \\
    \beta(t,x) & = \beta(x)\cos\Chi t,
  \end{align}
\end{subequations}
where $\psi_{+}(x)$ and $\psi_{-}(x)$ are both real functions.  This
is the most general ansatz allowing for separation of $t$ and $x$
dependence, making at the same same time the resulting system of
equations relatively simple (cf.~\cite{Buchel2013}).  Plugging the
(\ref{eq:421})-(\ref{eq:422}) into (\ref{eq:67})-(\ref{eq:71}) and
linearizing about $\mu=0$ we obtain a set of differential-algebraic
equations
\begin{align}
  \label{eq:423}
  \begin{split}
    \alpha & = -\sin{2x}\left\{\frac{\Omega}{\Chi}\,\f
      \left(\psi_{+}'-\psi_{-}'\right) \right. \\
    & \left. \quad + \,\f' \left[\left(1-\frac{\Omega
          }{\Chi}\right)\psi_{+} + \left(1 +
          \frac{\Omega}{\Chi}\right)\psi_{-}\right]\right\}\,,
  \end{split}
  \\[1.5ex]
  \label{eq:424}
  \beta' & = - \frac{d - 1 - \cos 2x}{\sin{x}\cos{x}}\frac{\alpha}{\a}\,,
  \\[1.5ex]
  \label{eq:425}
  \begin{split}
    \psi_{\pm}'' & = - \frac{d - 1 - \cos{2x}\,(1 -
      \a)}{\a\sin{x}\cos{x}}\psi_{\pm}' - \left(1 \mp
      \frac{\Chi}{\Omega}\right)^2
    \left(\frac{\Omega e^{\de}}{\a}\right)^2\psi_{\pm} \\
    & \quad - \frac{1}{2}\beta'\f' + \left(1 \mp \frac{\Chi}{2
        \Omega}\right) \left(\frac{\Omega
        e^{\de}}{\a}\right)^2\beta\,\f\,.
  \end{split}
\end{align}
This system supplied with the boundary conditions (inherited from
(\ref{eq:85}))
\begin{equation}
  \label{eq:426}
  \psi_{\pm}(\pi/2) = 0, \quad \alpha(\pi/2) = 0,
  \quad \beta'(\pi/2) = 0,
\end{equation}
and the regularity conditions at $x=0$
\begin{equation}
  \label{eq:427}
  \psi_{\pm}'(0) = 0, \quad \alpha(0)=\beta(0)=0,
\end{equation}
is a linear eigenvalue problem with $\Chi$ as an eigenvalue.  In
principle, knowing standing wave solution $\f(x)$, $\a(x)$, $\de(x)$,
we could integrate (\ref{eq:423})-(\ref{eq:425}) to obtain a solution
in a closed form.  Since this is not the case here, we again resort on
perturbative method (in principle we could solve the equations
numerically, using either a pseudospectral or shooting method, these
can be solved simultaneously with the equations determining a standing
wave but from a perturbative approach we expect to gain more insight).

Relaying on perturbative approach we expand the unknown functions
$\alpha(x)$, $\beta(x)$, $\psi_{\pm}(x)$ and frequency $\Chi$ in small
parameter $\ep$ (the same as in (\ref{eq:402})-(\ref{eq:405}) for the
standing wave solution)
\begin{align}
  \label{eq:428}
  \Chi(\ep) &= \sum_{\mbox{{\small even }}\lambda\geq
    0}\ep^{\lambda}\chi_{\lambda},
  \\
  \psi_{\pm}(x;\ep) &=
  \sum_{\mbox{{\small even }}\lambda\geq
    0}\ep^{\lambda}\psi_{\pm,\lambda}(x),
  \\
  \alpha(x;\ep) & = \sum_{\mbox{{\small odd }}\lambda\geq
    1}\ep^{\lambda}\alpha_{\lambda}(x),
  \\
  \label{eq:429}
  \beta(x;\ep) &=
  \sum_{\mbox{{\small odd }}\lambda\geq
    1}\ep^{\lambda}\beta_{\lambda}(x).
\end{align}
Plugging (\ref{eq:402})-(\ref{eq:405}) and
(\ref{eq:428})-(\ref{eq:429}) into (\ref{eq:423})-(\ref{eq:425}) we
demand that the equations are satisfied at each order of $\ep$.
Moreover, as for the standing wave solution we expand the unknown
functions in eigenbasis $e_{j}(x)$
\begin{subequations}
  \label{eq:430}
  \begin{align}
    \psi_{\pm,\lambda}(x) & = \sum_{j\geq
      0}\inner{e_{j}}{\psi_{\pm,\lambda}} e_{j}(x),
    \\
    \alpha_{\lambda}(x) & = \sum_{j\geq
      0}\hat{\alpha}_{\lambda,j}\left(e_{j}(x)-e_{j}(0)\right),
    \\
    \beta_{\lambda}(x) & = \sum_{j\geq 0}\hat{\beta}_{\lambda,j}
    \left(e_{j}(x)-e_{j}(0)\right).
  \end{align}
\end{subequations}
At the lowest order $\mathcal{O}(\ep^{0})$ the constraints
(\ref{eq:423}) and (\ref{eq:424}) are identically satisfied, while
from (\ref{eq:425}) we get two linear second order equations
\begin{equation}
  \label{eq:431}
  L\psi_{\pm,0} - (\chi_{0} \mp \omega_{\gamma})^{2}\psi_{\pm,0} = 0\,,
\end{equation}
Using decomposition of $\psi_{\pm,0}(x)$ and orthogonality of the
basis functions (\ref{eq:90}) we get the condition for the frequency
$\chi_{0}$
\begin{equation}
  \label{eq:432}
  \left\{
    \begin{aligned}
      \omega_{j}^{2} - (\chi_{0} - \omega_{\gamma})^{2}  & = 0, \\
      \omega_{k}^{2} - (\chi_{0} + \omega_{\gamma})^{2}  & = 0.
    \end{aligned}
  \right.
\end{equation}
This system is satisfied when: $\psi_{-,0}\equiv 0$,
$\psi_{+,0} = e_{j}(x)$, and $\chi_{0}=\omega_{\gamma}\pm\omega_{j}$
or $\psi_{+,0}\equiv 0$, $\psi_{-,0} = e_{k}(x)$, and
$\chi_{0}=-\omega_{\gamma}\pm\omega_{k}$ (there is also the case when
neither of $\psi_{\pm,0}(x)$ is zero, i.e.
$\psi_{+,0}(x) = e_{j}(x)$, $\psi_{-,0}(x) = e_{k}(x)$ with $k$, $j$
such that $d+2\gamma = |k-j|$ holds, but construction of solutions for
this choice breaks down at higher orders, thus we exclude this case).
Taking into account the form of the ansatz (\ref{eq:422}), due to its
symmetry: $\psi_{\pm}\ra\psi_{\mp}$ and $\Chi\ra-\Chi$, these two
seemingly different cases are in fact equivalent.  Therefore, it
suffices to consider the former case, so as a solution of the linear
system (\ref{eq:431}) we take
\begin{equation}
  \label{eq:433}
  \psi_{+,0}(x) = e_{\zeta}(x),
  \quad \psi_{-,0}(x) = 0,
  \quad \chi_{0}^{\pm} = \omega_{\gamma} \pm \omega_{\zeta}\,.
\end{equation}

Thus, at the lowest order in $\ep$, solution (\ref{eq:433}) specifies
a standing wave with $\gamma$ nodes perturbed by a single eigenmode
with $\zeta$ nodes.  Next, at each odd order $\lambda$ the constraints
are solved as follows.  The coefficients $\hat{\alpha}_{\lambda,j}$
are simply given in terms of the decomposition of the order $\lambda$
of the right hand side of the equation (\ref{eq:423}).  Next, we
rearrange Eq.~(\ref{eq:424}) at the order $\lambda$ to obtain the
linear system for the expansion coefficients of the
$\beta_{\lambda}(x)$ function
\begin{equation}
  \label{eq:434}
  \sum_{i}\hat{\beta}_{\lambda,i}\inner{e_{k}}{\sin x\cos x e_{i}'}
  = - \inner{e_{k}}{\coef{\lambda}(d-1-\cos 2x)\frac{\alpha}{\a}}.
\end{equation}
For any even $\lambda$ the system (\ref{eq:423})-(\ref{eq:425})
reduces to two inhomogeneous equations
\begin{equation}
  \label{eq:435}
  L\psi_{\pm,\lambda} - (\chi_{0} \mp \omega_{\gamma})^{2}
  \psi_{\pm,\lambda} = S_{\pm,\lambda}\,,
\end{equation}
with source terms $S_{\pm,\lambda}$ depending on the lower order
expansion coefficients in (\ref{eq:402})-(\ref{eq:405}) and
(\ref{eq:428})-(\ref{eq:429}).  Using the $\psi_{+,\lambda}(x)$
expansion formula (\ref{eq:430}) and projecting the first equation in
(\ref{eq:435}) onto the $e_{i}(x)$ mode we have
\begin{equation}
  \label{eq:436}
  \inner{e_{i}}{\psi_{+,\lambda}}
  = \frac{\inner{e_{i}}{S_{+,\lambda}}}{\omega_{i}^{2} - \omega_{\zeta}^{2}},
   \quad i\in\mathbb{N}_{0},\ i\neq\zeta\,,
\end{equation}
where we have used the definition of $\chi_{0}^{\pm}$ given in
(\ref{eq:433}).  For $i=\zeta$ the necessary condition
\begin{equation}
  \label{eq:437}
  \inner{e_{\zeta}}{S_{+,\lambda}} = 0\,,
\end{equation}
is satisfied by an appropriate choice of the parameter
$\chi_{\lambda}$, while the free coefficient
$\inner{e_{\zeta}}{\psi_{+,\lambda}}$ is fixed as follows.  We set the
value of $\psi_{+}(x)$ at the origin to unity (we use the fact that
governing equations are linear and we set
$\psi_{+,0}(x)=e_{\zeta}(x)/e_{\zeta}(0)$), then since
$\psi_{+,0}(0)=1$ we require that $\psi_{+,\lambda}(0)=0$ for
$\lambda\geq 2$ which corresponds to taking
\begin{equation}
  \label{eq:438}
  \inner{e_{\zeta}}{\psi_{+,\lambda}} =
  - \sum_{i\neq\zeta}\inner{e_{i}}{\psi_{+,\lambda}}e_{i}(0)\,.
\end{equation}
For a second equation in (\ref{eq:435}) after projection on $e_{k}(x)$
mode, we get
\begin{equation}
  \label{eq:439}
  \inner{e_{k}}{\psi_{-,\lambda}}
  = \frac{\inner{e_{k}}{S_{-,\lambda}}}{\omega_{k}^{2}
    - \left(2\omega_{\gamma}\pm\omega_{\zeta}\right)^{2}},
  \quad k\neq k_{*},
\end{equation}
where $\omega_{k_{*}} = |2\omega_{\gamma}\pm\omega_{\zeta}|$ and the
sign depends on the particular choice of $\chi_{0} =
\chi_{0}^{\pm}$.  For $\chi_{0} = \chi_{0}^{+} = \omega_{\gamma} +
\omega_{\zeta}$ the $k_{*}=d+2\gamma+\zeta>0$ and the condition
\begin{equation}
  \label{eq:440}
  \inner{e_{k_{*}}}{S_{-,\lambda}} = 0,
\end{equation}
can always be satisfied by an appropriate choice of a constant
$\inner{e_{k_{*}}}{\psi_{-,\lambda-2}}$ (it is remarkable that at the
lowest nontrivial order $\lambda=2$ the coefficient
$\inner{e_{k_{*}}}{S_{-,\lambda=2}}$ is always zero for any
combination of $\gamma$ and $\zeta$, so we can continue our
construction to arbitrary high order $\lambda$, having exactly one
undetermined constant after solving order $\lambda$, which will be
fixed at higher order $\lambda+2$).  On the other hand, for
$\chi_{0}=\chi_{0}^{-}=\omega_{\gamma}-\omega_{\zeta}$ we have $k_{*}
= \frac{1}{2}\left(\left|d + 2(2\gamma - \zeta)\right| - d\right)$
which can be either positive or negative.  For $k_{*}<0$ there are
always solutions to (\ref{eq:439}) since the denominator, on right
hand side, is always different from zero for any $k\geq 0$, and the
coefficient $\inner{e_{k}}{\psi_{-,\lambda}}$ will be determined by
the formula (\ref{eq:439}).  The $k_{*}\geq 0$ case is more involved
since there are two possibilities: either $d + 2(2\gamma - \zeta)\geq
0$ which gives $k_{*}=2\gamma - \zeta\geq 0$ and there are no
solutions to (\ref{eq:439}) since it turns out that the coefficient
$\inner{e_{k_{*}}}{S_{2,\lambda=2}}$ is nonzero, which leads to
contradiction, either $d + 2(2\gamma - \zeta)<0$ and for $k_{*} =
\zeta - 2\gamma - d\geq 0$ the coefficient
$\inner{e_{k_{*}}}{S_{-,\lambda=2}}$ is zero and the unknown
$\inner{e_{k_{*}}}{\psi_{-,\lambda=2}}$ will be fixed at higher order
$\lambda=4$ and we proceed just like for the $\chi_{0}=\chi_{0}^{+}$
case.  To sum up, for $\chi_{0}=\chi_{0}^{+}$ there are solutions for
any choice of $\gamma$ and $\zeta$, while for the
$\chi_{0}=\chi_{0}^{-}$ there exists solutions only for $\zeta >
2\gamma$.

In this way we construct a solution describing a standing wave with
dominant eigenmode $e_{\gamma}(x)$ peturbed (at the linear level) by a
dominant eigenmode $e_{\zeta}(x)$.  Note the (general) ansatz
(\ref{eq:421})-(\ref{eq:422}) allows us to perturb a fixed standing
wave with any eigenmode, as opposed to the analysis presented in
\cite{Buchel2013}.  The ansatz proposed in \cite{Buchel2013} restricts
the form of perturbations, such that it allows for a $\gamma$-node
standing wave to be perturbed by a solution with $\gamma$-nodes only.
For that reason it is not suitable to find the full spectrum of linear
perturbations.

\subsection{Results}
\label{sec:StandingResults}

\begin{figure}[!t]
  \centering
  \includegraphics[width=\swidth]{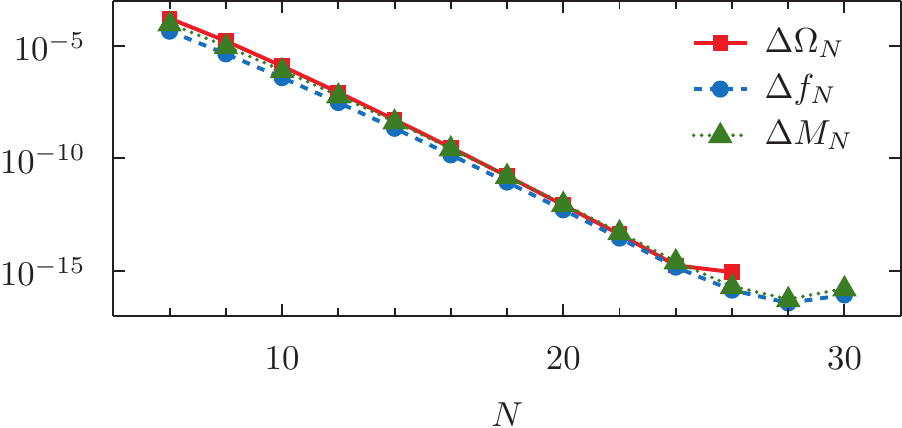}
  \\[2ex]
  \includegraphics[width=\swidth]{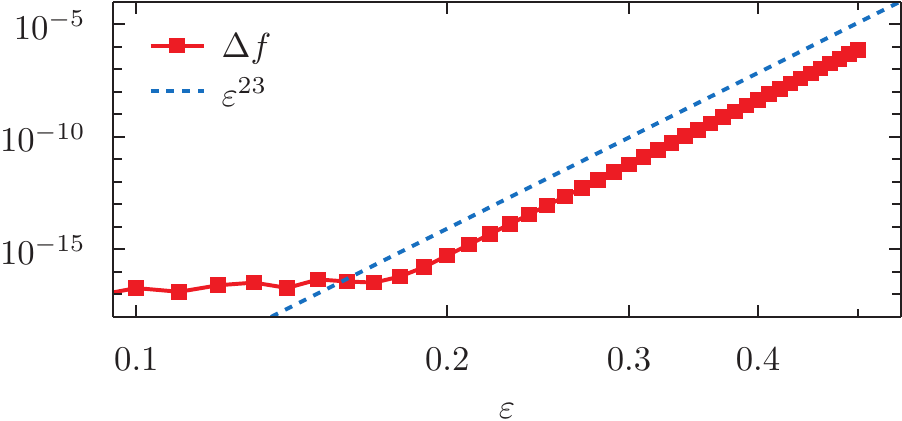}
  \caption{\textit{Top panel}.  The convergence test of numerical code
    for ground state standing wave solution with $\f(0)=3/10$
    ($\mbox{$\Omega\approx \num{4.56690}$}$).  The frequency error
    $\Delta\Omega_{N} := |\Omega_{N}-\Omega_{N=32}|$, scalar field
    profile error $\Delta\f_{N}:=\left\|\,\f_{N}-\f_{N=32}\right\|_{2}$,
    and total mass error $\Delta M_{N}:=|M_{N}-M_{N=32}|$ computed for
    increasing number of Fourier coefficients $N$ in
    (\ref{eq:415})-(\ref{eq:417}) compared with reference solution
    with $N=32$. \textit{Bottom panel}.  The comparison of numerical
    and analytical ground state standing wave solutions for varying
    value of $\f(0)=\ep$.  The scalar field absolute error
    $\Delta\f:=\left\|\,\f_{\mathrm{num}} -
      \f_{\mathrm{pert}}\right\|_{2}$ is computed for numerical
    solution with $N=48$ eigenmodes, the perturbative series was found
    up to $\mathcal{O}\left(\ep^{23}\right)$ order.  For small values
    of $\ep<0.2$ the rounding errors dominate.  The discrete
    $l^{2}$-norm $\left\|\,\cdot\,\right\|_{2}$ was computed on as set
    of equally spaced grid points $x_{i}=i\pi/800$, $i=1,\ldots,
    400$.}
  \label{fig:StandingNumericsVsPerturbative}
\end{figure}

In analysis of the results obtained by the methods presented in
previous sections, we restrict ourselves to $d=4$ case and present the
fundamental ($\gamma=0$) standing wave solution; properties of excited
solutions ($\gamma>0$) are qualitatively very similar.

\begin{figure}[!t]
  \centering
  \includegraphics[width=\swidth]{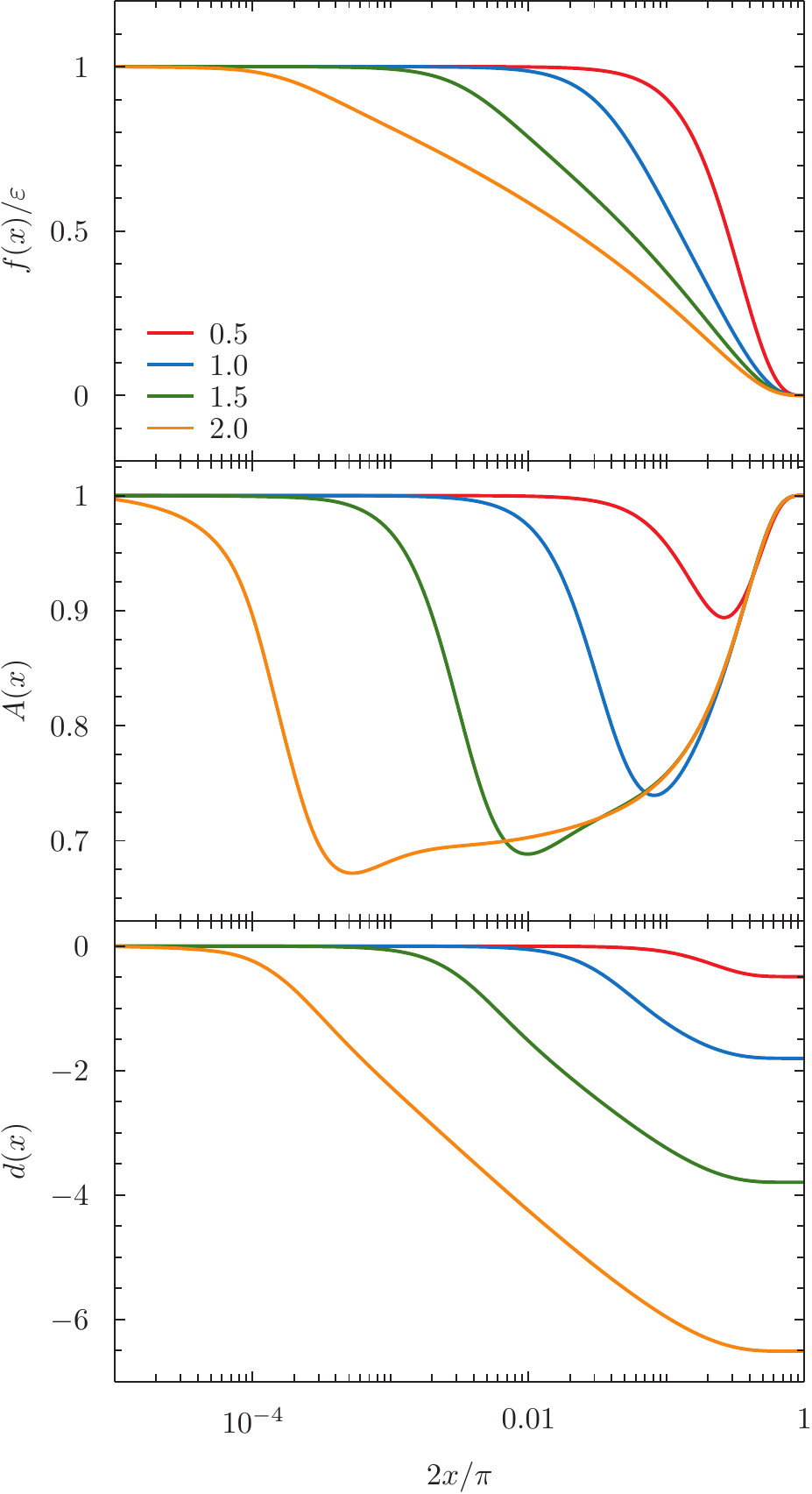}
  \caption{Dependence of standing wave soltuion scalar field profile
    and metric functions on the central density $\ep$ (color coded).
    Note the use of logarithmic scale on rescaled horizontal axis.}
  \label{fig:StandingNumericsLargeEpsilon}
\end{figure}

Fig.~\ref{fig:StandingNumericsVsPerturbative} shows both the
convergence rate of our numerical pseudospectral method and a
comparison with a perturbatively constructed solution.  A high order
perturbative series gives an accurate result which for small values of
$\mbox{$\ep\lesssim{}0.2$}$ is equivalent to the numerical solution up
to the roundoff errors.  On
Fig.~\ref{fig:StandingNumericsLargeEpsilon} we show a scalar field and
metric function profiles for few values of central density $\ep$.
With increasing value of central density the scalar field profile
$\f(x)$ concentrates at the origin but still has a polynomial tail.
Also the metric functions $\a(x)$ and $\de(x)$ exhibit a steep gradient
near the origin.
Since these compact configurations are difficult to accurately resolve
numerically with relatively small number of modes, the use of other
than spectral methods are preferred, as pointed out in the remark.
The results shown on Fig.~\ref{fig:StandingNumericsLargeEpsilon} where
obtained by shooting method.

\begin{figure}[!t]
  \centering
  \includegraphics[width=\swidth]{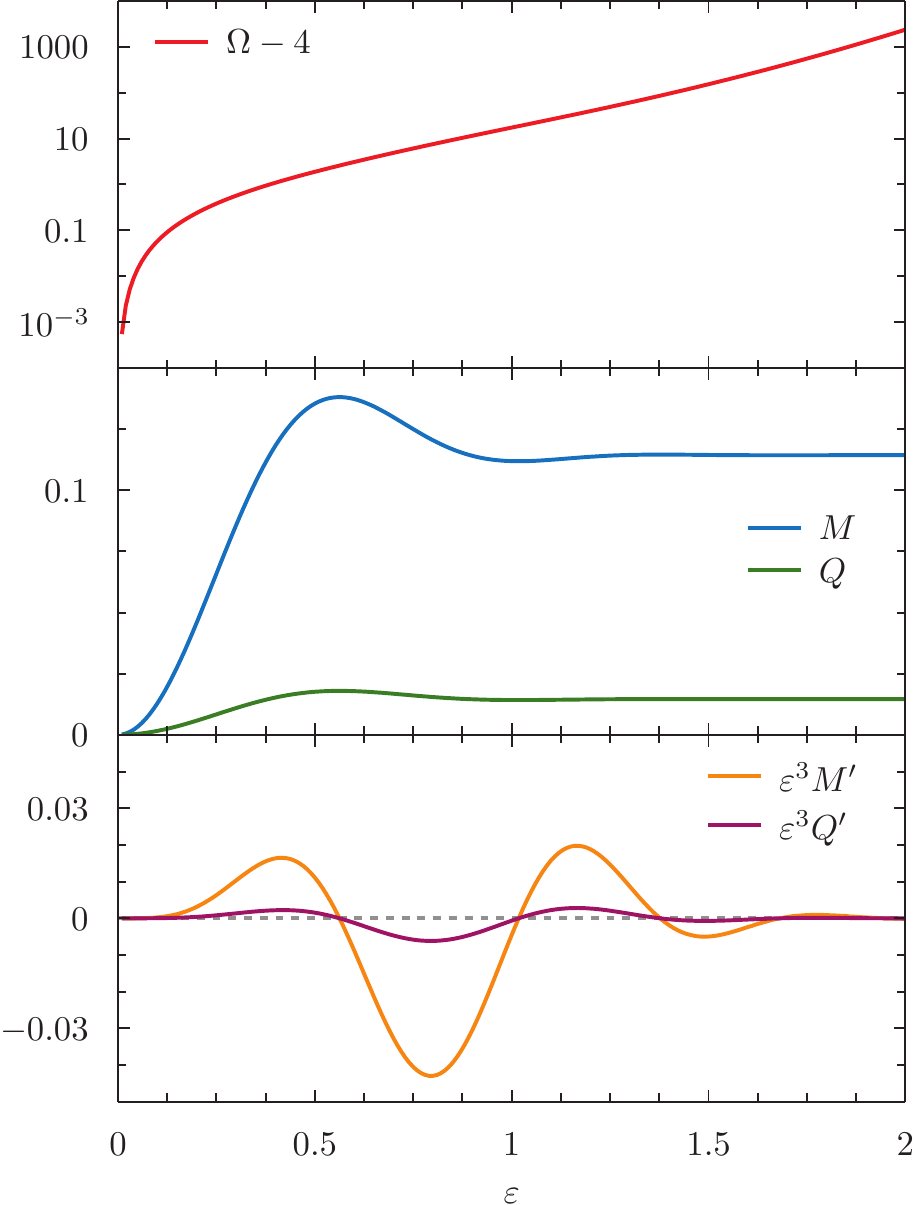}
  \caption{\textit{Top panel}.  The frequency $\Omega$ as a function
    of central density $\ep$ for fundamental standing wave solutions
    in $d=4$ space dimensions.  While for small values of $\ep$ the
    frequency has polynomial dependence on $\ep$ (in agreement with
    perturbative expansion) for large central densities it grows
    exponentially.  \textit{Middle and bottom panels}.  Dependence of
    mass $M$ and charge $Q$ (defined in (\ref{eq:400}) and
    (\ref{eq:401}) respectively) of standing wave solutions on the
    central density $\ep$.  The extremum points of $M(\ep)$ correspond
    to the extrema of $Q(\ep)$.  For presentation we multiply
    derivatives by $\ep^{3}$.}
  \label{fig:StandingNumericsMassCharge}
\end{figure}

With increasing $\ep$ the standing wave solution becomes more compact
and more massive (and at the same time more charged) while we note
from Fig.~\ref{fig:StandingNumericsMassCharge} that there exists a
maximum mass for the family of solutions.  Thus, in analogy to the
asymptotically flat solutions, this model also exhibits an analog of
the Chandrasekhar mass limit.  No stationary configurations exists
with masses greater than $M_{\ast}\equiv M(\ep_{\ast})$.  Moreover,
the evolution of configurations that lie to the left of the mass
maximum $\ep<\ep_{\ast}$ shows that they are stable with respect to
small perturbations while those to the right $(\ep>\ep_{\ast})$ are
unstable.  In asymptotically flat boson star models this type of
behaviour was verified by direct numerical evolution of unstable
configurations \cite{PhysRevD.42.384, PhysRevD.58.104004,
  PhysRevD.62.104024}, for self-interacting case see
\cite{Becerril2007263}.  The stability analysis of standing wave
solutions is discussed in more detail below.

The structure of solutions with different number of nodes $\gamma$ in
different number of space dimension $d$ is as follows.  For fixed
space dimension $d$ the location of first extremum of $M(\ep)$
decreases with increasing $\gamma$; similarly for fixed $\gamma$ (for
given family of solutions) the location of stability point,
$\ep_{\ast}$, decreases with increasing $d$.  For $d=4$ the maximum
value of mass of a ground state standing wave solution is
$M_{\ast}\approx\num{0.138065}$ and the first few consecutive
stationary points of mass are: $\ep_{\ast}\approx \num{0.563},
\num{1.018}, \num{1.377}$ and $\num{1.682}$.  Further, each such
extremum point corresponds to the appearance of a zero mode, i.e.
$\Chi(\ep)$ tends to zero as $\ep\ra\ep_{\ast}$ (this issue is
emphasized below).  Moreover each extremum point of $M(\ep)$
corresponds to the extremum of $Q(\ep)$, i.e.
$M'(\ep_{\ast})=Q'(\ep_{\ast})=0$.  These properties of standing wave
solutions in AdS are similar to the boson star models in
asymptotically flat case \cite{Gleiser1989733, LaiPhD,
  LaiChoptuik2007, PhysRevD.62.104024}.

To study the stability of small amplitude ($0<\left|\ep\right|\ll 1$)
standing wave solutions we have solved the higher orders of
perturbative equations (in terms of the $\ep$ expansion) to get
successive approximation to the solution of the system
(\ref{eq:423})-(\ref{eq:425}) and in particular for the
eigenfrequences $\Chi^{\pm}_{\zeta}(\ep)$.  Repeating this procedure
for successive values of $\zeta$ (the wave number) we can compute the
spectrum of linear perturbations around the standing wave (by deducing
a general expression for frequency corrections $\chi_{\lambda}$ in
perturbative series expansion (\ref{eq:428})).  A systematic analysis
of our results lead us to the observation that all of these
corrections are given in terms of the recurrence relation which is
easy to solve.\footnote{To be precise, these fall to the following
  class of recurrence relation: $a_{n}y_{n+1} + b_{n}y_{n} + c_{n}=0$,
  $n\in\mathbb{N}$, with coefficients $a_{n}$, $b_{n}$ and $c_{n}$
  being polynomials in $n$ of orders depending on and increasing with
  $\lambda$, $\gamma$ and $d$; these are solved case by case.}  Here
we present just a sample of our calculations for the ground state
solution $\gamma=0$ and first excited solution $\gamma=1$.  For
$\chi_{\lambda=0}^{+} = \omega_{\gamma=0} + \omega_{\zeta}$ the second
and fourth order coefficients in (\ref{eq:428}) read
\begin{equation}
  \label{eq:441}
  \begin{split}
    \chi_{2} &=
    \bigl(\num{1134}\,\zeta^6+\num{19003}\,\zeta^5+\num{124820}\,\zeta^4
    +\num{407705}\,\zeta^3+\num{688426}\,\zeta^2
    \\
    & \quad +\num{548112}\,\zeta
    +\num{146160}\bigr)\bigl(\num{448}\,\zeta^5+\num{5600}\,\zeta^4
    +\num{25760}\,\zeta^3+\num{53200}\,\zeta^2
    \\
    & \quad +\num{47292}\,\zeta +\num{13230}\bigr)^{-1},
  \end{split}
\end{equation}
\begin{equation}
  \label{eq:442}
  \begin{aligned}
    \chi_{4} & = \Bigl( \num{780065499136}\,\zeta^{21} +
    \num{41163366682624}\,\zeta^{20} +
    \num{1006051101954048}\,\zeta^{19}
    \\
    & \quad + \num{15102140468966400}\,\zeta^{18} +
    \num{155617295714301696}\,\zeta^{17}
    \\
    & \quad + \num{1164464261736365184}\,\zeta^{16} +
    \num{6520784612437551808}\,\zeta^{15}
    \\
    & \quad + \num{27691944373998004960}\,\zeta^{14} +
    \num{89026702549663104696}\,\zeta^{13}
    \\
    & \quad + \num{211802620660726920504}\,\zeta^{12} +
    \num{347405017468617340788}\,\zeta^{11}
    \\
    & \quad + \num{295005895048265576070}\,\zeta^{10} -
    \num{219816084830486177849}\,\zeta^9
    \\
    & \quad - \num{1183082173497216325931}\,\zeta^8 -
    \num{2011100419520146792317}\,\zeta^7
    \\
    & \quad - \num{1981235719849412491245}\,\zeta^6 -
    \num{1102026285670662519354}\,\zeta^5
    \\
    & \quad - \num{166769419982648034006}\,\zeta^4 +
    \num{213918880041409312548}\,\zeta^3
    \\
    & \quad + \num{161636443558413319440}\,\zeta^2 +
    \num{47492605169680204800}\,\zeta
    \\
    & \quad + \num{5337973362362256000} \Bigr)\Bigl( \num{1358280}
    (\zeta +3) (\zeta +4) (2 \zeta -3) (2 \zeta -1)
    \\
    & \quad \times (2 \zeta +1)^3 (2\zeta +3)^3 (2 \zeta +5)^3 (2
    \zeta +7)^3 (2 \zeta +9)^3 (2 \zeta +11) \Bigr)^{-1},
  \end{aligned}
\end{equation}
for $\zeta\in\mathbb{N}_{0}$, while in the $\chi_{\lambda=0}^{-} =
\omega_{\gamma=0} - \omega_{\zeta}$ case we get
\begin{equation}
  \label{eq:443}
  \begin{split}
    \chi_{2} &=
    \bigl(-\num{1134}\,\zeta^6-\num{8213}\,\zeta^5-\num{16920}\,\zeta^4
    -\num{455}\,\zeta^3+\num{28674}\,\zeta^2
    \\
    & \quad
    +\num{13168}\,\zeta-\num{15120}\bigr)\bigl(\num{448}\,\zeta^5
    +\num{3360}\,\zeta^4+\num{7840}\,\zeta^3 +\num{5040}\,\zeta^2
    \\
    & \quad -\num{1988}\,\zeta-\num{1470}\bigr)^{-1},
  \end{split}
\end{equation}
\begin{equation}
  \label{eq:444}
  \begin{aligned}
    \chi_{4} & = \Bigl( -\num{780065499136}\,\zeta^{21} -
    \num{24362135244800}\,\zeta^{20} -
    \num{334001844441088}\,\zeta^{19}
    \\
    & \quad -
    \num{2620283850820608}\,\zeta^{18}-\num{12724918886962944}\,\zeta^{17}
    \\
    & \quad -
    \num{37508263920951168}\,\zeta^{16}-\num{52331076299836096}\,\zeta^{15}
    \\
    & \quad +
    \num{49633095360308704}\,\zeta^{14}+\num{366300085095405128}\,\zeta^{13}
    \\
    & \quad +
    \num{630404345332416984}\,\zeta^{12}+\num{161482377313411596}\,\zeta^{11}
    \\
    & \quad -
    \num{1049533950802646634}\,\zeta^{10}-\num{1502573105590735255}\,\zeta^9
    \\
    & \quad - \num{244286507129741639}\,\zeta^8
    +\num{866752269056468573}\,\zeta^7
    \\
    & \quad + \num{103774233213806463}\,\zeta^6
    -\num{826669645928487918}\,\zeta^5
    \\
    & \quad - \num{328263009917069502}\,\zeta^4 +
    \num{300382014204958140}\,\zeta^3
    \\
    & \quad + \num{176773133899303200}\,\zeta^2 -
    \num{26938306783080000}\,\zeta
    \\
    & \quad -\num{27498788800560000} \Bigr)\Bigl( \num{1358280}\zeta
    (\zeta +1) (2 \zeta -3) (2 \zeta -1)^3
    \\
    & \quad \times (2 \zeta +1)^3 (2\zeta +3)^3 (2 \zeta +5)^3 (2
    \zeta +7)^3 (2 \zeta +9) (2 \zeta +11) \Bigr)^{-1},
  \end{aligned}
\end{equation}
for $\zeta\in\mathbb{N}$.  From this we can read off the asymptotic
expansion of the linear spectrum of perturbed standing wave
(\ref{eq:428}).  Up to fourth order in $\ep$, for large wave numbers
$\zeta$, the spectrum (of ground state standing wave solution
$\gamma=0$) reads
\begin{equation}
  \begin{split}
    \label{eq:445}
    \Chi^{+}_{\zeta}(\ep) & = \left(2+\fracn{81}{32} \ep^2 +
      \fracn{706663}{322560}\ep^4 + \ldots \right)\zeta
    \\
    & \quad +
    \left(8+\fracn{1207}{112}\ep^2+\fracn{908257501}{86929920}\ep^4 +
      \ldots \right)
    \\
    & \quad - \left(\fracn{105}{64}\ep^2+\fracn{29319}{28672}\ep^4 +
      \ldots \right)\zeta^{-1}
    \\
    & \quad + \left( \fracn{165}{16}\ep^2 + \fracn{472547}{28672}\ep^4
      + \ldots \right)\zeta^{-2}
    +\mathcal{O}\left(\zeta^{-3}\right)\,,
  \end{split}
\end{equation}
\begin{equation}
  \begin{split}
    \label{eq:446}
    \Chi^{-}_{\zeta}(\ep) & =
    -\left(2+\fracn{81}{32}\ep^2+\fracn{706663}{322560}\ep^4 + \ldots
    \right) \zeta
    \\
    & \quad +
    \left(\fracn{73}{112}\ep^2+\fracn{48824929}{28976640}\ep^4 +
      \ldots \right)
    \\
    & \quad + \left(\fracn{105}{64}\ep^2+\fracn{29319}{28672}\ep^4 +
      \ldots \right)\zeta^{-1}
    \\
    & \quad + \left(\fracn{15}{4}\ep^2 + \fracn{50753}{4096}\ep^4 +
      \ldots \right)\zeta^{-2} + \mathcal{O}\left(\zeta^{-3}\right)\,,
  \end{split}
\end{equation}
while for first excited solution $\gamma=1$ we get
\begin{equation}
  \begin{split}
    \label{eq:447}
    \Chi^{+}_{\zeta}(\ep) & = \left( 2 - \fracn{6579}{2048}\ep^{2} -
      \fracn{2518060221}{922746880}\ep^{4} + \ldots \right)\zeta
    \\
    & \quad + \left( 10 + \fracn{313641}{78848}\ep^{2} +
      \fracn{19711592741578761}{4231792395550720}\ep^{4} + \ldots
    \right)
    \\
    & \quad + \left( \fracn{14175}{8192}\ep^{2} +
      \fracn{8290232793}{5167382528}\ep^{4} + \ldots \right)\zeta^{-1}
    \\
    & \quad + \left( \fracn{24975}{2048}\ep^{2} +
      \fracn{171408520413}{5167382528}\ep^{4} + \ldots
    \right)\zeta^{-2} + \mathcal{O}\left(\zeta^{-3}\right)\,,
  \end{split}
\end{equation}
\begin{equation}
  \begin{split}
    \label{eq:448}
    \Chi^{-}_{\zeta}(\ep) & = - \left( 2 + \fracn{6579}{2048}\ep^{2} +
      \fracn{2518060221}{922746880}\ep^{4} + \ldots \right) \zeta
    \\
    & \quad + \left( 2 + \fracn{313641}{78848}\ep^{2}
      \fracn{19711592741578761}{4231792395550720}\ep^{4} + \ldots
    \right)
    \\
    & \quad + \left( \fracn{14175}{8192}\ep^{2} +
      \fracn{8290232793}{5167382528}\ep^{4} + \ldots \right)\zeta^{-1}
    \\
    & \quad + \left( \fracn{24975}{2048}\ep^{2} +
      \fracn{171408520413}{5167382528}\ep^{4} + \ldots
    \right)\zeta^{-2} + \mathcal{O}\left(\zeta^{-3}\right)\,,
  \end{split}
\end{equation}
for $\zeta\ra\infty$.  Thus the spectrum is manifestly dispersive (the
group velocity $\diff \Chi^{\pm}_{\zeta}/\diff\zeta$ depends on
$\zeta$) and is only asymptotically resonant for $\ep\neq{}0$ (the
$\ep=0$ corresponds to linear perturbations of AdS space and we
recover the resonant spectrum).  Additionally we note the following
$\lim_{\zeta\ra\infty}\diff\Chi^{+}_{\zeta}/\diff\zeta = -
\lim_{\zeta\ra\infty}\diff\Chi^{-}_{\zeta}/\diff\zeta$.  The
dispersive character of linear spectrum has a direct consequence on
the dynamics of perturbed standing wave solution, which we investigate
below.

\begin{table}[!t]
  \centering
  \begin{equation*}
    \begin{array}{cllll}
      \toprule
      \multirow{2}{*}{$\zeta$} & \multicolumn{2}{c}{\Chi^{+}} & \multicolumn{2}{c}{\Chi^{-}} \\
      & \multicolumn{1}{c}{\text{numeric}} & \multicolumn{1}{c}{\text{perturbative}} &
      \multicolumn{1}{c}{\text{numeric}} & \multicolumn{1}{c}{\text{perturbative}}
      \\ \midrule
      0 & \phantom{1}\num{8.111661210840} & \phantom{1}\num{8.111652} & \multicolumn{1}{c}{\text{---}} & \multicolumn{1}{c}{\text{---}} \\
      1 & \num{10.134321659884} & \num{10.134312} & \phantom{1}\num{-1.999578064801} & \phantom{1}\num{-1.999556} \\
      2 & \num{12.159077420529} & \num{12.159067} & \phantom{1}\num{-4.033347392183} & \phantom{1}\num{-4.0333488} \\
      3 & \num{14.184352284616} & \num{14.184340} & \phantom{1}\num{-6.062592316375} & \phantom{1}\num{-6.0625916} \\
      4 & \num{16.209810121153} & \num{16.20979} & \phantom{1}\num{-8.090070643161} & \phantom{1}\num{-8.090068} \\
      5 & \num{18.235341639390} & \num{18.235328} & \num{-10.116773090642} & \num{-10.116769} \\
      6 & \num{20.260904312594} & \num{20.26089} & \num{-12.143076829705} & \num{-12.1430714} \\
      7 & \num{22.286479760545} & \num{22.286463} & \num{-14.169151834984} & \num{-14.169145} \\
      8 & \num{24.312059525533} & \num{24.312041} & \num{-16.195084717138} & \num{-16.195076} \\
      9 & \num{26.337639597242} & \num{26.337620} & \num{-18.220923808978} & \num{-18.220914} \\
      \bottomrule
    \end{array}
  \end{equation*}
  \caption{The comparison of  perturbativelly and numerically derived the
    lowest eigenfrequencies (the linear spectrum) of fundamental
    ($\gamma=0$) standing wave solution with amplitude $\ep=1/10$ in
    $d=4$ space dimensions.  The numerical solutions where determined
    with $N=32$ modes (points).}
  \label{tab:StandingE0Spectrum}
\end{table}
\begin{figure}[!th]
  \centering
  \includegraphics[width=\swidth]
  {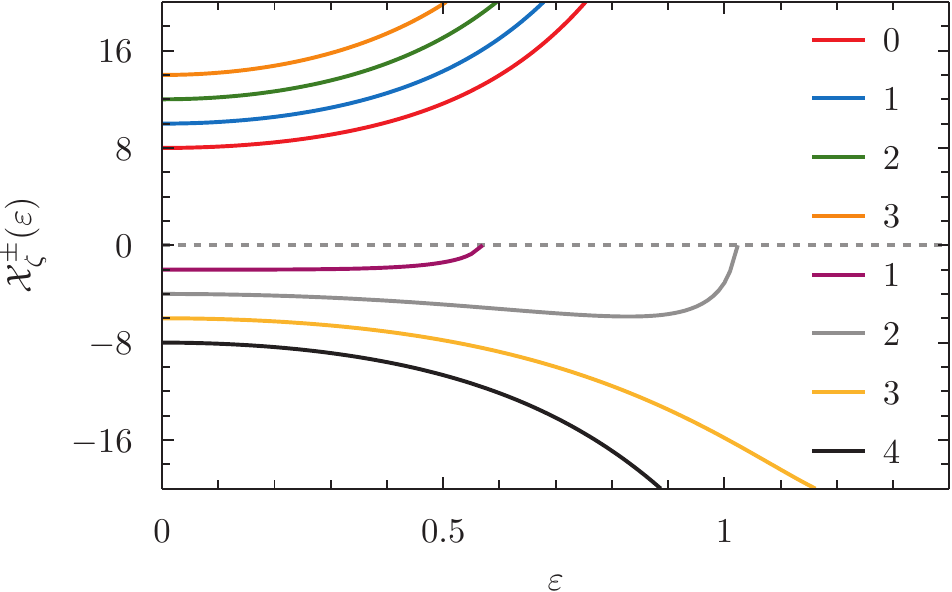}
  \caption{The linear perturbation spectrum $\Chi^{\pm}_{\zeta}(\ep)$
    of family of the fundamental standing wave solutions.  The first
    few eigenfrequencies are plotted with color coded wave number
    $\zeta$.  The consecutive negative oscillation frequencies tend to
    zero at stationary points of mass $M$ of the solutions
    parametrized by their central density $\ep$.}
  \label{fig:StandingE0Spectrum}
\end{figure}

In Tab.~\ref{tab:StandingE0Spectrum} the perturbativelly derived
eigenvalues are compared with the numerical solution of the system
(\ref{eq:423})-(\ref{eq:425}).  The fourth order accurate perturbative
series, with coefficients (\ref{eq:441})-(\ref{eq:442}) and
(\ref{eq:443})-(\ref{eq:444}), was evaluated for $\ep=1/10$ giving
result with precision of $6$ to $7$ significant digits.
\begin{figure}[!t]
  \centering
  \includegraphics[width=\swidth]{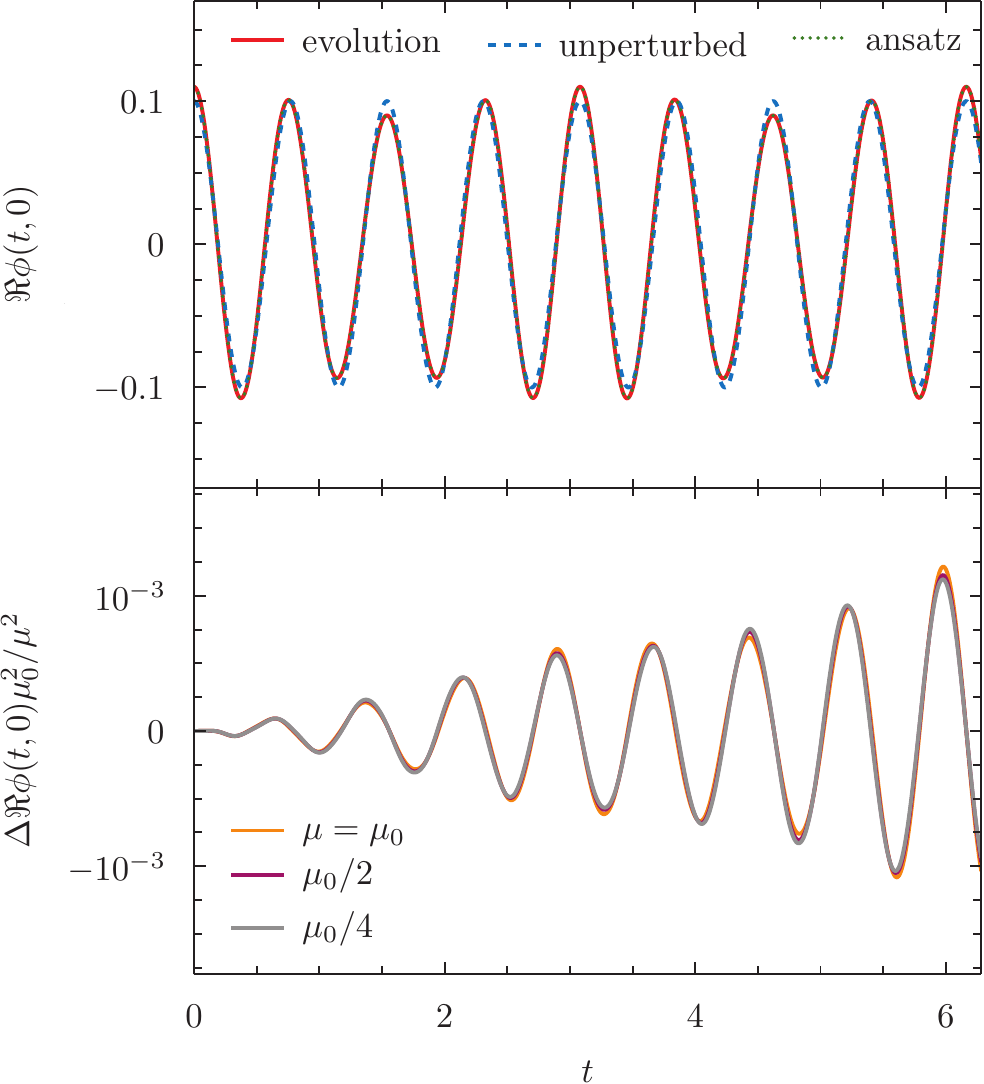}
  \caption{\textit{Top panel}.  The comparison of numerical solution
    (solid red line) of linear mode perturbed standing wave solution
    with the formula given by the ansatz (\ref{eq:421})-(\ref{eq:422})
    (dotted green line).  The $\gamma=2$, $\ep=1/10$
    ($\Omega\approx\num{8.155377}$) solution was perturbed with mode
    $\zeta=3$ (the plus case, $\Chi\approx\num{18.342540}$) of
    amplitude $\mu=10^{-2}$.  The numerical solutions were obtained by
    using $N=64$ eigenmodes. \textit{Bottom panel}.  The difference of
    numerical and analytical solution, denoted by
    $\Delta\Re\phi(t,0)$.  The rescaled error by $\mu^{2}$ is
    convergent with $\mu\ra 0$ which supports the consistency of the
    ansatz (\ref{eq:421})-(\ref{eq:422}).}
  \label{fig:StandingE2ModePerturbation}
\end{figure}
To verify these results further---the form of the ansatz and derived
solutions---we have solved the system (\ref{eq:67})-(\ref{eq:69})
subject to reflecting boundary conditions, using methods of
Section~\ref{sec:AdSEvolution}, with initial conditions derived from
(\ref{eq:421})-(\ref{eq:422}) for small values of $\mu$ and compared
numerical solution with analytical prediction (as given by the
perturbative ansatz) for different choices of $\gamma$ and
$\zeta$.  The results, of which a nontrivial example we present on
Fig.~\ref{fig:StandingE2ModePerturbation}, show both the consistency
and convergence with $\mu\ra 0$.  The solution exhibits harmonic
oscillation around stable standing wave solution, i.e. a pure
harmonic dependence of the scalar field at the origin is
modulated\footnote{For our normalization condition $\psi_{+}(0)=1$.}
\begin{equation}
  \label{eq:449}
  \phi(t,0) = \ep e^{i\Omega t} +
  \mu\left( e^{i(\Omega-\Chi)t} + \psi_{-}(0)e^{i(\Omega+\Chi)t} \right)
  + \mathcal{O}\left(\mu^{2}\right),
\end{equation}
which is depicted on Fig.~\ref{fig:StandingE2ModePerturbation} (with
dashed blue line for unperturbed solution ($\mu=0$) and with dotted
green line the real part of the formula (\ref{eq:449}); the $d=4$,
$\gamma=2$, $\zeta=3$ case with $\ep=1/10$, $\mu=1/100$,
$\Omega\approx\num{8.155377}$, $\Chi\approx\num{18.342540}$ and
$\psi_{-}(0)\approx\num{0.002804}$).

\begin{figure}[!t]
  \centering
  \includegraphics[width=\swidth]
  {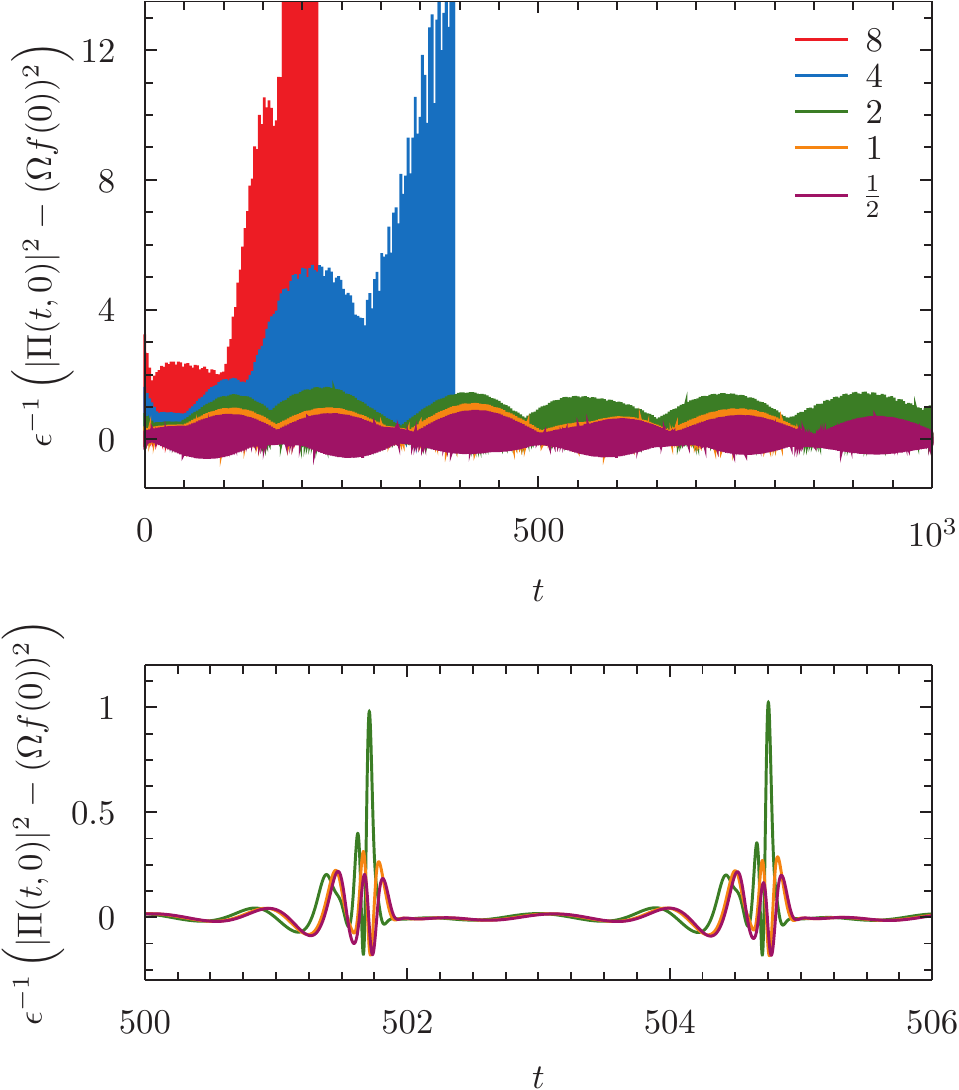}
  \caption{\textit{Top panel}.  The time evolution of the squared
    module of a scalar field $\Pi(t,x)$ at the origin ($x=0$) for a
    perturbed ground state standing wave solution with evaluated at
    the origin for perturbed standing wave with $\f(0)=0.16$
    ($\Omega\approx\num{4.15034}$) by a narrow Gaussian pulse
    (\ref{eq:450}) of decreasing amplitude (labeled with different
    line colors).  \textit{Bottom panel}. A close-up showing scaling
    with an amplitude of the perturbation $\epsilon$, which improves
    when $\epsilon\ra 0$.  Because of the nonlinearity of governing
    field equations, we cannot exactly separate contributions coming
    from a standing wave solution (a constant value) and a
    perturbation.}
  \label{fig:StandingE0GaussPerturbation}
\end{figure}
\begin{figure}[!t]
  \centering
  \includegraphics[width=\swidth]
  {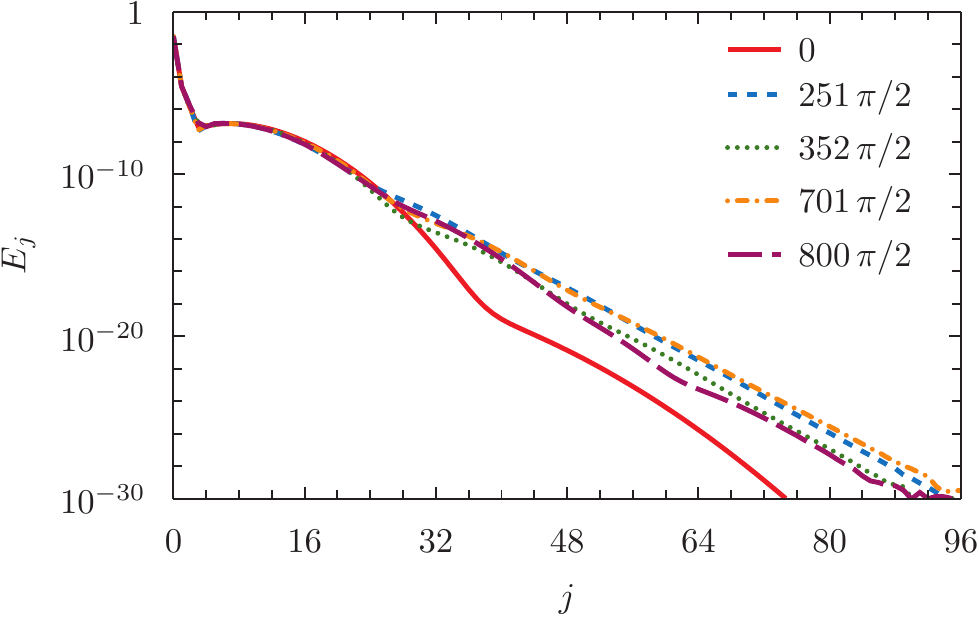}
  \caption{Plot of the energy spectrum defined as in
    Eqs.~(\ref{eq:326}) and (\ref{eq:327}) at initial and late times
    (labeled by different line types) for the solution of perturbed
    standing wave (\ref{eq:450}) with $\f(0)=0.16$
    ($\Omega\approx\num{4.15034}$) and amplitude of the Gaussian
    perturbation with $\epsilon=1/2$.  For late times the spectrum
    falls off exponentially with an almost constant slope (compare
    Fig.~2 in \cite{MRIJMPA} for perturbed AdS solution).}
  \label{fig:StandingE0GaussPerturbationEnergySpectra}
\end{figure}

The part of the linear oscillation perturbation spectrum
$\Chi^{\pm}_{\zeta}(\ep)$ and the change of its character with $\ep$,
for a family of fundamental solutions, is shown on
Fig.~\ref{fig:StandingE0Spectrum}.  While frequencies
$\Chi^{+}_{\zeta}(\ep)$ increase monotonically with $\ep$ the
$\Chi^{-}_{\zeta}(\ep)$ tent to zero when $\ep$ tends to consecutive
critical points $\ep_{\ast}$ of mass $M(\ep)$.  At these points a
standing wave solution lose stability, the zero mode appears (a
feature observed in asymptotically flat boson stars
\cite{Gleiser1989733}); moreover for solutions with central densities
close to $\ep_{\ast}$ and for $\ep>\ep_{\ast}$ our numerical procedure
ceases to find solution for Eqs.~(\ref{eq:423})-(\ref{eq:425}).  The
location of these points can be approximated with perturbative series
for $\Chi$.  Indeed, the fourth order formula (\ref{eq:443}) and
(\ref{eq:444}) for $\zeta=1$ gives a single positive real root
$\ep_{\ast}\approx\num{0.819208}$ which with inclusion of higher order
approximation to $\Chi^{-}_{\zeta=1}$ is expected to converge to the
numerical value.

Next, we solved the system (\ref{eq:67})-(\ref{eq:69}) subject to
generic initial conditions.  For purely real initial data as in
\cite{br} we reproduce the scaling
$\Pi(t,0)^{2}\ra\ep^{-2}\Pi(\ep^{2}t,0)^{2}$ (cf. Fig.~2 in
\cite{br}), which improves with decreasing amplitude of the
perturbation $\ep$, supporting the conjectured AdS instability for
reflecting boundary conditions.  On the other hand, for a perturbed
standing wave solution, i.e. for the initial data\footnote{Note use
  of different symbols for Greek epsilon to distinguish central
  density $\ep$ and the amplitude of initial perturbation $\epsilon$.}
\begin{equation}
  \label{eq:450}
  \begin{aligned}
    \phi(t,0) &= \f(x), \\
    \Pi(t,0) &= i\,\Omega\,\f(x)\frac{e^{\de(x)}}{\a(x)} +
    \epsilon\,\frac{2}{\pi} \exp\left(
      -\frac{4}{\pi^{2}}\frac{\tan^{2}{x}}{\sigma^{2}} \right),
  \end{aligned}
\end{equation}
(with $\sigma=1/16$), evolution is different (see
Fig.~\ref{fig:StandingE0GaussPerturbation} for a perturbed ground
state solution; we observe the same behaviour also for small amplitude
excited states).  While for large amplitudes of the Gaussian
perturbation, after several dozens of reflections, the modulus squared
of the scalar field $\Pi(t,0)$ starts to grow, indicating the
formation of the apparent horizon, the situation changes when the
perturbation becomes small.  For slightly perturbed standing wave
solution and for simulated time intervals, the evolution does not show
any sign of instability staying close to the stationary state at all
times.  Moreover, similarly to the cavity model with Neumann boundary
condition discussed in Section~\ref{sec:BoxNeumann}, the scalar field
$\Pi(t,x)$ evaluated at the origin exhibits linear scaling with
$\epsilon$ (so does, in the leading order, the squared modulus).  Here
to see this scaling we subtract the constant contribution of the
standing wave solution, namely the constant factor
$(\Omega\f(0))^{2}$.  The energy spectra of a noncollapsing solution
seems to equilibrate around some stationary distribution, with a small
fluctuation of energy between eigenmodes (see
Fig.~\ref{fig:StandingE0GaussPerturbationEnergySpectra}).  Moreover,
in contrast to perturbations of the pure AdS space, we do not observe
any scaling with the coordinate time $t$ in this case.  This picture
holds for standing wave solutions on a stable branch, i.e. solutions
with $\ep<\ep_{\ast}$ and for both initial perturbations conserving
total charge (e.g. such as given in (\ref{eq:450})) and these that
modify charge of a standing wave solution.

\begin{figure}[!t]
  \centering
  \includegraphics[width=\swidth]{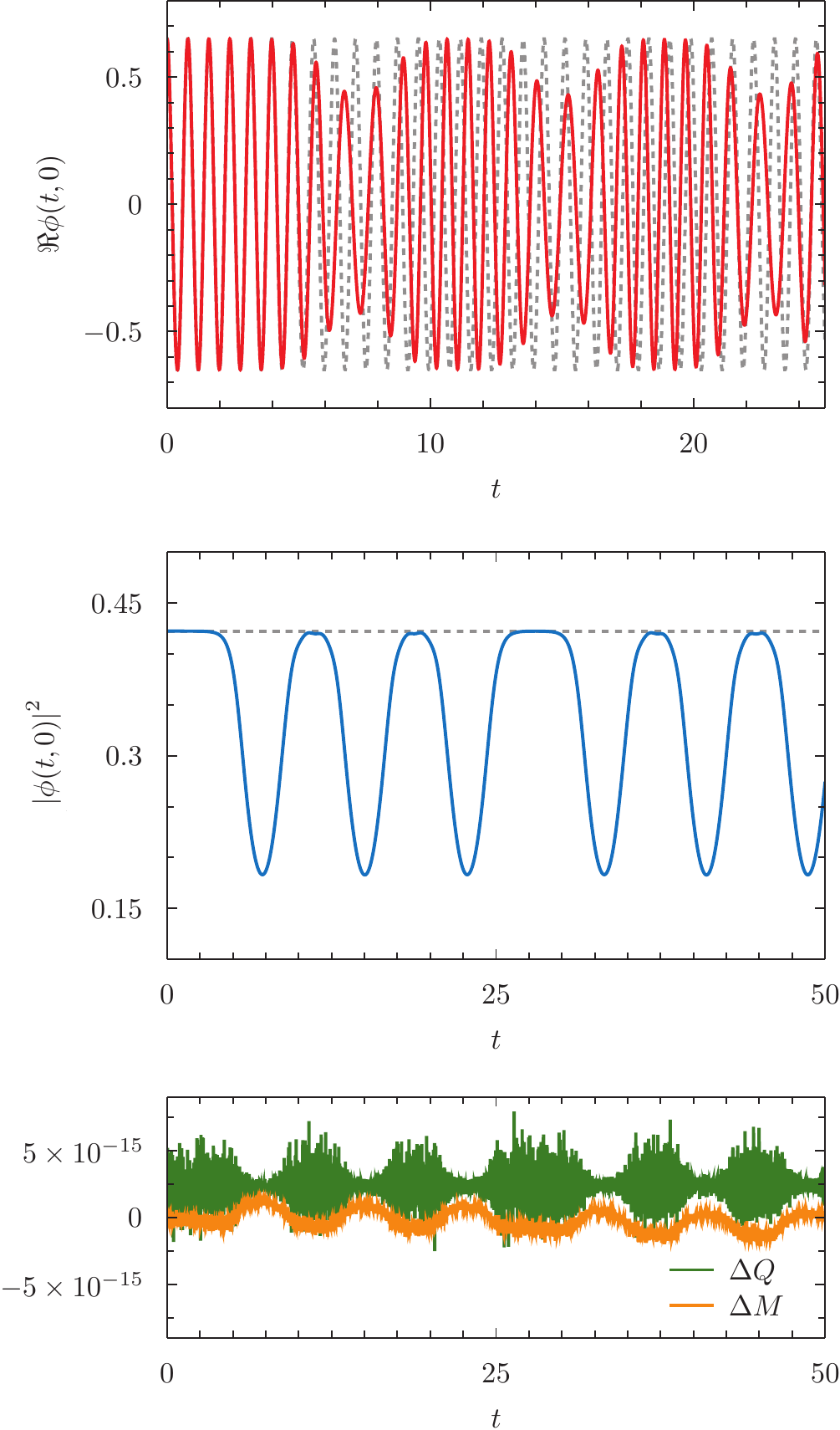}
  \caption{Time evolution of slightly perturbed (with Gaussian
    amplitude $\epsilon=10^{-4}$ set in (\ref{eq:450})) unstable
    fundamental standing wave solution $\ep=0.65$
    ($\Omega\approx \num{7.894722}$).  \textit{Top panel}.  The real
    part of scalar field evaluated at the origin (red solid line)
    compared with harmonic oscillation of unperturbed standing wave
    solution (dashed gray line).  \textit{Middle panel}.  The squared
    module of scalar field at the origin.  \textit{Bottom panel}.  The
    charge and total mass absolute errors during the evolution.}
  \label{fig:StandingE0GaussUnstable}
\end{figure}

This picture changes dramatically when we perturb an unstable standing
wave solution, such with $\ep>\ep_{\ast}$.  Starting with the same
initial conditions as in (\ref{eq:450}) with small positive values of
$\epsilon$ the solution inevitably collapses to a black hole.  For
negative $\epsilon$ the apparent horizon does not form and solution
stays regular all the future times but diverges away from a stationary
state in a sense that it exhibits large amplitude oscillations around
stationary state (this is in contrast with what we observe for the
unstable time-periodic solutions which undergo prompt or delayed
collapse independently on the sign of initial perturbation).  This is
illustrated on Fig.~\ref{fig:StandingE0GaussUnstable} where the
unstable solution with central density $\ep=0.65$
($\Omega\approx\num{7.894722}$) was perturbed with Gaussian profile
(\ref{eq:450}) of small amplitude $\epsilon=10^{-4}$.  The solution
undergoes high amplitude oscillations; a very similar phenomena was
observed for a long time evolution of unstable boson stars for
$\Lambda=0$ case \cite{LaiPhD}.  A use of sufficient number of
eigenmodes in numerical evolution together with symplectic integration
method guarantees the high accuracy which is confirmed by the mass and
charge conservation (see bottom panel of
Fig.~\ref{fig:StandingE0GaussUnstable}).  In contrast to the
asymptotically flat case, here the excess of mass and charge cannot
leave the system and no convergence to a stable solution is expected.
Since the oscillations are of large amplitude it makes it difficult to
precisely identify an effective background state.  A natural candidate
would be one of the standing solutions on stable branch.  But as for
the time-periodic solutions, we were unsuccessful to provide
satisfactory description for such evolutions.

\clearpage

\section[Cohomogenity-two biaxial Bianchi~IX ansatz]{Cohomogenity-two
  biaxial Bianchi~IX\newline ansatz}
\label{sec:BCS}

In this section we deal with time-periodic solutions of the pure
vacuum Einsteins equations---the system of equations introduced in
Section~\ref{sec:BCSModel}.  The construction and the methods follow
similar steps as for the EKG system.  The differences among these two
systems of PDEs lead to crucial modifications of perturbative
construction which we emphasize here.  We start with perturbative
construction in Section~\ref{sec:BCSPerturbative}.  Then we describe
the spatial discretization method (Section~\ref{sec:BCSEvolution}) and
finally we show the numerical construction of time-periodic solutions
(Section~\ref{sec:BCSNumerical}).  We verify our approaches in
Section~\ref{sec:BCSResults} where the properties of time-periodic
solutions are also discussed.

\subsection{Perturbative construction}
\label{sec:BCSPerturbative}

We follow the steps of Section~\ref{sec:AdSPeriodicPerturbative-Even}
when constructing time-periodic solutions to the model introduced in
Section~\ref{sec:BCSModel}.  We assume the following small amplitude
$\ep$ expansion for time-periodic solution bifurcating from a single
eigenmode $e_{\gamma}(x)$ ($\gamma\in\mathbb{N}_{0}$)
\begin{align}
  \label{eq:451}
  B(t,x;\ep) & = \ep\cos{\tau}\,e_{\gamma}(x) + \sum_{\lambda\geq
    2}\ep^{\lambda}B_{\lambda}(\tau,x),
  \\
  \label{eq:452}
  A(t,x;\ep) & = 1 - \sum_{\lambda\geq 2}\ep^{\lambda}A_{\lambda}(\tau,x),
  \\
  \label{eq:453}
  \delta(t,x;\ep) & = \sum_{\lambda\geq
    2}\ep^{\lambda}\delta_{\lambda}(\tau,x),
\end{align}
where $\tau = \Omega\,t$ is a rescaled time variable with
\begin{equation}
  \label{eq:454}
  \Omega(\ep)  = \omega_{\gamma} + \sum_{\lambda\geq
    1}\ep^{\lambda}\xi_{\lambda}.
\end{equation}
This form of perturbative ansatz reduces for $\ep\ra 0$ to a single
eigenmode $e_{\gamma}(x)$ oscillating with frequency
$\omega_{\gamma}$---the solution of linearized Einstein's equations
(\ref{eq:111}).  At the linear level this is a time-periodic solution
(refereed by other authors as oscillon), and with perturbative
construction of time-periodic solutions we retain periodicity of the
higher orders of functional series (\ref{eq:451})-(\ref{eq:453}) term
by term.

The strategy is a minor modification of methods given in
Section~\ref{sec:AdSPeriodicPerturbative-Even} and restricted to the
problem at hand.  One of the differences is the form of perturbative
expansion, given in Eqs.~(\ref{eq:451})-(\ref{eq:453}), where both
even and odd powers of $\ep$ are present, which is a direct
consequence of the dependence of Eqs.~(\ref{eq:97})-(\ref{eq:100}) on
sign of the squashing field $B$.  For completeness we present all the
steps in this construction, and point out the necessary modifications
we have to make in adapting the techniques developed for the scalar
field model.

We plug the expansion (\ref{eq:451})-(\ref{eq:454}) into the field
equations (\ref{eq:97})-(\ref{eq:99}), perform the expansion around
$\ep=0$ and require the resulting perturbative equations to be
satisfied at each perturbative order $\lambda$.  To reduce the
complexity of the solution procedure we decompose each of the metric
functions, at each order $\lambda$, as follow
\begin{align}
  \label{eq:455}
  B_{\lambda}(\tau, x) & = \sum_{j\geq 0}\hat{B}_{\lambda,j}(\tau)e_{j}(x),
  \\
  \label{eq:456}
  A_{\lambda}(\tau, x) & = \sum_{j \geq 0} \hat{A}_{\lambda,
    j}(\tau)\cos\left(2jx\right),
  \\
  \label{eq:457}
  \delta_{\lambda}(\tau, x) & = \sum_{j \geq 1} \hat{\delta}_{\lambda,
    j}(\tau)\left(\cos\left(2jx\right)-1\right),
\end{align}
where the upper limits of the sums are finite and depend on the
particular choice of $\gamma$ and the order of the $\ep$ expansion.
The choice of the basis functions decomposition is dictated by the
boundary conditions (\ref{eq:105}) and (\ref{eq:106}) and the
simplicity.\footnote{In spate the fact the boundary expansion for $A$
  and eigenbasis functions are consistent we decided to expand $A$ in
  cosine basis to reduce the number of integrals needed to be
  computed, see below.  For $\delta$ function we have no alternative
  for the cosine basis.}  Such decomposition also reduces the
perturbative differential equations to algebraic systems for expansion
coefficients of $A_{\lambda}$ and $\delta_{\lambda}$ function and the
PDE for $B_{\lambda}$ to a system of second order ODEs.

The exponential nonlinearity and structure of the field equations
causes that at each perturbative order we need to solve three
equations, as opposed to the scalar field case, the constraints and
the wave equation.  Therefore at each order $\lambda\geq 2$ we solve
first (\ref{eq:98}) and (\ref{eq:99}) for $\delta_{\lambda}$ and
$A_{\lambda}$ respectively, then from Eq.~(\ref{eq:97}) we derive
$B_{\lambda}$.  We do this in this particular order since the former
functions serve as source terms for the latter.  Starting with
$\delta$ function we plug (\ref{eq:453}) and (\ref{eq:457}) into
(\ref{eq:98}) and get the following
\begin{equation}
  \label{eq:458}
  - \sum_{j\geq 1} \hat{\delta}_{\lambda,j}(\tau) 2j \sin(2jx)
  = \coef{\lambda}\left( - 2\sin{x}\cos{x}\left(\beta(t,x)^{2}
      + \Pi(t,x)^{2}\right) \right).
\end{equation}
The right hand side of this equation is a finite combination of the
$\sin(2jx)$ terms, so using orthogonality property of the $\sin(2jx)$
functions we can read off directly the
$\hat{\delta}_{\lambda,j}(\tau)$ coefficients, which completely
determine the solution $\delta_{\lambda}(\tau,x)$.  The gauge
condition $\delta_{\lambda}(\tau, 0)=0$ is already enforced by the
form of decomposition (\ref{eq:457}).
Next, inserting the perturbative series (\ref{eq:452}) into
(\ref{eq:99}) and multiplying both sides by $\sin{x}\cos{x}$ we obtain
\begin{multline}
  \label{eq:459}
  -\sin{x}\cos{x}\,A_{\lambda}'(\tau,x) - (3 -
  \cos{2x})A_{\lambda}(\tau,x) =
  \\
  \coef{\lambda}\cos{x}\left( \sin{x}\,\delta' A + \frac{2}{3}\cos{x}
    \left(4e^{-2B(t,x)} - e^{-8B(t,x)} - 3\right)\right).
\end{multline}
The resulting right hand side is a finite combination of $\cos(2jx)$
with the time dependent coefficients.  Then we insert (\ref{eq:456})
into (\ref{eq:459}) and use the orthogonality of $\cos(2jx)$ functions
to obtain an algebraic system (of finite size; when expressed in
matrix form this system has a banded main matrix) for the coefficients
$\hat{A}_{\lambda,j}(\tau)$.  The boundary conditions
$A_{\lambda}(\tau, 0) = A_{\lambda}(\tau, \pi/2) = 0$ are then
automatically satisfied.  Finally, for $B_{\lambda}$ we have the
following wave equation to be solved at each order $\lambda\geq 2$
\begin{equation}
  \label{eq:460}
  \left(\omega_{\gamma}^{2}\,\partial_{\tau}^{2} + L\right) B_{\lambda}
  = S_{\lambda},
\end{equation}
(where by $S_{\lambda}$ we denote the source function at order
$\lambda$ resulting from the Taylor expansion of Eq.~(\ref{eq:97})
with (\ref{eq:451})-(\ref{eq:454}) substituted).  When we plug
(\ref{eq:455}) into Eq.~(\ref{eq:460}) and project onto $e_{k}(x)$
mode we get the system of second order ODEs which have the following
form
\begin{equation}
  \label{eq:461}
  \left(\omega_{\gamma}^{2}\,\partial_{\tau}^{2}
    + \omega_{k}^{2}\right)\hat{B}_{\lambda,k}
  = \inner{e_{k}}{S_{\lambda}}.
\end{equation}
These have to be solved with general initial conditions
\begin{equation}
  \label{eq:462}
  \hat{B}_{\lambda,k}(0) = c_{\lambda,k}, \quad
  \partial_{\tau}\hat{B}_{\lambda,k}(0) = \tilde{c}_{\lambda,k}.
\end{equation}
At each order we set two of them, $c_{\lambda,\gamma} =
\tilde{c}_{\lambda,\gamma} = 0$, to impose the amplitude of dominant
eigenmode $e_{\gamma}(x)$ to $\ep$ and its velocity to zero at time
$\tau=0$ ($t=0$)
\begin{equation}
  \label{eq:463}
  \inner{e_{\gamma}}{B}\big|_{\tau=0} = \ep, \quad
  \inner{e_{\gamma}}{\partial_{\tau}B}\big|_{\tau=0} = 0,
\end{equation}
(this freedom is related to the fact that we may pick one particular
solution from the whole family of time-periodic solutions build on a
given eigenmode $e_{\gamma}(x)$, while the second condition
corresponds to a freedom we have to fix the phase of periodic
solution).  It turns out that the choice
$\tilde{c}_{\lambda,\gamma}=0$ implies that all other modes tune in
phase with the dominant one, i.e. $\tilde{c}_{\lambda,k}=0$ for all
$k\geq 0$ and $\lambda\geq 2$.  Next, following the idea given in
Section~\ref{sec:MethodsPerturbative} we use the freedom we have in
specifying initial data $c_{\lambda,k\neq \gamma}$ (the remaining
integration constants) and the expansion parameter $\xi_{\lambda}$ to
remove the resonant terms $\cos(\omega_{k}/\omega_{\gamma}\tau)$,
naturally present in $\inner{e_{k}}{S_{\lambda}}$ in the case of fully
resonant system, which give rise to secular terms
$\tau\sin(\omega_{k}/\omega_{\gamma}\tau)$, which neither are periodic
nor bounded when $\tau\ra\infty$.  Thus all of the resonances have to
be removed by fixing the remaining integration constants and the
frequency corrections $\xi_{\lambda}$.  However, it turns out that
this is possible at any order $\lambda$ only if one solve the system
(\ref{eq:461}) in a proper way, starting at the lowest nontrivial
order $\lambda=2$.  The source function in (\ref{eq:460}) has the
following property
\begin{equation}
  \label{eq:464}
  \inner{e_{k}}{S_{\lambda=2}}\neq 0, \quad \text{for}\ k=0, 1, \ldots,
  2\gamma + 2,
\end{equation}
and for $\lambda\geq 3$
\begin{equation}
  \label{eq:465}
  \inner{e_{k}}{S_{\lambda}}\neq 0, \quad \text{for}\ k=0, 1, \ldots,
  \frac{1}{2}\left(7(\lambda - 1) - \frac{1 + (-1)^{\lambda}}{2}\right)
  + \gamma\lambda,
\end{equation}
so in particular the projection $\inner{e_{k}}{S_{\lambda=2}}$
vanishes for $k=k_{\ast}=2\gamma + 3$.  But the eigenmode
$e_{k_{\ast}}(x)$ is in resonance with $e_{\gamma}(x)$, that is its
frequency satisfies the following condition
\begin{equation}
  \label{eq:466}
  \frac{\omega_{k_{\ast}}}{\omega_{\gamma}} = 2,
\end{equation}
so $k_{\ast}\in O_{\gamma}$ (it is the lowest resonant mode with
$e_{\gamma}(x)$, cf.  (\ref{eq:27}) for definition of the resonant
set).  We include this eigenmode in the solution $B_{2}(\tau,x)$
resulting from inhomogenous system (\ref{eq:461}) by adding the term
\begin{equation}
  \label{eq:467}
  \widetilde{B}_{2,2\gamma + 3}\cos(2\tau)\,e_{2\gamma + 3}(x),
\end{equation}
being itself a solution of homogeneous wave equation (\ref{eq:461}).
The additional parameter, namely the amplitude
$\widetilde{B}_{2,2\gamma + 3}$, will be used to remove one resonant
term at higher order (specifically at order $\lambda=4$).  The
inclusion of (\ref{eq:467}) at order $\lambda=2$ allow us to continue
our construction up to arbitrary high order without need to explicitly
include further homogeneous solutions to (\ref{eq:461}) at any higher
order.  Furthermore, the number of integration constants
$c_{\lambda,k}$ together with the expansion parameter $\xi_{\lambda}$
is sufficient to remove all of the resonances (with an exception for
$\lambda=4$ where we also need the explicitly included parameter
$\widetilde{B}_{2,2\gamma + 3}$ at order $\lambda=2$), and all of
these parameters are fixed at higher order of perturbative calculation
leading to a unique solution.

In fact, the projections of the source to the inhomogeneous wave
equation (\ref{eq:460}) onto $e_{k}(x)$ mode, appearing at any order
of the perturbative procedure described above, can be reduced to just
a few inner products; the source functions $S_{\lambda}$ consists
mostly of the terms like: $\csc^2{x}\,e_{i}(x)e_{j}(x)$,
$\cos(2ix)e_{j}(x)$, $\sin(2ix)e_j'(x)$, $e_{i}(x)e_{j}(x)$ and
$\sin^{2}{x}\,e_{i}(x)$, where each term individually can be written
as a finite combination of the eigenmodes.  This property does not
hold for the $\csc^{2}{x}\cos(2jx)e_{i}(x)$ terms which come form the
products like $\csc^{2}{x}\,B_{i}\delta_{j}$ (where $i+j=\lambda$,
$i,j\geq 1$, at order $\lambda$), since each of them separately do not
have finite decomposition onto the eigenmodes $e_i(x)$.  For that
reason the straightforward procedure to decompose the source term by
term does not work here (as it does for a scalar field system in odd
spacetime dimensions, as analyzed in
Section~\ref{sec:AdSPeriodicPerturbative-Odd}).  To overcome this we
proceed as follow.  Multiplication of the source $S_{\lambda}$ by the
factor $\sin^{2}x$ removes troublesome terms and allow us to write (to
simplify notation we drop subscript of $S_{\lambda}$ and write $S$
instead here; this trick applies to any $\lambda\geq 2$)
\begin{equation}
  \label{eq:468}
  S \sin^2{x} = \sum_{j\geq 0}\tilde{S}_j e_j(x),
\end{equation}
(as a finite sum) with known coefficients $\tilde{S}_j$.  Since we
know that the source can be written as the finite sum of eigenmodes
$S=\sum_{i\geq 0}\hat{S}_{i}e_{i}(x)$ we can plug this expression into
the former one and use the orthogonality property of the $e_{j}(x)$
basis functions to get
\begin{equation}
  \label{eq:469}
  \sum_{i\geq 0}\inner{e_k}{\sin^{2}\!{x}\,e_i}\hat{S}_i = \tilde{S}_{k},
\end{equation}
the linear equation for the $\hat{S}_i$ coefficients.  The matrix with
elements $\inner{e_k}{\sin^{2}\!{x}\,e_i}$ has a tri-diagonal form and
is non-degenerate whence can be inverted to find the necessary
coefficients $\hat{S}_i$ easily.

The number of terms in the source function in (\ref{eq:459}) and in
(\ref{eq:460}) rapidly grows with the order $\lambda$, because of the
exponential nonlinearity of the field equations, so the construction
of solution to a given order is much more involved then for a scalar
field system discussed in
Section~\ref{sec:AdSPeriodicPerturbative-Even} (where the wave
equation for the scalar field is linear).  Therefore, in practice,
with the same computational resources available we can obtain
approximation to a time-periodic solution to
(\ref{eq:97})-(\ref{eq:99}) of lower order compared to the scalar
field case (at least for the current implementation).  The reason of
that is also the lack of symmetry of the perturbative expansion
(\ref{eq:451})-(\ref{eq:454}) where all powers of the $\ep$ are
present, whereby at each order the constraints and the dynamical
equation have to be solved.
\subsubsection{Integrals}
In contrast to the scalar field case here we decompose the metric
functions in cosine basis (\ref{eq:456}) and (\ref{eq:457}), therefore
for convenience the source functions in the corresponding equations
(\ref{eq:458}) and (\ref{eq:459}) are simply decomposed by expanding
the eigenbasis functions and all of appearing products at a given
perturbative order in the cosine series.\footnote{This step is carried
  automatically by the \mathematica{} itself, by simply using the
  built-in function \codenamestyle{TrigReduce}.}  This simplifies the
implementation of the algorithm but this step is both time and memory
consuming operation so possibly further improvements are needed, like
using eigenbasis expansion in (\ref{eq:452}).

The decomposition of the source function $S_{\lambda}$ of
Eq.~(\ref{eq:460}) into the eigenbasis $e_{i}(x)$ can be computed by
applying the following formulae
\begin{equation}
  \label{eq:470}
  \csc^2{x}\,e_{i}(x)e_{j}(x) =
  \sum_{k=\max(0,|i-j|-2)}^{i+j+2}\inner{e_{k}}{\csc^{2}{x}\,e_{i}\,e_{j}}e_{k}(x),
\end{equation}

\begin{equation}
  \label{eq:471}
  \cos (2ix) e_j(x) =
  \sum_{k=\max(0,j-i)}^{i+j}\inner{e_{k}}{\cos(2ix)e_{j}}e_{k}(x),
\end{equation}

\begin{equation}
  \label{eq:472}
  \sin(2ix)e_j'(x) =
  \sum_{k=\max(0,j-i)}^{i+j}\inner{e_{k}}{\sin(2ix)e_j'}e_{k}(x),
\end{equation}

\begin{equation}
  \label{eq:473}
  e_{i}(x) e_{j}(x) =
  \sum_{k=\max(0,|i-j|-3)}^{i+j+3}\inner{e_{k}}{e_{i}\,e_{j}}e_{k}(x),
\end{equation}

\begin{equation}
  \label{eq:474}
  \sin^{2}{x}\,e_{i}(x) =
  \sum_{k=\max(0,i-1)}^{i+1}\inner{e_{k}}{\sin^{2}{x}\,e_{i}}e_{k}(x),
\end{equation}
where the expansion coefficients in each of the sum are calculated as
described in Appendix~\ref{cha:inter-coeff}.  Denoting by
$\mathcal{N}_{j}$ the normalization constant in (\ref{eq:113})
\begin{equation}
  \label{eq:475}
  \mathcal{N}_{j} = 2\sqrt{\frac{(j+3)(j+4)(j+5)}{(j+1)(j+2)}},
  \quad j\in\mathbb{N}_{0},
\end{equation}
we have (with the inner product defined in Eq.~(\ref{eq:112}))
\begin{multline}
  \label{eq:476}
  \inner{e_{k}}{\csc^{2}{x}\,e_{i}\,e_{j}} =
  \mathcal{N}_{i}\,\mathcal{N}_{j}\, \mathcal{N}_{k} \sum_{q=0}^k
  \sum_{r=0}^j \sum_{s=0}^i \Biggl[ (-1)^{i+j+k-q-r-s}
  \\
  \times \binom{i+2}{i-s} \binom{i+3}{s} \binom{j+2}{j-r}
  \binom{j+3}{r} \binom{k+2}{k-q} \binom{k+3}{q}
  \\
  \times \frac{\Gamma(q+r+s+5) \Gamma(i+j+k-q-r-s+4)}{2
    \Gamma(i+j+k+9)} \Biggr],
\end{multline}
\begin{multline}
  \label{eq:477}
  \inner{e_{k}}{\cos(2ix)e_{j}} = \mathcal{N}_{j}\,\mathcal{N}_{k}
  \sum_{q=0}^k \sum_{r=0}^j \sum_{s=0}^i \Biggl[ (-1)^{i+j+k-q-r+s}
  \\
  \times \binom{2 i}{2 s} \binom{j+2}{j-r} \binom{j+3}{r}
  \binom{k+2}{k-q} \binom{k+3}{q}
  \\
  \times \frac{\Gamma(q+r+s+3) \Gamma(i+j+k-q-r-s+4)}{2
    \Gamma(i+j+k+7)} \Biggr],
\end{multline}
\begin{multline}
  \label{eq:478}
  \inner{e_{k}}{\sin(2ix)e_j'} = \mathcal{N}_{j}\,\mathcal{N}_{k}
  \Biggl\{ \sum_{q=0}^k \sum_{r=0}^j \sum_{s=1}^i \Biggl[
  (-1)^{i+j+k-q-r+s}
  \\
  \times \binom{2 i}{2 s-1} \binom{j+2}{j-r} \binom{j+3}{r}
  \binom{k+2}{k-q} \binom{k+3}{q}
  \\
  \times \frac{\Gamma(q+r+s+3)\Gamma(i+j+k-q-r-s+4)}{\Gamma (i+j+k+7)}
  \Biggr]
  \\
  + \sum_{q=0}^k \sum_{r=0}^{j-1} \sum_{s=1}^i \Biggl[
  (-1)^{i+j+k-q-r+s} (j+6)
  \\
  \times \binom{2i}{2 s-1} \binom{j+2}{j-r-1} \binom{j+3}{r}
  \binom{k+2}{k-q} \binom{k+3}{q}
  \\
  \times \frac{\Gamma(q+r+s+3) \Gamma(i+j+k-q-r-s+4)}{\Gamma(i+j+k+7)}
  \Biggr]
  \\
  - 2 \sum_{q=0}^k \sum_{r=0}^j \sum_{s=1}^i (-1)^{i+j+k-q-r+s}
  \\
  \times \binom{2 i}{2 s-1} \binom{j+2}{j-r} \binom{j+3}{r}
  \binom{k+2}{k-q} \binom{k+3}{q}
  \\
  \times \frac{\Gamma(q+r+s+2) \Gamma(i+j+k-q-r-s+5)}{\Gamma(i+j+k+7)}
  \Biggr] \Biggr\},
\end{multline}
\begin{multline}
  \label{eq:479}
  \inner{e_{k}}{e_{i}\,e_{j}} = \mathcal{N}_{i}\,\mathcal{N}_{j}\,
  \mathcal{N}_{k}\sum_{q=0}^k \sum_{r=0}^j \sum_{s=0}^i
  \Biggl[(-1)^{i+j+k-q-r-s}
  \\
  \times \binom{i+2}{i-s} \binom{i+3}{s} \binom{j+2}{j-r}
  \binom{j+3}{r} \binom{k+2}{k-q} \binom{k+3}{q}
  \\
  \times \frac{\Gamma(q+r+s+5) \Gamma(i+j+k-q-r-s+5)}{2
    \Gamma(i+j+k+10)} \Biggr],
\end{multline}
\begin{multline}
  \label{eq:480}
  \inner{e_{j}}{\sin^{2}{x}\,e_{i}} = \mathcal{N}_{i}\,
  \mathcal{N}_{j} \sum_{r=0}^j \sum_{s=0}^i \Biggl[ (-1)^{i+j-r-s}
  \\
  \times \binom{i+2}{i-s} \binom{i+3}{s} \binom{j+2}{j-r}
  \binom{j+3}{r}
  \\
  \times \frac{\Gamma(r+s+3) \Gamma(i+j-r-s+5)}{2 \Gamma(i+j+8)}
  \Biggr].
\end{multline}
Similarly to the EKG system
(Section~\ref{sec:AdSPeriodicPerturbative}) we stress the importance
of symmetries of the integrals (\ref{eq:476})-(\ref{eq:480}) which are
worth noting when performing calculations.

\subsection{Pseudospectral code for the time evolution}
\label{sec:BCSEvolution}

We apply the MOL approach with pseudospectral discretization in space
to solve the initial value problem of the system
(\ref{eq:97})-(\ref{eq:99}) using constrained evolution scheme.  We
expand dynamical fields $B(t,x)$ and $\Pi(t,x)$ into $N$ eigenmodes of
the linear problem
\begin{equation}
  \label{eq:481}
  B(t,x) = \sum_{j=0}^{N-1}\hat{B}_{j}(t)\,e_{j}(x), \quad
  \Pi(t,x) = \sum_{j=0}^{N-1}\hat{\Pi}_{j}(t)\,e_{j}(x).
\end{equation}
With the pseudospectral approach we choose a spatial grid of $N$ points
\begin{equation}
  \label{eq:482}
  x_{j} = \frac{\pi}{2}\frac{j}{N+1}, \quad j=1,\ldots,N,
\end{equation}
(here we prefer to use the analytical approximation to the zeros of
$e_{N+1}(x)$, as discussed in \cite{Boyd199935}) and require necessary
equations to be identically satisfied at these collocation points.
For convenience, instead of evolving in time the values of the
dynamical fields at discrete spatial grid, we evolve their Fourier
coefficients.  To calculate time derivatives of the coefficients
$\hat{B}_{j}(t)$ and $\hat{\Pi}_{j}(t)$, instead of Eq.~(\ref{eq:97}),
we use
\begin{align}
  \label{eq:483}
  \dot B &= Ae^{-\delta}\Pi\,,
  \\
  \label{eq:484}
  \dot\Pi &=
  \begin{multlined}[t]
    e^{-\delta}\Biggl[ A\left(\beta' + 2\cot{2x}\beta\right) +
    \beta\left(4\tan{x} +
      \frac{2}{3}\cot{x}\left(4e^{-2B}-e^{-8B}\right)\right)
    \\
    - \frac{4}{3}\frac{e^{-2B}-e^{-8B}}{\sin^{2}\!{x}} \Biggr],
  \end{multlined}
\end{align}
(where we have used the constraint equations to eliminate spatial
derivatives of $A$ and $\delta$ functions).  Since we perform
constrained time evolution we solve for the constraints at each
intermediate integration time step as follows.  We know that the
metric function $\delta$ and the integrand in (\ref{eq:109}) can be
approximated as
\begin{align}
  \label{eq:485}
  \delta(t,x) &= \sum_{j=0}^{N-1}\hat{\delta}_{j}(t)\cos(2jx),
  \\
  \label{eq:486}
  e^{-\delta}\left( \beta^{2} + \Pi^{2} - \frac{1}{3}\frac{4e^{-2B} -
      e^{-8B} - 3}{\sin^{2}{x}} \right) &= \sum_{j=0}^{N-1}
  \tilde{A}_{j}(t) \cos(2jx),
\end{align}
(they have compatible boundary behaviour to the expansion functions,
so for smooth dynamical fields $B$ and $\Pi$ the coefficients fall off
exponentially).  Plugging (\ref{eq:485}) into (\ref{eq:98}) we get
(after cancellation of common terms and multiplication by constant
trigonometric factor)
\begin{equation}
  \label{eq:487}
  \sum_{j=1}^{N-1}j\,\frac{\sin(2jx)}{\sin{x}\cos{x}}\,\hat{\delta}_{j}(t) =
  \beta^{2} + \Pi^{2},
\end{equation}
which evaluated at the set of collocation points and supplied with one
extra condition, for the remaining coefficient $\hat{\delta}_{0}(t)$,
fixing the gauge freedom $\delta(t,0) = \sum_{j=0}^{N-1}
\hat{\delta}_{j}(t)$ forms a linear system for the Fourier
coefficients of the $\delta(t,x)$ function.  Similarly we evaluate
(\ref{eq:486}) at the collocation points and solve the resulting
system for the $\tilde{A}_{j}(t)$ coefficients.  Thus the metric
function $A$ can be expressed as
\begin{equation}
  \label{eq:488}
  A(t,x) = 1 - 2 e^{\delta}\,\frac{\cos^{4}{x}}{\sin^{2}{x}}
  \sum_{j=0}^{N-1}\tilde{A}_{j}(t)w_{j}(x),
\end{equation}
where the weight functions $w_{j}(x)$ read\footnote{These are easy to
  obtain using trigonometric identities $\cos(2jy) = 4\cos^2{y}
  \cos(2(j-1)y) - \cos(2(j-2)y) - 2\cos(2(j-1)y)$ and $\cos(2jy) =
  2\cos{y}\cos((2j-1)y) - \cos(2(j-1)y)$ to derive recurrence
  equations $w_{j}(x)=4b_{j-1}(x)-w_{j-1}(x)-w_{j-2}(x)$ and
  $b_{j}(x)=2c_{j}(x)-b_{j-1}(x)$, with
  $b_{j}(x)=\int_{0}^{x}\frac{\sin^{3}{y}}{\cos{y}}\cos(2jy)\diff y$,
  and $c_{j}(x)$ defined in (\ref{eq:490}).  The solution to such
  recurrence yields the presented result.}
\begin{equation}
  \label{eq:489}
  \begin{split}
    w_{j}(x) & = \int_{0}^{x}\cos(2jy)\tan^{3}{y}\diff y = (-1)^j
    \Biggl[ \frac{\sec^{2}{x}}{2} + (2j^{2} + 1)\log(\cos{x})
    \\
    \quad & + \frac{1}{8}\biggl(-4 j (2 j-1) \cos(2x) + (j-1) j \cos(4
    x) + (j-1)(7j+4)\biggr)
    \\
    \quad & - 4 \sum_{k=2}^{j-1} (-1)^{k}(j-k)(j-k+1)c_{k}(x) \Biggr],
  \end{split}
\end{equation}
with
\begin{equation}
  \label{eq:490}
  \begin{split}
    c_{j}(x) & = \int_{0}^{x}\sin^{3}{y}\cos\left((2j-1)y\right)\diff
    y = \frac{1}{8}\Biggl(-\frac{\cos (2 (j-2) x)}{2 (j-2)}+\frac{3
      \cos (2 (j-1) x)}{2 (j-1)}
    \\
    \quad & -\frac{3 \cos (2 j x)}{2 j} + \frac{\cos (2 (j+1) x)}{2
      (j+1)} + \frac{3}{(j-2) (j-1) j (j+1)}\Biggr).
  \end{split}
\end{equation}
Now, substituting the expansions (\ref{eq:481}), (\ref{eq:485}) and
(\ref{eq:488}) into Eqs.~(\ref{eq:483}) and (\ref{eq:484}) and
evaluating both sides at the collocation points, we get the linear
system of equations to be solved for the time derivatives of
$\hat{B}_{j}(t)$ and $\hat{\Pi}_{j}(t)$ expansion coefficients.

The total mass of the system, given by the integral (\ref{eq:104}), we
compute as follows.  The integral in (\ref{eq:104}) can be
approximated by truncated expansion
\begin{equation}
  \label{eq:491}
  2\left[ \left(\beta^{2} + \Pi^{2}\right)A +
    \frac{3 + e^{-8B} - 4e^{-2B}}{3\sin^{2}{x}} \right] =
  \sum_{j=0}^{N-1}m_{j}e_{j}(x).
\end{equation}
With this we have
\begin{equation}
  \label{eq:492}
  M = \int_{0}^{\pi}\sum_{j=0}^{N-1}m_{j}e_{j}(x)\tan^{3}{x}\diff x =
  \sum_{j=0}^{N-1}m_{j}\int_{0}^{\pi}e_{j}(x)\tan^{3}{x}\diff x,
\end{equation}
where the weighted integral of the eigenfunctions can be calculated,
similarly as for the scalar field model, using the integral identity
of the Jacobi polynomials (\ref{eq:596}), which gives
\begin{equation}
  \label{eq:493}
  \int_{0}^{\pi}e_{j}(x)\tan^{3}{x}\diff x =
  \frac{2+(-1)^{j}(j+3)(j(j+6)+6)}{\sqrt{(j+1)(j+2)(j+3)(j+4)(j+5)}}.
\end{equation}

As in the EKG model, to advance the solution in time we use the
Gauss-Legendre implicit Runge-Kutta method, see
Section~\ref{sec:AppIRK}, with fixed time step, to preserve symplectic
structure of equations and at the same time have a total mass
conserving scheme.  Use of implicit time integrator does not change
the order of complexity of our algorithm (solving for time derivatives
of $\hat{B}_{j}$ and $\hat{\Pi}_{j}$ requires $O\left(N^{2}\right)$
floating points operations) thus the complexity of time evolution
scheme is $O\left(N^{3}\right)$ (for stability reasons time step size
must be of order $N^{-1}$).  The robustness of this approach is
reported in subsequent section.

\subsection{Numerical construction}
\label{sec:BCSNumerical}

Seeking for time-periodic solutions numerically it is convenient to
use rescaled time coordinate $\tau=\Omega\,t$ where, as in the
perturbative construction, $\Omega$ denotes the frequency of the
solution we are looking for.  Assuming that time-periodic solution
does exist, we decompose both $B(t,x)$ and $\Pi(t,x)$ functions into
eigenmodes of the linearized problem in space and Fourier coefficients
in time.  Choosing a grid with $N$ collocation points in space
(\ref{eq:482}) and $K$ collocation points in time $\tau_{k} =
\pi(k-1/2)/K$, $k=1,\ldots,K$ (suited for trigonometric expansion) we
truncate these expansions as follows
\begin{align}
  \label{eq:494}
  B(\tau, x) & = \sum_{k=0}^{K-1}\sum_{j=0}^{N-1}
  \hat{B}_{k,j}\cos(k\tau)\,e_j(x),
  \\
  \label{eq:495}
  \Pi(\tau, x) &= \sum_{k=1}^{K}\sum_{j=0}^{N-1}
  \hat{\Pi}_{k,j}\sin(k\tau)\,e_j(x).
\end{align}
Next, at each instant of time $\tau_{i}$ we calculate the coefficients
\begin{align}
  \label{eq:496}
  \hat{B}_j(\tau_i) & = \sum_{k=0}^{K-1}\hat{B}_{k,j}\cos(k\tau_n) \,,
  \\
  \label{eq:497}
  \hat{\Pi}_j(\tau_i) &= \sum_{k=1}^{K}\hat{\Pi}_{k,j}\sin(k\tau_n)
  \,,
\end{align}
and use them as an input for our time evolution procedure, getting as
the output their time derivatives, which we equate to the time
derivatives of (\ref{eq:494}) and (\ref{eq:495}) (remembering that
$\partial_t = \Omega\,\partial_{\tau}$) evaluated at the set of $K
\times N$ tensor product grid points $(\tau_k, x_j)$.  We supply this
system with one additional equation, the condition to pick one
solution from a continuous family of time-periodic solutions.  Either
we set the amplitude of the dominant mode $\gamma$ in the initial data
to $\ep$
\begin{equation}
  \label{eq:498}
  \left.\inner{e_{\gamma}}{B}\right|_{\tau=0} =
  \sum_{k=0}^{K-1}\hat{B}_{k,\gamma} = \ep,
\end{equation}
as we did in perturbative construction or we choose the condition
setting second spatial derivative at the origin at initial time to
$\ep$
\begin{equation}
  \label{eq:499}
  B''(0,0) = \sum_{k=0}^{K-1}\sum_{j=0}^{N-1}\hat{B}_{k,j}\,e_{j}''(0) = \ep,
\end{equation}
which corresponds to controlling the dynamical part of the Kretschmann
scalar evaluated at $x=0$, cf. Eq.~(\ref{eq:110}).  In this way we get
a closed nonlinear system of $2 \times K \times N + 1$ equations for
$2 \times K \times N + 1$ unknowns: $\hat{B}_{k,j}$, $\hat{\Pi}_{k,j}$
and $\Omega$ ($k=0,1,\ldots,K-1$, $j=0,1,\ldots,N-1$).  This system is
solved with the Newton-Raphson algorithm yielding the time-periodic
solution of the system (\ref{eq:97})-(\ref{eq:99}).  To initialize the
numerical root-finding algorithm we take (when using the normalization
condition (\ref{eq:498}))
\begin{align}
  \label{eq:500}
  \hat{B}_{1,\gamma} &= \ep,
  \\
  \label{eq:501}
  \hat{\Pi}_{1,\gamma} &= -\ep\,\omega_{\gamma},
  \\
  \label{eq:502}
  \Omega &= \omega_{\gamma},
\end{align}
or (when taking (\ref{eq:499}))
\begin{align}
  \label{eq:503}
  \hat{B}_{1,\gamma} &= \frac{\ep}{e_{\gamma}''(0)},
  \\
  \label{eq:504}
  \hat{\Pi}_{1,\gamma} &= -\frac{\ep\,\omega_{\gamma}}{e_{\gamma}''(0)},
  \\
  \label{eq:505}
  \Omega &= \omega_{\gamma},
\end{align}
and the remaining expansion coefficients in (\ref{eq:494}) and
(\ref{eq:495}) we set to zero.  This provides a good guess for small
values of $\ep$ only, as is for the EKG system, for larger absolute
values of $\ep$ amplitudes Newton's algorithm converges slowly or
ceases to converge starting from such initial conditions.  Thus, we
apply the same method as for the scalar matter model and use the local
polynomial extrapolation from the data corresponding to time-periodic
solutions of smaller amplitudes used to generate initial values for
Newton's iteration.

In fact, at the output of this procedure we get a finite dimensional
representation of the $B(t,x)$ and $\Pi(t,x)$ fields only, given as
truncated expansions (\ref{eq:494}) and (\ref{eq:494}), the remaining
metric functions $\delta(t,x)$ and $A(t,x)$ can be also determined, at
any instant of time, by solving the constraint equations (\ref{eq:98})
and (\ref{eq:99}), as described in previous section, with
time-periodic sources $B(t,x)$ and $\Pi(t,x)$.

\subsection{Results}
\label{sec:BCSResults}

The analysis of the outcomes of perturbative calculations shows that
the time-periodic solutions of the system (\ref{eq:97})-(\ref{eq:99})
have the following regular structure.  The perturbative expansion of
the metric function $B$ is
\begin{equation}
  \label{eq:506}
  B_{\lambda}(\tau,x) = \sum_{j=0}^{j_{\text{max}}}\hat{B}_{\lambda,j}(\tau)e_{j}(x),
\end{equation}
where the upper limit of the sum is finite (as was pointed out before)
and depends on both the dominant mode index $\gamma$ and the
perturbative order $\lambda$, that is
\begin{equation}
  \label{eq:507}
  j_{\text{max}} =
  \frac{1}{2}\left(7(\lambda - 1) - \frac{1 + (-1)^{\lambda}}{2}\right)
  + \gamma\lambda,
\end{equation}
cf. (\ref{eq:465}).  Each coefficient $\hat{B}_{\lambda,j}(\tau)$ is a
finite linear combination of cosines
\begin{equation}
  \label{eq:508}
  \hat{B}_{\lambda,j}(\tau) = \sum_{k=0}^{\lambda}\hat{B}_{\lambda,j,k}\cos{k\tau}.
\end{equation}
The remaining metric functions have very similar form, namely
\begin{align}
  \label{eq:509}
  \delta_{\lambda}(\tau,x) &= \sum_{j=0}^{j_{\text{max}}}
  \hat{\delta}_{\lambda,2j}(\tau)\cos{2jx},
  \\
  \label{eq:510}
  A_{\lambda}(\tau,x) &= \sum_{j=0}^{j_{\text{max}}}
  \hat{A}_{\lambda,2j}(\tau)\cos{2jx},
\end{align}
here $j_{\text{max}}=\frac{1}{2}\left( 7\lambda -
  \frac{1-(-1)^{\lambda}}{2} \right) + \gamma\lambda$, and
\begin{align}
  \label{eq:511}
  \hat{\delta}_{\lambda,2j}(\tau) &=
  \sum_{k=0}^{\lambda}\hat{\delta}_{\lambda,2j,k}\cos{k\tau},
  \\
  \label{eq:512}
  \hat{A}_{\lambda,2j}(\tau) &=
  \sum_{k=0}^{\lambda}\hat{A}_{\lambda,2j,k}\cos{k\tau}.
\end{align}
The frequency expansion $\Omega(\ep)$ contains both even and odd
powers of $\ep$.  Due to the exponential nonlinearity of the field
equations and the complexity of perturbative equations, with the
current version of the \mathematica{} script we were not able to
construct a very high order perturbative solutions.\footnote{The time
  limitation is one thing but the memory consumption of a running
  program is enormous, even for the least complicated case,
  $\gamma=0$, the memory usage exceeds 128GB RAM at order $\lambda=11$
  (which is our main limitation now).  This is a strong motivation for
  further improvements of proposed algorithm.}

\begin{figure}[!t]
  \centering
  \includegraphics[width=\swidth]{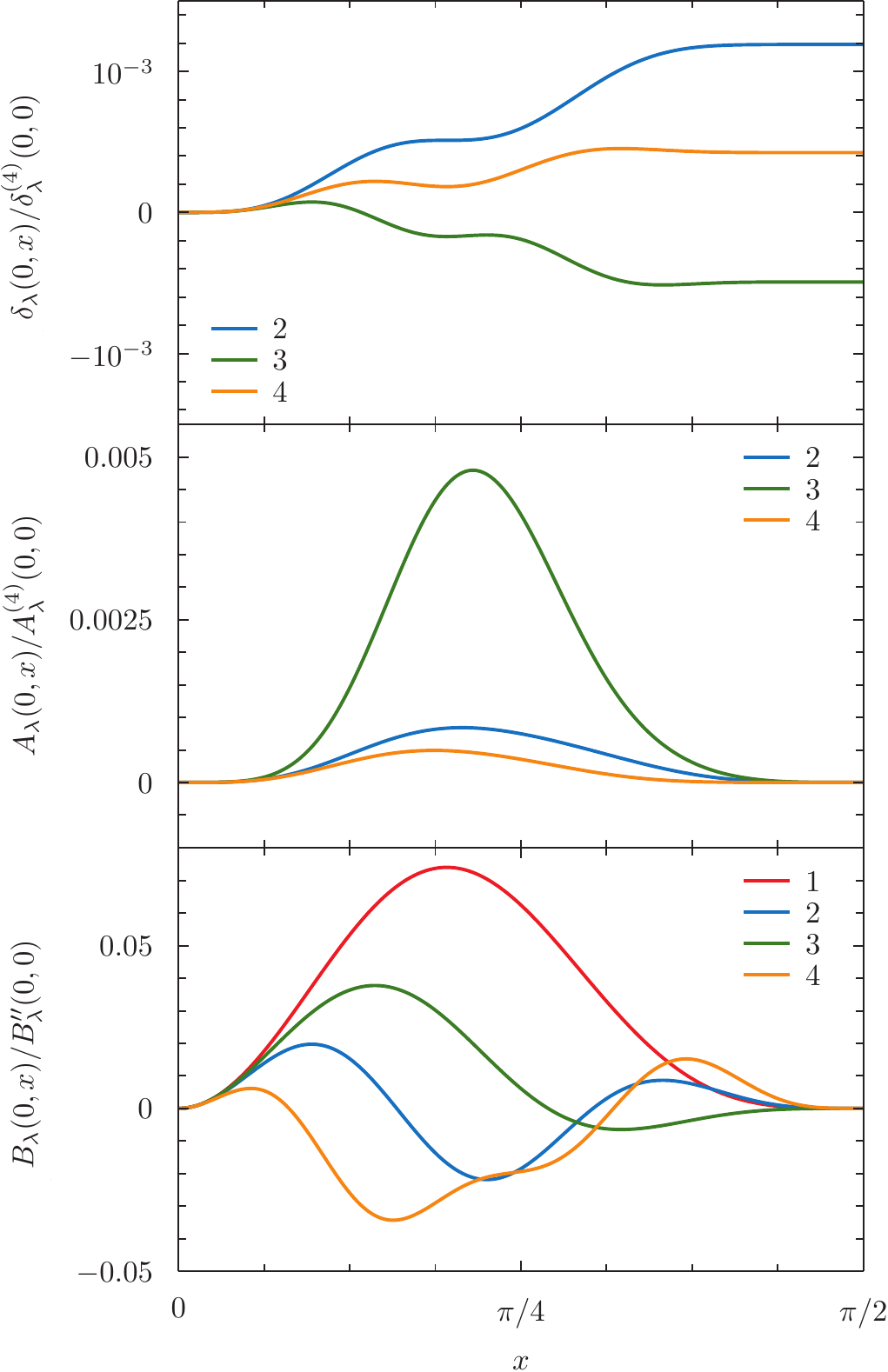}
  \caption{The perturbative profiles of the ground state ($\gamma=0$)
    time-periodic solution at time $t=0$. Successive order of
    perturbative approximations $\lambda$ are labeled and color
    coded. \textit{Top and middle panels}. The metric functions
    $\delta(0,x)$ and $A(0,x)$ are normalized by their fourth order
    derivative at the origin.  \textit{Bottom panel.} The
    corresponding profiles of the metric function $B(0,x)$ normalized
    by the slope at the origin.}
  \label{fig:BCSPerturbativeInitialProfiles}
\end{figure}

As an example of perturbative results we give an explicit form of the
frequency of time-periodic solutions bifurcating from the fundamental
mode ($\gamma=0$), expressed in terms of
$\ep=\inner{e_{\gamma}}{B(t=0,\,\cdot\,)}$
\begin{multline}
  \label{eq:513}
  \Omega(\ep) = 6 + \fracn{255520}{3003}\,\ep^{2}
  -\fracn{5110400}{63063}\sqrt{\frac{10}{3}}\,\ep^{3}
  \\[0.5ex]
  +
  \fracn{107529019665139827384118120}{14784728151271728210561}\,\ep^{4}
  \\[0.5ex]
  -\fracn{5194671967013678873387121856}{310479291176706292421781}
  \sqrt{\frac{10}{3}}\,\ep^{5}
  \\[0.5ex]
  \scalebox{0.95}{$ +
    \fracn{2814967732905508988971083255906420114422532599349512863
      7040470425757284}{304447658872843044986605891324506574364429
      00405674579718433434835}\,\ep^{6}$}
  \\
  \scalebox{0.82}{$ -
    \fracn{10791151457587107681786545030759576262227559
      46970978533253250354698196272}{658714025561242224607383655774
      84149726121911786823181572610522643}\sqrt{\frac{2}{15}}\,\ep^{7}$}
  \\
  \scalebox{0.52}{$ +
    \fracn{128452457711940337844909784683077365245488504914493
      6717987779052999954476099453106768376303533469242824001482580827
      82207374548949379594}{907688602853863710393337352764135094943701
      7883337835071744817865426425831403443119467095619561667345871234
      75200153484563288925}\,\ep^{8}$}
  \\
  \scalebox{0.5}{$ -
    \fracn{1319799259225578564690908463596706031125376945997255910
      78868001533610563183026797618097579407121829111034165654553
      7673474247185144320996288}{40355835282882780564087778703893446321
      19698150932001472897746022968588924641970810915070712457117
      30197435097073988239236838256055}\sqrt{\frac{2}{15}}\,\ep^{10}$}
  \\
  + \mathcal{O}\left(\ep^{11}\right),
\end{multline}
(note that $\xi_{9}\equiv 0$ in this expansion) and for solution
bifurcating from the first excited mode ($\gamma=1$)
\begin{multline}
  \label{eq:514}
  \Omega(\ep) = 8 + \fracn{8126280}{17017}\,\ep^{2} -
  \fracn{559810400}{1310309} \sqrt{5}\,\ep^{3}
  \\[0.5ex]
  +
  \fracn{24660035615444486661715203182184698}
  {170539264538251206633687202551}\,\ep^{4}
  \\[0.5ex]
  - \fracn{765219490326897319946768148832418981072}
  {2245490496175153637745759395989017}\sqrt{5}\,\ep^{5}
  \\[0.5ex]
  \scalebox{0.72}{$ +
    \fracn{1264528574989006273353048057864571823158
      830924295674846135208710817072318527577074413
      773009479340538}{1826250284722116053506617264
      523612421977382604219037949595481692953013662
      4436593376555377975}\,\ep^{6}$}
  \\
  \scalebox{0.52}{$ -
    \fracn{3119474612497936249223107682696235977392
      007646153665115483454298749412982003711002604
      47086157702721156376}{24046237498936102076521
      629521982404760176196749752072682323707451112
      3308933956624989104661796825}\sqrt{5}\,\ep^{7}$}
  \\
  \scalebox{0.38}{$ +
    \fracn{5879503137384970819961152533095432079777
      233512420449858057163834043523409774198645233
      722857851377553707206132924338495646546791099
      705013262303851763603057743273872624263451366
      81667030913}{14917335804180774894031565899835
      112371583805282719747531145478326680055890754
      338807798536228176484081362028850650480639747
      099793273961963052930264861020909348634010595
      953375000}\,\ep^{8}$}
  \\
  + \mathcal{O}\left(\ep^{9}\right).
\end{multline}

\begin{figure}[!t]
  \centering
  \includegraphics[width=\swidth]
  {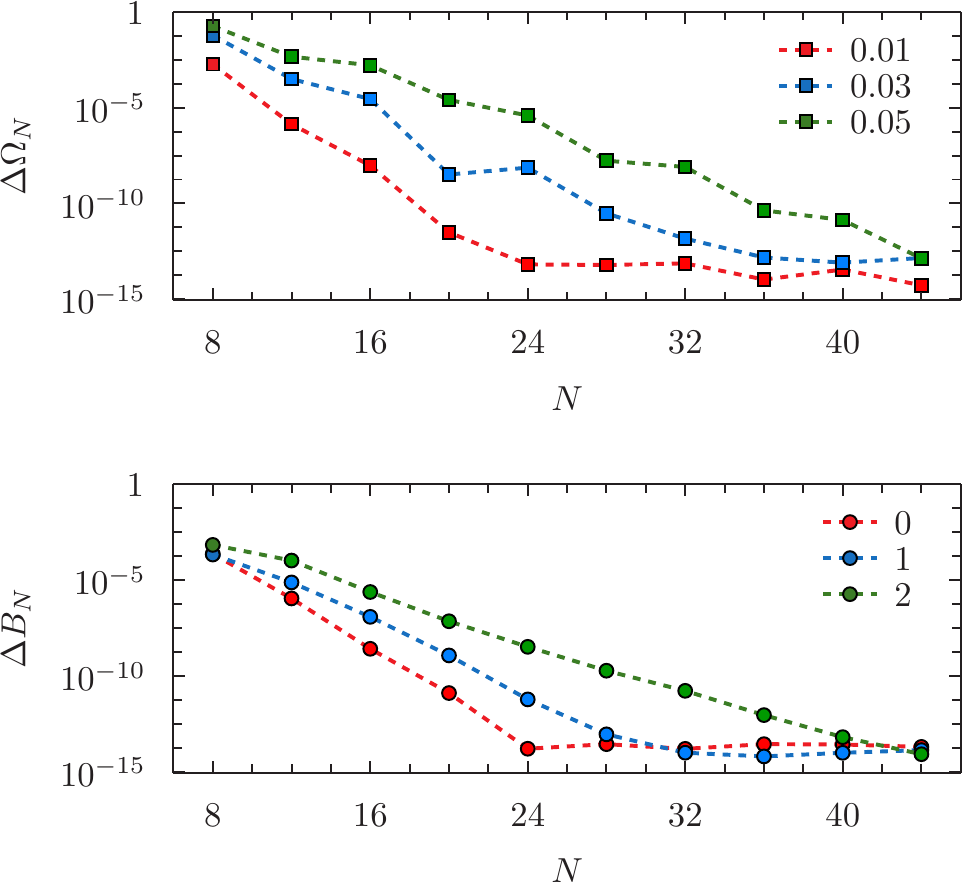}
  \caption{The results of convergence tests of the numerical procedure
    of Section~\ref{sec:BCSNumerical} used to find time-periodic
    solutions.  The number of Fourier modes $K$ of truncated expansion
    (\ref{eq:494}) and (\ref{eq:495}) was set to $N/2$.  For reference
    solution we have taken $N=44$.  \textit{Top panel}. The absolute
    frequency error
    $\Delta\Omega_{N}:=\left|\Omega_{N}-\Omega_{N=44}\right|$ for a
    fundamental mode solution $\gamma=0$.  The points color code
    different values of
    $\ep=\left.\inner{e_{\gamma}}{B}\right|_{\tau=0}$.  \textit{Bottom
      panel}. The $B(\tau,x)$ function absolute error $\Delta
    B_{N}:=\|B_{N}-B_{N=44}\|_{2}$ for solutions with $\ep=1/100$ and
    with varying $\gamma$.  The discrete $l^{2}$-norm was calculated
    on a set of equally spaced grid points $x_{i}=i\,\pi/128$,
    $i=1,\ldots,63$ and $\tau_{j}=j\,\pi/128$, $j=1,\ldots,127$.}
  \label{fig:BCSPeriodicNumericConvergence}
\end{figure}

The perturbative profiles of the metric functions for the $\gamma=0$
case at $t=0$ are shown on
Fig.~\ref{fig:BCSPerturbativeInitialProfiles}.  Derived solutions
share similar features as those of the scalar field model.  Higher
order perturbative expansions slightly modify the profile of dominant
mode.  A self-consistency test, of the perturbative construction,
includes computation of the total conserved mass of the solution,
either by using the integral (\ref{eq:104}) or by using expansion
(\ref{eq:106}).  This gives, as expected, the time independent
$\ep$-series which we list below, for $\gamma=0$
\begin{multline}
  \label{eq:515}
  M(\ep) = 72\,\ep^2 - \frac{160\sqrt{30}}{7}\,\ep^{3} +
  \fracn{2323060416265662648}{11476268992291103}\,\ep^4
  \\
  -\fracn{377668750201628764128}{80333882946037721}
  \sqrt{\frac{6}{5}}\,\ep^5
  \\
  \scalebox{0.72}{$+\fracn{7714049493666698396313208
      9188636318625246733261540632353240931802}{118208
      49933875753338504888235944694912301445854019
      50256449925}\,\ep^6$}
  \\
  \scalebox{0.72}{$-\fracn{92421374067020897159229756351914921
      32734310131865906674215728010552}{4964756972227816402172
      0530590967718631666072586881
      91077089685}\sqrt{\frac{2}{15}}\,\ep^7$}
  \\
  \scalebox{0.52}{$+\fracn{83884794866616251381183707817349583
      092772248524569086710647984936280836491055126056661186
      7137039832137024241667991722581446627184}{80851602698400
      558530039134979792198434405945551973
      305112365434463029107934757122543486858172213023941988
      088575361506116125}\,\ep^8$}
  \\
  \scalebox{0.38}{$-\fracn{4137888998461859967407475873
      945978000527429539513871885046484880644929229052905547
      32731981277112659513659718906370001719654288992228
    }{12903915790664729141394245942774834870131188910094
      939495933523340299445626387236757940502564285198621141
      29893662769637613355}\sqrt{\frac{2}{15}}\,\ep^9$}
  \\
  + \mathcal{O}\left(\ep^{10}\right),
\end{multline}
and for $\gamma=1$
\begin{multline}
  \label{eq:516}
  M(\ep) = 128\,\ep^2 - \frac{79360\sqrt{5}}{693}\,\ep^3 -
  \fracn{49021708911471544946735104}{88722504124676105028345}\,\ep^4
  \\
  - \fracn{24535320799809131378995801223168}{23364184236192
    2054981643723}\frac{1}{\sqrt{5}}\,\ep^5
  \\
  \scalebox{0.72}{$+\fracn{52092673515415257535135372906997172401
      07720292251257865697699905084289499388705463
      50170889118208}{199473671269575593939073428
      0000930809332179491361934033874229156361319
      85838883461001336875}\,\ep^6$}
  \\
  \scalebox{0.52}{$-\fracn{188758 0667797356873126815482336536013138
      68554324901360615411589478048328907946129118200
      5118934032384}{27647050837963177319955577120812901017344007750276
      405709496816107167893237269247694785290875}\frac{1}{\sqrt{5}}\,\ep^7$}
  \\
  \scalebox{0.38}{$+\fracn{83971526965991783575506316247174421650
      121362198774067971634420008972725327255849600666105480
      961066965756420937992332675476283846262220077633380546
      62991301458149772753380415173632}{63777115548378486509
      426491878868121769319408157661
      457402004031226516586938692998628208334806081591186447
      522740903708579428593574729183716272930696566869253351
      88271640625}\,\ep^8$}
  \\
  + \mathcal{O}\left(\ep^{9}\right).
\end{multline}

\begin{figure}[!t]
  \centering
  \includegraphics[width=\swidth]{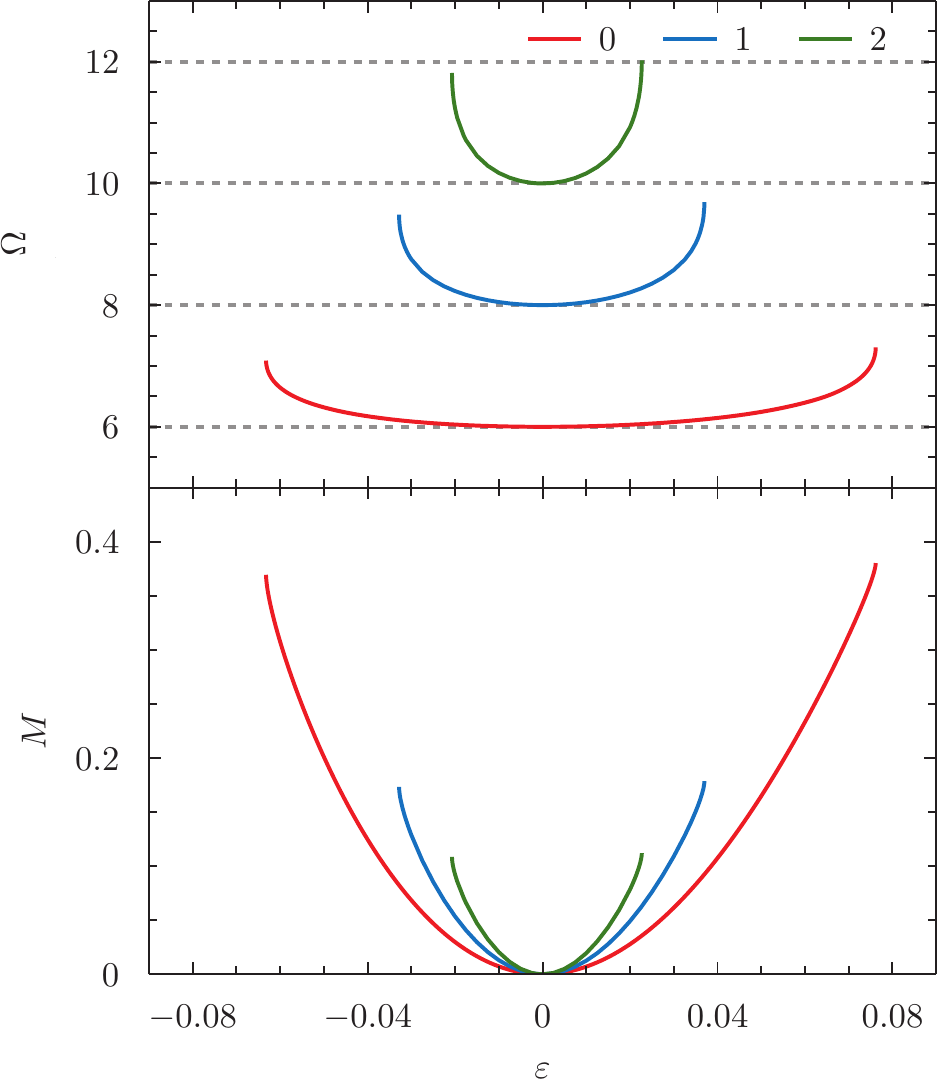}
  \caption{\textit{Top panel}. The frequency of time-periodic
    solutions bifurcating from the lowest eigenmodes (with $\gamma$
    color coded) parametrized by dominant mode amplitude
    $\ep=\left.\inner{e_{\gamma}}{B}\right|_{\tau=0}$. \textit{Bottom
      panel}.  The corresponding total mass of the solutions.}
  \label{fig:BCSPeriodicOmegaMass}
\end{figure}
\begin{table}[!t]
  \centering
  \begin{tabular}{c ccc}
    \toprule
    \multirow{2}{*}{$n$} & \multicolumn{3}{c}{$\gamma$} \\
    \cmidrule(r){2-4}
    & 0 & 1 & 2 \\ \midrule
    $2$ & $\{ \num{-0.100},\num{0.121} \}$
    & $\{ \num{-0.055},\num{0.061} \}$
    & $\{ \num{-0.036},\num{0.039} \}$ \\
    $4$ & $\{ \num{-0.075}, \num{0.089} \}$
    & $\{ \num{-0.040}, \num{0.044} \}$ & --- \\
    \bottomrule
  \end{tabular}
  \caption{The zeros of the denominator of the diagonal $[n/n]_{\Omega}$
    Pad\'e approximation of $\Omega(\ep)$, expressed in terms of
    $\ep=\left.\inner{e_{\gamma}}{B}\right|_{\tau=0}$, closest to 0
    for $\gamma=0,1$ and $2$. (For $\gamma=2$ the $n=4$ rational
    approximation is missing due to the insufficient number of terms in
    perturbative expansion.)}
  \label{tab:BCSOmegaPade}
\end{table}

The convergence of a numerical method used to find time-periodic
solutions numerically is presented in
Fig.~\ref{fig:BCSPeriodicNumericConvergence}.  The spectral
convergence is observed whenever number of eigenmodes $N$ and number
of trigonometirc polynomials $K$ is increased; the optimal results we
get for $K\approx N/2$ for considered amplitudes.  Using
pseudospectral method of Sections~\ref{sec:BCSEvolution} and
\ref{sec:BCSNumerical} we achieve exponential convergence for $\Omega$
and both dynamical fields $B$ and $\Pi$ (as a consequence the spectral
accuracy is achieved also for the remaining metric functions $\delta$
and $A$).  It is evident from these tests that in order to obtain an
accurate approximation of the time-periodic solution we need to
increase number of grid points (equivalently the number of basis
functions in truncated expansions (\ref{eq:494}) and (\ref{eq:495}))
both in space and time when either $\left|\ep\right|$ or $\gamma$ is
increased.  This is due to the fact that higher modes become
significant when $\left|\ep\right|$ and $\gamma$ are increasing, which
stays in agreement with the form of perturbative expansion
(\ref{eq:506})-(\ref{eq:511}).

\begin{figure}[!t]
  \centering
  \includegraphics[width=\swidth]{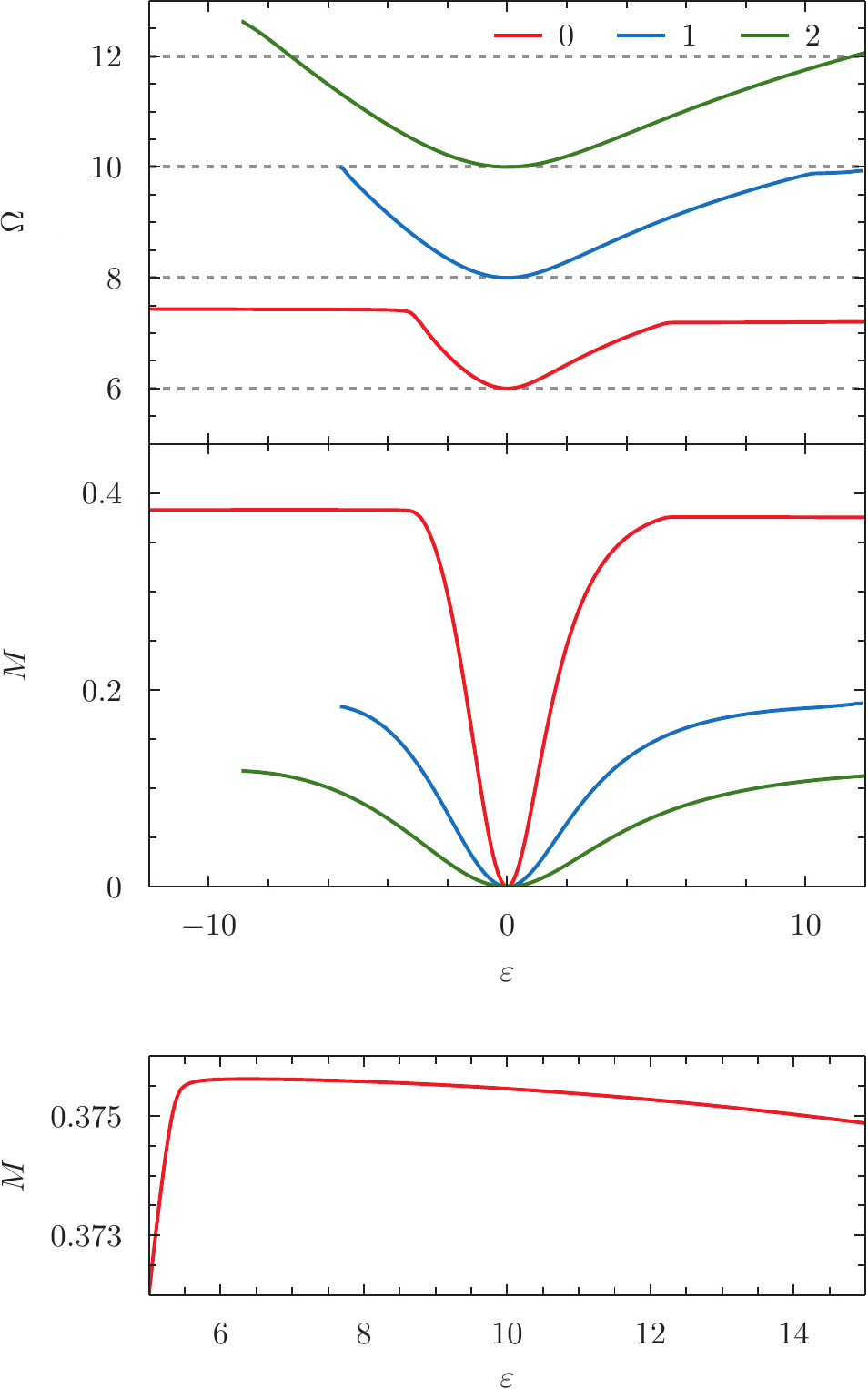}
  \caption{An analogue of Fig.~\ref{fig:BCSPeriodicOmegaMass} using
    different parametrization of time-periodic solutions.  \textit{Top
      panel}.  For large values of $|\ep|$ the frequency $\Omega$
    grows monotonically which is hidden on this scale.  \textit{Middle
      and bottom panels}.  The mass function on the other hand is
    bounded from above $M\leq M_{\ast}$.  At stationary points of mass
    function of $\gamma=0$ family of solutions:
    $M(\ep_{\ast}^{+})\approx\num{0.375619}$ at
    $\ep_{\ast}^{+}\approx\num{6.380482}$ and
    $M(\ep_{\ast}^{-})\approx\num{0.382 891}$ at
    $\ep_{\ast}^{-}\approx\num{-7.226 530}$, for numerical data with
    $12\times 32$ modes.}
  \label{fig:BCSPeriodicOmegaMassNew}
\end{figure}

On Fig.~\ref{fig:BCSPeriodicOmegaMass} we plot the bifurcation curves
for time-periodic solutions with $\gamma=0,1$ and $2$, showing both
the frequency $\Omega$ and the total mass $M$ as a function of
$\ep=\left.\inner{e_{\gamma}}{B}\right|_{\tau=0}$.  These numerical
solutions were found by taking $24\times 64$ modes for $\gamma=0,1$
and $32\times 72$ modes for $\gamma=2$.  Because of lack of symmetry
$B\ra -B$ in the system (\ref{eq:97})-(\ref{eq:100}), there are both
negative and positive $\ep$ branches of solutions for each family
$\gamma$.  These are not symmetric with respect to $\ep=0$, neither
$\Omega$ nor $M$, so is the range of amplitudes for which the
time-periodic solutions do exist.  Similarly to the EKG model (see
Section~\ref{sec:AdSPeriodicResults}) this is related to the fact that
$\left.\inner{e_{\gamma}}{B}\right|_{\tau=0}$ is bounded and no
time-periodic solutions exist with larger absolute values of dominant
mode amplitude than some maximal value.  This interval rapidly
shrinks---magnitude of maximal allowed amplitudes decrease---with
increasing $\gamma$.  Using the extrapolated initial guess for the
Newton method we were able to find time-periodic solutions staying
very close to the boundary of their existence.  The estimated ranges
of $\ep$ for which solutions exists are:
$-0.06322\lesssim\ep\lesssim 0.076175$ for $\gamma=0$,
$-0.032786\lesssim\ep\lesssim 0.037023$ for $\gamma=1$ and
$-0.0207\lesssim\ep\lesssim 0.02269$ for $\gamma=2$.  Near these
limiting values $\Omega$ rapidly increases while $M$ stays finite (so
is the Kretschmann scalar evaluated at the origin (\ref{eq:110})).
For still larger absolute values of $\ep$ the Newton method ceases to
converge to a true solution.

\begin{figure}[!t]
  \centering
  \includegraphics[width=\swidth]{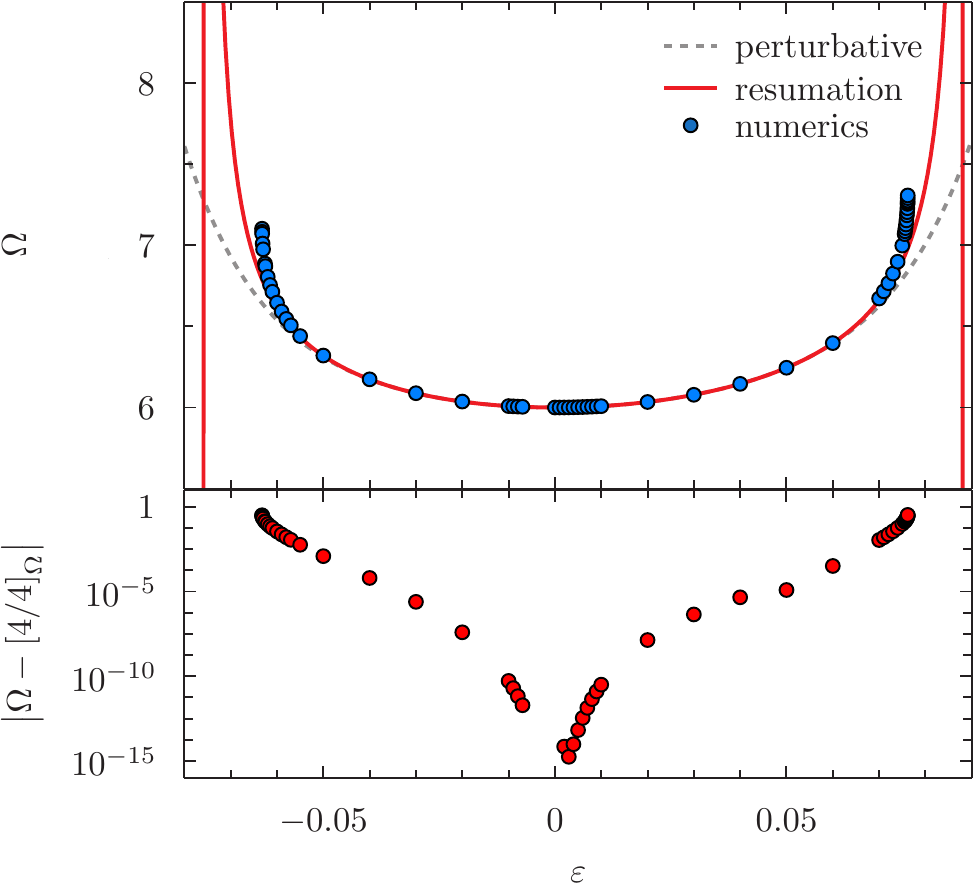}
  \caption{The comparison of oscillation frequency of time-periodic
    solution bifurcating from eigenmode $e_{0}(x)$ from numerical and
    from perturbative calculation.  \textit{Top panel}. The numerical
    data (points) align on a smooth curve which is well approximated
    by perturbative expansion (dashed gray line) only for small $\ep$.
    The Pad\'e resumation accelerates the convergence (solid red
    line). \textit{Bottom panel}.  The absolute difference between
    numerical data and $[4/4]_{\Omega}$ Pad\'e resumation of $\Omega$
    series (\ref{eq:513}).}
  \label{fig:BCSPeriodicOmegaPade}
\end{figure}

These limiting values can be estimated using perturbative series
expansion with Pad\'e resumation.  Since the frequency expansion
contains the most terms we analyze this series.  Computing a diagonal
$[n/n]_{\Omega}$ Pad\'e approximation we have found that for even $n$
it contains simple poles on the real line only.  These being closest
to zero can be taken as the upper bound of the convergence radius of
constructed perturbative series.  The approximate locations of these
poles are given in Tab.~\ref{tab:BCSOmegaPade}, and these agree with
numerical results given above.  Additionally, the Pad\'e resumation
can be used to accelerate convergence of perturbative series.  In
Fig.~\ref{fig:BCSPeriodicOmegaPade} we show such comparison for a
solution bifurcating from fundamental eigenmode ($\omega_{0}=6$); we
get similar results for solutions bifurcating from higher eigenmodes.

The limiting values of the parameter $\ep$ are absent when we use an
alternative definition of $\ep$, given in Eq.~(\ref{eq:499}), see
Fig.~\ref{fig:BCSPeriodicOmegaMassNew} where we plot frequency and
mass of large amplitude solutions as a function of $B''(0,0)$.  As for
the scalar field system (Sections~\ref{sec:AdSPeriodic} and
\ref{sec:Standing}), the mass function retains its maximal value, both
for branches $\ep>0$ and $\ep<0$.  The curves shown on
Fig.~\ref{fig:BCSPeriodicOmegaMassNew} are expected to be smoothly
continued for greater values of $|\ep|$ when taking larger number of
modes in the truncated approximation for $B$ and $\Pi$.

\begin{figure}[!t]
  \centering
  \includegraphics[width=\swidth]{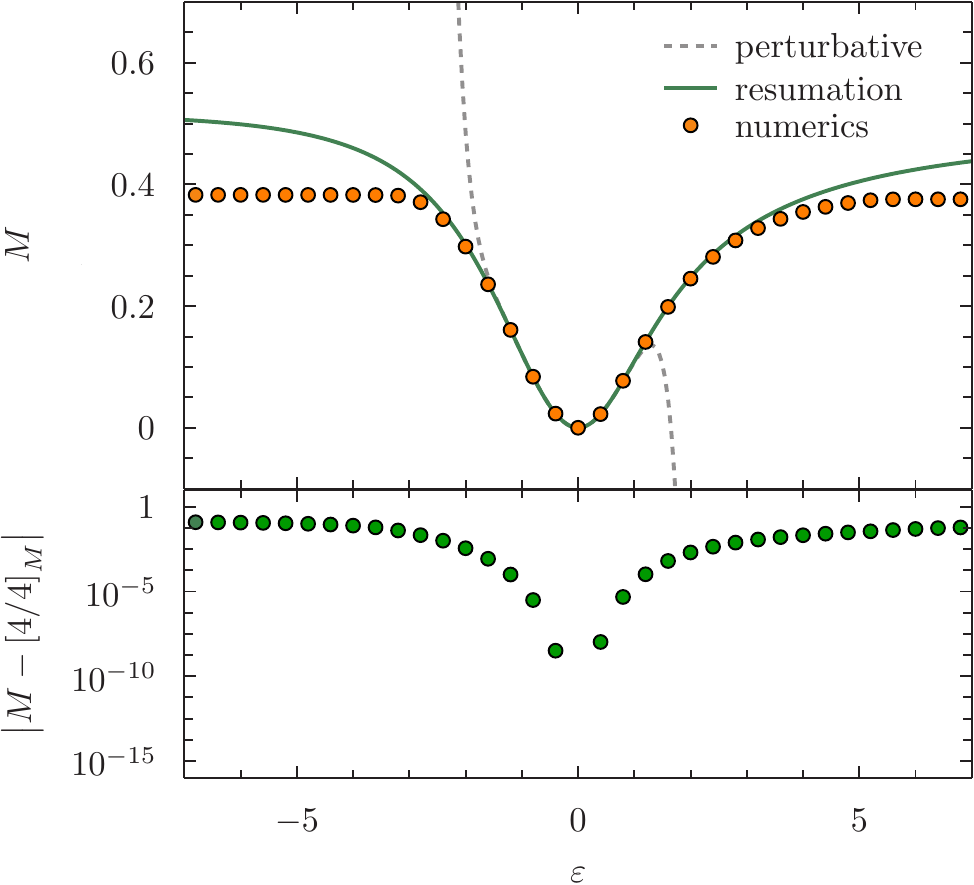}
  \caption{The comparison of mass of time-periodic solution
    bifurcating from eigenmode $e_{0}(x)$ from numerical and from
    perturbative calculation.  \textit{Top panel}. The Pad\'e
    resumation accelerates the convergence of the mass function series
    with $B''(0,0)$ as the expansion parameter (compare dashed and
    solid lines with points).  \textit{Bottom panel}. The absolute
    difference between numerical data and $[4/4]_{M}$ Pad\'e
    resumation of $M$ series (\ref{eq:515}).}
  \label{fig:BCSPeriodicMassPadeNew}
\end{figure}

Similarly, we can use the Pad\'e resumation to improve perturbative
series when using $B''(0,0)$ as the expansion parameter.  We
illustrate this on Fig.~\ref{fig:BCSPeriodicMassPadeNew}, where we
plot results for the $M(\ep)$ function. (A similar we get while
looking at $\Omega(\ep)$; this time though the Pad\'e approximation
does not have the poles on the real axis and their structure rapidly
changes with $n$.)  This illustrates the agreement of our two
independent methods used to find time-periodic solutions
(demonstrating their correctness) also shows the superiority of Pad\'e
resumation.

\begin{figure}[!t]
  \centering
  \begin{tabular}{lr}
  \includegraphics[width=0.45\textwidth]
  {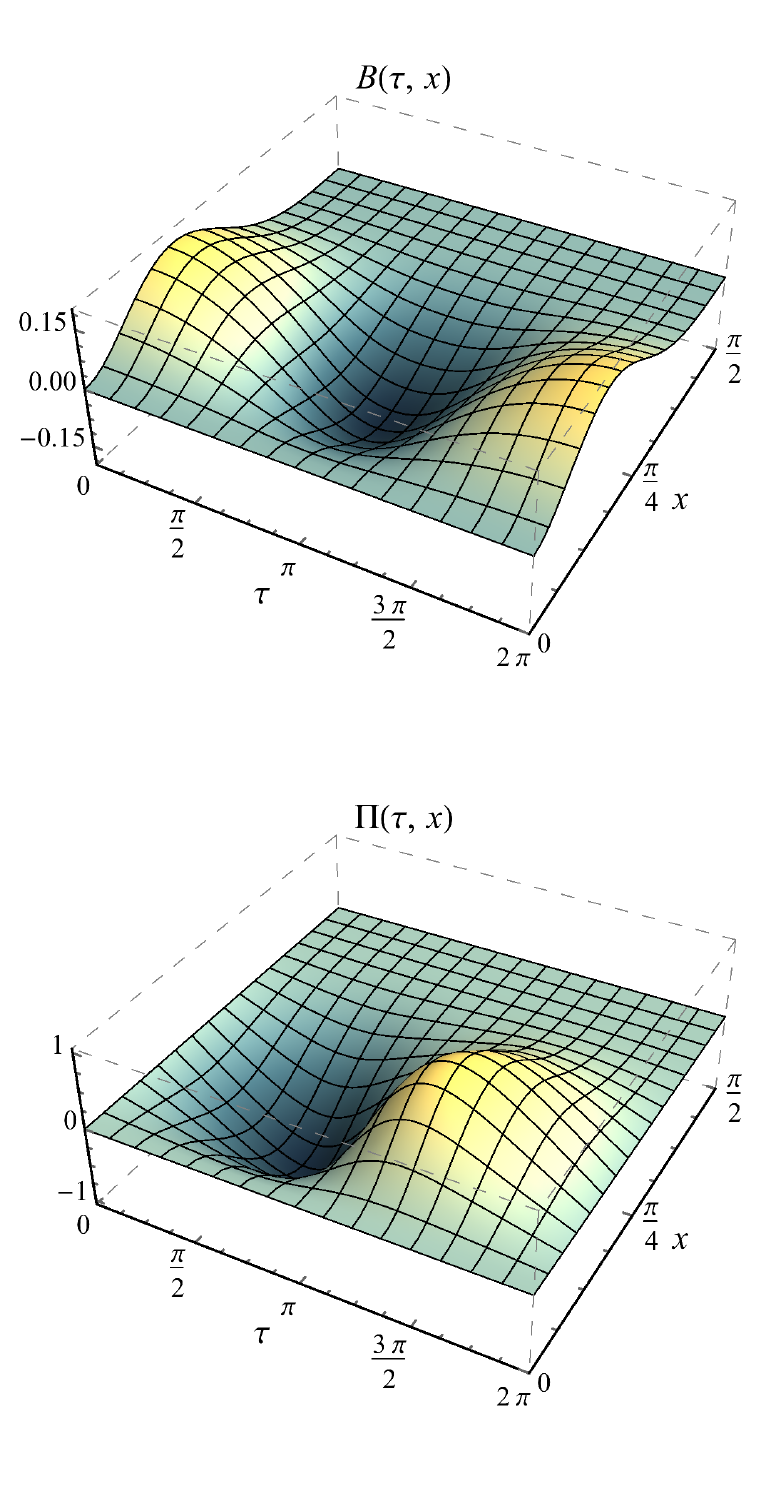} &
  \includegraphics[width=0.45\textwidth]
  {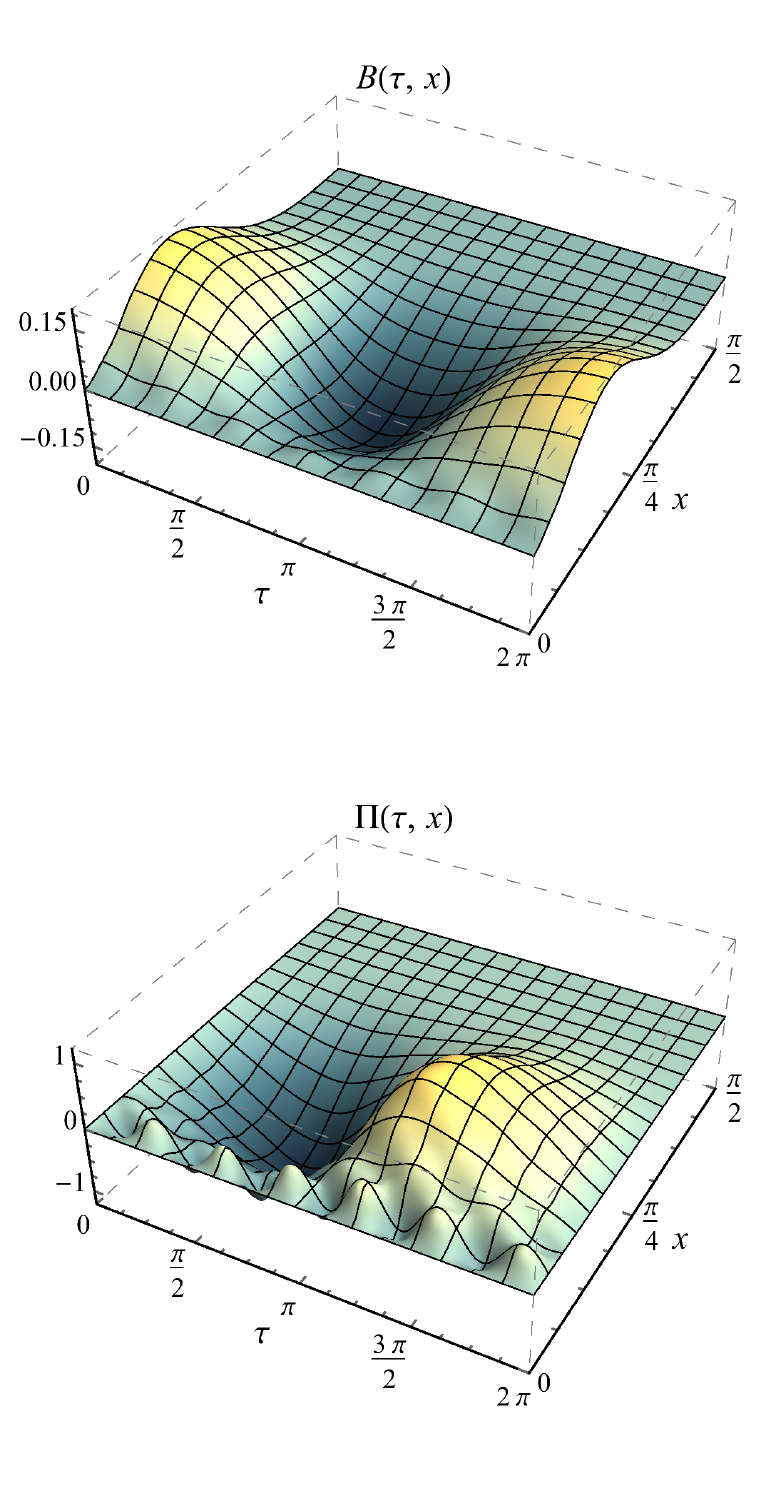}
  \end{tabular}
  \caption{The spatio-temporal plots of the time-periodic solutions
    bifurcating from fundamental eigenmode of different amplitude
    $\ep=B''(0,0)$.  \textit{Left}. The $\ep=4$ case
    ($\Omega\approx\num{6.929985}$, $M\approx\num{0.354816}$).  Both
    dynamical fields $B$ and $\Pi$ exhibit an almost harmonics
    oscillations with no significant nonlinear effects.  The numerical
    solution was derived on a grid with $16\times 96$ points.
    \textit{Right}. Large amplitude solution $\ep=10$ right to the
    mass stationary point ($\Omega\approx\num{7.198278}$,
    $M\approx\num{0.375623}$).  A clearly visible nonlinear
    oscillations required definitely more points to resolve the fast
    oscillations; here we present the results for the grid of
    $48\times 80$ points.}
  \label{fig:BCSPeriodicEvolutionE0Epsilon4Epsilon10}
\end{figure}

Looking at derived solutions we have found a notable change of their
profiles when moving along the branch of time-periodic family (with
fixed $\gamma$).  While all solutions with amplitudes\footnote{By
  $\ep_{\ast}^{\pm}$ we denote the smallest in absolute value critical
  points of the mass, $M'\left(\ep_{\ast}^{\pm}\right)\equiv 0$,
  $\ep_{\ast}^{+}>0$ and $\ep_{\ast}^{-}<0$.}
$\ep_{\ast}^{-}<\ep<\ep_{\ast}^{+}$ are dominated by the harmonic
oscillation with profile of bifurcating mode, those beyond the mass
extremum have much reacher structure and are of only slightly larger
amplitude.  For positive $\ep$ branch of $\gamma=0$ family this is
visualized on Fig.~\ref{fig:BCSPeriodicEvolutionE0Epsilon4Epsilon10},
similar we observe for negative values of $\ep$ and for other families
$\gamma>0$.

\begin{figure}[!pt]
  \centering
  \includegraphics[width=0.62\columnwidth]
  {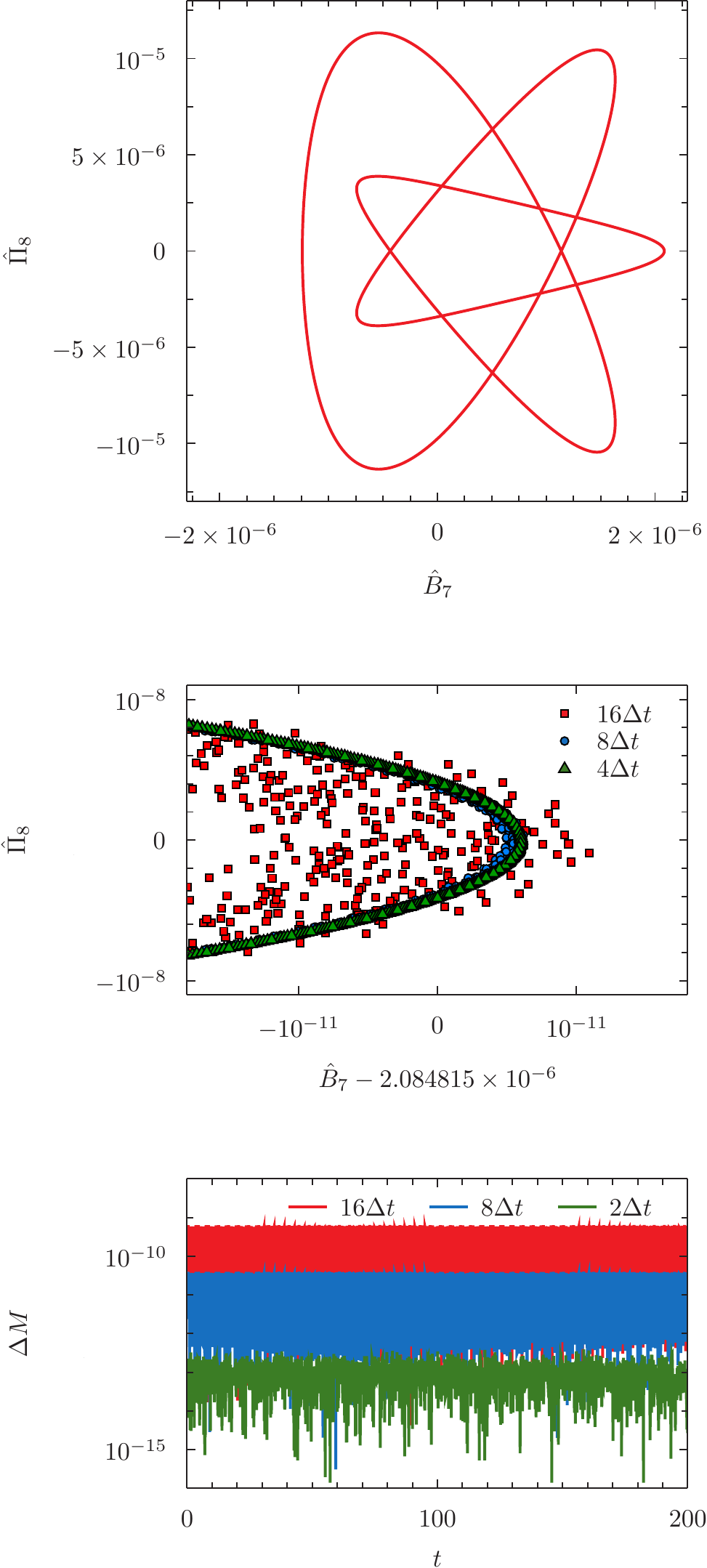}
  \caption{The numerical evolution scheme and stability test of
    time-periodic solution with $\gamma=0$ and mode amplitude
    $\ep=0.05$ ($B''(0,0)\approx\num{1.356402}$,
    $\Omega\approx\num{6.245130}$).  \textit{Top panel}.  The plot
    shows a parametric plot of $\hat{B}_{7}(t)$ and $\hat{\Pi}_{8}(t)$
    for $t\in[0,200T]$ ($T\approx\num{1.006094}$).  On this scale of
    the phase space section a numerically evolved solution appears as
    a closed curve.  \textit{Middle panel}. A zoom of a small region
    of the phase space.  Solutions were obtained by integrating in
    time (with decreasing step size) initial conditions corresponding
    to the time-periodic solution.  The Gauss-RK method of order $4$
    was used and we set $\Delta t=2^{-13}\pi\approx 3.8\cdot 10^{-4}$
    (in all runs $N=64$).  \textit{Bottom panel}.  The total mass
    conservation test showing the absolute error $\Delta M:=
    M(t)-M(0)$.  Due to the large rounding error a further decrease of
    $\Delta t$ does not reduce $\Delta M$ below $10^{-13}$.}
  \label{fig:BCSPerturbativeStabilityTestTimeStep}
\end{figure}

\begin{figure}[!pt]
  \centering
  \begin{tabular}{cc}
    \includegraphics[width=0.46\columnwidth]
    {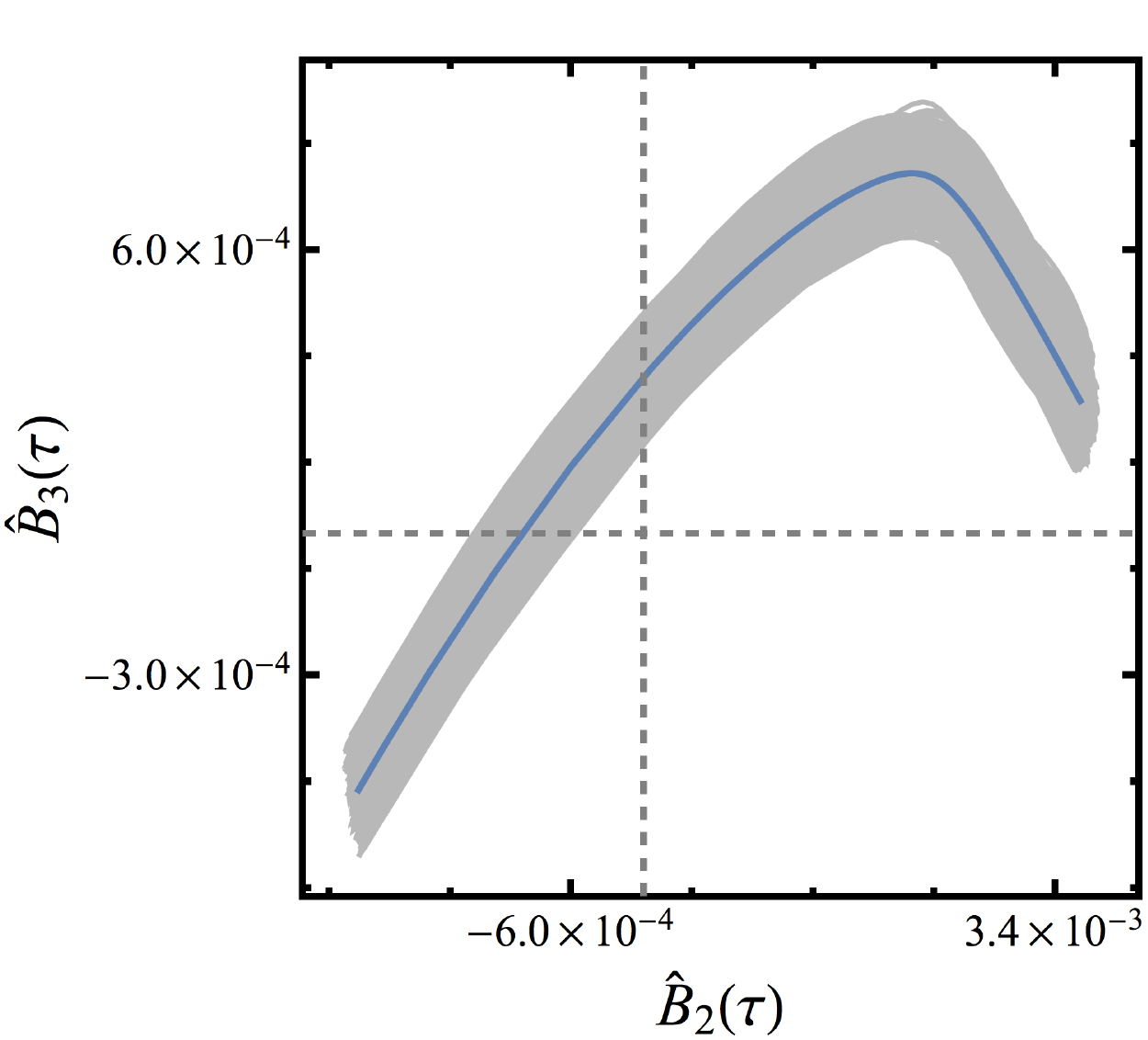} &
    \includegraphics[width=0.46\columnwidth]
    {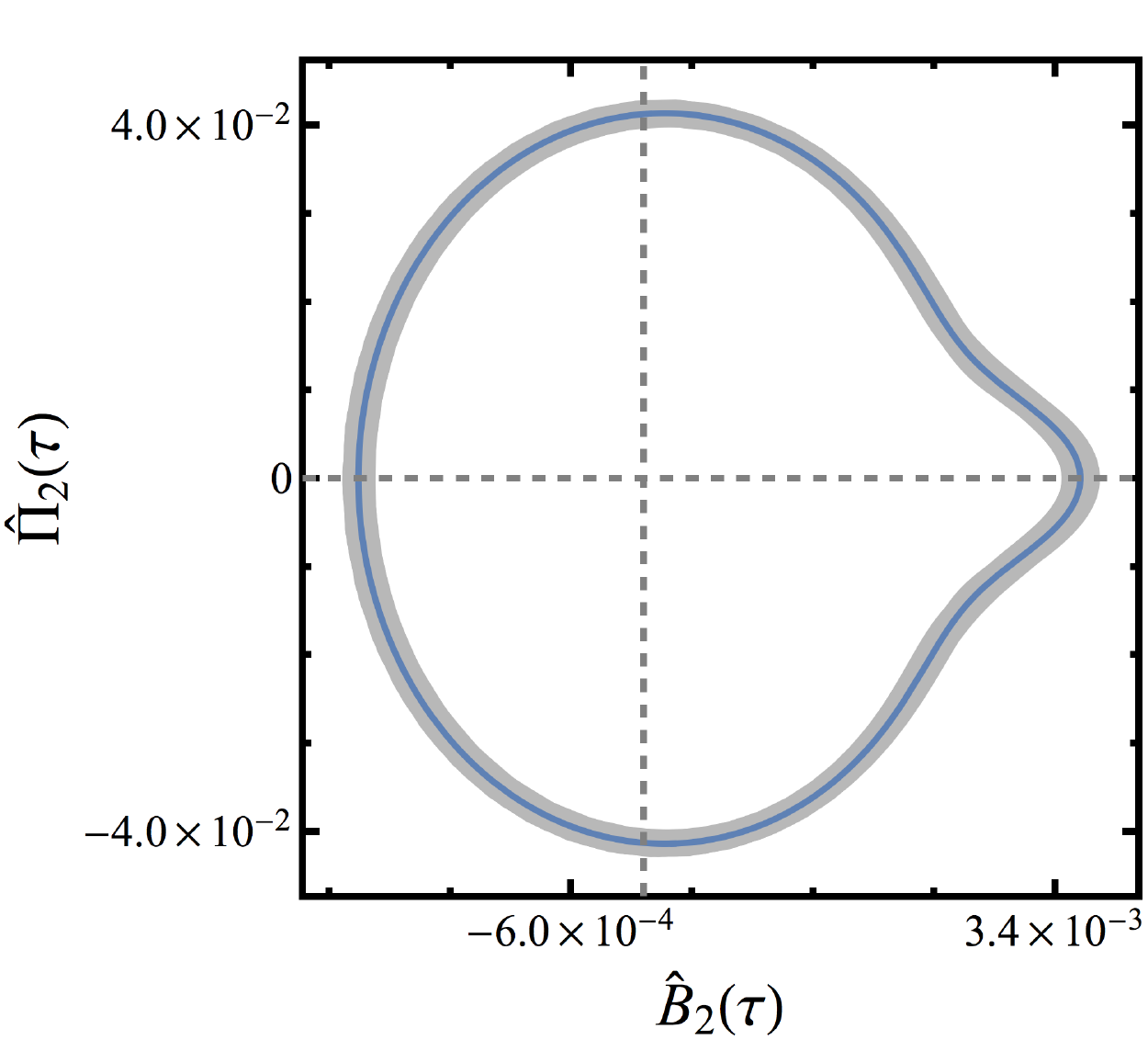}
    \\[4ex]
    \includegraphics[width=0.46\columnwidth]
    {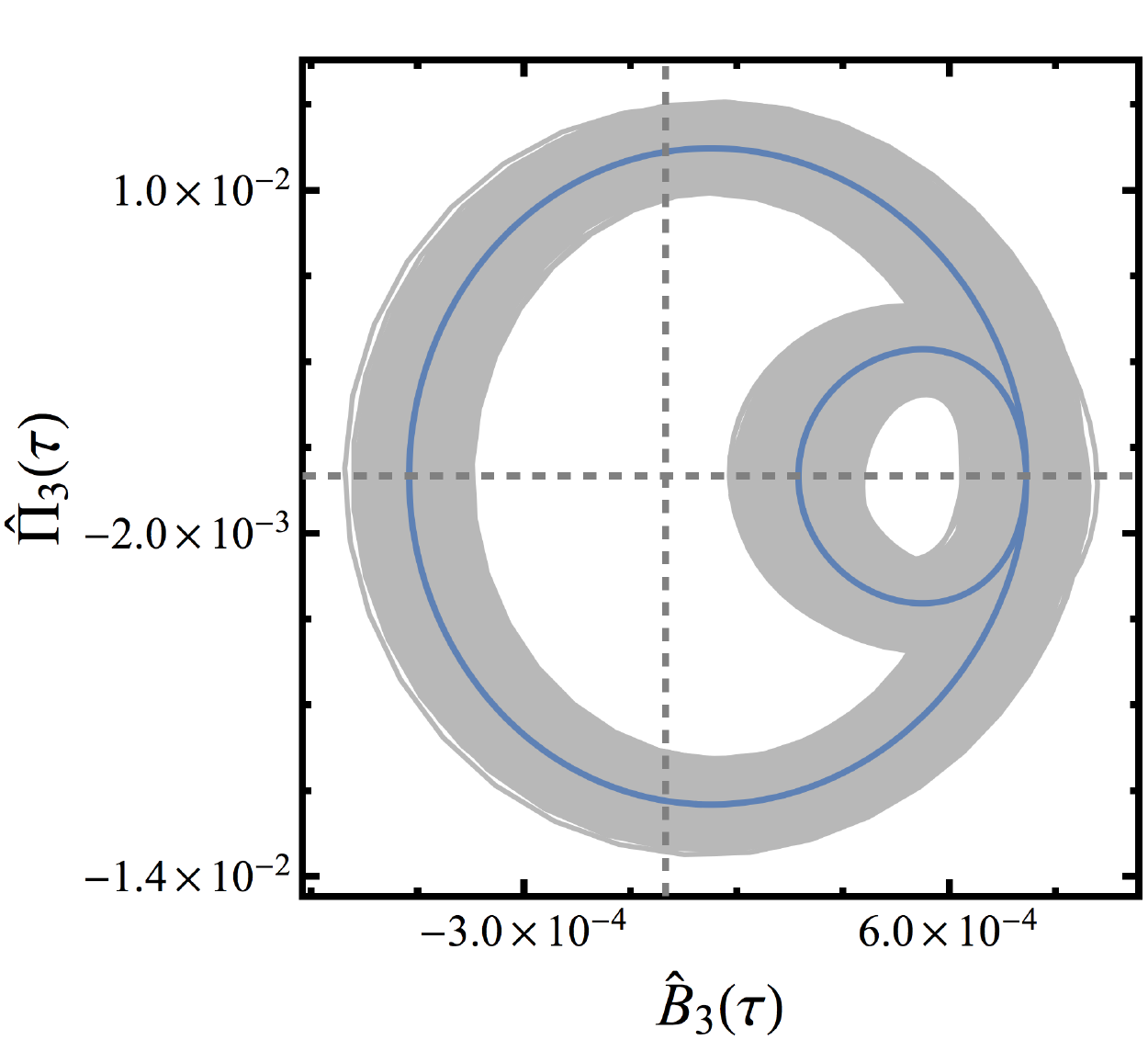} &
    \includegraphics[width=0.46\columnwidth]
    {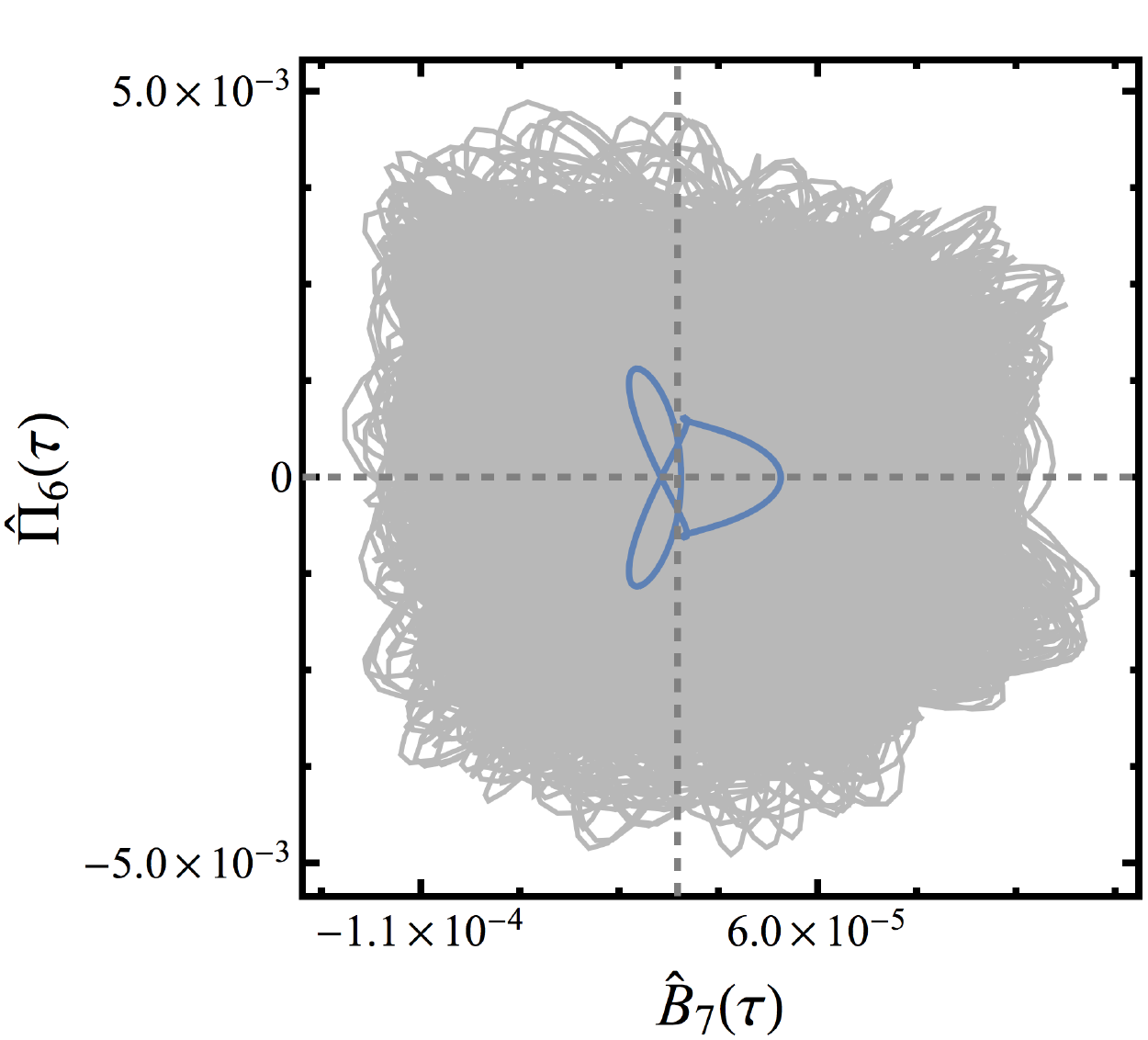}
    \\[4ex]
    \includegraphics[width=0.46\columnwidth]
    {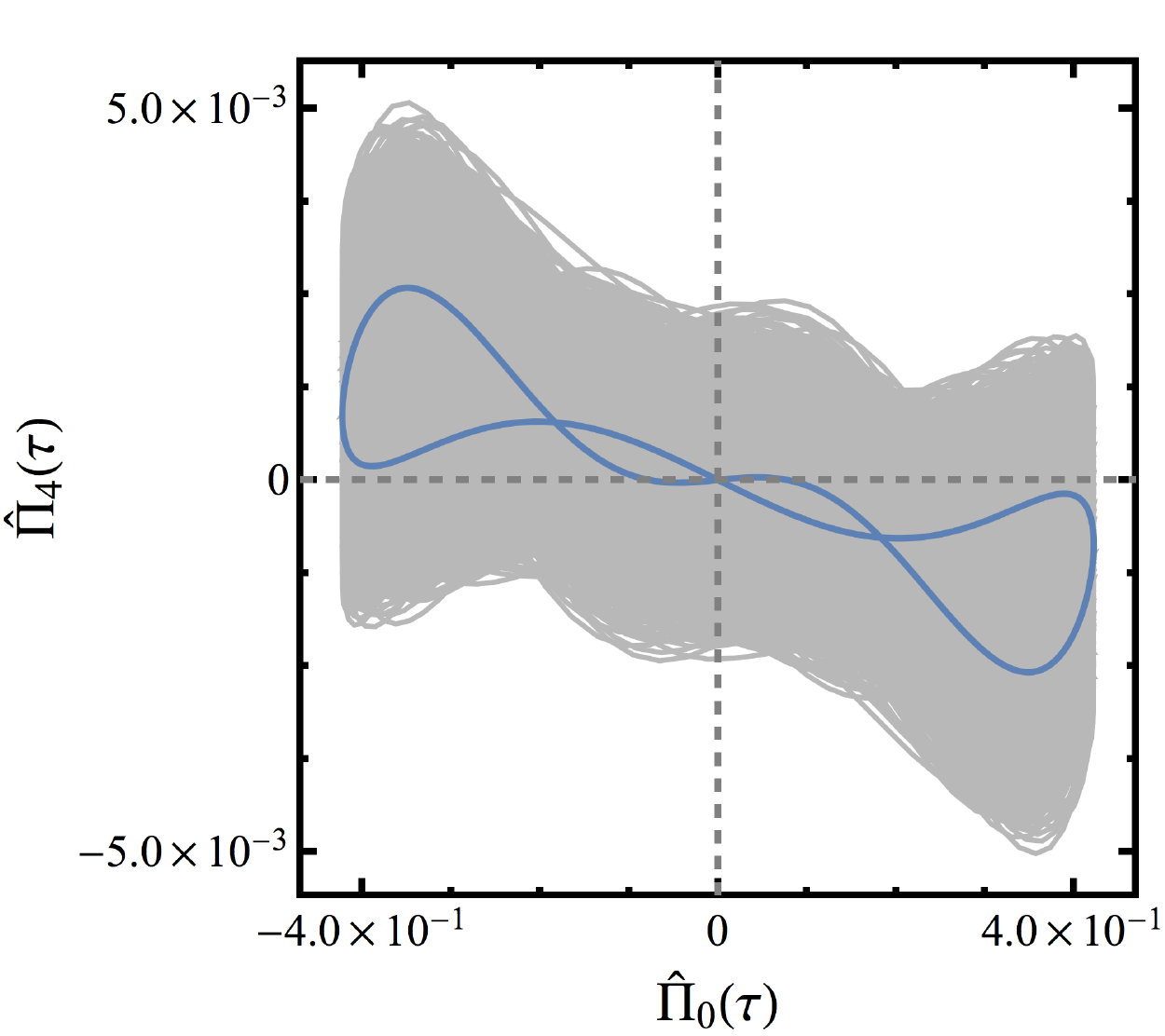} &
    \includegraphics[width=0.46\columnwidth]
    {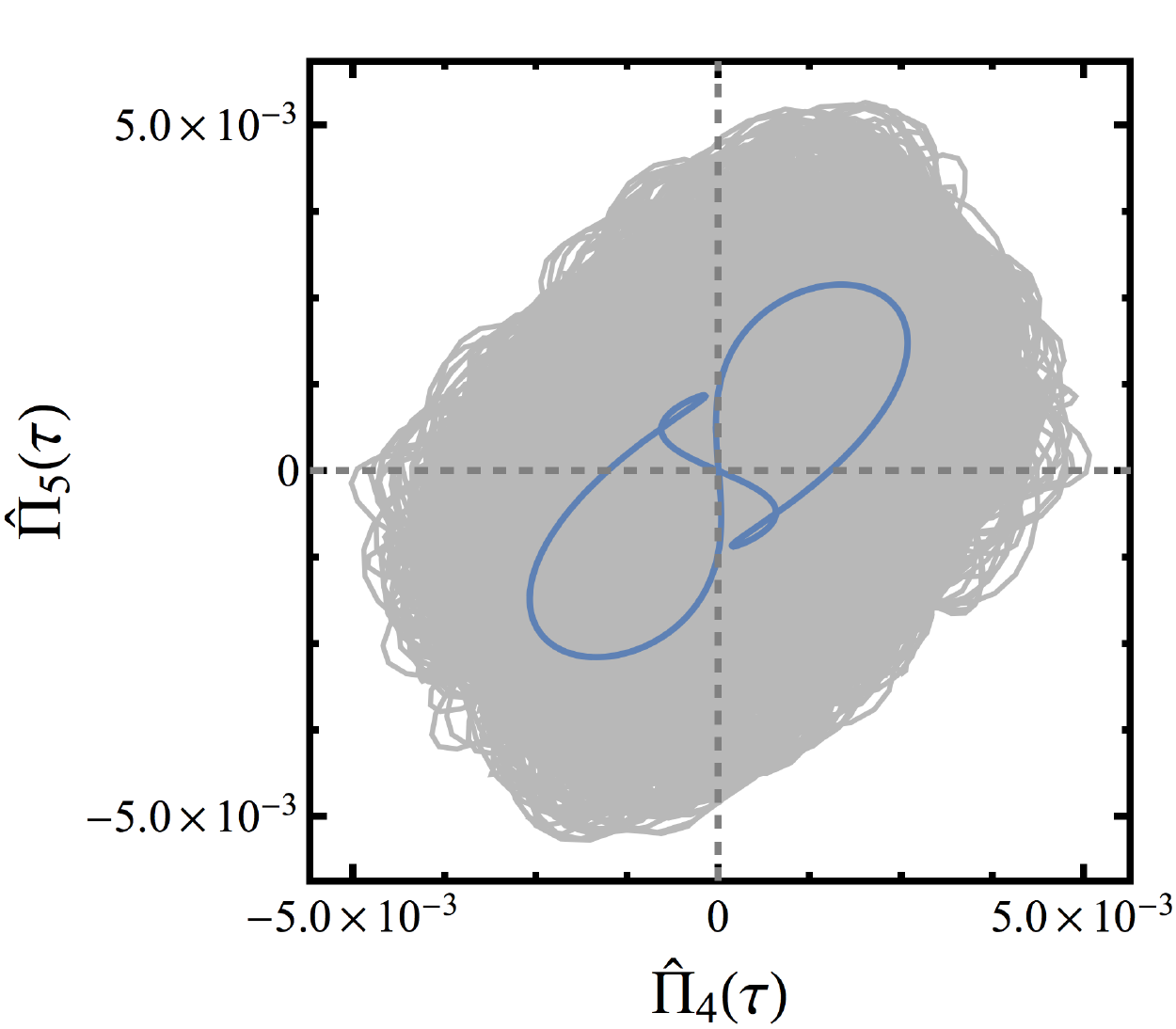}
  \end{tabular}
  \caption{The plots of projections of a phase space of perturbed, by
    a Gaussian pulse (\ref{eq:517}) with $\epsilon=70$, time-periodic
    solution of central amplitude $\ep=4$ (gray lines) with
    unperturbed trajectories overlaid (blue lines).  The shaded
    regions---the explored volume of a phase space---does not expand
    further during long time evolution; here $\tau\in[0,1700\pi]$.
    The evolution was performed using $N=128$ eigenmodes with sixth
    order Gauss-RK with time step $\Delta t=2^{-10}\pi$.  The larger
    the index is the larger the perturbation which is related to the
    fact that the initial data we have chosen (\ref{eq:517}) are wide
    in a Fourier space so these introduce relatively large
    perturbation into the higher modes.}
  \label{fig:BCSLoopsE0}
\end{figure}

Having numerically derived time-periodic solutions we have put them
into evolution code, as initial conditions, configuration read off
from (\ref{eq:494}) and (\ref{eq:495}) at $t=0$, and monitor their
periodicity.  On Fig.~\ref{fig:BCSPerturbativeStabilityTestTimeStep}
we show the results of such test performed for a time-periodic
solution bifurcating from fundamental mode with amplitude
$\ep=\left.\inner{e_{\gamma}}{B}\right|_{\tau=0}=0.05$ which was
constructed on a numerical grid with $24\times 64$ points.  Fixing the
order of time-integration algorithm (the $s=2$ stage Gauss-RK method
of order $4$) we varied the magnitude of integration step size (to
control the error of the time integration).  Since the initial
conditions contain some amount of error and the time-integration
itself gives only an approximate result and we do not observe any
signs of instability---still this may happen on a much larger scales.
The projections of a phase space, spanned by $\hat{B}_{i}(t)$ and
$\hat{\Pi}_{i}(t)$, show Lissajous curves.  These can be monitored
whether they close or not over one or more revolutions.  Decreasing an
error in the time-integration (by decreasing time step size) we
observe a solution to get closer to the periodic orbit.  Even for
relatively large step sizes, when the time-integration introduces
significant amount of numerical noise, the distance to the periodic
orbit stays bounded over considered time intervals.  The same is
observed on a total mass conservation plot.  But due to large rounding
errors we were not able to decrease the absolute error in total energy
below $10^{-13}$ by refining the time-integration method.\footnote{A
  possible source of such error sits in a way of computing the total
  energy (\ref{eq:491})-(\ref{eq:493}).}

\begin{figure}[!t]
  \centering
  \includegraphics[width=\swidth]
  {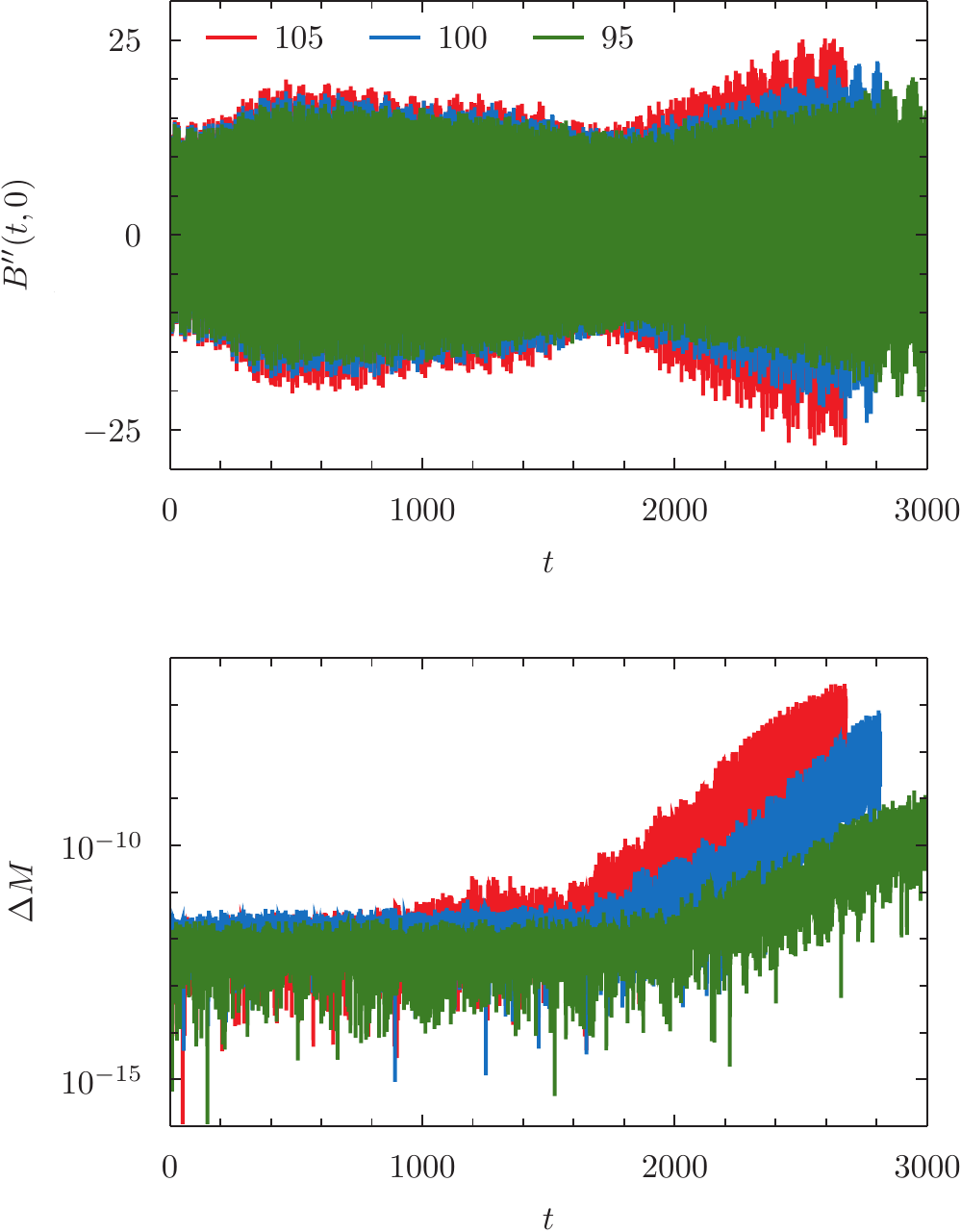}
  \caption{The long time evolution of a perturbed time-periodic
    solution ($\gamma=0$ case with $\ep=10$, right to the mass
    extremum, cf. bottom panel of
    Fig.~\ref{fig:BCSPeriodicOmegaMassNew}).  Curves are labelled by
    amplitude of initial momentum perturbation (\ref{eq:517}).
    \textit{Top panel}.  The Kretschmann scalar evaluated at the
    origin stays bounded for integrated intervals.  \textit{Bottom
      panel}.  The mass conservation error signals the onset of the
    instability roughly at the same time independently of the initial
    perturbation amplitude (see discussion in the text).}
  \label{fig:BCSEvolutionKretschmannE010}
\end{figure}

In order to investigate the issue of stability of constructed
time-periodic solutions we have considered the Cauchy problem with
initial conditions corresponding to perturbed time-periodic solution,
with perturbation in initial momenta given by
\begin{equation}
  \label{eq:517}
  \Pi(0,x) = \epsilon\,\frac{2}{\pi} \sin^2{x}
  \exp\left(-4\frac{\tan^2{x}}{\pi^2\sigma^2}\right),
\end{equation}
with $\sigma=2/25$.  We have chosen two representatives of the
fundamental family $\gamma=0$, namely those shown on
Fig.~\ref{fig:BCSPeriodicEvolutionE0Epsilon4Epsilon10} (with very
similar behaviour for the negative $\ep$ branch).  For perturbed
solution with $\ep<\ep_{\ast}^{+}$ the evolution is qualitatively the
same as we have seen in preceding sections.  For relatively small
amplitudes of initial perturbation solution stays within a bounded
distance from a periodic orbit.  This is illustrated for the $\ep=4$
and $\epsilon=70$ case on a series of parametric plots showing time
evolution of eigenbasis Fourier coefficients (\ref{eq:481}).  While
unperturbed trajectories (blue lines) stay periodic (up to numerical
errors which are under control---used numerical procedures are
consistent), the perturbed solution (gray lines) does not exhibit any
recurrences and stays in a bounded and finite distance (over simulated
times).

Remarkably this picture does not change as dramatically for solutions
with $\ep>\ep_{\ast}^{+}$ as it does for the EKG system, i.e. no
instant or delayed collapse of solutions $\ep>\ep_{\ast}^{+}$ is
observed.  On Fig.~\ref{fig:BCSEvolutionKretschmannE010} we plot the
time evolution of $B''(t,0)$ for time-periodic solution ($\ep=10$)
perturbed with Gaussian profiles of increasing amplitudes
$\epsilon=95, 100$ and $105$.  While for perturbations of larger
amplitudes ($\ep\gtrsim 115$) the focusing effect of gravity is strong
enough to focus the energy on sufficiently small scales and trigger
black hole formation in much shorten times than those shown on
Fig.~\ref{fig:BCSEvolutionKretschmannE010}. (In fact since we are
unable to perform evolution up to the black hole formation we identify
this process by noting continuous growth of the Kretschmann scalar at
the origin, by monitoring $B''(t,0)$.) For considered amplitudes the
evolutions stays smooth.  An observed modest growth of a Kretschmann
scalar for $t\gtrsim 1700$ in each of the runs, and connected with
that the increase of the mass conservation error, suggests an onset of
instability, of the unknown origin (whether this is a numerical
artifact, which we have not excluded at the time of writing, or just a
long time modulation being a real effect).\footnote{One possible
  explanation would be that, regardless of the size of the controlled
  perturbation, already the initial data corresponding to the
  time-periodic solution contains enough and significant error to
  trigger such growth.  In fact the $\ep=10$ solution was hardly
  derived using 'only' $48\times 80$ grid points, which noticeably is
  to small to accurately resolve its rough profile.}  Clearly, we
observe different behaviour than for unstable solutions of the EKG
system.  Therefore, if these solutions are unstable then their
unstable modes have small Lyapunov exponents.

\section{Spherical cavity model}
\label{sec:BoxPeriodic}

In this section we present and analyze procedures used to construct
time-periodic solutions of the system (\ref{eq:117})-(\ref{eq:117}).
The numerical procedure (Section~\ref{sec:BoxPeriodicNumeric}) is an
adaptation of Chebyshev pseudospectral spatial discretization, the
perturbative construction (Section~\ref{sec:BoxPeriodicPerturbative})
shows a different structure of the solutions (compared to AdS case).
The analysis of results, presented in
Section~\ref{sec:BoxPeriodicResults}, is restricted to the small
amplitude regime only and concentrates on verification and comparison
of these two approaches.

\subsection{Perturbative construction}
\label{sec:BoxPeriodicPerturbative}

We start the perturbative construction of time-periodic solutions to
the system (\ref{eq:118})-(\ref{eq:121}) by taking the following
ansatz
\begin{align}
  \label{eq:518}
  \phi(\tau,r;\ep) &= \sum_{\substack{\lambda\geq 1 \\
      \mathrm{odd}}}\ep^{\lambda}\phi_{\lambda}(\tau,r),
  \\
  \delta(\tau,r;\ep) &= \sum_{\substack{\lambda\geq 2 \\
      \mathrm{even}}}\ep^{\lambda}\delta_{\lambda}(\tau,r),
  \\
  A(\tau,r;\ep) &= 1 - \sum_{\substack{\lambda\geq 2 \\
      \mathrm{even}}}\ep^{\lambda}A_{\lambda}(\tau,r),
\end{align}
together with a time coordinate rescaling $\tau = \Omega\,t$, and the
perturbative $\ep$-expansion of the frequency
\begin{equation}
  \label{eq:519}
  \Omega(\ep) = \omega_{\gamma} +
  \sum_{\substack{\lambda\geq 2 \\ \mathrm{even}}}\ep^{\lambda}\xi_{\lambda}.
\end{equation}
We use the freedom we have in defining the perturbative parameter
$\ep$ and set it to the value of the second spatial derivative of a
scalar field at the origin at time $t=0$
\begin{equation}
  \label{eq:520}
  \phi''(0,0) = \ep.
\end{equation}
This choice is dictated by the ease of comparing the perturbative
results with the spectral code and its implementation used in
numerical construction.  As is discussed in Section~\ref{sec:BoxCheb},
in the spectral code we use the $\Phi$ field instead of $\phi$ itself
and since $\Phi$ vanishes at the origin (being the spatial derivative
of an even function) the natural choice is to control the second
derivative of $\phi$.  In this way, fixing $\gamma\in\mathbb{N}_{0}$
and specifying a real number $\ep$ we choose a single solution of the
one-parameter family of time-periodic solutions bifurcating from the
frequency $\omega_{\gamma}$.  The phase of solutions is fixed by the
requirement
$\left.\inner{e_{\gamma}}{\partial_{t}\phi}\right|_{t=0}=0$ (as in
previously considered models).

At the lowest (linear) order $\phi_{1}(\tau,r)$ has to be a solution
to the homogeneous wave equation (with $L$ defined in (\ref{eq:129}))
\begin{equation}
  \label{eq:BoxPertFirst}
  -\omega_{\gamma}^{2}\,\ddot{\phi}_{1} - L\phi_{1} = 0.
\end{equation}
The second order solution (the back-reaction on the metric) can be
easily written as the integrals
\begin{equation}
  \label{eq:BoxPertSecond}
  \begin{aligned}
    \delta_{2}(\tau,r) &= - \int_{0}^{r} s\left( \phi_{1}'(\tau,s)^{2}
      + \omega_{\gamma}^{2}\dot{\phi}_{1}(\tau,s)^{2} \right) \diff s,
    \\
    A_{2}(\tau,r) &= - \delta_{2}(\tau,r) +
    \frac{1}{r}\int_{0}^{r}\delta_{2}(\tau,s) \diff s.
  \end{aligned}
\end{equation}
At the third order one gets the inhomogeneous wave equation
\begin{multline}
  \label{eq:BoxPertThird}
  -\omega_{\gamma}^{2}\ddot{\phi}_{3} - L\phi_{3} = S_{3} =
  \bigl(A_{2}' + \delta_{2}'\bigr)\phi_{1}' +
  \omega_{\gamma}^{2}\bigl(\dot{A}_{2} + \dot{\delta}_{2}
  \bigr)\dot{\phi}_{1}
  \\
  + 2\omega_{\gamma}\bigl( \xi_{2} + \omega_{\gamma}\left( A_{2} +
    \delta_{2} \right) \bigr)\ddot{\phi}_{1}.
\end{multline}

Due to the incompatibility of eigenbasis functions (\ref{eq:130})
with the regularity conditions, for both reflecting boundary
conditions (\ref{eq:126}) and (\ref{eq:127}), the source function in
(\ref{eq:BoxPertThird}) cannot be written as a finite combination of
the eigenfunctions.  This makes the calculations much more involved
and these have to be done case by case for each $\gamma$ (half-full
automatization is still possible).  For that case we carry out only
the necessary calculations up to the third order to to demonstrate
main idea.  Despite the fact that the perturbative equations
(\ref{eq:BoxPertFirst})-(\ref{eq:BoxPertThird}) are independent of a
particular choice of the boundary condition, the construction
procedure is very different in each case, for that reason it is
considered separately.

\subsubsection{Dirichlet boundary condition}
\label{sec:BoxPeriodicPerturbative-Dirichlet}

\textit{A posteriori}, we know that for this choice of boundary
conditions the calculation are $\gamma$ dependent; in the presentation
we further restrict ourselves to the $\gamma=0$ case.  At the end of
this paragraph we comment on generalization to $\gamma\in\mathbb{N}$.

\begin{figure}[t]
  \centering
  \includegraphics[width=\swidth]
  {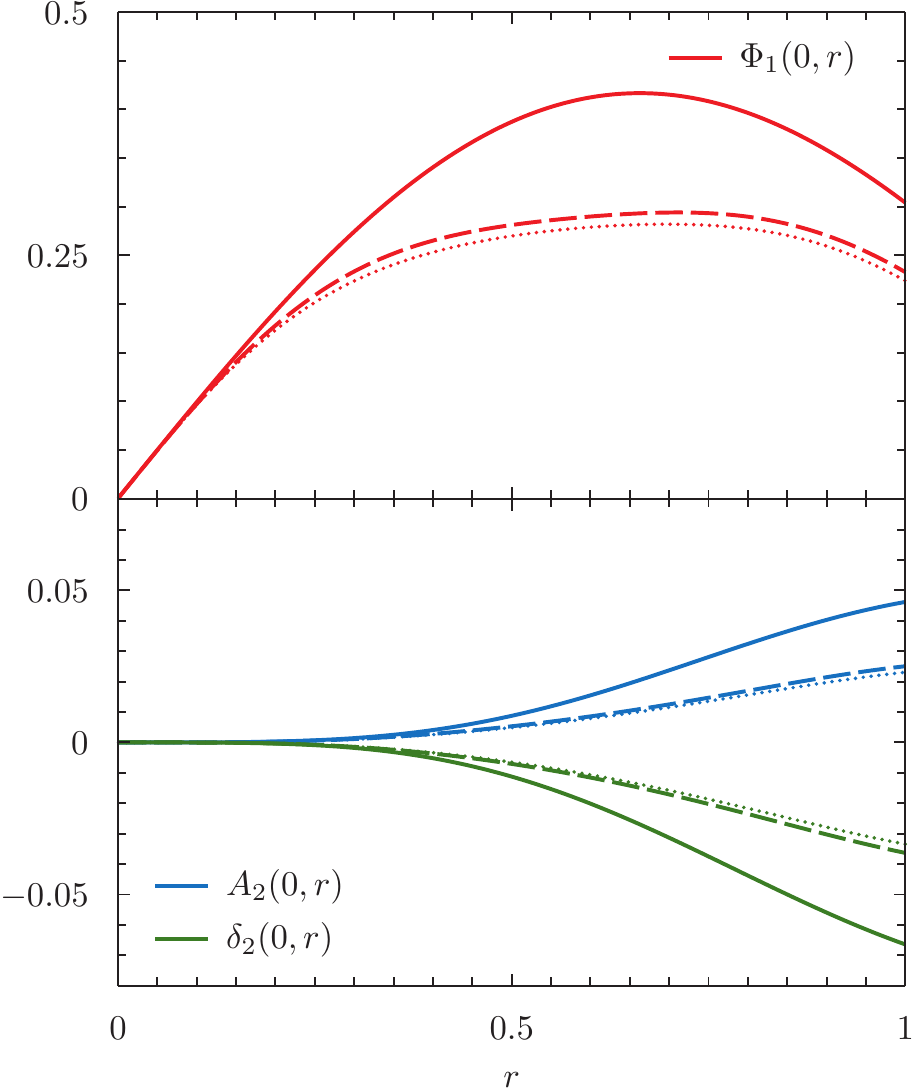}
  \caption{The perturbative profiles of the ground state ($\gamma=0$)
    time-periodic solution for Dirichlet boundary condition.  With
    different line types we plot the solution obtained by taking more
    terms in the first order solution $\phi_{1}(t,x)$
    (\ref{eq:BoxPertPhi1InfSum}) at $t=0$; solid line is just a single
    mode, with dashed line solution with two eigenmodes, dotted line
    denotes three mode approximation.  The convergence with the number
    of modes taken into account is clearly visible. \textit{Top
      panel}. The scalar field profile. \textit{Bottom panel.} The
    corresponding profiles of the metric functions.}
  \label{fig:BoxPertDirichlet}
\end{figure}

As for the solution to the linear wave equation
(\ref{eq:BoxPertFirst}) we take the single eigenmode (\ref{eq:130})
($\omega_{0}=\pi$)\footnote{Note the use of the sinc function
  $\sinc x=\frac{\sin x}{x}$.}
\begin{equation}
  \label{eq:521}
  \phi_{1}(\tau, r) = - \frac{3}{\omega_{0}^{2}}\cos{\tau}\,
  \sinc\left(\omega_{0}r\right),
\end{equation}
whose amplitude was set such that $\phi_{1}''(0,0)= 1$ holds.  Then,
it is easy to get the back-reaction (the second order solution) by
integrating (\ref{eq:BoxPertSecond}).  At the third order, equation
(\ref{eq:BoxPertThird}) has the following separable form
\begin{equation}
  \label{eq:522}
  -\omega_{\gamma}^{2}\ddot{\phi}_{3} - L\phi_{3} = S_{3,1}(r)\cos{\tau} +
  S_{3,3}(r)\cos{3\tau}.
\end{equation}
Using the orthogonality of cosine basis $\{\cos(i\tau)\,|\
i\in\mathbb{N}_{0}\}$ and setting
\begin{equation}
  \label{eq:523}
  \phi_{3}(\tau,r)=\phi_{3,1}(r)\cos{\tau} + \phi_{3,3}(r)\cos{3\tau},
\end{equation}
we reduce this PDE to a system of two independent second order ODEs
for $\phi_{3,1}(r)$ and $\phi_{3,3}(r)$.  These have to be solved with
the Dirichlet boundary condition at $r=1$ and regularity requirement
at $r=0$ (\ref{eq:520}).  The condition to meet the proper boundary
behaviour for the first of these ODEs fixes the second order frequency
correction parameter $\xi_{2}$
\begin{multline}
  \label{eq:524}
  \xi_{2} = \frac{9}{8\pi^4}\biggl[2\pi\bigl(-2\Ci(2\pi) - 5 +
  2\eulergamma + \log\left(4\pi^{2}\right)\bigr)
  \\
  + 10\Si(2\pi) - 5\Si(4\pi)\biggr] \simeq \num{0.06857053973},
\end{multline}
where $\eulergamma\simeq\num{0.577216}$ is the Euler-Mascheroni
constant \cite{Fun:EM}, $\Ci(z)$, $\Si(z)$ are the cosine
\cite{Fun:Ci} and sine \cite{Fun:Si} integral functions respectively,
together with one of the integration constants. The boundary condition
for $\phi_{3,3}(\tau,r)$ cannot be satisfied since only one of the
integration constants is fixed by imposing the Dirichlet condition at
$r=1$, while the regularity condition at $r=0$ is violated, even if
the second constant is still unspecified.  This is a direct
consequence of the presence of the resonance to the eigenmode
$e_{2}(r)$, i.e. the projection $\inner{e_{2}}{S_{3,3}}$ does not
vanish.  This is in contrast to the AdS case, where such cancellations
are present.  To overcome this difficulty we modify the ansatz for the
first order solution (\ref{eq:521}), and instead of just single mode
we take, in advance, an infinite combination of all of the eigenmodes
whose frequencies $\omega_{k}$ satisfy the condition
\begin{equation}
  \label{eq:525}
  \frac{\omega_{k}}{\omega_{0}} = 2i+1, \quad i,k\in\mathbb{N},
\end{equation}
explicitly
\begin{equation}
  \label{eq:BoxPertPhi1InfSum}
  \phi_{1}(\tau,r) = \sum_{i\geq 0}\hat{\phi}_{1,i}\cos\left((2i+1)\tau\right)
  \sinc\left(\omega_{2i}r\right),
\end{equation}
(we still assume that at the lowest order the frequency is
$\Omega=\omega_{0} + \mathcal{O}\left(\ep^{2}\right)$).  Note that
$k$'s fulfilling (\ref{eq:525}) is the subset of $O_{\gamma=0}$
(cf. (\ref{eq:27})).  In this way we introduce additional parameters,
the eigenmode amplitudes $\left\{\hat{\phi}_{1,i}\right\}_{i\geq 0}$,
into our procedure.  These will be used, together with $\xi_{2}$, to
remove all appearing resonances (or equivalently, to satisfy
regularity and boundary conditions) present at the third order.  By
including subsequent eigenmodes in (\ref{eq:BoxPertPhi1InfSum}) we
generate more resonances at the third order, so in fact the number of
terms in (\ref{eq:BoxPertPhi1InfSum}) needs to be infinite.  Such
modification of the first order solution will lead to the
inhomogeneous PDE with the source term composed of an infinite number
of Fourier modes (an analogous of (\ref{eq:522}))
\begin{equation}
  \label{eq:526}
  -\omega_{\gamma}^{2}\ddot{\phi}_{3} - L\phi_{3} =
  \sum_{i\geq 0}S_{3,2i+1}(r)\cos((2i+1)\tau).
\end{equation}
Then, the absence of resonances enforces an infinite system of
algebraic equations
\begin{equation}
  \label{eq:527}
  \inner{e_{2i+1}}{S_{3,2i+1}} = 0, \quad i\in\mathbb{N}_{0},
\end{equation}
for $\left\{\hat{\phi}_{1,i}\right\}_{i\geq 0}$ and $\xi_{2}$, with an
additional normalization condition (\ref{eq:520})
\begin{equation}
  \label{eq:528}
  \phi_{1}''(0,0) =
  - \frac{1}{3}\sum_{i\geq 0}\omega_{i}^{2}\hat{\phi}_{1,i} = 1.
\end{equation}
In practice, since we do not have tools (as for the AdS case in even
space dimensions) to produce and manipulate effectively the higher
order equations, and in particular to determine the conditions
(\ref{eq:527}), we truncate the sum in (\ref{eq:BoxPertPhi1InfSum}) at
some $i=i_{\text{max}}$.  To demonstrate that this is the proper way
to solve this problem we show the steps for $i_{\text{max}}=1$.
Approximating $\phi_{1}(\tau,r)$ by taking
\begin{equation}
  \label{eq:529}
  \phi_{1}(\tau, r) =
  \hat{\phi}_{1,0}\cos{\tau}\,\sinc\left(\pi r\right) +
  \hat{\phi}_{1,1}\cos{3\tau}\,\sinc\left(3\pi r\right),
\end{equation}
we find the $\delta_{2}(\tau,r)$, $A_{2}(\tau,r)$ and compute the
source in the wave equation (\ref{eq:BoxPertThird}).  Then we solve
the algebraic equations
\begin{equation}
  \label{eq:530}
  \begin{aligned}
    \inner{e_{0}}{S_{3,1}} &= 0,
    \\
    \inner{e_{2}}{S_{3,3}} &= 0,
  \end{aligned}
\end{equation}
(which are linear in $\xi_{2}$ and cubic in $\hat{\phi}_{1,0}$) and
get a nontrivial, real and unique solution
\begin{equation}
  \label{eq:531}
  \hat{\phi}_{1,0} \simeq \num{-0.2146299761}, \quad
  \xi_{2} \simeq \num{0.03440105174},
\end{equation}
(which is given by a lengthy formula in terms of $\Si(z)$ function, so
we do not present it here).  In addition to the $e_{0}(r)$ and
$e_{2}(r)$ eigenmode resonances (which are removed by imposing the
condition (\ref{eq:530})) there are higher source projections which
cannot be set to zero, since we do not have a sufficient number of
free parameters as was mentioned earlier.  But, as was seen already,
the coefficients in (\ref{eq:529}) are rapidly decreasing with their
mode index, so truncating the series (\ref{eq:BoxPertPhi1InfSum}) with
moderate number of terms would produce partial but sufficient
approximation to a time-periodic solution see
Fig.~\ref{fig:BoxPertDirichlet} (accurate up to the $\ep^{2}$ order
term with an approximate $\phi_{1}$ and back-reaction $\delta_{2}$ and
$A_{2}$).  To sum up, taking more terms in the initial sum
(\ref{eq:BoxPertPhi1InfSum}), one gets successively better
approximation to the first order solution $\phi_{1}(\tau,r)$ which is
determined by the lack of resonances at third order, at the same time
removes more and more resonances (among infinite number of them)
present in $S_{3}$.

Therefore, cancellation of all of the resonances is possible only if
$\phi_{1}(\tau,r)$ is a very special linear combination of an infinite
number of the eigenmodes,\footnote{This is also the case for the cubic
  NLW on a circle \cite{Khrustalev2001239, 10.1007/BF02557130}.} among
which the $e_{\gamma}(r)$ mode has the highest absolute value of the
amplitude relative to amplitudes of the other eigenmodes and the
frequency of time-periodic solution is
$\Omega(\ep=0)=\omega_{\gamma}$.  For a general
$\gamma\in\mathbb{N}_{0}$ the ansatz for the lowest order solution
$\phi_{1}(t,x)$ generalizing (\ref{eq:BoxPertPhi1InfSum}) would be
\begin{equation}
  \label{eq:532}
  \phi_{1}(\tau,r) = \sum_{i\geq 0}\hat{\phi}_{1,i}\cos\left((2i+1)\tau\right)
  \sinc\left(\omega_{(2i+1)(\gamma+1)-1}r\right),
\end{equation}
with parameters $\left\{\hat{\phi}_{1,i}\right\}_{i\geq 0}$ determined
at the third order in a similar way as for the fundamental mode
$\gamma=0$.
\subsubsection{Neumann boundary condition}
\label{sec:BoxPeriodicPerturbative-Neumann}

For the Neumann boundary condition, the dispersive case
(\ref{eq:133}), there is only one resonance present at each odd
perturbative order $\lambda\geq 3$.  This is due to the fact that
$O_{\gamma}=\{\gamma\}$, i.e. the equation
\begin{equation}
  \label{eq:533}
  \frac{\omega_{j}}{\omega_{\gamma}}=m, \quad \gamma,j\in\mathbb{N}_{0},
  \ \ m\in\mathbb{N},
\end{equation}
for eigenfrequencies (\ref{eq:133}) has only trivial solution
$j=\gamma$ and $m=1$ for any $\gamma$.\footnote{This is easy to show
  by the contradiction.  If we assume that (\ref{eq:533}) holds then
  from the definition (\ref{eq:133}) we get
  $\tan(m\omega_{\gamma})=m\tan{\omega_{\gamma}}$.  This condition can
  be reduced, using a trigonometric identity
  $\tan(n+1)x=(\tan{nx}+\tan{x})/(1-\tan{nx}\tan{x})$ for
  $n\in\mathbb{Z}$, to an algebraic equation whose root
  $\omega_{\gamma}$ for $m\geq 2$ is an algebraic number.  This
  contradicts the assumption that $\omega_{\gamma}$ is a
  transcendental number.}  Therefore this single resonance can be
removed be setting the value of a free parameter $\xi_{\lambda-1}$.
All of the integration constants are fixed by the boundary condition
$\phi_{\lambda}(\tau,1)=0$ and the normalization condition
(\ref{eq:520}).  Moreover, since the form of the basis functions
(\ref{eq:130}) is independent on $\gamma$, the construction can be
performed without specifying $\gamma$.

\begin{figure}[!t]
  \centering
  \includegraphics[width=\swidth]{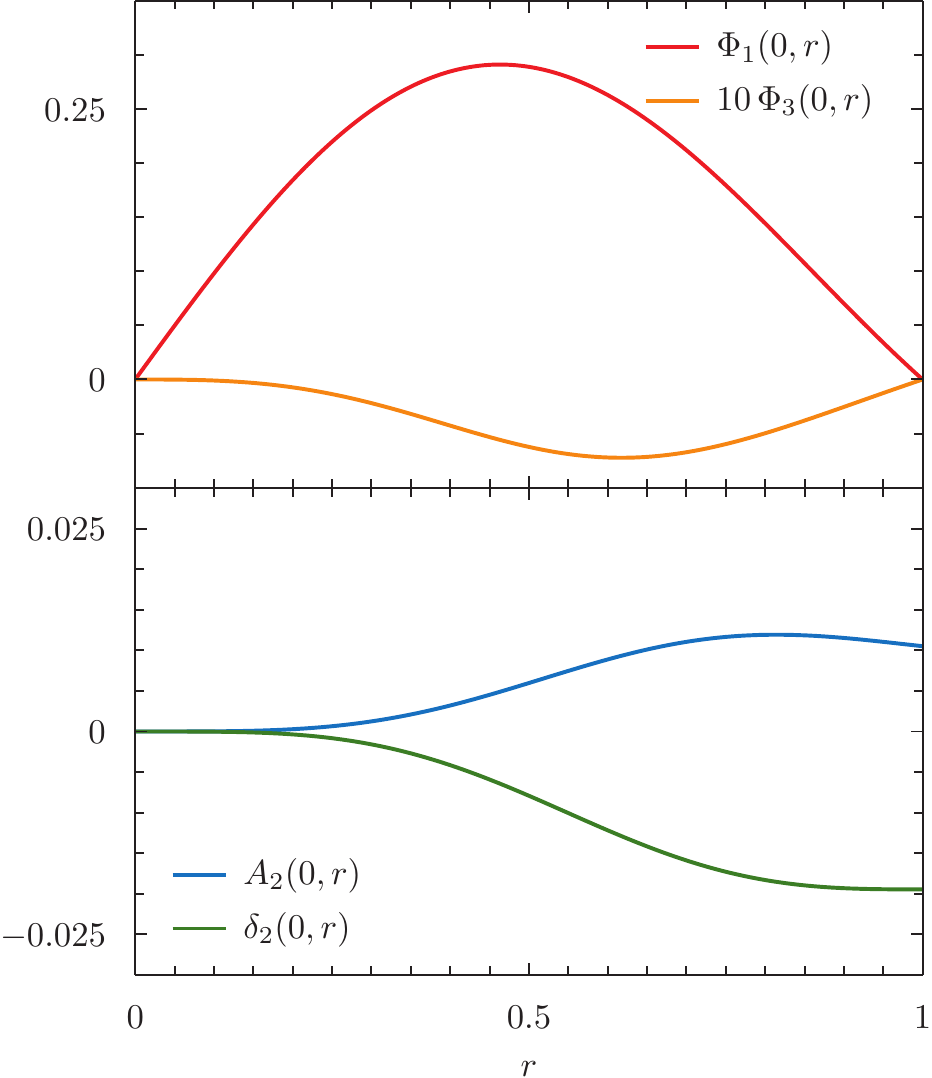}
  \caption{The perturbative profiles of the ground state ($\gamma=0$)
    time-periodic solution for Neumann boundary condition at
    $t=0$. \textit{Top panel}. The scalar field
    profiles. \textit{Bottom panel.} The corresponding metric
    functions.}
  \label{fig:BoxPertNeumann}
\end{figure}

For time-periodic solution with frequency bifurcating from
$\omega_{\gamma}$ as first order approximation we take an analogous of
(\ref{eq:521})
\begin{equation}
  \label{eq:534}
  \phi_{1}(\tau, r) = - \frac{3}{\omega_{\gamma}^{2}}\cos{\tau}\,
  \sinc\left(\omega_{\gamma}r\right),
\end{equation}
(with $\omega_{\gamma}$ given in Tab.~\ref{tab:BoxEigenval}) and
calculate the integrals (\ref{eq:BoxPertSecond}).  Next, since the
wave equation at order $\lambda=3$ (\ref{eq:BoxPertThird}), has
exactly the same structure as for the resonant case, i.e.
(\ref{eq:522}), we follow the same steps as in the previous paragraph.
Both of the solutions $\phi_{3,1}(\tau,r)$ and $\phi_{3,3}(\tau,r)$,
will contain two integration constants.  An additional parameter will
be the frequency correction $\xi_{2}$.  These constants we fix in a
following way.  The regularity conditions at the origin and the
Neumann condition at the cavity will uniquely determine the function
$\phi_{3,3}(\tau,r)$.  Imposing the same conditions on the
$\phi_{3,1}(\tau,r)$ function will fix one integration constant and
the $\xi_{2}$ parameter
\begin{multline}
  \label{eq:535}
  \xi_{2} = -\frac{9}{64\omega_{\gamma }^7} \left(\cos \left(2 \omega
      _{\gamma }\right)-5\right) + \frac{9}{64 \omega _{\gamma }^5}
  \left(9 \cos \left(2 \omega _{\gamma }\right)-4
    \csc^2\omega_{\gamma} - 13\right)
  \\
  +\frac{9}{32\omega_{\gamma}^3} \left(16 \log \left(2 \omega _{\gamma
      }\right)-11 \cos \left(2 \omega _{\gamma }\right)+3
    \csc^2\omega_{\gamma} - 16 \text{Ci}\left(2 \omega _{\gamma
      }\right)+16 \eulergamma -54\right)
  \\
  +\frac{45}{8\omega_{\gamma}^4} \csc^2\omega_{\gamma} \left(2
    \text{Si}\left(2 \omega _{\gamma }\right)-\text{Si}\left(4
      \omega_{\gamma}\right)\right).
\end{multline}
(see Tab.~\ref{tab:BoxXi2Neumann} for numerical values of $\xi_{2}$).
The remaining integration constant (multiplying the
$\inner{e_{\gamma}}{\phi_{3,1}}$ term) is a free parameter related to
the freedom to define the expansion parameter $\ep$, which we fix by
imposing the condition $\phi_{3}''(0,0)=0$.  In this way we obtain the
$\mathcal{O}\left(\ep^{3}\right)$ accurate approximation to the
time-periodic solution with $\gamma$ being the only parameter, in the
nonresonant case, see Fig.~\ref{fig:BoxPertNeumann}.  It is
straightforward to continue this construction with the higher order
approximation in $\ep$, with the only (but very serious) limitation
that the computations are much more involved and the resulting lengthy
formulas one gets are fairly complicated (at least as generated and
simplified by the \mathematica).  The analysis and manipulation of
these is cumbersome (they are not as easy to analyze as in the
AdS$_{d+1}$ case with $d$ even), which makes it hard to give some
general statements about the obtained time-periodic solutions.  For
this reason we stop this procedure at order $\ep^{3}$ and limit the
analysis to the comparison with numerically constructed solutions.
\begin{table}[!t]
  \centering
  \begin{tabular}[h]{c|ccccc}
    \toprule
    $\gamma$ & 0 & 1 & 2 & 3 & $\cdots$ \\ \midrule
    $\xi_{2}$ & \num{0.0319852} & \num{0.0101963} & \num{0.00460518} &
    \num{0.00249952} & $\cdots$ \\
    \bottomrule
  \end{tabular}
  \caption{The numerical values of the frequency corrections
    (\ref{eq:535}) of time-periodic solutions for the Neumann
    boundary condition $\phi'(t,1)=0$ case bifurcating from the
    eigenmode $e_{\gamma}(r)$.}
  \label{tab:BoxXi2Neumann}
\end{table}

\subsection{Numerical construction}
\label{sec:BoxPeriodicNumeric}

We construct the time-periodic solutions of the system
(\ref{eq:118})-(\ref{eq:121}) as follows.  We expand both functions
$\Phi(t,r)$ and $\Pi(t,r)$ in Fourier series in time and use the
following truncated approximation
\begin{align}
  \label{eq:536}
  \Phi(\tau,r) & =
  \sum_{k=0}^{K-1}\cos\left((2k+1)\tau\right)\hat{\Phi}_{k}(r),
  \\
  \label{eq:537}
  \Pi(\tau,r) &=
  \sum_{k=0}^{K-1}\sin\left((2k+1)\tau\right)\hat{\Pi}_{k}(r),
\end{align}
where we use the rescaled time coordinate $\tau=\Omega\,t$, with
$\Omega$ the frequency of the solution we are looking for.  Instead of
expanding the Fourier coefficients $\hat{\Phi}_{k}(r)$,
$\hat{\Pi}_{k}(r)$ in Chebyshev basis (we do not expand them in the
eigenbasis of the linear problem (\ref{eq:129}) since these
eigenfunctions do not have correct boundary expansion, see
Section~\ref{sec:BoxEigenvalue}, which would cause inefficient
polynomial decay of the expansion coefficients in that case) we use
the nodal representation, i.e. we operate on the function values on
the grid points $\hat{\Phi}_{ki}\equiv\hat{\Phi}_{k}(r_{i})$ and
$\hat{\Pi}_{ki}\equiv\hat{\Pi}_{k}(r_{i})$ as described in
Section~\ref{sec:BoxCheb} and Section~\ref{sec:AdSEvolutionCheb}.

With $N$ radial Chebyshev collocation points~(\ref{eq:635}) in space,
and $K$ collocation points in time $\tau_k = \pi(k-1/2)/K$,
$k=1,\ldots,K$, at each instant of time $\tau_k$ we calculate values
of the fields $\Phi(\tau_{k},r_{i})$ and $\Pi(\tau_{k},r_{i})$ at grid
points $r_{i}$ and use them as an input in our time evolution
procedure, getting as the output their time derivatives.  Equating
those to the time derivatives of (\ref{eq:536}) and (\ref{eq:537})
(remembering that $\partial_t = \Omega\,\partial_{\tau}$) at the set
of $K \times N$ Cartesian product grid points $(\tau_k, r_i)$,
together with the additional equation
\begin{equation}
  \label{eq:538}
  \Phi'(0, 0) = \ep,
\end{equation}
setting the center value of the dominant mode $\gamma$ in the initial
data to $\ep$, we get a nonlinear system of $2 \times K \times N + 1$
equations for $2 \times K \times N + 1$ unknowns: $\hat{\Phi}_{k,i}$,
$\hat{\Pi}_{k,i}$ and $\Omega$.  This system is solved with the
Newton-Raphson algorithm yielding the time-periodic solution of the
system (\ref{eq:118})-(\ref{eq:120}) (the corresponding geometry of
space-time given by metric functions $\delta(t,r)$ and $A(t,r)$ can be
determined, at each instant of time, by solving the constraint
equations (\ref{eq:119}) and (\ref{eq:120}) for the time-periodic data
(\ref{eq:536}) and (\ref{eq:537}) at each constant time slice).

As a starting point for the numerical root-finding algorithm we choose
the single eigenmode approximation, fulfilling the condition
(\ref{eq:538}), i.e. we set
\begin{align}
  \label{eq:539}
  \Phi(\tau,r) &= \ep\cos\tau\left(
    -\frac{3}{\omega_{\gamma}^{2}}\sinc'\left(\omega_{\gamma}r\right)\right),
  \\
  \Pi(\tau,r) &= \ep\sin\tau\left(
    \frac{3}{\omega_{\gamma}}\sinc\left(\omega_{\gamma}r\right)
  \right),
  \\
  \Omega &= \omega_{\gamma},
\end{align}
while looking for solution bifurcating from eigenmode $\gamma$.  Such
initial conditions, provide a good enough approximation so that the
Newton algorithm converges relatively fast even for moderate values of
amplitudes $\ep$.

We use exactly the same methods (with the same solution representation
and initial conditions) for both boundary conditions, Dirichlet and
Neumann, with only minor modification within the code, in the part
calculating the RHS of the wave equation (\ref{eq:118}) (in this case
the time derivatives of (\ref{eq:536}) and (\ref{eq:537})) as is
described in Section~\ref{sec:BoxCheb}.

\subsection{Results}
\label{sec:BoxPeriodicResults}

\begin{table}[!t]
  \centering
  \begin{tabular}{cllll} \toprule
    & \multicolumn{3}{c}{$i_{\text{max}}$} \\ \cmidrule(r){2-4}
    & \multicolumn{1}{c}{0} & \multicolumn{1}{c}{1} &
    \multicolumn{1}{c}{2} & \multicolumn{1}{c}{fit}
    \\ \midrule
    $\inner{e_{0}'}{\phi_{1}'}$ & $\num{-0.214935}$ & $\num{-0.158182}$ &
    $\num{-0.151766}$ & $\num{-0.150903}$
    \\
    $\inner{e_{2}'}{\phi_{1}'}$ & $\phantom{-}0$ & $\num{-0.00630589}$ &
    $\num{-0.00605718}$ & $\num{-0.00602276}$
    \\
    $\inner{e_{4}'}{\phi_{1}'}$ & $\phantom{-}0$ & $\phantom{-}0$ &
    $\num{-0.000346149}$ & $\num{-0.000344889}$
    \\
    $\xi_{2}$ & $\phantom{-}\num{0.0685705}$ &
    $\phantom{-}\num{0.0373695}$ & $\phantom{-}\num{0.0344011}$ &
    $\phantom{-}\num{0.0340109}$
    \\
    \bottomrule
  \end{tabular}
  \caption{The convergence of perturbative approximation of
    $\phi_{1}(\tau,r)$ evaluated at $\tau=0$ and frequency expansion
    parameter $\xi_{2}$ to the numerical data with increasing $i_{max}$
    in (\ref{eq:BoxPertPhi1InfSum}).}
  \label{tab:BoxDirichlet01Comp}
\end{table}

\begin{figure}[!t]
  \centering
  \includegraphics[width=\swidth]{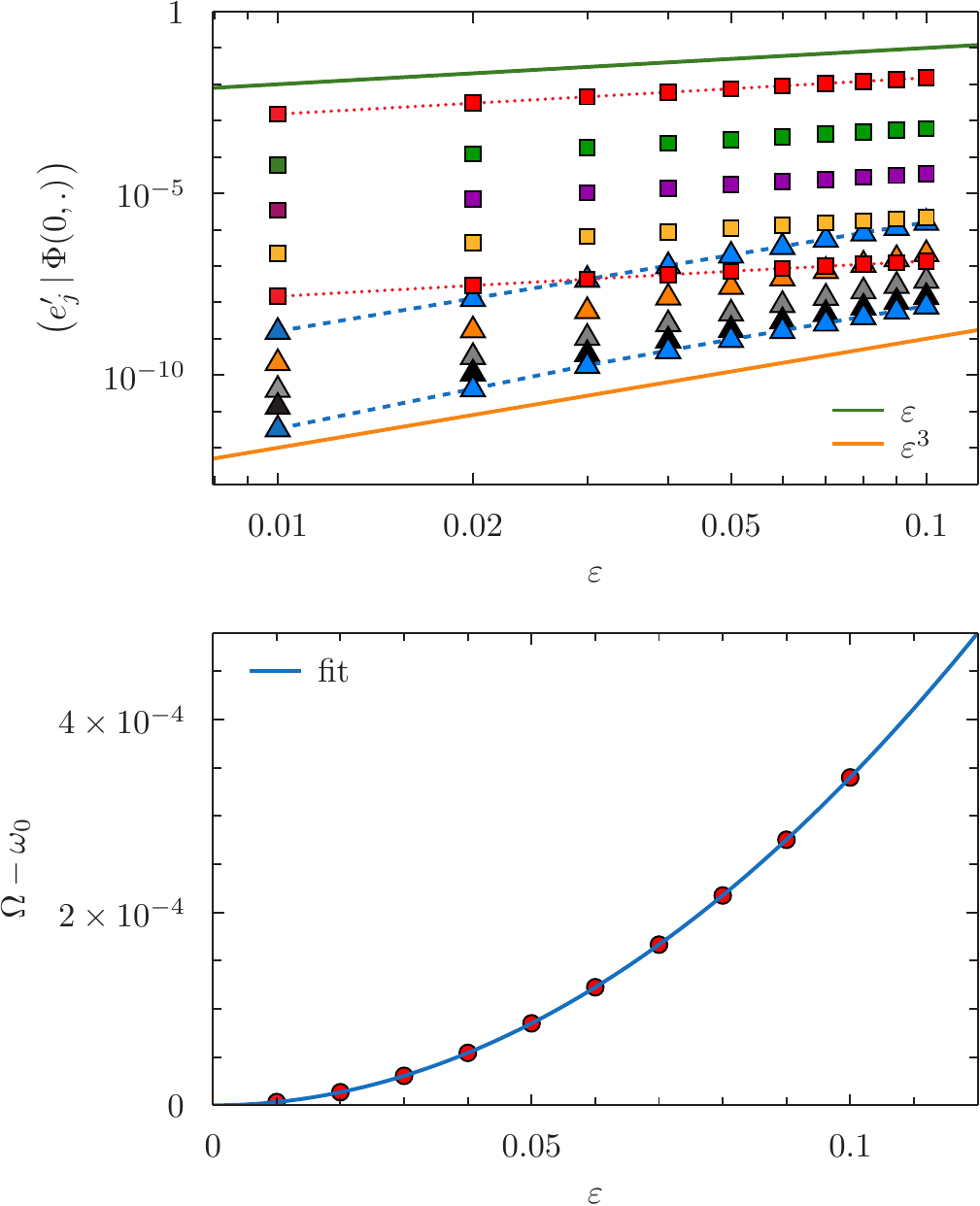}
  \caption{The numerical results showing time-periodic solutions
    bifurcating from fundamental ($\gamma=0$) eigenmode for Dirichlet
    boundary condition at cavity. \textit{Top panel.} The log-log plot
    showing the polynomial scaling of initial data $\Phi(0,x)$
    projections onto $e_{j}'(r)$ with $\ep$.  The data shows
    $\inner{e_{2j}'}{\Phi(0,\,\cdot\,)}\sim\ep$ (squares) and
    $\inner{e_{2j+1}'}{\Phi(0,\,\cdot\,)}\sim\ep^{3}$ (triangles) as
    predicted by perturbative calculation
    ($j\in\mathbb{N}_{0}$).  \textit{Bottom panel.} The frequency of
    the time-periodic solution as a function of $\ep$ with a
    polynomial fit (see Tab.~\ref{tab:BoxDirichlet01Comp}).}
  \label{fig:BoxNumericCoeffScalingDir01}
\end{figure}

We restrict the presentation of results to a comparison of our
perturbative construction with the numerical approach for small
amplitude solutions only.  Further studies should be devoted to large
amplitude solutions and their stability properties (as for in AdS).

To verify results of our methods we look at the profile of the scalar
field at $t=0$ (in fact we can make the comparison for any instant of
time, but taking $t=0$ is natural, since at that moment all eigenmodes
have equal phase since $\Pi(0,r)\equiv 0$ and this is also
computationally straightforward).  Instead of performing comparison in
the physical space, it is convenient to do this in a Fourier space.
Therefore we read off the initial data for the $\Phi(t,r)$ field
resulting from (\ref{eq:536}).  Then we project this function onto the
$e_{j}'(r)$ modes (the necessary integrals were computed numerically
using the Gauss-Legendre quadrature, see the
Appendix~\ref{sec:legendre-polynomials}), and using the relation
(\ref{eq:134}) we get the decomposition of initial data for the
$\phi'(0,r)$ function.  Repeating this procedure for the time-periodic
solutions with different values of the $\ep$ parameter (\ref{eq:538}),
we can perform a fit to the numerical data to get the coefficients of
polynomial dependence of the expansion coefficients
$\inner{e_{j}'}{\phi'(0,\,\cdot\,)}$ on the amplitude $\ep$.

\begin{figure}[!t]
  \centering
  \includegraphics[width=\swidth]{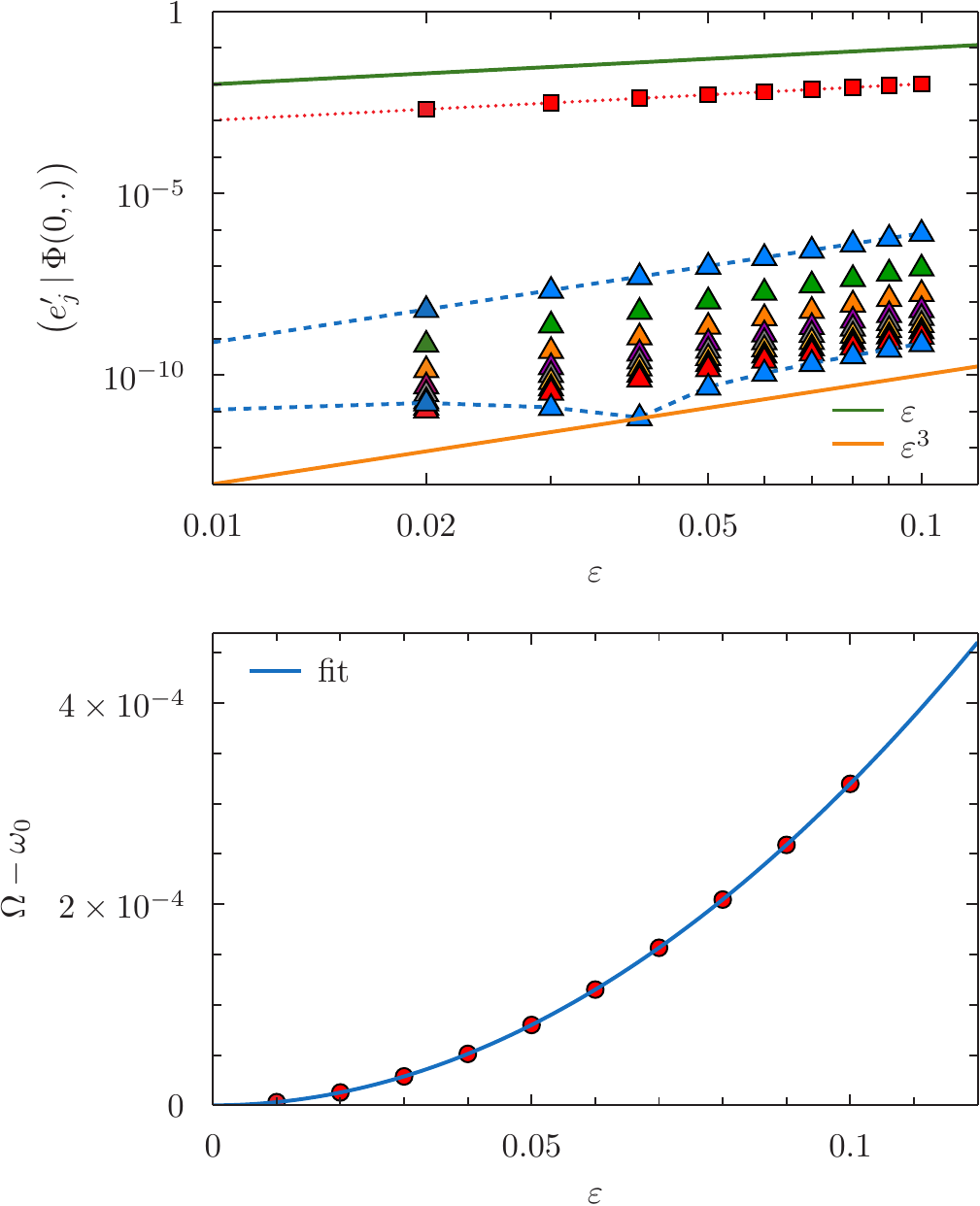}
  \caption{An analogue of Fig.~\ref{fig:BoxNumericCoeffScalingDir01}
    for Neumann boundary condition at cavity. \textit{Top panel.} The
    log-log plot showing the polynomial scaling of initial data
    $\Phi(0,x)$ projections onto $e_{j}'(r)$ with $\ep$.  The data
    shows $\inner{e_{0}'}{\Phi(0,\,\cdot\,)}\sim\ep$ (squares) and
    $\inner{e_{j}'}{\Phi(0,\,\cdot\,)}\sim\ep^{3}$, $j\in\mathbb{N}$
    (triangles) as predicted by perturbative
    calculation.  \textit{Bottom panel.} The frequency of the
    time-periodic solution as a function of $\ep$ with a polynomial
    fit (\ref{eq:540}).}
  \label{fig:BoxNumericCoeffScalingNeu01}
\end{figure}
\begin{table}[!t]
  \centering
  \begin{equation*}
    \begin{array}{ccl}
      \toprule
      j & \inner{e_{j}'}{\phi_{3}'} & \multicolumn{1}{c}{$\text{fit}$}
      \\ \midrule
 1 & -7.99298\times 10^{-4} & -7.992982\times 10^{-4} \\
 2 & -8.87371\times 10^{-5} & -8.873713\times 10^{-5} \\
 3 & \phantom{-}1.77219\times 10^{-5} & \phantom{-}1.772189\times 10^{-5} \\
 4 & \phantom{-}6.34559\times 10^{-6} & \phantom{-}6.345592\times 10^{-6} \\
 5 & -3.87437\times 10^{-6} & -3.87437\times 10^{-6} \\
 6 & \phantom{-}2.46376\times 10^{-6} & \phantom{-}2.463799\times 10^{-6} \\
 7 & -1.65876\times 10^{-6} & -1.656686\times 10^{-6} \\
 8 & \phantom{-}1.17002\times 10^{-6} & \phantom{-}1.223907\times 10^{-6} \\
 9 & -8.56511\times 10^{-7} & -1.01226\times 10^{-7} \\
 \bottomrule
    \end{array}
  \end{equation*}
  \caption{The comparison of perturbativelly and numerically obtained
    time-periodic fundamental solution $\gamma=0$ for the Neumann
    boundary condition.  The Fourier coefficients of the third order
    term of perturbative series (\ref{eq:518}) at $t=0$ are compared
    with the fitting to the numerical data.  The numerical solutions
    where determined with $N=16$, $K=24$ modes (points).}
  \label{tab:BoxPertNumCompNum}
\end{table}

For the Dirichlet boundary condition and the fundamental solution
($\gamma=0$) the results are summarized in the
Tab.~\ref{tab:BoxDirichlet01Comp}.  These results are in excellent
agreement, bearing in mind that for the first order solution
(\ref{eq:BoxPertPhi1InfSum}) we have taken only first few terms (up to
$i_{\text{max}}=2$).  Moreover, from the numerical data we get that
the coefficients $\inner{e_{2j}'}{\phi'(0,\,\cdot\,)}$,
$j\in\mathbb{N}_{0}$ scale (in the leading order) as $\ep$, while the
remaining modes behave as
$\inner{e_{2j+1}'}{\phi'(0,\,\cdot\,)}\sim \ep^{3}$,
$j\in\mathbb{N}_{0}$, see Fig.~\ref{fig:BoxNumericCoeffScalingDir01}.
This supports our assumption about the form of the $\phi_{1}(\tau,r)$
solution (\ref{eq:BoxPertPhi1InfSum})).  From this comparison it is
evident that including more terms in (\ref{eq:BoxPertPhi1InfSum}) will
lead to the successive decrease of coefficient $\hat{\phi}_{1,0}$ and
its convergence to the true value, while $\hat{\phi}_{1,i>0}$ will
have smaller but growing (in absolute value) numerical values with the
number of terms in (\ref{eq:BoxPertPhi1InfSum}).  This agreement is
strengthened when we compare frequencies of the solutions
Fig.~\ref{fig:BoxNumericCoeffScalingDir01}.  Fitting an even
polynomial in $\ep$ to the numerical data $\Omega(\ep)$ we read off
the quadratic term coefficient which has to be compared with the
frequency correction $\xi_{2}$ obtained by solving the algebraic
equations (see Tab.~\ref{tab:BoxDirichlet01Comp}).

These results validate our perturbative procedure for this model, in
particular the form of the first order term in the perturbative
expansion, given by the infinite sum (\ref{eq:BoxPertPhi1InfSum}),
together with the scheme which gives the unique values of the
coefficients $\left\{\hat{\phi}_{1,i}\right\}_{i\geq 0}$ and the
frequency expansion coefficient $\xi_{2}$.

For the Neumann case, we can perform similar comparison as for the
Dirichlet boundary condition.  However, having an exact form of
$\phi_{1}(\tau,r)$ and a third order solution $\phi_{3}(\tau,r)$
determined, we can perform more detailed comparison of perturbative
and numerical results (we limit the presentation to the $\gamma=0$
case).  First of all, we get the following scaling of the Fourier
coefficients $\inner{e_{0}}{\phi'(0,\,\cdot\,)}\sim\ep$ and
$\inner{e_{j}}{\phi'(0,\,\cdot\,)}\sim\ep^{3}$, for $j\in\mathbb{N}$
(see Fig.~\ref{fig:BoxNumericCoeffScalingNeu01}).  That is, the first
order term (\ref{eq:518}) consist of only one eigenmode, the one with
index $\gamma=0$, while $\phi_{3}(\tau,r)$ is a linear combination of
all eigenmodes.  Secondly, we get good agreement of the numerical
values of the Fourier coefficients which are shown in
Tab.~\ref{tab:BoxPertNumCompNum}.  For perturbative solution the
projections are exact numbers, while for the numerical solution we
have performed the least square fitting procedure.  Additionally,
fitting a polynomial function to the $\Omega(\ep)$ numerical data we
get
\begin{equation}
  \label{eq:540}
  \Omega(\ep) \approx \omega_{0} + \num{0.0319852}\ep^{2},
\end{equation}
which agrees with an exact perturbative frequency correction
(\ref{eq:535}) listed in Tab.~\ref{tab:BoxXi2Neumann} (the absolute
value of relative error is $4\times 10^{-11}$).  The same quality of
agreement we obtain when we analyze results for excited solutions
$\gamma>0$.

\section{Yang-Mills on Einstein Universe}
\label{sec:YMPeriodic}

In this section we continue studies of the YM system, initiated in
Section~\ref{sec:YMTurbulence}.  After discussing perturbative
construction of time-periodic solutions
(Section~\ref{sec:YMPeriodicPerturbative}) we review the numerical
method (Section~\ref{sec:YMPeriodicNumerical}).  Since the
perturbative analysis, in contrary to numerical procedure, differs
when we consider perturbations around different static solutions (in
different topological sectors) both construction and results analysis
is considered separately.  In Section~\ref{sec:YMPeriodicResults} we
present the results and verify them by comparing numerical and
perturbative construction.

\subsection{Perturbative construction}
\label{sec:YMPeriodicPerturbative}

The perturbative construction of time-periodic solutions is less
involved for the YM model when compared to previously analyzed
systems, since here we deal with a single PDE.  As in previous cases
this construction is based on the Poincar\'e-Lindstedt method,
therefore the initial steps we follow here are the same as taken in
Section~\ref{sec:poinc-lindst-appr}, where the perturbative
construction of single mode initial conditions where presented.  We
introduce new time coordinate through (\ref{eq:213}) and assume series
expansion in small parameter $\ep$ ($0<\left|\ep\right|\ll 1$) for
solution profile $u$ and corresponding frequency $\Omega$, as in
(\ref{eq:215}) and (\ref{eq:216}) respectively.  The Taylor series
expansion in $\ep$ of the Eq.~(\ref{eq:214}) with (\ref{eq:215}) and
(\ref{eq:216}) substituted gives the perturbative second order linear
PDEs for $u_{\lambda}(\tau,x)$.  The first four of them are given
explicitly in (\ref{eq:217})-(\ref{eq:220}); the higher order
equations contain much more complicated source terms.  As in
proceeding sections $\gamma$ refers to an index of dominant mode in
the solution, and since we are looking for bifurcating solutions we
assume for the leading order solution
\begin{equation}
  \label{eq:541}
  \lim_{\ep\rightarrow 0} \frac{1}{\ep}u(\tau,x;\ep) =
  u_{1}\left(\Omega(0)\,t,x\right) =
  \cos(\omega_{\gamma}t)\,e_{\gamma}(x), \quad   \Omega(0) = \omega_{\gamma},
\end{equation}
(because of used convention here $\gamma\in\mathbb{N}$).  We define
the perturbative parameter to be the amplitude of the $e_{\gamma}(x)$
eigenmode at the initial time, i.e. we set
\begin{equation}
  \label{eq:542}
  \left.\inner{e_{\gamma}}{u}\right|_{t=0} = \ep, \quad
  \left.\inner{e_{\gamma}}{\dot{u}}\right|_{t=0} = 0,
\end{equation}
which fixes also the phase of time-periodic solution.  Then all the
higher order equations are solved by assuming decomposition in
eigenbasis of $u_{\lambda}(\tau,x)$ as in (\ref{eq:222}), so the time
evolution of $\hat{u}_{\lambda,j}(\tau)$ is governed by
(\ref{eq:223}).

Here, when solving the perturbative equations for time-periodic
solutions, instead of enforcing (\ref{eq:224}), we relax the initial
conditions and for modes other than $\gamma$ we set
\begin{equation}
  \label{eq:543}
  \hat{u}_{\lambda,k}(0) = c_{\lambda,k}, \quad
  \frac{\diff \hat{u}_{\lambda,k}}{\diff\tau}(0) = 0,
  \quad k\in\mathbb{N},\ k\neq\gamma,
\end{equation}
while for $k=\gamma$
\begin{equation}
  \label{eq:544}
  \hat{u}_{\lambda,\gamma}(0) = 0, \quad
  \frac{\diff \hat{u}_{\lambda,\gamma}}{\diff\tau}(0) = 0.
\end{equation}
Then, free parameters $c_{\lambda,k}$ and $\xi_{\lambda}$ are used to
remove all of the appearing resonant terms and to force the
$2\pi$-periodicity of $\hat{u}_{\lambda,k}(\tau)$ (some of
$c_{\lambda,k}$ will be used to remove terms with rational frequencies
in $\tau$, which produce spurious secular terms appearing at higher
orders).  As it turns out, at some orders of the perturbative
calculation there may not be enough free parameters available to
remove all of the resonances.  Then, as for the model of
Section~\ref{sec:BCSModel}, we modify the solution by adding the
homogeneous solution whose amplitude would serve as a missing
parameter.  Since the construction and the structure of governing
equations depends on whether we construct time-periodic solutions
around the vacuum static solution or around the kink we discuss these
cases separately.
\subsubsection{Vacuum sector}
\label{sec:vacuum-sector}
The construction of time-periodic solutions in the vacuum topological
sector, with $S=1$, is analogous for the previously considered cases
because of the character of the linear spectrum.  As before, we need
to ensure that at each perturbative order there will be enough
parameters availiable to remove all resonances.  It turns out that for
any $\lambda$ the projection $\inner{e_{k}}{s_{\lambda}}$ in
(\ref{eq:223}) vanishes when $k>k_{\ast}=\lambda(\gamma+1)-3$.  The
resonant frequencies to $\omega_{\gamma}$ present in the source to the
wave equation are
\begin{equation}
  \label{eq:545}
  \{k \omega_{\gamma}\,|\ k=1,2,\ldots, K\},
\end{equation}
where $K$ is the largest positive integer such that
\begin{equation}
  \label{eq:546}
  K\omega_{\gamma} \leq \omega_{k_{\ast}},
\end{equation}
holds, so $K\omega_{\gamma}$ is the largest resonant frequency present
in $s_{\lambda}$.  From this we get $K=\lambda-1$.  Thus, at order
$\lambda\geq 2$ we will have $K-1=\lambda-2$ free integration
constants $c_{\lambda,k}$ (since one of them is used to satisfy the
normalization condition (\ref{eq:542})).  With these and with
frequency expansion parameter $\xi_{\lambda+1}$ we can remove at most
$\lambda-1$ resonances at order $\lambda+2$ (the resonances at order
$\lambda+1$ are removed by using the integration constants left at
order $\lambda-1$ and parameters $\xi_{\lambda}$).  However, for
$\lambda\geq 2$ there are exactly $\lambda$ resonances present at
order $\lambda+2$ (at order $\lambda=2$ there is only one resonance
which is removed by setting $\xi_{1}\equiv 0$). Therefore this scheme
would break down already at the fourth order where there are two
resonances present and we have only one parameter available, namely
$\xi_{3}$.  Consequently, at each perturbative order $\lambda\geq 2$
me modify the solution to the wave equation by adding the term with
the frequency being the $(K+1)$-th multiple of $\omega_{\gamma}$, i.e,
the term
\begin{equation}
  \label{eq:547}
  \widetilde{u}_{\lambda,k_{\ast}+2,K+1}
  \cos\left((K+1)\tau\right)\,e_{k_{\ast}+2}(x),
\end{equation}
which is itself a solution to the homogeneous equation with
$k=k_{\ast}+2=\lambda(\gamma+1)-1$ in (\ref{eq:223}).  Thus, an
arbitrary amplitude of (\ref{eq:547}) serve as a remaining parameter
to remove all of the resonances appearing at order $\lambda+2$.
Repeating this reasoning with $\lambda$ replaced by $\lambda+1$ we see
that in this way we get a unique solution, which be constructed up to
arbitrary high order.

\subsubsection{Kink sector}
\label{sec:kink-sector}
As we pointed out already the solution to Eq.~(\ref{eq:223}) will
contain terms $\cos(\omega_{k}/\omega_{\gamma}\tau)$ which in general
are not $2\pi$-periodic functions (especially for frequencies
$\omega_{j}=\sqrt{(j+1)^{2}-3}$), with obvious exception for
$k=\gamma$.  Therefore we can choose the integration constants
$c_{\lambda,k}$ appearing at order $\lambda$ to eliminate all such
terms (for $k=\gamma$ we have parameter $\xi_{\lambda-1}$ which is
used to cancel resonant term---the $\omega_{\gamma}$ is resonant to
itself---then the constant $c_{\lambda,\gamma}$ is fixed by the
normalization condition).  If we have left such terms at order
$\lambda$, they would generate secular terms in the solution at order
$\lambda+2$, through spurious resonances (this is the reason why the
Poincar\'e-Lindstedt method failed to give a uniformly bounded
solution for a single mode initial data, as was emphasized in
Section~\ref{sec:poinc-lindst-appr}) and the condition for their
absence would fix the parameters $c_{\lambda,k}$.  This is equivalent,
but the former way of fixing integration constants is the optimal
approach.  In this way we can proceed to higher order up to appearance
of another proper resonance.  Even though the eigenfrequencies are
irrational numbers (with an exception of the fundamental mode) there
is an infinite number of resonances, but they are irregular (as
opposed to previous cases, where the eigenfrequencies are
equidistant).  In fact the resonant set $O_{\gamma}$, defined in
(\ref{eq:27}), is of measure greater than one for any $\gamma\geq 2$,
since the equation for $k$
\begin{equation}
  \label{eq:548}
  \frac{(k+1)^{2}-3}{(\gamma+1)^{2}-3} = m^{2},
  \quad k,\gamma,m\in\mathbb{N}, \ k\geq\gamma,
\end{equation}
has infinitely many solutions; for $\gamma=1$ there is only a trivial
solution and $\mbox{$O_{\gamma=1}=\{1\}$}$ is a singleton, and this
makes the $\gamma=1$ case special.  As an example we give below an
explicit form of solution to (\ref{eq:548}) for $\gamma=2$
\begin{multline}
  \label{eq:549}
  O_{\gamma=2} =
  \\
  \bigg\{ \frac{1}{2} \left[\left(3+\sqrt{6}\right) \left(5-2
      \sqrt{6}\right)^i + \left(3-\sqrt{6}\right) \left(5+2
      \sqrt{6}\right)^i-2\right]\, \bigg| \ i\in\mathbb{N} \bigg\} =
  \\
  \big\{2,\, 26,\, 266,\, \num{2642},\, \num{26162},\, \num{258986},\,
  \num{2563706},\, \num{25378082},\, \ldots \big\}.
\end{multline}
We see that the resonant frequencies are irregularly distributed,
hence the construction of time-periodic solutions is not so systematic
and has to be carried case by case.  Besides that, the strategy of
resonance removing is identical to the previous cases.  If the
resonance to the eigenmode $k\in O_{\gamma}$ appears at order
$\lambda_{\text{res}}$ then we move back to the order
$\lambda_{\text{res}}-2$ and modify the solution
$u_{\lambda_{\text{res}}-2}(\tau,x)$ by adding to it the term
\begin{equation}
  \label{eq:550}
  \widetilde{u}_{\lambda_{\text{res}}-2,k,m}\cos(m\tau)\,e_{k}(x),
\end{equation}
with $m=\omega_{k}/\omega_{\gamma}$, whose amplitude will be used at
remove this resonance at order $\lambda_{\text{res}}$.  Since then, at
any order $\lambda\geq\lambda_{\text{res}}+1$, there will be two
resonances present (up to the appearance of the next resonance for the
eigenmode with next index from the set $O_{\gamma}$).  These will be
removed by utilizing frequency expansion parameter $\xi_{\lambda-1}$
and free integration constant for resonant mode $k$ the
$c_{\lambda-1,k}$.  In this way the whole procedure can be continued
indefinitely with growing number of proper resonances and number of
availiable parameters to ensure absence of secular terms.
\begin{table}[!t]
  \centering
  \begin{equation*}
    \begin{array}{cccc}
      \toprule
      \gamma & k & m=\omega_{k}/\omega_{\gamma} & \lambda_{\text{res}} \\
      \midrule
      2 & 26 & 11 & 13 \\
      3 & 255 & 71 & 85 \\
      4 & 60 & 13 & 15 \\
      5 & 269 & 47 & 54 \\
      6 & \num{5296} & 781 & 883 \\
      7 & \num{335159611} & \num{42912791} & \num{47879945} \\
      8 & 944 & 107 & 118 \\
      \bottomrule
    \end{array}
  \end{equation*}
  \caption{The lowest resonant frequency indices $k$ for the frequency
    $\omega_{\gamma}$ of bifurcating time-periodic solution constructed
    on a kink static solution for $\gamma=2,\ldots,8$.  The order
    $\lambda_{\text{res}}$ at which given resonance occurs is listed in
    the last column.}
  \label{tab:YMResonances}
\end{table}
In general the appearance of a given proper resonance would be hard to
predict, nevertheless the order at which the first nontrivial
resonance occurs can be precisely predicted.  It turns out that before
that happens the solution at orders
$2\leq\lambda\leq\lambda_{\text{res}}$ is composed of eigenmodes among
which the highest one has index $\gamma\lambda$.  Whence the first
resonance will occur at order
$\lambda_{\text{res}}= \lceil k/\gamma \rceil$ since then the resonant
mode $e_{k}(x)$ will be present in $s_{\lambda_{\text{res}}}(\tau,x)$,
and at the same time $s_{\lambda_{\text{res}}}(\tau,x)$ would contain
Fourier mode $\cos{m\tau}$ (because $\lambda_{\text{res}}>m$).  The
first resonant eigenmodes and their order of occurrence in
perturbative calculation for first few dominant modes $\gamma$ are
listed in Table.~\ref{tab:YMResonances}.  It is evident that the first
proper resonance appears at relatively high perturbative order,
especially for $\gamma=7$ it would be impossible to go to
$\lambda_{\text{res}}$ in perturbative construction, in order to check
that indeed the terms like (\ref{eq:550}) are necessary.  However one
cannot neglect the resonances, even if they appear at relatively high
perturbative order, since they modify the solution in a significant
manner.  The two lowest order cases, namely those with $\gamma=2$ and
$\gamma=4$, which are possible to obtain are discussed in the
following section when also the properties of constructed solutions
are discussed.

\subsubsection{Integrals}

The products appearing in the perturbative equations are expressed in
terms of the following finite sums
\begin{equation}
  \label{eq:551}
  \csc^{2}{x}\,e_i(x)e_j(x) =
  \sum_{\substack{k=\max(|i-j|,1) \\ i+j+k\ \mathrm{odd}}}^{i+j}
  \inner{e_{k}}{\csc^{2}{x}\,e_{i}\,e_{j}} e_k(x),
\end{equation}

\begin{equation}
  \label{eq:552}
  \cos{x}\csc^{2}{x}\,e_i(x)e_j(x) =
  \sum_{\substack{k=\max(|i-j|,1) \\ i+j+k\ \mathrm{even}}}^{i+j}
  \inner{e_{k}}{\cos{x}\csc^{2}{x}\,e_{i}\,e_{j}} e_k(x),
\end{equation}

\begin{equation}
  \label{eq:553}
  e_i(x)e_j(x) =
  \sum_{\substack{k=\max(|i-j|-1,1) \\ i+j+k\ \mathrm{odd}}}^{i+j+1}
  \inner{e_{k}}{e_{i}\,e_{j}} e_k(x),
\end{equation}

\begin{equation}
  \label{eq:554}
  \csc^{2}{x}\,e_{i}(x)e_{j}(x)e_{k}(x) = \csc^{2}{x}\,e_{i}(x)
  \sum_{\substack{l=\max(|j-k|,-1,1) \\ j+k+l\ \mathrm{odd}}}^{j+k+1}
  \inner{e_{l}}{e_{j}\,e_{k}}e_{l}(x),
\end{equation}
where the expansion coefficients were derived using the approach given
in Appendix~\ref{cha:inter-coeff}.  Those are
\begin{multline}
  \label{eq:555}
  \inner{e_{k}}{\csc^{2}{x}\,e_{i}\,e_{j}} =
  \frac{1}{4}\mathcal{N}_{i}\,\mathcal{N}_{j}\,\mathcal{N}_{k}
  \sum_{s=0}^{i-1} \sum_{r=0}^{j-1} \sum_{q=0}^{k-1} \Biggl[ (-1)^{i+j+k-q-r-s+1}
  \\
  \times \binom{i+\frac{1}{2}}{i-s-1} \binom{i+\frac{1}{2}}{s}
  \binom{j+\frac{1}{2}}{j-r-1} \binom{j+\frac{1}{2}}{r}
  \binom{k+\frac{1}{2}}{k-q-1} \binom{k+\frac{1}{2}}{q}
  \\
  \times \frac{ \Gamma\left(q+r+s+\frac{5}{2}\right)
    \Gamma\left(i+j+k-q-r-s-\frac{1}{2}\right)}{\Gamma(i+j+k+2)}
  \Biggr],
\end{multline}

\begin{multline}
  \label{eq:556}
  \inner{e_{k}}{\cos{x}\csc^{2}{x}\,e_{i}\,e_{j}} =
  \frac{1}{4}\mathcal{N}_{i}\,\mathcal{N}_{j}\,\mathcal{N}_{k}
  \sum_{s=0}^{i-1} \sum_{r=0}^{j-1} \sum_{q=0}^{k-1} \Bigg[
  (-1)^{i+j+k-q-r-s+1}
  \\
  \times \binom{i+\frac{1}{2}}{i-s-1} \binom{i+\frac{1}{2}}{s}
  \binom{j+\frac{1}{2}}{j-r-1} \binom{j+\frac{1}{2}}{r}
  \binom{k+\frac{1}{2}}{k-q-1} \binom{k+\frac{1}{2}}{q}
  \\
  \times \frac{1}{\Gamma(i+j+k+3)}
  \Biggl(\Gamma\left(q+r+s+\frac{7}{2}\right)
  \Gamma\left(i+j+k-q-r-s-\frac{1}{2}\right)
  \\
  - \Gamma\left(q+r+s+\frac{5}{2}\right)
  \Gamma\left(i+j+k-q-r-s+\frac{1}{2}\right)\Biggr) \Biggr],
\end{multline}

\begin{multline}
  \label{eq:557}
  \inner{e_{k}}{e_{i}\,e_{j}} =
  \mathcal{N}_{i}\,\mathcal{N}_{j}\,\mathcal{N}_{k} \sum_{s=0}^{i-1}
  \sum_{r=0}^{j-1} \sum_{q=0}^{k-1} \Biggl[ (-1)^{i+j+k-q-r-s+1}
  \\
  \times \binom{i+\frac{1}{2}}{i-s-1} \binom{i+\frac{1}{2}}{s}
  \binom{j+\frac{1}{2}}{j-r-1} \binom{j+\frac{1}{2}}{r}
  \binom{k+\frac{1}{2}}{k-q-1} \binom{k+\frac{1}{2}}{q}
  \\
  \times \frac{\Gamma\left(q+r+s+\frac{7}{2}\right)
    \Gamma\left(i+j+k-q-r-s+\frac{1}{2}\right)}{\Gamma(i+j+k+4)}
  \Biggr],
\end{multline}
where we use the shorthand notation for the normalization constant of
eigenfunctions (\ref{eq:160})
\begin{equation}
  \label{eq:558}
  \mathcal{N}_{j}=
  \frac{(j+1)\sqrt{2j(j+2)}\Gamma(j)}{\Gamma\left(j+\frac{3}{2}\right)}.
\end{equation}
The \mathematica{} assisted guess gives a closed form for the
expressions (\ref{eq:555}) and (\ref{eq:557}) in a following form
\begin{equation}
  \label{eq:559}
  \inner{e_{k}}{\csc^{2}{x}\,e_{i}\,e_{j}} =
  \begin{cases}
    \ {\displaystyle \frac{1}{4\sqrt{2\pi}}\,\tilde{t}_{ijk}} &
    \text{for}\ (i+j+k)\ \text{odd number and}
    \\
    & \bigl(-i+j+k\ge 0\,\wedge\,i-j+k\ge 0
    \\
    & \,\wedge\,i+j-k\ge 0\bigr),
    \\
    \ 0 & \text{otherwise},
  \end{cases}
\end{equation}
where the nonzero elements are
\begin{equation}
  \label{eq:560}
  \tilde{t}_{ijk} = \frac{(-i+j+k+1)(i+j-k+1)(i-j+k+1)(i+j+k+3)}
  {\sqrt{i(i+2) j(j+2) k(k+2)}}.
\end{equation}
Similarly we get
\begin{multline}
  \label{eq:561}
  \inner{e_{k}}{e_{i}\,e_{j}} = \frac{1}{\sqrt{2\pi
      i(i+2)j(j+2)k(k+2)}} \bigl( -(i+2)jk\delta_{i,j+k+1} \\ -
  i(j+2)k\delta_{j,i+k+1} - ij(k+2)\delta_{k,i+j+1} + \tilde{s}_{ijk}
  \bigr),
\end{multline}
with
\begin{equation}
  \label{eq:562}
  \tilde{s}_{ijk} =
  \begin{cases}
    \ i(i+2)+j(j+2)+k(k+2)+1 & \text{for}\ (i+j+k)\
    \text{odd number and}
    \\
    & \bigl(-i+j+k\geq 1\,\wedge\,i-j+k\geq 1
    \\
    & \,\wedge\,i+j-k\geq 1\bigr),
    \\
    \ 0 & \text{otherwise},
  \end{cases}
\end{equation}
and $\delta_{ij}$ denoting the Kronecker delta.  The use of
(\ref{eq:559}) and (\ref{eq:561}) in place of (\ref{eq:555}) and
(\ref{eq:557}) respectively greatly reduces time that \mathematica{}
spends on computation of the expansions and thus allows us to derive a
very high order approximation to the time-periodic solutions.

\subsection{Numerical construction}
\label{sec:YMPeriodicNumerical}

To find a time-periodic solutions of (\ref{eq:153}) we follow the
steps given in Section~\ref{sec:MethodsNumeric} of introductory
chapter.  In this case, where we have a single second order wave
equation to solve, the complexity of the overall algorithm greatly
simplifies, compared to the case of Einstein's equations, since the
lack of constraints and in addition the number of dynamical variables
is reduced here by the factor of two since the velocity and the field
are no longer independent (see (\ref{eq:198}) and (\ref{eq:199})).

As in the previous cases, we use the rescaled time coordinate
$\tau=\Omega\,t$, with $\Omega$ standing for the frequency of
time-periodic solution.  We assume the following truncated double
expansion in the trigonometric series and in the eigenbasis of the
linear operator (\ref{eq:155}) of the time-periodic solution
\begin{equation}
  \label{eq:563}
  u(t,x) = \sum_{k=0}^{K-1}\sum_{j=1}^{N}
  \hat{u}_{k,j}\cos\left(k\tau\right)e_{j}(x),
\end{equation}
(note that this already fixes the phase
$\left.\partial_{t}u\right|_{t=0}=0$). The unknown $K\times N$
expansion coefficients $\hat{u}_{k,j}$ are then determined, similarly
to the previous cases, by constructing suitable number of algebraic
equations.  These are taken to be the requirement for the governing
equation, namely (\ref{eq:153}), to be identically satisfied at the
discrete grid points.  Thus, we take a cartesian product grid with $K$
collocation points of time coordinate $\tau_{k}=\pi(k-1/2)/K$,
$k=1,\ldots,K$ and $N$ collocation points in radial direction
$x_{n}=\pi(n-1/2)/N$, $n=1,\ldots,N$, exactly the same set used in the
spatial discretization procedure discussed in
Section~\ref{sec:YMEvolution}, suited to the expansion (\ref{eq:563}).
Next, at each instant of time $\tau_{i}$ we calculate the coefficients
\begin{equation}
  \label{eq:564}
  \hat{u}_{j}\left(\tau_{i}\right) = \sum_{k=0}^{K-1}
  \widehat{u}_{k,j}\cos\left(k\tau_{i}\right), \quad j=1,\ldots,N,
\end{equation}
and use them as an input in our spectral procedure, which produces as
an output their second time derivatives
$\ddot{\hat{u}}_{j}\left(\tau_{i}\right)$.  These values are equated
to the second time derivatives calculated directly from (\ref{eq:563})
(taking into account the change of independent variables and the
transformation $\partial_t=\Omega\,\partial_{\tau}$) again at the set
of $K\times N$ grid points $\left(\tau_{k},\,x_{n}\right)$.  We choose
the convenient normalization condition as for the time-periodic
solutions in AdS, i.e. we control the amplitude of the dominant mode
$\gamma$ by setting
\begin{equation}
  \label{eq:565}
  \sum_{k=0}^{K-1}\hat{u}_{k,\gamma} = \ep,
\end{equation}
for the solution bifurcating from the eigenmode $e_{\gamma}(x)$ (of
eigenvalue $\omega_{\gamma}$).  This closes the system of
$K\times N+1$ nonlinear equations for the expansion coefficients and
the frequency of the time-periodic solution which are found by
applying the Newton-Raphson algorithm.  As in previous models, we
initialize this iterative procedure by setting a single mode
conditions, i.e, we take
\begin{equation}
  \label{eq:566}
  u(\tau,x) = \ep\,\cos{\tau}\,e_{\gamma}(x), \quad \Omega=\omega_{\gamma}.
\end{equation}

\subsection{Results}
\label{sec:YMPeriodicResults}

\subsubsection{Vacuum sector}
The time-periodic solutions for $S=1$ case admit a regular structure.
For $\gamma$ odd and for any $\lambda\geq 2$ the solutions have the
following Fourier decomposition
\begin{equation}
  \label{eq:567}
  u_{\lambda}(\tau,x) =
  \sum_{j=1}^{(\gamma+1)\lambda/2}\hat{u}_{\lambda,2j-1}(\tau)e_{2j-1}(x),
\end{equation}
\begin{equation}
  \label{eq:568}
  \hat{u}_{\lambda,2j-1}(\tau) = \sum_{k=0}^{\lambda}
  \hat{u}_{\lambda,2j-1,k} \cos{k\tau},
\end{equation}
together with the frequency expansion $\Omega(\ep)$ containing terms
with even and odd powers of $\ep$.  Since the time-periodic solution
is composed of odd eigenmodes only, the following symmetry holds
\begin{equation}
  \label{eq:569}
  u(\tau,x;\ep) = u(\tau,\pi-x;\ep),
  \quad \tau\in[0,2\pi],\ x\in[0,\pi],
\end{equation}
because of the identity $e_{j}(\pi-x)=(-1)^{j+1}e_{j}(x)$,
$j\in\mathbb{N}$, for the eigenbasis functions (\ref{eq:161}).

\begin{figure}[!t]
  \centering
  \begin{tabular}{rl}
    \includegraphics[width=0.465\textwidth]
    {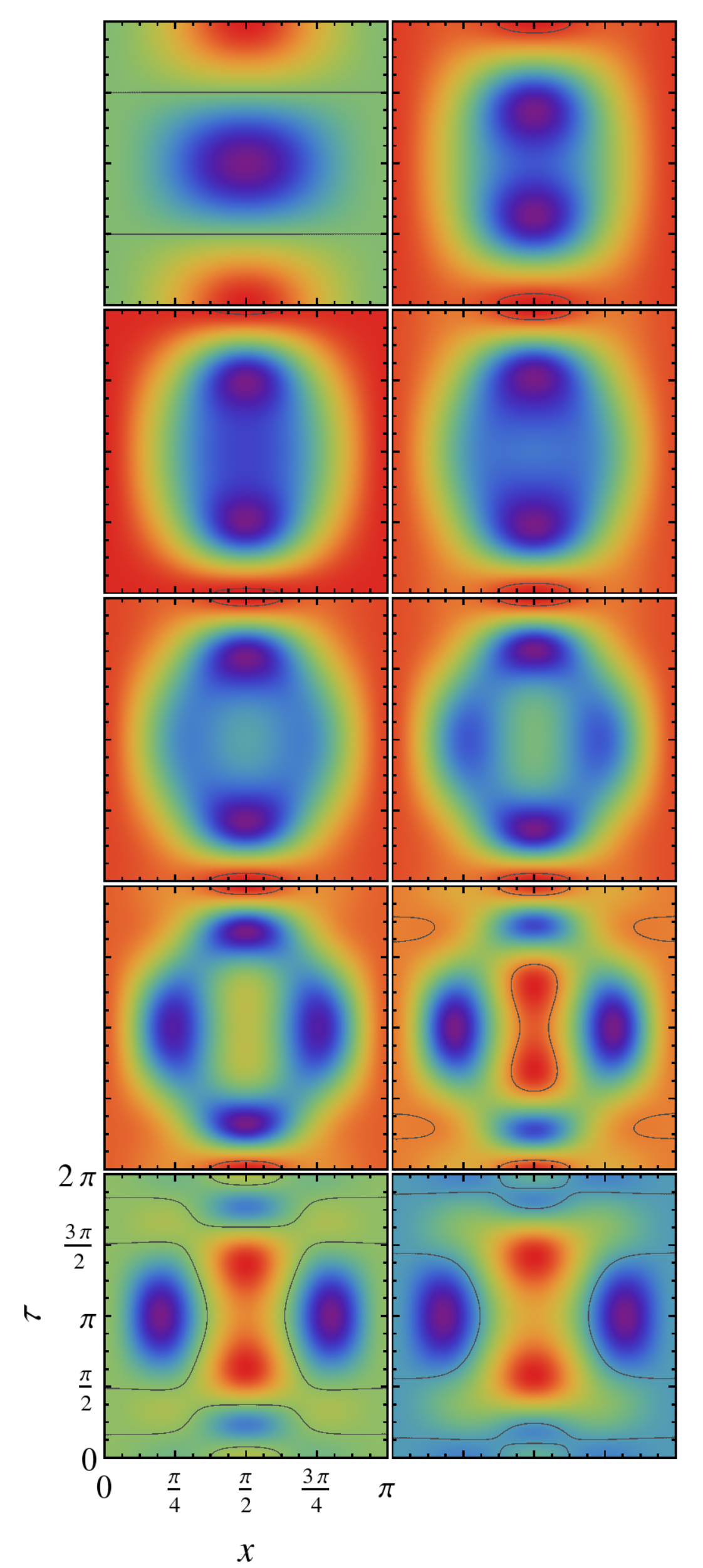} &
    \includegraphics[width=0.465\textwidth]
    {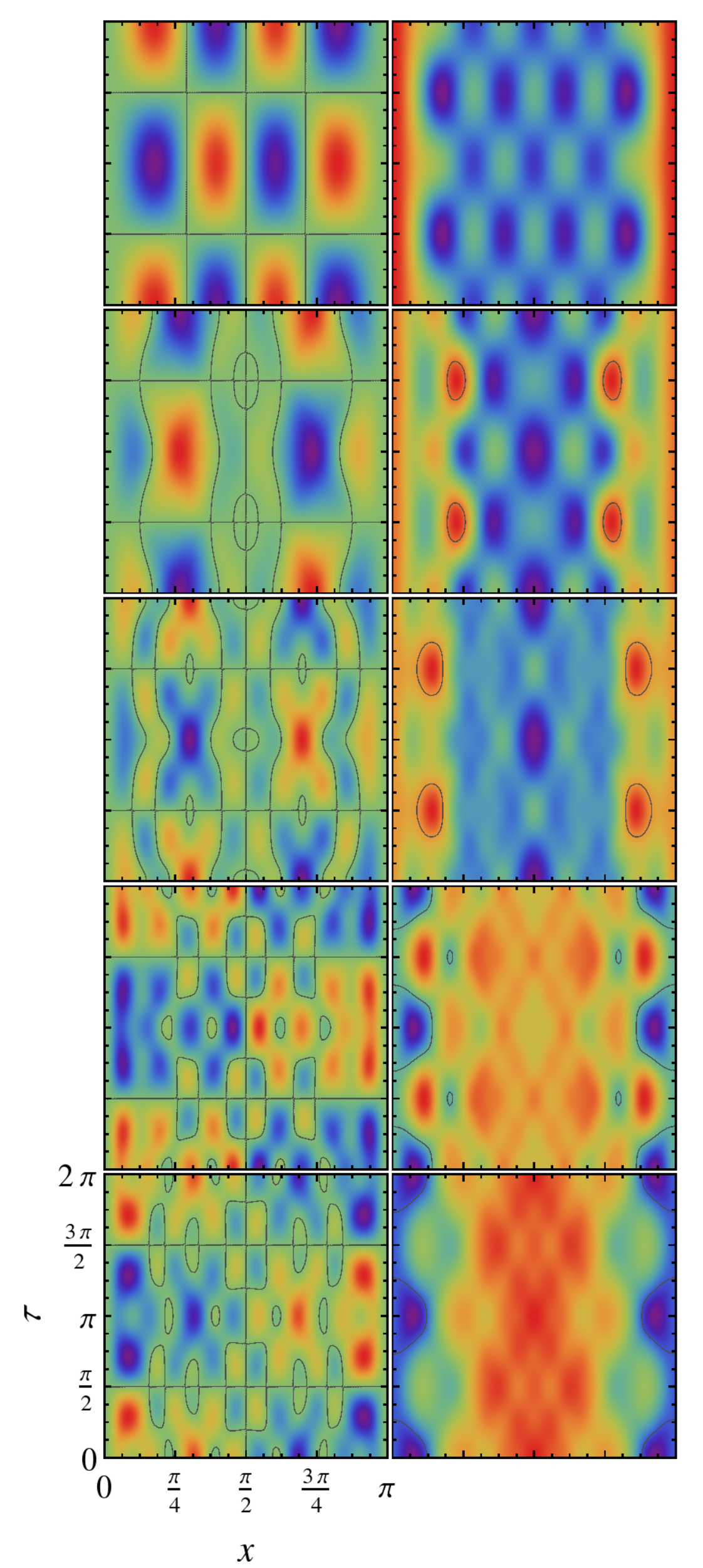} \\
    \multicolumn{2}{c}{\includegraphics[width=0.5\textwidth]
      {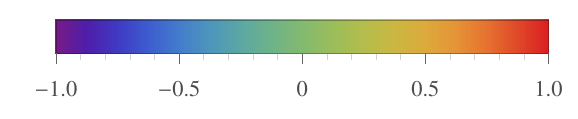}}
  \end{tabular}
  \caption{The density plots of perturbative profiles
    $u_{\lambda}(\tau,x)$ of time-periodic solution for $S=1$.  The
    plots show consecutive perturbative approximations from
    $\lambda=1$ (top left) up to $\lambda=10$ (bottom right).  At each
    plot the range of $u_{\lambda}(\tau,x)$ was normalized to $[-1,1]$
    for better presentation.  As $\gamma$ (the index of dominant mode)
    increases the solution has finer structure, both in space and
    time, as the number of modes contributing to solution grow both
    with $\gamma$ and $\lambda$ as given in
    (\ref{eq:567})-(\ref{eq:568}) and (\ref{eq:570})-(\ref{eq:573}).
    The gray contours are zeros of $u_{\lambda}(\tau,x)$. \textit{Left
      panel}. The $\gamma=1$ ($\omega_{1}=2$) case. \textit{Right
      panel}. The $\gamma=4$ ($\omega_{4}=5$) case.}
  \label{fig:YMCosDensityMaps}
\end{figure}

The structure of solutions with $\gamma$ even is different, and these
solutions have more symmetries.  There are both even and odd
eigenmodes present in the solution and the Fourier decomposition of
time-periodic solution depends on the parity of $\lambda$; for
$\lambda$ even
\begin{equation}
  \label{eq:570}
  u_{\lambda}(\tau,x) = \sum_{j=1}^{(\gamma+1)\lambda/2}
  \hat{u}_{\lambda,2j-1}(\tau)e_{2j-1}(x),
\end{equation}
and
\begin{equation}
  \label{eq:571}
  \hat{u}_{\lambda,2j-1}(\tau) = \sum_{k=0}^{\lambda/2}
  \hat{u}_{\lambda,2j-1,2k}\cos\left(2k\tau\right),
\end{equation}
whereas for $\lambda$ odd
\begin{equation}
  \label{eq:572}
  u_{\lambda}(\tau,x) = \sum_{j=1}^{\left[(\gamma+1)\lambda-1\right]/2}
  \hat{u}_{\lambda,2j}(\tau)e_{2j}(x),
\end{equation}
and
\begin{equation}
  \label{eq:573}
  \hat{u}_{\lambda,2j}(\tau) =
  \sum_{k=0}^{(\lambda+1)/2}\hat{u}_{\lambda,2j,2k-1}
  \cos\left(\left(2k-1\right)\tau\right).
\end{equation}
The frequency expansion for $\gamma$ even contains only even powers of
$\ep$, so $\Omega(\ep)=\Omega(-\ep)$.  From
(\ref{eq:570})-(\ref{eq:573}) we see that the solution profile
exhibits the following symmetries
\begin{equation}
  \label{eq:574}
  u(\tau+\pi,x;-\ep) = u(\tau,x;\ep), \quad
  u(\tau,\pi-x;-\ep) = u(\tau,x;\ep),
\end{equation}
for $\tau\in[0,2\pi]$ and $x\in[0,\pi]$.

To visualize results of perturbative construction we plot on
Fig.~\ref{fig:YMCosDensityMaps} density maps of successive terms of
the $\ep$ expansion, namely the $u_{\lambda}(\tau,x)$ functions.
Clearly, when both the index of dominant mode $\gamma$ and the order
of perturbative expansion $\lambda$ increase the solution oscillates
on smaller scales, both in space and time.  Note also that the
symmetry of solution with respect to the equator of the three sphere,
$x=\pi/2$, for $\gamma=1$ is retained at any perturbative order, while
for $\gamma=4$ only even order terms (right column) remain symmetric,
the odd order terms are antisymmetric (left column).

\begin{figure}[!t]
  \centering
  \includegraphics[width=\swidth]
  {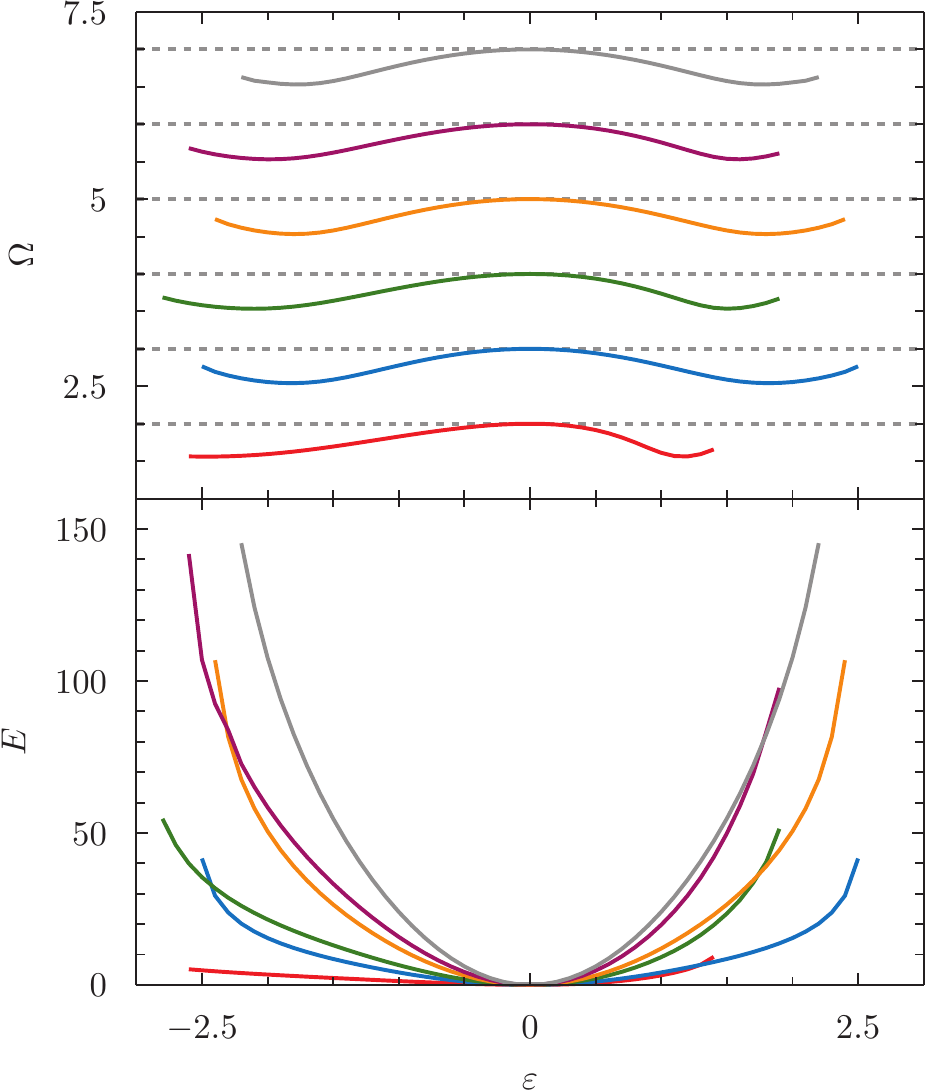}
  \caption{The frequency and energy of numerically computed
    time-periodic solutions as a function of $\ep$ (the $S=1$ case).
    Plot shows solutions bifurcating from first six eigenmodes (with
    $\gamma$ increasing from bottom to top).  The numerical
    calculations were performed on a grid with $32\times 64$ points.
    Beyond the plotted ranges of $\ep$ the numerical procedure ceases
    to converge indicating finite range of $\ep$ for which
    time-periodic solutions exist.}
  \label{fig:YMOneFrequencyEnergy}
\end{figure}

The change of energy and frequency of time-periodic solutions with the
amplitude $\ep$ of dominant mode is presented on
Fig.~\ref{fig:YMOneFrequencyEnergy}.  The nonlinearity causes the
decrease of frequency (the period of oscillation increases) and
increase of energy (with $|\ep|$).  Though continuous, this dependence
on amplitude is not monotonic---for any $\gamma$ there exist different
solutions having the same period and energy.

We derived numerically solutions on a mesh of $24\times 48$ points,
which produces accurate results for small values of $\ep$, but gives
only rough approximation to true solutions for larger amplitudes.  The
Newton procedure used to find the time-periodic solutions converges
rapidly for small and moderate values of $\ep$.  For larger values its
convergence is degraded, while for still larger values the algorithm
ceases to converge even when we provide better initial conditions than
a single mode approximation (\ref{eq:566}), e.g. by using
extrapolation from solutions with smaller values of $\ep$ for which
the algorithm converged.  The borderline of existence of time-periodic
solutions, the limiting values of $\ep$ are computationally difficult
to determine.  The high amplitude time-periodic solutions have spectra
which fall off exponentially but with very small slope, hence to
accurately approximate a solution we would need very large number of
eigenmodes present in truncated series approximation (\ref{eq:563}).
Even using finer grids and taking better initial guess for the Newton
root-finding algorithm it would not be possible to greatly extend the
bifurcating curves shown on
Fig.~\ref{fig:YMOneFrequencyEnergy}.\footnote{As we have seen earlier,
  when discussing time-periodic solutions of other systems, this may
  be related with the \textit{bad} definition of parametrization
  variable.  Since the dominant mode amplitude
  $\left.\inner{e_{\gamma}}{u}\right|_{t=0}$ may be bounded there
  still may exist time-periodic solutions of greater
  \textit{amplitude}, e.g. defined by $u''(0,0)$.  We left this issue
  for further studies, and continue discussion using the
  parametrization (\ref{eq:542}). }

\begin{figure}[!t]
  \centering
  \includegraphics[width=\swidth]
  {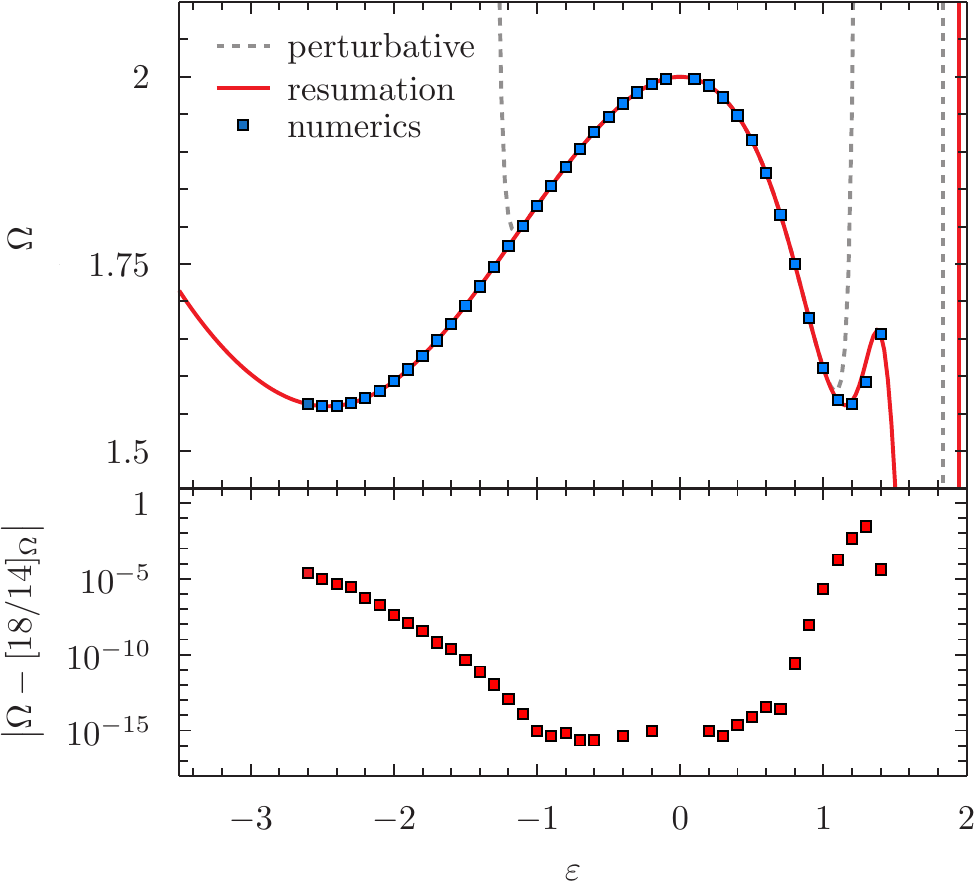}
  \caption{The frequency of time-periodic solution bifurcating from
    fundamental mode ($\gamma=1$).  The perturbative series (gray
    dashed line), despite of being of high order (here we take into
    account first forty terms of $\Omega(\ep)$ Taylor expansion),
    converges slowly but reaches amplitudes outside the linear regime.
    The Pad\'e resumation, of order $[18/14]$, gives qualitatively
    better approximation to numerical data (especially for $\ep<0$).}
  \label{fig:YMOneE1Pade}
\end{figure}

However, for any $\gamma$ there seems to be a finite range of $\ep$
for which time-periodic solutions exist.  Constructing solutions for a
few lowest dominant modes we observe that the length of this interval
slowly decreases with increasing $\gamma$ (separately for $\gamma$
even and $\gamma$ odd).  Using the Pad\'e resummation, as in previous
Sections, we were unable to find a reasonable approximation to the
limiting values of $\ep$.  In this case the structure of poles of
$[m/n]_{\Omega}$ either with $m=n$ or $m\neq n$ rapidly changes when
we change order parameters $m$ and $n$.  Different also is the
character of a frequency bifurcation curve.
\begin{figure}[!t]
  \centering
  \includegraphics[width=\mwidth]
  {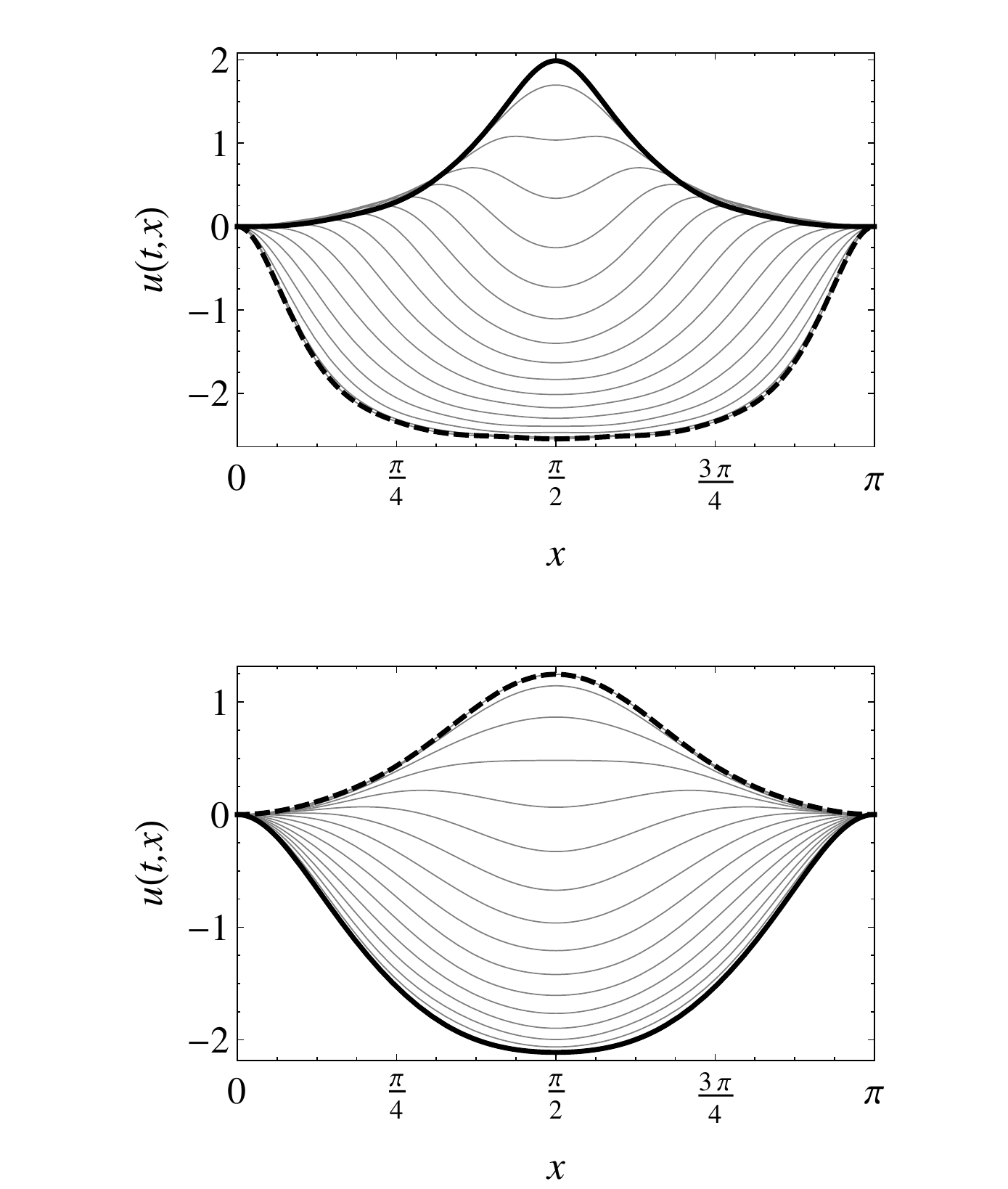}
  \caption{The snapshots of the time-periodic solution in vacuum
    topological sector ($S=1$) bifurcating from mode $e_{1}(x)$.  The
    large amplitude (with marginal values, near the existence border)
    was chosen to emphasize the nonlinear effects.  The numerical
    calculations were performed on grid with $32\times 64$
    (\ref{eq:563}).  The profile at initial time $t=0$ is plotted with
    thick black line, the dashed line is the profile at $t=T/2$, while
    gray lines show profile at intervals $T/32$. \textit{Top
      panel}.  Positive amplitude solution with $\ep=1.45$ and
    $\Omega\approx\num{1.714833}$. \textit{Bottom panel}.  Negative
    amplitude solution with $\ep=-2.62$ and
    $\Omega\approx\num{1.563661}$.}
  \label{fig:YMOneE1Evolution}
\end{figure}
\begin{figure}[!t]
  \centering
  \includegraphics[width=\swidth]
  {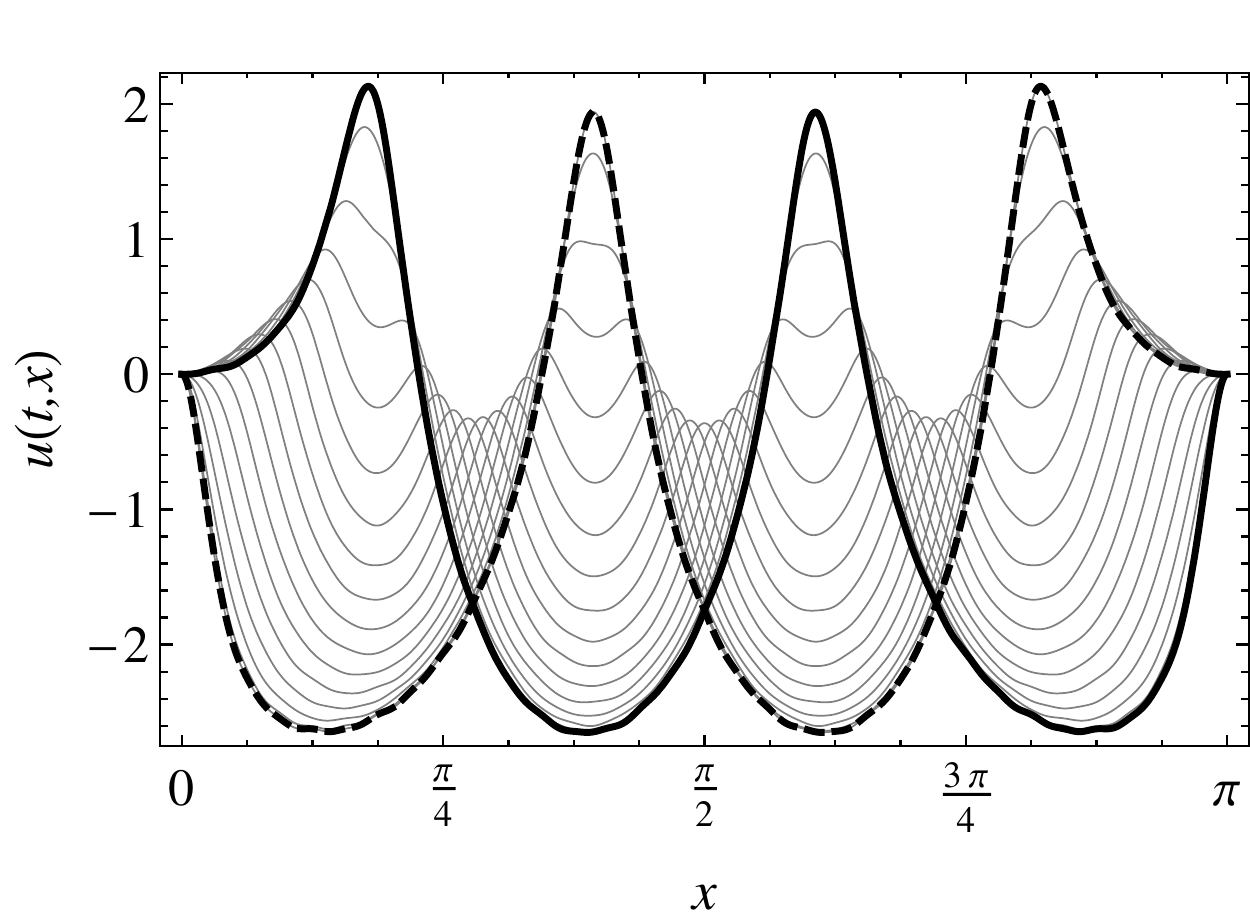}
  \caption{An analogue of Fig.~\ref{fig:YMOneE1Evolution} for solution
    bifurcating from eigenmode $e_{4}(x)$ with amplitude $\ep=2.42$
    and frequency $\Omega\approx\num{4.750961}$.  The solution with
    negative value of $\ep$ is not shown these can be obtained from
    $\ep>0$ solution using symmetry (\ref{eq:574}), as for other
    $\gamma$ even.}
  \label{fig:YMOneE4Evolution}
\end{figure}
Nevertheless, the resumation greatly improves the convergence of
perturbative series.  The result showing the comparison of numerical
data (points) with perturbative expansion of $\Omega$ and its
resumation is shown on Fig.~\ref{fig:YMOneE1Pade} (the $\gamma=1$
case).  The off-diagonal $[18/14]_{\Omega}$ Pad\'e approximation,
chosen as the best qualitative approximation, greately extends the
range of applicability of perturbative result (especially for
$\ep<0$).  This shows also the agreement of the results of two
independent methods used to construct time-periodic solutions (for
relatively small $\left|\ep\right|$ the difference of these two
methods is dominated by the numerical precision).

Since for small values of the expansion parameter $\ep$ the solutions
do not significantly differ from the single mode approximation (as the
name suggests the dominant mode is a main component of the solution)
we present the results for relatively large values of $\ep$ to make
the nonlinear effect clearly visible.  On
Figs.~\ref{fig:YMOneE1Evolution} and \ref{fig:YMOneE4Evolution} we
present the results of numerical calculations for $\gamma=1$ and
$\gamma=4$ respectively.  For $\gamma$ odd the solutions for positive
and negative values of $\ep$ are distinct, which is clearly visible on
Fig.~\ref{fig:YMOneE1Evolution}, where we present the fundamental
solutions.  The initial profiles of time-periodic solutions preserve
the number of zeros of dominant modes, even when other modes have
significant amplitudes (especially for large values of $\ep$).  This
is no longer true for other times, during the evolution the number of
zeros changes, in contrast to the linear solutions.  This
characteristic change of shape in time of function profile, visible
also on Fig.~\ref{fig:YMOneE4Evolution}, alongside with frequency
change, is the nonlinear effect.  It is worth to mention that
symmetries, given in Eqs.~(\ref{eq:569}) and (\ref{eq:574}), which are
apparent on Figs.~\ref{fig:YMOneE1Evolution} and
\ref{fig:YMOneE4Evolution}, are not forced in numerical code, for
given $\gamma$ and initial guess for Newton's algorithm, the parity is
preserved through successive iterations.

\subsubsection{Kink sector}

Up to the occurrence of the first proper resonance, i.e. for
$\lambda\leq\lambda_{\text{res}}$, the time-periodic solutions have
the following structure.  For $\gamma$ even the solution is composed
of even eigenmodes only
\begin{equation}
  \label{eq:575}
  u_{\lambda}(\tau,x) =
  \sum_{j=1}^{\gamma\lambda/2}\hat{u}_{\lambda,2j}(\tau)e_{2j}(x),
\end{equation}
\begin{equation}
  \label{eq:576}
  \hat{u}_{\lambda,2j}(\tau) =
  \sum_{k=0}^{\lambda}\hat{u}_{\lambda,2j,k}\cos(k\tau),
\end{equation}
so it exhibits the following symmetry
\begin{equation}
  \label{eq:577}
  u(\tau,\pi-x;\ep) = - u(\tau,x;\ep).
\end{equation}
The frequency $\Omega(\ep)$ contains both even and odd powers of $\ep$
in its Taylor expansion.  For $\gamma$ odd the eigenmode decomposition
depends on parity of $\lambda$, i.e. there is for even $\lambda$
\begin{equation}
  \label{eq:578}
  u_{\lambda}(\tau,x) =
  \sum_{j=1}^{\gamma\lambda/2}\hat{u}_{\lambda,2j}(\tau)e_{2j}(x),
\end{equation}
\begin{equation}
  \label{eq:579}
  \hat{u}_{\lambda,2j}(\tau) =
  \sum_{k=0}^{\lambda/2}\hat{u}_{\lambda,2j,2k}\cos(2k\tau),
\end{equation}
while for odd $\lambda$
\begin{equation}
  \label{eq:580}
  u_{\lambda}(\tau,x) =
  \sum_{j=1}^{(\gamma\lambda+1)/2}\hat{u}_{\lambda,2j-1}(\tau)e_{2j-1}(x),
\end{equation}
\begin{equation}
  \label{eq:581}
  \hat{u}_{\lambda,2j-1}(\tau) =
  \sum_{k=0}^{(\lambda+1)/2}\hat{u}_{\lambda,2j-1,2k-1}
  \cos\left((2k-1)\tau\right),
\end{equation}
and the frequency expansion contains only even powers of $\ep$.

\begin{figure}[!t]
   \centering
   \includegraphics[width=\swidth]
   {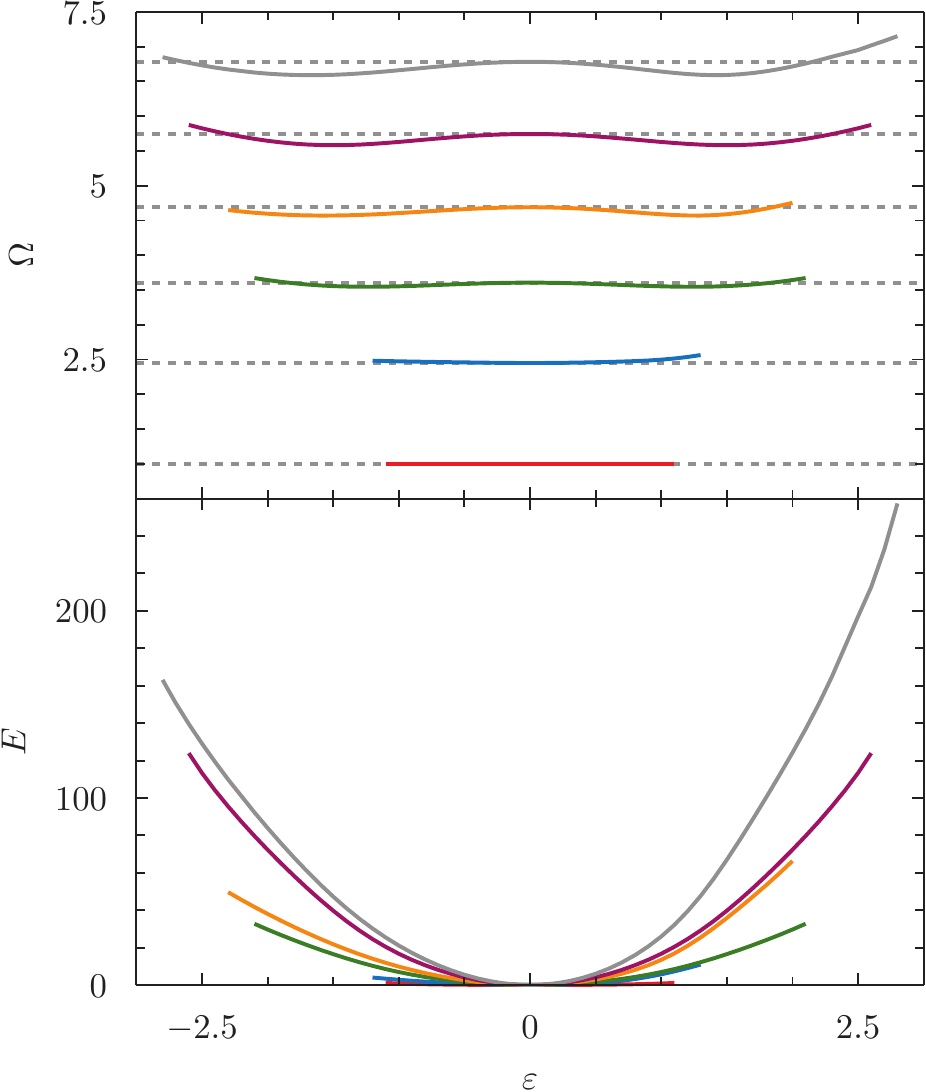}
   \caption{An analogue of Fig.~\ref{fig:YMOneFrequencyEnergy} for
     $S=\cos{x}$ case.}
  \label{fig:YMCosFrequencyEnergy}
\end{figure}

Since by adding the term (\ref{eq:550}) we modify the solution at
order $\lambda_{\text{res}}-2$ these decomposition formulae hold only
for $\lambda\leq\lambda_{\text{res}}-3$ (but in practice the order at
which first proper resonance appears $\lambda_{\text{res}}$ is very
large).  Then the eigenmode decomposition at higher perturbative
orders greately depends on $\gamma$, since the resonant frequencies do
not seem to have any regular structure (cf. \ref{eq:549}). (The
exception is the $\gamma=1$, which is the \textit{fully nonresonant}
(contrary to fully resonant) case and the formulae
(\ref{eq:578})-(\ref{eq:581}) holds for arbitrary high order
$\lambda$.)  However, the resonances for $\gamma>1$ do not affect the
symmetry of the solution, so the following holds
\begin{equation}
  \label{eq:582}
  u(\tau+\pi,x;-\ep) = u(\tau,x;\ep), \quad
  u(\tau,\pi-x;-\ep) = -u(\tau,x;\ep),
\end{equation}
for $\tau\in[0,2\pi]$ and $x\in[0,\pi]$, in analogy to (\ref{eq:574}).

\begin{figure}[!t]
  \centering
  \includegraphics[width=\swidth]
  {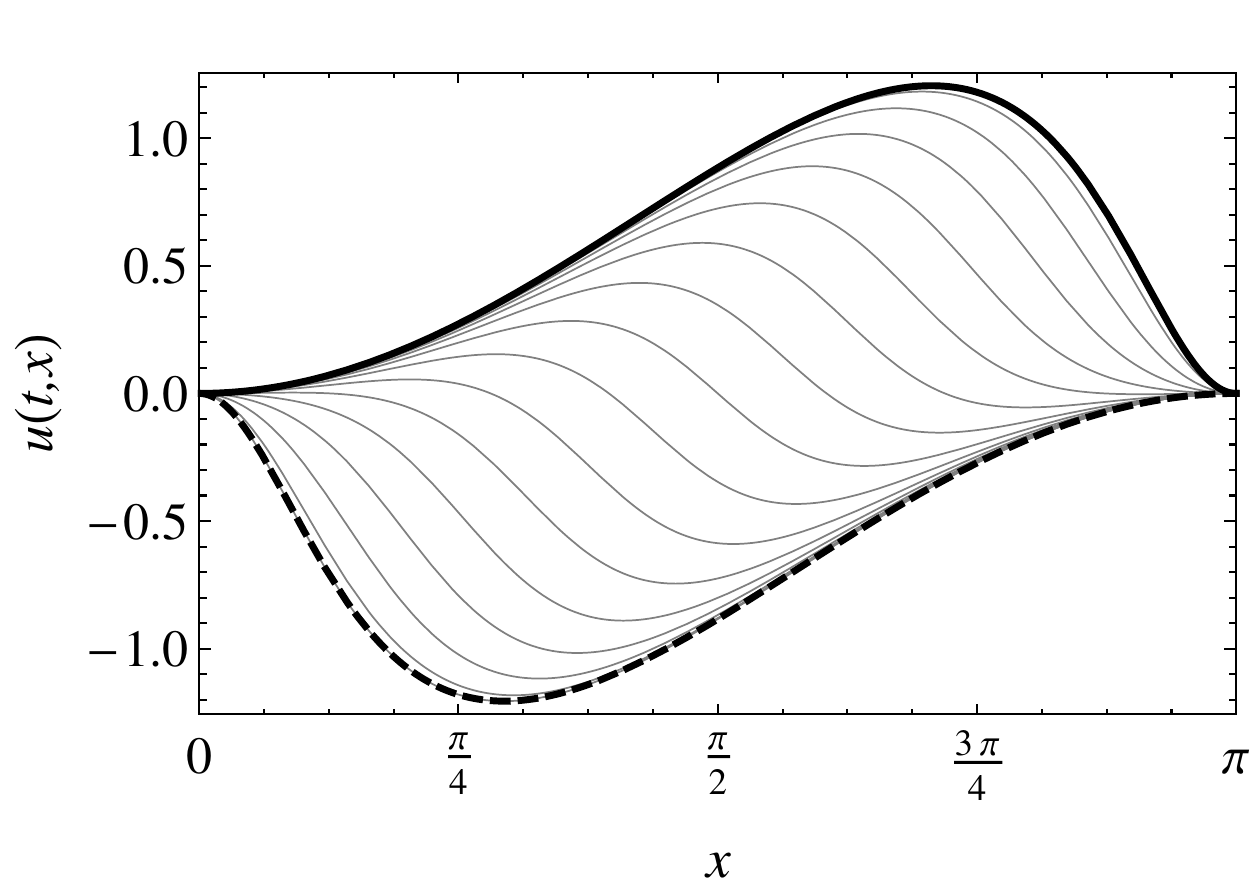}
  \caption{An analogue of Fig.~\ref{fig:YMOneE1Evolution} in kink
    topological sector showing solution bifurcating from fundamental
    mode $e_{1}(x)$ with amplitude $\ep=1.15$ and frequency $\Omega=1$
    (up to numerical precision).  The negative $\ep$ time-periodic
    solution is reconstructed by taking into account the symmetry of
    all odd $\gamma$ solutions (\ref{eq:582}).}
  \label{fig:YMCosE1Evolution}
\end{figure}

As was mentioned already, there are two things which distinguish the
time-periodic solutions in the kink and vacuum topological sectors.
The first characteristic concerns the frequency of time-periodic
solutions.  For the family of solutions bifurcating from fundamental
mode ($\gamma=1$) the frequency does not depend on $\ep$.  In other
words, all time-periodic solutions which as a dominant mode have the
$e_{1}(x)$ eigenmode oscillate with the same frequency $\Omega(\ep)=1$
even if they are far from the linear regime.  This independence on
amplitude is shown on Fig.~\ref{fig:YMCosFrequencyEnergy}, where we
show results from numerical calculations, this is also confirmed by
high order perturbative expansion.

The frequency of solutions based on higher modes depends on amplitude.
As for the vacuum topological sector this is not a monotonic function,
which is illustrated on Fig.~\ref{fig:YMCosFrequencyEnergy}.  But most
importantly, the numerical procedure is able to find the solutions
with frequencies both smaller and larger than the corresponding
frequency of bifurcation point.  Also the range of existence is
increasing with the index of dominant mode, in contrast to the vacuum
case, demonstrated on Fig.~\ref{fig:YMOneFrequencyEnergy}, where the
range of existence shrinks with increasing $\gamma$.  This makes the
kink topological sector different from the vacuum case.

\begin{figure}[!t]
  \centering
  \includegraphics[width=\mwidth]
  {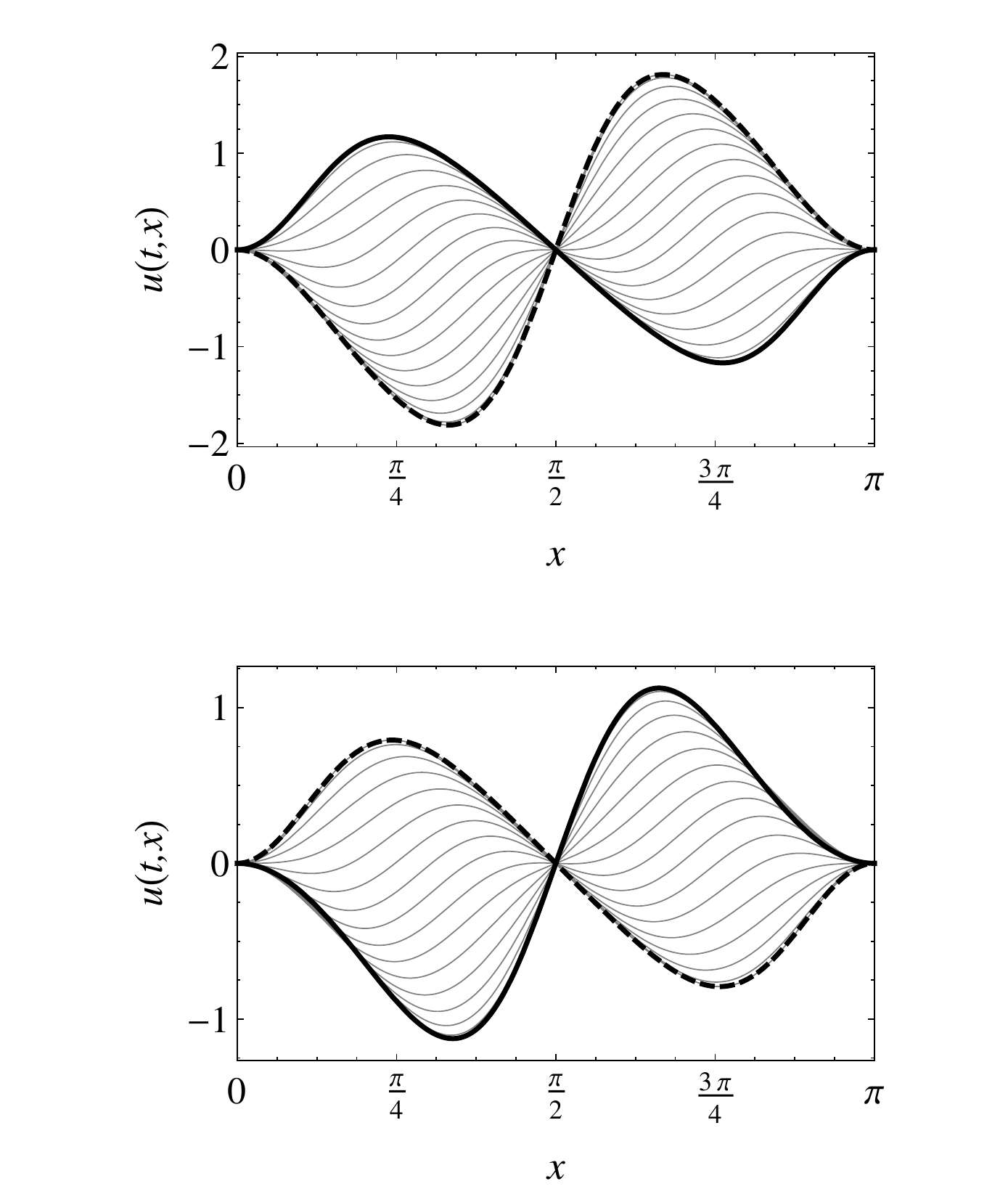}
  \caption{An analogue of Fig.~\ref{fig:YMOneE1Evolution} in kink
    topological sector showing solutions bifurcating from eigenmode
    $e_{2}(x)$.  \textit{Top panel}. Positive amplitude solution with
    $\ep=1.31$ and $\Omega\approx\num{2.567217}$. \textit{Bottom
      panel}.  Negative amplitude solution with $\ep=-1.24$ and
    $\Omega\approx\num{2.484884}$.}
  \label{fig:YMCosE2Evolution}
\end{figure}

Profiles of solutions for $\gamma=1$ and $\gamma=2$ are shown on
Figs.~\ref{fig:YMCosE1Evolution} and \ref{fig:YMCosE2Evolution}
respectively.  Again, note that the symmetries (\ref{eq:577}) and
(\ref{eq:582}) are respected by the numerical code.

Another feature which distinguishes these two topological sectors is
the resonant structure.  To deal with the proper resonances appearing
in perturbative construction we modify the solution by adding some
extra terms, which are homogeneous solutions to the governing
equations, with amplitude which subsequently serves as a parameter to
cancel otherwise occurring secular terms.  This was emphasized over
current chapter.  While for $S=1$ these resonances have regular
structure, in the kink topological sector $S=\cos{x}$ resonances are
irregular.  Additionally they appear at relatively high perturbative
orders.  Therefore it is interesting to validate perturbatively
constructed solutions with numerical calculations.

\begin{figure}[!t]
  \centering
  \includegraphics[width=\swidth]
  {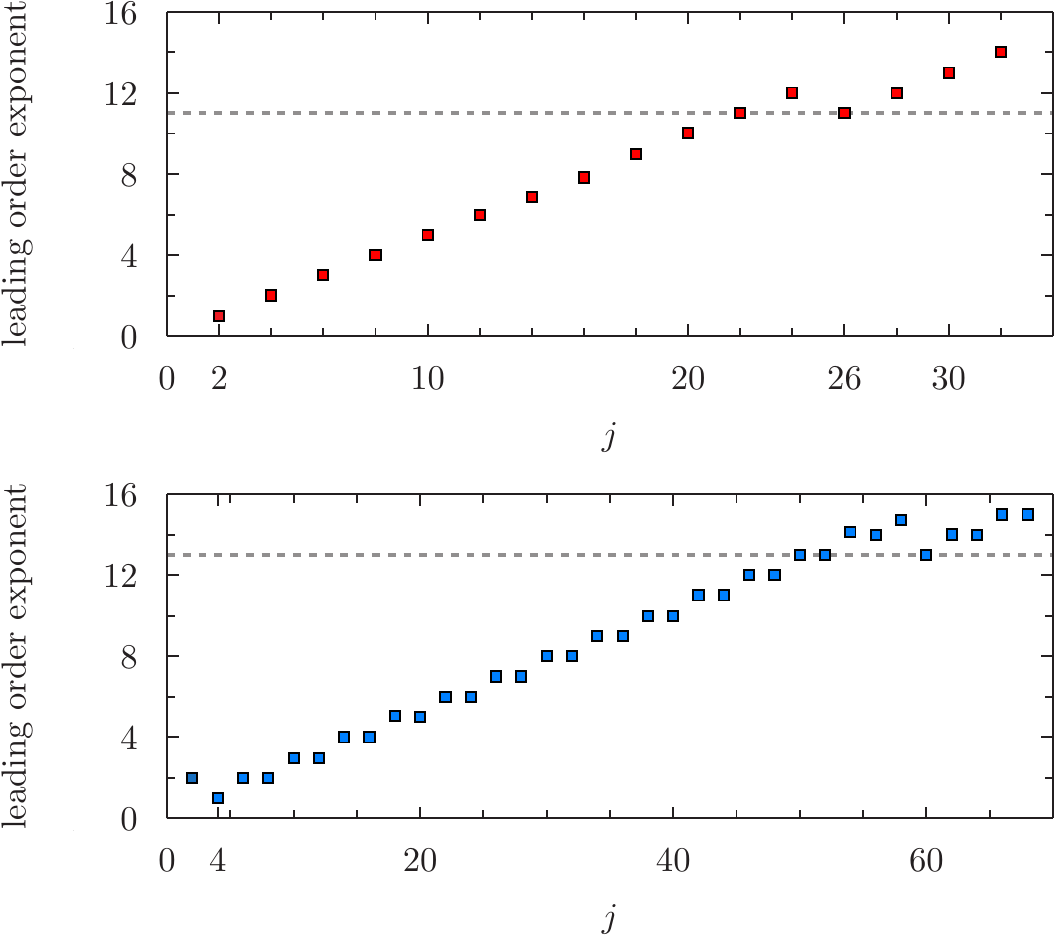}
  \caption{The leading order scaling with $\ep$ of eigenmode
    decomposition of initial data $u(0,x\,;\ep)=\sum_{j\geq
      1}\hat{u}_{j}(0\,;\ep)\,e_{j}(x)$ corresponding to time-periodic
    solution (results obtained by fitting to numerical
    data). \textit{Top panel}. The $\gamma=2$ case. To remove proper
    resonance at order $\lambda=13$ the eigenmode $e_{26}(x)$ was
    added at order $\lambda=11$.  Thus the expected leading order
    behaviour $\hat{u}_{26}(0\,;\ep)\sim\ep^{11}$. \textit{Bottom
      panel}. The $\gamma=4$ case. Here the resonance to $e_{60}(x)$
    was removed by adding this mode at order $\lambda=13$, and
    $\hat{u}_{60}(0\,;\ep)\sim\ep^{13}$.}
  \label{fig:YMCosE2E4Scaling}
\end{figure}

The two lowest order and tractable cases are the $\gamma=2$ and
$\gamma=4$, see Tab.~\ref{tab:YMResonances}.  For $\gamma=2$ the
resonance for the eigenmode $e_{26}(x)$ appears at order $\lambda=13$
and we claim that inclusion of the term (\ref{eq:550}) at order
$\lambda=11$ removes the resonance, for $\gamma=4$ the resonance
appears at order $\lambda=15$ and the solution modification is done at
order $\lambda=13$.  Solving higher order perturbative equations we
supported this procedure, since the first proper resonance can be
successfully removed and the construction can be continued further.
These additional parameters are, for $\gamma=2$
\begin{multline}
  \label{eq:583}
  \hat{u}_{11,26,11} = -\frac{\num{118242196219}}
  {\num{280553932619513856}\pi^5}\,\sqrt{\frac{7}{13}}
  \\[0.5ex]
  \approx \num{-1.010611200842075}\times 10^{-9},
\end{multline}
and for $\gamma=4$
\begin{multline}
  \label{eq:584}
  \hat{u}_{13,60,13} = \frac{\num{124000385740925526359933}}
  {\num{12339744967221655227732493520486400000}\pi^6}
  \,\sqrt{\frac{31}{5}}
  \\[0.5ex]
  \approx \num{2.602636168242766}\times 10^{-17}.
\end{multline}
Further analysis of perturbative series and comparison with
numerically obtained solutions validates our approach.  In order to
perform quantitative comparison of both approaches it is convenient to
rearrange the perturbative series given in Eqs.~(\ref{eq:215}) and
(\ref{eq:222}) as
\begin{equation}
  \label{eq:585}
  u(\tau,x;\ep) = \sum_{j\geq 1}\hat{u}_{j}(\tau;\ep)e_{j}(x),
\end{equation}
with the eigenmode expansion coefficients being $\ep$ dependent.
Computing numerical solutions for few values of $\ep$ we can determine
the dependence on $\ep$ by fitting polynomial in $\ep$ to the
numerical data.  In particular we can check a leading order scaling
with $\ep$ of decomposition coefficients $\hat{u}_{j}(\tau;\ep)$ at
some instant of time, for convenience we take $\tau=0$. These results,
with the resonant modes marked, are presented on
Fig.~\ref{fig:YMCosE2E4Scaling}.  This is in agreement with
perturbative series given in Eq.~(\ref{eq:575}) up to
$\lambda=\lambda_{\text{res}}-1$.
\begin{table}[!t] \centering
  \begin{equation*}
    \begin{array}{cll}
      \toprule
      \multirow{2}{*}{$\lambda$} &
      \multicolumn{2}{c}{\coef{\lambda}\hat{u}_{26}(0;\ep)}
      \\
      & \multicolumn{1}{c}{\text{numerical data}} &
      \multicolumn{1}{c}{\text{perturbative data}}
      \\
      \midrule
      11 &
      \num{-1.010611200842076}\times 10^{-9} &
      \num{-1.010611200842075}\times 10^{-9}
      \\
      12 &
      \num{-1.565630367129}\times 10^{-9} & \num{-1.565630367147342}\times
      10^{-9}
      \\
      13 & \phantom{-}\num{1.138758143}\times 10^{-8} &
      \phantom{-}\num{1.138758153956382}\times 10^{-8}
      \\
      14 & \phantom{-}\num{2.41603}\times
      10^{-8} & \phantom{-}\num{2.415999875082757}\times 10^{-8}
      \\
      15 & \num{-2.342}\times 10^{-8} &
      \num{-2.313753657153318}\times 10^{-8}
      \\
      \bottomrule
    \end{array}
  \end{equation*}
  \caption{The comparison of numerically and perturbatively
    constructed time-periodic solutions on a branch bifurcating from
    $\gamma=2$ eigenmode.  The $e_{26}(x)$ mode projection of initial
    conditions where compared.  The coefficients of polynomial dependence
    of $\hat{u}_{26}(0;\ep)$ on $\ep$ where read off from perturbative
    series, while with numerical method these were obtained by
    the least square fitting to a discrete series of  solutions with
    small amplitudes $\ep$ (ranging from $10^{-4}$ to $2^{10}\times
    10^{-4}$) on a mesh with $16\times 32$ points.}
  \label{tab:YMPeriodicCosE2NumercPerturbative}
  \centering
  \begin{equation*}
    \begin{array}{cll}
      \toprule
      \multirow{2}{*}{$\lambda$} &
      \multicolumn{2}{c}{\coef{\lambda}\hat{u}_{60}(0;\ep)} \\
      & \multicolumn{1}{c}{\text{numerical data}} &
      \multicolumn{1}{c}{\text{perturbative data}} \\
      \midrule
      13 & \phantom{-}\num{2.6026361682427657}\times 10^{-17} &
      \phantom{-}\num{2.602636168242766}\times 10^{-17}
      \\
      14 & \phantom{-}\num{2.2522448521489787}\times 10^{-17} &
      \phantom{-}\num{2.252244852148980}\times 10^{-17}
      \\
      15 & \phantom{-}\num{1.481102}\times 10^{-16} &
      \phantom{-}\num{1.481130035912182}\times 10^{-16}
      \\
      16 & \phantom{-}\num{8.337571}\times 10^{-16} &
      \phantom{-}\num{8.337598936082316}\times 10^{-16}
      \\
      17 & \phantom{-}\num{2.60569}\times 10^{-15} &
      \phantom{-}\num{2.605917516055920}\times 10^{-15}
      \\
      18 & \phantom{-}\num{6.9}\times 10^{-16} &
      \phantom{-}\num{7.110705447625604}\times 10^{-16}
      \\
      \bottomrule
    \end{array}
  \end{equation*}
  \caption{Same as Tab.~\ref{tab:YMPeriodicCosE2NumercPerturbative}
    for $\gamma=4$.  The numerical solutions were calculated on mesh
    $18\times 68$ with $\ep$ ranging from $10^{-5}$ to
    $2^{10}\times 10^{-4}$.}
  \label{tab:YMPeriodicCosE4NumercPerturbative}
\end{table}
Moreover the analysis can be even more detailed.  Performing the
polynomial fit to the resonant mode $e_{26}(x)$ coefficient
$\hat{u}_{26}(0;\ep)$ (in $\gamma=2$ case) we can compare this with
the perturbative calculation.  The straightforward approach would fail
because of limited numerical precision.  This is because for small
amplitudes of dominant mode $\ep$ the amplitudes od resonant modes are
hidden below the level set by double-floating point precision (which
is of order $10^{-16}$).  In order to be above this threshold we need
to take $\ep\gtrsim 0.23$ and $\ep\gtrsim 1.1$ for $\gamma=2$ and
$\gamma=4$ respectively.  This if far beyond the linear regime, where
the nonlinearities are expected to dominate, and the perturbative
approximation does not work.  The resolution to this remedy is to use
the extended precision arithmetic excellent facilities of
\mathematica{}.  Setting the arithmetic precision to $80$ significant
digits we were able to determine solutions for very small $\ep$ (with
residuals not exceeding $10^{-80}$ in absolute value).  The results of
the fitting procedure and the coefficients extracted from perturbative
series are compared in
Tab.~\ref{tab:YMPeriodicCosE2NumercPerturbative}.  The agreement for
lower order coefficients of a polynomial fit is astonishing, while the
higher modes are moderately accurate.  This would converge both with
increasing the number of eigenmodes taken in truncated series
expansion (\ref{eq:563}) and with denser probing in $\ep$ (especially
for relatively small amplitudes where the perturbative series
converge).  The same quality of the result we obtain in the $\gamma=4$
case by performing the same analysis, see
Tab.~\ref{tab:YMPeriodicCosE4NumercPerturbative}.

\section{Conclusions}
\label{sec:PeriodicConclusions}

In this chapter we presented the perturbative and numerical methods
used to find time-periodic solutions to specific mixed
elliptic-hyperbolic systems of PDEs on bounded domains especially to
the Einstein equations with negative cosmological constant.
\vspace{0.4ex}

The construction follows a general scheme given in
Chapter~\ref{cha:Methods} but the structure of considered equations
enforces various modifications.  This mainly concerns the perturbative
construction where certain additional resonant terms may appear when
building successive approximation, and their removal is nontrivial
(apart from proper choice of basis functions and the efficient
solution to the perturbative equations).  As a part of the numerical
construction of time-periodic solutions, which itself is rather
universal, we have described spatial discretization and in particular
we demonstrated how to effectively solve the constraint equations.
Finally, we have shown that our approach leads to a stable and
convergent numerical scheme.  For time evolution problems we have
demonstrated superior properties (near preservation of constants of
motion) of symplectic time-integration algorithms in long-time
simulations.
\vspace{0.4ex}

Extending the results of \cite{MRPRL}, we have shown the steps of
perturbative construction of time-periodic solutions to the EKG system
in any space dimension $d\geq{}2$, especially for the physically most
interesting case $d=3$.  While for odd $d$ the construction procedure
(general approach) is challenging (it requires manipulation of long
formulae of a very complex structure) it can be greatly simplified in
even $d$ by using the eigenbasis expansion (algorithmic approach).
For even $d$ we can reduce the problem of solving coupled PDEs to
algebraic equations and a system of second order ODEs.  Thus the
equations are much easier to solve and the whole procedure is much
more efficient.
\vspace{0.4ex}

We presented a way to express the products of eigenfunctions in terms
of their (finite in even $d$) linear combinations, which proved to be
a necessary element of the efficient algorithm.  In both odd and even
$d$ cases we give an argument by counting the number of appearing
resonant terms, that at each perturbative order we are able to cancel
all of the resonances and that produced expressions give a unique
solution up to the definition of the expansion parameter.  The
difference between even and odd $d$ also manifests in the numerical
construction.  For studies of even dimensional cases we have used
improved numerical scheme of \cite{MRPRL, MR2014} based on the
eigenbasis expansion.  For odd dimensional cases, where the former
code cannot be used, we employed Chebyshev pseudospectral spatial
discretization, which additionally is universal in terms of the space
dimension $d$.
\vspace{0.4ex}

Using two independent approaches to the construction of time-periodic
solutions we verified our results directly comparing the perturbative
and numerical data, performing convergence tests, and finally by the
time evolution of the time-periodic solutions (supporting their
periodicity).  We stressed the importance of appropriate choice of the
continuous parameter identifying solutions of a given family
(solutions bifurcating from given linear mode).  While this choice is
not particularly important in the case of small amplitude solutions,
the use of proper parametrization is crucial in studies of large
amplitude solutions (we noted this in all of the considered problems).
Using a center value of the scalar field as an expansion parameter we
found that there are no solutions with arbitrarily large mass---the
mass function attains its maximum for large amplitude time-periodic
solutions.  The critical amplitude separates stable branch and
unstable branch of solutions. Time evolution of perturbed
configurations on the stable branch gives evidence on their nonlinear
stability (solutions stays all the time in finite distance to the
periodic orbit).  In contrast, solutions on the unstable branch
unstable branch when perturbed collapse to a black hole almost
immediately.
\vspace{0.4ex}

These techniques were applied also to a complex scalar field in the
studies of standing waves (counterparts of time-periodic solutions).
Simple form of standing wave solutions (static metric and purely
harmonic oscillation of matter field) allowed us to study their linear
stability problem.  In particular, using analytical techniques we were
able to solve the eigenvalue problem, proving their linear stability.
Using numerical methods we verified our analytical results, in
particular consistency of the ansatz we used in the perturbation
analysis.  Moreover, we determined explicit perturbative (in size of
the standing wave solution) expression for the frequency of the
oscillatory modes.  This linear spectrum turned out, not surprisingly,
to be dispersive which explains the nonlinear evolution of perturbed
standing waves.  These behave in a similar fashion to perturbed
(stable) time-periodic solutions and scalar perturbations in spherical
perfectly reflecting cavity model with the Neumann boundary condition.
\vspace{0.4ex}

Next we examined, by means of the same methods, time-periodic
solutions for vacuum Einstein's equations using the cohomogenity-two
biaxial Bianchi IX ansatz.  Explicit exponential nonlinearity of the
field equations, in terms of the squashing field makes the
perturbative construction rather involved.  The perturbative expansion
contains both odd and even terms and all of the equations need to be
solved at each perturbative order.  Among that one additional
resonance---unremovable by the straightforward approach---appears at
the fifth perturbative order.  This induces a modification of lower
order solution by inclusion of a homogeneous term.  After
demonstrating the convergence of the numerical method and the
conformity of numerical and perturbative results, we studied the
stability problem of time-periodic solutions (here also the mass of
the solutions is bounded from above).  We confirmed the stability of
small amplitude time-periodic solutions.  However, large amplitude
solutions, exhibits more complex structure than less massive
solutions, also behave differently than their scalar 'counterparts'.
They do not collapse rapidly when perturbed slightly, and the
Kretschmann scalar monitored at the origin does not give any evidence
for the instability during long time evolution, but due to numerical
difficulties, stability of such configurations remains an open problem
requiring further investigation.
\vspace{0.4ex}

In a similar manner we analyzed two remaining models, the perfectly
reflecting spherical cavity model and the YM system, aimed mainly to
exploit various structures of small amplitude time-periodic solutions.
(The issue of reparametrization and detailed stability analysis of
constructed solutions are left for further studies.)  The linearized
spectra of these models can have different character---this affects
the form of the solutions (again, this concerns the perturbative
expansion, the numerical approach stays unaltered).  The analytical
construction relies heavily on the fact that resonances appear and
their successive removal fixes free parameters (either those in
homogeneous terms or the ones appearing as integration constants) in
successive perturbative orders.  One may naively think that for
dispersive linear spectrum resonances are absent and therefore no
time-periodic solutions exist then---we show that this is not the
case.
\vspace{0.4ex}

For the YM equation the resonant structure require homogeneous terms
to be included in solution at more than only one perturbative order
(for the kink topological sector with irregular structure of the
resonant set) or even at each perturbative order (for the vacuum
topological sector with regular structure of the resonant set) during
the construction.  Without such modifications the number of available
parameters will be insufficient to remove all of the appearing
resonances and so the procedure will break.  (We have used this
modification already for time-periodic solutions of the vacuum
Einstein equations.)  A notable exception is a family of solutions
bifurcating from the fundamental mode in the kink topological sector
(in this case the bifurcation frequency is an integer, while higher
mode frequencies are irrational numbers).  For that case the resonant
set is a singleton and so there is only one resonance present at each
perturbative order.  Surprisingly, the oscillation frequency for this
family of time-periodic solutions does not depend on the amplitude.
This is due to the fact that the structure of perturbative equations,
in that particular case, enforces the frequency corrections to vanish.
This special case and the presence, in other families of time-periodic
solutions, of homogeneous terms (especially their amplitudes) are
confirmed, with great accuracy, in the numerically constructed
solutions with extended precision arithmetics (in some cases it was
necessary to use higher than machine precision in order to verify
perturbative prediction).
\vspace{0.4ex}

In all of the above systems the first order approximation to the
time-periodic solution is a single eigenmode only.  This is also the
case for the cavity model with Neumann boundary condition.  However,
with Dirichlet boundary condition there are more resonances present
already at the third (first nontrivial) perturbative order and a
single free parameter (the frequency correction) does not suffice to
construct a bounded solution with just one mode at the linear level.
For that reason the first order solution has to be very specific
superposition of eigenmodes such that all of the resonances are
absent.  We verified this by truncating this infinite series and
comparing results (in principle convergence with the truncation order)
with the numerics.
\vspace{0.4ex}

To summarize, in this chapter we have presented details of
construction of time-periodic solutions to nonlinear systems of PDEs,
in particular to the Einstein equations with negative cosmological
constant (in the $1+1$ dimensional settings).  For each of the
considered models we used two independent methods which give
consistent results which demonstrated their correctness and give
credence to the existence of time-periodic solutions.  Additionally,
we analyzed constructed solutions and described their properties
(including in few cases the stability analysis).  These lead us to the
conclusion that the existence of time-periodic
solutions\footnote{Proved by explicit construction.}  to the nonlinear
nondissipative evolution equations on bounded domains seems to be a
rule rather than exception.  In such class of PDEs systems the
time-periodic solutions are dynamical counterparts of the static
solutions, thus their derivation (we demonstrated that this is not
very difficult) should be the first step in studies of nonlinear wave
equations on bounded domains.
\vspace{0.4ex}

In considered models the time-periodic solutions bifurcating from
linear modes can be found by means of general methods, both numerical
and perturbative.  Proposed techniques proved to be extremely
efficient; spectral methods in numerical calculations give accurate
results with moderate effort,\footnote{When the spectral convergence
  rate is observed.} the perturbative expansion procedure (preferably
with use of eigenbasis functions of associated linear operator),
assisted with CAS, produces high order approximation.  In addition, we
have shown how to discretize equations with the spectral
decomposition, and in particular how to solve the constraint equations
effectively.  The numerical code was used not only to find
time-periodic solutions but was also central part of the
time-evolution code.  Presented methods with use of symplectic
integrators lead to stable near energy preserving numerical scheme
allowing for long time evolutions.

\chapter{Summary and Outlook}
\label{cha:Summary}

In this closing chapter we provide summary of the thesis and give
several remarks on future work in the field.

\section{Summary}
\label{sec:SummarySummary}

In this thesis we studied dynamics of nonlinear waves on bounded
domains.  Because of our primary motivation, (in)stability problem of
AdS, we have devoted most space to the analysis of equations
describing perturbations of aAdS spaces.  We have demonstrated the
existence of (strictly) time-periodic solutions within each of
analyzed systems thus showing that time-periodic solutions are common
to the nondissipative nonlinear PDEs on bounded domains.  In
particular, we constructed time-periodic solutions to the vacuum
Einstein equations with negative cosmological constant under
cohomogenity-two biaxial Bianchi IX ansatz. Even though the analyzed
systems differ in details, time-periodic solutions can be constructed
by means of general procedures: the perturbative Poincar\'e-Lindstedt
method and Newton's root-finding numerical algorithm based on
pseudospectral discretization, as described in
Chapter~\ref{cha:Methods}.

We made extensive tests and detailed comparisons of our two
independent constructions and direct numerical evolution of initial
data corresponding to time-periodic solution.  All of them corroborate
the correctness of our approaches.  We analyzed the properties
constructed solutions, in particular we demonstrated their stability.
For the complex scalar field case of the EKG system we investigated in
more detail the linear stability of standing wave solutions. We showed
the dispersive character of the spectrum of their linear perturbations
from which we also infer the change of stability at the turning point
for any family of solutions.

Studying perfectly reflecting spherical cavity model we gave evidence
that turbulent dynamics is not an exclusive domain of aAdS spaces but
can be also observed in confined geometries with zero $\Lambda$ (see
also \cite{Okawa2014}).  Concerning the question how the character of
the linear spectrum affect the nonlinear dynamics, we analyzed by
weakly nonlinear perturbative expansion small amplitude initial data
and highly nonlinear regime by means of numerical solution of initial
or initial-boundary value problems.  In all considered cases with
dispersive spectra we observed spreading of the initial perturbation
which for small amplitude was preventing solution to collapse (for
self-gravitating models).  Moreover, we observed studying the YM
system (in perturbative calculations) that the resonances are equally
common in dispersive and nondispersive cases and that in both cases in
the long time evolution the energy cascade eventually stops.

Above that we have developed numerical methods intended for long and
stable time evolution of Einstein equations with negative cosmological
constant and related problems.  These methods are based on MOL
approach with pseudospectral discretization in space and RK time
integration.  We verified their robustness and in particular we
demonstrated superiority of symplectic ODE integrators.

Although our results are not rigorous (neither we provide proof of
existence of time-periodic solutions nor we prove the convergence of
derived perturbative series) we hope that the presented evidence leave
little doubt.

\section{Future work}
\label{sec:SummaryFutureWork}

This thesis focuses on only a few particular aspects of a very
interesting and broad area lying on the interface of General
Relativity and turbulence.  There are many further directions to
pursue or questions to answer, let alone attempts to prove the
existence of time-periodic solutions we have constructed.  Below we
enumerate some of them:

\begin{enumerate}[\itshape i\upshape)]
\item Do all time-periodic solutions of the EKG model on the unstable
  branch collapse to a black hole? (We have confirmed this only for
  several points on the unstable branch.)  Why do we not observe a
  similar behaviour (large amplitude oscillations) as in the complex
  scalar field case?

\item The results of stability studies of time-periodic solutions
  within the cohomogenity-two biaxial Bianchi IX ansatz are not
  conclusive; it would be particularly valuable to refine these
  results.

\item What is the long-time dynamics of perturbed unstable standing
  wave solutions?

\item Concerning the problem of stability of the time-periodic
  solutions, it would be interesting to see whether these solutions
  are stable with respect to perturbations outside the ansatz, e.g.
  by considering generalizations of the YM field to the sphaleron
  sector \cite{0264-9381-30-9-095009, PhysRevD.60.124011}, or by
  considering coupled Einstein-Yang-Mills system, or by relaxing the
  cohomogenity-two biaxial Bianchi IX ansatz to include triaxial case
  \cite{PhysRevLett.97.131101}.

\item Full understanding of the nonlinear evolution of solutions to
  the YM equation is certainly beyond the reach of our currently
  developed mathematical technology.  Therefore, it would be valuable
  to study the system (\ref{eq:237}) as a particular low dimensional
  approximation of the original equation.

\item Recently, there has been considerable interest in studies of the
  EKG system with a massive scalar field \cite{Holzegel20142436,
    Holzegel2013}.  In the context of AdS instability this seems to be
  'the next step' towards answering the question how the boundary
  conditions, imposed on the conformal boundary of the AdS space,
  affect global dynamics.  Under particular conditions, for massive
  scalar field, there is a freedom to consider more general boundary
  data \cite{10.1007/s10714-014-1724-0}, which require an extensive
  exploration.

\item Exploring the perfectly reflecting spherical cavity model one
  can study more systematically the effect of dispersive character of
  the linear spectrum on nonlinear dynamics either by considering
  system with angular momentum \cite{PhysRevD.76.124014} (which by a
  centrifugal barrier term introduces the discrete parameter
  controlling dispersion relation) or by going to higher dimensions
  (strikingly the $d=3$ case is special in a sense that with Dirichlet
  boundary condition the linear spectrum is equidistant
  (nondispersive), which is not the case for $d\neq 3$ both for
  Dirichlet and Neumann boundary conditions).

\item Since the analysis of the Einstein equations is particularly
  involved (both analytically and numerically) its seems useful to
  analyze simpler models.  The YM system considered here turned out to
  not to be the best model of the AdS dynamics (the energy transfer
  eventually stops, even though the spectrum is nondispersive).
  Alternative toy model would be semi-linear wave equation derived
  from the cohomogenity-two biaxial Bianchi IX ansatz by 'ignoring'
  the metric functions $A$ and $\delta$ (Section~\ref{sec:BCSModel}).
  The resulting PDE for the squashing field has a nondispersive
  spectrum and the nonlinearity of geometric nature.  Preliminary
  studies have shown the existence of time-periodic solutions and a
  blow-up phenomena for relatively small initial data.

\item Finally, with methods presented in this thesis one can
  effectively analyze (in particular perform the time evolution) cases
  when the turbulence is 'not active'.  For field configurations
  developing steep gradients, for which black hole formation or
  blow-up is expected, current methods are fairly inefficient.  Still
  more involved techniques like moving mesh methods \cite{ANU:5537300}
  may be required.

\end{enumerate}


\appendix
\chapter{Orthogonal polynomials}
\label{cha:AppOrthogonalPolynomials}

In this chapter we collect the most important facts about orthogonal
polynomials which appear in this thesis.  We refer to general
resources \cite{abramowitz1972handbook, olver2010nist, nistDigital,
  Fun} for more details; in connection with spectral methods we
especially recommend references \cite{boyd2001chebyshev,
  shen2011spectral}.

\section{The Jacobi polynomials}
\label{sec:AppJacobiPolynomials}

The Jacobi polynomials $P^{(\alpha,\beta)}_{n}(x)$ of degree
$n\in\mathbb{N}_{0}$, where both parameters $\alpha$, $\beta$ $>-1$,
are defined as regular solutions to the second order differential
equation
\begin{multline}
  \label{eq:586}
  (1-x^{2})P_{n}''(x) + \left[\left(\beta-\alpha\right) -
    \left(\alpha+\beta+2\right)x\right]P_{n}'(x)
  \\
  + n\left(n+\alpha+\beta+1\right)P_{n}(x) = 0\,.
\end{multline}
They form an important class of orthogonal polynomials in numerical
analysis.
\begin{enumerate}[\itshape i\upshape)]
\item Weight function
  \begin{equation}
    \label{eq:587}
    w(x) = (1-x)^{\alpha}(1+x)^{\beta}.
  \end{equation}
\item The orthogonality relation
  \begin{multline}
    \label{eq:588}
    \int_{-1}^{1}P^{(\alpha,\beta)}_{n}P^{(\alpha,\beta)}_{m}
    (1-x)^{\alpha}(1+x)^{\beta}\diff x
    = \\
    \frac{2^{\alpha+\beta+1}}{2n+\alpha+\beta+1}
    \frac{\Gamma(n+\alpha+1)\Gamma(n+\beta+1)}
    {n!\Gamma(n+\alpha+\beta+1)}\delta_{nm},
  \end{multline}
  $n,m\in\mathbb{N}_{0}$.
\item Customary normalization
  \begin{equation}
    \label{eq:589}
    P^{(\alpha,\beta)}_{n}(1) = \binom{n+\alpha}{n}.
  \end{equation}
\item First few polynomials
  \begin{subequations}
    \label{eq:590}
    \begin{align}
      P^{(\alpha,\beta)}_{0}(x) & = 1,
      \\
      P^{(\alpha,\beta)}_{1}(x) & = \frac{1}{2}(2+\alpha+\beta) x +
      \frac{1}{2}(\alpha-\beta),
      \\
      P^{(\alpha,\beta)}_{2}(x) & = \frac{1}{8} x^2 (\alpha +\beta +3)
      (\alpha +\beta +4)+\frac{1}{4} x \left(\alpha ^2+3 \alpha -\beta
        (\beta +3)\right)
      \\ \nonumber
      & \quad + \frac{1}{8} \left(\alpha ^2-\alpha (2 \beta +1)+\beta
        ^2-\beta -4\right).
    \end{align}
  \end{subequations}
  General formula
  \begin{equation}
    \label{eq:591}
    P^{(\alpha,\beta)}_{n}(x) = 2^{-n}\sum_{k=0}^{n}
    \binom{n+\alpha}{k}\binom{n+\beta}{n-k}(x-1)^{n-k}(x+1)^{k}\,.
  \end{equation}
\item Three-term recursion formula
  \begin{equation}
    \label{eq:592}
    P^{(\alpha,\beta)}_{n}(x) =
    c_{1}^{-1}\left(\left(c_{2}x+c_{3}\right)P^{(\alpha,\beta)}_{n-1}(x)
      - c_{4}P^{(\alpha,\beta)}_{n-2}(x)\right),
  \end{equation}
  for $n\geq 2$, with $n=0,1$ given by (\ref{eq:590}), where
  \begin{subequations}
    \label{eq:593}
    \begin{align}
      c_{1} &= 2n(n+\alpha+\beta)(2n+\alpha+\beta-2),
      \\
      c_{2} &=
      (2n+\alpha+\beta-1)(2n+\alpha+\beta-2)(2n+\alpha+\beta),
      \\
      c_{3} &= (2n+\alpha+\beta-1)(\alpha^{2}-\beta^{2}),
      \\
      c_{4} &= 2(n+\alpha-1)(2n+\alpha+\beta)(n+\beta-1)\,.
    \end{align}
  \end{subequations}
\item Derivative
  \begin{equation}
    \label{eq:594}
    \frac{\diff}{\diff x}P^{(\alpha,\beta)}_{n}(x) =
    \frac{1}{2}(n+\alpha+\beta+1) P^{(\alpha+1,\beta+1)}_{n-1}(x)\,.
  \end{equation}
\item Relation with hypergeometric functions
  \begin{equation}
    \label{eq:595}
    \, _2F_1(a,b;c;z) = \frac{\Gamma(c) \Gamma(b-c+1)}{\Gamma(b)}
    (1-z)^{c-a-b} P_{b-c}^{(c-1,c-a-b)}(1-2 z)\,.
\end{equation}
\item Integral \cite{fun:PIntegral}
  \begin{multline}
    \label{eq:596}
    \int (z-1)^{c}P^{(a,b)}_{n}(z) \diff z = (z-1)^{c+1}
    \frac{\Gamma (c+1) \Gamma (a+n+1)}{\Gamma (n+1)}
    \times
    \\
    \,
    _3\tilde{F}_2\left(-n,a+b+n+1,c+1;a+1,c+2;\frac{1-z}{2}\right),
  \end{multline}
  where $_3\tilde{F}_2\left(a_{1},a_{2},a_{3};b_{1},b_{2};z\right)$ is
  regularized hypergeometric function \cite{Fun:tildeF}.
\item Zeros of $P^{(\alpha,\beta)}_{N+1}(x)$ are the eigenvalues of
  the symmetric tridiagonal matrix (see
  \cite[p.~55,~84]{shen2011spectral})
  \begin{equation}
    \label{eq:597}
    \begin{pmatrix}
      a_{0} & b_{1} & & & \\
      b_{1} & a_{1} & b_{2} & & \\
      & \ddots & \ddots & \ddots & \\
      & & b_{N-1} & a_{N-1} & b_{N} \\
      & & & b_{N} & a_{N}
    \end{pmatrix},
  \end{equation}
  where
  \begin{align}
    a_{n} &= \frac{\beta ^2-\alpha ^2}{(\alpha +\beta +2 n) (\alpha
      +\beta +2 n+2)}\,,
    \\
    b_{n} &= \frac{2}{(\alpha +\beta +2 n)} \sqrt{\frac{n (\alpha +n)
        (\beta +n) (\alpha +\beta +n)} {(\alpha +(\beta -1)+2
        n)(\alpha +\beta +2 n+1)}}\,,
  \end{align}
  are related to the three-term recursion coefficients (\ref{eq:592}).
\end{enumerate}

\section{The Chebyshev polynomials}
\label{sec:AppChebyshevPolynomials}

Chebyshev polynomials $T_{n}(x)$ are orthogonal polynomials with
respect to the weight function $w(x)=(1-x^{2})^{-1/2}$ solving the
following differential equation
\begin{equation}
  \label{eq:598}
  \left(1-x^{2}\right)T_{n}''(x) - x T_{n}'(x) + n^{2}T_{n}(x) = 0\,.
\end{equation}
\begin{enumerate}[\itshape i\upshape)]
\item Orthogonality property
  \begin{equation}
    \label{eq:599}
    \int_{-1}^{1}T_{n}(x)T_{m}(x) \frac{1}{\sqrt{1-x^{2}}}\diff x =
    \begin{cases}
      {\displaystyle  \frac{\pi}{2}\delta_{nm} } & \text{for}
      \ n,m\neq 0, \\[2ex]
      {\displaystyle \pi} & \text{for}\ n=m=0\,.
    \end{cases}
  \end{equation}
\item Trigonometric definition
  \begin{equation}
    \label{eq:600}
    T_{n}(x) = \cos(n\arccos{x})\,.
  \end{equation}
\item First few polynomials
  \begin{subequations}
    \label{eq:601}
    \begin{align}
      T_{0}(x) &= 1,
      \\
      T_{1}(x) &= x,
      \\
      T_{2}(x) &= 2x^{2} - 1\,.
    \end{align}
  \end{subequations}
\item Special values of $T_{n}(x)$
  \begin{equation}
    \label{eq:602}
    T_{n}(1) = 1,
  \end{equation}
  \begin{equation}
    \label{eq:603}
    T_{2n}(0) = (-1)^{n}, \quad T_{2n+1}(0) = 0,
  \end{equation}
  \begin{equation}
    \label{eq:604}
    T_{n}(-1) = (-1)^{n}\,.
  \end{equation}
\item Integration of Chebyshev polynomials
  \begin{equation}
    \label{eq:605}
    \int T_{0}(x) \diff x = T_{1}(x),
  \end{equation}
  \begin{equation}
    \label{eq:606}
    \int T_{1}(x) \diff x = \frac{T_{2}(x)}{4},
  \end{equation}
  \begin{equation}
    \label{eq:607}
    \int T_{n}(x) \diff x = \frac{1}{2}\left(\frac{T_{n+1}(x)}{n+1}
      -\frac{T_{n-1}(x)}{n-1}\right), \quad n \geq 2,
  \end{equation}
  \begin{equation}
    \label{eq:608}
    \int_{0}^{1} T_{2n}(x) \diff x = \frac{1}{1-4n^{2}}\,.
  \end{equation}
\item For an even function $f:[-1,1]\mapsto\mathbb{R}$ which has the
  expansion in terms of even Chebyshev polynomials only
  \begin{equation}
    \label{eq:609}
    f(x) = \sum_{n\geq 0} \hat{f}_{n}T_{2n}(x),
  \end{equation}
  a value of $f(x)$ at $x=0$ can be computed referring to the
  expansion coefficients $\hat{f}_{n}$ only, from (\ref{eq:603}) we
  have
  \begin{equation}
    \label{eq:610}
    f(0) = \sum_{n\geq 0}\hat{f}_{2n} - \sum_{n\geq 0}\hat{f}_{2n+1}\,.
  \end{equation}
\item Zeros $\{x_{j}\,|\,\,T_{N+1}(x_{j})=0,\ j=0,1,\ldots,N\}$
  \begin{equation}
    \label{eq:611}
    x_{k} = \cos\left(\frac{k+1/2}{N+1}\pi\right),
    \ k=0,1,\ldots,N,
  \end{equation}
  (known as Chebyshev points of the first kind or Chebyshev Gauss
  points).
\item Extrema $\{x_{j}\cup\{-1,1\}\,|\,\,T_{N}'(x_{j})=0,\
  j=1,\ldots,N-1\}$
  \begin{equation}
    \label{eq:612}
    x_{k} = \cos\left(\frac{k}{N}\pi\right), \quad k=0,1,\ldots,N,
  \end{equation}
  (these are known as Chebyshev points of the second kind or
  Chebyshev-Gauss-Lobatto points).
\item Relation with Jacobi polynomials
  \begin{equation}
    \label{eq:613}
    P^{(-1/2,-1/2)}_{n}(x) = \frac{1}{4^{n}}\binom{2n}{n}T_{n}(x)\,.
  \end{equation}
\end{enumerate}

\section{The Legendre polynomials}
\label{sec:legendre-polynomials}

Legendre polynomials $P_{n}(x)$ are regular solutions to the
Legendre's differential equation
\begin{equation}
  \label{eq:614}
  (1-x^{2})P_{n}''(x) -2xP_{n}'(x) + n(n+1)P_{n}(x) = 0\,.
\end{equation}
They are a special case of the Jacobi polynomials with
$\alpha=\beta=0$ (subclass of the ultraspherical polynomials).

\begin{enumerate}[\itshape i\upshape)]
\item Orthogonality property
  \begin{equation}
    \label{eq:615}
    \int_{-1}^{1}P_{n}(x)P_{m}(x) \diff x = \frac{2}{2n+1}\delta_{nm}
    \quad n,m\in\mathbb{N}_{0}\,.
  \end{equation}
\item Rodrigues representation
  \begin{equation}
    \label{eq:616}
    P_{n}(x) = \frac{1}{2^{n}n!}\frac{\diff^{n}}{\diff x^{n}}
    \left[(x^{2}-1)^{n}\right]\,.
  \end{equation}
\item First few polynomials
  \begin{subequations}
    \label{eq:617}
    \begin{align}
      P_{0}(x) & = 1,
      \\
      P_{1}(x) & = x,
      \\
      P_{2}(x) & = \frac{1}{2}\left(3x^{2}-1\right)\,.
    \end{align}
  \end{subequations}
  General formula
  \begin{equation}
    \label{eq:618}
    P_{n}(x) = \frac{1}{2^{n}}\sum_{k=0}^{\left\lfloor n/2\right\rfloor}
    (-1)^{k}\binom{n}{k}\binom{2n-2k}{n}x^{n-2k}\,.
  \end{equation}
\item Gauss-Legendre quadrature approximates the integral
  \begin{equation}
    \label{eq:619}
    I(f) = \int_{-1}^{1}f(x)\diff x,
  \end{equation}
  for smooth function $f:[-1,1]\mapsto\mathbb{R}$, by the finite sum
  \begin{equation}
    \label{eq:620}
    S_{N+1}(f) = \sum_{k=0}^{N}w_{k} f(x_{k}),
  \end{equation}
  where the weights $w_{k}$ are
  \begin{equation}
    \label{eq:621}
    w_{k} = - \frac{2}{(N+2)P_{N+2}(x_{k})P_{N+1}'(x_{k})},
  \end{equation}
  and the abscissas (quadrature points) are the zeros of $P_{N+1}(x)$.
  The error of Gauss-Legendre quadrature is
  \begin{equation}
    \label{eq:622}
    I(f) - S_{N+1}(f) = \frac{2^{2N+3}\left[\left(N+1\right)!\right]^{4}}
    {(2N+3)\left[\left(2N+2\right)!\right]^{3}}f^{(2N+3)}(\xi),
    \quad -1<\xi<1\,.
  \end{equation}
  From this it follows that the Gauss-Legendre quadrature is exact for
  polynomials of order $2N+1$.
\end{enumerate}

\chapter{Polynomial pseudospectral methods in spherical symmetry}
\label{cha:polyn-pseud-meth}

In this chapter we point out difficulties of using spectral algorithms
with spherical coordinates and give the standard solution using
Chebyshev polynomials which avoids singularity at the origin.

\section{Chebyshev method}
\label{sec:spectr-chebysh-meth}

The Chebyshev polynomials (see
Section~\ref{sec:AppChebyshevPolynomials}) are one of the most used
class of orthogonal polynomials in spectral methods.  They are not
only distinguished by the analytic form of the quadrature nodes and
weights but also by their simple form and the connection with the
Fourier series makes them particularly important in the context of
spectral methods.

We assume that $u(x)$ is a smooth function (defined on interval
$x\in[-1,1]$), which is well approximated in the finite $(N+1)$
dimensional subspace of
$\mathsf{B}_{N}=\text{span}\left\{x^{j}\,\left.\right|\
  j=0,1,\ldots,N\right\},$
that satisfy prescribed boundary conditions.  Taking Chebyshev
polynomials as basis functions of $\mathsf{B}_{N}$, any function can
be approximated as
\begin{equation}
  \label{eq:623}
  \mathcal{I}_{N}u(x) = \sum_{j=0}^{N}\hat{u}_{j}T_{j}(x)\,.
\end{equation}
The expansion coefficients $\hat{u}_{j}$ can be effectively computed
using the trigonometric relation (\ref{eq:600}) with Fast Fourier
Transform (FFT) algorithm, see e.g. \cite{boyd2001chebyshev}.
Alternatively these can be computed using either quadrature formulae
or by solving the linear algebraic system
\begin{equation}
  \label{eq:624}
  u_{i} = \sum_{j=0}^{N}\hat{u}_{j}T_{j}(x_{i}), \quad i=0,1,\ldots,N\,,
\end{equation}
where $u_{i}=u(x_{i})$ and $x_{i}$ are $N+1$ suitably chosen grid
points.

The expansion (\ref{eq:623}) can be written using the Lagrange
interpolation polynomial
\begin{equation}
  \label{eq:625}
  \mathcal{I}_{N}u(x) = \sum_{j=0}^{N}u_{j}\ell_{j}(x)\,,
\end{equation}
which is mathematically equivalent to (\ref{eq:623}) when the grid
points are chosen appropriately \cite{hesthaven2007spectral}.  This
polynomial interpolation representation is particularly convenient
while working with (anti)symmetric functions (as is the case in
spherical symmetry, see discussion in the following section).  For any
choice of $(N+1)$ nodes $x_i$ the Lagrange interpolating polynomials
are
\begin{equation}
  \label{eq:626}
  \ell_{j}(x) = \frac{{\displaystyle
      \prod_{\substack{k=0\\k\neq j}}^{N}\left(x-x_{k}\right)}}
  {{\displaystyle
      \prod_{\substack{j=0\\j\neq k}}^{N}\left(x_{j}-x_{k}\right)}}\,,
\end{equation}
with the property
\begin{equation}
  \label{eq:627}
  \ell_{j}(x_{k}) = \delta_{jk}, \quad j,k=0,1,\ldots,N\,.
\end{equation}
The formulae (\ref{eq:625}) and (\ref{eq:626}) (referred to as the
Lagrange form) are usually mentioned in numerical analysis literature,
but (\ref{eq:626}) is neither computationally efficient nor
numerically stable \cite{doi:10.1137/S0036144502417715}.  Therefore,
following \cite{doi:10.1137/S0036144502417715} we prefer the more
optimal form of interpolating polynomial.  Introducing weights
\begin{equation}
  \label{eq:628}
  w_{j} = \frac{1}
  {{\displaystyle
      \prod_{\substack{j=0\\j\neq k}}^{N}\left(x_{j}-x_{k}\right)}}\,,
\end{equation}
we rearrange (\ref{eq:625}) and (\ref{eq:626}) to the following
symmetric form
\begin{equation}
  \label{eq:629}
  \mathcal{I}_{N}u(x) = \frac{{\displaystyle
      \sum_{j=0}^{N} \frac{w_{j}}{x-x_{j}}}u_{j}}
  {{\displaystyle
      \sum_{j=0}^{N} \frac{w_{j}}{x-x_{j}}}}\,,
\end{equation}
which is referred as barycentric formula.

The weights (\ref{eq:628}) can be calculated analytically on few sets
of grid points.  Specifically for Chebyshev points of the second kind
(\ref{eq:612}), used in our numerical codes, these are given
explicitly
\begin{equation}
  \label{eq:630}
  w_{0}=\frac{c}{2},\quad w_{N}=(-1)^{N}\frac{c}{2},
  \quad  w_{k}=(-1)^{k}c, \quad k=1,2,\ldots,N-1,
\end{equation}
(note that these are not unique, but due to the special form of
(\ref{eq:629}) the common factor $c$ cancels out).  We refer to
(\ref{eq:630}) as Chebyshev (barycentric) weights.

Further, using the interpolating representation (\ref{eq:629}),
derivatives of the interpolant at the grid points can be computed
using the differentiation matrices (as in the standard pseudospectral
method), so
\begin{equation}
  \label{eq:631}
  \frac{\diff^{n}}{\diff x^{n}} \mathcal{I}_{N}u(x_{j}) =
  \sum_{k=0}^{N}D^{(n)}_{jk} u_{k}\,.
\end{equation}
The element $(j,k)$-th of $n$-th order differentiation matrix
$D^{(n)}$ ($n\geq 1$) can be expressed by the following recurrence
relation (the hybrid formula)
\begin{equation}
  \label{eq:632}
  D_{jk}^{(n)} =
  \begin{cases}
    {\displaystyle
     \ \frac{n}{x_{j}-x_{k}}\left(\frac{w_{k}}{w_{j}}D_{jj}^{(n-1)} -
        D_{jk}^{(n-1)}\right)}& \text{if $j\neq k$}, \\
    & \\
    {\displaystyle
      \ - \sum_{\substack{l=0 \\ l\neq j}}^{N}D_{jl}^{(n)}}
    & \text{if $j=k$},
  \end{cases}
\end{equation}
for $j,k=0,1,\ldots,N$, where $D_{jk}^{(0)}=\id_{jk}$ is the identity
matrix.

\section{Spherical symmetry}
\label{sec:spherical-symmetry}

Each of the problems encountered in this thesis suffer for the
singularity of the spherical coordinate system.  In most cases we use
the basis functions (the eigenbasis associated with the linear
operator) for the expansion of approximated functions.  These
functions are regular at the origin and also convenient to use.  But
not always they are compatible with the boundary expansion at the
outer boundary for the nonlinear problem.  Therefore other set of
functions should be used and these typically are Chebyshev polynomials
\cite{boyd2001chebyshev}.

The difficulty in using polynomial spectral methods in spherical
coordinate systems is twofold.  First, the equations written in
spherical coordinate system are typically singular at the origin
(especially those containing Laplacian operator).  Second, for
polynomial interpolation methods the grid points are clustering near
the domain boundaries which for the time-dependent problems severely
restricts time steps taken (because of stability restrictions).  These
issues can be easily resolved by noting that the singularity of the
Laplace operator written in spherical coordinates is just an apparent
singularity.  The solution to a differential equation is usually
smooth at the origin $r=0$ and it should be taken into account in
numerical calculations.  Additionally, more grid points in the region
where the approximated function is smooth is usually unnecessary.

While many resolutions for the spectral methods in spherical (and
related) coordinate systems have been proposed for various problems
(see, e.g. \cite[Chapter~18]{boyd2001chebyshev}) we prefer to use the
'double covering' method \cite{trefethen2000spectral}.  We demonstrate
this approach by approximating and computing derivatives of smooth
function defined on the unit radial interval $r\in[0,1]$.  To weaken
the coordinate singularity at the origin $r=0$ we extend the radial
coordinate to $r\in [-1,1]$ (without rescaling) and use the
(anti)symmetry (with respect to $r=0$) of approximated functions to
reduce actual calculations to within $[0,1]$.  We assume
$u:[-1,1]\mapsto\mathbb{R}$ be a smooth function such that
\begin{equation}
  \label{eq:633}
  u(-r)=(-1)^{p}u(r)\,,
\end{equation}
holds, with $p=0$ or $p=1$ for even or odd cases respectively.  Then
we take $(2N+1)$ Chebyshev grid points (\ref{eq:612})
\begin{equation}
  \label{eq:634}
  \tilde{x}_{k} = \cos\left(\frac{k\pi}{2N+1}\right), \quad
  k=0,1,\ldots,2N+1,
\end{equation}
covering the interval $[-1,1]$ and composing the computational grid
such that, there are exactly $N+1$ points of the 'physical' part of
the grid ($r\geq 0$), i.e.
\begin{equation}
  \label{eq:635}
  x_{k} = \tilde{x}_{k} = \cos\left(\frac{k\pi}{2N+1}\right), \quad
  k=0,1,\ldots,N.
\end{equation}
(Here we adapt a convention that the quantities defined on a
computational grid are denoted by over tilde while these on the
'physical' part of the grid are not.)  In this way we exclude the
origin from our scheme, since $x_{N}=\cos(N\pi/(2N+1))\ra 0$ only
asymptotically $N\ra\infty$ and there is no clustering at $r=0$, so we
do not need to impose 'pole conditions' to ensure regularity.

A finite-dimensional polynomial interpolation of $u(r)$ is
(cf. (\ref{eq:629}))
\begin{equation}
  \label{eq:636}
  \mathcal{I}_{2N+2}u(r) = \sum_{k=0}^{2N+1}u(\tilde{x}_{k})\ell_{k}(r)\,.
\end{equation}
and the $n$-th derivative of the interpolant of $u(r)$ at the
computational nodes can be computed using the differentiation matrices
(\ref{eq:632}).  Denoting by $\tilde{u}_{k}=u(\tilde{x}_{k})$ and
$\tilde{u}^{(n)}_{k}=u^{(n)}(\tilde{x}_{k})$ the values of function
and its $n$-th derivative at the grid nodes $\tilde{x}_{k}$
respectively, we have
\begin{equation}
  \label{eq:637}
  \begin{pmatrix}
    \tilde{u}^{(n)}_{0} \\
    \vdots \\
    \tilde{u}^{(n)}_{N} \\
    \tilde{u}^{(n)}_{N+1} \\
    \vdots \\
    \tilde{u}^{(n)}_{2N+1}
  \end{pmatrix}
  =
  \begin{pmatrix}
    \\
    & \tilde{D}^{(n)}_{\,++} & & \tilde{D}^{(n)}_{\,+-} & \\
    \\
    & \tilde{D}^{(n)}_{\,-+} & & \tilde{D}^{(n)}_{\,--} & \\
    & & & &
  \end{pmatrix}
  \begin{pmatrix}
    \tilde{u}_{0} \\
    \vdots \\
    \tilde{u}_{N} \\
    \tilde{u}_{N+1} \\
    \vdots \\
    \tilde{u}_{2N+1}
  \end{pmatrix},
\end{equation}
where we explicitly divide the $2(N+1)\times 2(N+1)$ matrix
$\tilde{D}^{(n)}$ (composed of $\tilde{x}_{k}$ and the barycentric
weights: $\tilde{w}_{0}=1/2$, $\tilde{w}_{2N+1}=-1/2$,
$\tilde{w}_{k}=(-1)^{k}$, $k=1,2,\ldots,2N$) into four
$(N+1)\times(N+1)$ blocks.  Since we are interested in the values of
the derivative at the 'physical' part of the grid only, i.e. in
$\tilde{u}_{k}$, $k=0,\ldots,N$, we rewrite (\ref{eq:637}) as
\begin{equation}
  \label{eq:638}
  \begin{pmatrix}
    \tilde{u}^{(n)}_{0} \\
    \vdots \\
    \tilde{u}^{(n)}_{N}
  \end{pmatrix}
  =
  \left(\tilde{D}^{(n)}_{\,++} + (-1)^{p}\tilde{D}^{(n)}_{\,+-}\right)
  \begin{pmatrix}
    \tilde{u}_{0} \\
    \vdots \\
    \tilde{u}_{N}
  \end{pmatrix},
\end{equation}
where we have used (anti)symmetry of the function $u(r)$.  In this way
we reduce the number of floating-point operations (thus the
computational complexity) by the factor of two when computing the
derivatives (the necessary matrices in (\ref{eq:638}) are calculated
only once at the initialization phase).  To simplify notation we adapt
the following convention to the differentiation matrices of
(anti)symmetric functions
\begin{equation}
  \label{eq:639}
  D^{(n,\pm)}_{i,j} = \tilde{D}^{(n)}_{++,ij} \pm \tilde{D}^{(n)}_{+-,ij} =
  \tilde{D}^{(n)}_{ij} \pm \tilde{D}^{(n)}_{i,N+1+j},
  \quad i,\,j=0,1,\ldots,N,
\end{equation}
with plus sign for $p=0$ and minus sign for $p=1$ in (\ref{eq:633}).

Whenever the value of the function $u(r)$ at $r$ which is not a grid
node ($r\neq \tilde{x}_{k}$, for any $k=0,1,\ldots,2N+1$) is needed we
use the barycentric interpolation formula (\ref{eq:629}).  For that we
use the computational grid (\ref{eq:635}) with corresponding
barycentric weights and the function values for $r<0$ at the grid
nodes given by the (anti)symmetry.  Then the barycentric formula reads
\begin{equation}
  \label{eq:640}
  u(r) =
  \frac{{\displaystyle
      \sum_{k=0}^{2N+1}\frac{\tilde{w}_{k}}{r-\tilde{x}_{k}}\tilde{u}_{k}}}
  {{\displaystyle\sum_{k=0}^{2N+1}\frac{\tilde{w}_{k}}{r-\tilde{x}_{k}}}}\,.
\end{equation}
For example, the value of the symmetric function at the origin $r=0$
is
\begin{equation}
  \label{eq:641}
  u(0) = \frac{{\displaystyle
      \sum_{k=0}^{2N+1}\frac{\tilde{w}_{k}}{\tilde{x}_{k}}\tilde{u}_{k}}}
  {{\displaystyle\sum_{k=0}^{2N+1}\frac{\tilde{w}_{k}}{\tilde{x}_{k}}}} =
  \frac{{\displaystyle \sum_{k=0}^{N}\frac{w_{k}}{x_{k}}u_{k}}}
  {{\displaystyle\sum_{k=0}^{N}\frac{w_{k}}{x_{k}}}}\,,
\end{equation}
since $w_{k}=\tilde{w}_{k}=-\tilde{w}_{2N+1-k}$ and
$x_{k}=\tilde{x}_{k}=-\tilde{x}_{2N+1-k}$ holds for $k=0,1,\ldots,N$
and from symmetry of $u(r)$ we have $u_{k} = \tilde{u}_{k} =
\tilde{u}_{2N+1-k}$ for $k=0,1,\ldots,N$.

Note that this approach is equivalent to (\ref{eq:610}) where the
coefficients of even Chebyshev polynomials expansion are used.  In the
computations presented in this thesis we prefer to use the barycentric
formulation with the differentiation matrices approach because this is
convenient in solving the constraints (elliptic PDEs), moreover for
relatively small grids we can reduce the complexity of the algorithm
computing derivatives by the factor $4\log{4}$ compared to commonly
used FFT based differentiation.

\chapter{Runge-Kutta methods}
\label{cha:runge-kutta-methods}

In this chapter we consider one-step methods for the systems of ODEs.
In particular we discuss the Runge-Kutta methods.  Here we give only
necessary definitions and state key theorems characterizing the
methods.  We motivate the use of specialized methods for the problems
considered in this thesis.

For proofs and more details concerning methods for ODE systems we
refer to specialized series of books \cite{hairer2008solving,
  hairer1996solving, hairer2006geometric} and references therein to
research papers.

\section{Definition}
\label{sec:DefinitionRK}

\begin{mydef}[The Runge-Kutta methods]
  Let $b_{i}$, $a_{ij}$ ($i,j=1,\ldots,s$) be real numbers and let
  $c_{i}$ be defined by
  \begin{equation}
    \label{eq:642}
    c_{i} = \sum_{j=1}^{s}a_{ij}\,.
  \end{equation}
  The method
  \begin{equation}
    \label{eq:643}
    \begin{aligned}
      k_{i} &= f\left(x_{0} + c_{i}h, y_{0} +
        h\sum_{j=1}^{s}a_{ij}k_{j}\right),
      \quad i=1,\ldots,s,
      \\
      y_{1} &= y_{0} + h\sum_{i=1}^{s}b_{i}k_{i},
    \end{aligned}
  \end{equation}
  is called an s-stage Runge-Kutta method for ODE system
  \begin{equation}
    \label{eq:644}
    y'=f(x,y), \quad y(x_{0})=y_{0},
  \end{equation}
  with sufficiently well behaved
  $f: [x_{0},\infty)\times\mathbb{R}^{n}\mapsto\mathbb{R}^{n}$
  ($\,'\equiv \diff/\diff x$).  When $a_{ij}=0$ for $i\leq j$ we have
  an explicit (ERK) method. If $a_{ij}=0$ for $i<j$ and at least one
  $a_{ii}\neq 0$, we have an diagonal implicit Runge-Kutta method
  (DIRK). If in addition all diagonal elements are identical
  ($a_{ii}=\gamma$ for $i=1,\ldots,s$), we speak of a singly diagonal
  implicit (SDIRK) method. In all other cases we speak of an implicit
  Runge-Kutta method (IRK).
\end{mydef}

The coefficients $b_{i}$, $a_{ij}$ and $c_{i}$ of RK methods are
commonly listed in a table (the Butcher tableau)
\begin{equation}
  \label{eq:645}
  \begin{array}{c|cccc}
    c_{1}   & a_{11} & a_{12} & \cdots & a_{1s} \\
    c_{2}   & a_{21} & a_{22} & \cdots & a_{2s} \\
    \vdots & \vdots & \vdots & \ddots & \vdots \\
    c_{s}   & a_{s1} & a_{s2} & \cdots & a_{ss} \\[1ex] \hline
    \rule{0pt}{2,5ex}
            & b_{1} & b_{2} & \cdots & b_s \\
  \end{array}
\end{equation}

\begin{mydef}
  A Runge-Kutta method (\ref{eq:643}) has order $p$ if for
  sufficiently smooth problems (\ref{eq:644}), the Taylor series for
  the exact solution $y(x_{0}+h)$ and for $y_{1}$ coincide up to (and
  including) the term $h^{p}$, i.e.
  \begin{equation}
    \label{eq:646}
    \left\| y(x_{0}+h)-y_{1}\right\| \leq C h^{p+1},
  \end{equation}
  holds with some $C\in\mathbb{R}$.
\end{mydef}

\section{Explicit methods}
\label{sec:AppERK}

The ease of use of explicit methods makes them particularly popular
(this does not mean this is always the optimal choice).  One of the
best known ERK method is the classical fourth order $p=4$
\begin{equation}
  \label{eq:ButcherRK4}
  \begin{array}{c|cccc}
    0 & & & & \\[1ex]
    \frac{1}{2} & \frac{1}{2} & & & \\[1ex]
    \frac{1}{2} & 0 & \frac{1}{2} & & \\[1ex]
    1 & 0 & 0 & 1 & \\[1ex]
    \hline
    \rule{0pt}{3,3ex}
    & \frac{1}{6} & \frac{1}{3} & \frac{1}{3} & \frac{1}{6}
  \end{array}
\end{equation}
known from its balance between the cost and accuracy.  Many higher
order methods where constructed over the years of research, but they
are necessary much more computationally costly, since the number of
stages ($s$ in formulas (\ref{eq:642}) and (\ref{eq:643})) rapidly
increases with $p$ (in fact for $p\geq 5$ no ERK method exists of
order $p$ with $s=p$ stages).

It is often advantageous to continue integration with dynamically
changing step size $h$.  This is either done by using Richardson
extrapolation or by using RK formulae which contain two numerical
approximations for $y(x_{0}+h)$ where their difference yields an
estimate of the local error which is then used for step size control.
The embedded ERK methods are characterized by the Butcher tableau
\begin{equation}
  \label{eq:647}
  \begin{array}{c|ccccc}
    0 & & & & & \\
    c_{2} & a_{21} & & & & \\
    c_{3} & a_{31} & a_{32} & & & \\
    \vdots & \vdots & \ddots & & & \\
    c_{s} & a_{s1} & a_{s2} & \cdots & a_{s,s-1} & \\[1ex] \hline
    \rule{0pt}{2.7ex} & b_{1} & b_{2} & \cdots & b_{s-1} & b_{s} \\
    \rule{0pt}{2.7ex}
    & \hat{b}_{1} & \hat{b}_{2} & \cdots & \hat{b}_{s-1} & \hat{b}_{s}
  \end{array}
\end{equation}
such that
\begin{equation}
  \label{eq:648}
  y_{1} = y_{0} + h \sum_{i=1}^{s}b_{i}k_{i},
\end{equation}
is of order $p$, and
\begin{equation}
  \label{eq:649}
  \hat{y}_{1} = y_{0} + h \sum_{i=1}^{s}\hat{b}_{i}k_{i},
\end{equation}
is of order $\hat{p}$ (usually $\hat{p}=p-1$ or $\hat{p}=p+1$).

The ERK method used in this thesis is the Dormand-Prince $s=7$ method
with $p=5$, $\hat{p}=4$ (known as DOPRI5), which has the following
coefficients
\begin{equation}
  \label{eq:ButcherTabDormandPrince}
  \begin{array}{c|ccccccc}
    0 & & & & & & & \\[1ex]
    \fracn{1}{5} & \fracn{1}{5} & & & & & & \\[1ex]
    \fracn{3}{10} & \fracn{3}{40} & \fracn{9}{40} & & & & & \\[1ex]
    \fracn{4}{5} & \fracn{44}{45} & -\fracn{56}{15} & \fracn{32}{9} & & & & \\[1ex]
    \fracn{8}{9} & \fracn{19372}{6561} & -\fracn{25360}{2187} & \fracn{64448}{6561} & -\fracn{212}{729} & & & \\[1ex]
    1 & \fracn{9017}{3168} & -\fracn{355}{33} & \fracn{46732}{5247} & \fracn{49}{176} & -\fracn{5103}{18656} & &  \\[1ex]
    1 & \fracn{35}{384} & 0 & \fracn{500}{1113} & \fracn{125}{192} & -\fracn{2187}{6784} & \fracn{11}{84} & \\[1ex]
    \hline
    \rule{0pt}{3,3ex} & \fracn{35}{384} & 0 & \fracn{500}{1113} & \fracn{125}{192} & -\fracn{2187}{6784} & \fracn{11}{84} & 0 \\
    \rule{0pt}{3,3ex} & \fracn{5179}{57600} & 0 & \fracn{7571}{16695} & \fracn{393}{640} & -\fracn{92097}{339200} & \fracn{187}{2100} & \fracn{1}{40}
  \end{array}
\end{equation}
The characteristic feature of the Dormand-Price methods
\cite{Dormand198019} is that they have a minimal error coefficients of
the higher order result ($p=5$ here) which is then used as numerical
solution (as opposed to Fehlberg methods \cite{fehlberg1969low} which
use the lower order approximation as an initial value for the next
step).

The part of the computations in this thesis using
(\ref{eq:ButcherTabDormandPrince}) where obtained by adapting the
\codenamestyle{DOPRI5} routine of the \prognamestyle{FORTRAN} code
\footnote{The source codes including also few usage examples are
  available at the webpage \url{http://goo.gl/oDPoN9}.}  implementing
the adaptive step method (\ref{eq:ButcherTabDormandPrince}).

\section{Implicit methods}
\label{sec:AppIRK}

For implicit methods, the $k_{1},\ldots,k_{s}$ in (\ref{eq:643}) are
not given explicitly (as is for ERK).  The fully implicit method of
$s$ stages constitute of $s\times n$ equations to be solved at each
step (for ODE system of size $n$).  This makes the IRK more
complicated to implement and simultaneously more expensive to use than
the ERK methods.  Nevertheless, the properties like larger domains of
stability make IRK especially useful for stiff equations, where the
possibility to make larger step sizes compensates the cost of solving
nonlinear system for $k_{i}$'s.  Moreover, the importance of some IRK
for Hamiltonian systems, where quality is more important than
accuracy, is discussed later.

One of the class of IRK are the collocation methods, and in particular
Gauss methods (known also as Gauss-Legendre RK methods), which are
collocation methods based on the Gaussian quadrature formulae, i.e.
the $c_{1},\ldots,c_{s}$ in (\ref{eq:656}) are the zeros of the
shifted Legendre polynomial (see
Section~\ref{sec:legendre-polynomials}) of degree $s$
\begin{equation}
  \label{eq:650}
  P_{s}(2x-1) \sim
  \frac{\diff^{s}}{\diff x^{s}}\bigl(x^{s}\left(x-1\right)^{s}\bigr).
\end{equation}

The simplest and probably the best known implicit method is the
implicit midpoint rule
\begin{equation}
  \label{eq:ButcherTabGauss2}
  \begin{array}{c|c}
    \frac{1}{2} & \frac{1}{2} \\[1ex]
    \hline
    \rule{0pt}{3,3ex} & 1
  \end{array}
\end{equation}
which is the lowest order $p=2$ Gauss method.  The higher order
implicit Gauss RK schemes are for orders $p=4$ (the
Hammer-Hollingsworth method)
\begin{equation}
  \label{eq:ButcherTabGauss4}
  \begin{array}{c|cc}
    \frac{1}{2} - \frac{\sqrt{3}}{6} & \frac{1}{4} & \frac{1}{4} - \frac{\sqrt{3}}{6} \\[1ex]
    \frac{1}{2} + \frac{\sqrt{3}}{6} & \frac{1}{4} + \frac{\sqrt{3}}{6} & \frac{1}{4} \\[1ex]
    \hline
    \rule{0pt}{3,3ex} & \frac{1}{2} & \frac{1}{2}
  \end{array}
\end{equation}
and $p=6$ (the Kuntzmann-Butcher method)
\begin{equation}
  \label{eq:ButcherTabGauss6}
  \begin{array}{c|ccc}
    \frac{1}{2} - \frac{\sqrt{15}}{10} & \frac{5}{36} & \frac{2}{9} - \frac{\sqrt{15}}{15} & \frac{5}{36} - \frac{\sqrt{15}}{30} \\[1ex]
    \frac{1}{2} & \frac{5}{36} + \frac{\sqrt{15}}{24} & \frac{2}{9} & \frac{5}{36} - \frac{\sqrt{15}}{24} \\[1ex]
    \frac{1}{2} + \frac{\sqrt{15}}{10} & \frac{5}{36} + \frac{\sqrt{15}}{30} & \frac{2}{9} + \frac{\sqrt{15}}{15} & \frac{5}{36} \\[1ex]
    \hline
    \rule{0pt}{3,3ex} & \frac{5}{18} & \frac{4}{9} & \frac{5}{18}
  \end{array}
\end{equation}
Schemes of order $p=2s$ can be constructed for any $s\geq 1$ (see
\cite{Butcher1964} where also methods of orders $p=8$ and $p=10$ are
explicitly given).  Their stability domains are precisely the left
half-plane (these methods are A-stable).  However, our main interest
of these methods is due to their properties when applied to the
Hamiltonian systems.

Hamiltonian systems is given by
\begin{equation}
  \label{eq:651}
  \dot{p}_{i}= -\frac{\partial H}{\partial q_{i}}, \quad
  \dot{q}_{i}= \frac{\partial H}{\partial p_{i}}, \quad i=1,\ldots,n,
\end{equation}
($\,\,\dot{}\equiv \diff/\diff t$), where the Hamiltonian function
$H(q,p)=H(q_{1},\ldots,q_{n},p_{1},\ldots,p_{n})$ is the first
integral.  Moreover, the flow corresponding to (\ref{eq:651}) is
symplectic, it preserves the differential 2-form
\begin{equation}
  \label{eq:652}
  \omega^{2} = \sum_{i=1}^{n}\diff p_{i}\wedge \diff q_{i}.
\end{equation}
Special properties of Hamiltonian systems motivated studies of
numerical methods suitable for the ODEs of this special form
(\ref{eq:651}).

\begin{mydef}
  A one-step method is called symplectic if for every smooth
  Hamiltonian $H$ and for every step size $h$ the mapping $\psi_{h}$
  (the transformation defined by the method)
  \begin{equation}
    \label{eq:653}
    \psi_{h}: \mathbb{R}^{2n}\ni (p_{0},q_{0})
    \mapsto (p_{1},q_{1})\in\mathbb{R}^{2n},
  \end{equation}
  is symplectic, i.e. preserves the differential 2-form
  (\ref{eq:652}).
\end{mydef}

An important property of the Gauss methods is stated in the following
theorem
\begin{mythe}[{\cite[p.~315, Theorem II.16.5]{hairer2008solving}}]
  The implicit $s$-stage Gauss methods of order $2s$ are symplectic
  for all $s$.
\end{mythe}
The characteristic feature of all of symplectic RK methods is
summarized in the following theorem
\begin{mythe}[{\cite[p.~316, Theorem II.16.6]{hairer2008solving}}]
  If the $s\times s$ matrix $M$ with elements
  $m_{ij}=b_{i}a_{ij}+b_{j}a_{ji}-b_{i}b_{j}$, ($i,j=1,\ldots,s$)
  satisfies $M=0$, then the Runge-Kutta (\ref{eq:643}) method is
  symplectic.
\end{mythe}
An important implication of this theorem is that the ERK methods are
never symplectic.  Moreover the most familiar classes of IRK (Radau IA
and IIA, or Lobatto IIIA, IIIB and IIIC methods) are also not
symplectic. Although, the Lobatto IIIA-IIIB pair, composed into
partitioned Runge-Kutta method (see below) play important role as it
generalizes the St\"ormer-Verlet scheme \cite{hairer2006geometric}.

The important property of symplectic methods is preservation of the
Hamiltonian and other first integrals.

\begin{mythe}[{\cite[p.~319, Theorem II.16.7]{hairer2008solving}}]
  Denote $y=(p,q)$ and let $G$ be a symmetric $2n\times 2n$ matrix.  A
  symplectic Runge-Kutta method leaves all quadratic first integrals
  \begin{equation}
    \label{eq:654}
    \left<y,y\right>_{G}:=y^{T}Gy,
  \end{equation}
  of the system (\ref{eq:651}) invariant.
\end{mythe}

\section{Partitioned methods}
\label{sec:AppPRK}

Let us consider differential equations in the partitioned form
($'\equiv \diff/\diff x$)
\begin{equation}
  \label{eq:655}
  y'=f(y,z), \quad z'=g(y,z),
\end{equation}
$f:\mathbb{R}^{n}\times\mathbb{R}^{m}\mapsto\mathbb{R}^{n}$,
$g:\mathbb{R}^{n}\times\mathbb{R}^{m}\mapsto\mathbb{R}^{m}$ (with
$n,m\in\mathbb{N}$, not necessarily equal) with special case including
(\ref{eq:651}).
\begin{mydef}[Partitioned Runge-Kutta method]
  Let $b_{i}$, $a_{ij}$ and $\hat{b}_{i}$, $\hat{a}_{ij}$ be the
  coefficient of two Runge-Kutta methods. A partitioned Runge-Kutta
  method (PRK) for the solution of (\ref{eq:655}) is given by
  \begin{equation}
    \label{eq:656}
    \begin{aligned}
      k_{i} &= f\left(y_{0} + h\sum_{j=1}^{s}a_{ij}k_{j}, z_{0} +
        h\sum_{j=1}^{s}\hat{a}_{ij}\ell_{j}\right),
      \\
      \ell_{i} &= g\left(y_{0} + h\sum_{j=1}^{s}a_{ij}k_{j}, z_{0} +
        \sum_{j=1}^{s}\hat{a}_{ij}\ell_{j} \right),
      \\
      y_{1} &= y_{0} + h\sum_{i=1}^{s}b_{i}k_{i},
      \\
      z_{1} &= z_{0} + h\sum_{i=1}^{s}\hat{b}_{i}\ell_{i}.
    \end{aligned}
  \end{equation}
\end{mydef}
The idea behind (\ref{eq:656}) is to take two RK methods, and to treat
the $y$-variables with the first method (with coefficients $a_{ij}$,
$b_{i}$), and the $z$-variables with the second method (coefficients
$\hat{a}_{ij}$, $\hat{b}_{i}$).  The following theorem gives the
condition for symplecticity of (\ref{eq:656})
\begin{mythe}[{\cite[p.~326, Theorem II.16.10]{hairer2008solving}}]
  \begin{enumerate}[(a)]
  \item If the coefficients of (\ref{eq:656}) satisfy
    \begin{align}
      \label{eq:657}
      b_{i} &= \hat{b}_{i},
      \\
      \label{eq:658}
      b_{i}\hat{a}_{ij} + \hat{b}_{j}a_{ji} - b_{i}\hat{b}_{j} &= 0,
    \end{align}
    (i,j=1,\ldots,s) then the method (\ref{eq:656}) is symplectic.
  \item If the Hamiltonian is separable then the condition
    (\ref{eq:658}) alone implies symplecticity of the method.
  \end{enumerate}
\end{mythe}
In particular, for separable Hamiltonian
systems
\begin{equation}
  \label{eq:659}
  H(p,q) = T(p) + V(q),
\end{equation}
it is possible to obtain explicit symplectic methods.  If the PRK
method (\ref{eq:656}) consist of diagonally implicit and explicit
methods
\begin{equation}
  \label{eq:660}
  \begin{aligned}
    a_{ij}&=0, \quad \text{for} \ i<j,
    \\
    \hat{a}_{ij}&=0, \quad \text{for} \ i\leq j,
  \end{aligned}
\end{equation}
respectively, then the resulting scheme is explicit.  Moreover,
assuming $b_{i}\neq 0$ and $\hat{b}_{i}\neq 0$ ($i=1,\ldots,s$)
without loss of generality, then the symplecticity condition
(\ref{eq:658}) becomes
\begin{equation}
  \label{eq:661}
  \begin{aligned}
    a_{ij} &= b_{j}, \quad \text{for}\ i\geq j,
    \\
    \hat{a}_{ij} &= \hat{b}_{j}, \quad \text{for}\ i\geq j,
  \end{aligned}
\end{equation}
so that the method (\ref{eq:656}) is characterized by
\begin{equation}
  \label{eq:662}
  \begin{array}{c|cccc}
    & b_1 & \cdots  & 0 & 0 \\
    & b_1 & b_2 & \cdots  & \vdots  \\
    & \vdots  & \vdots  & \ddots & \vdots  \\
    & b_1 & b_2 & \cdots  & b_s \\[1ex] \hline
    \rule{0pt}{2.7ex} & b_1 & b_2 & \cdots  & b_s \\
  \end{array} \qquad
  \begin{array}{c|cccc}
    & 0 & \cdots  & 0 & 0 \\
    & \hat{b}_1 & 0 & \cdots  & \vdots  \\
    & \vdots  & \ddots & \ddots & \vdots  \\
    & \hat{b}_1 & \cdots  & \hat{b}_{s-1} & 0 \\[1ex] \hline
    \rule{0pt}{2.7ex} & \hat{b}_1 & \cdots  & \hat{b}_{s-1} & \hat{b}_s \\
  \end{array}
\end{equation}

For example, the fourth order $s=4$ method given in
\cite{Yoshida1990262} has the following set of coefficients, of the
first Butcher tableau
\begin{equation}
  \label{eq:663}
  \begin{aligned}
    b_1&=\frac{1}{6} \left(2+\frac{1}{\sqrt[3]{2}}+\sqrt[3]{2}\right),
    \\
    b_2&=\frac{1}{12} \left(2-2 \sqrt[3]{2}-2^{2/3}\right),
    \\
    b_{3}&=b_{2},
    \\
    b_{4}&=b_{1},
  \end{aligned}
\end{equation}
and of the second one
\begin{equation}
  \label{eq:664}
  \begin{aligned}
    \hat{b}_{1}&=\frac{1}{3}
    \left(2+\frac{1}{\sqrt[3]{2}}+\sqrt[3]{2}\right),
    \\
    \hat{b}_2&=-\frac{1}{3} \left(1+\sqrt[3]{2}\right)^2,
    \\
    \hat{b}_{3}&=\hat{b}_{1},
    \\
    \hat{b}_{4}&=0.
  \end{aligned}
\end{equation}

\chapter{Interaction coefficients}
\label{cha:inter-coeff}

In this chapter we briefly review methods we use to calculate
integrals (interaction coefficients) appearing in perturbative
calculations.

\section{General formula}
\label{sec:general-formula}

In perturbative calculations we need to expand different quantities
into a linear combination of eigenfunctions themselves (the
interaction coefficients).  This requires computation of integrals of
products of eigenfunctions.  Let assume (for generality) that there is
a given set of orthogonal functions $e_{j}(x)$ on a interval
$[\alpha,\beta]\subset\mathbb{R}$ with a weight function $w(x)>0$
\begin{equation}
  \label{eq:665}
  \inner{e_{i}}{e_{j}} \equiv
  \int_{\alpha}^{\beta}e_{i}(x)e_{j}(x)w(x)\diff x=\delta_{ij},
  \quad i,j\in\mathbb{N}_{0},
\end{equation}
($\delta_{ij}$ is the Kronecker delta).  We want to calculate an
integral of the form
\begin{multline}
  \label{eq:666}
  \inner{e_{k}}{f(\,\cdot\,)\,e_{i_{1}}\,\cdots\,e_{i_{n}}
    e_{j_{1}}'\,\cdots\,e_{j_{m}}'} \equiv
  \\
  \int_{\alpha}^{\beta}e_{k}(x)f(x)e_{i_{1}}(x)\,\cdots\,e_{i_{n}}(x)
  e_{j_{1}}'(x)\,\cdots\,e_{j_{m}}'(x) \diff x,
\end{multline}
for $i_{1},\ldots, i_{n},j_{1},\ldots,j_{m}\in \mathbb{N_{0}}$, where
$f:[\alpha,\beta]\mapsto\mathbb{R}$ is some well behaved function.
Because the eigenbasis functions $e_{i}(x)$ which we encounter here
are polynomials in $\cos{x}$ we reduce the integrand (\ref{eq:666}) to
the sum of the cosine and sine products.  Then if this integral is
convergent (which we assume is true) we make use of the Euler beta
function $B(a,b)$ \cite{Fun:B}, i.e.
\begin{equation}
  \label{eq:667}
  \int_{0}^{\pi}\cos^{a}\left(\frac{x}{2}\right)
  \sin^{b}\left(\frac{x}{2}\right) \diff x =
  B\left(\frac{a+1}{2},\frac{b+1}{2}\right)=\frac{\Gamma
    \left(\frac{a+1}{2}\right) \Gamma
    \left(\frac{b+1}{2}\right)}{\Gamma \left(\frac{1}{2}
      (a+b+2)\right)},
\end{equation}
which is valid for integer powers $a,\,b>-1$.  Reducing of each the
products in (\ref{eq:666}) to the sum (allowing for use of
(\ref{eq:667})), and changing variables (by rescaling the interval
$[\alpha,\beta]$ to $[0,\pi]$) the integral (\ref{eq:666}) can be
written as
\begin{multline}
  \label{eq:668}
  \inner{e_{k}}
  {f(\,\cdot\,)\,e_{i_{1}}\,\cdots\,e_{i_{n}}e_{j_{1}}'\,\cdots\,e_{j_{m}}'}
  =
  \\
  \int_{\alpha}^{\beta}
  \sum_{\mathbf{J}}C_{ki_{1}\,\cdots\,i_{n}j_{1}\,\cdots\,j_{m}}\left(\mathbf{J}\right)
  \cos^{a(\mathbf{J})}\left(\frac{\pi}{2}\frac{x-\alpha}{\beta-\alpha}\right)
  \sin^{b(\mathbf{J})}\left(\frac{\pi}{2}\frac{x-\alpha}{\beta-\alpha}\right)
  \diff x =
  \\
  \frac{\beta-\alpha}{\pi}
  \sum_{\mathbf{J}}C_{ki_{1}\,\cdots\,i_{n}j_{1}\,\cdots\,j_{m}}\left(\mathbf{J}\right)
  B\left(\frac{a(\mathbf{J})+1}{2},\frac{b(\mathbf{J})+1}{2}\right),
\end{multline}
where we use the sum rule, assuming that this integral is convergent
whence both $a(\mathbf{J})$ and $b(\mathbf{J})>-1$, for any
$\mathbf{J}\in\mathbb{N_{0}}^{M}$.  The dimension $M$ of summation
multi-index $\mathbf{J}$ and real coefficients
$C_{ki_{1}\,\cdots\,i_{n}j_{1}\,\cdots\,j_{m}}\left(\mathbf{J}\right)$
depend on a specific form of the integral (\ref{eq:666}).  In exactly
the same way we can obtain the value of the integrals
$\inner{e_{k}'}{f(\,\cdot\,)\,
  e_{i_{1}}\,\cdots\,e_{i_{n}}e_{j_{1}}'\,\cdots\,e_{j_{m}}'}$.

\section{Example}
\label{sec:example}

Let us consider the YM model (see Section \ref{sec:YMModel}) to
illustrate the above.  For this model we have $w(x)=1$,
$[\alpha,\beta]=[0,\pi]$ and the eigenfunctions (\ref{eq:161}) are
\begin{equation}
  \label{eq:669}
  e_j(x) =
  \frac{(j+1)\sqrt{j(j+2)}\Gamma(j)}
  {2\sqrt{2}\Gamma\left(j+\frac{3}{2}\right)}
  \sin^{2}{x}\,P_{j-1}^{\left(3/2,3/2\right)}(\cos{x}),\quad j\in\mathbb{N},
\end{equation}
(we rewrite (\ref{eq:160}) for convenience).  In order to use the
formula (\ref{eq:668}) we express the eigenfunctions using the series
representation of the Jacobi functions (\ref{eq:591}) to get
\begin{multline}
  \label{eq:670}
  e_{j}(x) = \frac{(j+1)\sqrt{2j(j+2)}\Gamma(j)}
  {\Gamma\left(j+\frac{3}{2}\right)} \sum _{k=0}^{j-1} \Bigg[
  (-1)^{j-k+1} \binom{j+\frac{1}{2}}{j-k-1} \binom{j+\frac{1}{2}}{k}
  \\
  \times \cos^{2k+2}\left(\frac{x}{2}\right)
  \sin^{2(j-k)}\left(\frac{x}{2}\right) \Bigg].
\end{multline}
As an example let us calculate the projection
$\inner{e_{k}}{\csc^{2}{x}\,e_{i}\,e_{j}}$.  Denoting by
$\mathcal{N}_{j}$ the normalization constant in (\ref{eq:670})
\begin{equation}
  \label{eq:671}
  \mathcal{N}_{j}=\frac{(j+1)\sqrt{2j(j+2)}\Gamma(j)}
  {\Gamma\left(j+\frac{3}{2}\right)},
\end{equation}
we have
\begin{multline}
  \label{eq:672}
  \inner{e_{k}}{\csc^{2}{x}\,e_{i}\,e_{j}} =
  \int_{0}^{\pi}e_{k}(x)\frac{e_{i}(x)e_{j}(x)}
  {4\sin^{2}\left(\frac{x}{2}\right)\cos^{2}\left(\frac{x}{2}\right)}
  \diff x =
  \\
  \frac{1}{4}\mathcal{N}_{i}\,\mathcal{N}_{j}\mathcal{N}_{k}
  \int_{0}^{\pi}\Biggl\{
  \sin^{-2}\left(\frac{x}{2}\right)\cos^{-2}\left(\frac{x}{2}\right)
  \\
  \times \sum_{s=0}^{i-1} \Bigg[ (-1)^{i-s+1}
  \binom{i+\frac{1}{2}}{i-s-1} \binom{i+\frac{1}{2}}{s}
  \cos^{2s+2}\left(\frac{x}{2}\right)
  \sin^{2(i-s)}\left(\frac{x}{2}\right) \Bigg]
  \\
  \times \sum_{r=0}^{j-1} \Bigg[ (-1)^{j-r+1}
  \binom{j+\frac{1}{2}}{j-r-1} \binom{j+\frac{1}{2}}{r}
  \cos^{2r+2}\left(\frac{x}{2}\right)
  \sin^{2(j-r)}\left(\frac{x}{2}\right) \Bigg]
  \\
  \times \sum_{q=0}^{k-1} \Bigg[ (-1)^{k-q+1}
  \binom{k+\frac{1}{2}}{k-q-1} \binom{k+\frac{1}{2}}{q}
  \cos^{2q+2}\left(\frac{x}{2}\right)
  \sin^{2(k-q)}\left(\frac{x}{2}\right) \Bigg] \Biggr\}\diff x =
  \\
  \frac{1}{4}\mathcal{N}_{i}\,\mathcal{N}_{j}\mathcal{N}_{k}
  \int_{0}^{\pi}\Biggl\{
  \sum_{s=0}^{i-1} \sum_{r=0}^{j-1} \sum_{q=0}^{k-1} \Biggl[
  (-1)^{i+j+k-s-r-q+1}
  \\
  \times \binom{i+\frac{1}{2}}{i-s-1} \binom{i+\frac{1}{2}}{s}
  \binom{j+\frac{1}{2}}{j-r-1} \binom{j+\frac{1}{2}}{r}
  \binom{k+\frac{1}{2}}{k-q-1} \binom{k+\frac{1}{2}}{q}
  \\
  \times \cos^{2(s+r+q+2)}{\left(\frac{x}{2}\right)}
  \sin^{2(i+j+k-s-r-q-1)}{\left(\frac{x}{2}\right)} \Biggr] \Bigg\} \diff x
  =
  \\
  \frac{1}{4}\mathcal{N}_{i}\,\mathcal{N}_{j}\mathcal{N}_{k}
  \sum_{s=0}^{i-1} \sum_{r=0}^{j-1} \sum_{q=0}^{k-1} \Biggl[
  (-1)^{i+j+k-s-r-q+1}
  \\
  \times \binom{i+\frac{1}{2}}{i-s-1} \binom{i+\frac{1}{2}}{s}
  \binom{j+\frac{1}{2}}{j-r-1} \binom{j+\frac{1}{2}}{r}
  \binom{k+\frac{1}{2}}{k-q-1} \binom{k+\frac{1}{2}}{q}
  \\
  \times \frac{\Gamma\left(s+r+q+\frac{5}{2}\right)
    \Gamma\left(i+j+k-s-r-q-\frac{1}{2}\right)}{\Gamma\left(i+j+k+2\right)}
  \Biggr],
\end{multline}
since for any $i,j,k\geq 1$ the powers of $\cos(x/2)$ and $\sin(x/2)$
in this integral are always greater than $-1$ as is required for
convergence.  This result is a special case of (\ref{eq:668}) with
$f(x)=\csc^{2}{x}$, $n=2$, $m=0$ and $[\alpha,\beta]=[0,\pi]$.  Then,
calculating (\ref{eq:672}) for specific values of indices we have,
e.g.
\begin{align}
  \label{eq:673}
  \csc^{2}{x}\,e_{1}(x)^{2} &= 2 \sqrt{\frac{2}{3 \pi }} e_1(x),
  \\
  \csc^{2}{x}\,e_{1}(x)e_{2}(x) &= 2 \sqrt{\frac{2}{3 \pi }} e_2(x),
  \\
  \csc^{2}{x}\,e_{2}(x)^{2} &= 2 \sqrt{\frac{2}{3 \pi }}
  e_1(x)+\sqrt{\frac{10}{3 \pi }} e_3(x),
  \\
  \csc^{2}{x}\,e_3(x)e_4(x) &= 3 \sqrt{\frac{2}{5 \pi }} e_2(x)+7
  \sqrt{\frac{2}{15 \pi }} e_4(x)+\frac{8}{\sqrt{15 \pi }} e_6(x),
  \\
  \csc^{2}{x}\,e_3(x)e_{60}(x) &= \frac{1}{5} \sqrt{\frac{1798}{15 \pi
    }} e_{58}(x)+\frac{413}{155} \sqrt{\frac{6}{5 \pi }}
  e_{60}(x)+\frac{48}{31} \sqrt{\frac{2}{\pi }} e_{62}(x).
\end{align}
Therefore in general we have
\begin{equation}
  \label{eq:674}
  \csc^{2}{x}\,e_{i}(x)e_{j}(x) =
  \sum_{k=|i-j|+1}^{i+j-1} \inner{e_{k}}{\csc^{2}{x}\,e_{i}\,e_{j}}e_{k}(x).
\end{equation}


\backmatter
\begingroup
\sloppy
\cleardoublepage
\phantomsection
\addcontentsline{toc}{chapter}{Bibliography}
\printbibliography
\endgroup

\newpage
\mbox{}
\newpage

\pagestyle{plain}

\newpage
\mbox{}
\newpage

\newpage
\mbox{}
\newpage

\newpage
\mbox{}
\newpage

\pagestyle{plain}
\topskip0pt \vspace*{\fill}
\begin{center}
  \begin{minipage}{.667\textwidth}
    \begin{flushleft}
      ``Quite simple, my dear Watson.''
    \end{flushleft}
    \vspace{-2ex}
    \begin{flushright}
      ---Sir Arthur Conan Doyle, \textit{The adventure of the Retired
        Colourman}
    \end{flushright}
  \end{minipage}
\end{center}
\vspace*{\fill}

\end{document}